%% file: hdr.tex
\pdfoutput=1
\documentclass[a4paper,11pt,twoside,fancyhdr]{report}
\usepackage{epsfig,amsmath,amssymb,graphicx,fancyheadings,indentfirst,%
  undertilde,color,multirow,slashed}
\usepackage[english]{babel}
\usepackage[colorlinks=true,urlcolor=black]{hyperref}

\usepackage{feynmf} %for drawing Feynman diagrams
\usepackage{empheq}

\hypersetup{backref = true,pagebackref = true,hyperindex = true, colorlinks = true,
  breaklinks = true, urlcolor = blue, linkcolor = blue, bookmarks = true,
  bookmarksopen = true, citecolor=red}

\graphicspath{{figures/}}
\renewcommand{\chaptermark}[1]{\markboth{Chapter \thechapter\ - #1}{}}

\input{commands}

\allowdisplaybreaks

\begin{document}
\pagestyle{empty} 
\input{Title/title2}

\cleardoublepage 

\pagestyle{fancy}\pagenumbering{roman}

\chapter*{Acknowledgements} \input{Acknowledgements/acknowledgements}

\cleardoublepage

\chapter*{Abstract}\input{Abstract/abstract}

\tableofcontents \cleardoublepage

\chapter{Introduction} \pagenumbering{arabic} \input{Chapter1/intro}

\chapter{Multi-loop calculations} \input{Chapter2/methods}

\chapter{Reduced QED} \input{Chapter3/rqed}

\chapter{Critical properties of QED$_3$ and QED$_{4,3}$} \input{Chapter4/dchisb}

\chapter{Optical conductivity of graphene} \input{Chapter5/graphene}

\chapter{Outlook}\input{Conclusions/conclusions}

\appendix
\renewcommand{\chaptermark}[1]{\markboth{Appendix \thechapter\ - #1}{}}

\chapter{Notations and conventions} \input{Appendix/conventions}

\chapter{More on Gegenbauer polynomials} \input{Appendix/gegenbauer}

\cleardoublepage
\phantomsection
\addcontentsline{toc}{chapter}{Bibliography}
\bibliographystyle{utphys}
\footnotesize{\bibliography{biblio}}

\end{document}

%% file: commands.tex
\definecolor{lightgrey}{gray}{0.9}

\newcommand{\N}{\mathbb{N}}

\newcommand{\R}{\mathbb{R}}

\def\B{\langle }
\def\K{\rangle }

\def\Tr{\mbox{Tr}}

\def\al{\alpha}
\def\eps{\epsilon}

\def\om{\omega}
\def\veps{\varepsilon}
\def\ep{\varepsilon}

\def\be{\begin{equation}}
\def\ee{\end{equation}}
\def\bea{\begin{eqnarray}}
\def\eea{\end{eqnarray}}
\def\bse{\begin{subequations}}
\def\ese{\end{subequations}}
\def\bc{\begin{center}}
\def\ec{\end{center}}

\def\ms{\medskip}
\def\bs{\bigskip}

\def\ni{\noindent}
\def\ra{\rightarrow}
\def\Ra{\Rightarrow}
\def\lra{\leftrightarrow}

\def\nonum{\nonumber}

\def\E{{\rm e}}
\def\I{{\rm i}}

\def\D{{\rm d}}
\def\Ord{{\rm O}}

\newcommand{\comment}[1]{}

\newcommand{\ie}{{\it i.e.}}
\newcommand{\eg}{{\it e.g.}}

\newcommand{\psibar}{{\bar\psi}}

\def\Sp{{\slashed p}}
\def\Sq{{\slashed q}}
\def\Sk{{\slashed k}}

\newtheorem{theorem}{Theorem}

\newcommand*\pFqskip{8mu}
\catcode`,\active
\newcommand*\pFq{\begingroup
        \catcode`\,\active
        \def ,{\mskip\pFqskip\relax}%
        \dopFq
}
\catcode`\,12
\def\dopFq#1#2#3#4#5{%
        {}_{#1}F_{#2}\biggl(\genfrac..{0pt}{}{#3}{#4} \bigg | #5\biggr)%
        \endgroup
}

%% file: Title/title2.tex
\thispagestyle{empty}
%%%%%%%%%%%%%%%%%%%%%%%%%%%%%%%%%%%%%%%%%%%%%%%%%%%%%

%\begin{minipage}{\linewidth}
%\begin{minipage}{.4\linewidth}
%\begin{flushleft}
%%\ifpdf
%  \includegraphics[width=40mm]{Figs/logop6Bord}
%%\else
%%  \includegraphics[width=40mm]{PolyFigs/logop6Bord}
%%\fi
%\end{flushleft}
%\end{minipage}
%\hfill
%\begin{minipage}{.6\linewidth}
%\begin{flushright}
%\bfseries
%\sffamily
%\large
%Universit\'e Pierre et Marie Curie\\
%Paris~VI
%\end{flushright}
%\end{minipage}
%\end{minipage}

%%%%%%%%%%%%%%%%%%%%%%%%%%%%%%%%%%%%%%%%%%%%%%%%%%%%%
\vspace{2cm}
%%%%%%%%%%%%%%%%%%%%%%%%%%%%%%%%%%%%%%%%%%%%%%%%%%%%%
\begin{center}
{\Huge

Field theoretic study of electron-electron interaction effects in Dirac liquids

}

\vspace{3cm}

{%
%\sffamily
\Large

{\bf Sofian Teber}

\vspace{1cm}

{\large \sffamily
Sorbonne Universit\'e, CNRS, Laboratoire de Physique Th\'eorique et Hautes Energies, LPTHE, F-75005 Paris, France

%\vspace{0.1cm}

%CNRS, UMR 7589, LPTHE, F-75005, Paris, France.
	}
}

\vspace{7cm}

{%
\sffamily
\Large

{\bfseries Habilitation \`a Diriger des Recherches}

\bigskip

{Defended on 12/12/2017}

}

\vspace{2cm}

\begin{tabular}{lcr}
\begin{minipage}[l]{0.4\textwidth}
{%\vspace{-0.6in}
%\hspace*{-1.1in}
\includegraphics[width=4.5cm,keepaspectratio]{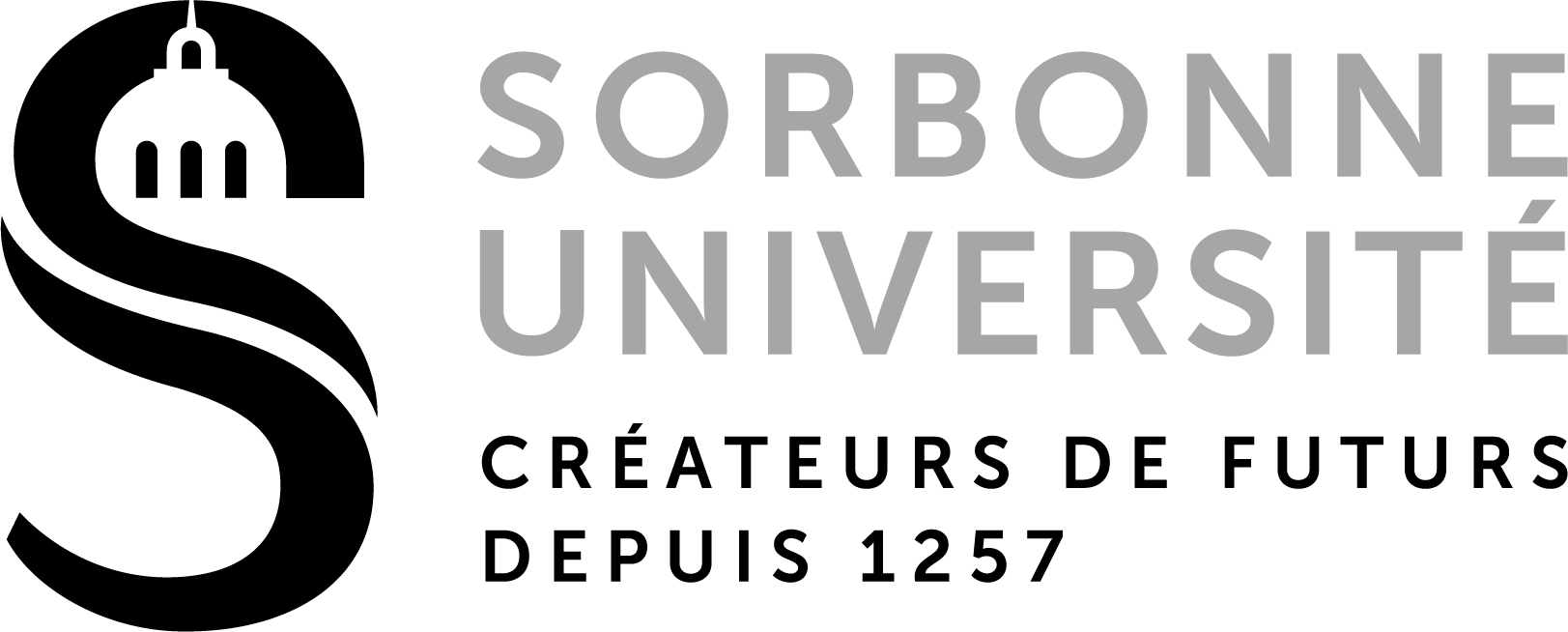}
}
\end{minipage}
&
%%\vspace{-0.4in}
\begin{minipage}[c]{0.2\textwidth}
{$\qquad$%\vspace{-0.6in}
%\includegraphics[width=4cm,keepaspectratio]{upmc_logo}
%}
%\vskip 0.2cm

%\vspace{-0.3cm}
%    \baselineskip 11pt\largeheadfont
%\centerline{
%Universit\'e Pierre et Marie Curie - Paris 6 % Laboratoire de Physique Th\'eorique et Hautes \'Energies
}
\end{minipage}
&
\begin{minipage}[r]{0.4\textwidth}
{%\vspace{-0.6in}
%\hspace*{0.7in}
\includegraphics[width=4.5cm,keepaspectratio]{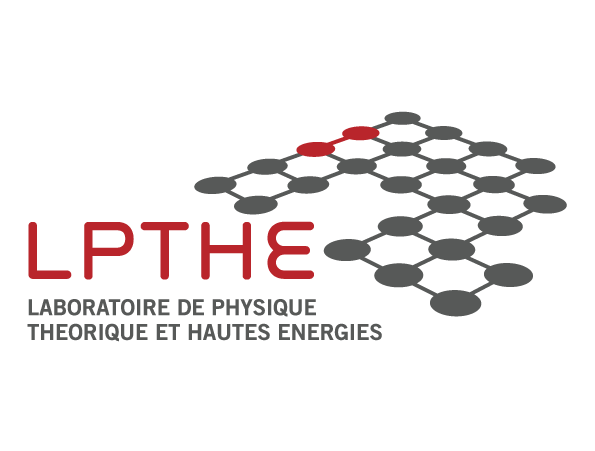}
}
\end{minipage}
\end{tabular}

\end{center}

%% file: Acknowledgements/acknowledgements.tex
I am deeply grateful to Anatoly Kotikov, my collaborator, master and friend, for the beautiful moments we had over the last years working on the problems reviewed in this manuscript;
his dedication to science, extraordinary computational skills as well as true kindness were crucial in helping us go through our arduous investigations.

I warmly thank all my colleagues at Laboratoire de Physique Th\'eorique et Hautes Energies (LPTHE) for exchanges and enjoyable moments over the years. In particular, I thank Olivier Babelon and Beno\^it
Dou\c cot, respective directors of LPTHE since I joined the lab in 2007, for giving to me complete freedom in choosing my research topics as well as for their support. 
 Special thanks also go to Fran\c coise Got and Isabelle Nicolai, secretaries of LPTHE, 
for their constant kind help and support. Teaching and administrative duties at Sorbonne Universit\'e (former Universit\'e Pierre et Marie Curie - UPMC) 
have given to me opportunities to meet and interact with many colleagues whom I warmly thank.
In particular, I would like to express my deep gratitude to Claude Aslangul from Laboratoire de Physique Th\'eorique de la Mati\`ere Condens\'ee (LPTMC) 
and Jacques Lefr\`ere from Laboratoire Atmosph\`eres, Milieux, Observations Spatiales (LATMOS) for wonderful years spent teaching together. 
I consider Claude Aslangul as my Godfather at UPMC; his dedication to both teaching and research are exemplary to me. 
I am also very grateful to Nicolas Dupuis (LPTMC) for so kindly accepting to relief me from some of my teaching duties over the last two years through an 
ABER (``Accord de Bin\^ome d'Enseignement et de Recherche'') program. 
I am also grateful to Benjamin Basso from Laboratoire de Physique Th\'eorique at Ecole Normale Sup\'erieure (LPTENS) for all the generous help and high scientific expertise he 
provided in the organization of the Multi-Loop conference held at UPMC in June 2017. And to Johannes Braathen and Hugo Ricato, two brilliant PhD students at LPTHE, for technical assistance during the workshop.
It is my pleasure to thank Mikhail Kompaniets for sharing his expertise on multi-loop techniques through many discussions during the last years, for accurate reading and important comments on various 
Chapters of this manuscript and for very kind hospitality during each of my visits to St Petersbourg. 
I would like to express my deep gratitude to the referees (John Gracey from the University of Liverpool, Valery Gusynin from BITP Kiev and Dmitry Kazakov from JINR Dubna) 
for accepting to review the manuscript as well as for their encouragements  and stimulating comments (illuminating discussions with Valery Gusynin were highly appreciated). I am very grateful to the other 
members of the jury (Jean-No\"el Fuchs from LPTMC, Thierry Jolicoeur from the University of Orsay and Dominique Mouhanna from LPTMC) for kindly accepting to examine my work. 

Last but not least, I thank with all my heart my family for true love and continuous encouragements, help and support. 
I thereby dedicate this manuscript to my wife Ratka, to my two sons Aleksandar and Marko and to our Baba.

%% file: Abstract/abstract.tex
L'objectif de cette th\`ese d'habilitation %\`a diriger des recherches 
 est de
pr\'esenter des r\'esultats r\'ecents, obtenus dans la p\'eriode 2012-2017, concernant
l'effet des interactions dans les liquides planaires de type Dirac, \eg, le graph\`ene, les \'etats de surface de certains isolants topologiques 
voire m\^eme les syst\`emes \`a effet Hall quantique fractionnaire \`a demi-remplissage (pour les fermions composite de Dirac). 
Ces liquides sont caracteris\'es par des bandes sans gap, de fortes interactions \'electron-\'electron et une invariance de
Lorentz \'emergeant dans l'infra-rouge profond. Les probl\`emes abord\'es
prennent leur source dans des exp\'eriences portant sur ces syst\`emes et
pr\'esentent un large spectre d'int\'er\^et aussi bien en physique de basse
(physique de la mati\`ere condens\'ee) que de haute (physique des particules) \'energie. %et se situent \`a la fronti\`ere entre physique de la mati\`ere condens\'ee et physique des particules. 
On traitera en particulier de l'influence subtile des interactions sur les propriet\'es de transport
ainsi que de leur r\^ole d\'eterminant dans la potentielle g\'en\'eration dynamique d'une masse.
La r\'esolution de ces probl\`emes %pr\'esente un large spectre d'int\'er\^et en physique aussi bien de basse
%que de haute \'energie. Elle 
nous guidera de l'examen approfondi de la structure perturbative de
th\'eories de champs de jauge au d\'eveloppement et l'application de m\'ethodes non-perturbatives issues de l'\'electro/chromo-dynamique quantique pour aborder le r\'egime de couplage fort.

\vspace{1cm}

The aim of this habilitation thesis is to present recent results, obtained during the period 2012-2017, related to interaction effects in 
condensed matter physics systems such as planar Dirac liquids, \eg, graphene and graphene-like systems, the surface states of some topological insulators 
and possibly half-filled fractional quantum Hall systems (for their Dirac composite fermions). These liquids are characterized by gapless bands, 
strong electron-electron interactions and emergent Lorentz invariance deep in the infra-red.
We address a number of important issues raised by experiments on these systems covering subjects of wide current interest in low-energy (condensed matter) 
as well as high-energy (particle) physics.
%at the boundary between condensed matter and particle physics raised by experiments on graphene.
We shall consider in particular the subtle influence of interactions on transport properties and their supposedly crucial influence on a potential
dynamical mass generation. The resolution of these problems will guide us from the thorough examination of the perturbative structure of 
gauge field theories to the development and application of non-perturbative approaches known from quantum electro/chromo-dynamics to address
strong coupling issues.
%The project is of inter-disciplinary nature and covers subjects of wide current interest in low-energy as well as high-energy physics.

%% file: Chapter1/intro.tex
\label{chap1}
 
The study of quantum systems composed of a large number of interacting particles represents one of the most fascinating and challenging
aspects of modern physics. Its study dates back to the birth of the theory of quantum fields. Throughout the decades, progress in understanding such systems
resulted from deep exchanges, both at the technical and conceptual levels, between various fields of physics, see the monograph \cite{Anderson1984basic} for a beautiful non-technical illustration.
In this manuscript, we shall summarize a series of works applying, and to some extent adapting, powerful analytic techniques, originally developed
in the framework of (four-dimensional) relativistic gauge field theories, to study the effects of electron-electron interactions in condensed matter physics systems
such as planar Dirac liquids, \eg, graphene and graphene-like systems, the surface states of some topological insulators and possibly half-filled fractional quantum Hall systems.
The use of such techniques should be viewed as an attempt to reach high precision quantitative understanding of some of the most intriguing properties, \eg, 
related to transport and dynamical gap generation, of these peculiar (low-dimensional) electronic systems.
This chapter will briefly review some of the historical background and modern context surrounding these developments with an emphasize on field theory as
a thread between low and high-energy physics. 

\section{Historical background}

\subsection{High-energy physics}

Since the early works of James Clerk Maxwell, % to the discovery of asymptotic freedom, 
the study of gauge theories has had a profound impact on physics in general. Contemporary elementary particle physics is dominated by
gauge theories \cite{tHooft:1995wad} the prototype of which is quantum electrodynamics (QED).
This theory is an abelian gauge theory with a U$(1)$ gauge group.
The gauge field mediating the interaction between two spin-$1/2$ fields is the electromagnetic field.
Physically, particles interact by exchanging photons.
At textbook level, \cite{Berestetskii82,itzykson2012quantum,peskin1995introduction,Weinberg:1995mt}, relativistic quantum field theories such as QED are born from the fact that they are the only way to reconcile
quantum mechanics to special relativity. Their study has led to great achievements.
In the 40s, the first perturbative techniques using covariance and gauge invariance were independently developed by Tomonaga \cite{Tomonaga:1946zz}, Schwinger \cite{Schwinger:1948yk+Schwinger:1948yj+Schwinger:1949ra} 
and Feynman \cite{Feynman:1948fi+Feynman:1949zx+Feynman:1950ir} and unified by Dyson \cite{PhysRev.75.486}.
This led to the discovery of the concept of renormalization as an attempt to give a meaning to divergent integrals appearing in the perturbative series. 
Specific ideas of the field theoretic renormalization group (RG) were then developed in the 50s 
\cite{Stueckelberg:1953dz,GellMann:1954fq,Bogolyubov:1956gh,Bogoliubov:1957gp,Hepp:1966eg,Zimmermann1969,Bogolyubov:1980nc}.
 %Soon, the renormalizability of a quantum field theory appeared as a fundamental concept and a general criterion for the acceptability of this theory.
The success of QED came from the unprecedented agreement between high precision experiments and high precision theoretical computations of measurable quantities (anomalous magnetic moment, Lamb's shift, etc...).
In the 50s, non-abelian gauge theories were discovered %by Yang and Mills 
\cite{Yang:1954ek} and, in the 60s, the weak interaction was unified with electromagnetism (electro-weak interaction) \cite{Glashow61+Salam:1964ry+Weinberg:1967tq}.
%Glashow, Salam and Weinberg \cite{Glashow61Salam64Weinberg67} unified the weak interaction with electromagnetism (electro-weak interaction).
  %by Sheldon Glashow, Abdus Salam and Steven Weinberg.
%Despite this success, there where important conceptual problems that made these theories look quite suspicious even to its own creators.
%For some, the existence of the so-called ``Landau poles'' (divergence of the coupling constant at high energies or small distances) totally discredited such theories.
%For others, the renormalization procedure was itself suspicious as it gave the feeling that difficulties, the divergences, were ``swept under the rug''.
%During the 50s and 60s, it was generally believed that the real world could actually not be described by a quantum field theory and perturbative gauge theories
%were then even banned from some institutions to the benefit of other approaches based on symmetries ($S$ matrix, current algebras, dualities).
%Nevertheless, fundamental works appeared during this period. In particular, the discovery in 1954 of non-abelian gauge theories by Chen Ning Yang and Robert Mills
%and, in the 60s, the unification of the weak interaction with electromagnetism (electro-weak interaction)  by Sheldon Glashow, Abdus Salam and Steven Weinberg.
In the 70s, 't Hooft and Veltman \cite{'tHooft:1971rn,'tHooft:1972fi} proved the renormalizability of non-abelian gauge field theories and, to this purpose, invented 
the dimensional regularization (DR) technique which was also independently discovered in \cite{Bollini:1972ui,Cicuta:1972jf,Ashmore:1972uj}.
Combined with the minimal subtraction (MS) scheme \cite{'tHooft:1973mm} within an RG framework, this regularization technique is particularly well suited to 
compute radiative corrections and we shall use it extensively later on.
In the wake of these far-reaching developments, asymptotic freedom has then been discovered \cite{Politzer:1973fx,Gross:1973id}.
It eliminates the problem of the Landau poles and allows for the existence of theories which are well defined at arbitrarily low energies.
The first example of such a theory is quantum chromodynamics (QCD) which describes quarks as well as their interactions (strong interaction).
Contrarily to QED, a perturbative approach to QCD is possible only at high energy.
At low energies, a strong coupling regime takes place corresponding to quark confinement. All these developments led to the Standard Model (SM) of particle
physics which provides a unified field theoretic description of three elementary forces (electromagnetism, weak and strong). The latter allows 
a classification of all elementary particles presently known in a remarkable agreement with experiments. Still, some phenomena remain unexplained such as the origin of the mass, 
of dark matter and energy or gravity. Such questions are commonly addressed in what is now known as BSM (beyond the SM) physics. 
The latter includes, among other things, supersymmetric extensions of the SM \cite{Dimopoulos:1981zb}. A radically different approach (though also in need
of supersymmetry to be consistent) includes string theory which was developed to explain the strong interaction in the pre-QCD era but later evolved into a potential theory of quantum gravity.  
At this point, let us recall that, back in the 70s, 't Hooft discovered a simplification of U$(N)$ gauge theories in the limit of large $N$ \cite{'tHooft:1973jz}.
This work initiated the large-$N$ study of gauge theories, see the monograph \cite{Brezin:1994eb} for a review.
Recently, it played an important role in examining the conjecture relating string theories in anti-de Sitter spaces to (the 't Hooft limit of) superconformal field theories
in one less space-time dimension (AdS/CFT correspondence) \cite{Maldacena:1997re+Gubser:1998bc+Witten:1998qj}. The most prominent example of such a field theory is the 
$3+1$-dimensional $\mathcal{N}=4$ super-Yang-Mills theory at the conformal point which now serves as a kind of theoretical laboratory to gain insight on the beautiful and complex 
structure of quantum field theories (see below for a little more). Even though this theory is not QCD, the correspondence suggests that string theory might be, after all, not too far from a theory of the strong 
interaction, see \cite{Nicolai:2007zza} for a very nice short review.

The fundamental developments which led to the elaboration of such theories in particle/high-energy physics have had profound consequences
in other fields such as statistical mechanics. Formally, a $D$-dimensional quantum system is equivalent to a $D+1$-dimensional (in Euclidean space) 
statistical mechanics one.
Physically, complex systems composed of a large number of interacting particles are subject to emergent phenomena such as 
 %the most striking and common example being that of a 
phase transitions. It is Wilson that first realized that the vicinity of a second order phase transition may be described by a
continuous QFT  and formulated the so-called momentum-shell RG \cite{Wilson:1971bg,Wilson:1971dh}.
This led to the development of the small-$\veps$ expansion technique to compute critical exponents by Wilson and Fisher \cite{Wilson:1971dc} 
and brought up the important notion of (non-trivial) infra-red (IR) fixed points.\footnote{According to \cite{Brezin:1994eb}, the large-$N$ expansion also first appeared
in the context of statistical mechanics through the work of Stanley on spin systems \cite{Stanley:1968gx}.} The works of the Saclay group, see, \eg, \cite{Brezin:1974eb}, led to the development of the field-theoretic Wilsonian RG
culminating in the book \cite{zinn2002quantum}.
Within particle physics and statistical mechanics, the ability to access high order corrections, and therefore achieve high precision calculations, arose in the 80s with the developments
of powerful methods to compute renormalization group functions, {\it i.e.}, $\beta$-functions and anomalous dimensions of fields.
These methods include, \eg, the method of infra-red rea-rangement (IRR) \cite{Vladimirov:1979zm} and the
$R^*$-operation \cite{Chetyrkin:1982nn,Chetyrkin:1983wh,Chetyrkin:1984xa,Smirnov:1986me,Chetyrkin:2017ppe}, the method of uniqueness \cite{D'Eramo:1971zz,Vasiliev:1981dg,Usyukina:1983gj,Kazakov:1983ns,Kazakov:1986mu} 
and (Vasil'ev's) conformal bootstrap technique \cite{Vasiliev:1982dc}, 
integration by parts (IBP) \cite{Vasiliev:1981dg,Tkachov:1981wb,Chetyrkin:1981qh}, the Gegenbauer polynomial technique \cite{Chetyrkin:1980pr}
and the combination of these methods with symmetry arguments \cite{Broadhurst:1986bx,Barfoot:1987kg}. These techniques were applied to four-dimensional relativistic theories (and their classical analogues)
using RG in DR within the MS scheme which is incomparably more efficient computationally  than the more physically
appealing momentum-shell RG, see the classic monographs: \cite{Collins:1984xc,kleinert2001critical} and especially \cite{Vasil'evbook} for a beautiful historical introduction; see also the 
more recent books \cite{grozin2007lectures,smirnov2013analytic} devoted to Feynman diagram calculations. 
Among their greatest early achievements, let us mention, \eg, the computation of the $3$-loop $\beta$-function of QCD \cite{Tarasov:1980au}
and the computation of the $5$-loop $\beta$-function of $\Phi^4$-theory \cite{Kazakov:1984km,Gorishnii:1983gp,Kazakov:1983pk}. The following decades witnessed further ground-breaking developments, \eg,
in the exact evaluation of massless Feynman diagram, \eg, \cite{Gracey:1992ew,Kotikov:1995cw,Broadhurst:1996ur,Panzer:2013cha}, in developing and applying new techniques to deal 
with massive ones, \eg, \cite{Kotikov:1990kg,Boos:1990rg,Kotikov:1991hm,Kotikov:1991pm},  in discovering dimensional recurrence relations for Feynman diagrams \cite{Tarasov:1996br,Lee:2009dh},
in performing high order $\veps$-expansions together with understanding some intriguing relations with number theoretical issues, 
\eg, \cite{Broadhurst:1996ur,Broadhurst:1996yc,Broadhurst:2002gb,Bierenbaum:2003ud,Brown:2008um,Brown:2009ta},
in applying such techniques to numerous models, \eg, \cite{Kleinert:1991rg,Gracey:1993sn,Vasiliev:1992wr,Gracey:1993kx,Kivel:1993wq,Gracey:1993iu}, 
in the Hopf-algebraic interpretation of the renormalization group \cite{Kreimer:1997dp,Connes:1998qv}, in the notion of a cosmic Galois group \cite{Cartier:2001,Kontsevich:1999,Connes:2004zi}... 
(the list is not exhaustive).

%see, \eg, \cite{Panzer:2014kia} for a recent review (the list is not exhaustive). 

Throughout the years, there has been a dramatic increase in the complexity of the calculations.
Modern challenges often require the manipulation of thousands of diagrams each one of them eventually breaking in hundreds of integrals. 
While some of the most complicated integrals, the so called ``master integrals'' \cite{Broadhurst:1987ei}, require human assistance for their analytic evaluation, other tasks are rather systematic
in nature as well as highly symbolic: from generating the diagrams, to performing eventual gamma matrix algebra, to reducing large numbers of
diagrams to a few master integrals... This led to the automation of such tasks via the developments of powerful computer algebra systems, 
\eg, from SCHOONSHIP \cite{Veltman}, to REDUCE \cite{Hearn}, FORM \cite{Vermaseren:2000nd}, GiNac \cite{Bauer:2000cp} and the commercial MATHEMATICA \cite{wolfram1991mathematica}, see \cite{Weinzierl:2002cg} for a short review.
Specific algorithms and tools were developed to generate Feynman diagrams, \eg, QGRAF \cite{Nogueira:1991ex}, GRAPHSTATE/GRAPHINE \cite{Batkovich:2014bla} and EXP \cite{Seidensticker:1999bb}. 
Others to deal with the reduction problem such as Laporta's algorithm \cite{Laporta:2001dd}, Baikov's method \cite{Baikov:1996rk} 
as well as computer codes combining several algorithms to achieve this task such as REDUZE \cite{Studerus:2009ye,vonManteuffel:2012np}, 
FIRE \cite{Smirnov:2008iw}, KIRA \cite{Maierhoefer:2017hyi} and LiteRed \cite{Lee:2013mka}. %and CRUSCHER (Marquard). 
Some algorithms are devoted to the subtle problems of dealing with subdivergences and generating the Laurent expansion 
such as, \eg, the sector decomposition technique \cite{Binoth:2000ps,Bogner:2007cr}, see also the dissertation of Bogner \cite{Bogner2009}, 
parametric integration using hyperlogarithms \cite{Panzer:2014caa}, see also the dissertation of Panzer \cite{Panzer:2015ida}, the recently discovered 
method of graphical functions \cite{Schnetz:2013hqa} and its combination with parametric 
integration \cite{Golz:2015rea} together with the automation of the $R$ and $R^*$ operators by Batkovich and Kompaniets \cite{Batkovich:2014rka}, see also the very recent \cite{Herzog:2017bjx}.
The computation of massless propagator-type Feynman integrals has been fully automated by the MINCER program at three loops \cite{Gorishnii:1989gt} and, quite recently, 
by the FORCER program at five loops \cite{Ueda:2016yjm}, see also \cite{Ruijl:2017cxj} for more details. Though computer assisted, all these remarkable developments 
may be considered as analytical as opposed to the numerical methods which may eventually be used at the final stage of the procedure in order to extract a numerical value for the coefficients 
of the Laurent series associated with a given diagram.
Often, they involve advanced mathematical concepts from, \eg, graph theory, algebraic geometry and number theory.
Nowadays, computer algebra systems combined with appropriate algorithms are an integral part of the field of multi-loop calculations.
For the years 2016/2017 alone, they allowed breakthrough achievements such as, {\it e.g.}, the 4-loop $\beta$-function calculation for the Gross-Neveu \cite{Gracey:2016mio}~\footnote{Notice that the lower number of loops  
presently achieved for the Gross-Neveu model with respect to other models is related to the loss of multiplicative renormalizability of 4-fermion operators in dimensional regularization and the generation of 
evanescent operators; so calculations for this model are less straightforward than in other models.}, see also \cite{Gracey:2018qba,Gracey:2018fwq} for very recent large-$N$ calculations, 
 and Gross-Neveu-Yukawa models \cite{Mihaila:2017ble,Zerf:2017zqi},
the 5-loop $\beta$-function calculation for QCD \cite{Baikov:2016tgj}, its generalization to an arbitrary gauge group \cite{Luthe:2016xec,Luthe:2017ttc,Herzog:2017ohr} and
gauge fixing parameter \cite{Chetyrkin:2017bjc,Luthe:2017ttg},
the $6$-loop calculations of the $\Phi^4$-model renormalization group functions \cite{Batkovich:2016jus,Kompaniets:2016hct,Kompaniets:2017yct}, 
the 7-loop anomalous field dimension calculation for the $\Phi^4$ model  \cite{Schnetz:2016fhy}
and the 7-loop anomalous dimension calculation of twist-2 operators in planar $\mathcal{N}=4$ super-Yang-Mills theory with the help of integrability arguments \cite{Marboe:2016igj}.

As a result of these enormous developments, mainly concerning particle physics, statistical mechanics and (supersymmetric) gauge field theories, there is by now a wide panel of advanced techniques available  
which may also have potentially successful applications to a broad range of problems in other fields, \eg, such as condensed matter physics.

\subsection{Interplay between high- and low-energy physics}

In this manuscript, we will be interested in condensed-matter physics systems consisting of a large number of interacting electrons. Such systems may be described in terms of 
effective QFTs at low energies which, in many cases, correspond to gauge field theories and which rival in complexity with the most sophisticated models of high energy physics as 
they are subject to their own peculiar complications due to, {\it e.g.},
strong non-linearities, reduced symmetries due to the underlying presence of a lattice, the highly non-trivial role played by the Fermi surface, possible enhanced effects of interactions in low dimensions...
%Since the 50s there were numerous fruitful exchanges of concepts and methods between condensed matter physics, statistical mechanics and particle physics.
Early developments consisted in applying Feynman diagram, or Green's function, technique to solid state physics systems, see the notorious \cite{abrikosov1975methods} for a monograph.
Conversely, condensed matter physics concepts such as symmetry breaking have deeply influenced particle/high-energy physics.
In order to illustrate this with concrete examples, let us recall that Landau's theory of Fermi liquids \cite{landau1956,landau1957,landau1958} is the standard framework to understand
the properties of Galilean invariant band metals in three space dimensions (3D). 
In a Fermi liquid, interacting fermions are in one-to-one correspondence with non-interacting fermionic quasi-particles
albeit with renormalized parameters (mass, velocity,...); see \cite{PinesN66} for a classic monograph, \cite{abrikosov1975methods} for a Green's function approach and 
\cite{Shankar:1993pf,Polchinski:1992ed} for more recent RG approaches to Fermi liquid theory.  
Several types of instabilities may profoundly change the ground-state.
As known from Bardeen, Cooper and Schrieffer (BCS) \cite{Bardeen:1957mv}, see also Bogoliubov \cite{bogoliubov1958} and then Nambu \cite{Nambu60.PhysRev.117.648} and Eliashberg \cite{eliashberg1959} 
for refined theories of superconductivity,  an attractive interaction due to electron-phonon coupling near the Fermi surface
is responsible for the effective (Cooper) pairing of fermions. This leads to the formation of a fermion condensate which in turn breaks the $U(1)$ gauge symmetry (charge conservation) and opens
a gap in the single particle fermionic spectrum. 
The BCS model is a remarkable example of an effective field theory whereby integrating out phonons yields an effective four-fermion interaction. It is also the first
model where the dynamical breaking of a symmetry takes place.\footnote{Following \cite{Miransky:1994vk}, we distinguish spontaneous and dynamical symmetry breaking in that the latter
is characterized by the non-vanishing vacuum expectation value of a composite operator, \eg, $\psi^\dagger(x) \psi^\dagger(x)$ in the case of superconductivity where $\psi(x)$ is a fermionic particle-field operator.}
Inspired by the Cooper instability, Nambu realized that, within the particle physics context, the pairing mechanism might provide a mass for elementary particles \cite{Nambu:1960xd}.
Together with Jona-Lasinio, he constructed a relativistic counterpart of the BCS model \cite{Nambu:1961tp}.\footnote{As noticed in \cite{Miransky:1994vk}, independently, and at the same time, 
a similar work was carried out by Vaks and Larkin \cite{VaksL61}.} 
The Nambu-Jona-Lasinio (NJL) model describes Dirac fermions interacting via four-fermion interactions. It has brought to attention the fundamental phenomenon of dynamical
mass generation and dynamical chiral symmetry breaking (D$\chi$SB) in particle physics together with the notion of a critical coupling constant above which such phenomena could take place.
 %This effective field theory (non-renormalizable in four dimension) has QCD as its ultra-violet (UV) completion. 
Many subsequent models of BSM physics whereby particles are supposed to dynamically acquire a mass,
{\it e.g.}, the so-called technicolor theories, see, \eg, \cite{Cohen:1988sq} for a paper somehow related to the subjects we will study, 
were inspired by the NJL model, see the book \cite{Miransky:1994vk} for a beautiful and detailed review on dynamical symmetry breaking in four-dimensional
theories.  Another type of instability, in the case of repulsive interactions and provided the Fermi surface is ``nested'', is towards the formation of an excitonic 
insulator\footnote{There are also Peierls insulators resulting from the Peierls instability and leading to (charge or spin) density waves.
The Peierls instability originates from electron-phonon interactions. In the case of the excitonic insulator, the driving mechanism is the Coulomb interaction between electrons and holes in overlapping bands. 
In both cases some form of nesting, \ie, different parts of the Fermi surface have to coincide when shifted in momentum space, is required for the instability to take place. 
In the case of excitonic insulators, the nesting condition is that electron and hole Fermi surfaces coincide modulo some wave-vector. 
See the textbook \cite{khomskii2010basic} for a nice account.} 
\cite{KeldyshK64+HalperinR68}, see also \cite{RevModPhys.40.755} for an early review.  
In this case, the pairing is between electrons and holes leading to the formation of a so-called excitonic condensate 
which breaks lattice symmetry and opens a gap in the single particle fermionic spectrum.
Contrary to the BCS case, the pairs are neutral and the ground-state insulating.\footnote{There is no sliding of the density-wave as a consequence of commensurability effects and pinning by disorder.} 
Notice that, in the case of the Cooper and excitonic instabilities, the underlying non-interacting fermion system had a finite density of state (DOS) at the Fermi surface. 
In contrast, in the case of relativistic models such as the NJL model, 
the DOS vanishes at the Dirac points. As reviewed in \cite{Miransky:1994vk}, the difference between these two cases results in a non-zero critical coupling constant 
above which dynamical breaking takes place in relativistic theories while it takes place for an arbitrary small coupling constant in systems with a Fermi surface. 
Moreover, in the case of band-touching, \ie, the semi-metallic case with zero DOS at the Fermi surface, an early study of Abrikosov and Beneslavskii (devoted to 3D systems) \cite{abrikosov1971}
showed that quadratic band touching is always unstable towards an excitonic instability once the interaction is turned on while linear band touching is stable, at least at weak coupling. 
We shall come back on these important facts below in relation with the newly discovered ``Dirac materials''.

In the 80s, new and very intriguing electronic phases of matter were discovered: high-temperature superconductivity \cite{HTc-1986}
and fractional quantum Hall effect (FQHE) \cite{PhysRevLett.48.1559}.
These phases appear in (quasi) planar condensed matter physics systems such as layered copper oxide (``cuprate'') materials for high-temperature superconductors\footnote{Interestingly,
conventional (BCS-like) superconductivity at $203$K at high pressures has been recently discovered in the sulfur hydride system (hydrogen-based superconductor) \cite{H2S-2015}. The highest
transition temperature obtained to date in cuprates is $133$K. So, since 2015, the highest-temperature superconductors are actually conventional ones!}
 and two-dimensional electron gases (2DEG) realized in transistor-like structures made from semiconductors for the FQHE.
 %They are non-Fermi liquid like and the microscopic origin of their ground state is still not known. 
Cuprate superconductors display a very rich phase diagram as a function of doping and temperature: superconductivity, antiferromagnetism, pseudogap and strange metal phases... 
They are the subject of extensive experimental and theoretical works for more than thirty years now. 
In the absence of doping, it is generally agreed that the system is already in a very unusual phase: a Mott insulator.\footnote{The driving mechanism 
for the formation of a Mott insulators is the strong electron-electron interaction.} All the richness (and complexity) 
of cuprates seems to arise from doping such a strongly correlated insulator, see \cite{RevModPhys.78.17} for a rather recent and detailed review.
There is now good evidence that their superconducting state is not conventional BCS-like; they are d-wave superconductors (dSC) but the precise microscopic pairing mechanism between charge carriers is still not known. 
Even the properties of the metallic phases above the transition temperature, pseudogap or strange metal depending on doping,
are still not well understood. Nevertheless, gauge-field approaches provided a lot of qualitative insights into the rich physics of these systems. 
Very soon after their discovery, it was indeed realized that various regions of the phase diagram of dSC could be described by $(2+1)$-dimensional lattice gauge theories where the gauge fields are artificially introduced
in order to take into account of occupation constraints \cite{PhysRevB.37.580,Marston:1989zz,Ioffe:1989zz,PhysRevLett.64.2450,PhysRevB.46.5621}.  In the strange metal phase, gauge forces explain qualitatively the 
linear temperature dependence of the resistivity \cite{PhysRevLett.65.653}; such linear dependence contrasts with the typical $T^2$ scaling in usual Fermi liquids and is typical of the so-called marginal Fermi 
liquids, see \cite{PhysRevLett.63.1996} for a phenomenological approach. %\footnote{In a Fermi liquid the resistivity scales as $T^2$.} 
In the normal phase of the underdoped region, the so-called algebraic spin liquid phase \cite{PhysRevLett.86.3871,Rantner2002}, 
low-energy excitations were found to have a gapless linear, relativistic-like, spectrum \cite{Marston:1989zz,Ioffe:1989zz}. Depending on the precise spin liquid phase, the effective
gauge-field theory description may take the form of $(2+1)$-dimensional QED (or QED$_3$), see also \cite{PhysRevLett.87.257003,PhysRevB.66.054535,Herbut:2002yq} for QED$_3$-like theories
of dSC, or even of $(2+1)$-dimensional QCD (or QCD$_3$) \cite{PhysRevB.65.165113}. In the particle physics context, odd dimensional field theories can be viewed as 
a high temperature limit of even dimensional models \cite{Appelquist:1981vg,Appelquist:1981sf}. The study of QED$_3$ in particular was advocated by Pisarski \cite{Pisarski:1984dj}
because of its similarities with $(3+1)$-dimensional QCD (or QCD$_4$) and the fact that D$\chi$SB may be studied systematically in this simpler model. 
The study of the critical properties of QED$_3$ was soon after refined by Appelquist et al.~\cite{Appelquist:1985vf,Appelquist:1986fd,Appelquist:1986qw,Appelquist:1988sr,Nash:1989xx} 
who later on also considered the case of QCD$_3$ \cite{PhysRevLett.64.721}.  
Since then, QED$_3$ has been extensively studied with motivations coming from both particle and condensed matter physics, see, \eg, 
\cite{Pennington:1990bx,Curtis:1992gm,Pisarski:1991kg,Atkinson:1989fp,Karthik:2015sgq,Karthik:2016ppr,Dagotto:1988id,Dagotto:1989td,Hands:2004bh,%
Strouthos:2008kc,Azcoiti:1993fb,Azcoiti:1995mi,Appelquist:2004ib,Raya:2013ina,Appelquist:1999hr,Giombi:2015haa,Giombi:2016fct,Braun:2014wja,Janssen:2016nrm,Kotikov:1993wr,Kotikov:2011kg,Gusynin:1995bb,Bashir:2005wt,%
Bashir:2009fv,Kubota:2001kk,DiPietro:2015taa,Herbut:2016ide,Gusynin:2016som}. Notice that the first application of the massless Feynman diagram techniques (such as those described in the previous section) 
to D$\chi$SB in QED$_3$ was carried out by Kotikov \cite{Kotikov:1993wr,Kotikov:2011kg}. We shall come back in detail on D$\chi$SB in QED$_3$ later on. Notice also that we have so-far considered
the so-called non-compact QED$_3$. It seems that for a small number of fermion flavours, the compactness of $U(1)$ cannot be neglected and that monopole
configurations proliferate leading to the confinement of electric charge \cite{Polyakov:1975rs,Polyakov:1976fu}, see, \eg, Refs.~\cite{PhysRevB.70.214437,Pufu:2013vpa,Chester:2015wao} for
interesting recent developments. In the following, only the non-compact case will be considered. 
%Moreover, recent developments have revealed the existence of powerful dualities
%in $(2+1)$-dimensional QFTs (including QED$_3$), see, \eg, \cite{Son:2015xqa,Hsiao:2017lch,Xu:2015lxa,Cheng:2016pdn,Seiberg:2016gmd}. 
%These dualities may provide exact results at specific values of the parameters of the model under consideration, \eg, at self-duality.  
%We shall come back briefly on some of these results in the next chapters.

 %{\it Following \cite{Kotikov:2016wrb,Kotikov:2016-2}, the problem of D$\chi$SB in QED$_3$ will be carefully analysed later on in this manuscript}.
Gauge field theory approaches are also sometimes used for a phenomenological description of the FQHE.~\footnote{In a way somehow similar to superconductivity, the physics of the quantum Hall 
effect is essentially about understanding the appearance of gaps in the single particle electron spectrum. The situation is quite clear in the integer case with a crucial role played by disorder 
(broadening of plateaus). The fractional case is more mysterious with gaps appearing at fractional fillings of the Landau levels. It is believed that these gaps are opened due to electron-electron interactions
but a microscopic theory is still missing. One fruitful idea in FQH physics was Jain's idea of composite fermions \cite{Jain:1989tx}, see also the textbook \cite{Jain:book}, 
which are quasi-particles formed by attaching an even number of magnetic flux quanta to an electron.
It turns out that, at a mean-field level, the composite fermion is subject to an effective magnetic field which is such that the 
FQH problem for electrons can be mapped to an integer QH for the composite fermion. A special case is at half-filling where the effective magnetic field is zero and the composite fermion becomes gapless.
According to Son, composite fermions at half-filling are Dirac like \cite{Son:2015xqa} and the corresponding low-energy effective theory for these Dirac composite fermions 
(at the self-dual point) is reduced QED$_{4,3}$ that will be analyzed in detail in this manuscript. Notice that, experimentally, it is quite clear that the $1/2$-filled QH state is a compressible Fermi liquid state  
as there are no plateaus in the Hall conductivity at this filling, see, \eg, Refs.~\cite{PfeiferCF:1993,JainCF:1994,PfeifferCF:1996} 
for experimental confirmation of the existence of the Fermi surface and \cite{RezayiCF:1994} for a numerical study. 
Recently, there has been experimental confirmation of the Dirac nature of the composite fermion, see \cite{PanCF:2017} and more references therein. 
Finally, the existence of a Fermi surface for the Dirac composite fermions implies that we should in principle add a chemical potential
to our model; in this manuscript, we will only study reduced QED$_{4,3}$ at zero chemical potential.}   In this case, they are very different from those of Maxwell, 
namely of the Chern Simons type. The $(2+1)$-dimensional Chern-Simons theories provide a good description of the phenomenology of quantum Hall fluids, \eg, 
incompressibility, fractional Hall conductance with odd denominators and fractional charges and statistics of the quasiparticles \cite{Zhang:1988wy,Fradkin:1991wy}. 
There is a correspondence \cite{Witten:1988hf} between these $(2+1)$-dimensional topological field theories and
conformal field theories (CFT) in $1+1$ dimensions which are believed to be able to describe either the edges of the quantum Hall sample or even some of its bulk properties \cite{Moore:1991ks}.
The heavy machinery of CFT is now extensively applied to investigate the formal properties of quantum Hall states. %(with, probably, little connexion with experimentally relevant issues). 
 %It is not clear, however, that such approaches may give any insight in understanding
 %(experimentally relevant) properties of FQHE systems. 
 Moreover, recent developments have revealed the existence of powerful dualities
in $(2+1)$-dimensional QFTs (essentially of the Chern-Simons type), see, \eg, \cite{Son:2015xqa,Hsiao:2017lch,Xu:2015lxa,Cheng:2016pdn,Seiberg:2016gmd}.
These dualities may provide exact results at specific values of the parameters of the model under consideration, \eg, at self-duality.
We shall come back briefly on some of these results in the next chapters.
 Let's note that only in FQHE systems, and so far only those FQHE systems made in 2DEG, did experiments reveal the existence of quasiparticles with fractional charges \cite{PhysRevLett.79.2526,picciotto1997fqhe}.
In particle physics, quarks are assumed to have fractional charges but they are confined.
The concept of fractional excitations first appeared in the study of (one-dimensional) relativistic QFTs as topological solitons polarizing the Dirac vacuum \cite{PhysRevD.13.3398}, see 
also \cite{0038-5670-30-5-R02} for a beautiful review at the interface between QFT and solid-state physics.  
In the latter, solitons as topological defect structures of (quasi-one dimensional) Peierls insulators act as charge 
(or spin) carriers \cite{PhysRevLett.36.432,brazovskii78,brazovskii1980,PhysRevLett.42.1698,rice1979}. There are some experimental indications that solitons exist, \eg,  in conducting polymers, see
\cite{RevModPhys.60.781} for a review, but the clearest evidence has only been given recently in \cite{PhysRevLett.104.227602} where it was shown that spinless charged solitons are the constituents 
of ferroelectric domain walls, see \cite{0953-8984-13-18-311} for an earlier theoretical anticipation  and \cite{Karpov16} for a more recent study.
Still, there is no direct experimental evidence that they carry fractional charges.
Actually, the practical motivations to study one-dimensional (1D) and quasi-1D systems were based on Little's theoretical proposal
that some conducting polymers might be superconducting at room 
temperature \cite{PhysRev.134.A1416}. This is far from being achieved yet.\footnote{Presently, the highest achieved critical temperature for an organic superconductor is 
about $33$K in alkali-doped fullerene RbCs$_2$C$_60$.}
Certainly, the structural complexity of all the above considered materials makes them difficult to control experimentally and difficult to understand (beyond phenomenological considerations) theoretically.
% and controlling systems such as cuprates or conducting polymers might be due, in part, 
%to the fact that they are extremely complicated at the structural level 
%(polymers look more like a bunch of entangled spaghettis than nice quasi-one dimensional structures!).

Starting in the 90s, the quest for simplicity arose with the study of carbon-based materials.
This is the case of carbon nanotubes: filaments made of carbon atoms with nanometer thickness and lengths which can be more than micrometers. 
The date of their discovery is not too clear (maybe in the 50s or even before) but their extensive experimental study was carried out in the 90s.
These systems are amongst the rare 1D structures known (others being quantum wires and edge states in Hall systems). They have many remarkable electrical, thermal and mechanical properties, see the book \cite{saito1998physical}. 
Theoretically, interacting electrons in such 1D structures are supposed to form the so-called Tomonaga-Luttinger (TL) liquid, see the textbooks \cite{giamarchi2003quantum,gogolin2004bosonization} 
for detailed theoretical accounts, especially \cite{gogolin2004bosonization} for a nice historical introduction.
According to \cite{egger2001}, it is probably in metallic carbon nanotubes that the cleanest experimental evidence of TL liquid behaviour was found, see, \eg, \cite{nanotubestomonaga1998,nanotubestomonaga2003}. 
Historically, the Tomonaga model \cite{Tomonaga:1950zz} was the first theoretical realization
of Bloch's idea \cite{Bloch1933,Bloch1934} that the low-energy excitations of an assembly of Fermi particles could be described in terms of ``sound waves''.
Tomonaga proved that low-energy excitations in interacting 1D Fermi systems are not fermionic quasi-particles but collective (bosonic) ones. 
That one of the founding fathers of QED is at the origin of such an important condensed matter physics model may be related, to put it simply, to the fact that the 
emergence of Lorentz invariance at low-energies is generic to 1D systems as the Fermi surface always reduces to 2 points in this dimensionality. 
As we already alluded to above, when mentioning solitons, it is probably so far within the  study of 1D quantum systems that relativistic QFTs have had the most important input.
The TL model, see \cite{Luttinger:1963zz,Mattis:1964wp} for complementary early studies by Luttinger as well as Mattis and Lieb, is a $(1+1)$-dimensional effective theory describing 
interacting fermions with gapless linear spectrum. The related TL liquid is the prime example of a non-Fermi liquid.
In physical terms, this is related to phase space constraints in 1D which lead to dramatic changes in the ground-state of non-interacting fermions once an, even very small, interaction is turned on.
The sensitivity of the ground-state to interactions as well as the large fluctuations in 1D, make it delicate to use conventional perturbative techniques.
For example, both Cooper and density-wave channels compete in 1D and, soon after Little's proposal, specific RG methods were developed in order to resum the singularities in both channels, 
the so-called ``parquet'' summation \cite{bychkov1966,dzyaloshinskii1972}, see also \cite{solyom1979} for an early review on g-ology.\footnote{The parquet summation first appeared
in particle physics \cite{diatlov1957} and was first introduced in the West by Nozi\`eres et al., see \cite{PhysRev.178.1072,PhysRev.178.1084} for a very nice presentation of the method
and application to the X-ray absorption edge problem.} It was then shown that, in 1D, both instabilities
cancel each other and that, from this cancellation, the TL liquid arises as a non-trivial metallic ground-state, see, \eg, \cite{PhysRevB.55.3200} for a review. 
Complementary early fermionic approaches confirmed that this unusual liquid is characterized by power-law correlation functions \cite{dzyaloshinskii1973,efetov1975}. 
The subsequent developments gave rise to radically new methods which are specific to relativistic $(1+1)$-dimensional (and 2D statistical) models allowing sometimes to find exact solutions. These include, \eg, 
bosonization \cite{Coleman:1974bu,PhysRevD.11.3026,PhysRevB.9.2911}, CFT techniques \cite{Belavin:1984vu}, see also \cite{Dotsenko:1984nm} for the link between CFT and bosonization,
and methods of quantum integrable systems, see \eg, the textbook \cite{korepin1997quantum}. 
All these developments, see, \eg, \cite{Haldane:1981zza}, confirmed the typical power-law correlations of TL liquids originally obtained via fermionic techniques.
They also suggest that the absence of (usual) fermionic quasi-particles might be interpreted as a separation of the electron into a spinon and a holon.
It is the power-law correlations, with exponents in good agreement with theoretical predictions, which are the experimental signatures of TL behaviour, \eg, \cite{nanotubestomonaga1998,nanotubestomonaga2003}.
Apparently, there is also some experimental evidence for spin-charge charge separation, \eg, \cite{jompol2009}, and charge fractionalization \cite{steinberg2008frac}. 
Let's also note that there were numerous attempts to adapt these non-perturbative
techniques to higher dimensional systems, in particular $(2+1)$-dimensional, with applications to high-temperature superconductors, see, \eg, \cite{PhysRevB.55.3200} for an application of the parquet summation to this case
as well as a discussion of other methods such as higher dimensional bosonization. Without entering any detail, it seems that such rigorous approaches are extremely difficult to implement in higher dimensions 
especially for systems such as cuprates where, to begin with, the complicated Fermi surface is always taken into account in some approximate way.  

Another interesting theoretical approach developed in the 90s is that of the functional renormalization group (FRG), see \cite{Delamotte:2007pf} for a pedagogical review. 
The basis of the FRG is an exact (one-loop like) flow equation for an (effective) action functional \cite{Wetterich:1992yh}. Because it is the flow of the full functional which is examined, the FRG does not make any 
distinction between renormalizable, marginal and super-renormalizable couplings as in the field-theoretic approach.  There is by now a vast literature on FRG and many applications. 
However, only a few of them are devoted to interacting fermions, see, \eg, the review of Metzner et al.~\cite{RevModPhys.84.299}, and gauge field theories, see, \eg, the review of Gies \cite{Gies:2006wv} 
as well as the very nice dissertation of Janssen \cite{Janssen:2012oca} where topics close to the ones considered in this manuscript are studied with the help of the FRG.~\footnote{Anticipating the next section, we may also from now
on mention the work of Katanin \cite{Katanin16.PhysRevB.93.035132} who studied dynamical gap generation in graphene with the help of a combination of FRG and Bethe-Salpeter approaches.}

Summarizing this paragraph, we have, roughly speaking, considered two broad categories of interacting systems:
\begin{itemize}
\item those where interactions lead to spectacular effects well observed experimentally (dSC, FQHE which are $(2+1)$-dimensional) but seem to still challenge theoretical understanding,%\footnote{Which may also in part be due 
%to the fact that papers which have explained correctly some of their phenomena got rapidly flooded by the ever-growing literature on the subject}
\item others where interaction effects are rather well understood theoretically even at a fully non-perturbative level (($1+1$)-dimensional systems) but where experiments seem to be very difficult to carry out.
\end{itemize}
In the next paragraph we will in some sense consider another (somehow intermediate between the two above) type of system: the Dirac materials. 
They are strongly coupled $(2+1)$ or $(3+1)$-dimensional systems and therefore, {\it a priori}, not amenable to exact solutions.
They are however considered to be very clean systems in which to experiment. And, though interaction effects are not (so far) as spectacular as in dSC, their understanding raises a number of well defined 
issues awaiting for a quantitative theoretical analysis. Our approach, as we shall motivate it in more detail in the next section, will then rely on a field-theoretic renormalization study making use of modern multi-loop techniques
to address these issues. As we mentioned before, such an approach is quite efficient and rigorous (within its range of applicability) to capture analytically many interaction effects. 
For example, within the framework of Schwinger-Dyson equations, such an approach allows to study strong-coupling phenomena such as D$\chi$SB. Hopefully, our results will serve as a basis for future (fully) non-perturbative techniques
which have yet to be developed. 

\section{Electron-electron interactions in Dirac liquids}

\subsection{Overview}

In 2004, another allotrope of carbon was isolated \cite{novoselov2004,novoselov2005,Novoselov:2005kj}: graphene which is a 2D crystal of carbon atoms forming a 
honeycomb lattice (or lattice of benzene rings). Since then, it has been the subject of an enormous amount of studies related to its rather extraordinary electrical, optical, thermal and mechanical 
properties, see \cite{Gusynin:2007ix,CastroNeto:2009zz} for early reviews as well as the textbook \cite{katsnelson2012graphene}. As a matter of fact, graphene is the most bi-dimensional 
system presently known: it has a one-atom thickness, that is about $0.5$nm, and more than micrometer lengths. Despite being so thin, a graphene sheet still absorbs a significant
amount of visible light \cite{Nair:2008zz,PhysRevLett.101.196405}:
\be
\pi \al \approx 2.3 \%, \qquad \al = \frac{e^2}{4 \pi \hbar c} = \frac{1}{137}\, ,
\label{chap1:abs}
\ee
where $\al$ is the QED fine structure constant (we shall come back on this remarkable result below). 
From the structural point of view, graphene is much simpler than most other planar condensed matter physics systems such as, 
\eg, the cuprates mentioned above. It is characterized by a high crystal quality and is a clean system in which to experiment; electron mobility as well as thermal conductivity are
exceptionally high in graphene. It is stronger than steel and yet stretchable. There is by now a whole graphene industry with many promissing applications... 
In this manuscript, we shall focus on fundamental aspects of graphene physics. Actually, long before the isolation of graphene, it has been recognized that, from its peculiar lattice structure, an effective low-energy
description (below $1$eV which is a rather large energy scale for a condensed matter system) emerges in terms of massless Dirac fermions \cite{PhysRev.71.622,Semenoff:1984dq}. 
In the absence of doping,\footnote{In this manuscript, we shall focus only on the case of intrinsic graphene. 
Doped graphene shares many similarities with Fermi liquids and will not be considered. Notice that the intrinsic case is some kind of 
idealization because the system is never, strictly speaking, at the Dirac point due to
the presence of charge fluctuations or electron-hole puddles. Nevertheless, experiments manage to get rather close to the Dirac points, \eg, down to
$\mu \sim 0.1$meV. For simplicity, and as a first approximation, we shall therefore take $\mu=0$ in the following.} the 
so-called intrinsic case, graphene is a semimetal (or zero-band semiconductor) with a Fermi surface reducing to two points 
(the Dirac points or valleys), linear single-particle spectrum and an electron wave function taking the form of a four-component spinor (in sublattice$\otimes$valley space) with two additional 
species ($N_F=2$) corresponding to the real spin. Such peculiarities strongly differentiate graphene from usual Galilean invariant band metals and semi-conductors 
and are reminiscent of the 1D conductors, d-wave superconductors and 3D semimetals mentioned above. In condensed matter physics, 
the emergence of Lorentz invariance at low energies is a generic feature of systems with two stable Fermi points, see the book \cite{Volovik2009book}. 
Experimentally, the existence of gapless relativistic-like excitations in graphene has been first 
clearly confirmed by the observation of an anomalous integer quantum Hall effect \cite{Novoselov:2005kj} in accordance with theoretical studies \cite{Schakel:1991af,Gusynin:2005pk}. The 
low-energy effective description of graphene may therefore very well be a simple continuous $U(1)$ QED-like gauge-field theory as anticipated long ago by Semenoff \cite{Semenoff:1984dq}.
There are, however, important differences with respect to usual relativistic QEDs.
Firstly, electrons in graphene propagate at the Fermi velocity: $v = \sqrt{3} a t / (2\hbar)$, which depends on the lattice spacing $a =2.5$\AA{} and on
the hopping integral: $t = 3$eV. Such lattice parameters yield a (bare) velocity which is much smaller than the velocity of light: $v = c/300$.
In the presence of interactions (coupling to a gauge field) this implies that Lorentz invariance is {\it a priori} lost and we shall refer to this system as being pseudo-relativistic.
 %As a first step, we may altogether neglect relativistic effects (coupling to vector photons) which corresponds to assume that the interaction between electrons
%is the instantaneous long-range Coulomb interaction. Such system is therefore not relativistic. But because it also differs from usual, low-energy galilean invariant, parabolic band metals and
%semi-conductors, it will be referred to as a pseudo-relativistic system.
A second difference with respect to usual QED is that we are dealing with fermions in $D_e=2$ space dimensions interacting via a gauge field in $D_\gamma = 3$ space dimensions.
This system may therefore be thought of as a physical realization of a ``brane''-like universe (``brane'' coming from membrane) such as those which are often
evoked in particle physics for a larger (and unphysical) number of space dimensions.
Thirdly, we are dealing with a strong coupling theory. Indeed, the QED fine structure constant is replaced by: 
\be
\alpha_g = \frac{e^2}{4 \pi \hbar \kappa v} = \frac{2.2}{\kappa}\, ,
\label{chap1:alg}
\ee
where the dielectric constant $\kappa \approx 1$ in the case of suspended graphene, that is, without substrate. Moreover, there is no screening of the interaction (in the intrinsic case) which is therefore
long-ranged. %A priori, physical properties of graphene should be strongly affected by such interactions. 
A fourth peculiarity, in the case of suspended graphene, is that the sheet is actually subject to some out-of-plane deformations, the so-called ripples, which may reach $1$nm height \cite{meyer2007ripples}.
The microscopic origin of these ripples is not too clear yet but there is a common belief that they allow the planar system to overcome the Landau-Peierls-Mermin-Wagner theorem which stipulates
that long-range crystalline order cannot exist in 2D. %Suspended (intrinsic) graphene should therefore be viewed as a soft semi-metallic crystalline membrane with strong (long-ranged) electron-electron interactions.
At this point, let's note that around 2008 topological materials were experimentally discovered, see, \eg, \cite{hsieh2008topological,xia2009topological},
and were since then the subject of extensive studies. It turns out that the surface electronic structure of these topological insulators consists of Dirac cones similarly to graphene, see the 
review \cite{RevModPhys.82.3045} for more details. There are also, by now, many artificial graphene-like materials under study. The hope is that they may offer a tunable platform to study the properties
of 2D pseudo-relativistic systems, see \cite{polini2013artificial} for more details. More recently, 3D Dirac (described in terms of $4$-spinors), %(predicted to be stable in \cite{abrikosov1971} as we already mentioned), 
see, \eg, \cite{liu2014dirac3d,neupane2013dirac3d,PhysRevLett.113.027603}, and Weyl ($2$-spinors) semimetals, see, \eg, \cite{huang2014weyl,xu2015weyl}, were experimentally discovered, 
see the review \cite{RevModPhys.88.021004,Wehling2014Dirac} for more details on these three-dimensional analogues of graphene. 
%Finally, let us note that the gapless half-filled fractional quantum Hall states may also be descibed in terms of relativisic-like low energy excitations  \cite{Son:2015xqa}. So, in the following, even though we shall essentially focus on graphene for which many experiments are available, 
%our theoretical developpements will concern all of these planar Dirac liquids, including 3D ones for which our approach should be rather straightforward to adapt.% (and are left to interested students as an exercise!).

From the above presentation, suspended intrinsic graphene appears as a soft semi-metallic crystalline membrane with strong (long-ranged) electron-electron interactions.
Issues commonly addressed are then broadly related to the effects of interactions, disorder (extrinsic or intrinsic such as ripples) as well as their interplay.
Gauge-field theory approaches to graphene are very common. Some very nice developments are related to working in curved space in order to take into account of the ripples 
and introducing random gauge fields in order to take into account of disorder, see \cite{vozmediano2010gauge} for a review. 
The combined effect of disorder and interactions seems to be non-trivial and interactions may be enhanced when disorder is present, 
see \cite{RevModPhys.84.1067} which briefly reviews some of these aspects. The interplay between ripples and electron-electron interactions is, as far as we know, still not well understood, 
see however Ref.~\cite{Herbut08.PhysRevLett.100.046403} for an early study.~\footnote{Let's also note that the study of the elastic and curvature degrees of freedom of graphene 
as a membrane, \ie, in the absence of a coupling to electronic degrees of freedom, is rather involved, see, \eg, the recent \cite{Coquand:2017wak} and references therein}.
 %Such studies reveal, \eg, a topological stability of the Dirac point for weak Coulomb and %lattice deformations. 
In the following, we shall neglect disorder as well as flexural distortions and focus on the sole effect of electron-electron interactions on some of the properties of an ideally 
perfectly flat and clean undoped graphene sheet. Moreover, we shall exclusively focus on the (marginal or marginally irrelevant) long-range Coulomb interaction and neglect possible 
({\it a priori} irrelevant) short-range (Hubbard-like) interactions.~\footnote{There are very nice studies of the effect of short-range interactions %(that preserve relativistic invariance) 
	and the instabilities they might trigger by Igor Herbut and co.\ starting from Ref.~\cite{Herbut:2006cs}. It seems that the instabilities are in the Gross-Neveu universality class. 
As previously mentioned, the computation of RG functions of this model in $(3+1)$-dimensions has recently reached  the impressive 4-loop level \cite{Gracey:2016mio,Mihaila:2017ble,Zerf:2017zqi}.} 
Finally, as we mentioned above, graphene is now part of a larger class of materials that we may refer to as planar Dirac liquids. So we shall keep in mind that similar issues related
to interaction effects may also arise in these systems, \eg, surface states of topological insulators, possibly half-filled fractional quantum Hall systems and others which are yet to be discovered (our
approach can also be adapted to $(3+1)$-dimensional variants of these systems).
 % (experiments we shall refer to in what follows are done with ultra-clean suspended graphene samples). 
In this respect, three questions are commonly raised:
\begin{itemize}
\item are there any specific effects of interactions on Dirac materials such as graphene that cannot be found in usual Galilean invariant band metals?
\item are there any clear experimental signatures of these effects?
\item is the present theoretical understanding of these effects satisfactory?
\end{itemize}
The answer to the first question is yes and given the large strength and long-range character of the Coulomb interaction it is {\it a priori} expected that there will be
clear experimental signatures of these effects. The surprising fact is that interaction effects turn out to be difficult to observe experimentally in graphene and, up to now, 
only few experiments were devoted to analyzing interaction effects {\it per se}. It seems that more pronounced effects are found in recently available 
ultra-clean high-mobility samples, see, \eg, \cite{PhysRevLett.114.126804}. A similar feature is expected to take place in (tunable) artificial materials as well as
new and more strongly coupled Dirac materials \cite{Hirata2017}. 
As for the last question, as mentioned above, the theoretical study of interactions started long before the actual isolation of graphene. Since 2004, 
a lot of progress has been made, see \cite{RevModPhys.84.1067} for a review. However, the reason why the effects of interactions are so difficult to observe experimentally is still not yet clearly explained. Moreover, there is often no
definitive agreement on the precise value of important quantities directly related to interaction effects ($\mathcal{C}$, $\al_c$, $N_c$, ... see below).   
Let us turn to some concrete examples to explain these statements in more detail.

We have already mentioned several times the emergence of Lorentz invariance at low energies in systems with stable Fermi points. In the case of graphene, 
this has been proven theoretically by Gonz\'ales, Guinea and Vozmediano in \cite{Gonzalez:1993uz} with the help of an RG study. 
The later revealed the running of the Fermi velocity which flows towards the velocity of light deep in the infrared, 
$v \ra v^*=c$. This goes along with a flow of the coupling constant to the QED one, $\al_g \ra \al_g^* = \al=1/137$. The fixed point corresponds to a zero of the beta function associated with the velocity at $v^*$.
There is experimental evidence for the existence of this fixed point coming from the observation of a reshaping of the Dirac cones via measurements of the effective cyclotron mass \cite{elias2011cones}
and high-resolution angle-resolved photoemission spectroscopy \cite{siegel2011cones}.
Of course, the real system is no where close to the deep IR fixed point. Nevertheless, the observed reshaping is compatible with an increase of the velocity by a factor of $3$ %(and hence a decrease of
 %the coupling constant by the same factor) 
upon reducing as much as possible the carrier density. The observed increase of the velocity is logarithmic in accordance with the theoretical prediction 
\cite{Gonzalez:1993uz} (a similar feature was predicted to happen for 3D Dirac materials in \cite{abrikosov1971}). Up to now, the reshaping of the Dirac 
cones, as observed in \cite{elias2011cones,siegel2011cones}, 
 is considered as one of the most important signature of the effect of interactions in graphene. Other experiments find evidence for a quasi-particle decay rate having a linear dependence 
in energy, \eg, \cite{bostwick2007,PhysRevLett.102.176804}, in accordance with theory \cite{PhysRevLett.77.3589}. This is similar to what happens in marginal Fermi liquids. 
However, it is theoretically understood that, despite a strong renormalization of the quasi-particle residue, $Z_\psi$, 
the latter does not vanish in the IR because of the corresponding decrease of $\al_g$ and a quasi-particle description is still believed to hold, see \cite{PhysRevB.59.R2474} for an early paper, 
\cite{PhysRevLett.113.105502} for a more recent one and \cite{RevModPhys.84.1067} for a review. 
In this respect, marginal Fermi liquid features are most pronounced in the UV while in the IR the system seems to behave as a weakly coupled Fermi liquid.

Another expected difference between Dirac materials and usual metals comes from the non applicability of Kohn's theorem \cite{Kohn61PhysRev.123.1242} to systems with non-parabolic bands.
Kohn's theorem stipulates that, in Galilean invariant systems, the long wave-length ($k=0$ where $k$ is the wave-number) properties are not affected by interactions. 
This is the case of the cyclotron frequency and optical conductivity as well as other properties probed by applying homogeneous electric or magnetic fields.\footnote{The 
original paper of Kohn \cite{Kohn61PhysRev.123.1242} deals only with the cyclotron frequency. But the argument can be easily generalized to
the case of the optical conductivity, see, \eg , \cite{PhysRevB.14.4439} for an early paper.} Indeed, for a Galilean invariant system the current operator is proportional to the total momentum which is conserved
while interactions affect only the motion of the relative particle. The theorem fails in the case where Galilean invariance is broken. Typically, this is the case of disordered systems. 
In the absence of disorder, it is also the case of graphene and other pseudo-relativistic systems. One therefore expects that, in Dirac liquids, the cyclotron 
frequency and the optical conductivity will be affected by interactions (see \cite{Throckmorton:2018} for a recent paper). 
As we shall not consider any applied magnetic field in the following let us first focus on the 
case of the optical conductivity (the case with magnetic field will be briefly considered next). 
In this case, the available experimental results \cite{Nair:2008zz,PhysRevLett.101.196405},\footnote{These experiments do not directly probe the homogeneous optical conductivity (conductivity at zero momentum and
frequencies which are in the visible range of the spectrum) but rather the optical properties of graphene and in particular the above mentioned opacity, see Eq.~(\ref{chap1:abs}), which is related to the optical conductivity.}
see also \cite{Peres:2010mx} for a review, are well understood within a free fermion model with deviations of the order of $2\%$ only. These deviations were not clearly attributed to interactions by the 
experimentalists.\footnote{Indeed, many other factors affect the conductivity such as, \eg, temperature, chemical potential, substrate, scattering rates etc... Even in the case of free fermions their effect on the optical conductivity 
is rather non-trivial and lead to results, see, \eg, \cite{Gusynin:2006ym,Gusynin:2009-ac}, which already fit well the experimental data \cite{Nair:2008zz}.}
Nevertheless, there have been extensive theoretical attempts to understand interaction corrections to the optical conductivity, see, {\it e.g.},
\cite{PhysRevB.83.195401,Herbut08.PhysRevLett.100.046403,Mishchenko2008,Juricic:2010dm,Sheehy09.PhysRevB.80.193411,Abedinpour11.PhysRevB.84.045429,Sodemann12.PhysRevB.86.115408,Gazzola13.0295-5075-104-2-27002,%
Rosenstein13.PhysRevLett.110.066602,PhysRevB.87.205445,Link16.PhysRevB.93.235447,Boyda:2016emg,Stauber17.PhysRevLett.118.266801}.
It was shown that there is no correction from short-range interactions among the fermions \cite{PhysRevB.83.195401}. 
No exact result is available in the case of long-range interactions. Analytically, the problem is generally considered with the help of perturbation theory:~\footnote{At this point, 
the reader may be horrified by the fact that we use perturbation theory for $\al_g \approx 2.2$. As we shall comment on briefly in Sec.~\ref{chap1:sec:approach} and in Chap.~\ref{chap5}, 
the perturbation theory needs to be properly ``optimized'', see, \eg, see \cite{PhysRevLett.113.105502}. Our feeling is that, before doing so, 
agreement has to be found for the value of $\mathcal{C}_1$ and this will be our focus in this manuscript.}
\be
\sigma(\om) = \sigma_0\,\big( 1 + \mathcal{C}_1 \al_g + \mathcal{C}_2 \al_g^2 + \cdots \big)\, ,
\label{chap1:sigma}
\ee
where $\sigma_0 = e^2/(4\hbar)$ is the free fermion conductivity \cite{Fradkin86.PhysRevB.33.3263,Lee93.PhysRevLett.71.1887,Ludwig94.PhysRevB.50.7526} and $\mathcal{C}_i$ are interaction correction coefficients.
Despite the apparent simplicity of the task, the precise value of the first order interaction correction coefficient, $\mathcal{C}_1$, to the conductivity was subject to some debate.
As we shall see in the following, it seems that the most plausible analysis is that of Mishchenko \cite{Mishchenko2008}.
He showed that vertex and self-energy contributions almost completely cancel eachother yielding $\mathcal{C}_1 \approx 0.01$; this implies an only weak violation of Kohn's theorem 
compatible with the $\mathcal{C}_1 \al_g \approx 2\%$ deviation from free fermion result observed experimentally.
The question of whether such cancellation takes place at each order of perturbation theory is still open (even $\mathcal{C}_2$ is still unknown).
In the presence of a magnetic field, it was assumed for some time that Kohn's theorem is approximately valid with no clear sign of interaction effects on Landau levels, see \cite{0268-1242-25-6-063001} for an early review. 
Recently, however, more refined experiments on ultra-clean high mobility samples have revealed subtle deviations from the free fermion case that were attributed to interactions \cite{PhysRevLett.114.126804}.

A final example concerns the existence of an excitonic instability and related dynamical gap generation in graphene and graphene-like materials.
As we argued in previous sections, for relativistic-like systems, such an instability takes place only if the coupling constant $\al_g$ is
larger than a critical value, $\al_{c}$. In the case of graphene, even though the (bare) coupling constant is of the order of $2.2$ and the interaction is supposed to be
long-ranged, there is no experimental evidence that a gap of more that $0.1$meV is open at the Dirac point \cite{elias2011cones}. 
Theoretically, the computation of $\al_{c}$ has been the subject of extensive work,\footnote{Again, we mention only works focusing on the effect of the long-range Coulomb interactions.
Some studies indicate that the dynamical symmetry breaking is essentially driven by point-like interactions and that the long-range interaction makes only weak modifications, see, \eg, 
\cite{Herbut:2006cs,Juricic:2009px} and \cite{Semenoff:2011jf} for a very nice review and more references (including to other approaches such as, \eg, holography).
As discussed in the text, the situation is not yet clarified for long-range interactions alone. So, for simplicity, we shall focus only on the latter leaving
the addition of short-range interactions for future studies. Notice that, in the relativistic limit and for QED$_3$, the situation seems to have been clarified recently, 
see \cite{Gusynin:2016som}.} see, \eg, \cite{Khveshchenko:2001zz,Gorbar:2002iw,Leal:2003sg,Son07.PhysRevB.75.235423,VafekCase08,Khveshchenko:2008ye,Liu09.PhysRevB.79.205429,%
Gamayun:2009em,Drut:2008rg,Drut:2009aj,Drut:2009zi,Gonzalez12.PhysRevB.85.085420,Wang2012,Buividovich12.PhysRevB.86.245117,Ulybyshev:2013swa,%
Popovic13.PhysRevB.88.205429,Gonzalez15.PhysRevB.92.125115,Katanin16.PhysRevB.93.035132,Carrington:2017hlc}. 
All these works seem to agree that $\al_c$ should be of order $1$. 
However, presently, there is still no general agreement on the precise value of $\al_c$ and whether or not it is larger
that $\al_g$.~\footnote{Again, we focus on model-independent estimates of $\al_c$. For applications to real materials other effects may have to be taken into account such as, \eg, screening by residual charge carriers.}
  Clearly, an accurate computation of $\al_c$ is a major theoretical challenge in order to gain a  precise knowledge of the phase diagram of graphene and related Dirac liquids. 
One hopes that, for example, the precise knowledge of $\al_c$ may enable a fine tuning of (artificial) graphene-like materials in order to open a gap in a controlled way in these systems. 
This  would be of extreme practical importance, \eg, for the development of graphene(-like)-based transistors \cite{nevius2015} see also \cite{neto2009} for a nice early review attributing the possibility that
graphene could be an insulator due to electron-electron interactions to Linus Pauling. Notice that evidence for an excitonic instability was recently found 
in quasi-two-dimensional organic conductors which seem to possess an extraordinarily strong Coulomb interaction among their anisotropic Weyl fermions \cite{Hirata2017}.
Notice also that a magnetic field is expected to catalyze the dynamical generation, see \cite{Miransky:2015ava,Gusynin:2013noa} for reviews and, \eg, \cite{DeTar:2016vhr,DeTar:2016dmj} 
for recent lattice studies.  The study of dynamical gap generation in Dirac liquids, with or without magnetic field, is a very active field of research and 
we shall come back on it later (in the case of zero magnetic field).
 %{\it Following \cite{Kotikov2016gap}, the problem of dynamical gap generation in graphene will be considered at zero magnetic field}. 

%{\it Following \cite{Teber:2012de,Kotikov:2013kcl,Teber:2014ita}, we will consider the problem 
%of computing interaction corrections to the conductivity of graphene both at the IR fixed point and away from it}.

Understanding the above issues requires a {\it quantitative} theoretical understanding of the effect of electron-electron interactions in 
Dirac materials.
Due to the simplicity of the models describing these systems, which goes along with the structural simplicity of the materials themselves, 
there is some hope that such a task may be achieved analytically.  In this respect, the focus of this manuscript will be on the computation of the interaction correction to 
the optical conductivity, $\mathcal{C}$, as well as the critical coupling and flavour number, $\al_c$ and $N_c$, respectively.

\subsection{Approach and brief overview of some results}
\label{chap1:sec:approach}

The most general low-energy effective action (model I) that we wish to consider reads (in Minkowski space):
\bea
S =&& \int \D t\, \D^{D_e} x\, \left[ \bar{\psi}_\sigma \left( \I \gamma^0 \partial_t + \I v \vec{\gamma} \cdot \vec{\nabla}\,\right) \psi^\sigma - e\bar{\psi}_\sigma \,\gamma^0 A_0\, \psi^\sigma
+ e \frac{v}{c}\,\bar{\psi}_\sigma\, \vec{\gamma} \cdot \vec{A}\, \psi^\sigma \right ]
\nonum \\
&&+\, \int \D t\, \D^{D_\gamma} x\,\left[ - \frac{1}{4}\,F^{\mu \nu}\,F_{\mu \nu} - \frac{1}{2\xi}\left(\partial_{\mu}A^{\mu}\right)^2 \right]\, ,
\label{chap1:model-general}
%&&\int \D t\, \D^3 x\, \left[ - \frac{1}{4}\,F^{\mu \nu}\,F_{\mu \nu} -\frac{1}{2a}\,(\partial_\mu A^\mu)^2 \right]\, ,
\eea
where $\psi^\sigma \equiv \psi^\sigma(t,\vec x\,)$ is a four component spinor field of spin index $\sigma$ which varies from $1$ to $N_F$ ($N_F=2$ for graphene),
$v$ is the Fermi velocity, $c$ is the velocity of light which is also implicitly contained in the gauge field action through $\partial_\mu = (\frac{1}{c}\partial_t,\vec \nabla\,)$,
$\xi$ is the gauge fixing parameter and $\gamma^\mu$ is a $4\times 4$ Dirac matrix satisfying the
usual algebra: $\{ \gamma^\mu,\gamma^\nu \} = 2 g^{\mu \nu}$
where $g^{\mu \nu} = {\rm diag}(1,-1,-1,\cdots,-1)$ is the metric tensor in $D_e+1$-dimensions.
The action (\ref{chap1:model-general}) describes the coupling of a fermion field in $d_e=D_e+1$-dimensions with a $U(1)$ gauge field in $d_\gamma = D_\gamma + 1$-dimensions.
In the case of graphene, we have: $D_e=2$ and $D_\gamma = 3$, \ie, fermions in the plane and gauge field in the bulk.
Because of the running of $v$ all the way up to $c$, any complete renormalization group analysis of Dirac materials should in principle
be based on (\ref{chap1:model-general}). %Moreover, this model allows in principle the computation of relativistic corrections, {\it i.e.}, corrections
%in $v/c$, to any observable. It is expected that such corrections are weak because of the smallness of the ratio $v/c \approx 1/300$. Nevertheless, 
It turns out that such a task is rather involved and, presently, very few results are available, see \eg, \cite{Gonzalez:1993uz,PhysRevLett.116.116803}.

In the literature, the overwhelming majority of works focus on the non-relativistic limit where $v/c \ra 0$ (instantaneous interactions).
This is of course, a very realistic assumption given the smallness of the ratio $v/c \approx 1/300$ at the experimentally accessible scales.
In this limit, there is no coupling to vector photons and Eq.~(\ref{chap1:model-general}) simplifies as (model II):
\be
S = \int \D t\, \D^{D_e} x\, \bar{\psi}_\sigma \left[ \gamma^0 \left( \I \partial_t -eA_0 \right) + \I v \vec{\gamma} \cdot \vec{\nabla}\,\right] \psi^\sigma
+\, \frac{1}{2}\,\int \D t\, \D^{D_\gamma} x\, \left( \vec{\nabla} A_0 \right)^2 \, ,
\label{chap1:model-inst}
\ee
where the Coulomb gauge is used. Most of the theoretical results reviewed in the previous section were derived on the basis of (\ref{chap1:model-inst}). 
They are mainly perturbative with expansions in the (bare) coupling constant reaching two-loop accuracy (some partial results are available at three-loop \cite{PhysRevB.89.235431}). 
Of course, given the strength of the interaction in this limit ($\al_g \approx 2.2$) such expansions may not be reliable and a non-perturbative treatment of 
the interactions seems to be required. Such treatments are in general limited to an RPA-like resummation or leading order (LO) in the $1/N$-expansion, see \cite{PhysRevLett.113.105502} for an attempt to compute NLO corrections. 
Often, even LO results are approximate (using the so-called static approximation, neglecting Fermi velocity renormalization, etc...). So, despite the fact that 
(\ref{chap1:model-inst}) is simpler than (\ref{chap1:model-general}), calculations are difficult to carry out in a rigorous way in this limit. This often results in a rather confusing
situation where even the simplest quantities are subject to theoretical uncertainties as was reviewed in the last section for $\mathcal{C}_1$, $\al_c$ and $N_c$.

In this manuscript, we will follow an alternative non-conventional route. We will study interaction effects starting from the IR Lorentz invariant
fixed point where $v/c \ra 1$ and the interaction is fully retarded. In this limit, Eq.~(\ref{chap1:model-general}) can be written in covariant form as (model III):
\bea
S = \int \D^{d_e} x\, \bar{\psi}_\sigma \I \slashed D  \psi^\sigma + \int \D^{d_\gamma} x\,\left[ - \frac{1}{4}\,F^{\mu \nu}\,F_{\mu \nu} - \frac{1}{2\xi}\left(\partial_{\mu}A^{\mu}\right)^2 \right]\, ,
\label{chap1:rqed}
\eea
where $D_\mu = \partial_\mu + \I e A_\mu$ is the covariant derivative. For $d_e=d_\gamma =d$, Eq.~(\ref{chap1:rqed}) simply reduces to QED$_d$. In the case $d_e < d_\gamma$, this model is sometimes referred to in the literature
as the so-called reduced \cite{Gorbar:2001qt} or pseudo \cite{Marino:1992xi} or even very recently mixed-dimensional \cite{Hsiao:2017lch} QED; it is a very natural model describing low-dimensional quantum systems and 
certainly appeared in many other works, see, \eg, \cite{Dorey:1991kp,Kovner:1990zz,Kaplan:2009kr} and references therein. In the following, 
the notation reduced QED$_{d_\gamma,d_e}$ (or RQED$_{d_\gamma,d_e}$ or even QED$_{d_\gamma,d_e}$) will be used; the peculiar case of QED$_{4,3}$ 
describes graphene at its Lorentz invariant fixed point as was first pointed out in \cite{Teber:2012de}. In this respect, model II (\ref{chap1:model-inst}) corresponds to a non-relativistic reduced QED$_{d_\gamma,d_e}$ 
(NRRQED$_{d_\gamma,d_e}$). From the field theoretic point of view, the model of Eq.~(\ref{chap1:rqed}) (and similarly for the two previous ones) is characterized by an
effective free gauge-field action with fractional d'Alembertian.\footnote{The appearance of fractional d'Alembertian in Eq.~(\ref{chap1:rqed-int}) for $\veps_e>0$ implies that the reduced theory is non-local.
In the terminology of Gracey, see \cite{Gracey:2006jc}, the action of the reduced gauge field seems to be a ``localizable non-locality'' because it can be written as a finite number of local operators as in Eq.~(\ref{chap1:rqed}).
Notice also that the gauge fixing condition for the reduced gauge-field in (\ref{chap1:rqed-int}) appears to be non-local.}$^,\,$\footnote{Fractional d'Alembertians (or Laplacians) appear in the field of fractional calculus, see
\cite{samko1993fractional} for an extended monograph. In mathematics, there is a trick apparently due to Caffarelli and Silvestre \cite{Caffarelli07} which amounts to re-write the fractional field theory in $d$-dimensional space 
as a local theory in a $d+1$-dimensional space; in our frame, this is nothing else but simply going from (\ref{chap1:rqed-int}) back to (\ref{chap1:rqed})  in the peculiar case $\veps_e=1/2$. 
I thank M.~Rajabpour for pointing these references to me. See also \cite{Rajabpour:2011qr} for a nice account on the conformal invariance of (``localizable'') non-local field theories and \cite{Herzog:2017xha} for a recent
work on QED$_{4,3}$ and other ``mixed dimensional'' QFTs as boundary CFTs. See also \cite{Limtragool:2016gnl,LaNave:2017lwf,LaNave:2017nex} for recent references on non-local QFTs 
which make explicit use of the Caffarelli-Silvestre trick and applications to cuprates.}    
 The later can be derived from Eq.~(\ref{chap1:rqed}) by integrating out the gauge degrees of freedom
transverse to the $d_e$-dimensional manifold. Including fermions, the result reads:
\be
S =  \int \D^{d_e} x\, \Bigg [ \bar{\psi}_\sigma \I \bigg( \slashed{\partial} + \I e \tilde{\slashed{A}} \bigg)  \psi^\sigma 
- \frac{1}{4}\,\tilde{F}^{\mu \nu}\,\frac{(4\pi)^{\veps_e}}{\Gamma(1-\veps_e)\,[-\Box \,]^{\veps_e}}\,\tilde{F}_{\mu \nu} 
+ \frac{1}{2\tilde{\xi}}\,\tilde{A}^\mu \frac{(4\pi)^{\veps_e}\,\partial_{\mu} \partial_\nu}{\Gamma(1-\veps_e)\,[-\Box \,]^{\veps_e}}\,\tilde{A}^{\nu} \Bigg]\, ,
\label{chap1:rqed-int}
\ee
where we used the notation $\tilde{A}^\mu$ to emphasize the fact that it is a reduced gauge field (in $d_e$-dimensional space), $\veps_e = (d_\gamma-d_e)/2$ and $\tilde{\xi} = \veps_e + (1-\veps_e)\, \xi$, see
Chap.~\ref{chap3} for more.  Though {\it a priori} mainly of academic interest, the general motivation to consider reduced QED models is that relativistic invariance allows a rigorous and systematic study of interaction effects.
Indeed, all the powerful multi-loop machinery combined with the field-theoretic renormalization technique described in the beginning of this chapter, and originally developed in particle physics and statistical mechanics, 
can be rather straightforwardly applied to models such as Eq.~(\ref{chap1:rqed}). Interestingly, and as we will see in the following, the odd dimensionality of space-time together with the 
(related) presence of Feynman diagrams with non-integer indices brings a lot of novelties (as well as highly non-trivial additional complications) with respect to what is usually known from the study of $(3+1)$-dimensional theories. 
As will be shown in the following, the study of (\ref{chap1:rqed}) has reached two-loop accuracy for perturbative expansions (for photon and fermion self-energy with application to the computation of $\mathcal{C}_1^*$, the interaction 
correction coefficient at the fixed point) 
and next-to-leading order (NLO) in the $1/N$-expansion (within a Schwinger-Dyson approach to dynamical mass generation and computation of $\al_c^*$ and $N_c^*$). 
This may not seem very high, but all of these results are exact and constitute a robust base from which higher order corrections may be computed. 
Indeed, a full three-loop accuracy for photon, \eg, $\mathcal{C}_2^*$, and fermion self-energies seems to be reachable and, perhaps, next-to-next-to-leading order (NNLO) corrections to the dynamical mass. 
The study of the fixed point with the help of (\ref{chap1:rqed}) also offers a robust base from which the physics away from the fixed point 
(which is closer to the experimental situation) may be explored. As will be seen in the following, multi-loop techniques have been applied to (\ref{chap1:model-inst}) with two-loop accuracy for the photon self-energy; an application
to the optical conductivity of graphene helped clarifying the theoretical controversy surrounding the first order interaction correction coefficient, $\mathcal{C}_1$. 
A striking feature of the results obtained so far (in the period 2012-2017) is that there seems to be a
good {\it quantitative} agreement between the fixed point physics (relativistic limit with fully retarded interactions) and physics far away from it (non-relativistic limit with instantaneous interactions). 
This can be seen at the level of the first order interaction correction coefficient which reads in the two extreme limits (see Chap.~\ref{chap5} for more):
\be
\mathcal{C}_1 = \frac{19-6\pi}{12} \approx 0.013, \qquad \mathcal{C}_1^* = \frac{92-9\pi^2}{18\pi} \approx 0.056\, .
\label{chap1:C-res}
\ee
Similarly, the critical coupling constant in the two extreme limits reads (see Chap.~\ref{chap4} for more):
\be
0.833 < \al_c < 7.65, \qquad \al_c^* = 1.22 \, ,
\label{chap1:alc-res}
\ee
where the range of values in the non-relativistic limit has been taken from the literature on the subject (see Chap.~\ref{chap4}) and, in the ultra-relativistic limit, the result has NLO precision and is gauge-invariant.
Extensions to model (\ref{chap1:model-general}) remain challenging and are currently under study.

\section{Outline of the manuscript}

The outline of the manuscript is the following. 
In Chap.~\ref{chap2}, we shall give a basic presentation of some of the multi-loop techniques that will be used in subsequent chapters as well as some results for non-trivial
massless two-loop propagator-type integrals. This chapter is partly based on the following papers:
{\small
\begin{quote}
\cite{Kotikov:2013kcl} \textbf{A.\ V.~Kotikov and S.~Teber}, ``{\sl Note on an application of the method of uniqueness to reduced quantum electrodynamics}'', Phys.\ Rev.\ D {\bf 87} (2013) 087701,
\end{quote}}
{\small
\begin{quote}
\cite{Kotikov:2013eha} \textbf{A.\ V.~Kotikov and S.~Teber}, ``{\sl Two-loop fermion self-energy in reduced quantum electrodynamics and application to the ultra-relativistic limit of graphene}'',
Phys.\ Rev.\ D {\bf 89} (2014) 065038,
\end{quote}}
{\small
\begin{quote}
\cite{Teber:2016unz} \textbf{S.~Teber and A.\ V.~Kotikov}, ``{\sl The method of uniqueness and the optical conductivity of graphene: new application of a powerful technique for multi-loop calculations}'', Theoretical and Mathematical Physics {\bf 190} (2017) 446
[Teor.\ Mat.\ Fiz.\  {\bf 190} (2017) no.3,  519],
\end{quote}}
{\small
\begin{quote}
\cite{Kotikov:2016rgs} \textbf{A.\ V.~Kotikov and S.~Teber}, ``{\sl New results for a two-loop massless propagator-type Feynman diagram}'', arXiv:1611.07240 [hep-th].
\end{quote}}
%
%{\small 
%\begin{quote}
%\cite{Kotikov2017review} \textbf{A.\ V.~Kotikov and S.~Teber}, ``{\sl Review on multi-loop techniques for massless Feynman diagram calculations}''  (2017).
%\end{quote}}
%

\ms

Chap.~\ref{chap3} will then provide a brief presentation of the IR Lorentz-invariant fixed point on the basis of (\ref{chap1:model-general}) for graphene ($D_e=2$, $D_\gamma=3$). 
This will motivate the general study of reduced QED$_{d_\gamma,d_e}$. The perturbative structure of the later will be studied up to two loops for the fermion self-energy and
polarization operators with the help of the multi-loop methods of Chap.~\ref{chap2}. The results of this chapter will be used in the two subsequent ones.  They are based on:
{\small
\begin{quote}
\cite{Teber:2012de} \textbf{S.~Teber}, ``{\sl Electromagnetic current correlations in reduced quantum electrodynamics}'', Phys.\ Rev.\ D {\bf 86} (2012) 025005,
\end{quote}}
{\small
\begin{quote}
\cite{Kotikov:2013kcl} \textbf{A.\ V.~Kotikov and S.~Teber}, ``{\sl Note on an application of the method of uniqueness to reduced quantum electrodynamics}'', Phys.\ Rev.\ D {\bf 87} (2013) 087701,
\end{quote}}
{\small
\begin{quote}
\cite{Kotikov:2013eha} \textbf{A.\ V.~Kotikov and S.~Teber}, ``{\sl Two-loop fermion self-energy in reduced quantum electrodynamics and application to the ultra-relativistic limit of graphene}'', 
Phys.\ Rev.\ D {\bf 89} (2014) 065038,
\end{quote}}
{\small
\begin{quote}
\cite{Teber:2014hna} \textbf{S.~Teber}, ``{\sl Two-loop fermion self-energy and propagator in reduced QED$_{3,2}$}'', Phys.\ Rev.\ D {\bf 89} (2014) 067702.
\end{quote}}

\ms

In Chap.~\ref{chap4}, we study the critical properties of both the large-$N$ limit of QED$_3$ and reduced QED$_{4,3}$ focusing in particular on D$\chi$SB in these models and the possible dynamical generation of a mass. 
With the help of the methods of Chap.~\ref{chap2}, we will present a solution of the Schwinger-Dyson equations for the fermion self-energy of QED$_3$
and extract from it the critical fermion flavour number for QED$_3$. %and critical coupling constant for QED$_{4,3}$. 
We will show then that  the large-$N$ limit of QED$_3$ and reduced QED$_{4,3}$ are actually related via a mapping. The later will allow us to derive results for reduced QED$_{4,3}$ (in particular for the value
of the critical coupling constant) from those obtained for the large-$N$ limit of QED$_3$ without further complicated calculations. This chapter is based on:
{\small
\begin{quote}
\cite{Kotikov:2016wrb} \textbf{A.\ V.~Kotikov, V.~Shilin and S.~Teber}, ``{\sl Critical behaviour of ($2+1$)-dimensional QED: $1/N_f$-corrections in the Landau gauge}'',
Phys.\ Rev.\ D {\bf 94} (2016) 056009,
\end{quote}}
{\small
\begin{quote}
\cite{Kotikov:2016prf} \textbf{A.\ V.~Kotikov and S.~Teber}, ``{\sl Critical behaviour of ($2+1$)-dimensional QED: $1/N_f$-corrections in an arbitrary non-local gauge}'',
 Phys.\ Rev.\ D {\bf 94} (2016) no.11,  114011,
\end{quote}}
{\small
\begin{quote}
\cite{Kotikov:2016yrn} \textbf{A.\ V.~Kotikov and S.~Teber}, ``{\sl Critical behaviour of reduced QED$_{4,3}$ and dynamical fermion gap generation in graphene}'',
Phys.\ Rev.\ D {\bf 94} (2016) no.11,  114010.
\end{quote}}

\ms

In Chap.~\ref{chap5}, we depart from relativistic models and consider graphene in the non-relativistic limit ($v/c \ra 0$). The focus is on the computation of the interaction correction 
to the optical conductivity of graphene. Such a calculation has actually been performed in the ultra-relativistic limit ($v/c \ra 1$) in Chap.~\ref{chap3} on reduced QED.
We shall recall the result and go over in applying the multi-loop methods of Chap.~\ref{chap2} to the non-relativistic case. Broken Lorentz invariance
will be shown to increase the calculational complexity with semi-massive diagrams (fortunately only of the tadpole type for the optical conductivity) to be computed.
A short discussion comparing results obtained in the two extreme limits will be included. The results presented in this chapter are based on:
{\small
\begin{quote}
\cite{Teber:2014ita} \textbf{S.~Teber and A.\ V.~Kotikov}, ``{\sl Interaction corrections to the minimal conductivity of graphene via dimensional regularization}'', 
Europhys.\ Lett.\  {\bf 107} (2014) 57001,
\end{quote}}
{\small
\begin{quote}
\cite{Teber:2016unz} \textbf{S.~Teber and A.\ V.~Kotikov}, ``{\sl The method of uniqueness and the optical conductivity of graphene: new application of a powerful technique for multi-loop calculations}'', Theoretical and Mathematical Physics {\bf 190} (2017) 446
[Teor.\ Mat.\ Fiz.\  {\bf 190} (2017) no.3,  519].
\end{quote}}

\ms

A brief summary of the results as well as some future perspectives will be given in the Outlook \ref{chap6}. Some notations and conventions are given in Appendix~\ref{app:conv} and useful formulas in relation with the Gegenbauer polynomial technique are given in Appendix~\ref{app:gegen}.

\bs \bs

\textbf{Note:} after defending this habilitation, the following papers were published which are partly based on the results presented in this manuscript.
In relation with Chap.~\ref{chap2}:
{\small
\begin{quote}
\cite{Kotikov:2018wxe} \textbf{A.~V.~Kotikov and S.~Teber}, ``{\sl Multi-loop techniques for massless Feynman diagram calculations}'',
arXiv:1805.05109 [hep-th] (accepted for publication in Phys.\ Part.\ Nucl.).
\end{quote}}
In relation with Chap.~\ref{chap3}:
{\small
\begin{quote}
\cite{Teber:2018goo} \textbf{S.~Teber and A.~V.~Kotikov},
	``{\sl Field theoretic renormalization study of reduced quantum electrodynamics and applications to the ultrarelativistic limit of Dirac liquids}'',
Phys.\ Rev.\ D {\bf 97} (2018) no.7,  074004.
\end{quote}}
In relation with Chap.~\ref{chap5}:
{\small
\begin{quote}
\cite{Teber:2018qcn} \textbf{S.~Teber and A.~V.~Kotikov}, ``{\sl Field theoretic renormalization study of interaction corrections to the universal ac conductivity of graphene}'',
JHEP {\bf 1807} (2018) 082.
\end{quote}}

\cleardoublepage

%% file: Chapter2/methods.tex
\label{chap2}

This Chapter introduces the general notations, concepts and methods related to multi-loop calculations that will be used throughout the text. 
The focus will be on some of the methods that we consider as being the most efficient from the point of view of the analytic computation of ``master integrals''. 
These integrals may be viewed as basic building blocs at the {\it core} of multi-loop calculations. Their evaluation is therefore a fundamental
task prior to any automation. Their importance is witnessed by the recent appearance of ``Loopedia'' \cite{Bogner:2017xhp} which attempts at providing
a database for all known loop integrals. Many of such integrals related to four-dimensional models were known before the enormous developments mentioned in the Introduction
concerning particle physics, statistical mechanics and (supersymmetric) gauge field theories.
In other cases such as, \eg, in application to odd dimensional theories relevant to condensed matter physics systems that we are interested in, non-standard master integrals appear the systematic evaluation of which is more recent,
see, \eg, \cite{Kotikov:2013kcl,Kotikov:2013eha,Teber:2014ita,Teber:2016unz,Kotikov:2016wrb,Kotikov:2016prf,Kotikov:2016rgs}.
The main emphasis will then be on the standard rules of perturbation theory for massless Feynman diagrams as described in the pioneering work of Kazakov \cite{Kazakov:1984km}, see
also the beautiful lectures \cite{Kazakov:1984bw}. A brief (historically oriented) overview of some of the results obtained for the massless 2-loop propagator-type diagram is also provided.
In addition, renormalization methods, which give a meaning to the regularized integrals computed with the help of multi-loop techniques, will be reviewed at the end of this Chapter.

\section{Basics of Feynman diagrams}
\label{chap2:basics}
% only in one file to avoid multiple inclusions
%\nocite{*}

\begin{fmffile}{fmf-chap2a}

\subsection{Notations}
\label{chap2:notations}

We consider an Euclidean space-time of dimensionality $D$. Throughout this manuscript we shall use dimensional regularization (DR) which has many advantages over other regularizations 
such as, \eg, introducing a cut-off, as it preserves the symmetries (gauge invariance, Lorentz symmetry, chiral/flavour symmetry,...) of the model as well as (dimensional) power counting 
(it is a ``mass independent'' regularization), see the textbook \cite{Collins:1984xc} for a more complete account on DR, 
the classic \cite{Leibbrandt:1975dj} for a detailed early review together with \cite{Narison:1980ti} for an early review with applications to QED and QCD 
as well as the more recent \cite{Kilgore:2011ta} for a very instructive review on 
{\bf conventional dimensional regularization} that will be used throughout the text as well as other schemes.~\footnote{In this manuscript, 
we will not have to deal with, \eg, completely antisymmetric tensors, and other objects which pose difficulties to DR.}
 We shall therefore set, {\it e.g.}, $D=4-2\veps$ in the case of $(3+1)$-dimensional theories, where $\veps \ra 0$ is the
regularization parameter. Then, the infinitesimal volume element, {\it e.g.}, in momentum space, can be written as:~\footnote{Warning: it should be kept in mind that the correct prescription to introduce the
renormalization scale is via the replacement of (dimensionful) bare coupling constants by (dimensionless) renormalized one. For our purposes we shall introduce $\mu$ at the level of the 
measure of integration which, in our case, is strictly equivalent to the former prescription.}
\be
\D^4 k = \mu^{2\eps}\,\D^D k\, ,
\label{chap2:d4k}
\ee
where $\mu$ is the so-called renormalization scale in the minimal subtraction ($\text{MS}$) scheme 
which is related to the corresponding parameter $\overline{\mu}$ in the modified minimal subtraction ($\overline{\text{MS}}$) scheme with the help of:
\be
\overline{\mu}^{\,2} = 4\pi e^{-\gamma_E} \mu^2\, ,
\label{chap2:muMSbar}
\ee
where $\gamma_E$ is Euler's constant. 
Our main focus will be on {\bf massless Feynman diagrams of the propagator type} (the so called p-integrals \cite{Chetyrkin:1981qh}).~\footnote{Integrals with many legs 
are much more complicated, see the recent \cite{Chawdhry:2018awn} and references therein, and their
consideration is beyond the scope of this manuscript.}
Diagrams will be analyzed mainly in momentum space but position space will also be considered in some cases. 
In momentum space, Feynman diagrams are defined as integrals over dummy momentum variables or loop momenta.
The dependence of these integrals on the external momentum follows from dimensional analysis~\footnote{This statement is valid for relativistic QFTs. When Lorentz invariance is broken things
get more complicated. In this Chapter and the next two ones, only relativistic QFTs will be considered (unless explicitly indicated). The non-relativistic case
will be considered in Chapter \ref{chap5}.} and is power-like. The goal
of the calculations is then to compute the dimensionless coefficient function, $C_D$, associated with a given diagram (see below).
In some peculiar cases this function can be computed exactly. In most cases, only an approximate solution can be found. Often,
it takes the form of a Laurent series in $\veps$ and of great interest are the coefficients of negative power of $\veps$
which may be related to $\beta$-functions and anomalous dimensions of fields.

In this Chapter, for simplicity, we shall assume that all algebraic manipulations related to gamma matrices 
	such as, \eg, contraction of Lorentz indices, calculations of traces, etc... have been already performed.~\footnote{In some cases, \eg, for $n$-point functions, a
	tensorial reduction, the so-called Passarino-Veltman reduction scheme \cite{Passarino:1978jh}, see also \cite{Denner:1991kt} for a review, allows to express a tensor integral in terms 
	of scalar ones with tensor coefficients depending on the external
	kinematic variables and eventually the metric tensor. We assume that such a reduction has been performed and essentially focus on the computation of the scalar integrals. Notice that, at
	one-loop, the Passarino-Veltman reduction has been automated in FeynCalc \cite{Shtabovenko:2016sxi,Mertig:1990an}.}
The diagrams we shall be mainly interested in are therefore expressed in terms of scalar integrals; for completeness, and because they are of practical interest in concrete calculations,
we shall also consider diagrams with simple numerators such as traceless symmetric tensors. 
In reciprocal space, momentum is conserved at each vertex and integrations are over all internal
momenta. This has to be contrasted with calculations in real space where integrations are over the coordinates of all vertices.
In both spaces, the lines of such diagrams correspond to scalar propagators and are simple power laws. In momentum space, they take the form: $1/k^{2\al}$ where $\al$ is the so-called {\bf index} of the line.
A line with an arbitrary index $\al$ can be represented graphically as:
\be
\frac{1}{k^{2\al}}: \quad \parbox{10mm}{
  \begin{fmfgraph*}(10,10)
    \fmfleft{i}
    \fmfright{o}
    \fmf{plain,label=$\al$,l.s=left}{i,o}
  \end{fmfgraph*}
} \quad , \qquad
\frac{k^\mu}{k^{2\al}}: \quad \parbox{10mm}{
  \begin{fmfgraph*}(10,10)
    \fmfleft{i}
    \fmfright{o}
    \fmf{fermion,label=$\al^\mu$,l.s=left}{i,o}
  \end{fmfgraph*}
} \quad  \, , \qquad
\frac{k^{\mu_1 \cdots \mu_n}}{k^{2\al}}: \quad \parbox{10mm}{
  \begin{fmfgraph*}(10,10)
    \fmfleft{i}
    \fmfright{o}
    \fmf{fermion,label=$\al^{\mu_1 \cdots \mu_n}$,l.s=left}{i,o}
  \end{fmfgraph*}
} \quad  \, ,
\label{chap2:def:lines}
\ee
where the absence of arrow implies a scalar propagator while arrows indicate the presence of a non-trivial numerator the tensorial structure of which is displayed on the index for clarity.
In momentum space, ordinary lines have index $1$. The link with coordinate space is given by the Fourier transform:
\be
\int [\D^D p]\,\frac{e^{\I p x}}{[p^2]^\al} = \frac{2^{D-2\al}}{(4\pi)^{D/2}}\,\frac{a(\al)}{[\,x^2\,]^{D/2-\al}}, \qquad a(\al) = \frac{\Gamma(D/2-\al)}{\Gamma(\al)} \, ,
\label{chap2:FT-line}
\ee
where  $\al \not= D/2, D/2+1, ...$, $\Gamma(x)$ is Euler's gamma function and, in the following, we shall often use the notation 
\be
[\D^D p] = \frac{\D^D p}{(2 \pi)^D}\, .
\ee
From Eq.~(\ref{chap2:FT-line}), we see that, in coordinate space, ordinary lines have dimension $D/2-1$.
For a line of arbitrary index $\al$, the indices $\al$ and $D/2-\al$ are said to be {\bf dual} to each other in the sense of Fourier transform.
The inverse Fourier transform reads:
\be
\frac{1}{[\,p^2\,]^\al} = \frac{a(\al)}{\pi^{D/2}\,2^{2\al}}\,\int \D^D x \,\frac{e^{-\I p x}}{[\,x^2\,]^{D/2-\al}}\, .
\label{chap2:FT-inv-line}
\ee
In $p$-space, a zero index means that the corresponding line should shrink to a point while in 
$x$-space it means that the line has to be deleted. 

Let us then consider a general $L$-loop propagator-type Feynman diagram with $n$-internal lines.
Schematically, this diagram can be represented as:
\bea
\parbox{15mm}{
  \begin{fmfgraph*}(15,13)
      \fmfleft{i}
      \fmfright{o}
      \fmf{plain}{i,v}
      \fmfblob{.75w}{v}
      \fmf{plain}{v,o}
  \end{fmfgraph*}
}
\quad = \quad 
\frac{(p^2)^{\frac{LD}{2} - \sum_{i=1}^n\,\al_i}}{(4\pi)^{\frac{LD}{2}}}\,\text{C}_D(\vec{\al})\, ,
\label{chap2:def:p-integral}
\eea
where $\vec{\al} = (\al_1,\,\al_2,\, \cdots,\,\al_n)$ and the index of the diagram corresponds to the sum of the indices of its constituent lines: $\sum_{i=1}^n\,\al_i$. 
Eq.~(\ref{chap2:def:p-integral}) defines the dimensionless coefficient function, $\text{C}_D(\vec{\al})$, 
of the propagator-type diagram. This function depends on the indices, $\vec \al$, and the dimension of space-time, $D$. 
In the following, we shall extensively study the one-loop and two-loop p-integrals. At this point, we briefly consider vacuum-type diagrams (the so-called v-integrals).
A general multi-loop v-integral can be represented as:
\bea
\parbox{15mm}{
  \begin{fmfgraph*}(15,13)
      \fmfleft{v}
      \fmfblob{.75w}{v}
  \end{fmfgraph*}
} 
\eea
and can be obtained from the p-integral by putting the external momenta to zero. Such diagrams are therefore scaleless.

\subsection{Massless vacuum diagrams (and Mellin-Barnes transformation)}

In the frame of dimensional regularization, the one-loop massless vacuum diagram obeys the following identity \cite{Gorishnii:1984te}:
\be
\parbox{10mm}{
  \begin{fmfgraph*}(10,10)
      \fmfleft{i}
      \fmfright{o}
      \fmf{plain,left,label=$\al$,l.s=right}{i,o}
      \fmf{plain,left}{o,i}
  \end{fmfgraph*}
} 
\quad = \quad
\int \frac{\D^D k}{(k^2)^\al} \quad = \quad i \pi \, \Omega_D \, \delta(\al-D/2)\, ,
\label{chap2:one-loop-v-int}
\ee
where $\Omega_D = 2 \pi^{D/2} / \Gamma(D/2)$ is the surface of the unit hypersphere in $D$-dimensional Euclidean space-time.
One way to check the consistency of this relation is to consider the Mellin-Barnes transformation of a massive scalar propagator \cite{Boos:1990rg}:
\be
\frac{1}{(k^2+m^2)^\al} = \frac{1}{2\I \pi \Gamma(\al)}\,\int_{-\I \infty}^{\I \infty} \D s \,\frac{(m^2)^s}{(k^2)^{s+\al}}\,\Gamma(-s) \Gamma(\al+s)\, .
\label{chap2:Mellin-Barnes}
\ee
Interestingly, this transformation allows to express a massive propagator in terms of a contour integral involving a massless one.
As first noticed by Boos and Davydychev \cite{Boos:1990rg}, this suggests that techniques for computing massless Feynman diagrams may be of importance to compute massive ones. We shall come back on this in Chap.~\ref{chap5}.
For our present purpose, only the expression of the massive one-loop tadpole integral will be useful. It reads:
\be
\int \frac{[\D^D k]}{(k^2 + m^2)^\al} = \frac{(m^2)^{D/2-\al}}{(4\pi)^{D/2}}\,\frac{\Gamma(\al - D/2)}{\Gamma(\al)}\, ,
\label{chap2:massive-one-loop-tadpole}
\ee
where the mass dependence follows from dimensional analysis and the dimensionless factor $\Gamma(\al - D/2)/\Gamma(\al)$ corresponds to the coefficient function of this simple diagram. 
Eq.~(\ref{chap2:massive-one-loop-tadpole}) can be straightforwardly obtained using the standard parametric integration technique, see Sec.~\ref{chap2:sec:meth:param-int}. 
It can also be obtained using the Mellin-Barnes transformation (\ref{chap2:Mellin-Barnes}) upon assuming that
Eq.~(\ref{chap2:one-loop-v-int}) holds. This proves the consistency of Eq.~(\ref{chap2:one-loop-v-int}).

There is a connection between v-integrals and p-integrals, see Ref.~\cite{Gorishnii:1984te}. The latter may be derived by turning a p-integral
into a v-integral upon multiplying it by $(p^2)^{-\sigma}$ and integrating over $p$.~\footnote{This is also known as gluing, see Ref.~\cite{Chetyrkin:1981qh} and \cite{Baikov:2010hf} for a recent review.} 
From Eq.~(\ref{chap2:def:p-integral}), such a procedure yields:
\begin{subequations}
\bea
\parbox{15mm}{
  \begin{fmfgraph*}(15,13)
      \fmfleft{i}
      \fmfright{o}
      \fmf{plain}{i,v}
      \fmfblob{.75w}{v}
      \fmf{plain}{v,o}
      \fmf{plain,right,label=$\sigma$,l.s=right}{i,o}
  \end{fmfgraph*}
}
\quad & = & \quad
\frac{C_D(\vec{\al})}{(4\pi)^{\frac{LD}{2}}}\,\int \frac{\D^D p}{(p^2)^{\sigma + \sum_{i=1}^n\, \al_i - \frac{LD}{2}}} 
\nonum \\
& = & \quad 
\frac{i \pi}{(4\pi)^{\frac{LD}{2}}} \, \Omega_D \, C_D(\vec{\al})\,\delta(\sigma + \sum_{i=1}^n\, \al_i - \frac{LD}{2}) 
\\
& = & \quad 
\frac{i \pi}{(4\pi)^{\frac{LD}{2}}} \, \Omega_D \, X_D(\vec{\al},\sigma)\,\delta(\sigma + \sum_{i=1}^n\, \al_i - \frac{LD}{2}) \, .
\eea
\end{subequations}
Hence, the coefficient functions of the v-type diagram, $X_D$, and the p-type diagram, $C_D$, are related by:
\be
C_D(\vec{\al}) = X_D(\vec{\al},\sigma)|_{\sigma = \frac{LD}{2} - \sum_{i=1}^n\, \al_i}\, .
\label{chap2:v-vs-p}
\ee

As can be noticed from Eq.~(\ref{chap2:one-loop-v-int}), vacuum diagrams are rather ambiguously defined within dimensional regularization. Their scaleless
nature does not provide any clue of what their actual value might be a priori: it can be zero, infinity or even some finite number, 
see also the review \cite{Leibbrandt:1975dj}. This results from a subtle interplay between infrared and ultraviolet divergences of massless
diagrams. Following t'Hooft and Veltman, we shall often assume that the continuous dimension regularizes such highly divergent integrals to {\bf zero}. 
Similarly, whenever a diagram contains a scaleless subdiagram, {\it e.g.}, such as the massless tadpole diagram,
its value will be set to zero. In principle, however, care must be taken in the case where $D=2\al$. 
On the other hand, the consequence of Eq.~(\ref{chap2:one-loop-v-int}) is unambiguous for integrals over polynomials corresponding to the case where $\al<0$ in (\ref{chap2:one-loop-v-int});
within dimensional regularization such integrals vanish identically. Summarizing, in the following, we shall always assume that:~\footnote{In the case where $\al=D/2$ is encountered, it is
also possible to use the following trick: introduce a regulator $\delta \ra 0$ shifting the index $\al$, \eg, $\al \ra \al+\delta$. The limit $\delta \ra 0$ is taken at the end of the calculation. 
See Ref.~\cite{Kotikov:2013kcl} for an example.}
\be
\int \frac{\D^D k}{(k^2)^\al} = 0 \qquad (\al \not= D/2)\, .
\label{chap2:one-loop-v-int2}
\ee

\subsection{Massless one-loop propagator-type diagram}
\label{chap2:sec:one-loop-p-diag}

The one-loop (scalar) p-type massless integral is defined as:
\be
J(D,p,\al,\beta) = \int \frac{[\D^D k]}{k^{2\al} (p-k)^{2\beta}}\, ,
\label{chap2:def:one-loop-p-int}
\ee
where $p$ is the external momentum and $\al$ and $\beta$ are arbitrary indices.
In graphical form it is represented as:
\bea
J(D,p,\al,\beta) =
\qquad
\parbox{13mm}{
    \begin{fmfgraph*}(15,13)
      \fmfleft{i}
      \fmfright{o}
      \fmfleft{ve}
      \fmfright{vo}
      \fmffreeze
      \fmfforce{(-0.3w,0.5h)}{i}
      \fmfforce{(1.3w,0.5h)}{o}
      \fmfforce{(0w,0.5h)}{ve}
      \fmfforce{(1.0w,0.5h)}{vo}
      \fmffreeze
      \fmf{fermion,label=$p$}{i,ve}
      \fmf{plain,left=0.6,label=$\al$,l.d=0.1h}{ve,vo}
      \fmf{plain,left=0.6,label=$\beta$,l.d=0.05h}{vo,ve}
      \fmf{plain}{vo,o}
      \fmffreeze
      \fmfdot{ve,vo}
    \end{fmfgraph*}
} \quad \qquad .
\eea
In Eq.~(\ref{chap2:def:one-loop-p-int}), the momentum dependence is easily extracted from dimensional analysis which allows to write the diagram in the following form:
\be
J(D,p,\al,\beta) =  \frac{(p^2)^{D/2 - \al - \beta}}{(4\pi)^{D/2}}\,G(D,\al,\beta)\, ,
\label{chap2:def:one-loop-G-func}
\ee
where $G(D,\al,\beta)$ is the (dimensionless) coefficient function of the diagram. In graphical form, the latter is represented by a diagram similar to the one for $J(D,p,\al,\beta)$ but with amputated external legs:
\be
G(D,\al,\beta) = \text{C}_D\bigg[\int \frac{[\D^D k]}{k^{2\al} (p-k)^{2\beta}}\bigg] \quad =
\quad
  \parbox{13mm}{
  \begin{fmfgraph*}(15,13)
      \fmfleft{i}
      \fmfright{o}
      \fmf{plain,right=0.6,label=$\al$,l.d=0.1h}{i,o}
      \fmf{plain,right=0.6,label=$\beta$,l.d=0.05h}{o,i}
      \fmfdot{i,o}
  \end{fmfgraph*}
}\qquad  .
\ee
In the one-loop case, the so-called $G$-function is known exactly and reads:
\be
G(D,\al,\beta) = \frac{a(\al) a(\beta)}{a(\al + \beta -D/2)}, \qquad a(\al) = \frac{\Gamma(D/2 - \al)}{\Gamma(\al)}\, .
\label{chap2:one-loop-G}
\ee
All these results may be generalized to integrands with numerators. In particular:
\be
J^{\mu_1 \cdots  \mu_n}(D,p,\al,\beta) = \int [\D^D k] \frac{k^{\mu_1 \cdots  \mu_n}}{k^{2\al} (p-k)^{2\beta}} = \frac{(p^2)^{D/2 - \al - \beta}}{(4\pi)^{D/2}}\,p^{\mu_1 \cdots \mu_n}\,G^{(n,0)}(D,\al,\beta)\, ,
\label{chap2:def:one-loop-p-int+tp}
\ee
where $k^{\mu_1 \cdots  \mu_n}$ denotes the traceless symmetric tensor, see App.~\ref{app:gegen} for more, and
\be
G^{(n,0)}(D,\al,\beta) = \frac{a_n(\al) a_0(\beta)}{a_n(\al + \beta -D/2)}, \qquad a_n(\al) = \frac{\Gamma(n+D/2 - \al)}{\Gamma(\al)}\, ,
\label{chap2:one-loop-Gn}
\ee
where $G(D,\al,\beta) \equiv G^{(0,0)}(D,\al,\beta)$ and, for simplicity, the argument $D$ is also sometimes dropped unless an ambiguity may arise. Graphically, a one-loop p-type 
diagram with numerator can be represented as:
%
%\begin{subequations}
%\label{chap2:master-1loop-Gn-simple}
%\bea
%G^{(n,0)}(\al,\beta) &\equiv& G^{(n,0)}(D,\al,\beta)\, ,
%\\
%G(\al,\beta) &\equiv& G^{(0,0)}(D,\al,\beta), \qquad a_0(\al) = a(\al) \, .
%\label{chap2:int-G0}
%\eea
%\end{subequations}
%%
%Similarly:
%
\be
G^{(1,0)}(D,\al,\beta) = \text{C}_D\bigg[\int \frac{[\D^D k]~~k^\mu}{k^{2\al} (p-k)^{2\beta}}\bigg] \quad =
\quad
  \parbox{13mm}{
  \begin{fmfgraph*}(15,13)
      \fmfleft{i}
      \fmfright{o}
      \fmf{fermion,left=0.6,label=$\al^\mu$,l.d=0.2h}{i,o}
      \fmf{plain,left=0.6,label=$\beta$,l.d=0.05h}{o,i}
      \fmfdot{i,o}
  \end{fmfgraph*}
} \quad = \quad
-\quad 
  \parbox{13mm}{
  \begin{fmfgraph*}(15,13)
      \fmfleft{i}
      \fmfright{o}
      \fmf{fermion,right=0.6,label=$\al^\mu$,l.d=0.2h}{o,i}
      \fmf{plain,right=0.6,label=$\beta$,l.d=0.05h}{i,o}
      \fmfdot{i,o}
  \end{fmfgraph*}
} \quad\,~~ .
\ee
Whenever the integrand contains a scalar product, the corresponding lines are arrowed. As an example:
\begin{flalign}
\text{C}_D\bigg[\int \frac{[\D^D k]~~(k,p-k)}{k^{2\al} (p-k)^{2\beta}}\bigg] \quad = \quad
  \parbox{13mm}{
  \begin{fmfgraph*}(15,13)
      \fmfleft{i}
      \fmfright{o}
      \fmf{fermion,left=0.6,label=$\al^\mu$,l.d=0.15h}{i,o}
      \fmf{fermion,right=0.6,label=$\beta_\mu$,l.d=0.1h}{i,o}
      \fmfdot{i,o}
  \end{fmfgraph*}
} \quad = \quad
  \parbox{13mm}{
  \begin{fmfgraph*}(15,13)
      \fmfleft{i}
      \fmfright{o}
      \fmf{fermion,right=0.6,label=$\al^\mu$,l.d=0.15h}{o,i}
      \fmf{fermion,left=0.6,label=$\beta_\mu$,l.d=0.1h}{o,i}
      \fmfdot{i,o}
  \end{fmfgraph*}
} \quad = \quad - \quad
  \parbox{13mm}{
  \begin{fmfgraph*}(15,13)
      \fmfleft{i}
      \fmfright{o}
      \fmf{fermion,left=0.6,label=$\al^\mu$,l.d=0.15h}{i,o}
      \fmf{fermion,left=0.6,label=$\beta_\mu$,l.d=0.1h}{o,i}
      \fmfdot{i,o}
  \end{fmfgraph*}
} \qquad ,
\end{flalign}
\vskip 2mm

\ni where the notation $(k,p)=k^\mu p_\mu$ denotes the scalar product of the $D$-dimensional momenta $k$ and $p$.
When such a diagram is encountered, it can be evaluated as:
\vspace{2mm}
\begin{flalign}
  \parbox{13mm}{
  \begin{fmfgraph*}(15,13)
      \fmfleft{i}
      \fmfright{o}
      \fmf{fermion,left=0.6,label=$\al^\mu$,l.d=0.15h}{i,o}
      \fmf{fermion,right=0.6,label=$\beta_\mu$,l.d=0.1h}{i,o}
      \fmfdot{i,o}
  \end{fmfgraph*}
} \quad = \quad - \quad
  \parbox{13mm}{
  \begin{fmfgraph*}(15,13)
      \fmfleft{i}
      \fmfright{o}
      \fmf{plain,left=0.6,label=$\al-1$,l.d=0.15h}{i,o}
      \fmf{plain,right=0.6,label=$\beta$,l.d=0.1h}{i,o}
      \fmfdot{i,o}
  \end{fmfgraph*}
} \quad + \qquad
\parbox{13mm}{
    \begin{fmfgraph*}(15,13)
      \fmfleft{i}
      \fmfleft{ve}
      \fmfright{vo}
      \fmffreeze
      \fmfforce{(-0.3w,0.5h)}{i}
      \fmfforce{(0w,0.5h)}{ve}
      \fmfforce{(1.0w,0.5h)}{vo}
      \fmffreeze
      \fmf{fermion,label=$p_\mu$}{i,ve}
      \fmf{fermion,left=0.6,label=$\al^\mu$,l.d=0.15h}{ve,vo}
      \fmf{plain,left=0.6,label=$\beta$,l.d=0.1h}{vo,ve}
      \fmffreeze
      \fmfdot{ve,vo}
    \end{fmfgraph*}
} \qquad = \quad -G(\al-1,\beta) + G^{(1,0)}(\al,\beta)\, ,
\label{chap2:oneloop+arrows}
\end{flalign}
\vskip 2mm

\ni where, in the third diagram, the internal momentum is dotted with an external one. Hence:
\be
\int \frac{[\D^D k]~~(k,p-k)}{k^{2\al} (p-k)^{2\beta}} = \frac{(p^2)^{D/2+1-\al-\beta}}{(4\pi)^{D/2}}\,
\bigg( G^{(1,0)}(D,\al,\beta) - G(D,\al-1,\beta)\bigg)\, .
\label{chap2:oneloop+arrows:res}
\ee

The expression of $G(D,\al,\beta)$ given above may be derived by using parametric integration, see Sec.~\ref{chap2:sec:meth:param-int}.
Following Ref.~\cite{Kazakov:1984bw}, an alternative simple derivation consists in first going to $x$-space using Eq.~(\ref{chap2:FT-inv-line})
and then going back to $p$-space:
\bea
\int \frac{\D^D k}{k^{2\al} (p-k)^{2\beta}}
&=&  \frac{a(\al) a(\beta)}{\pi^D\,2^{2(\al + \beta)}}\,\int \frac{\D^D x\, \D^D y\,\D^D k\, e^{-\I k x-\I (p-k)y}}{[\,x^2\,]^{D/2-\al}\,[\,y^2\,]^{D/2-\beta}}
\nonum \\
&=&  \frac{a(\al) a(\beta)}{2^{2(\al + \beta)-D}}\,\int \frac{\D^D x\, e^{-\I p x}}{[\,x^2\,]^{D-\al-\beta}}
\nonum \\
&=&  \pi^{D/2}\,\frac{a(\al) a(\beta)}{a(\al + \beta - D/2)}\,\frac{1}{[\,p^2\,]^{\al + \beta - D/2}} \, .
\label{chap2:def:loop_calc}
\nonum
\eea
This leads to Eq.~(\ref{chap2:one-loop-G}).

An important property of the $G$-function is that it vanishes whenever one (or more) of the indices is zero or a negative {\it integer}\,:
\be
G(D,n,m) = 0, \qquad n \leq 0, \quad m \leq 0\, .
\ee
This property follows from the fact that a massless one-loop p-integral with a zero or negative integer index corresponds to a massless vacuum
diagram (possibly with a polynomial numerator) which vanishes according to Eqs.~(\ref{chap2:one-loop-v-int2}) 
(provided the case $D=2\al$ is not encountered). 

Finally, the $G$-function informs about the nature of the singularities in Eq.~(\ref{chap2:def:one-loop-p-int}). The latter can be either ultraviolet or infrared. In both cases, they will appear 
as $1/\veps$ poles in the expression of the $G$-function as dimensional regularization treats both types of singularities on an equal footing. 
In order to see that, let's note that from dimensional analysis Eq.~(\ref{chap2:def:one-loop-p-int}) has an ultraviolet singularity ($k \ra \infty$)
for $\al + \beta \leq D/2$; on the other hand, it has an infrared singularity ($k \ra 0$) for $\al \geq D/2$ and/or $\beta \geq D/2$. Then, from the explicit expression of $G(\al, \beta)$ 
in terms of $\Gamma$-functions:
\be
G(D,\al,\beta) = \frac{\Gamma(D/2-\al) \Gamma(D/2-\beta)\Gamma(\al + \beta -D/2)}{\Gamma(\al)\Gamma(\beta)\Gamma(D-\al-\beta)}\, ,
\label{chap2:G:IR+UV}
\ee
we see that poles coming from either one of the first two $\Gamma$-functions in the numerator are IR poles while those coming from the last $\Gamma$-function in the numerator are UV poles.

\section{Massless two-loop propagator-type diagram}

\subsection{Basic definition}

Central to this manuscript is the massless two-loop propagator-type diagram which is 
defined as:
\be
J(D,p,\al_1,\al_2,\al_3,\al_4,\al_5) = \int \frac{[\D^Dk][\D^Dq]}{(k-p)^{2\al_1}\,(q-p)^{2\al_2}\,q^{2\al_3}\,k^{2\al_4}\,(k-q)^{2\al_5}}\, ,
\label{chap2:def:two-loop-p-int}
\ee
where $p$ is the external momentum and the $\al_i$, $i=1-5$, are five arbitrary indices.
In graphical form, it is represented as:
\be
J(D,p,\al_1,\al_2,\al_3,\al_4,\al_5) 
\quad = \qquad
\parbox{16mm}{
    \begin{fmfgraph*}(16,14)
      \fmfleft{i}
      \fmfright{o}
      \fmfleft{ve}
      \fmfright{vo}
      \fmftop{vn}
      \fmftop{vs}
      \fmffreeze
      \fmfforce{(-0.3w,0.5h)}{i}
      \fmfforce{(1.3w,0.5h)}{o}
      \fmfforce{(0w,0.5h)}{ve}
      \fmfforce{(1.0w,0.5h)}{vo}
      \fmfforce{(.5w,0.95h)}{vn}
      \fmfforce{(.5w,0.05h)}{vs}
      \fmffreeze
      \fmf{fermion,label=$p$}{i,ve}
      \fmf{plain,left=0.8}{ve,vo}
      \fmf{phantom,left=0.7,label=$\al_1$,l.d=-0.1w}{ve,vn}
      \fmf{phantom,right=0.7,label=$\al_2$,l.d=-0.1w}{vo,vn}
      \fmf{plain,left=0.8}{vo,ve}
      \fmf{phantom,left=0.7,label=$\al_3$,l.d=-0.1w}{vo,vs}
      \fmf{phantom,right=0.7,label=$\al_4$,l.d=-0.1w}{ve,vs}
      \fmf{plain,label=$\al_5$,l.d=0.05w}{vs,vn}
      \fmf{plain}{vo,o}
      \fmffreeze
      \fmfdot{ve,vn,vo,vs}
    \end{fmfgraph*}
} \qquad \, .
\label{chap2:graph:J}
\ee
Similarly to the one-loop case, the momentum dependence of Eq.~(\ref{chap2:def:two-loop-p-int}) follows from dimensional analysis which allows to write this diagram in the form:
\be
J(D,p,\al_1,\al_2,\al_3,\al_4,\al_5) = \frac{(p^2)^{D-\sum_{i=1}^5 \al_i}}{(4\pi)^D}\, G(D,\al_1,\al_2,\al_3,\al_4,\al_5)\, ,
\label{chap2:def:two-loop-G-func}
\ee
where the (dimensionless) coefficient function of the diagram, $G(D,\{\al_i\})$, has been defined according to the general case Eq.~(\ref{chap2:def:p-integral}).
Graphically, the latter is represented as:
\be
G(D,\al_1,\al_2,\al_3,\al_4,\al_5) = \text{C}_D[J(D,p,\al_1,\al_2,\al_3,\al_4,\al_5)]
\quad = \quad
\parbox{16mm}{
    \begin{fmfgraph*}(16,14)
      \fmfleft{ve}
      \fmfright{vo}
      \fmftop{vn}
      \fmftop{vs}
      \fmffreeze
      \fmfforce{(0w,0.5h)}{ve}
      \fmfforce{(1.0w,0.5h)}{vo}
      \fmfforce{(.5w,0.95h)}{vn}
      \fmfforce{(.5w,0.05h)}{vs}
      \fmffreeze
      \fmf{plain,left=0.8}{ve,vo}
      \fmf{phantom,left=0.7,label=$\al_1$,l.d=-0.1w}{ve,vn}
      \fmf{phantom,right=0.7,label=$\al_2$,l.d=-0.1w}{vo,vn}
      \fmf{plain,left=0.8}{vo,ve}
      \fmf{phantom,left=0.7,label=$\al_3$,l.d=-0.1w}{vo,vs}
      \fmf{phantom,right=0.7,label=$\al_4$,l.d=-0.1w}{ve,vs}
      \fmf{plain,label=$\al_5$,l.d=0.05w}{vs,vn}
      \fmffreeze
      \fmfdot{ve,vn,vo,vs}
    \end{fmfgraph*}
}\quad \, .
\label{chap2:graph:G2loop}
\ee
\begin{figure}
  \begin{center}
    \begin{fmfgraph*}(20,17.5)
      \fmfleft{i}
      \fmfright{o}
      \fmfleft{ve}
      \fmfright{vo}
      \fmftop{vn}
      \fmftop{vs}
      \fmffreeze
      \fmfforce{(-0.3w,0.5h)}{i}
      \fmfforce{(1.3w,0.5h)}{o}
      \fmfforce{(0w,0.5h)}{ve}
      \fmfforce{(1.0w,0.5h)}{vo}
      \fmfforce{(.5w,0.95h)}{vn}
      \fmfforce{(.5w,0.05h)}{vs}
      \fmffreeze
      \fmf{fermion,label=$p$}{i,ve}
      \fmf{plain,left=0.8}{ve,vo}
      \fmf{phantom,left=0.7,label=$\al_1$,l.d=-0.01w}{ve,vn}
      \fmf{phantom,right=0.7,label=$\al_2$,l.d=-0.01w}{vo,vn}
      \fmf{plain,left=0.8}{vo,ve}
      \fmf{phantom,left=0.7,label=$\al_3$,l.d=-0.01w}{vo,vs}
      \fmf{phantom,right=0.7,label=$\al_4$,l.d=-0.01w}{ve,vs}
      \fmf{plain,label=$\al_5$,l.d=0.05w}{vs,vn}
      \fmf{plain}{vo,o}
      \fmffreeze
      \fmfdot{ve,vn,vo,vs}
    \end{fmfgraph*}
  \caption{\label{chap2:fig:J}
  Two-loop scalar massless propagator-type diagram.}
  \end{center}
\end{figure}
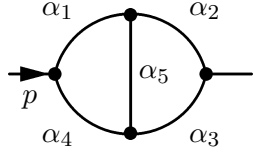

As in the one-loop case, whenever the integrand contains a scalar product, the corresponding lines are arrowed. As an example:
\be
\text{C}_D\bigg[\int \frac{[\D^Dk][\D^Dq]\,(k-q,q-p)}{(k-p)^{2\al_1}\,(q-p)^{2\al_2}\,q^{2\al_3}\,k^{2\al_4}\,(k-q)^{2\al_5}} \bigg]
\quad = \quad
\parbox{16mm}{
    \begin{fmfgraph*}(16,14)
      \fmfleft{ve}
      \fmfright{vo}
      \fmftop{vn}
      \fmftop{vs}
      \fmffreeze
      \fmfforce{(0w,0.5h)}{ve}
      \fmfforce{(1.0w,0.5h)}{vo}
      \fmfforce{(.5w,0.95h)}{vn}
      \fmfforce{(.5w,0.05h)}{vs}
      \fmffreeze
      \fmf{plain,left=0.8}{ve,vo}
      %\fmf{phantom_arrow,right=0.4}{ve,vs}
      \fmf{phantom_arrow,right=0.4}{vo,vn}
      %\fmf{phantom_arrow,right=0.4}{vn,ve}
      %\fmf{phantom_arrow,right=0.4}{vs,vo}
      \fmf{phantom,left=0.7,label=$\al_1$,l.d=-0.01w}{ve,vn}
      \fmf{phantom,right=0.7,label=$\al_{2 \mu}$,l.d=-0.01w}{vo,vn}
      \fmf{plain,left=0.8}{vo,ve}
      \fmf{phantom,left=0.7,label=$\al_3$,l.d=-0.01w}{vo,vs}
      \fmf{phantom,right=0.7,label=$\al_4$,l.d=-0.01w}{ve,vs}
      \fmf{fermion,label=$\al_5^\mu$,l.d=0.05w}{vs,vn}
      \fmffreeze
      \fmfdot{ve,vn,vo,vs}
    \end{fmfgraph*}
} \quad \, .
\label{chap2:2loop-arrows-1}
\ee
As another example, we may consider the case where the scalar product involves the external momentum:
\be
\text{C}_D\bigg[\int \frac{[\D^Dk][\D^Dq]\,(p,q)}{(k-p)^{2\al_1}\,(q-p)^{2\al_2}\,q^{2\al_3}\,k^{2\al_4}\,(k-q)^{2\al_5}} \bigg]
\qquad = \qquad
\parbox{16mm}{
    \begin{fmfgraph*}(16,14)
      \fmfleft{i}
      \fmfleft{ve}
      \fmfright{vo}
      \fmftop{vn}
      \fmftop{vs}
      \fmffreeze
      \fmfforce{(-0.3w,0.5h)}{i}
      \fmfforce{(0w,0.5h)}{ve}
      \fmfforce{(1.0w,0.5h)}{vo}
      \fmfforce{(.5w,0.95h)}{vn}
      \fmfforce{(.5w,0.05h)}{vs}
      \fmffreeze
      \fmf{fermion,label=$p^\mu$}{i,ve}
      \fmf{plain,left=0.8}{ve,vo}
%      \fmf{phantom_arrow,right=0.4}{ve,vs}
%      \fmf{phantom_arrow,right=0.4}{vo,vn}
%      \fmf{phantom_arrow,right=0.4}{vn,ve}
      \fmf{phantom_arrow,right=0.4}{vs,vo}
      \fmf{phantom,left=0.7,label=$\al_1$,l.d=-0.01w}{ve,vn}
      \fmf{phantom,right=0.7,label=$\al_2$,l.d=-0.01w}{vo,vn}
      \fmf{plain,left=0.8}{vo,ve}
      \fmf{phantom,left=0.7,label=$\al_{3 \mu}$,l.d=-0.01w}{vo,vs}
      \fmf{phantom,right=0.7,label=$\al_{4}$,l.d=-0.01w}{ve,vs}
      \fmf{plain,label=$\al_5$,l.d=0.05w}{vs,vn}
      \fmffreeze
      \fmfdot{ve,vn,vo,vs}
    \end{fmfgraph*}
}\quad \, .
\label{chap2:2loop-arrows-2}
\ee

Notice that that there is a single topological class of two-loop propagator-type diagrams represented in Fig.~\ref{chap2:fig:J}.
In some cases, two-loop diagrams may be reduced to products of one-loop diagrams (and hence products of $\Gamma$-functions) and are said to be {\bf primitively one-loop}, or {\bf recursively one-loop}, 
diagrams \cite{Chetyrkin:1981qh};  some examples of the later are given in Fig.~\ref{chap2:fig:primitive}. In general, however, no simple expression is known for the diagram of Fig.~\ref{chap2:fig:J}.

\subsection{Symmetries}
\label{chap2:sec:symmetries}

Symmetries are important because they yield identities among the coefficient functions with changed indices. 
We shall introduce a number of other such identities which follow from non-trivial transformations 
in the following sections. An appropriate use of identities among diagrams with different indices is central to multi-loop calculations 
and very often significantly reduces the amount of computations which has to be done. As a matter of fact,
these identities, when used in an appropriate way, may reduce a considerable number of two-loop diagrams to primitively
one-loop ones leaving only a small set of truly two-loop diagrams. Following Broadhurst, the set of irreducible integrals (at $1$, $2$ or higher loop order) 
is refereed to as the {\bf master integrals} \cite{Broadhurst:1987ei}. We shall come back on the computation of master integrals and their use all along this manuscript.

We start with some basic symmetries of the diagram which follow from changing the integration variables in Eq.~(\ref{chap2:def:two-loop-p-int}):
\begin{itemize}
\item the invariance of the integral upon changing $k \lra q$ implies that the diagram is invariant under the change $1 \leftrightarrow 2$
and $3 \leftrightarrow 4$. Geometrically, this can be viewed as an invariance of the diagram in a reflection 
through the plane perpendicular to the diagram and
containing the line of index $\al_5$. At the level of the coefficient function, Eq.~(\ref{chap2:def:two-loop-G-func}), this yields the following trivial identity:
\be
G(D,\al_1,\al_2,\al_3,\al_4,\al_5) = G(D,\al_2,\al_1,\al_4,\al_3,\al_5).
\label{chap2:transf1}
\ee
\item the invariance of the integral upon changing $k \lra k-p$ and $q \lra q-p$ implies that the diagram is 
invariant under the change $1 \leftrightarrow 4$ and $2 \leftrightarrow 3$. 
Geometrically, this can be viewed as an invariance of the diagram in a reflection through the plane perpendicular to the diagram and
to the line of index $\al_5$. This yields the following trivial identity among the coefficient functions with changed indices:
\be
G(D,\al_1,\al_2,\al_3,\al_4,\al_5) = G(D,\al_4,\al_3,\al_2,\al_1,\al_5).
\label{chap2:transf2}
\ee
\end{itemize}
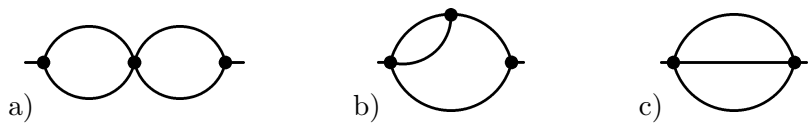
\begin{figure}
\begin{center}
a)
  \begin{fmfgraph*}(24,14)
    \fmfleft{i}
    \fmfright{o}
    \fmfleft{ve}
    \fmfright{vo}
    \fmftop{v}
    \fmffreeze
    \fmfforce{(-0.1w,0.5h)}{i}
    \fmfforce{(1.1w,0.5h)}{o}
    \fmfforce{(0w,0.5h)}{ve}
    \fmfforce{(1.0w,0.5h)}{vo}
    \fmfforce{(.5w,0.5h)}{v}
    \fmffreeze
    \fmf{plain}{i,ve}
    \fmf{plain,left=0.8}{ve,v}
    \fmf{plain,left=0.8}{v,ve}
    \fmf{plain,left=0.8}{v,vo}
    \fmf{plain,left=0.8}{vo,v}
    \fmf{plain}{vo,o}
    \fmffreeze
    \fmfdot{ve,v,vo}
  \end{fmfgraph*}
  \qquad \qquad
b)
  \begin{fmfgraph*}(16,14)
    \fmfleft{i}
    \fmfright{o}
    \fmfleft{ve}
    \fmfright{vo}
    \fmftop{v}
    \fmffreeze
    \fmfforce{(-0.1w,0.5h)}{i}
    \fmfforce{(1.1w,0.5h)}{o}
    \fmfforce{(0w,0.5h)}{ve}
    \fmfforce{(1.0w,0.5h)}{vo}
    \fmfforce{(.5w,0.95h)}{v}
    \fmffreeze
    \fmf{plain}{i,ve}
    \fmf{plain,left=0.8}{ve,vo}
    \fmf{plain,left=0.8}{vo,ve}
    \fmf{plain,left=0.5}{v,ve}
    \fmf{phantom,right=0.1}{vo,v}
    \fmf{plain}{vo,o}
    \fmffreeze
    \fmfdot{ve,v,vo}
  \end{fmfgraph*}
          \qquad \qquad
c)
  \begin{fmfgraph*}(16,14)
    \fmfleft{i}
    \fmfright{o}
    \fmfleft{ve}
    \fmfright{vo}
    \fmftop{v}
    \fmffreeze
    \fmfforce{(-0.1w,0.5h)}{i}
    \fmfforce{(1.1w,0.5h)}{o}
    \fmfforce{(0w,0.5h)}{ve}
    \fmfforce{(1.0w,0.5h)}{vo}
    \fmfforce{(.5w,0.95h)}{v}
    \fmffreeze
    \fmf{plain}{i,ve}
    \fmf{plain,left=0.8}{ve,vo}
    \fmf{plain,left=0.8}{vo,ve}
    \fmf{plain}{ve,vo}
    \fmf{phantom,right=0.1}{vo,v}
    \fmf{plain}{vo,o}
    \fmffreeze
    \fmfdot{ve,vo}
  \end{fmfgraph*}
  \caption{\label{chap2:fig:primitive}
  Two-loop primitive, or recursively one-loop, diagrams. Diagram a) corresponds to $\al_5=0$. Diagram b) corresponds to $\al_4=0$. Diagram c) corresponds to $\al_1=\al_3=0$.}
\end{center}
\end{figure}

It turns out that the 2-loop diagram is invariant under a very large group: $Z_2 \times S_6$ which has $1,440$ elements \cite{Broadhurst:1986bx,Barfoot:1987kg}.
Historically, some of the symmetry properties of the diagram were observed by the St Petersburg group \cite{Vasiliev:1981dg} (see also the textbook \cite{Vasil'evbook}). A few years later, 
the study of Gorishny and Isaev~\cite{Gorishnii:1984te} clearly revealed that the diagram has a full tetrahedral symmetry. 
To demonstrate this, they used the relation between the coefficient functions of the 2-loop p-integral 
and the related 3-loop v-integral. From Eq.~(\ref{chap2:v-vs-p}), 
this relation reads:\footnote{This is again the gluing (or glue and cut) method,
see Ref.~\cite{Chetyrkin:1981qh} and \cite{Baikov:2010hf} for a recent review.}
\be
G(D,\al_1,\al_2,\al_3,\al_4,\al_5) = X_D(\vec{\al},\sigma)|_{\sigma = D - \sum_{i=1}^5\, \al_i} \, .
\label{chap2:pv}
\ee
The 3-loop v-integral has the full tetrahedral symmetry. 
The coefficient function $X_D$ is therefore invariant under the permutations of the indices corresponding to the elements of the full
tetrahedral group (rotations and reflections) which is isomorphic to $S_4$, the symmetric group of 4 elements (the vertices of the tetrahedron).
This group has $4!=24$ elements and 3 generators;~\footnote{There are two possible sets of generators for the symmetric group $S_n$:
\begin{itemize}
\item $n-1$ generators formed by the transpositions $(1\,2),\,\,(2\,3),\,\, \cdots (n\,n-1)$,
\item 2 generators formed by a transposition $(1\,2)$ and an n-cycle: $(1\,2\,\cdots\,n)$.
\end{itemize}} 
for example: rotations around 2 axes passing by some vertex and the center
of the opposite side and one reflection. The generating elements can, for example, be chosen as: the rotation axes passing through the vertices $(\al_1,\,\al_5,\,\al_2)$
and $(\al_1,\,\sigma,\,\al_4)$ and a reflection corresponding to the permutation of the two vertices $(\al_1,\,\al_2,\,\al_5)$ and $(\al_3,\,\al_4,\,\al_5)$ (similar
to the transformation $(1 \leftrightarrow 4,\,\,2 \leftrightarrow 3)$).
For the v-integral, this yields:
\begin{subequations}
\label{chap2:S4generators}
\bea
X_D(\al_1,\al_2,\al_3,\al_4,\al_5,\sigma) & = & X_D(\al_4,\sigma,\al_2,\al_5,\al_1,\al_3)
%it can also be: X_D(\al_2,\al_5,\al_4,\sigma,\al_1,\al_3),
\label{chap2:S4generators-1} \\
& = & X_D(\al_4,\al_5,\al_2,\sigma,\al_3,\al_1),
\label{chap2:S4generators-2} \\
& = & X_D(\al_4,\al_3,\al_2,\al_1,\al_5,\sigma).
\label{chap2:S4generators-3}
\eea
\end{subequations}
Then, cutting one line of the v-integral to transform it into a p-integral yields:
\begin{subequations}
\label{chap2:transf3}
\bea
G(D,\al_1,\al_2,\al_3,\al_4,\al_5) & = & G(D,\al_2,\al_5,\al_4,\frac{3D}{2}-\sum_i \al_i,\al_1),
\label{chap2:transf3-1} \\
& = & G(D,\al_4,\al_5,\al_2,\frac{3D}{2}-\sum_i \al_i,\al_3),
\label{chap2:transf3-2} \\
& = & G(D,\al_4,\al_3,\al_2,\al_1,\al_5).
\label{chap2:transf3-3}
\eea
\end{subequations}
The last transformation in Eq.~(\ref{chap2:transf3}) is the same as the one of Eq.~(\ref{chap2:transf2}).
Combining the 3 transformations of Eq.~(\ref{chap2:transf3}) one generates all the possible transformations of indices
of the 2-loop diagram compatible with the tetrahedral symmetry.

Gorishny and Isaev further noted that the 2-loop diagram is invariant under another transformation which follows
from the uniqueness relation (see Sec.~\ref{chap2:sec:meth:uniqueness} for more on uniqueness):
\bea
&&C_D[J(D,p,\al_1,\al_2,\al_3,\al_4,\al_5)] 
\nonum \\
&&= a(\al_2)a(\al_3)a(\al_5)a(D-t_2)\,C_D[J(D,p,\al_1,\tilde{\al}_3,\tilde{\al}_2,\al_4,t_2-D/2)],
\label{chap2:transf4}
\eea
where $\tilde{\al} = D/2-\al$ and $t_2=\al_2+\al_3+\al_5$.
The existence of this additional transformation suggests that the symmetry group of the 2-loop diagram is larger than $S_4$.
As a matter of fact, instead of the 3 generators of Eq.~(\ref{chap2:S4generators}) we could have chosen one transposition and a $6$-cycle:
\begin{subequations}
\label{chap2:S6generators}
\bea
X_D(\al_1,\al_2,\al_3,\al_4,\al_5,\sigma) & = & X_D(\al_2,\al_1,\al_3,\al_4,\al_5,\sigma),
\label{chap2:S6generators-1} \\
& = & X_D(\sigma,\al_1,\al_2,\al_3,\al_4,\al_5).
\label{chap2:S6generators-2}
\eea
\end{subequations}
The transformations of Eq.~(\ref{chap2:S6generators}) generate the group $S_6$. Furthermore, the transformation of Eq.~(\ref{chap2:transf4})
involves dual indices $\al \rightarrow D/2-\al$ suggesting that there might be an additional $Z_2$ symmetry.
That the actual group is indeed $Z_2 \times S_6$, which has $2 \times 6!=1,440$ elements, was realized by Broadhurst~\cite{Broadhurst:1986bx} 
soon after the work of Gorishny and Isaev. 
The 3 generators of the $Z_2 \times S_6$ group are: a transposition and a $6$-cycle which generate $S_6$; 
and the dual transformation $\al_i \ra D/2-\al_i$ which generates $Z_2$.
Broadhurst~\cite{Broadhurst:1986bx} and then Barfoot and Broadhurst~\cite{Barfoot:1987kg} not only defined the symmetry 
group but also gave the set of 10 group invariants. 
As we shall review in the next paragraph, this allowed a more
accurate expansion of the massless 2-loop propagator-type diagram. 
%We shall not review the group theoretical results of 
%Broadhurst and Barfoot and Broadhurst. Let's simply note that they have defined the following 3 generators of the
%$Z_2 \times S_6$ group: a transposition and a $6$-cycle which generate $S_6$; and the dual transformation $\al_i \ra D/2-\al_i$ which generates $Z_2$.

\subsection{Brief historical overview of some results}
\label{Sec:chap2:hist}

The importance of the 2-loop massless propagator-type diagram of Fig.~\ref{chap2:fig:J} comes from the fact that it is a basic building block of many 
multi-loop calculations. As such, it has been extensively studied during the last three decades, see Ref.~\cite{Grozin:2012xi,Bierenbaum:2003ud} for historical reviews.
In this section, we shall review some results obtained for this diagram over the years keeping in mind our interest in three-dimensional theories. 

\begin{figure}
  \begin{center}
    a) \qquad
    \begin{fmfgraph*}(20,17.5)
      \fmfleft{i}
      \fmfright{o}
      \fmfleft{ve}
      \fmfright{vo}
      \fmftop{vn}
      \fmftop{vs}
      \fmffreeze
      \fmfforce{(-0.3w,0.5h)}{i}
      \fmfforce{(1.3w,0.5h)}{o}
      \fmfforce{(0w,0.5h)}{ve}
      \fmfforce{(1.0w,0.5h)}{vo}
      \fmfforce{(.5w,0.95h)}{vn}
      \fmfforce{(.5w,0.05h)}{vs}
      \fmffreeze
      \fmf{fermion,label=$p$}{i,ve}
      \fmf{plain,left=0.8}{ve,vo}
      \fmf{phantom,left=0.7,label=$\al$,l.d=-0.01w}{ve,vn}
      \fmf{phantom,right=0.7,label=$1$,l.d=-0.01w}{vo,vn}
      \fmf{plain,left=0.8}{vo,ve}
      \fmf{phantom,left=0.7,label=$1$,l.d=-0.01w}{vo,vs}
      \fmf{phantom,right=0.7,label=$\beta$,l.d=-0.01w}{ve,vs}
      \fmf{plain,label=$1$,l.d=0.05w}{vs,vn}
      \fmf{plain}{vo,o}
      \fmffreeze
      \fmfdot{ve,vn,vo,vs}
    \end{fmfgraph*}
    \qquad \qquad \qquad
    b) \qquad
    \begin{fmfgraph*}(20,17.5)
      \fmfleft{i}
      \fmfright{o}
      \fmfleft{ve}
      \fmfright{vo}
      \fmftop{vn}
      \fmftop{vs}
      \fmffreeze
      \fmfforce{(-0.3w,0.5h)}{i}
      \fmfforce{(1.3w,0.5h)}{o}
      \fmfforce{(0w,0.5h)}{ve}
      \fmfforce{(1.0w,0.5h)}{vo}
      \fmfforce{(.5w,0.95h)}{vn}
      \fmfforce{(.5w,0.05h)}{vs}
      \fmffreeze
      \fmf{fermion,label=$p$}{i,ve}
      \fmf{plain,left=0.8}{ve,vo}
      \fmf{phantom,left=0.7,label=$\al$,l.d=-0.01w}{ve,vn}
      \fmf{phantom,right=0.7,label=$\gamma$,l.d=-0.01w}{vo,vn}
      \fmf{plain,left=0.8}{vo,ve}
      \fmf{phantom,left=0.7,label=$1$,l.d=-0.01w}{vo,vs}
      \fmf{phantom,right=0.7,label=$1$,l.d=-0.01w}{ve,vs}
      \fmf{plain,label=$1$,l.d=0.05w}{vs,vn}
      \fmf{plain}{vo,o}
      \fmffreeze
      \fmfdot{ve,vn,vo,vs}
    \end{fmfgraph*}
  \caption{\label{chap2:fig:triangle}
  Some simple two-loop massless propagator diagrams which satisfy the rule of triangle and can be computed exactly using IBP identities.}
  \end{center}
\end{figure}
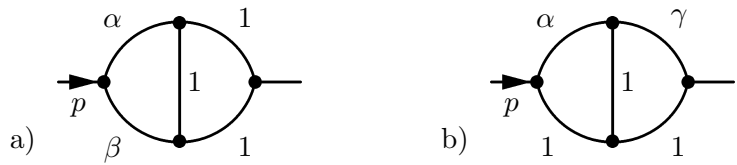

Generally speaking, when all indices are integers the 2-loop massless propagator-type diagram is easily computed. 
One of the earliest and most well-known result is the one due to Chetyrkin, Kataev and Tkachov \cite{Chetyrkin:1980pr} who found an exact expression
for $G(D,1,1,1,1,1)$ with the help of the Gegenbauer polynomial technique (see Sec.~\ref{chap2:sec:meth:Gegenbauer} for an introduction to the latter).
Soon after, the exact result for $G(D,1,1,1,1,1)$ could be re-obtained in a much more simple and straightforward way
by Vasiliev et al.\ \cite{Vasiliev:1981dg}, Tkachov \cite{Tkachov:1981wb} as well as Chetyrkin and Tkachov \cite{Chetyrkin:1981qh} using integration by parts (see Sec.~\ref{chap2:sec:meth:IBP} for more on IBP).
The result reads:
\be
G(D,1,1,1,1,1) =  \frac{2}{D-4}\, \left[ \,\, G(D,1,1) G(D,1,2) - G(D,1,1) G(D,2,3-D/2) \right ]\, ,
\label{chap2:G(1,1,1,1,1)}
\ee
where $G(D,\al,\beta)$ is the coefficient function of the one-loop p-type integral, Eq.~(\ref{chap2:one-loop-G}).  The fact that
$G(D,1,1,1,1,1)$ reduces to products of $G$-functions implies that this peculiar 2-loop diagram  is actually primitively one-loop.
It can therefore be expressed in terms of $\Gamma$-functions:
\be
G(4-2\veps,1,1,1,1,1) = \frac{\Gamma(1+2\veps)}{\veps^3\,(1-2\veps)}\, \left[  \frac{\Gamma^4(1-\veps)\Gamma^2(1+\veps)}{\Gamma^2(1-2\veps)\Gamma(1+2\veps)}
- \frac{\Gamma^3(1-\veps)}{\Gamma(1-3\veps)} \right ]\, ,
\label{chap2:G(1,1,1,1,1)-G4d}
\ee
where the case $D=4-2\veps$ has been considered. From dimensional analysis, $G(4-2\veps,1,1,1,1,1)$ is expected to be UV finite with no $1/\veps$ poles.
This can be checked explicitly by writing it in expanded form and keeping only the first few terms for short:~\footnote{Notice that in Eq.~(\ref{chap2:G(1,1,1,1,1)-exp4d}), we have used
a scheme in which $\gamma_E$ and $\zeta_2$ were subtracted from the remaining $\veps$-expansion. There are several other such schemes, \eg, the $G$-scheme \cite{Chetyrkin:1980pr}, see Eq.~(\ref{chap2:G-scheme}), where
a factor of $G^l(\veps)$ is extracted from every $l$-loop diagram and may be absorbed in a redefinition of the renormalization scale $\mu$. 
As they resum part of the $\veps$-expansion, these schemes appear to converge faster than the $\overline{\text{MS}}$ scheme.}
\begin{flalign}
G(4-2\veps,1,1,1,1,1) =
\frac{e^{-2\gamma_E \veps - \zeta_2 \veps^2}}{1-2\veps}\,
\left( 6 \zeta_3 + 9 \zeta_4 \veps + 42 \zeta_5 \veps^2 + (90 \zeta_6 - 46 \zeta_3^2)\,\veps^3 + \Ord(\veps^4) \right)\, ,
\label{chap2:G(1,1,1,1,1)-exp4d}
\end{flalign}
where the expansion formula for Gamma functions has been used:
\be
\Gamma(1+x) = \exp \left(-\gamma_E x + \sum_{n=2}^\infty\,\frac{(-1)^n}{n}\,\zeta_n\,x^n \right)\, ,
\label{chap2:gamma-expansion1}
\ee
and $\zeta_n$ is the Riemann zeta function which is defined as:
\be
\zeta_s = \zeta(s) = \sum_{n=1}^{\infty}\frac{1}{n^s} \qquad (\Re{s}>1)\, .
\label{chap2:def:zeta}
\ee
In Eq.~(\ref{chap2:G(1,1,1,1,1)-exp4d}), the $\Ord(1)$ term reduces to the celebrated $\zeta_3$ and all coefficients of higher order $\veps$ terms can be expressed in terms
of zeta functions.  The authors of \cite{Chetyrkin:1981qh} also noticed that the functions $G(D,\al,1,1,\beta,1)$ as well as
$G(D,\al,\gamma,1,1,1)$, where $\al$ and $\beta$ are arbitrary indices, can also be computed exactly using IBP.
This follows from the so-called {\bf rule of triangles} \cite{Chetyrkin:1981qh} whereby the 2-loop diagram can be exactly computed with the help of IBP
whenever three adjacent lines have integer indices, see Fig.~\ref{chap2:fig:triangle}.
For completeness, we give their expressions \cite{Chetyrkin:1981qh}:
\begin{subequations}
\label{chap2:IBP:other-applications}
\bea
G(D,\al,1,1,\beta,1)  &=&  \frac{G(D,1,1)}{D-\al-\beta-2}\,\bigg[ \al G(D,1+\al,\beta) - \al G(D,1+\al,2+\beta-D/2) \bigg .
\nonum \\
&&\bigg. + \beta G(D,\al,1+\beta) - \beta G(D,2+\al-D/2,1+\beta) \bigg]\, ,
\label{sec:IBP:application1}
\\
G(D,\al,\gamma,1,1,1) &=&  \frac{2(1+\al+\gamma-D/2)}{(D-3)(D-4)}\, \bigg[ \al \, G(D,1,1+\al)G(D,1,\al+\gamma+2-D/2) + \{ \gamma \leftrightarrow \al\}  \bigg]
\nonum \\
&&- \frac{2\al \gamma}{(D-3)(D-4)}\,G(D,1,1+\al)G(D,1,1+\gamma) \, .
\label{sec:IBP:application2}
\eea
\end{subequations}
These functions can be expanded for arbitrary indices $\al$, $\beta$ and $\gamma$ in $D=n-2\veps$ ($n \in \N^*$) with the help of:
\be
\Gamma(x+\veps) = \Gamma(x)\,\exp \Big[ \,\,\sum_{k=1}^{\infty} \psi^{(k-1)}(x) \frac{\veps^k}{k!} \,\,\Big]\, ,
\label{chap2:gamma-expansion2}
\ee
where $\psi^{(k)}$ is the polygamma function of order $k$:
\be
\psi(x) = \psi^{(0)}(x) = \frac{\Gamma'(x)}{\Gamma(x)}, \quad  \psi^{(k)}(x) = \frac{\D^k}{\D x^k} \,\psi(x)\, ,
\label{chap2:def:polygamma}
\ee
$\psi(x)$ being the digamma function and $\psi'(x) = \psi^{(1)}(x)$ the trigamma function.

On the other hand, for arbitrary (non-integer) values of all the indices, it's evaluation is highly non-trivial (even for the lowest order coefficients of the $\veps$-expansion) and
peculiar cases have to be considered, see, {\it e.g.},
Refs.~\cite{Chetyrkin:1980pr,Vasiliev:1981dg,Chetyrkin:1981qh,Kazakov:1983ns,Kazakov:1984km,Kazakov:1983pk,Broadhurst:1986bx,Barfoot:1987kg,Gracey:1992ew,Kivel:1993wq,Kotikov:1995cw,Broadhurst:1996ur,Broadhurst:1996yc,Broadhurst:2002gb,Bierenbaum:2003ud}. 
In the case where all the indices take the form:\footnote{Indices of this kind appear when considering multi-loop Feynman diagrams with integer indices. 
Upon integrating some of the subdiagrams using, \eg, IBP or another technique, the diagram transforms into a diagram with less loops but having lines where the integer indices are shifted by $\veps$ quantities.} 
$\al_i = n_i + a_i \veps$, where the $n_i$ are positive integers and the $a_i$ are non-negative real numbers,
the diagram is known only in the form of an $\veps$-expansion. For $n_i=1$ and $D=4-2\veps$, an expansion to order $\veps^3$ could be 
carried out in the seminal paper of Kazakov using the method of uniqueness \cite{Kazakov:1983ns}. Using functional identities among complicated diagrams, 
Kazakov further managed to extend his computation to order $\veps^4$ \cite{Kazakov:1984km,Kazakov:1983pk}. 
Then, using the symmetry arguments outlined in Sec.~\ref{chap2:sec:symmetries}, Broadhurst~\cite{Broadhurst:1986bx} and then
Barfoot and Broadhurst~\cite{Barfoot:1987kg} managed to extend the computation to order $\veps^5$ and then $\veps^6$, respectively. 
%As noted by Broadhurts: ``Our $\Ord(\veps^5)$ results subsums Kazakov's $\Ord(\veps^4)$ \cite{Kazakov84} result whose apparently enormous complexity is now reduced
%to the specification of merely 10 numbers''. In addition, Broadhurst found at $\Ord(\veps^5)$ a 
%term which takes the form of an alternating double sum:
%
%\be
%U_{6,2} = \sum_{n>m>0} \frac{(-1)^{n-m}}{n^6 m^2}\, .
%\ee
%
%Such a term cannot be reduced to a single (or products of) Riemann zeta function(s). 
%This suggests that, starting at $\Ord(\veps^5)$, some coefficients of the $\veps$-expansion may not be obtained by simple expansion of Gamma functions:
%
Subsequently, the orders $\veps^7$ and $\veps^8$ were computed by Broadhurst, Gracey and Kreimer~\cite{Broadhurst:1996ur}. After two decades of calculations, 
an expansion to order $\veps^9$ was achieved in Ref.~\cite{Broadhurst:2002gb}. A this point, number theoretical issues were raised, see, \eg, \cite{Broadhurst:2002gb} and references therein. It was known 
since the early days of quantum field theory that the Riemann zeta function, Eq.~(\ref{chap2:def:zeta}), often arises in Feynman diagram computations. More complicated diagrams
depending on an additional scale, such as massive propagator-type Feynman diagrams, were also known to be expressed in terms of the polylogarithm. This
function is a generalization of the zeta function and is defined as:
\be
\mbox{Li}_s(z) = \sum_{n=1}^{\infty} \frac{z^n}{n^s}\, ,
\label{chap2:def:polylog}
\ee
where $\mbox{Li}_1(z) = - \log(1-z)$ and $\mbox{Li}_s(1) = \zeta_s$ with $s \geq 2$. 
Other generalizations relevant to multi-loop massless Feynman diagrams include multiple zeta functions, also known as multiple zeta values (MZV) or Euler sums; they are defined as:
\be
\zeta_{s_1,s_2, \cdots, s_l} = \zeta(s_1,s_2, \cdots, s_l) = \sum_{n_1>n_2> \cdots n_l>0} \frac{1}{n_1^{s_1} n_2^{s_2} \cdots n_l^{s_l}} \qquad (s_1>1,~~s_2,\cdots,s_l \in \N)\, ,
\label{chap2:def:MZV}
\ee
where the integer $l$ is referred to as the length of the multiple zeta value and $s = \sum_{i=1}^l s_i$ to its weight. In general they reduce
to zeta functions, \eg, $\zeta(2,2) = (3/4)\zeta(4)$, $\zeta(3,2) = 3\zeta(2) \zeta(3) - (11/2) \zeta(5)$, etc... 
In some cases they are irreducible, \eg, at length $2$ and weight $8$, $\zeta(6,2)$ cannot be written in terms of zeta functions.
Multiple zeta functions are themselves a peculiar case of the multiple polylogarithm:
\be
\mbox{Li}_{s_1,s_2, \cdots, s_l}(z_1, z_2, \cdots, z_l) = \sum_{n_1>n_2> \cdots n_l>0} \frac{z_1^{n_1} z_2^{n_2} \cdots z_l^{n_l}}{n_1^{s_1} n_2^{s_2} \cdots n_l^{s_l}} \, ,
\label{chap2:def:mpl}
\ee
which appears in multi-scale Feynman diagrams. The important question that is then asked is whether Feynman diagrams (and in particular the coefficients of the $\veps$ expansion) 
can be fully expressed in terms of the zeta functions, Eq.~(\ref{chap2:def:zeta}), or generalization of these functions such as the multiple zeta functions and multiple polylogarithms?
In the general case, this question is very difficult to answer.

Progress appeared in 2003 with the work of Bierenbaum and Weinzierl \cite{Bierenbaum:2003ud} on the massless two-loop p-type diagram $G(D,\al_1,\al_2,\al_3,\al_4,\al_5)$ 
with $\al_i = n_i + a_i \veps$. For $D=2m-2\veps$ ($m \in \N$), they managed to automate its $\veps$-expansion; in principle, their computer assisted method allows an expansion to arbitrary order in $\veps$ 
the only restrictions arising from hardware constraints. Furthermore, they proved the following important theorem \cite{Bierenbaum:2003ud}:\footnote{Some generalization of the theorem appeared in 
Refs.~\cite{Brown:2008um,Brown:2009ta}.} 

\begin{theorem}[Bierenbaum and Weinzierl (2003)]
Multiple zeta values are sufficient for the Laurent expansion of the two-loop integral, $G(D,\al_1,\al_2,\al_3,\al_4,\al_5)$, with $D=2m-2\veps$ ($m \in \N$) if all powers of the propagators are of the form
$\al_i = n_i + a_i \veps$ where the $n_i$ are positive integers and the $a_i$ are non-negative real numbers.
\end{theorem}

Let's also note that, on a more ``phenomenological'' level, a principle of ``maximum weight'' or ``maximal transcendentality''
was discovered in Refs.~\cite{Broadhurst:1996ur,Fleischer:1998nb,Kotikov:2001sc,Kotikov:2002ab}.\footnote{I was informed by David Broadhurst that this principle
appears to be first due to John Gracey in an example of supersymmetric nonlinear sigma model preceding Ref.~\cite{Broadhurst:1996ur}.}
In very simple terms, this property can be checked at the level of the elementary example provided by Eq.~(\ref{chap2:G(1,1,1,1,1)-exp4d}). 
For this, let's assume that the transcendentality level of $\zeta_s$ is $s$ and the one of $\veps$ is $-1$. 
Then, we see that all displayed terms of the $\veps$-expansion have transcendentality $3$. Such an observation strongly constrains the coefficients of the series and, when judiciously used, sometimes allows
to reconstruct a whole series from the knowledge of the first few terms, see examples in \cite{Fleischer:1998nb}. 

It turns out that, unfortunately, none these powerful theorems and beautiful observations hold 
in the case of odd-dimensional field theories and related expansions in the vicinity of non-integer indices that we will be concerned with in the following.  
%These are precisely the cases that will be of interest to us in the following from the point of view of applications to reduced QED$_{4,3}$ and QED$_3$.
From the few existing studies, one may anticipate that odd-dimensional theories are ``transcendentally'' more complex that even dimensional ones %and that multiple zeta functions are insufficient 
\cite{Broadhurst:1996yc}. As a simple example, let's reconsider Eq.~(\ref{chap2:G(1,1,1,1,1)}) in $D=3-2\veps$. In this case:
\begin{flalign}
G(3-2\veps,1,1,1,1,1) = -2\,\frac{\Gamma(1+2\veps)}{1+2\veps}\,\left[  2\veps\,\frac{\Gamma^4(1/2-\veps)\Gamma^2(1/2+\veps)}{\Gamma^2(1-2\veps)\Gamma(1+2\veps)} 
+ \frac{1+6\veps}{\veps\,(1+2\veps)}\, \frac{\Gamma^3(1/2-\veps)}{\Gamma(1/2-3\veps)} \right ]\, ,
\label{chap2:G(1,1,1,1,1)-G3d}
\end{flalign}
where the first argument in $G$ emphasizes that we work near $3$ dimensions. From Eq.~(\ref{chap2:G(1,1,1,1,1)-G3d}), we first see the appearance of a $1/\veps$-pole. The latter is of IR nature
and arises from the last $G$-function in Eq.~(\ref{chap2:G(1,1,1,1,1)}). Moreover, as $G$-functions in $D$-dimensions are expressed in terms of $\Gamma$-functions with arguments depending on $D/2$, 
we see the appearance of half-integer indices in Eq.~(\ref{chap2:G(1,1,1,1,1)-G3d}) around which the expansion has to be made. With the help of Eq.~(\ref{chap2:gamma-expansion2}), this leads to:
\begin{flalign}
&G(3-2\veps,1,1,1,1,1) = -2\pi\,\frac{e^{-2\gamma_E \veps + \zeta_2 \veps^2}}{1+2\veps}\,
\left[ \frac{1}{\veps} + 4 + \big(2\pi^2 - 9\zeta_2 -8 \big) \veps 
\right .
\nonum \\
&\left . \qquad \qquad \qquad \qquad \qquad \qquad  
+ \frac{4}{3}\,\big( 12 - 44 \zeta_3 -27\zeta_2 + 6\pi^2\,\log 2 \big) \, \veps^2 + \Ord(\veps^3) \right]\, ,
\label{chap2:G(1,1,1,1,1)-exp3d}
\end{flalign}
where we have used the fact that $\psi(1/2)=-\gamma_E - 2 \log 2$, $\psi'(1/2) = 3 \zeta_2$ and $\psi''(1/2)=-14\zeta_3$.\footnote{We have: $\psi^{(n)}(1/2)=(-1)^{n+1}\,n!\,(2^{n+1}-1)\,\zeta_{n+1}$ for $n \in \N^*$.}
As anticipated, Eq.~(\ref{chap2:G(1,1,1,1,1)-exp3d}) has no property of ``maximal transcendentality''. Moreover, it features numbers such as $\pi$ and $\log 2$
which arise from derivatives of the $\Gamma$-function at half-integer argument. We shall encounter such numbers when studying reduced QED at two-loops.~\footnote{In principle, 
according to \cite{Broadhurst:1996yc}, even more complicated (beyond zeta values) numbers may appear, the first of which is:
\be
U(3,1) = \sum_{m>n>0} \frac{(-1)^{m+n}}{m^3 n} = \frac{1}{2}\,\zeta(4) -2 \left \{ \text{Li}_4(1/2) +\frac{1}{24}\,\log^2 2 \, \left( \log^2 2 - \pi^2 \right) \right \} \, ,
\ee
with a non-trivial $\text{Li}_4(1/2)$. I thank David Broadhurst for pointing out to me that such non-trivial numbers may appear in reduced QED$_{4,3}$ at higher loops.
}
%This can be readily seen at the level of the simple examples given above and will appear again in the following. In $D$ dimensions, Eqs.~(\ref{chap2:G(1,1,1,1,1),chap2:IBP:other-applications}) 
%express the diagrams in terms of $\Gamma$-functions of arguments depending on $D/2$. The $\veps$-expansion leads, \eg, with the help of Eq.~(\ref{chap2:gamma-expansion2}), 
%to coefficients which may be expressed in terms of $\psi$-functions. 
%If $D=4$, the latter have integer arguments such as, \eg, $\psi(1) = -\gamma_E$  where $\gamma_E$ does not appear in the $\overline{\text{MS}}$, Eq.~(\ref{chap2:muMSbar}).
%On the other hand, if $D=3$, the $\psi$-functions have half-integer arguments such as, \eg, $\psi(1/2) = -\gamma_E - 2 \log 2$ where again $\gamma_E$ may be cancelled but the $\log 2$ remains.
%Upon going further along this ``transcendentality'' line, let's also note that  

\begin{figure}
  \begin{center}
    a) \qquad
    \begin{fmfgraph*}(20,17.5)
      \fmfleft{i}
      \fmfright{o}
      \fmfleft{ve}
      \fmfright{vo}
      \fmftop{vn}
      \fmftop{vs}
      \fmffreeze
      \fmfforce{(-0.3w,0.5h)}{i}
      \fmfforce{(1.3w,0.5h)}{o}
      \fmfforce{(0w,0.5h)}{ve}
      \fmfforce{(1.0w,0.5h)}{vo}
      \fmfforce{(.5w,0.95h)}{vn}
      \fmfforce{(.5w,0.05h)}{vs}
      \fmffreeze
      \fmf{fermion,label=$p$}{i,ve}
      \fmf{plain,left=0.8}{ve,vo}
      \fmf{phantom,left=0.7,label=$1$,l.d=-0.01w}{ve,vn}
      \fmf{phantom,right=0.7,label=$1$,l.d=-0.01w}{vo,vn}
      \fmf{plain,left=0.8}{vo,ve}
      \fmf{phantom,left=0.7,label=$1$,l.d=-0.01w}{vo,vs}
      \fmf{phantom,right=0.7,label=$1$,l.d=-0.01w}{ve,vs}
      \fmf{plain,label=$\al$,l.d=0.05w}{vs,vn}
      \fmf{plain}{vo,o}
      \fmffreeze
      \fmfdot{ve,vn,vo,vs}
    \end{fmfgraph*}
    \qquad \qquad \qquad
    b) \qquad
    \begin{fmfgraph*}(20,17.5)
      \fmfleft{i}
      \fmfright{o}
      \fmfleft{ve}
      \fmfright{vo}
      \fmftop{vn}
      \fmftop{vs}
      \fmffreeze
      \fmfforce{(-0.3w,0.5h)}{i}
      \fmfforce{(1.3w,0.5h)}{o}
      \fmfforce{(0w,0.5h)}{ve}
      \fmfforce{(1.0w,0.5h)}{vo}
      \fmfforce{(.5w,0.95h)}{vn}
      \fmfforce{(.5w,0.05h)}{vs}
      \fmffreeze
      \fmf{fermion,label=$p$}{i,ve}
      \fmf{plain,left=0.8}{ve,vo}
      \fmf{phantom,left=0.7,label=$\al$,l.d=-0.01w}{ve,vn}
      \fmf{phantom,right=0.7,label=$1$,l.d=-0.01w}{vo,vn}
      \fmf{plain,left=0.8}{vo,ve}
      \fmf{phantom,left=0.7,label=$\beta$,l.d=-0.01w}{vo,vs}
      \fmf{phantom,right=0.7,label=$1$,l.d=-0.01w}{ve,vs}
      \fmf{plain,label=$1$,l.d=0.05w}{vs,vn}
      \fmf{plain}{vo,o}
      \fmffreeze
      \fmfdot{ve,vn,vo,vs}
    \end{fmfgraph*}
  \caption{\label{chap2:fig:beyondtriangle}
   Examples of two-loop massless p-type diagrams which are beyond IBP identities and uniqueness ($\al$ and $\beta$ are arbitrary indices).}
%  Diagram b) is part of the more general class of diagram with 2 adjacent lines having indecies equal to 1, see Fig.~\ref{sec:mou:fig:beyondtriangle+}.}
  \end{center}
\end{figure}
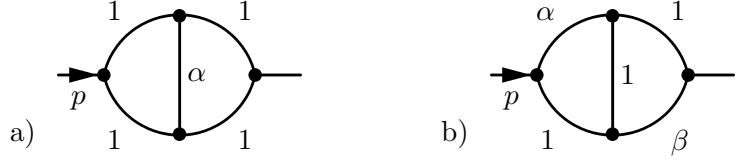

Pursuing with our historical overview, there are some specific cases where an exact evaluation of the diagram can be found. 
One of the simplest non-trivial, {\it i.e.}, beyond the rules of triangle see Fig.~\ref{chap2:fig:beyondtriangle},
diagram which may be considered is the one with an arbitrary index on the central line:
\begin{flalign}
G(D,1,1,1,1,\al) \quad = \quad
\parbox{16mm}{
    \begin{fmfgraph*}(16,14)
      \fmfleft{ve}
      \fmfright{vo}
      \fmftop{vn}
      \fmftop{vs}
      \fmffreeze
      \fmfforce{(0w,0.5h)}{ve}
      \fmfforce{(1.0w,0.5h)}{vo}
      \fmfforce{(.5w,0.95h)}{vn}
      \fmfforce{(.5w,0.05h)}{vs}
      \fmffreeze
      \fmf{plain,left=0.8}{ve,vo}
      \fmf{phantom,left=0.7,label=$1$,l.d=-0.01w}{ve,vn}
      \fmf{phantom,right=0.7,label=$1$,l.d=-0.01w}{vo,vn}
      \fmf{plain,left=0.8}{vo,ve}
      \fmf{phantom,left=0.7,label=$1$,l.d=-0.01w}{vo,vs}
      \fmf{phantom,right=0.7,label=$1$,l.d=-0.01w}{ve,vs}
      \fmf{plain,label=$\al$,l.d=0.05w}{vs,vn}
      \fmffreeze
      \fmfdot{ve,vn,vo,vs}
    \end{fmfgraph*}
}\quad \, .
\label{chap2:def:I(al)}
\end{flalign}
\vskip 2mm

\ni Early calculations by Vasil'ev, Pis'mak and Khonkonen~\cite{Vasiliev:1981dg} focused on the case where the index $\al$ is related to the dimensionality of the system as follows:
$\al = \lambda = D/2 - 1$. Using the method of uniqueness in real space, they have shown that~\cite{Vasiliev:1981dg} (see also discussions in Refs.~\cite{Vasiliev:1992wr,Kivel:1993wq}):
\be
G(D=2\lambda+2,1,1,1,1,\lambda) = 3\,\frac{\Gamma(\lambda)\Gamma(1-\lambda)}{\Gamma(2\lambda)} \,\Big[ \psi'(\lambda) - \psi'(1) \Big] \, ,
\label{chap2:res:I(lambda)}
\ee
where $\psi'(x)$ is the trigamma function. Notice that this result has been recently recovered, hopefully in a simpler way, in Ref.~\cite{Kotikov:2013kcl} using the method of uniqueness
in momentum space. For an arbitrary index $\al$, the diagram is beyond IBP as well as uniqueness. A one-fold series representation of Eq.~(\ref{chap2:def:I(al)}) has first been given by Kazakov in Ref.~\cite{Kazakov:1983pk}.
His expression reads:
%
%\begin{flalign}
%&G(D,1,1,1,1,\al) =
%-2\, \frac{\Gamma^2(1-\veps)\Gamma(\veps)\Gamma(-\veps-\al) \Gamma(\al+2\veps)}{\Gamma(2-2\veps)} \Biggl[
%\frac{1}{\Gamma(1+\al)\Gamma(1-3\veps-\al)}
%%\times
%\label{chap2:res:I(al):Kazakov}\\
%&\qquad \times
%%\left[ \frac{1}{\Gamma(1+\al)\Gamma(1-3\veps-\al)}\,
%\sum_{n=1}^{\infty}\,(-1)^n \frac{\Gamma(n+1-2\veps)}{\Gamma(n+\veps)}\,\left(\frac{1}{n+\al+\veps} + \frac{1}{n-\al-2\veps}\right)
%+\cos [\pi \veps] \Biggr]\, ,
%\nonum
%\end{flalign}
%
%
\begin{flalign}
&G(2\lambda+2,1,1,1,1,\al) =
-2\, \frac{\Gamma^2(\lambda)\Gamma(1-\lambda)\Gamma(\lambda-\al) \Gamma(1-2\lambda+\al)}{\Gamma(2\lambda)} \Biggl[
\frac{1}{\Gamma(\al)\Gamma(3\lambda-\al-1)}
%\times
\nonum\\
&\qquad \times
%\left[ \frac{1}{\Gamma(1+\al)\Gamma(1-3\veps-\al)}\,
\sum_{n=1}^{\infty}\,(-1)^n \frac{\Gamma(n+2\lambda-1)}{\Gamma(n+1-\lambda)}\,\left(\frac{1}{n-\lambda+\al} + \frac{1}{n-1+2\lambda-\al}\right)
-\cos [\pi \lambda] \Biggr]\, ,
\label{chap2:res:I(al):Kazakov}
\end{flalign}
where the one-fold series can be represented as a combination of two ${}_3F_2$-hypergeometric functions of argument $-1$.

Later, a whole class of complicated diagrams where two adjacent indices are integers and the three others are arbitrary, see Fig.~\ref{chap2:fig:beyondtriangle+},
could be computed exactly by Kotikov~\cite{Kotikov:1995cw} on the basis of a new development of the Gegenbauer polynomial technique.
For this class of diagrams, similar results have been obtained in Ref.~\cite{Broadhurst:1996ur} using an Ansatz to solve the
recurrence relations arising from IBP for the 2-loop diagram.
All these results are expressed in terms of (combinations of) generalized hypergeometric functions, ${}_3F_2$ with argument $1$.
As a matter of fact, from \cite{Kotikov:1995cw}, the diagram of Eq.~(\ref{chap2:def:I(al)}) could be expressed as:
\begin{flalign}
&G(2\lambda+2,1,1,1,1,\al) =
-2\, \Gamma(\lambda)\Gamma(\lambda-\al) \Gamma(1-2\lambda+\al) \times
\label{chap2:res:I(al):Kotikov}\\
&\qquad \times \left [ \frac{\Gamma(\lambda)}{\Gamma(2\lambda)\Gamma(3\lambda-\al-1)}\,
\sum_{n=0}^{\infty}\,\frac{\Gamma(n+2\lambda)\Gamma(n+1)}{n!\,\Gamma(n+1+\al)}\,\frac{1}{n+1-\lambda+\al}
+\frac{\pi \cot \pi (2\lambda-\al)}{\Gamma(2\lambda)} \right ]\, .
\nonum
\end{flalign}
Notice that the equality of the two representations (\ref{chap2:res:I(al):Kazakov}) and (\ref{chap2:res:I(al):Kotikov}) was proven only recently \cite{Kotikov:2016rgs}, see also 
this reference for other representations of this diagram. This proof provides the following relation, conjectured in Ref.~\cite{Kotikov:1995cw},  
between two ${}_3F_2$-hypergeometric functions of argument $-1$ and a single  ${}_3F_2$-hypergeometric function of argument $1$:
%
%\bea
\begin{flalign}
&{}_3F_2(2A,B,1;B+1,2-A;-1)+ \frac{B}{1+A-B} \, {}_3F_2(2A,1+A-B,1;2+A-B,2-A;-1) \nonumber \\
&= B \cdot \frac{\Gamma(2-A)\Gamma(B+A-1)\Gamma(B-A)\Gamma(1+A-B)}{\Gamma(2A)\Gamma(1+B-2A)} - \frac{1-A}{B+A-1} \, {}_3F_2(2A,B,1;B+1,A+B;1) \, ,
\label{chap2:id:F32}
\end{flalign}
%\eea
%
where $A$, $B$ and $C$ are arbitrary. As far as we know, such a relation does not appear in standard textbooks.

On the basis of Ref.~\cite{Kotikov:1995cw}, the case of:
\be
G(D,\al,1,\beta,1,1) \quad = \quad
\parbox{16mm}{
    \begin{fmfgraph*}(16,14)
      \fmfleft{ve}
      \fmfright{vo}
      \fmftop{vn}
      \fmftop{vs}
      \fmffreeze
      \fmfforce{(0w,0.5h)}{ve}
      \fmfforce{(1.0w,0.5h)}{vo}
      \fmfforce{(.5w,0.95h)}{vn}
      \fmfforce{(.5w,0.05h)}{vs}
      \fmffreeze
      \fmf{plain,left=0.8}{ve,vo}
      \fmf{phantom,left=0.7,label=$\al$,l.d=-0.01w}{ve,vn}
      \fmf{phantom,right=0.7,label=$1$,l.d=-0.01w}{vo,vn}
      \fmf{plain,left=0.8}{vo,ve}
      \fmf{phantom,left=0.7,label=$\beta$,l.d=-0.01w}{vo,vs}
      \fmf{phantom,right=0.7,label=$1$,l.d=-0.01w}{ve,vs}
      \fmf{plain,label=$1$,l.d=0.05w}{vs,vn}
      \fmffreeze
      \fmfdot{ve,vn,vo,vs}
    \end{fmfgraph*}
}\quad \, ,
\label{chap2:def:I(al,beta)}
\ee
\vskip 2mm

\ni could also be computed explicitly \cite{Kotikov:2013eha}. It was shown to be expressed in terms of two generalized hypergeometric functions, ${}_3F_2$ with argument $1$~\cite{Kotikov:2013eha}. 
The result reads \cite{Kotikov:2013eha}:
\be
G(D,\al,1,\beta,1,1) =
%\frac{1}{(4\pi)^D}
%\frac{1}{\Gamma(\lambda)}
\frac{1}{\tilde{\al}-1} \frac{1}{1-\tilde{\beta}} \,
\frac{\Gamma(\tilde{\al})\Gamma(\tilde{\beta})
\Gamma(3-\tilde{\al}-\tilde{\beta})}{\Gamma(\al)
\Gamma(\lambda-2+\tilde{\al}+\tilde{\beta})}
\frac{\Gamma(\lambda)}{\Gamma(2\lambda)} \, I(\tilde{\al},\tilde{\beta}) \, ,
\label{chap2:res:Gab}
\ee
where $\tilde{\al} = D/2-\al$, $\lambda=D/2-1$, $D=4-2\veps$ and the function $I(\tilde{\al},\tilde{\beta})$ can be written in four different forms; for example:
\bea
&&I(\tilde{\al},\tilde{\beta}) =
%\biggl\{
\frac{\Gamma(1+\lambda-\tilde{\al})}{
%\Gamma(2\lambda)\Gamma(\lambda)
\Gamma(3-\tilde{\al}-\tilde{\beta})}
\frac{\pi \sin[\pi(\tilde{\beta}- \tilde{\al}+\lambda)]}{\sin[\pi (\lambda-1+\tilde{\beta})]\sin[\pi \tilde{\al}]}
%\nonumber \\&&
+ \sum_{n=0}^{\infty} \frac{\Gamma(n+2\lambda)}{n!(n+\lambda+\tilde{\al}-1)}
\label{chap2:res:Gab:I1} \\
&& \times \biggl( \frac{\Gamma(n+1)}{\Gamma(n+2+\lambda-\tilde{\beta})}
%\nonumber \\&&
- \frac{\Gamma(n-2+\lambda +\tilde{\al}+\tilde{\beta})\Gamma(2-\tilde{\beta}) \Gamma(\lambda)
}{\Gamma(n-1+2\lambda+\tilde{\al})\Gamma(3-\tilde{\al}-\tilde{\beta})\Gamma(\lambda+\tilde{\al}-1)}
\frac{\sin[\pi(\tilde{\beta}+\lambda-1)]}{\sin[\pi \tilde{\al}]}
\biggl)
%\biggr\}
\, ,
\nonum
\eea
\bea
&&I(\tilde{\al},\tilde{\beta}) =
%\biggl\{
\frac{\Gamma(1+\lambda - \tilde{\al})}{
%\Gamma(2\lambda)\Gamma(\lambda)
\Gamma(3-\tilde{\al}-\tilde{\beta})}
\frac{\pi \sin[\pi \tilde{\al}]}{\sin[\pi (\lambda-1+\tilde{\beta})] \sin[\pi (\tilde{\al}+\tilde{\beta}+\lambda-1) ]}
\label{chap2:res:Gab:I2} \\&&
+ \sum_{n=0}^{\infty} \frac{\Gamma(n+2\lambda)}{n!
%(n+\lambda+\tilde{\al}-1)
}
%\nonumber \\&& \times
\biggl(\frac{1}{n+\lambda+\tilde{\al}-1}
\frac{\Gamma(n+1)}{\Gamma(n+2+\lambda-\tilde{\beta})}
\nonumber \\&&
+ \frac{1}{n+\lambda+1-\tilde{\al}}
\frac{\Gamma(n+2-\tilde{\al})\Gamma(2-\tilde{\beta})\Gamma(\lambda)
}{\Gamma(n+3+\lambda-\tilde{\al}-\tilde{\beta})\Gamma(3-\tilde{\al}-\tilde{\beta})\Gamma(\lambda+\tilde{\al}-1)}
\frac{\sin[\pi(\tilde{\beta}+\lambda-1)]}{\sin[\pi (\tilde{\al}+\tilde{\beta}+\lambda-1)]}
\biggl)
%\biggr\}
\, ,
\nonum
\eea
see \cite{Kotikov:2013eha} for other representations in terms of $\psi$-functions. Of course, Eq.~(\ref{chap2:res:Gab}) allows to recover all previously known cases
for integer indices. 

\begin{figure}
  \begin{center}
    a) \qquad
    \begin{fmfgraph*}(20,17.5)
      \fmfleft{i}
      \fmfright{o}
      \fmfleft{ve}
      \fmfright{vo}
      \fmftop{vn}
      \fmftop{vs}
      \fmffreeze
      \fmfforce{(-0.3w,0.5h)}{i}
      \fmfforce{(1.3w,0.5h)}{o}
      \fmfforce{(0w,0.5h)}{ve}
      \fmfforce{(1.0w,0.5h)}{vo}
      \fmfforce{(.5w,0.95h)}{vn}
      \fmfforce{(.5w,0.05h)}{vs}
      \fmffreeze
      \fmf{fermion,label=$p$}{i,ve}
      \fmf{plain,left=0.8}{ve,vo}
      \fmf{phantom,left=0.7,label=$\al$,l.d=-0.0w}{ve,vn}
      \fmf{phantom,right=0.7,label=$\beta$,l.d=-0.0w}{vo,vn}
      \fmf{plain,left=0.8}{vo,ve}
      \fmf{phantom,left=0.7,label=$\gamma$,l.d=-0.0w}{vo,vs}
      \fmf{phantom,right=0.7,label=$1$,l.d=-0.0w}{ve,vs}
      \fmf{plain,label=$1$,l.d=0.05w}{vs,vn}
      \fmf{plain}{vo,o}
      \fmffreeze
      \fmfdot{ve,vn,vo,vs}
    \end{fmfgraph*}
    \qquad \qquad \qquad
    b) \qquad
    \begin{fmfgraph*}(20,17.5)
      \fmfleft{i}
      \fmfright{o}
      \fmfleft{ve}
      \fmfright{vo}
      \fmftop{vn}
      \fmftop{vs}
      \fmffreeze
      \fmfforce{(-0.3w,0.5h)}{i}
      \fmfforce{(1.3w,0.5h)}{o}
      \fmfforce{(0w,0.5h)}{ve}
      \fmfforce{(1.0w,0.5h)}{vo}
      \fmfforce{(.5w,0.95h)}{vn}
      \fmfforce{(.5w,0.05h)}{vs}
      \fmffreeze
      \fmf{fermion,label=$p$}{i,ve}
      \fmf{plain,left=0.8}{ve,vo}
      \fmf{phantom,left=0.7,label=$\al$,l.d=-0.0w}{ve,vn}
      \fmf{phantom,right=0.7,label=$\beta$,l.d=-0.0w}{vo,vn}
      \fmf{plain,left=0.8}{vo,ve}
      \fmf{phantom,left=0.7,label=$1$,l.d=-0.0w}{vo,vs}
      \fmf{phantom,right=0.7,label=$1$,l.d=-0.0w}{ve,vs}
      \fmf{plain,label=$\gamma$,l.d=0.05w}{vs,vn}
      \fmf{plain}{vo,o}
      \fmffreeze
      \fmfdot{ve,vn,vo,vs}
    \end{fmfgraph*}
    \qquad \qquad \qquad
    c) \qquad
    \begin{fmfgraph*}(20,17.5)
      \fmfleft{i}
      \fmfright{o}
      \fmfleft{ve}
      \fmfright{vo}
      \fmftop{vn}
      \fmftop{vs}
      \fmffreeze
      \fmfforce{(-0.3w,0.5h)}{i}
      \fmfforce{(1.3w,0.5h)}{o}
      \fmfforce{(0w,0.5h)}{ve}
      \fmfforce{(1.0w,0.5h)}{vo}
      \fmfforce{(.5w,0.95h)}{vn}
      \fmfforce{(.5w,0.05h)}{vs}
      \fmffreeze
      \fmf{fermion,label=$p$}{i,ve}
      \fmf{plain,left=0.8}{ve,vo}
      \fmf{phantom,left=0.7,label=$\al$,l.d=-0.0w}{ve,vn}
      \fmf{phantom,right=0.7,label=$1$,l.d=-0.0w}{vo,vn}
      \fmf{plain,left=0.8}{vo,ve}
      \fmf{phantom,left=0.7,label=$1$,l.d=-0.0w}{vo,vs}
      \fmf{phantom,right=0.7,label=$\gamma$,l.d=-0.0w}{ve,vs}
      \fmf{plain,label=$\beta$,l.d=0.05w}{vs,vn}
      \fmf{plain}{vo,o}
      \fmffreeze
      \fmfdot{ve,vn,vo,vs}
    \end{fmfgraph*}
  \caption{\label{chap2:fig:beyondtriangle+}
  Examples of more complicated two-loop massless p-type diagrams ($\al$, $\beta$ and $\gamma$ are arbitrary indices).}
%  A class of complicated two-loop massless propagator diagrams computable with the technique of Ref.~\cite{Kotikov:1995cw}.}
  \end{center}
\end{figure}

As an application, we may consider the peculiar case where $\al=\veps$ and $\beta=2\veps$ which appears in the computation of the renormalization group function of the $\Phi^4$-model at 
6 loops \cite{Kompaniets:2016hct,Kompaniets:2017yct}. Using Eqs.~(\ref{chap2:res:Gab}) and (\ref{chap2:res:Gab:I2}), the first terms of the expansion are easily found and read:
%
%\be
%G(\veps,1,2\veps,1,1) = \frac{e^{-2\gamma \veps - \zeta_2 \veps^2}}{1-2\veps}\,\Bigg( - \frac{7}{30 \veps} - \frac{34}{15} - \frac{503}{30}\,\veps
%+ \left( \frac{766}{45} \zeta_3 - \frac{338}{3} \right)\,\veps^2 + \left( \frac{305}{12} \zeta_4 + \frac{7094}{45}\zeta_3 - \frac{21791}{30} \right)\, \veps^3 + \Ord(\veps^4) \Bigg)\, .
%\label{chap2:exp:G1e2e}
%\ee
%
%In the $G$-scheme:
%
\begin{flalign}
&G(4-2\veps,\veps,1,2\veps,1,1) = G^2(\veps) \, \bigg [  - \frac{7}{30 \veps} - \frac{9}{5} - \frac{367}{30}\,\veps + \left( \frac{239}{15} \zeta_3 - \frac{1187}{15} \right)\,\veps^2 \bigg .
\nonum \\
&\qquad \qquad \qquad \qquad \qquad \qquad \qquad \bigg .  + \left( \frac{239}{10} \zeta_4 + \frac{576}{5}\zeta_3 - \frac{15031}{30} \right)\, \veps^3 + \Ord(\veps^4) \bigg ] \, ,
\label{chap2:exp:G1e2e}
\end{flalign}
where the so-called $G$-scheme \cite{Chetyrkin:1980pr} has been used where:
\be
G(\veps) = \veps\,G(1,1) = \frac{\Gamma^2(1-\veps)\Gamma(1+\veps)}{\Gamma(2-2\veps)}\, .
\label{chap2:G-scheme}
\ee
In numerical form, the terms in the brackets of Eq.~(\ref{chap2:exp:G1e2e}) read:
\begin{flalign}
&G(4-2\veps,\veps,1,2\veps,1,1) = G^2(\veps) \,\bigg [- \frac{0.23333333333333333333}{\veps} - 1.8 - 12.2333333333333333336\,\veps \bigg .
\nonum \\
&\qquad \bigg . -59.98056000965713105\,\veps^2 -336.68885280365186888\, \veps^3 + \Ord(\veps^4)  \bigg ]\, ,
\label{chap2:exp:G1e2e:num}
\end{flalign}
and are in good agreement with estimates from the sector decomposition technique% with %2.5M and 25M points
:~\footnote{Unpublished result from M.~Kompaniets reproduced with his kind permission.}
%
%\bea
%  -0.2333331(12) * e**-1 +  -1.799941(34) +  -12.23342(18) * e**1 +  -59.9800(11) * e**2 +  -336.694(5) * e**3 +  -1550.789(29) * e**4 +  -8523.32(13) * e**5 
%\\
\begin{flalign}
&G(4-2\veps,\veps,1,2\veps,1,1) = G^2(\veps) \, \bigg [-\frac{0.23333324(12)}{\veps}  -1.7999970(31) -12.233338(24) \,\veps  \bigg .
\nonum \\ 
&\qquad \bigg . - 59.98056(12) \,\veps^2 - 336.6893(7) \, \veps^3 + \Ord(\veps^4)  \bigg ]\, . %- 1550.8017(34) \, \veps^4 - 8523.191(18) \, \veps^5\, .
\end{flalign}

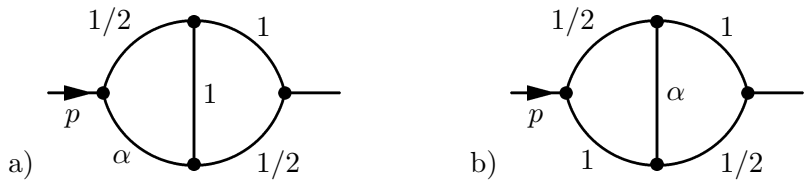
\begin{figure}
  \begin{center}
    a) \qquad
    \begin{fmfgraph*}(24,21)
      \fmfleft{i}
      \fmfright{o}
      \fmfleft{ve}
      \fmfright{vo}
      \fmftop{vn}
      \fmftop{vs}
      \fmffreeze
      \fmfforce{(-0.3w,0.5h)}{i}
      \fmfforce{(1.3w,0.5h)}{o}
      \fmfforce{(0w,0.5h)}{ve}
      \fmfforce{(1.0w,0.5h)}{vo}
      \fmfforce{(.5w,0.95h)}{vn}
      \fmfforce{(.5w,0.05h)}{vs}
      \fmffreeze
      \fmf{fermion,label=$p$}{i,ve}
      \fmf{plain,left=0.8}{ve,vo}
      \fmf{phantom,left=0.5,label=$1/2$,l.d=-0.01w}{ve,vn}
      \fmf{phantom,right=0.5,label=$1$,l.d=-0.01w}{vo,vn}
      \fmf{plain,left=0.8}{vo,ve}
      \fmf{phantom,left=0.5,label=$1/2$,l.d=-0.01w}{vo,vs}
      \fmf{phantom,right=0.5,label=$\alpha$,l.d=-0.01w}{ve,vs}
      \fmf{plain,label=$1$,l.d=0.05w}{vs,vn}
      \fmf{plain}{vo,o}
      \fmffreeze
      \fmfdot{ve,vn,vo,vs}
    \end{fmfgraph*}
    \qquad \qquad \qquad
    b) \qquad
    \begin{fmfgraph*}(24,21)
      \fmfleft{i}
      \fmfright{o}
      \fmfleft{ve}
      \fmfright{vo}
      \fmftop{vn}
      \fmftop{vs}
      \fmffreeze
      \fmfforce{(-0.3w,0.5h)}{i}
      \fmfforce{(1.3w,0.5h)}{o}
      \fmfforce{(0w,0.5h)}{ve}
      \fmfforce{(1.0w,0.5h)}{vo}
      \fmfforce{(.5w,0.95h)}{vn}
      \fmfforce{(.5w,0.05h)}{vs}
      \fmffreeze
      \fmf{fermion,label=$p$}{i,ve}
      \fmf{plain,left=0.8}{ve,vo}
      \fmf{phantom,left=0.5,label=$1/2$,l.d=-0.01w}{ve,vn}
      \fmf{phantom,right=0.5,label=$1$,l.d=-0.01w}{vo,vn}
      \fmf{plain,left=0.8}{vo,ve}
      \fmf{phantom,left=0.5,label=$1/2$,l.d=-0.01w}{vo,vs}
      \fmf{phantom,right=0.5,label=$1$,l.d=-0.01w}{ve,vs}
      \fmf{plain,label=$\alpha$,l.d=0.05w}{vs,vn}
      \fmf{plain}{vo,o}
      \fmffreeze
      \fmfdot{ve,vn,vo,vs}
    \end{fmfgraph*}
  \caption{\label{chap2:fig:morecomplicateddiags} Examples of complicated diagrams appearing in Refs.~\cite{Kotikov:2016wrb,Kotikov:2016prf,Kotikov:2016yrn}.}
  \end{center}
\end{figure}

The more complicated two-loop massless p-type diagrams of Fig.~\ref{chap2:fig:beyondtriangle+} are also in principle computable with the technique of Ref.~\cite{Kotikov:1995cw}.\footnote{I was
informed by M.~Kompaniets that, for even space dimensions, it is also possible to compute these integrals with the help of HyperInt \cite{Panzer:2014caa} using the technique developed in \cite{Kompaniets:2016hct}.}
The explicit computation has not been carried out yet but it is expected that they will also be expressed in terms of a linear combination of generalized hypergeometric functions, ${}_3F_2$ with argument $1$.
In some cases, one may expect that simpler expressions may be obtained, \eg, in the case of the diagrams of Fig.~\ref{chap2:fig:morecomplicateddiags} which appear in the study of D$\chi$SB in QED$_3$. 
In Chap.~\ref{chap4}, on the basis of the Gegenbauer polynomial technique, see Sec.~\ref{chap2:sec:meth:Gegenbauer}, we will provide a series representation for 
these diagrams which is a convenient starting point to compute them numerically for some specific values of $\al$.
%The numerical re

\section{Methods of calculations}
\label{chap2:sec:meth}

In this section we provide an overview of some useful methods to compute Feynman diagrams. Among these methods, some of them (parametric integration, Gegenbauer polynomial technique) involve explicit integrations. 
Other methods are {\bf algebraic} and involve identities between different diagrams which are conveniently expressed in graphical form. 
These identities transform a diagram into another one (with different indices) and sometimes allow its exact computation without performing any explicit integration (assuming we know the one-loop $G$-functions). 
We shall refer to them as the {\bf standard rules of perturbation theory for massless Feynman diagrams} \cite{Kazakov:1984km} and they will be extensively used in the next chapters.
%In the following, we will mainly use the algebraic methods which are very convenient to compute massless Feynman diagrams. 

\subsection{Parametric integration}
\label{chap2:sec:meth:param-int}

This is probably one of the oldest techniques known in Feynman diagram calculation, see, \eg, the textbooks \cite{grozin2007lectures,smirnov2013analytic,peskin1995introduction,itzykson2012quantum}. 
%It is also known as $\al$-parametrization and rests on the so-called {\bf Schwinger-trick} (see below). 
 It is based on so-called {\bf Schwinger-trick} (see below) and  amounts to represent a diagram, originally expressed in position or momentum space, in the space of {\bf Feynman parameters} (or parametric space).
The method is very useful in the case of massive Feynman diagrams. Many recent developments, even for massless multi-loop diagrams, are based on this technique, see, \eg,  
\cite{Binoth:2000ps,Bogner:2007cr,Brown:2008um,Brown:2009ta,Kompaniets:2016hct} and also the nice dissertations \cite{Bogner2009,Panzer:2015ida}. 

The Schwinger-trick is based on the integral representation of the $\Gamma$-function:
\be
\frac{1}{A_j^{\al_j}} = \frac{1}{\Gamma(\al_j)}\,\int_0^{+\infty}\D t e^{-A_j t}\,t^{\al_j-1}\, ,
\label{chap2:al-parametriz}
\ee
where $A_j=k_j^2+m_j^2$ and $\al_j$ is the index of the propagator. It immediately allows to compute the massive tadpole diagram, Eq.~(\ref{chap2:massive-one-loop-tadpole}).
It's generalization to the product of an arbitrary number of propagators with arbitrary exponents can be written as:
\be
\frac{1}{A_1^{\al_1} \cdots A_N^{\al_N}} = \frac{\Gamma(\al)}{\Gamma(\al_1) \cdots \Gamma(\al_N)} \, 
\int_{0}^1 \D u_1 \cdots \int_{0}^1 \D u_N \frac{\delta(1- \sum_{j=1}^N u_j) u_1^{\al_1-1} \cdots u_N^{\al_N-1}}{(u_1 A_1 + \cdots + u_N A_N)^{\al}}\, ,
\label{chap2:feyn-param}
\ee
where $\al=\sum_{j=1}^N \al_j$ and the $u_i$ are the so-called Feynman parameters.  The one-loop p-type integral can be computed straightforwardly using Eq.~(\ref{chap2:feyn-param}) yielding Eq.~(\ref{chap2:one-loop-G}).

Let's consider an $L$-loop diagram with external momenta collected in the vector $\vec p$ and $N$ internal propagators whose indices are collected in the vector $\vec \al$. In momentum space, this diagram reads:
\be
J_L(D,\vec p,\vec \al\,) = \int \frac{[\D^D k_1] \cdots [\D^D k_L]}{A_1^{\al_1} \cdots A_N^{\al_N}}\, .
\label{chap2:JL}
\ee
Using Eq.~(\ref{chap2:feyn-param}), Eq.~(\ref{chap2:JL}) can then be represented in parametric space under the general form:
\be
J_L(D,\vec p, \vec \al\,) = (4\pi)^{-LD}\,\frac{\Gamma(\al)}{\Gamma(\al_1) \cdots \Gamma(\al_N)}\,\int_0^1 \D^N u \, \delta \left(1- \sum_{j=1}^N u_j \right)\,\Pi_{j=1}^N u_j^{\al_j-1} \, 
\frac{\mathcal{U}^{\al - (L+1)D/2}}{\mathcal{F}(\vec{p})^{\al - LD/2}}\, ,
\label{chap2:expr:parametric}
\ee
where $\mathcal{U}$ and $\mathcal{F}$ are polynomials in the Feynman parameters and $\mathcal{F}$ also depends on the external momenta and masses.
For an arbitrary Feynman diagram, the Symansik polynomials $\mathcal{U}$ and $\mathcal{F}$ are not so easy to compute starting from the momentum representation.
Some more efficient derivations are based on the topology of the diagram under consideration, see, \eg, \cite{Bogner:2007cr}. Let's mention that in the case of the massless
2-loop p-type integral, these polynomials read:
\begin{subequations}
\label{chap2:symanzikU+F}
\bea
&&\mathcal{U} = (u_1 + u_2 + u_3 + u_4) u_5 + (u_1+u_4)(u_2+u_3)\, ,
\label{chap2:symanzikU}\\
&&\mathcal{F} = \bigg( (u_1+u_4)u_2u_3 +(u_2+u_3)u_1u_4 + (u_1+u_2)(u_3+u_4)u_5 \bigg) p^2\, . 
\label{chap2:symanzikF}
\eea
\end{subequations}

\subsection{Integration by parts}
\label{chap2:sec:meth:IBP}

Thanks to it's simplicity and efficiency, integration by parts (IBP) is one of the most widely used methods in multi-loop calculations. It has been introduced by Vasil'ev, Pis'mak and Khonkonen~\cite{Vasiliev:1981dg},
Tkachov \cite{Tkachov:1981wb} and Chetyrkin and Tkachov~\cite{Chetyrkin:1981qh}. It allows to reduce a complicated Feynman diagram in terms of a limited number of so-called ``master integrals'' \cite{Broadhurst:1987ei};
such reduction is now automated via it's implementation in computer programs with the help of various algorithms, see, \eg, \cite{Laporta:2001dd,Baikov:1996rk,Smirnov:2008iw,Lee:2013mka}.
As reviewed in Sec.~\ref{Sec:chap2:hist}, in some simple cases the master integrals themselves can be computed from IBP alone. In general, however, other methods have to be used often in combination with IBP (see next items).

IBP recurrence relations in momentum-space are essentially based on the translational invariance of dimensionally regularized integrals:
\be
\int\,\D^D k\,f(k) = \int\,\D^D k\,f(k+q) \quad \Ra \quad 0 = \int\,\D^D k\,\frac{\partial}{\partial k^\mu}\,f(k) \, .
\ee
In the following, we shall mainly be concerned with the application of the IBP procedure to the 2-loop massless p-type diagram of Eq.~(\ref{chap2:def:two-loop-p-int}).~\footnote{The original references 
\cite{Tkachov:1981wb,Chetyrkin:1981qh} were actually focusing on IBP relations for 3-loop massless p-type diagrams.}
The IBP relations for this diagram then follow from:
\be
0 = (\partial_{\mathcal{C}} \cdot P)\,J(D,p,\al_1,\al_2,\al_3,\al_4,\al_5)\, ,
\label{chap2:sec:meth:IBP:general}
\ee
where $\mathcal{C}$ is a closed oriented circuit and $P$ some momentum. In the above identity it is understood that differentiation goes before integration.
Let's define $p_i$ as the momentum carried by the line of index $\al_i$:
\be
p_1=k-p,\quad p_2=q-p,\quad p_3=q,\quad p_4=k,\quad p_5=k-q\, ,
\label{chap2:sec:meth:IBP:parametrization}
\ee
where $p$ is the external momentum. Different IBP relations come from different choices for the contour $\mathcal{C}$ and the momentum $P$. According to Chetyrkin and Tkachov \cite{Chetyrkin:1981qh}, 
for an $L$-loop p-integral one can write down $(L+1)^2$ independent IBP identities. This comes from the fact that there are $L+1$ possible choices for $\mathcal{C}$ ($L$ internal and $1$ external)
and a similar number for $P$ ($L$ loop momenta and one external momentum). So, for the 2-loop massless p-type diagram there are {\it a priori} $9$ IBP relations.

Let's first consider the case corresponding to $P=p_5$ and $\mathcal{C}=\{+p_1,\,+p_4,\,+p_5\}$. Along this contour the derivative reads:
\be
\partial_{\mathcal{C}} = + \frac{\partial}{\partial p_1} + \frac{\partial}{\partial p_4} + \frac{\partial}{\partial p_5}\, ,
\ee
where the sign before $p_i$ or $\partial / \partial p_i$ is plus if $p_i$ flows in the direction of $\mathcal{C}$ and minus otherwise.
With the help of Eq.~(\ref{chap2:sec:meth:IBP:general}), this leads to:
\be
\int \int \D^D k \, \D^D q\, \frac{\partial}{\partial k^\mu} \left \{ \frac{1}{(k-p)^{2\al_1}\,(q-p)^{2\al_2}\,q^{2\al_3}\,k^{2\al_4}}
\,\frac{(k-q)^\mu}{(k-q)^{2\al_5}} \right \} = 0 \, .
\label{chap2:sec:IBP:example}
\ee
Using equalities of the type:
\bea
\frac{\partial k^\nu}{\partial k^\mu} = \delta_\mu^\nu \quad (\delta_\mu^\mu = D) \, ,
\qquad
\frac{\partial (k-p)^{-2\al}}{\partial k^\mu} = - 2\al \,\frac{(k-p)_\mu}{(k-p)^{2\al+2}} \, ,
%\quad
%2(k-p_a)\cdot(k-p_b) = (k-p_a)^2+(k-p_b)^2-(p_b-p_a)^2 \, ,
\label{chap2:sec:meth:IBP:equalities}
\eea
and canceling squared combinations of momentum in the numerator and denominator yields, in graphical form:
%Central line, left triangle:
%
\begin{flalign}
(D-\al_1-\al_4-2\al_5) \quad
\parbox{16mm}{
    \begin{fmfgraph*}(16,14)
      \fmfleft{i}
      \fmfright{o}
      \fmfleft{ve}
      \fmfright{vo}
      \fmftop{vn}
      \fmftop{vs}
      \fmffreeze
      \fmfforce{(-0.1w,0.5h)}{i}
      \fmfforce{(1.1w,0.5h)}{o}
      \fmfforce{(0w,0.5h)}{ve}
      \fmfforce{(1.0w,0.5h)}{vo}
      \fmfforce{(.5w,0.95h)}{vn}
      \fmfforce{(.5w,0.05h)}{vs}
      \fmffreeze
      \fmf{plain}{i,ve}
      \fmf{plain,left=0.8}{ve,vo}
      \fmf{phantom,left=0.5,label=$\al_1$,l.d=-0.01w}{ve,vn}
      \fmf{phantom,right=0.5,label=$\al_2$,l.d=-0.01w}{vo,vn}
      \fmf{plain,left=0.8}{vo,ve}
      \fmf{phantom,left=0.5,label=$\al_3$,l.d=-0.01w}{vo,vs}
      \fmf{phantom,right=0.5,label=$\al_4$,l.d=-0.01w}{ve,vs}
      \fmf{plain,label=$\al_5$,l.d=0.05w}{vs,vn}
      \fmf{plain}{vo,o}
      \fmffreeze
      \fmfdot{ve,vn,vo,vs}
    \end{fmfgraph*}
}
\quad & =  \,\, \al_1 \left[\quad
\parbox{16mm}{
    \begin{fmfgraph*}(16,14)
      \fmfleft{ve}
      \fmfright{vo}
      \fmftop{vn}
      \fmftop{vs}
      \fmffreeze
      \fmfforce{(0w,0.5h)}{ve}
      \fmfforce{(1.0w,0.5h)}{vo}
      \fmfforce{(.5w,0.95h)}{vn}
      \fmfforce{(.5w,0.05h)}{vs}
      \fmffreeze
      \fmf{plain,left=0.8}{ve,vo}
      \fmf{phantom,left=0.5,label=$+$,l.d=-0.01w}{ve,vn}
      \fmf{phantom,right=0.4}{vo,vn}
      \fmf{plain,left=0.8}{vo,ve}
      \fmf{phantom,left=0.5}{vo,vs}
      \fmf{phantom,right=0.5}{ve,vs}
      \fmf{plain,label=$-$,l.d=0.05w}{vs,vn}
      \fmffreeze
      \fmfdot{ve,vn,vo,vs}
    \end{fmfgraph*}
} \qquad - \qquad
\parbox{16mm}{
    \begin{fmfgraph*}(16,14)
      \fmfleft{ve}
      \fmfright{vo}
      \fmftop{vn}
      \fmftop{vs}
      \fmffreeze
      \fmfforce{(0w,0.5h)}{ve}
      \fmfforce{(1.0w,0.5h)}{vo}
      \fmfforce{(.5w,0.95h)}{vn}
      \fmfforce{(.5w,0.05h)}{vs}
      \fmffreeze
      \fmf{plain,left=0.8}{ve,vo}
      \fmf{phantom,left=0.4,label=$+$,l.d=-0.01w}{ve,vn}
      \fmf{phantom,right=0.4,label=$-$,l.d=-0.01w}{vo,vn}
      \fmf{plain,left=0.8}{vo,ve}
      \fmf{phantom,left=0.5}{vo,vs}
      \fmf{phantom,right=0.5}{ve,vs}
      \fmf{plain}{vs,vn}
      \fmffreeze
      \fmfdot{ve,vn,vo,vs}
    \end{fmfgraph*}
} \quad
\right]
\nonum
\\ &\qquad \nonum
\\ & +  \,\,
\al_4 \left[\quad
\parbox{16mm}{
    \begin{fmfgraph*}(16,14)
      \fmfleft{ve}
      \fmfright{vo}
      \fmftop{vn}
      \fmftop{vs}
      \fmffreeze
      \fmfforce{(0w,0.5h)}{ve}
      \fmfforce{(1.0w,0.5h)}{vo}
      \fmfforce{(.5w,0.95h)}{vn}
      \fmfforce{(.5w,0.05h)}{vs}
      \fmffreeze
      \fmf{plain,left=0.8}{ve,vo}
      \fmf{phantom,left=0.5}{ve,vn}
      \fmf{phantom,right=0.5}{vo,vn}
      \fmf{plain,left=0.8}{vo,ve}
      \fmf{phantom,left=0.4}{vo,vs}
      \fmf{phantom,right=0.5,label=$+$,l.d=-0.01w}{ve,vs}
      \fmf{plain,label=$-$,l.d=0.05w}{vs,vn}
      \fmffreeze
      \fmfdot{ve,vn,vo,vs}
    \end{fmfgraph*}
} \qquad - \qquad
\parbox{16mm}{
    \begin{fmfgraph*}(16,14)
      \fmfleft{ve}
      \fmfright{vo}
      \fmftop{vn}
      \fmftop{vs}
      \fmffreeze
      \fmfforce{(0w,0.5h)}{ve}
      \fmfforce{(1.0w,0.5h)}{vo}
      \fmfforce{(.5w,0.95h)}{vn}
      \fmfforce{(.5w,0.05h)}{vs}
      \fmffreeze
      \fmf{plain,left=0.8}{ve,vo}
      \fmf{phantom,left=0.5}{ve,vn}
      \fmf{phantom,right=0.5}{vo,vn}
      \fmf{plain,left=0.8}{vo,ve}
      \fmf{phantom,left=0.4,label=$-$,l.d=-0.01w}{vo,vs}
      \fmf{phantom,right=0.4,label=$+$,l.d=-0.01w}{ve,vs}
      \fmf{plain}{vs,vn}
      \fmf{plain}{vs,vn}
      \fmffreeze
      \fmfdot{ve,vn,vo,vs}
    \end{fmfgraph*}
} \quad
\right]\, ,
\label{chap2:def:IBP-5-left}
\end{flalign}
where $\pm$ on the right-hand side of the equation denotes the increase or decrease of a line index by $1$ with respect to its value on the left-hand side.
Similarly, for $P=p_5$ but a contour running along the right triangle, the following identity is obtained:
\begin{flalign}
(D-\al_2-\al_3-2\al_5) \quad
\parbox{16mm}{
    \begin{fmfgraph*}(16,14)
      \fmfleft{i}
      \fmfright{o}
      \fmfleft{ve}
      \fmfright{vo}
      \fmftop{vn}
      \fmftop{vs}
      \fmffreeze
      \fmfforce{(-0.1w,0.5h)}{i}
      \fmfforce{(1.1w,0.5h)}{o}
      \fmfforce{(0w,0.5h)}{ve}
      \fmfforce{(1.0w,0.5h)}{vo}
      \fmfforce{(.5w,0.95h)}{vn}
      \fmfforce{(.5w,0.05h)}{vs}
      \fmffreeze
      \fmf{plain}{i,ve}
      \fmf{plain,left=0.8}{ve,vo}
      \fmf{phantom,left=0.5,label=$\al_1$,l.d=-0.01w}{ve,vn}
      \fmf{phantom,right=0.5,label=$\al_2$,l.d=-0.01w}{vo,vn}
      \fmf{plain,left=0.8}{vo,ve}
      \fmf{phantom,left=0.5,label=$\al_3$,l.d=-0.01w}{vo,vs}
      \fmf{phantom,right=0.5,label=$\al_4$,l.d=-0.01w}{ve,vs}
      \fmf{plain,label=$\al_5$,l.d=0.05w}{vs,vn}
      \fmf{plain}{vo,o}
      \fmffreeze
      \fmfdot{ve,vn,vo,vs}
    \end{fmfgraph*}
}
\quad & =  \,\, \al_2 \left[\quad
\parbox{16mm}{
    \begin{fmfgraph*}(16,14)
      \fmfleft{ve}
      \fmfright{vo}
      \fmftop{vn}
      \fmftop{vs}
      \fmffreeze
      \fmfforce{(0w,0.5h)}{ve}
      \fmfforce{(1.0w,0.5h)}{vo}
      \fmfforce{(.5w,0.95h)}{vn}
      \fmfforce{(.5w,0.05h)}{vs}
      \fmffreeze
      \fmf{plain,left=0.8}{ve,vo}
      \fmf{phantom,left=0.5}{ve,vn}
      \fmf{phantom,right=0.4,label=$+$,l.d=-0.01w}{vo,vn}
      \fmf{plain,left=0.8}{vo,ve}
      \fmf{phantom,left=0.5}{vo,vs}
      \fmf{phantom,right=0.5}{ve,vs}
      \fmf{plain,label=$-$,l.d=0.05w}{vs,vn}
      \fmffreeze
      \fmfdot{ve,vn,vo,vs}
    \end{fmfgraph*}
} \qquad - \qquad
\parbox{16mm}{
    \begin{fmfgraph*}(16,14)
      \fmfleft{ve}
      \fmfright{vo}
      \fmftop{vn}
      \fmftop{vs}
      \fmffreeze
      \fmfforce{(0w,0.5h)}{ve}
      \fmfforce{(1.0w,0.5h)}{vo}
      \fmfforce{(.5w,0.95h)}{vn}
      \fmfforce{(.5w,0.05h)}{vs}
      \fmffreeze
      \fmf{plain,left=0.8}{ve,vo}
      \fmf{phantom,left=0.4,label=$-$,l.d=-0.01w}{ve,vn}
      \fmf{phantom,right=0.4,label=$+$,l.d=-0.01w}{vo,vn}
      \fmf{plain,left=0.8}{vo,ve}
      \fmf{phantom,left=0.5}{vo,vs}
      \fmf{phantom,right=0.5}{ve,vs}
      \fmf{plain}{vs,vn}
      \fmffreeze
      \fmfdot{ve,vn,vo,vs}
    \end{fmfgraph*}
} \quad
\right]
\nonum
\\ &\qquad \nonum
\\ & +  \,\,
\al_3 \left[\quad
\parbox{16mm}{
    \begin{fmfgraph*}(16,14)
      \fmfleft{ve}
      \fmfright{vo}
      \fmftop{vn}
      \fmftop{vs}
      \fmffreeze
      \fmfforce{(0w,0.5h)}{ve}
      \fmfforce{(1.0w,0.5h)}{vo}
      \fmfforce{(.5w,0.95h)}{vn}
      \fmfforce{(.5w,0.05h)}{vs}
      \fmffreeze
      \fmf{plain,left=0.8}{ve,vo}
      \fmf{phantom,left=0.5}{ve,vn}
      \fmf{phantom,right=0.5}{vo,vn}
      \fmf{plain,left=0.8}{vo,ve}
      \fmf{phantom,left=0.4,label=$+$,l.d=-0.01w}{vo,vs}
      \fmf{phantom,right=0.5}{ve,vs}
      \fmf{plain,label=$-$,l.d=0.05w}{vs,vn}
      \fmffreeze
      \fmfdot{ve,vn,vo,vs}
    \end{fmfgraph*}
} \qquad - \qquad
\parbox{16mm}{
    \begin{fmfgraph*}(16,14)
      \fmfleft{ve}
      \fmfright{vo}
      \fmftop{vn}
      \fmftop{vs}
      \fmffreeze
      \fmfforce{(0w,0.5h)}{ve}
      \fmfforce{(1.0w,0.5h)}{vo}
      \fmfforce{(.5w,0.95h)}{vn}
      \fmfforce{(.5w,0.05h)}{vs}
      \fmffreeze
      \fmf{plain,left=0.8}{ve,vo}
      \fmf{phantom,left=0.5}{ve,vn}
      \fmf{phantom,right=0.5}{vo,vn}
      \fmf{plain,left=0.8}{vo,ve}
      \fmf{phantom,left=0.4,label=$+$,l.d=-0.01w}{vo,vs}
      \fmf{phantom,right=0.4,label=$-$,l.d=-0.01w}{ve,vs}
      \fmf{plain}{vs,vn}
      \fmf{plain}{vs,vn}
      \fmffreeze
      \fmfdot{ve,vn,vo,vs}
    \end{fmfgraph*}
} \quad
\right]\, .
\label{chap2:def:IBP-5-right}
\end{flalign}
Actually, Eqs.~(\ref{chap2:def:IBP-5-right}) and Eq.~(\ref{chap2:def:IBP-5-left}) are related to each other by using the symmetries of the diagram.
As another example, we may take $P=p_4$ and a contour running along the left triangle; this yields:
\begin{flalign}
(D-\al_1-\al_5-2\al_4) \quad
\parbox{16mm}{
    \begin{fmfgraph*}(16,14)
      \fmfleft{ve}
      \fmfright{vo}
      \fmftop{vn}
      \fmftop{vs}
      \fmffreeze
      \fmfforce{(0w,0.5h)}{ve}
      \fmfforce{(1.0w,0.5h)}{vo}
      \fmfforce{(.5w,0.95h)}{vn}
      \fmfforce{(.5w,0.05h)}{vs}
      \fmffreeze
      \fmf{plain,left=0.8}{ve,vo}
      \fmf{phantom,left=0.5,label=$\al_1$,l.d=-0.01w}{ve,vn}
      \fmf{phantom,right=0.5,label=$\al_2$,l.d=-0.01w}{vo,vn}
      \fmf{plain,left=0.8}{vo,ve}
      \fmf{phantom,left=0.5,label=$\al_3$,l.d=-0.01w}{vo,vs}
      \fmf{phantom,right=0.5,label=$\al_4$,l.d=-0.01w}{ve,vs}
      \fmf{plain,label=$\al_5$,l.d=0.05w}{vs,vn}
      \fmffreeze
      \fmfdot{ve,vn,vo,vs}
    \end{fmfgraph*}
}
\quad & =  \,\, \al_1 \left[\quad
\parbox{16mm}{
    \begin{fmfgraph*}(16,14)
      \fmfleft{ve}
      \fmfright{vo}
      \fmftop{vn}
      \fmftop{vs}
      \fmffreeze
      \fmfforce{(0w,0.5h)}{ve}
      \fmfforce{(1.0w,0.5h)}{vo}
      \fmfforce{(.5w,0.95h)}{vn}
      \fmfforce{(.5w,0.05h)}{vs}
      \fmffreeze
      \fmf{plain,left=0.8}{ve,vo}
      \fmf{phantom,left=0.5,label=$+$,l.d=-0.01w}{ve,vn}
      \fmf{phantom,right=0.4}{vo,vn}
      \fmf{plain,left=0.8}{vo,ve}
      \fmf{phantom,left=0.5}{vo,vs}
      \fmf{phantom,right=0.5,label=$-$,l.d=-0.01w}{ve,vs}
      \fmf{plain}{vs,vn}
      \fmffreeze
      \fmfdot{ve,vn,vo,vs}
    \end{fmfgraph*}
} \qquad - \quad (p^2)\times ~
\parbox{16mm}{
    \begin{fmfgraph*}(16,14)
      \fmfleft{ve}
      \fmfright{vo}
      \fmftop{vn}
      \fmftop{vs}
      \fmffreeze
      \fmfforce{(0w,0.5h)}{ve}
      \fmfforce{(1.0w,0.5h)}{vo}
      \fmfforce{(.5w,0.95h)}{vn}
      \fmfforce{(.5w,0.05h)}{vs}
      \fmffreeze
      \fmf{plain,left=0.8}{ve,vo}
      \fmf{phantom,left=0.4,label=$+$,l.d=-0.01w}{ve,vn}
      \fmf{phantom,right=0.4}{vo,vn}
      \fmf{plain,left=0.8}{vo,ve}
      \fmf{phantom,left=0.5}{vo,vs}
      \fmf{phantom,right=0.5}{ve,vs}
      \fmf{plain}{vs,vn}
      \fmffreeze
      \fmfdot{ve,vn,vo,vs}
    \end{fmfgraph*}
} \quad
\right]
\nonum
\\ &\qquad \nonum
\\ & +  \,\,
\al_5 \left[\quad
\parbox{16mm}{
    \begin{fmfgraph*}(16,14)
      \fmfleft{ve}
      \fmfright{vo}
      \fmftop{vn}
      \fmftop{vs}
      \fmffreeze
      \fmfforce{(0w,0.5h)}{ve}
      \fmfforce{(1.0w,0.5h)}{vo}
      \fmfforce{(.5w,0.95h)}{vn}
      \fmfforce{(.5w,0.05h)}{vs}
      \fmffreeze
      \fmf{plain,left=0.8}{ve,vo}
      \fmf{phantom,left=0.5}{ve,vn}
      \fmf{phantom,right=0.5}{vo,vn}
      \fmf{plain,left=0.8}{vo,ve}
      \fmf{phantom,left=0.4}{vo,vs}
      \fmf{phantom,right=0.5,label=$-$,l.d=-0.01w}{ve,vs}
      \fmf{plain,label=$+$,l.d=0.05w}{vs,vn}
      \fmffreeze
      \fmfdot{ve,vn,vo,vs}
    \end{fmfgraph*}
} \qquad - \qquad
\parbox{16mm}{
    \begin{fmfgraph*}(16,14)
      \fmfleft{ve}
      \fmfright{vo}
      \fmftop{vn}
      \fmftop{vs}
      \fmffreeze
      \fmfforce{(0w,0.5h)}{ve}
      \fmfforce{(1.0w,0.5h)}{vo}
      \fmfforce{(.5w,0.95h)}{vn}
      \fmfforce{(.5w,0.05h)}{vs}
      \fmffreeze
      \fmf{plain,left=0.8}{ve,vo}
      \fmf{phantom,left=0.5}{ve,vn}
      \fmf{phantom,right=0.5}{vo,vn}
      \fmf{plain,left=0.8}{vo,ve}
      \fmf{phantom,left=0.4,label=$-$,l.d=-0.01w}{vo,vs}
      \fmf{phantom,right=0.4}{ve,vs}
      \fmf{plain,label=$+$,l.d=0.05w}{vs,vn}
      \fmf{plain}{vs,vn}
      \fmffreeze
      \fmfdot{ve,vn,vo,vs}
    \end{fmfgraph*}
} \quad
\right]\, .
\label{chap2:def:IBP-4-left}
\end{flalign}
Other identities can be obtained from Eq.~(\ref{chap2:def:IBP-4-left}) by using the symmetries of the diagrams and will not be displayed.
%\subsection{Integration by parts identities (diagrams with arrows)}

%Another set of IBP identities envolves integrals containing a tensor in the numerator. Graphically, such tensors
%are displayed by using arrows on the corresponding lines as explained in App.~\ref{appen-loops}. Below, we shall
%give some identities at the level of triangles wich are themselves elements of the 2-loop diagrams.

Another set of useful identities, %comes from choosing an external circuit such as $\mathcal{C}=\{p,\,p_1,\,p_2\}$. 
the so-called ``homogeneity'' relations \cite{Chetyrkin:1981qh}, follow from the fact that the dimensionality $d_F$ of $J$ (in units of momentum) is known in terms of the $\al_i$ and $D$ from simple dimensional analysis:
\be
d_F = 2\left( D - \sum_{j=1}^5 \al_j \right)\, .
\ee
These relations read \cite{Chetyrkin:1981qh}:
\begin{subequations}
\label{chap2:sec:IBP:2loop-homog}
\bea
d_F J &=& \left( p \cdot \frac{d}{d p} \right)\,J = - p \cdot \left( \frac{\partial}{\partial p_1} + \frac{\partial}{\partial p_2} \right)\,J \, ,
\label{chap2:sec:IBP:2loop-homoga} \\
(D+d_F)\,\frac{p_1 \cdot p}{p^2}\, J &=& \left( \frac{d}{d p} \cdot p_1 \right)\,J = -\left( \frac{\partial}{\partial p_1} + \frac{\partial}{\partial p_2} \right) \cdot p_1 \,J \, ,
\label{chap2:sec:IBP:2loop-homogb} \\
(D+d_F)\,\frac{p_2 \cdot p}{p^2}\, J &=& \left( \frac{d}{d p} \cdot p_2 \right)\,J = -\left( \frac{\partial}{\partial p_1} + \frac{\partial}{\partial p_2} \right) \cdot p_2 \,J \, ,
\label{chap2:sec:IBP:2loop-homogc} 
\eea
\end{subequations}
where the chosen circuit is $\mathcal{C}=\{p,\,p_1,\,p_2\}$ and successive relations correspond to $P=p$, $P=p_1$ and $P=p_2$, respectively.
In graphical notations, Eq.~(\ref{chap2:sec:IBP:2loop-homogc}) reads:
\begin{flalign}
&\left( \frac{D}{2}+\al_1-\al_3-\al_4-\al_5 \right) \quad
\parbox{16mm}{
    \begin{fmfgraph*}(16,14)
%      \fmfleft{i}
%      \fmfright{o}
      \fmfleft{ve}
      \fmfright{vo}
      \fmftop{vn}
      \fmftop{vs}
      \fmffreeze
%      \fmfforce{(-0.1w,0.5h)}{i}
%      \fmfforce{(1.1w,0.5h)}{o}
      \fmfforce{(0w,0.5h)}{ve}
      \fmfforce{(1.0w,0.5h)}{vo}
      \fmfforce{(.5w,0.95h)}{vn}
      \fmfforce{(.5w,0.05h)}{vs}
      \fmffreeze
%      \fmf{plain}{i,ve}
      \fmf{plain,left=0.8}{ve,vo}
      \fmf{phantom,left=0.5,label=$\al_1$,l.d=-0.01w}{ve,vn}
      \fmf{phantom,right=0.5,label=$\al_2$,l.d=-0.01w}{vo,vn}
      \fmf{plain,left=0.8}{vo,ve}
      \fmf{phantom,left=0.5,label=$\al_3$,l.d=-0.01w}{vo,vs}
      \fmf{phantom,right=0.5,label=$\al_4$,l.d=-0.01w}{ve,vs}
      \fmf{plain,label=$\al_5$,l.d=0.05w}{vs,vn}
%      \fmf{plain}{vo,o}
      \fmffreeze
      \fmfdot{ve,vn,vo,vs}
    \end{fmfgraph*}
} 
 \quad  = \quad
\al_2 \left[ \quad
\parbox{16mm}{
    \begin{fmfgraph*}(16,14)
%      \fmfleft{i}
%      \fmfright{o}
      \fmfleft{ve}
      \fmfright{vo}
      \fmftop{vn}
      \fmftop{vs}
      \fmffreeze
%      \fmfforce{(-0.1w,0.5h)}{i}
%      \fmfforce{(1.1w,0.5h)}{o}
      \fmfforce{(0w,0.5h)}{ve}
      \fmfforce{(1.0w,0.5h)}{vo}
      \fmfforce{(.5w,0.95h)}{vn}
      \fmfforce{(.5w,0.05h)}{vs}
      \fmffreeze
%      \fmf{plain}{i,ve}
      \fmf{plain,left=0.8}{ve,vo}
      \fmf{phantom,left=0.5}{ve,vn}
      \fmf{phantom,right=0.4,label=$+$,l.d=-0.01w}{vo,vn}
      \fmf{plain,left=0.8}{vo,ve}
      \fmf{phantom,left=0.5}{vo,vs}
      \fmf{phantom,right=0.5}{ve,vs}
      \fmf{plain,label=$-$,l.d=0.05w}{vs,vn}
%      \fmf{plain}{vo,o}
      \fmffreeze
      \fmfdot{ve,vn,vo,vs}
    \end{fmfgraph*}
} \quad - \quad
\parbox{16mm}{
    \begin{fmfgraph*}(16,14)
%      \fmfleft{i}
%      \fmfright{o}
      \fmfleft{ve}
      \fmfright{vo}
      \fmftop{vn}
      \fmftop{vs}
      \fmffreeze
%      \fmfforce{(-0.1w,0.5h)}{i}
%      \fmfforce{(1.1w,0.5h)}{o}
      \fmfforce{(0w,0.5h)}{ve}
      \fmfforce{(1.0w,0.5h)}{vo}
      \fmfforce{(.5w,0.95h)}{vn}
      \fmfforce{(.5w,0.05h)}{vs}
      \fmffreeze
%      \fmf{plain}{i,ve}
      \fmf{plain,left=0.8}{ve,vo}
      \fmf{phantom,left=0.4,label=$-$,l.d=-0.01w}{ve,vn}
      \fmf{phantom,right=0.4,label=$+$,l.d=-0.01w}{vo,vn}
      \fmf{plain,left=0.8}{vo,ve}
      \fmf{phantom,left=0.5}{vo,vs}
      \fmf{phantom,right=0.5}{ve,vs}
      \fmf{plain}{vs,vn}
%      \fmf{plain}{vo,o}
      \fmffreeze
      \fmfdot{ve,vn,vo,vs}
    \end{fmfgraph*}
}
\quad \right]
\label{chap2:sec:IBP:2loop-homogc-graph}
\\
& \quad
\nonum \\
& \qquad \qquad + \left(\frac{3D}{2} - \sum_{i=1}^5\al_i \right) \,(k^2)^{-1}\, \left[ \quad
\parbox{16mm}{
    \begin{fmfgraph*}(16,14)
%      \fmfleft{i}
%      \fmfright{o}
      \fmfleft{ve}
      \fmfright{vo}
      \fmftop{vn}
      \fmftop{vs}
      \fmffreeze
%      \fmfforce{(-0.1w,0.5h)}{i}
%      \fmfforce{(1.1w,0.5h)}{o}
      \fmfforce{(0w,0.5h)}{ve}
      \fmfforce{(1.0w,0.5h)}{vo}
      \fmfforce{(.5w,0.95h)}{vn}
      \fmfforce{(.5w,0.05h)}{vs}
      \fmffreeze
%      \fmf{plain}{i,ve}
      \fmf{plain,left=0.8}{ve,vo}
      \fmf{phantom,left=0.5}{ve,vn}
      \fmf{phantom,right=0.4}{vo,vn}
      \fmf{plain,left=0.8}{vo,ve}
      \fmf{phantom,left=0.5}{vo,vs}
      \fmf{phantom,right=0.5,label=$-$,l.d=-0.01w}{ve,vs}
      \fmf{plain}{vs,vn}
%      \fmf{plain}{vo,o}
      \fmffreeze
      \fmfdot{ve,vn,vo,vs}
    \end{fmfgraph*}
} \quad - \quad
\parbox{16mm}{
    \begin{fmfgraph*}(16,14)
%      \fmfleft{i}
%      \fmfright{o}
      \fmfleft{ve}
      \fmfright{vo}
      \fmftop{vn}
      \fmftop{vs}
      \fmffreeze
%      \fmfforce{(-0.1w,0.5h)}{i}
%      \fmfforce{(1.1w,0.5h)}{o}
      \fmfforce{(0w,0.5h)}{ve}
      \fmfforce{(1.0w,0.5h)}{vo}
      \fmfforce{(.5w,0.95h)}{vn}
      \fmfforce{(.5w,0.05h)}{vs}
      \fmffreeze
%      \fmf{plain}{i,ve}
      \fmf{plain,left=0.8}{ve,vo}
      \fmf{phantom,left=0.4,label=$-$,l.d=-0.01w}{ve,vn}
      \fmf{phantom,right=0.4}{vo,vn}
      \fmf{plain,left=0.8}{vo,ve}
      \fmf{phantom,left=0.5}{vo,vs}
      \fmf{phantom,right=0.5}{ve,vs}
      \fmf{plain}{vs,vn}
%      \fmf{plain}{vo,o}
      \fmffreeze
      \fmfdot{ve,vn,vo,vs}
    \end{fmfgraph*}
} \quad \right] \, .
\nonum 
\end{flalign}
This identity is particularly useful in order to express a diagram as a function of another diagram with one index decreased.

Other relations follow from double differentiation with respect to the external momentum:
\be
d_F(d_F+D-2)\,J = p^2 \frac{d^2 \,J}{dp_\mu\,dp^\mu} = p^2 \, \left( \frac{\partial}{\partial p_1} + \frac{\partial}{\partial p_2} \right)^2 \,J\, .
\label{chap2:sec:IBP:2loop-dblext}
\ee
In graphical form, this reads:
\bea
& & d_F(d_F+D-2) \quad
\parbox{16mm}{
    \begin{fmfgraph*}(16,14)
%      \fmfleft{i}
%      \fmfright{o}
      \fmfleft{ve}
      \fmfright{vo}
      \fmftop{vn}
      \fmftop{vs}
      \fmffreeze
%      \fmfforce{(-0.1w,0.5h)}{i}
%      \fmfforce{(1.1w,0.5h)}{o}
      \fmfforce{(0w,0.5h)}{ve}
      \fmfforce{(1.0w,0.5h)}{vo}
      \fmfforce{(.5w,0.95h)}{vn}
      \fmfforce{(.5w,0.05h)}{vs}
      \fmffreeze
%      \fmf{plain}{i,ve}
      \fmf{plain,left=0.8}{ve,vo}
      \fmf{phantom,left=0.5,label=$\al_1$,l.d=-0.01w}{ve,vn}
      \fmf{phantom,right=0.5,label=$\al_2$,l.d=-0.01w}{vo,vn}
      \fmf{plain,left=0.8}{vo,ve}
      \fmf{phantom,left=0.5,label=$\al_3$,l.d=-0.01w}{vo,vs}
      \fmf{phantom,right=0.5,label=$\al_4$,l.d=-0.01w}{ve,vs}
      \fmf{plain,label=$\al_5$,l.d=0.05w}{vs,vn}
%      \fmf{plain}{vo,o}
      \fmffreeze
      \fmfdot{ve,vn,vo,vs}
    \end{fmfgraph*}
} 
\quad =  \,\, -  \quad
4 \al_1 \al_2 \, \quad
\parbox{16mm}{
    \begin{fmfgraph*}(16,14)
%      \fmfleft{i}
%      \fmfright{o}
      \fmfleft{ve}
      \fmfright{vo}
      \fmftop{vn}
      \fmftop{vs}
      \fmffreeze
%      \fmfforce{(-0.1w,0.5h)}{i}
%      \fmfforce{(1.1w,0.5h)}{o}
      \fmfforce{(0w,0.5h)}{ve}
      \fmfforce{(1.0w,0.5h)}{vo}
      \fmfforce{(.5w,0.95h)}{vn}
      \fmfforce{(.5w,0.05h)}{vs}
      \fmffreeze
%      \fmf{plain}{i,ve}
      \fmf{plain,left=0.8}{ve,vo}
      \fmf{phantom,left=0.5,label=$+$,l.d=-0.01w}{ve,vn}
      \fmf{phantom,right=0.5,label=$+$,l.d=-0.01w}{vo,vn}
      \fmf{plain,left=0.8}{vo,ve}
      \fmf{phantom,left=0.4}{vo,vs}
      \fmf{phantom,right=0.5}{ve,vs}
      \fmf{plain,label=$-$,l.d=0.05w}{vs,vn}
%      \fmf{plain}{vo,o}
      \fmffreeze
      \fmfdot{ve,vn,vo,vs}
    \end{fmfgraph*}
} \quad (k^2) 
\label{chap2:sec:IBP:dblext-graph} \\
& & \quad
\nonum \\
& & \,\, +  \quad
2(2\al_1+2\al_2+2-D)\, \left[\quad
\al_1 \quad
\parbox{16mm}{
    \begin{fmfgraph*}(16,14)
%      \fmfleft{i}
%      \fmfright{o}
      \fmfleft{ve}
      \fmfright{vo}
      \fmftop{vn}
      \fmftop{vs}
      \fmffreeze
%      \fmfforce{(-0.1w,0.5h)}{i}
%      \fmfforce{(1.1w,0.5h)}{o}
      \fmfforce{(0w,0.5h)}{ve}
      \fmfforce{(1.0w,0.5h)}{vo}
      \fmfforce{(.5w,0.95h)}{vn}
      \fmfforce{(.5w,0.05h)}{vs}
      \fmffreeze
%      \fmf{plain}{i,ve}
      \fmf{plain,left=0.8}{ve,vo}
      \fmf{phantom,left=0.5,label=$+$,l.d=-0.01w}{ve,vn}
      \fmf{phantom,right=0.4}{vo,vn}
      \fmf{plain,left=0.8}{vo,ve}
      \fmf{phantom,left=0.5}{vo,vs}
      \fmf{phantom,right=0.5}{ve,vs}
      \fmf{plain}{vs,vn}
%      \fmf{plain}{vo,o}
      \fmffreeze
      \fmfdot{ve,vn,vo,vs}
    \end{fmfgraph*}
} \quad + \quad \al_2\,\,\,\,\,\,
\parbox{16mm}{
    \begin{fmfgraph*}(16,14)
%      \fmfleft{i}
%      \fmfright{o}
      \fmfleft{ve}
      \fmfright{vo}
      \fmftop{vn}
      \fmftop{vs}
      \fmffreeze
%      \fmfforce{(-0.1w,0.5h)}{i}
%      \fmfforce{(1.1w,0.5h)}{o}
      \fmfforce{(0w,0.5h)}{ve}
      \fmfforce{(1.0w,0.5h)}{vo}
      \fmfforce{(.5w,0.95h)}{vn}
      \fmfforce{(.5w,0.05h)}{vs}
      \fmffreeze
%      \fmf{plain}{i,ve}
      \fmf{plain,left=0.8}{ve,vo}
      \fmf{phantom,left=0.4}{ve,vn}
      \fmf{phantom,right=0.4,label=$+$,l.d=-0.01w}{vo,vn}
      \fmf{plain,left=0.8}{vo,ve}
      \fmf{phantom,left=0.5}{vo,vs}
      \fmf{phantom,right=0.5}{ve,vs}
      \fmf{plain}{vs,vn}
%      \fmf{plain}{vo,o}
      \fmffreeze
      \fmfdot{ve,vn,vo,vs}
    \end{fmfgraph*}
} \quad
\right] \quad (k^2)\, .
\nonum
\eea

Finally, let us mention the IBP relations also apply to diagrams with numerator \cite{Kazakov:1986mu} and/or with mass \cite{Kotikov:1990kg}.

\subsection{The method of uniqueness}
\label{chap2:sec:meth:uniqueness}

The method of uniqueness is a powerful (but not very well known) technique devoted to the computation of {\bf massless} multi-loop Feynman diagrams.
This method owes its name to the so-called {\bf uniqueness relation}, otherwise known as the {\bf star-triangle} or {\bf Yang-Baxter
relation}, which is used in theories with conformal symmetry. Historically,
such relation was probably first used to compute three-dimensional integrals by D'Eramo, Peleti and Parisi \cite{D'Eramo:1971zz}.
Within the framework of multi-loop calculations, the method has first been introduced by Vasil'ev, Pis'mak and Khonkonen \cite{Vasiliev:1981dg}.
It allows, in principle, the computation of complicated Feynman diagrams using sequences of
simple transformations (including integration by parts) without performing any explicit integration.
A diagram is straightforwardly integrated once the appropriate sequence is found. In a sense, the method greatly simplifies
multi-loop calculations \cite{Vasiliev:1981dg,Usyukina:1983gj,Kazakov:1983ns,Kazakov:1984km,Kazakov:1983pk}.
As a matter of fact, the first analytical expression for the
five-loop $\beta$-function of the $\Phi^4$-model was derived by Kazakov using this technique \cite{Kazakov:1984km,Kazakov:1983pk}.
For a given diagram, the task of finding the sequence of transformations is, however, highly non-trivial. 
In the following, we will briefly present the method in momentum space in very close analogy with the beautiful lectures of Kazakov \cite{Kazakov:1984bw}
where the method was presented in coordinate space, see also \cite{Isaev:2003tk} for a recent short review. As mentioned previously, the set of all 
algebraic transformations (IBP, uniqueness, Fourier transform, duality, ...) is sometimes referred to as {\bf the 
standard rules of perturbation theory for massless Feynman diagrams} \cite{Kazakov:1984km}. In this review, only of a few of these rules are presented. For more, see \cite{Kazakov:1984bw}.

In momentum space, a triangle made of scalar propagators with three arbitrary indices is defined as:
\vspace{3mm}
\be
\parbox{16mm}{
  \begin{fmfgraph*}(16,16)
    \fmfleft{l}
    \fmfleft{vl}
    \fmfright{r}
    \fmfright{vr}
    \fmftop{t}
    \fmftop{vt}
    \fmffreeze
    \fmfforce{(0.0w,0.0h)}{l}
    \fmfforce{(0.25w,0.25h)}{vl}
    \fmfforce{(1.0w,0.0h)}{r}
    \fmfforce{(0.75w,0.25h)}{vr}
    \fmfforce{(.5w,1.0h)}{t}
    \fmfforce{(.5w,0.75h)}{vt}
    \fmffreeze
    \fmf{plain,label=$\al_2$,label.side=left}{vr,vl}
    \fmf{plain,label=$\al_3$,label.side=left}{vl,vt}
    \fmf{plain,label=$\al_1$,label.side=left}{vt,vr}
    \fmf{plain}{vl,l}
    \fmf{plain}{vr,r}
    \fmf{plain}{vt,t}
    \fmffreeze
    \fmfdot{vl,vt,vr}
    \fmflabel{$p_1$}{l}
    \fmflabel{$p_2$}{t}
    \fmflabel{$p_3=-p_1-p_2$}{r}
  \end{fmfgraph*}
} \,\,\, = \,\,\, \int\, \frac{[\D^D k]}{k^{2\al_2} (k-p_1)^{2\al_3} (k-p_1-p_2)^{2\al_1}} \, .
\label{chap2:def:triangle}
\ee
\vskip 4mm

\ni On the other hand, a vertex made of scalar propagators with three arbitrary indices is defined as:
\vspace{3mm}
\be
\parbox{16mm}{
  \begin{fmfgraph*}(16,16)
    \fmfleft{l}
    \fmfright{r}
    \fmftop{t}
    \fmfleft{v}
    \fmffreeze
    \fmfforce{(0.0w,0.0h)}{l}
    \fmfforce{(1.0w,0.0h)}{r}
    \fmfforce{(.5w,1.0h)}{t}
    \fmfforce{(.5w,0.5h)}{v}
    \fmffreeze
    \fmf{plain,label=$\beta_2$,label.side=right}{v,t}
    \fmf{plain,label=$\beta_3$,label.side=left}{v,r}
    \fmf{plain,label=$\beta_1$,label.side=right}{v,l}
    \fmffreeze
    \fmfdot{v}
  \end{fmfgraph*}
} \qquad = \qquad \frac{1}{p_1^{2\beta_1} p_2^{2\beta_2} (p_1+p_2)^{2\beta_3}}\, .
\label{chap2:def:star}
\ee
Both triangle and vertex diagrams have indices $\sum_{j=1}^3 \al_j$.
In momentum space, ordinary triangles and vertices, that is triangle and vertices made of ordinary lines, have index $3$.
Of great importance in the following will be the notion of a ``unique'' triangle and a ``unique'' vertex.
In momentum space, these diagrams are said to be ``unique'' if their indices are equal
to $D$ and $D/2$, respectively; see table \ref{chap2:meth:tab:indices} for a summary.

\begin{center}
\renewcommand{\tabcolsep}{1.2cm}
\renewcommand{\arraystretch}{1.5}
\begin{table}
    \begin{tabular}{  l || c | c | c }
      \hline
        ~~      &       {\bf Line}                      &       {\bf Triangle}          &       {\bf Vertex} \\
      \hline \hline
      Arbitrary &       $\al$                           & $\sum_{j=1}^3 \al_j$          &       $\sum_{j=1}^3 \al_j$    \\
      \hline
      Ordinary  &       $1$                             & $3$                           &       $3$     \\
      \hline
      Unique    &       $D/2$                           & $D$                           &       $D/2$  \\
      \hline
    \end{tabular}
    \caption{Indices of lines, triangles and vertices in $p$-space.}% ($\veps=2-D/2$, $\lambda = D/2-1$)}
    \label{chap2:meth:tab:indices}
\end{table}
\end{center}

The uniqueness (or star-triangle) relation connects a unique triangle to a unique vertex and reads:
\be
\parbox{16mm}{
  \begin{fmfgraph*}(16,16)
    \fmfleft{l}
    \fmfright{r}
    \fmftop{t}
    \fmfleft{v}
    \fmffreeze
    \fmfforce{(-0.1w,0.142265h)}{l}
    \fmfforce{(1.1w,0.142265h)}{r}
    \fmfforce{(.5w,1.1h)}{t}
    \fmfforce{(.5w,0.488675h)}{v}
    \fmffreeze
    \fmf{plain,label=${\al}_2$,label.side=left,l.d=0.05w}{v,t}
    \fmf{plain,label=${\al}_3$,label.side=right,l.d=0.05w}{r,v}
    \fmf{plain,label=${\al}_1$,label.side=left,l.d=0.05w}{v,l}
    \fmffreeze
    \fmfdot{v}
  \end{fmfgraph*}
}\, \qquad \underset{\underset{j}{\sum} \al_j = D/2}{=} \qquad \frac{(4\pi)^{D/2}}{G(\al_1,\al_2)}\, \quad
\parbox{16mm}{
  \begin{fmfgraph*}(16,16)
    \fmfleft{l}
    \fmfleft{vl}
    \fmfright{r}
    \fmfright{vr}
    \fmftop{t}
    \fmftop{vt}
    \fmffreeze
    \fmfforce{(-0.1w,0.042265h)}{l}
    \fmfforce{(0.1w,0.2h)}{vl}
    \fmfforce{(1.1w,0.042265h)}{r}
    \fmfforce{(0.9w,0.2h)}{vr}
    \fmfforce{(.5w,1.1h)}{t}
    \fmfforce{(.5w,0.966025h)}{vt}
    \fmffreeze
    \fmf{plain,label=$\tilde{\al}_2$,label.side=left,l.d=0.05w}{vr,vl}
    \fmf{plain,label=$\tilde{\al}_3$,label.side=left,l.d=0.05w}{vl,vt}
    \fmf{plain,label=$\tilde{\al}_1$,label.side=left,l.d=0.05w}{vt,vr}
    \fmf{plain}{l,vl}
    \fmf{plain}{r,vr}
    \fmf{plain}{t,vt}
    \fmffreeze
    \fmfdot{vl,vt,vr}
  \end{fmfgraph*}
}\, \qquad ,
\label{chap2:def:star-triangle}
\ee
where $\tilde{\al}_i=D/2-\al_i$ is the index dual to $\al_i$ and the condition $\sum_{j=1}^3 \al_j= D/2$ implies that the vertex is unique.
This relation can be proved by performing an inversion of all integration variables in the triangle: $k_\mu \ra k_\mu/(k)^2$,
keeping the external momenta fixed. Upon using the fact that the triangle is unique,  $\sum_j \al_j =D$, the integral simplifies and reduces to a simple vertex.

In order to illustrate the power of the method of uniqueness, let us derive Eq.~(\ref{chap2:transf4}) which was first obtained by Gorishny and Isaev \cite{Gorishnii:1984te}. Considering the 
massless two-loop p-type diagram with arbitrary indices, we start by replacing the central line by a loop~\footnote{In coordinate space, it corresponds to the insertion of a
point into this line (see the table of such transformations in Ref.~\cite{Vasiliev:1981dg} and also Ref.~\cite{Kazakov:1984bw} for a review).} in such
a way that the right triangle is unique. This yields:
\bea
\parbox{16mm}{
    \begin{fmfgraph*}(16,14)
      \fmfleft{i}
      \fmfright{o}
      \fmfleft{ve}
      \fmfright{vo}
      \fmftop{vn}
      \fmftop{vs}
      \fmffreeze
      \fmfforce{(-0.3w,0.5h)}{i}
      \fmfforce{(1.3w,0.5h)}{o}
      \fmfforce{(0w,0.5h)}{ve}
      \fmfforce{(1.0w,0.5h)}{vo}
      \fmfforce{(.5w,0.95h)}{vn}
      \fmfforce{(.5w,0.05h)}{vs}
      \fmffreeze
      \fmf{plain}{i,ve}
      \fmf{plain,left=0.8}{ve,vo}
      \fmf{phantom,left=0.5,label=$\al_1$,l.d=-0.01w}{ve,vn}
      \fmf{phantom,right=0.5,label=$\al_2$,l.d=-0.01w}{vo,vn}
      \fmf{plain,left=0.8}{vo,ve}
      \fmf{phantom,left=0.5,label=$\al_3$,l.d=-0.01w}{vo,vs}
      \fmf{phantom,right=0.5,label=$\al_4$,l.d=-0.01w}{ve,vs}
      \fmf{plain,label=$\al_5$,l.d=0.05w}{vs,vn}
      \fmf{plain}{vo,o}
      \fmffreeze
      \fmfdot{ve,vn,vo,vs}
    \end{fmfgraph*}
}  \qquad = \qquad \frac{(4\pi)^{D/2}}{G(\beta,\gamma)} \qquad ~
\parbox{18mm}{
    \begin{fmfgraph*}(18,16)
     \fmfleft{i}
      \fmfright{o}
      \fmfleft{ve}
      \fmfright{vo}
      \fmftop{vn}
      \fmftop{vs}
      \fmffreeze
      \fmfforce{(-0.3w,0.5h)}{i}
      \fmfforce{(1.3w,0.5h)}{o}
      \fmfforce{(0w,0.5h)}{ve}
      \fmfforce{(1.0w,0.5h)}{vo}
      \fmfforce{(.5w,0.95h)}{vn}
      \fmfforce{(.5w,0.05h)}{vs}
      \fmffreeze
      \fmf{plain}{i,ve}
      \fmf{plain,left=0.8}{ve,vo}
      \fmf{phantom,left=0.7,label=$\al_1$,l.d=-0.0w}{ve,vn}
      \fmf{phantom,right=0.7,label=$\al_2$,l.d=-0.0w}{vo,vn}
      \fmf{plain,left=0.8}{vo,ve}
      \fmf{phantom,left=0.7,label=$\al_3$,l.d=-0.0w}{vo,vs}
      \fmf{phantom,right=0.7,label=$\al_4$,l.d=-0.0w}{ve,vs}
      \fmf{plain,left=0.3,label=$\beta$,l.d=0.05w}{vs,vn}
      \fmf{plain,left=0.3,label=$\gamma$,l.d=0.05w}{vn,vs}
      \fmf{plain}{vo,o}
      \fmffreeze
      \fmfdot{ve,vn,vo,vs}
    \end{fmfgraph*}
}
\label{chap2:GI1}
\eea
where $\beta = \al_2 + \al_3 + \al_5 -D/2$ and $\gamma=D-\al_2-\al_3$. The right triangle being unique, we can use
Eq.~(\ref{chap2:def:star-triangle}) to simplify the diagram:
\bea
\parbox{16mm}{
    \begin{fmfgraph*}(16,14)
      \fmfleft{i}
      \fmfright{o}
      \fmfleft{ve}
      \fmfright{vo}
      \fmftop{vn}
      \fmftop{vs}
      \fmffreeze
      \fmfforce{(-0.3w,0.5h)}{i}
      \fmfforce{(1.3w,0.5h)}{o}
      \fmfforce{(0w,0.5h)}{ve}
      \fmfforce{(1.0w,0.5h)}{vo}
      \fmfforce{(.5w,0.95h)}{vn}
      \fmfforce{(.5w,0.05h)}{vs}
      \fmffreeze
      \fmf{plain}{i,ve}
      \fmf{plain,left=0.8}{ve,vo}
      \fmf{phantom,left=0.5,label=$\al_1$,l.d=-0.01w}{ve,vn}
      \fmf{phantom,right=0.5,label=$\al_2$,l.d=-0.01w}{vo,vn}
      \fmf{plain,left=0.8}{vo,ve}
      \fmf{phantom,left=0.5,label=$\al_3$,l.d=-0.01w}{vo,vs}
      \fmf{phantom,right=0.5,label=$\al_4$,l.d=-0.01w}{ve,vs}
      \fmf{plain,label=$\al_5$,l.d=0.05w}{vs,vn}
      \fmf{plain}{vo,o}
      \fmffreeze
      \fmfdot{ve,vn,vo,vs}
    \end{fmfgraph*}
}\qquad = \qquad \frac{G(\al_2,\gamma)}{G(\beta,\gamma)} \,\,(p^2)^{D/2-\al_2-\al_3}\qquad %\qquad
\parbox{16mm}{
    \begin{fmfgraph*}(16,14)
      \fmfleft{i}
      \fmfright{o}
      \fmfleft{ve}
      \fmfright{vo}
      \fmftop{vn}
      \fmftop{vs}
      \fmffreeze
      \fmfforce{(-0.3w,0.5h)}{i}
      \fmfforce{(1.3w,0.5h)}{o}
      \fmfforce{(0w,0.5h)}{ve}
      \fmfforce{(1.0w,0.5h)}{vo}
      \fmfforce{(.5w,0.95h)}{vn}
      \fmfforce{(.5w,0.05h)}{vs}
      \fmffreeze
      \fmf{plain}{i,ve}
      \fmf{plain,left=0.8}{ve,vo}
      \fmf{phantom,left=0.5,label=$\al_1$,l.d=-0.01w}{ve,vn}
      \fmf{phantom,right=0.5,label=$D/2-\al_3$,l.d=-0.01w}{vo,vn}
      \fmf{plain,left=0.8}{vo,ve}
      \fmf{phantom,left=0.5,label=$D/2-\al_2$,l.d=-0.01w}{vo,vs}
      \fmf{phantom,right=0.5,label=$\al_4$,l.d=-0.01w}{ve,vs}
      \fmf{plain,label=$\beta$,l.d=0.05w}{vs,vn}
      \fmf{plain}{vo,o}
      \fmffreeze
      \fmfdot{ve,vn,vo,vs}
    \end{fmfgraph*}
} \qquad \qquad .
\label{chap2:GI2}
\eea
Focusing for simplicity on the coefficient functions, all the dependence on the external momentum disappears.
Together with the simplification of the $G$-functions this yields:
\bea
C_D[J(D,p,\al_1,\al_2,\al_3,\al_4,\al_5)] = a(\al_2)a(\al_3)a(\al_5)a(D-t_2)\,C_D[J(D,p,\al_1,\tilde{\al}_3,\tilde{\al}_2,\al_4,t_2-D/2)],
\nonum
\eea
where $\tilde{\al} = D/2-\al$ and $t_2=\al_2+\al_3+\al_5$ and corresponds to the advertised Eq.~(\ref{chap2:transf4}).

\subsection{Fourier transform and duality}
\label{chap2:sec:meth:FT+Du}

Up to now, all diagrams were expressed in momentum space. Useful identities can be obtained by relating $p$-space and $x$-space diagrams.
Following \cite{Kazakov:1984bw}, we briefly present them in this paragraph.

Recall, from Eq.~(\ref{chap2:def:two-loop-p-int}), that the 2-loop massless propagator-type diagram
in $p$-space was defined as:
\be
J(D,p,\{\al_i\}) = \int \frac{[\D^Dk][\D^Dq]}{[(k-p)^2]^{\al_1}[(q-p)^2]^{\al_2}[q^2]^{\al_3}[k^2]^{\al_4}[(k-q)^2]^{\al_5}}
\quad = \qquad
\parbox{16mm}{
    \begin{fmfgraph*}(16,14)
      \fmfleft{i}
      \fmfright{o}
      \fmfleft{ve}
      \fmfright{vo}
      \fmftop{vn}
      \fmftop{vs}
      \fmffreeze
      \fmfforce{(-0.3w,0.5h)}{i}
      \fmfforce{(1.3w,0.5h)}{o}
      \fmfforce{(0w,0.5h)}{ve}
      \fmfforce{(1.0w,0.5h)}{vo}
      \fmfforce{(.5w,0.95h)}{vn}
      \fmfforce{(.5w,0.05h)}{vs}
      \fmffreeze
      \fmf{fermion,label=$p$}{i,ve}
      \fmf{plain,left=0.8}{ve,vo}
      \fmf{phantom,left=0.7,label=$\al_1$,l.d=-0.1w}{ve,vn}
      \fmf{phantom,right=0.7,label=$\al_2$,l.d=-0.1w}{vo,vn}
      \fmf{plain,left=0.8}{vo,ve}
      \fmf{phantom,left=0.7,label=$\al_3$,l.d=-0.1w}{vo,vs}
      \fmf{phantom,right=0.7,label=$\al_4$,l.d=-0.1w}{ve,vs}
      \fmf{plain,label=$\al_5$,l.d=0.05w}{vs,vn}
      \fmf{plain}{vo,o}
      \fmffreeze
      \fmfdot{ve,vn,vo,vs}
    \end{fmfgraph*}
} \qquad \,.
\label{chap2:def:two-loop-p-int2}
\ee
Equivalently, all calculations may be done in position space. In $x$-space, the 2-loop massless propagator-type diagram
is defined as:
\be
J(D,z,\{\overline{\al}_i\}) = \int \frac{[\D^D x][\D^D y]}{[y^2]^{\overline{\al}_1}[(y-z)^2]^{\overline{\al}_2}[(z-x)^2]^{\overline{\al}_3}[x^2]^{\overline{\al}_4}[(x-y)^2]^{\overline{\al}_5}}
\quad = \qquad
\parbox{16mm}{
    \begin{fmfgraph*}(16,14)
      \fmfleft{i}
      \fmfright{o}
      \fmfleft{ve}
      \fmfright{vo}
      \fmftop{vn}
      \fmftop{vs}
      \fmftop{vt}
      \fmfbottom{vb}		
      \fmffreeze
      \fmfforce{(-0.3w,0.5h)}{i}
      \fmfforce{(1.3w,0.5h)}{o}
      \fmfforce{(0w,0.5h)}{ve}
      \fmfforce{(1.0w,0.5h)}{vo}
      \fmfforce{(.5w,0.95h)}{vn}
      \fmfforce{(.5w,0.05h)}{vs}
      \fmfforce{(.5w,1.25h)}{vt}
      \fmfforce{(.5w,-0.3h)}{vb}
      \fmffreeze
      \fmf{plain,label=$0$,l.s=left}{i,ve}
      \fmf{plain,left=0.8}{ve,vo}
      \fmf{phantom,left=0.7,label=$\overline{\al}_1$,l.d=-0.1w}{ve,vn}
      \fmf{phantom,right=0.7,label=$\overline{\al}_2$,l.d=-0.1w}{vo,vn}
      \fmf{plain,left=0.8}{vo,ve}
      \fmf{phantom,left=0.7,label=$\overline{\al}_3$,l.d=-0.1w}{vo,vs}
      \fmf{phantom,right=0.7,label=$\overline{\al}_4$,l.d=-0.1w}{ve,vs}
      \fmf{plain,label=$\overline{\al}_5$,l.d=0.05w}{vs,vn}
      \fmf{plain,label=$z$,l.s=left}{vo,o}
      \fmf{phantom,label=$y$,l.d=-0.1w}{vn,vt}
      \fmf{phantom,label=$x$,l.d=-0.1w}{vs,vb}
      \fmffreeze
      \fmfdot{ve,vn,vo,vs}
    \end{fmfgraph*}
} \qquad \, ,
\label{chap2:def:two-loop-p-int-x}
\ee
where $0$ denotes the so-called ``root vertex'' and the $\overline{\al}_i$ are arbitrary indices. 
It is actually straightforward to show that the  $p$-space and $x$-space diagrams are related provided that $\overline{\al}_i=\tilde{\al}_i$ where $\tilde{\al} = D/2 -\al$ 
is the index which is dual (in the sense of Fourier transform) to $\al$. This follows from the Fourier transform, Eq.~(\ref{chap2:FT-inv-line}), that we reproduce here for clarity:
\be
\frac{1}{[k^2]^{\al}} = \frac{a(\al)}{(2\pi)^{D/2}}\,\int \D^D x \,\frac{e^{\I k x}}{[x^2]^{D/2-\al}}, \qquad a(\al) = \frac{\Gamma(D/2-\al)}{\Gamma(\al)}\, .
\ee
Hence, for a given diagram, the Fourier transform allows to relate its $p$-space and $x$-space coefficient functions and the relation reads:  
\be
\text{C}_D[\,J(D,p,\al_1,\al_2,\al_3,\al_4,\al_5)\,] \underset{(\text{FT})}{=}
\frac{\prod_{j=1}^5\,a(\al_j)}{a\left(\sum_{j=1}^5\al_j-D\right)}\,\text{C}_D[\,J(D,z,\tilde{\al}_1,\tilde{\al}_2,\tilde{\al}_3,\tilde{\al}_4,\tilde{\al}_5)\,].
\label{chap2:FT-inv-line2}
\ee
Graphically, this can be represented as:
\bea
\parbox{16mm}{
    \begin{fmfgraph*}(16,14)
%      \fmfleft{i}
%      \fmfright{o}
      \fmfleft{ve}
      \fmfright{vo}
      \fmftop{vn}
      \fmftop{vs}
      \fmffreeze
 %     \fmfforce{(-0.3w,0.5h)}{i}
 %     \fmfforce{(1.3w,0.5h)}{o}
      \fmfforce{(0w,0.5h)}{ve}
      \fmfforce{(1.0w,0.5h)}{vo}
      \fmfforce{(.5w,0.95h)}{vn}
      \fmfforce{(.5w,0.05h)}{vs}
      \fmffreeze
%      \fmf{fermion,label=$p$}{i,ve}
      \fmf{plain,left=0.8}{ve,vo}
      \fmf{phantom,left=0.7,label=$\al_1$,l.d=-0.1w}{ve,vn}
      \fmf{phantom,right=0.7,label=$\al_2$,l.d=-0.1w}{vo,vn}
      \fmf{plain,left=0.8}{vo,ve}
      \fmf{phantom,left=0.7,label=$\al_3$,l.d=-0.1w}{vo,vs}
      \fmf{phantom,right=0.7,label=$\al_4$,l.d=-0.1w}{ve,vs}
      \fmf{plain,label=$\al_5$,l.d=0.05w}{vs,vn}
%      \fmf{plain}{vo,o}
      \fmffreeze
      \fmfdot{ve,vn,vo,vs}
    \end{fmfgraph*}
} \quad = \quad \frac{\prod_{j=1}^5\,a(\al_j)}{a\left(\sum_{j=1}^5\al_j-D\right)}\,\qquad
\parbox{16mm}{
    \begin{fmfgraph*}(16,14)
      \fmfleft{i}
      \fmfright{o}
      \fmfleft{ve}
      \fmfright{vo}
      \fmftop{vn}
      \fmftop{vs}
      \fmffreeze
      \fmfforce{(-0.3w,0.5h)}{i}
      \fmfforce{(1.3w,0.5h)}{o}
      \fmfforce{(0w,0.5h)}{ve}
      \fmfforce{(1.0w,0.5h)}{vo}
      \fmfforce{(.5w,0.95h)}{vn}
      \fmfforce{(.5w,0.05h)}{vs}
      \fmffreeze
      \fmf{phantom,label=$0$,l.d=-0.1w,l.s=left}{i,ve}
      \fmf{plain,left=0.8}{ve,vo}
      \fmf{phantom,left=0.7,label=$\tilde{\al}_1$,l.d=-0.1w}{ve,vn}
      \fmf{phantom,right=0.7,label=$\tilde{\al}_2$,l.d=-0.1w}{vo,vn}
      \fmf{plain,left=0.8}{vo,ve}
      \fmf{phantom,left=0.7,label=$\tilde{\al}_3$,l.d=-0.1w}{vo,vs}
      \fmf{phantom,right=0.7,label=$\tilde{\al}_4$,l.d=-0.1w}{ve,vs}
      \fmf{plain,label=$\tilde{\al}_5$,l.d=0.05w}{vs,vn}
      \fmf{phantom,label=$z$,l.d=-0.1w,l.s=left}{vo,o}
      \fmffreeze
      \fmfdot{ve,vn,vo,vs}
    \end{fmfgraph*}
} \qquad \, ,
\eea
where all external legs were amputated as the diagrams correspond to coefficient functions
but we have explicitly indicated the location of the external vertices in the $x$-space function to distinguish it from its $p$-space counterpart.

Another useful transformation is the so-called duality transformation. It is based on the fact that the loop momenta are dummy integration variables.
They can therefore be replaced by dummy coordinate integration variables. Such an innocent looking change of
variables yields a dual diagram with some indices exchanged with respect to the original diagram. At the level of coefficient functions, the relation is given by:
\bea
\text{C}_D[\,J(D,p,\al_1,\al_2,\al_3,\al_4,\al_5)\,] & \underset{(\text{Du})}{=} & \text{C}_D[\,J(D,z,\al_1,\al_4,\al_3,\al_2,\al_5)\,]
\label{duality-def} \\
& \underset{(\text{Du})}{=} & \text{C}_D[\,J(D,z,\al_3,\al_2,\al_1,\al_4,\al_5)\,]
\nonum \\
& \underset{(\text{Du})}{=} & \text{C}_D[\,J(D,z,\al_2,\al_3,\al_4,\al_1,\al_5)\,].
\nonum
\eea
Notice that in the first line, the duality transformation exchanges indices 2 and 4. The two other equalities follow from the symmetries of the diagram
($1 \leftrightarrow 2$, $3 \leftrightarrow 4$ and $1 \leftrightarrow 4$, $2 \leftrightarrow 3$).

Both Fourier transform and duality transform relate diagrams which are in different spaces with different integration rules. By combining them,
it is possible to relate the coefficient functions of two $p$-space diagrams with changed indices:
\be
\text{C}_D[J(D,p,\al_1,\al_2,\al_3,\al_4,\al_5)] \underset{(\text{FT}+\text{Du})}{=} \frac{\Pi_i a(\al_i)}{a(\sum_i \al_i - D)}\,\text{C}_D[J(D,p,\tilde{\al}_2,\tilde{\al}_3,\tilde{\al}_4,\tilde{\al}_1,\tilde{\al}_5)].
\label{Fourier+duality-def}
\ee
Other similar transformations can be obtained from the symmetries of the diagram.

\subsection{Gegenbauer polynomial technique}
\label{chap2:sec:meth:Gegenbauer}

The $x$-space Gegenbauer polynomial technique is a powerful technique allowing to compute many complicated dimensionally regularized Feynman integrals.
In its modern form, it has been introduced by Chetyrkin, Kataev and Tkachov \cite{Chetyrkin:1980pr}, see references therein for earlier contributions. 
Later, subtle and important improvements were brought up by Kotikov \cite{Kotikov:1995cw} and we shall follow this reference 
in our brief review of the technique, see also App.~\ref{app:gegen} for more details on useful formulas.

The basic motivation for this technique lays in the fact that, in multi-loop computations, the complicated part of the integration is often the one over the angular variables.
This task is considerably simplified by expanding some of the propagators in the integrand in terms of the Gegenbauer polynomials (the so-called multipole expansion):
\bea
\frac{1}{(x_1 - x_2)^{2\lambda}}  =
\sum_{n=0}^\infty\,C_n^\lambda(\hat{x}_1 \cdot \hat{x}_2)\,\Bigg[ \frac{(x_1^2)^{n/2}}{(x_2^2)^{n/2+\lambda}}\,\Theta(x_2^2-x_1^2) + (x_1^2 \longleftrightarrow x_2^2) \Bigg]\, ,
\label{chap2:gegenbauer:multipole}
\eea
where $C_n^\lambda$ is the Gegenbauer polynomial of degree $n$ and $\hat{x}=x/\sqrt{x^2}$, and then using the orthogonality relation of Gegenbauer polynomials on the unit $D$-dimensional sphere: 
\be
\frac{1}{\Omega_D}\, \int\,\D_D\, \hat{x}\,C_n^\lambda(\hat{z} \cdot \hat{x})C_m^\lambda(\hat{x} \cdot \hat{z})=\delta_{n,m}\,\frac{\lambda\,\Gamma(n+2\lambda)}{\Gamma(2\lambda)\,(n+\lambda)\,n!},
\qquad \lambda = \frac{D}{2} -1\, ,
\label{chap2:gegenbauer:orth}
\ee
where $\D_D\,\hat{x}$ is the surface element of the unit $D$-dimensional sphere and $\Omega_D = 2 \pi^{D/2} / \Gamma(D/2)$. The Gegenbauer polynomials can be defined from their generating function:

\be
\frac{1}{(1-2xw+w^2)^\beta} = \sum_{k=0}^\infty\,C_k^\beta(x)\,w^k, \quad \qquad C_n^\beta(1) = \frac{\Gamma(n+2\beta)}{\Gamma(2\beta)\,n!}\, ,
\label{chap2:gegenbauer:def}
\ee
with some additional particular values given by:
\be
C_0^\lambda (x) = 1, \quad C_1^\lambda(x) = 2\lambda x, \quad C_2^\lambda(x) = 2\lambda(\lambda+1)x^2-\lambda\, .
\label{chap2:gegenbauer:values}
\ee
For our purpose, it is convenient to express the Gegenbauer polynomials in terms of traceless symmetric tensors \cite{Kotikov:1995cw}:
\be
C_n^\lambda(\hat{x}\cdot \hat{z})\,(x^2\,z^2)^{n/2} = S_n(\lambda)\,x^{\mu_1 \mu_2 \cdots \mu_n}\,z^{\mu_1 \mu_2 \cdots \mu_n}, \qquad S_n(\lambda) =  \frac{2^n \Gamma(n+\lambda)}{n!\,\Gamma(\lambda)}\, .
\label{chap2:gegenbauer:TSTGegen}
\ee
From Eq.~(\ref{chap2:gegenbauer:TSTGegen}) for $x=z$ and the last equation in (\ref{chap2:gegenbauer:def}), we deduce the following equation for products of traceless tensors:
\be
S_n(\lambda)\,z^{\mu_1 \mu_2 \cdots \mu_n}\,z^{\mu_1 \mu_2 \cdots \mu_n} = \frac{\Gamma(n+2\lambda)}{\Gamma(2\lambda)\,n!}\,z^{2n}\, .
\label{chap2:gegenbauer:TSTP}
\ee
With the help of Eq.~(\ref{chap2:gegenbauer:TSTGegen}), Eq.~(\ref{chap2:gegenbauer:multipole}) can be rewritten as:
\be
\frac{1}{(x_1 - x_2)^{2\lambda}}  =
\sum_{n=0}^\infty\,S_n(\lambda)\,x_1^{\mu_1 \cdots \mu_n}\,x_2^{\mu_1 \cdots \mu_n}\,
\Bigg[ \frac{1}{(x_2^2)^{n+\lambda}}\,\Theta(x_2^2-x_1^2) + (x_1^2 \longleftrightarrow x_2^2) \Bigg]\, .
\label{chap2:gegenbauer:multipole2}
\ee
Notice that, for a propagator with arbitrary index, Eq.~(\ref{chap2:gegenbauer:multipole}) can be generalized as:
\bea
\frac{1}{(x_1 - x_2)^{2\beta}}  =
\sum_{n=0}^\infty\,C_n^\beta(\hat{x}_1 \cdot \hat{x}_2)\,\Bigg[ \frac{(x_1^2)^{n/2}}{(x_2^2)^{n/2+\beta}}\,\Theta(x_2^2-x_1^2) + (x_1^2 \longleftrightarrow x_2^2) \Bigg]\, ,
\label{chap2:gegenbauer:multipole3}
\eea
where $C_n^\beta(x)$ can then be related to $C_{n-2k}^\lambda(x)$ ($0 \leq k \leq [n/2]$) with the help of:
\be
C_n^\delta(x) =  \sum_{k=0}^{[n/2]}\,C_{n-2k}^\lambda(x)\,\frac{(n-2k+\lambda)\Gamma(\lambda)}{k!\,\Gamma(\delta)}\,
\frac{\Gamma(n+\delta-k)\Gamma(k+\delta-\lambda)}{\Gamma(n-k+\lambda+1) \Gamma(\delta-\lambda)}\, .
\label{chap2:gegenbauer:diffind}
\ee
Moreover, the series appearing upon expanding the propagators and after performing all integrations may sometimes be resummed in the form of a  generalized
hypergeometric function ${}_3F_2$ of unit argument. There is a very useful transformation property relating such hypergeometric functions.
Even though not directly connected with Gegenbauer polynomials, we mention it here:
\bea
&& {}_3F_2(a,b,c;e,f;1) = \frac{\Gamma(1-a)\Gamma(e)\Gamma(f)\Gamma(c-b)}{\Gamma(e-b)\Gamma(f-b)\Gamma(1+b-a)\Gamma(c)}
\nonum 
\\
&& \quad \times {}_3F_2(b,b-e+1,b-f+1;1+b-c,1+b-a;1) + \big( b \longleftrightarrow c \big)\, .
\label{chap2:gegenbauer:hyperg-tranf0}
\eea
Of peculiar importance is the case where $e=b+1$ in which case the ${}_3F_2$ function can be expressed in terms of another ${}_3F_2$ plus a term involving only products of Gamma functions:
\bea
&&\sum_{p=0}^\infty \frac{\Gamma(p+a) \Gamma(p+c)}{p!\,\Gamma(p+f)}\,\frac{1}{p+b} = \frac{\Gamma(a) \Gamma(1-a)\Gamma(b)\Gamma(c-b)}{\Gamma(f-b)\Gamma(1+b-a)}
\nonum 
\\
\quad && - \frac{\Gamma(1-a)\Gamma(a)}{\Gamma(f-c)\Gamma(1+c-f)} \, \sum_{p=0}^\infty \frac{\Gamma(p+c-f+1) \Gamma(p+c)}{p!\,\Gamma(p+1+c-a)}\,\frac{1}{p+c-b}\, .
\label{chap2:gegenbauer:hyperg-tranf}
\eea

Let's consider some simple examples in order to illustrate the method. We start with the one-loop massless p-type diagram with two arbitrary indices in $x$-space (transformation rules between $x$-space and $p$-space are provided
in Sec.~\ref{chap2:sec:meth:FT+Du}):
\bea
J(D,z,\al,\beta)
= \int \frac{\D^D x}{x^{2\al} (x-z)^{2\beta}}\, , \qquad \D^D x = \frac{1}{2}\,x^{2\lambda}\,d x^2\,d_D\hat{x}\, .
\label{chap2:def:A00}
\eea
Combining Eqs.~(\ref{chap2:gegenbauer:multipole3}) and (\ref{chap2:gegenbauer:diffind}),
the integral can be separated into a radial and an angular part as follows:
\begin{flalign}
J(D,z,\al,\beta) &= \frac{1}{2}\, \sum_{n=0}^\infty \sum_{k=0}^{[n/2]} \int_0^\infty \D x^2 \, (x^2)^{\lambda - \al}\,
\bigg[\frac{(x^2)^{n/2}}{(z^2)^{n/2+\beta}}\,\Theta(z^2-x^2) + (x^2 \leftrightarrow y^2) \Bigg]
\nonum \\
&\times \underbrace{\int \D_D\, \hat{x}\, C_{n-2k}^\lambda(\hat{x} \cdot \hat{z})}_{\Omega_D\,\delta_{n,2k}}\,\frac{(n-2k+\lambda)\,\Gamma(\lambda)}{k!\,\Gamma(\beta)}\,
\frac{\Gamma(n+\beta-k)\Gamma(k+\beta-\lambda)}{\Gamma(n+\lambda+1-k)\Gamma(\beta-\lambda)}\, ,
\end{flalign}
where the orthogonality relation, Eq.~(\ref{chap2:gegenbauer:orth}) has been used to compute the angular part. It then follows that $n$ must be an even integer: $n=2p$ and $k=[n/2]=p$.
The remaining radial integrals are easily performed. The resulting expression can be conveniently written as a sum of two one-fold series:
\begin{flalign}
J(D,z,\al,\beta) &= \frac{\pi^{D/2}}{(z^2)^{\al+\beta-\lambda-1}}\,\frac{1}{\Gamma(\beta)\Gamma(\beta-\lambda)}\,
\nonum \\
&\times \, \sum_{p=0}^\infty\,\frac{\Gamma(p+\beta)\Gamma(p+\beta-\lambda)}{p!\,\Gamma(p+\lambda+1)}\,
\bigg[ \frac{1}{p+\al+\beta-1-\lambda} + \frac{1}{p-\al+\lambda+1} \bigg]\, .
\end{flalign}
This expression can be further simplified by transforming the first sum with the help of Eq.~(\ref{chap2:gegenbauer:hyperg-tranf}) with $a=\beta-\lambda$, $b=\al+\beta-1-\lambda$, $c=\beta$
and $f=\lambda+1$. Indeed, this yields:
\begin{flalign}
& \sum_{p=0}^\infty\,\frac{\Gamma(p+\beta)\Gamma(p+\beta-\lambda)}{p!\,\Gamma(p+\lambda+1)}\,\frac{1}{p+\al+\beta-1-\lambda} =
\\
&\frac{\Gamma(\beta-\lambda)\Gamma(1+\lambda-\al) \Gamma(1+\lambda-\beta) \Gamma(\al + \beta - 1 -\lambda)}{\Gamma(\al) \Gamma(2+2\lambda-\al-\beta)}
%\nonum \\ && \qquad 
- \sum_{p=0}^\infty\,\frac{\Gamma(p+\beta)\Gamma(p+\beta-\lambda)}{p!\,\Gamma(p+\lambda+1)}\,\frac{1}{p-\al+1+\lambda}\, ,
\nonum
\end{flalign}
and the sum on the lhs is simply the opposite of the second sum in $J(D,z,\al,\beta)$. Hence, the sum of the two one-fold series reduces to a product of $\Gamma$-functions
and we recover the well-known result:
\bea
J(D,z,\al,\beta) = \frac{\pi^{D/2}}{(z^2)^{\al+\beta-\lambda-1}}\,G(D,\al,\beta), \qquad G(D,\al,\beta) = \frac{a(\al) a(\beta)}{a(\al+\beta-1-\lambda)}\, ,% \qquad a(\al) = \frac{\Gamma(D/2-\al)}{\Gamma(\al)}\, ,
\eea
where $a(\al) = \Gamma(D/2-\al)/\Gamma(\al)$ and which was given in Eq.~(\ref{chap2:def:one-loop-G-func}) in $p$-space.

We may next generalize this result to the case where a traceless symmetric tensor, see App.~\ref{app:gegen} for more, appears in the numerator:
\bea
J^{\mu_1 \cdots \mu_n}(D,z,\al,\beta)
= \int \D^D x\, \frac{x^{\mu_1 \cdots \mu_n}}{x^{2\al} (x-z)^{2\beta}}\, , \qquad \D^D x = \frac{1}{2}\,x^{2\lambda}\,d x^2\,d_D\hat{x}\, .
\label{chap2:def:An0}
\eea
Dimensional analysis suggests that this integral should have the form:
\be
J^{\mu_1 \cdots \mu_n}(D,z,\al,\beta) = \pi^{D/2}\,\frac{z^{\mu_1 \cdots \mu_n}}{(z^2)^{\al+\beta-\lambda-1}}\,G^{(n,0)}(\al,\beta)\, ,
\label{chap2:def:An0-form}
\ee
where the coefficient function, $G^{(n,0)}(\al,\beta)$, is yet to be determined.
In order to do so, we consider the scalar function:
\be
z^{\mu_1 \cdots \mu_n}\,J^{\mu_1 \cdots \mu_n}(D,z,\al,\beta) =\pi^{D/2}\,\frac{z^{2n}}{(z^2)^{\al+\beta-\lambda-1}}\,\frac{\Gamma(\lambda) \Gamma(n+2\lambda)}{2^n \Gamma(2\lambda) \Gamma(n+\lambda)}\,G^{(n,0)}(\al,\beta)\, ,
\ee
where Eqs.~(\ref{chap2:def:An0-form}) and (\ref{chap2:gegenbauer:TSTP}) have been used.
The corresponding integral can be evaluated by using the relation between traceless symmetric tensors and Gegenbauer polynomials, Eq.~(\ref{chap2:gegenbauer:TSTGegen}):
\begin{flalign}
z^{\mu_1 \cdots \mu_n}\,J^{\mu_1 \cdots \mu_n}(D,z,\al,\beta) = \int \D^D x\, \frac{z^{\mu_1 \cdots \mu_n} x^{\mu_1 \cdots \mu_n}}{x^{2\al} (x-z)^{2\beta}}
= \frac{n! \Gamma(\lambda)}{2^n \Gamma(n+\lambda)}\,\int \D^D x\, \frac{C_n(\hat{z} \cdot \hat{x})\,(x^2 z^2)^{n/2}}{x^{2\al} (x-z)^{2\beta}}\, ,
\end{flalign}
and then expanding the propagator in Gegenbauer polynomials as before. This yields:
\begin{flalign}
&z^{\mu_1 \cdots \mu_n}\,J^{\mu_1 \cdots \mu_n}(D,z,\al,\beta) = \frac{n! \Gamma(\lambda)}{2^n \Gamma(n+\lambda)}\,\frac{1}{2}\,\sum_{p=0}^\infty \sum_{k=0}^{[p/2]}\,
\int dx^2\,(x^2)^{\lambda-\al}
\bigg[\frac{(x^2)^{\frac{p+n}{2}}}{(y^2)^{\frac{p-n}{2}+\beta}}\,\Theta(z^2-x^2) + (x^2 \leftrightarrow y^2) \Bigg]
\nonum \\
& \times \int \D_D\, \hat{x}\,C_n(\hat{z} \cdot \hat{x}) C_{p-2k}^\lambda(\hat{x} \cdot \hat{z})\,\frac{(p-2k+\lambda)\,\Gamma(\lambda)}{k!\,\Gamma(\beta)}\,
\frac{\Gamma(p+\beta-k)\Gamma(k+\beta-\lambda)}{\Gamma(p+\lambda+1-k)\Gamma(\beta-\lambda)}\, .
\end{flalign}
The angular integral is non-zero for $2k=p-n$ which implies that $p$ must have the same parity as $n$ and $p \geq n$. Separate analysis of the even and odd $n$ cases
yield, after some simple manipulations:
\bea
&& z^{\mu_1 \cdots \mu_n}\,J^{\mu_1 \cdots \mu_n}(D,z,\al,\beta) =  \pi^{D/2}\,\frac{z^{2n}}{(z^2)^{\al+\beta-\lambda-1}}\,\frac{\Gamma(\lambda) \Gamma(n+2\lambda)}{2^n \Gamma(2\lambda) \Gamma(n+\lambda)}
\times \nonum \\
&& \qquad \times \, \sum_{m=0}^\infty B(m,n| \beta,\lambda) \left( \frac{1}{m+\al+\beta-1-\lambda} + \frac{1}{m+n-\al+\lambda+1} \right)\, ,
\label{chap2:def:An0-calc}
\eea
where:
\be
B(m,n| \beta,\lambda) = \frac{\Gamma(m+n+\beta)}{m! \Gamma(m+n+\lambda+1) \Gamma(\beta)}\,\frac{\Gamma(m+\beta-\lambda)}{\Gamma(\beta-\lambda)}\, .
\label{chap2:gegenbauer-B-def}
\ee
Comparing Eqs.~(\ref{chap2:def:An0-calc}) and (\ref{chap2:def:An0-form}), we see that the coefficient function
equals the sum of two one-fold series:
\be
G^{(n,0)} (D,\al,\beta) = \sum_{m=0}^\infty B(m,n| \beta,\lambda) \left( \frac{1}{m+\al+\beta-1-\lambda} + \frac{1}{m+n-\al+\lambda+1} \right)\, .
\label{chap2:master-1loop-Gn-series}
\ee
Such a series representation reduces to a product of $\Gamma$-functions upon using the transformation properties of hypergeometric functions:
\be
G^{(n,0)} (D,\al,\beta) = \frac{a_n(\al) a_0(\beta)}{a_n(\al+\beta-\lambda-1)}\, , \qquad a_n(\al) = \frac{\Gamma(n+D/2-\al)}{\Gamma(\al)}\, ,
\ee
in accordance with Eq.~(\ref{chap2:one-loop-Gn}).

The above results yield {\bf integration rules for Feynman integrals involving traceless symmetric tensors and Heaviside functions} which were given in Ref.~\cite{Kotikov:1995cw}. 
From Eq.~(\ref{chap2:def:An0-calc}) we indeed recover the basic results of this reference:
%
%This expansion has then to be combined with the following rules for integrating one-loop massless integrals with $\Theta$-terms:
%
\bea
&&\int \D^D x\, \frac{x^{\mu_1 \cdots \mu_n}}{x^{2\al} (x-y)^{2\beta}}\,\Theta(x^2 - y^2) = \pi^{D/2}\,\frac{y^{\mu_1 \cdots \mu_n}}{(y^2)^{\al+\beta-\lambda-1}}\,
\sum_{m=0}^\infty \frac{B(m,n| \beta,\lambda)}{m+\al+\beta-1-\lambda}\, 
\nonum \\
&& \qquad \stackrel{(\beta = \lambda)}{=} \pi^{D/2}\,\frac{y^{\mu_1 \cdots \mu_n}}{(y^2)^{\al-1}}\,\frac{1}{\Gamma(\lambda)}\,\frac{1}{(\al-1) (n+\lambda)}\, ,
\eea
and
\bea
&&\int \D^D x\, \frac{x^{\mu_1 \cdots \mu_n}}{x^{2\al} (x-y)^{2\beta}}\,\Theta(y^2 - x^2) = \pi^{D/2}\,\frac{y^{\mu_1 \cdots \mu_n}}{(y^2)^{\al+\beta-\lambda-1}}\,
\sum_{m=0}^\infty \frac{B(m,n| \beta,\lambda)}{m+n-\al+1+\lambda}\, 
\nonum \\
&& \qquad \stackrel{(\beta = \lambda)}{=} \pi^{D/2}\,\frac{y^{\mu_1 \cdots \mu_n}}{(y^2)^{\al-1}}\,\frac{1}{\Gamma(\lambda)}\,\frac{1}{(n+\lambda + 1 - \al) (n+\lambda)}\, ,
\eea
where the peculiar case $\beta = \lambda$ has been explicitly displayed. The following more complicated cases are also useful:
\begin{flalign}
&\int \D^D x\, \frac{x^{\mu_1 \cdots \mu_n}}{x^{2\al} (x-y)^{2\beta}}\,\Theta(x^2 - z^2) = \pi^{D/2}\,y^{\mu_1 \cdots \mu_n}\, \left [ \frac{\Theta(y^2-z^2)}{(y^2)^{\al+\beta-\lambda-1}}\,G^{n,0} (\al,\beta) 
\right .
\nonum \\
& \left . \quad + \sum_{m=0}^\infty \frac{B(m,n| \beta,\lambda)}{(z^2)^{\al+\beta-\lambda-1}}\,\left( \left(\frac{y^2}{z^2} \right)^m\,\frac{\Theta(z^2-y^2)}{m+\al+\beta-1-\lambda}
- \left(\frac{z^2}{y^2} \right)^{m+n+\beta}\,\frac{\Theta(y^2-z^2)}{m-\al+n+1+\lambda}\right) \right]\, 
\nonum \\
& \quad \stackrel{(\beta = \lambda)}{=} \pi^{D/2}\,\frac{1}{\Gamma(\lambda)}\,y^{\mu_1 \cdots \mu_n}\, \left [ \frac{\Theta(y^2-z^2)}{(y^2)^{\al-1}}\,\frac{1}{(\al-1)(n+\lambda+1-\al)}
\right .
\nonum \\
& \left . \quad + \frac{1}{(z^2)^{\al-1}}\,\frac{1}{n+\lambda}\,\left( \frac{\Theta(z^2-y^2)}{\al-1}
- \left(\frac{z^2}{y^2} \right)^{n+\lambda}\,\frac{\Theta(y^2-z^2)}{n+1+\lambda-\al}\right) \right]\, ,
\end{flalign}
and
\begin{flalign}
&\int \D^D x\, \frac{x^{\mu_1 \cdots \mu_n}}{x^{2\al} (x-y)^{2\beta}}\,\Theta(z^2 - x^2) = \pi^{D/2}\,y^{\mu_1 \cdots \mu_n}\, 
\left [ \frac{\Theta(z^2-y^2)}{(y^2)^{\al+\beta-\lambda-1}}\,G^{n,0} (\al,\beta) \right .
\nonum \\
& \left . \quad - \sum_{m=0}^\infty \frac{B(m,n| \beta,\lambda)}{(z^2)^{\al+\beta-\lambda-1}}\,
\left( \left(\frac{y^2}{z^2} \right)^m\,\frac{\Theta(z^2-y^2)}{m+\al+\beta-1-\lambda}
- \left(\frac{z^2}{y^2} \right)^{m+n+\beta}\,\frac{\Theta(y^2-z^2)}{m-\al+n+1+\lambda}\right) \right]\, 
\nonum \\
& \quad \stackrel{(\beta = \lambda)}{=} \pi^{D/2}\,\frac{1}{\Gamma(\lambda)}\,y^{\mu_1 \cdots \mu_n}\, \left [ \frac{\Theta(z^2-y^2)}{(y^2)^{\al-1}}\,\frac{1}{(\al-1)(n+\lambda+1-\al)}
\right .
\nonum \\
& \left . \quad - \frac{1}{(z^2)^{\al-1}}\,\frac{1}{n+\lambda}\,\left( \frac{\Theta(z^2-y^2)}{\al-1}
- \left(\frac{z^2}{y^2} \right)^{n+\lambda}\,\frac{\Theta(y^2-z^2)}{n+1+\lambda-\al}\right) \right]\, .
\end{flalign}
%
%The result of the technique is an expression of the diagram in terms of a (multiple) series. For example:
%
%\be
%G^{n,0} (\al,\beta) = \sum_{m=0}^\infty B(m,n| \beta,\lambda) \left( \frac{1}{m+\al+\beta-1-\lambda} + \frac{1}{m+n-\al+\lambda+1} \right)\, .
%\label{master-1loop-Gn-series}
%\ee
%

\section{Renormalization (case of QED$_4$)}

The multi-loop techniques presented above offer a very powerful machinery that allows to compute complicated Feynman diagrams
arising within the perturbative study of a quantum field theory model. Dimensional regularization offered to us a very convenient way to regularize divergences which often occur
in such diagrams while preserving the symmetries of the theory. The renormalization procedure
enters at the last stage of the study in order to give a meaning to the divergent integrals: they are absorbed in some multiplicative constants, 
the so-called {\bf renormalization constant}, thereby allowing to define finite physical quantities such as, \eg, the renormalized mass and couplings.
The renormalization constants are themselves related to {\bf renormalization group functions}, $\ie$, $\beta$-function and anomalous dimensions, which enter as coefficients
of {\bf renormalization group equations}. The solution of these equations allows to access the non-trivial (UV or IR) asymptote of correlation functions. There are many excellent textbooks
devoted to renormalization and the renormalization group technique, see, \eg, \cite{Collins:1984xc,kleinert2001critical,Vasil'evbook,itzykson2012quantum,peskin1995introduction,Weinberg:1995mt}. 
In order to setup some of the notations and conventions that will be used in the next chapters, we present here a brief overview of this technique 
with massless QED$_4$ as an example.

\subsection{Feynman rules, Schwinger-Dyson equations and Ward identities}

The Lagrangian of massless QED$_4$ in a linear covariant gauge reads, in Minkowski space:
\be
L_{\text{QED}_4} = \psibar_\sigma \gamma^\mu \bigg( \I \partial_\mu - e A_\mu \bigg)  \psi^{\sigma} - \frac{1}{4}\,F^{\mu \nu}\,F_{\mu \nu} - \frac{1}{2\xi}\left(\partial_{\mu}A^{\mu}\right)^2\, , 
\label{chap2:model:QED_d}
\ee
where $\xi$ is the gauge fixing parameter, $\psi_\sigma$ a four component spinor of flavour $\sigma=1,\cdots,N_F$, 
the $\gamma_\mu$ are $4\times4$ Dirac matrices satisfying the usual algebra: $\{ \gamma^\mu,\gamma^\nu \} = 2 g^{\mu \nu}$
where $g^{\mu \nu} = {\rm diag}(1,-1,-1,\cdots,-1)$ is the metric tensor in $d$ space-time dimensions and all parameters and fields appearing in Eq.~(\ref{chap2:model:QED_d}) are the bare ones.
The most common gauge choices are:
\be
\xi = 0 ~~~(\text{Landau gauge}), \qquad  \xi=1~~~(\text{in Feynman gauge})\, .
\label{chap2:gauges}
\ee
The Feynman rules associated with Eq.~(\ref{chap2:model:QED_d}) together with their graphical representations are given by:
\begin{subequations}
\label{chap2:QED4:FR}
\bea
S_0(p) &=& \frac{\I}{\Sp} \qquad = \qquad
      \parbox{20mm}{
        \begin{fmfgraph*}(20,10)
          %\fmfpen{thick}
          \fmfleft{in}
          \fmfright{out}
          \fmf{fermion,label=$p$}{in,out}
        \end{fmfgraph*}
      } \quad \, ,
      \label{chap2:QED4:FR:S} \\
-\I e \Gamma_0^\mu &=& -ie\gamma^\mu  \qquad = \qquad
      \parbox{20mm}{
        \begin{fmfgraph*}(20,10)
          %\fmfpen{thick}
          \fmfleft{p}
          \fmflabel{$\mu$}{p}
          \fmfright{ei,eo}
          \fmf{boson}{p,v}
          \fmf{vanilla}{ei,v}
          \fmf{fermion}{v,eo}
          \fmfdot{v}
        \end{fmfgraph*}
      }\quad \, ,
      \label{chap2:QED4:FR:Gamma} \\
D_0^{\mu \nu}(q) &=& \frac{-\I}{q^2}\left( g^{\mu \nu} - (1-\xi) \frac{q^{\mu} q^{\nu}}{q^2} \right) \qquad = \qquad
      \parbox{20mm}{
        \begin{fmfgraph*}(20,10)
          %\fmfpen{thick}
          \fmfleft{in}
          \fmfright{out}
          \fmflabel{$\mu$}{in}
          \fmflabel{$\nu$}{out}
          \fmf{boson,label=$q$}{in,out}
        \end{fmfgraph*}
      } \qquad \, .
      \label{chap2:QED4:FR:D}
\eea
\end{subequations}
Notice that the photon propagator consists of a longitudinal and a transverse part which are defined and represented graphically as:
\begin{subequations}
\label{chap2:QED4:FR:Dperp+Dpara}
\begin{flalign}
D_{0 \perp}^{\mu \nu}(q) &= d_{0 \perp}(q^2)\,\left( g^{\mu \nu} - \frac{q^{\mu} q^{\nu}}{q^2} \right) \quad = \qquad
      \parbox{15mm}{
        \begin{fmfgraph*}(15,10)
          %\fmfpen{thick}
          \fmfleft{in}
          \fmfright{out}
          \fmflabel{$\mu$}{in}
          \fmflabel{$\nu$}{out}
          \fmf{boson,label=$\perp$}{in,out}
        \end{fmfgraph*}
      } \qquad \quad, \qquad d_{0 \perp}(q^2) = \frac{-\I}{q^2} \, ,
\label{chap2:QED4:FR:Dperp} \\
D_{0 \parallel}^{\mu \nu}(q) &= \qquad \,d_{0 \parallel}(q^2)\, \frac{q^{\mu} q^{\nu}}{q^2} \quad \qquad ~~= \qquad
      \parbox{15mm}{
        \begin{fmfgraph*}(15,10)
          %\fmfpen{thick}
          \fmfleft{in}
          \fmfright{out}
          \fmflabel{$\mu$}{in}
          \fmflabel{$\nu$}{out}
          \fmf{boson,label=$\parallel$}{in,out}
        \end{fmfgraph*}
      } \qquad \quad, \qquad ~~ d_{0 \parallel}(q^2) = \frac{-\I \xi}{q^2} \, .
\label{chap2:QED4:FR:Dpara}
\end{flalign}
\end{subequations}

The dressed Green's functions satisfy a set of coupled Schwinger-Dyson equations. The (unrenormalized) fermion Schwinger-Dyson equation reads:
\be
S(p) \quad =  \quad
      \parbox{17mm}{
        \begin{fmfgraph*}(17,10)
          %\fmfpen{thick}
          \fmfleft{in}
          \fmfright{out}
          \fmf{heavy,label=$p$}{in,out}
%         \marrow{a}{down}{bot}{$p$}{i1,v1}
        \end{fmfgraph*}
      }
\quad = \quad
 S_0(p) + S_0(p) \left( -i \Sigma(p) \right) S(p)\, ,
\label{chap2:SD:S}
\ee
where the fermion self-energy is given by:
\vspace{6mm}
\be
-\I \Sigma(p) \quad =
\quad
      \parbox{20mm}{
    \begin{fmfgraph*}(20,15)
      \fmfleft{vi}
      \fmfright{vo}
%      \fmf{double}{in,vi}
      \fmf{heavy,tension=0.2,label=$k$}{vi,vo}
      \fmf{dbl_wiggly,left,tension=0.2,label=$p-k$}{vi,vo}
%      \fmf{double}{vo,out}
      \fmfdot{vi}
      \fmfv{decor.shape=circle,decor.filled=shaded,decor.size=7thick}{vo}
    \end{fmfgraph*}
      }
 \quad = \quad
\int [\D^4 k] \,(-\I e \gamma^\mu)\,D_{\mu \nu}(p-k)\,S(k)\,(-\I e \Gamma^\nu(p,k))\, ,
\label{chap2:SD:Sigma}
\ee
and $D_{\mu \nu}(q)$ and $\Gamma^\nu(p,k)$ are the dressed photon propagator and vertex part, respectively.
The solution of Eq.~(\ref{chap2:SD:S}) can be written in the form:
\be
-\I S(p) = \frac{1}{\Sp}\frac{1}{1-\Sigma_V(p^2)}, \qquad \Sigma(p) = \Sp \Sigma_V(p^2)\, ,
\label{chap2:Sint}
\ee
where the parametrization of the self-energy is suited to the massless case.
Similarly, the (unrenormalized) Schwinger-Dyson equation for the photon propagator reads:
\bea
D^{\mu \nu}(q) \quad = \quad
      \parbox{20mm}{
        \begin{fmfgraph*}(20,10)
          %\fmfpen{thick}
          \fmfleft{in}
          \fmfright{out}
          \fmf{dbl_wiggly,label=$q$}{in,out}
%         \fmf{crossed}{v,v}
        \end{fmfgraph*}
      }
\quad = \quad
D_0^{\mu \nu}(q) +
D_0^{\mu \sigma}(q)\,i\Pi_{\sigma \rho}(q)\,D^{\rho \nu}(q) \, ,
\label{chap2:SD:D}
\eea
where the photon self-energy is given by:
\be
\I \Pi^{\mu \nu}(q) \quad =
\quad
  \parbox{20mm}{
  \begin{fmfgraph*}(20,18)
      \fmfleft{vi}
      \fmfright{vo}
      \fmf{heavy,right=0.6,label=$k$,l.d=0.1h}{vi,vo}
      \fmf{heavy,right=0.6,label=$k+q$,l.d=0.05h}{vo,vi}
      \fmfdot{vi}
      \fmfv{decor.shape=circle,decor.filled=shaded,decor.size=7thick}{vo}
  \end{fmfgraph*}
}
 \quad = \quad
-\int [\D^4 k]\, \Tr \bigg[ (-\I e \gamma^\mu)\,S(k)\,(-\I e \Gamma^\nu(k+q,k))\,S(k+q) \bigg]\, .
\label{chap2:SD:Pi}
\ee
Because of current-conservation, the following Ward identity holds: $k_\mu \Pi^{\mu \nu} (q)= k_\nu \Pi^{\mu \nu} (q)= 0$, as
a consequence of which the photon-self-energy is purely transverse:
\be
\Pi^{\mu \nu} (q)=\left( g^{\mu \nu}q^2 - q^{\mu} q^{\nu} \right)\,\Pi(q^2)\, .
\label{chap2:def:Pimunu}
\ee
Solving Eq.~(\ref{chap2:SD:D}), only the transverse part gets affected by interactions:
\begin{subequations}
\label{chap2:Dint}
\bea
&&D^{\mu \nu}(q) = d_\perp(q^2)\,\left( g^{\mu \nu} - \frac{q^{\mu} q^{\nu}}{q^2} \right) + d_\parallel(q^2)\frac{q^{\mu} q^{\nu}}{q^2},
\label{chap2:Dint-a} \\ 
&&d_\perp(q^2) = \frac{- \I}{q^2\,(1-\Pi(q^2))}, \qquad d_\parallel(q^2) = \frac{-\I \xi}{q^2}\, . 
\label{chap2:Dint-b}
\eea
\end{subequations}
Finally, the (unrenormalized) Schwinger-Dyson equation for the fermion-photon vertex reads:
\be
\Gamma^\mu(p,p') = \gamma^\mu + \Lambda^\mu (p,p')\, ,
\label{chap2:SD:Gamma}
\ee
where 
\be
\Lambda^\mu (p,p') = \int [\D^4 k]\, S(p+k) \Gamma^\mu(p+k,p'+k) S(p'+k)\, K(p+k,p'+k,k)\, ,
\label{chap2:SD:Lambda}
\ee
and $K$ is the (2-particle irreducible) fermion-antifermion scattering kernel. Notice that vertex part and fermion self-energy
are related to each other with the help of the following Ward identity:
\be
q_\mu \Gamma^\mu(p,p') = S^{-1}(p') - S^{-1}(p)\, ,
\label{chap2:Ward:Sigma}
\ee
where $q = p'-p$ is the exchanged momentum.

\subsection{Renormalization constants, $\beta$-function and anomalous dimensions}
\label{chap2:subsec:ren}

\begin{figure}
  \begin{center}
  a)
    \begin{fmfgraph*}(30,30)
      %\fmfpen{thick}
      \fmfleft{in}
      \fmflabel{$\mu$}{in}
      \fmfright{out}
      \fmflabel{$\nu$}{out}
      \fmf{boson}{in,ve}
      \fmf{fermion,right,tension=0.2,label=$k$}{ve,vw}
      \fmf{fermion,right,tension=0.2,label=$k+q$}{vw,ve}
      \fmf{boson}{vw,out}
      \fmfdot{ve,vw}
    \end{fmfgraph*}
  \qquad \quad
  b)
    \begin{fmfgraph*}(30,30)
      %\fmfpen{thick}
      \fmfleft{in}
      \fmflabel{$p$}{out}
      \fmfright{out}
      \fmflabel{$p$}{out}
      \fmf{plain}{in,vi}
      \fmf{fermion,tension=0.2,label=$k$}{vi,vo}
      \fmf{boson,left,tension=0.2,label=$p-k$}{vi,vo}
      \fmf{plain}{vo,out}
      \fmfdot{vi,vo}
    \end{fmfgraph*}
  \qquad \quad
  c)
    \begin{fmfgraph*}(30,30)
      %\fmfpen{thick}
      \fmfleft{in}
      \fmflabel{$\mu$}{in}
      \fmfright{e1,e2}
      \fmflabel{$p$}{e1}
      \fmflabel{$p'$}{e2}
      \fmf{boson,label=$q$}{in,vi}
      \fmf{fermion}{e1,v1}
      \fmf{fermion,tension=0.7,label=$p+k$,label.side=left}{v1,vi}
      \fmf{fermion,tension=0.7,label=$p'+k$,label.side=left}{vi,v2}
      \fmf{fermion}{v2,e2}
      \fmffreeze
      \fmf{boson,right,tension=0.7,label=$k$}{v1,v2}
      \fmfdot{vi,v1,v2}
    \end{fmfgraph*}
  \caption{\label{chap2:fig:one-loop}
  One-loop diagrams: a) gauge field self-energy, b) fermion self-energy and c) fermion-gauge field vertex.}
  \end{center}
\end{figure}
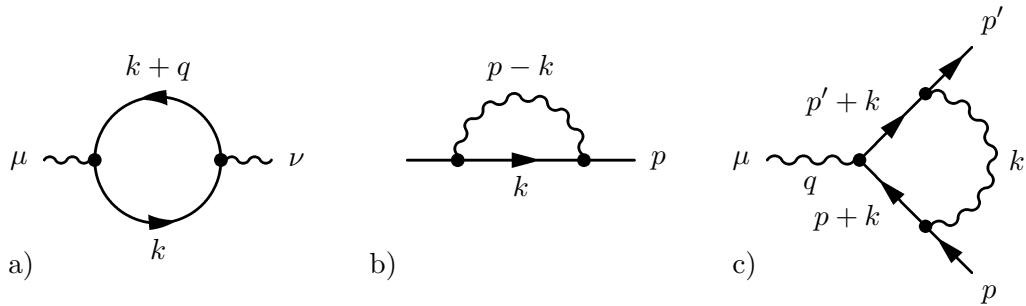

Dimensional analysis shows that all three integrals above are divergent in $d=4$. To see this in a little bit more detailed way, let's work in $d=4-2\veps$ and
perform the dimensional analysis at the level of (\ref{chap2:model:QED_d}). Then, in units of momentum (or mass):
\be
[\psi] = \frac{d-1}{2}= \frac{3}{2}-\veps, \qquad [A^\mu] = \frac{d-2}{2} = 1 -\veps, \qquad [e]=2 - \frac{d}{2} = \veps\, ,
\label{chap2:QED4:dim}
\ee
which shows that the coupling constant is dimensionless in $d=4$ while it has positive mass dimension in $d<4$ and negative mass dimension in $d>4$. Accordingly, the theory
is renormalizable in $d=4$, super-renormalizable in $d<4$ (as in the case of QED$_3$) and non-renormalizable in $d>4$. Similar power counting arguments
yield the {\bf superficial degree of divergence} (SDD) of an arbitrary multi-leg $L$-loop Feynman graph $G$:
\be
\om(G) = d - \frac{d-2}{2}\,N_\gamma - \frac{d-1}{2}\,N_e + \frac{d-4}{2}\,V\, ,
\label{chap2:QED4:w(G)}
\ee
where $N_\gamma$ is the number of external photon lines, $N_e$ the number of external fermion lines and $V$ the number of vertices.
In $d=4$, the SDD does not depend on the number of vertices which implies that there is a finite number of divergent structures again in accordance with the renormalizability of the theory.
Super-renormalizable theories, on the other hand, have a finite number of divergent diagrams. And non-renormalizable ones have an infinite number of divergent structures.
Going back to the three integrals above, Eq.~(\ref{chap2:QED4:w(G)}) shows that they are indeed all divergent in $d=4$: the vertex part diverges logarithmically ($\om(\Gamma)=0$),
the fermion self-energy linearly ($\om(\Sigma) =1$) and the photon-self-energy quadratically ($\om(\Pi) =2$). Keeping in mind the tensorial structure of $\Sigma(p)$ and $\Pi^{\mu \nu}(q)$,
the effective degree of divergence (of $\Sigma_V(p)$ and $\Pi(q^2)$) is zero. Hence, all three structures diverge logarithmically. At one-loop order, they are displayed on Fig.~\ref{chap2:fig:one-loop}.
At this point, let's notice that it is not because a graph has a negative SDD ($\om(G)<0$) that it is necessarily ``finite''.
Indeed, according to Weinberg's theorem \cite{Weinberg:1959nj}:

\begin{theorem}[Weinberg (1959)]
A Feynman diagram $G$ is absolutely convergent if its superficial degree of divergence, $\om(G)$, is negative and if the superficial degrees of divergence, $\om(\gamma)$, associated to all of its 
subgraphs $\gamma$ are also negative. 
\end{theorem}

\ni So $\om(G)$ of Eq.~(\ref{chap2:QED4:w(G)}) corresponds to the {\bf overall SDD} of graph $G$. When considering multi-loop diagrams, one often encounters diagrams with divergent subgraphs 
of the type shown on Fig.~\ref{chap2:fig:one-loop}. Dealing with these {\bf subdivergences} (nested or overlapping) is one of the central aspect of renormalization theory. We shall come back on this later in this section.

The Lagrangian (\ref{chap2:model:QED_d}) can then be re-written in the same form in terms of renormalized quantities:
\be
L_{\text{QED}_4} = Z_\psi\,\psibar_{r}\I {\slashed \partial} \psi_{r}(x)  
- Z_\Gamma \,e_r \mu^{\veps}\,\psibar_{r}(x) {\slashed A}_r  \psi_{r}(x) 
- \frac{1}{4}\,Z_A\,F_r^{\mu \nu}\,F_{r,\mu \nu} - \frac{1}{2\xi_r} Z_g\,\left(\partial_{\mu}A_r^{\mu}\right)^2\, ,
\label{chap2:model:QED_d-R}
\ee
involving a finite number of renormalization constants that absorb all singularities:
\be
\psi = Z_{\psi}^{1/2} \psi_r, \quad A = Z_A^{1/2} A_r, \quad e = Z_e e_r \mu^{\veps} = \frac{Z_\Gamma}{Z_\psi Z_A^{1/2}}\, e_r \mu^\veps, \quad \xi = Z_\xi \xi_r = \frac{Z_A}{Z_g} \xi_r\, ,
\label{chap2:model:QED_d:Z}
\ee
where the subscript $r$ denotes renormalized quantities and the renormalization scale, $\mu$, has been introduced in such a way that $e_r$ is dimensionless in $d=4-2\veps$ dimensions.
In the $\text{MS}$ scheme, these constants take the simple form:
\be
Z_x(\al_r,\xi_r) = 1 + \delta Z_x (\al_r,\xi_r) = 1 + \sum_{l=1}^\infty \sum_{j=1}^l Z_x^{(l,j)}(\xi_r)\,\frac{\al_r^l}{\veps^j} \qquad (x \in \{ \psi,A,e,\xi,\Gamma,g \})\, ,
\label{chap2:model:QED_d:Z-exp}
\ee
where $\al_r = e_r^2/(4\pi)$~\footnote{The true expansion parameter is actually: $\bar{\al} = \al/(4\pi) = e_r^2/(16\pi^2)$.} and
$l$ runs over the number of loops at which UV singularities are subtracted. The attractive feature of the $\text{MS}$ scheme is that the $Z_x$ do not depend on momentum or mass; furthermore, 
the dependence on $\mu$ is only through $\al_r$ and/or $\xi_r$. So the $Z_x$ depend only on $\al_r(\mu)$, $\veps$ and eventually $\xi_r(\mu)$. %(and not on momentum or mass). 
From the renormalization constants, it is possible to compute the $\beta$-function:
\be
\beta(\al_r) = \mu \, \frac{\partial \al_r}{\partial \mu}\bigg|_B \qquad (Z_\al = Z_e^2)\, ,
\label{chap2:model:QED_d:beta}
\ee
where the subscript $B$ indicates that bare parameters, which do not depend on $\mu$, are fixed. Eq.~(\ref{chap2:model:QED_d:beta}) encodes the dependence of the coupling constant on the energy scale.
As we shall see in the following, $Z_\al$ does not depend on $\xi_r$ (it is actually related to $Z_A$ which is gauge-invariant).
The $\beta$-function can then be re-written as:
\bea
\beta(\al_r) &=& \mu \, \frac{\partial \bigg( \mu^{-2\veps} Z_{\al}^{-1} \al \bigg)}{\partial \mu} 
= -2\veps \al_r + \al_r Z_\al \mu \, \frac{\partial Z_{\al}^{-1}}{\partial \mu} 
= -2\veps \al_r - \beta(\al_r) \al_r \frac{\partial \log Z_\al}{\partial \al_r}
\nonum \\
&=& \frac{-2\veps \al_r}{1+\al_r \frac{\partial \log Z_\al}{\partial \al_r}}\, .
\label{chap2:model:QED_d:beta2}
\eea
From this equation together with the general expression of $Z_\al$, Eq.~(\ref{chap2:model:QED_d:Z-exp}), the $\beta$-function can be written as a multiple-series:
\be
\beta(\al_r) = - 2\veps \al_r\,\sum_{n=0}^{+\infty} (-1)^n \, \bigg[ \al_r \frac{\partial}{\partial \al_r}\,\sum_{m=1}^{\infty} \frac{(-1)^{m+1}}{m}\,
\left(  \sum_{l=1}^\infty \sum_{j=1}^l Z_\al^{(l,j)}\,\frac{\al_r^l}{\veps^j} \right)^m  \bigg]^n\, .
\ee
Then, asking for the $\beta$-function to be finite in the limit $\veps \ra 0$, yields the very simple expression:~\footnote{Notice that, for the beta-function and anomalous dimensions, 
the vanishing of higher order terms in $1/\veps^k$ ($k>1$) yields constraints whereby the coefficients $Z_x^{(l,k)}$ may be expressed in terms of coefficients of lower $k$. 
In high order calculations, such relations provide an important check of the validity of the results. Anticipating the next Chapter, 
it was for example shown in Ref.~\cite{Kotikov:2013eha} that such relations hold at two-loop order for the coefficients of the wave-function renormalization constant in reduced QED.}
%~\footnote{Using the conventions of, \eg,  Grozin \cite{grozin2007lectures}, 
%only changes the result by simple global factors:
%%
%\be
%\tilde{\beta}(\al_r) = \mu \, \frac{\partial \al_r}{\partial \mu} =  2 \beta(\al_r), \qquad \tilde{\gamma}_X(\al_r) = \mu \, \frac{\D \log Z_X }{\D \mu} = -2\gamma_X(\al_r)\, .
%\ee 
%%
%}
%
\be
\beta(\al_r) = -2\veps \al_r + \sum_{l=0}^\infty \beta_l \al_r^{l+2}, \qquad \beta_l = 2(l+1)\,Z_\al^{(l+1,1)} = - 2(l+1)\,Z_A^{(l+1,1)}\, ,
\label{chap2:model:QED_d:beta3}
\ee
where the coefficients $\beta_l$ are completely determined by the simple $1/\veps$ poles in $Z_\al$ and the last equality is based on the Ward identity $Z_\al = Z_A^{-1}$, see Eq.~(\ref{chap2:WI:Zpsi}) below. 
Similarly, one may compute the field anomalous dimensions:
\be
\gamma_x(\al_r,\xi_r) = - \mu \,\frac{\D \log Z_x(\al_r,\xi_r)}{\D \mu}\bigg|_B \qquad (x \in \{ \psi,A \})\, .
\label{chap2:model:QED_d:anomalousdims}
\ee
In the case of the photon anomalous dimension, $Z_A$ does not depend on $\xi_r$. 
Proceeding along the same lines as for the $\beta$-function above, its expression in the limit $\veps \ra 0$ reads:
\be
\gamma_A(\al_r) = -\beta(\al_r)\,\frac{\partial \log Z_A(\al_r)}{\partial \al_r} = \sum_{l=0}^\infty \gamma_{A,l} \al_r^{l+1}, \qquad \gamma_{A,l}= 2(l+1)\,Z_A^{(l+1,1)}\, .
\label{chap2:model:QED_d:anomalousdims-A}
\ee
In the case of the fermion anomalous dimension, we have:
\be
\gamma_\psi(\al_r,\xi_r) = - \beta(\al_r)\,\frac{\partial \log Z_\psi(\al_r,\xi_r)}{\partial \al_r} - \xi_r \gamma_A(\al_r)\,\frac{\partial \log Z_\psi(\al_r,\xi_r)}{\partial \xi_r}\, ,
\ee
where the second term contains only singular contributions in the limit $\veps \ra 0$. Hence, in this limit:
\be
\gamma_\psi(\al_r,\xi_r) = \sum_{l=0}^\infty \gamma_{\psi,l}(\xi_r) \al_r^{l+1}, \qquad \gamma_{\psi,l}(\xi_r)= 2(l+1)\,Z_\psi^{(l+1,1)}(\xi_r)\, .
\label{chap2:model:QED_d:anomalousdims-psi}
\ee

The renormalization constants also relate renormalized and bare propagators as follows:
\begin{subequations}
\label{chap2:renormalized-propagators}
\bea
S(p;\al,\xi) &=& Z_\psi(\al_r) S_r(p;\al_r,\xi_r,\mu)\, , 
\label{chap2:renormalized-S} \\
D^{\mu \nu} (q;\al,\xi) &=& Z_A(\al_r) D_r^{\mu \nu} (q;\al_r,\xi_r,\mu)\, ,
\label{chap2:renormalized-D}\\
\Gamma^\mu(p,p'; \al,\xi) &=& Z_\Gamma^{-1} (\al_r) \Gamma_r^\mu(p,p'; \al_r,\xi_r, \mu)  \, ,
\label{chap2:renormalized-Gamma}
\eea
\end{subequations}
where the bare propagators do not depend on $\mu$. This allows to write renormalized Schwinger-Dyson equations where the renormalized photon and fermion self-energies are defined as:
\begin{subequations}
\label{chap2:renormalized-self-energies}
\bea
\Pi_r(q^2;\al_r,\mu) &=& Z_A(\al_r)\, \Pi(q^2;\al_r,\mu) - \big( Z_A(\al_r)-1 \big)\, ,
\label{chap2:renormalized-Pi}\\
\Sigma_{Vr}(p^2;\al_r,\xi_r, \mu) &=& Z_\psi(\al_r,\xi_r) \,\Sigma_{V}(p^2; \al_r,\xi_r, \mu) - \big( Z_\psi(\al_r,\xi_r) -1  \big)\, ,
\label{chap2:renormalized-Sigma}\\
\Lambda_r^{\mu}(p,p'; \al_r,\xi_r, \mu) &=& Z_\Gamma(\al_r,\xi_r)\, \Lambda^\mu(p,p') + \big( Z_\Gamma(\al_r,\xi_r) - 1 \big)\,\gamma^\mu\, .
\label{chap2:renormalized-Lambda} 
\eea
\end{subequations}
The Ward identities, together with Eqs.~(\ref{chap2:renormalized-propagators}), imply that the renormalization constants are constrained and therefore not all independent.
As a matter of fact, substituting the bare propagators for the renormalized ones in (\ref{chap2:Ward:Sigma}) and requiring that the final expression be finite, yields a renormalized Ward identity
together with the constraint:
\be
Z_\Gamma = Z_\psi \quad \Ra \quad e = Z_A^{-1/2} e_r \mu^{\veps}  \qquad (Z_e = Z_A^{-1/2})\, ,
\label{chap2:WI:Zpsi}
\ee
so that the renormalization of the charge is completely determined by the renormalization of the gauge field. Similarly, substituting the bare fields and parameters in the photon propagator 
Eq.~(\ref{chap2:Dint}) yields:
\bea
Z_g=1 \quad \Ra \quad \xi = Z_A \xi_r \qquad (Z_\xi = Z_A)\, ,
\label{chap2:WI:Zxi}
\eea
so that the renormalization of the gauge fixing parameter is also completely determined by the renormalization of the gauge field.\footnote{Combining Eqs.~(\ref{chap2:WI:Zpsi}) and (\ref{chap2:WI:Zxi}) we see that:
$\xi \al = Z_A Z_e^2 \xi_r \al_r = \xi_r \al_r$, so that the product $\xi \al$ is not renormalized.} With these important simplifications, the Lagrangian (\ref{chap2:model:QED_d-R}) can be written as:
\be
L_{\text{QED}_4} = Z_\psi\,\psibar_{r} \bigg( \I {\slashed \partial} - e_r \mu^{\veps}{\slashed A}_r  \bigg) \psi_{r}(x)
- \frac{1}{4}\,Z_A\,F_r^{\mu \nu}\,F_{r,\mu \nu} - \frac{1}{2\xi_r}\,\left(\partial_{\mu}A_r^{\mu}\right)^2\, ,
\label{chap2:model:QED_d-R2}
\ee
where only $Z_\psi$ and $Z_A$ appear as independent renormalization constants.  

Finally, let us note that from Eq.~(\ref{chap2:renormalized-propagators}) we may write renormalization group equations:
\begin{subequations}
\label{chap2:RG-eqns}
\bea
\mu^2\,\frac{\D S_r(p;\al_r,\xi_r,\mu)}{\D \mu^2} &=& \gamma_\psi(\al_r,\xi_r) S_r(p;\al_r,\xi_r,\mu)\, ,
\label{chap2:RG-eqn-S} \\
\mu^2\,\frac{\D D_r^{\mu \nu} (q;\al_r,\xi_r,\mu)}{\D \mu^2} &=& \gamma_A(\al_r) D_r^{\mu \nu} (q;\al_r,\xi_r,\mu)\, ,
\label{chap2:RG-eqn-D}
\eea
\end{subequations}
where the total derivative with respect to $\mu^2$ can be written in partial differential form as:
\be
\mu^2\,\frac{\D }{\D \mu^2} = \mu^2\,\frac{\partial }{\partial \mu^2} + \beta(\al_r)\,\frac{\partial }{\partial \al_r} + \gamma_A(\al_r)\,\xi_r\,\frac{\partial }{\partial \xi_r}\, .
\ee
The solution to Eqs.~(\ref{chap2:RG-eqns}) yields the asymptotic forms of the propagators which are determined by the $\beta$-function and anomalous dimensions that were defined above.

\end{fmffile}

\begin{fmffile}{fmf-chap2b}

\subsection{Renormalization methods}

From the last paragraph, we see that the main goal of multi-loop calculations is to determine the renormalization constants and, in particular, the coefficients of the simple $1/\veps$ poles
which determine the $\beta$-function and anomalous dimensions of fields. There are essentially two ways to do this.

The simplest way is to consider the (unrenormalized) Feynman rules Eq.~(\ref{chap2:QED4:FR}) and compute the various self-energies order by order in perturbation theory. This is the so-called
{\bf conventional renormalization}. The propagators can then be written as a double series in $\al$ and $\veps$. 
Upon substituting the bare parameters and fields for the renormalized ones, the renormalization constants, which are constrained to have the form (\ref{chap2:model:QED_d:Z-exp}) in the 
$\text{MS}$ scheme, are determined from the fact that they subtract all singular contributions.~\footnote{In the $\text{MS}$ and related schemes, no additional ``renormalization conditions'' are needed.} 
This simple method works fine at low orders (two-loop), see the textbook \cite{grozin2007lectures} where the method is applied to QED$_4$. 

A more efficient method is the well known {\bf method of counter-terms} due to Bogoliubov and Shirkov \cite{Bogolyubov:1956gh} and Bogoliubov and Parasiuk~\cite{Bogoliubov:1957gp}. In this method, the Lagrangian 
(\ref{chap2:model:QED_d}), or equivalently (\ref{chap2:model:QED_d-R2}), is written in terms of renormalized fields and parameters as follows:
\begin{subequations}
\label{chap2:model:QED_d:BPH-L}
\bea
L_{\text{QED}_4} &=& L_r + L_{\text{ct}}\, ,
\label{chap2:model:QED_d:BPH-L:decomp}
\\
L_r &=& \psibar_{r} \bigg( \I {\slashed \partial} - e_r \mu^{\veps} {\slashed A}_r  \bigg) \psi_{r}(x)
- \frac{1}{4}\,F_r^{\mu \nu}\,F_{r,\mu \nu} - \frac{1}{2\xi_r}\,\left(\partial_{\mu}A_r^{\mu}\right)^2
\label{chap2:model:QED_d:BPH-L:Lr}
\\
L_{\text{ct}} &=& \delta Z_\psi\,\psibar_{r} \bigg( \I {\slashed \partial} - e_r \mu^{\veps} {\slashed A}_r  \bigg) \psi_{r}(x)
- \frac{1}{4}\,\delta Z_A\,F_r^{\mu \nu}\,F_{r,\mu \nu} \, ,
\label{chap2:model:QED_d:BPH-L:Lct}
\eea
\end{subequations}
where $L_r$ is the renormalized Lagrangian and $L_{\text{ct}}$ is the counterterm Lagrangian. An important property of (\ref{chap2:model:QED_d:BPH-L:Lct}) is that all counterterms 
are local, \ie,  polynomial in momentum. This fact has been proved by Collins \cite{Collins:1974bg} (see also the textbook \cite{Collins:1984xc}):

\begin{theorem}[Collins (1974)]
A Feynman diagram $G$ of superficial degree of divergence $\om(G)$ has a (UV) counterterm which is polynomial in the external momenta of $G$ and in the massive parameters of the Lagrangian.
The dimensions of the terms in the polynomial are at most $\om(G)$.
\end{theorem}
 
\ni Because of this, if a non-local contribution is added to the free field action, it will {\bf not} be renormalized, \ie, $\delta Z_{\text{non-local}} =0$ (see Ref.~\cite{Vasil'evbook} page 223); this will play an important role
in discussing reduced QED and models of planar condensed matter systems. Moreover, the counterterm Lagrangian is invariant under a $U(1)$ gauge transformation:
\be
\psi \ra e^{\I e_r \mu^{\veps} \phi} \psi, \qquad \psibar \ra  e^{-\I e_r \mu^{\veps} \phi} \psibar, \qquad A_\mu \ra A_\mu + \partial_\mu \phi\, .
\label{chap2:model:QED4:U(1)-GT}
\ee
If gauge non-invariant counterterms were added, they would vanish, \ie, $\delta Z_{\text{gauge-non-invariant}} =0$ (see Ref.~\cite{Collins:1984xc} page 294); this is the case of the gauge-fixing term.
Finally, the renormalization constants in the $\text{MS}$ scheme, which have the form (\ref{chap2:model:QED_d:Z-exp}), do not depend on masses and external momenta (and, in the $\overline{\text{MS}}$ scheme,
on $\gamma_E$ and $\zeta_2$ as well) \cite{Vladimirov:1979zm}. This observation is at the basis of the so-called infrared rearrangement (IRR) method \cite{Chetyrkin:1980pr} that we shall come back on at the end of this section.

The basic method of counterterms then amounts to use renormalized Feynman rules (rules identical to those of Eq.~(\ref{chap2:QED4:FR}) with bare parameters replace by renormalized ones: 
$e \ra e_r$, $\xi \ra \xi_r$) together with three counterterm vertices:
\begin{subequations}
\label{chap2:QED4:ct-vertices}
\bea
      \parbox{20mm}{
        \begin{fmfgraph*}(20,10)
	\fmfcmd{
    		% Please let me know if there?s a more efficient way to do this
	    	path quadrant, q[], otimes;
    		quadrant = (0, 0) -- (0.5, 0) & quartercircle & (0, 0.5) -- (0, 0);
    		for i=1 upto 4: q[i] = quadrant rotated (45 + 90*i); endfor
    		otimes = q[1] & q[2] & q[3] & q[4] -- cycle;
	}
	\fmfwizard
          %\fmfpen{thick}
          \fmfleft{in}
          \fmfright{out}
          \fmf{fermion,label=$p$}{in,v}
	  \fmf{plain}{v,out}
	  %\fmfv{decor.shape=circle,decor.filled=empty,decor.size=5thick}{v}
	  %\fmfv{decor.shape=cross,decor.filled=empty,decor.size=5thick}{v}
	  \fmfv{d.sh=otimes,d.f=empty,d.si=.15w}{v}
        \end{fmfgraph*}
      } \qquad &=& \quad \I \Sp \, \delta Z_\psi
      \label{chap2:QED4:FR:Sct} \\
      \parbox{20mm}{
        \begin{fmfgraph*}(20,10)
        \fmfcmd{
                % Please let me know if there?s a more efficient way to do this
                path quadrant, q[], otimes;
                quadrant = (0, 0) -- (0.5, 0) & quartercircle & (0, 0.5) -- (0, 0);
                for i=1 upto 4: q[i] = quadrant rotated (45 + 90*i); endfor
                otimes = q[1] & q[2] & q[3] & q[4] -- cycle;
        }
        \fmfwizard
          %\fmfpen{thick}
          \fmfleft{p}
          \fmflabel{$\mu$}{p}
          \fmfright{ei,eo}
          \fmf{boson}{p,v}
          \fmf{vanilla}{ei,v}
          \fmf{fermion}{v,eo}
          \fmfv{d.sh=otimes,d.f=empty,d.si=.15w}{v}
        \end{fmfgraph*}
      }\qquad &=& \quad -\I e_r \mu^{\veps} \gamma^\mu \,\delta Z_\Gamma \, ,
      \label{chap2:QED4:FR:Gammact} \\
      \parbox{20mm}{
        \begin{fmfgraph*}(20,10)
        \fmfcmd{
                % Please let me know if there?s a more efficient way to do this
                path quadrant, q[], otimes;
                quadrant = (0, 0) -- (0.5, 0) & quartercircle & (0, 0.5) -- (0, 0);
                for i=1 upto 4: q[i] = quadrant rotated (45 + 90*i); endfor
                otimes = q[1] & q[2] & q[3] & q[4] -- cycle;
        }
          %\fmfpen{thick}
          \fmfleft{in}
          \fmfright{out}
          \fmflabel{$\mu$}{in}
          \fmflabel{$\nu$}{out}
          \fmf{boson,label=$q$}{in,v}
	  \fmf{boson}{v,out}
	  \fmfv{d.sh=otimes,d.f=empty,d.si=.15w}{v}
        \end{fmfgraph*}
      } \qquad &=& \quad -\I (g^{\mu \nu} q^2 -  q^{\mu} q^{\nu})\,\delta Z_A\, .
      \label{chap2:QED4:FR:Dct}
\eea
\end{subequations}
At one-loop, these rules immediately yield the renormalized self-energies:
\begin{subequations}
\label{chap2:RPT:Pi1+Sigma1}
\begin{flalign}
\I \Pi_{1r}^{\mu \nu}(q;\bar{\al}_r,\overline{\mu}) \quad &= \qquad 
      \parbox{15mm}{
    \begin{fmfgraph*}(15,15)
      %\fmfpen{thick}
      \fmfleft{in}
      \fmflabel{$\mu$}{in}
      \fmfright{out}
      \fmflabel{$\nu$}{out}
      \fmf{boson,label=$q$}{in,ve}
      \fmf{fermion,right,tension=0.2}{ve,vw}
      \fmf{fermion,right,tension=0.2}{vw,ve}
      \fmf{boson}{vw,out}
      \fmfdot{ve,vw}
    \end{fmfgraph*}
} \qquad + \qquad 
      \parbox{15mm}{
        \begin{fmfgraph*}(15,10)
        \fmfcmd{
                % Please let me know if there?s a more efficient way to do this
                path quadrant, q[], otimes;
                quadrant = (0, 0) -- (0.5, 0) & quartercircle & (0, 0.5) -- (0, 0);
                for i=1 upto 4: q[i] = quadrant rotated (45 + 90*i); endfor
                otimes = q[1] & q[2] & q[3] & q[4] -- cycle;
        }
          %\fmfpen{thick}
          \fmfleft{in}
          \fmfright{out}
          \fmflabel{$\mu$}{in}
          \fmflabel{$\nu$}{out}
          \fmf{boson,label=$q$}{in,v}
          \fmf{boson}{v,out}
          \fmfv{d.sh=otimes,d.f=empty,d.si=.15w}{v}
        \end{fmfgraph*}
      } \qquad \, ,
\label{chap2:RPT:Pi1}\\
-\I \Sigma_{1r}(p;\bar{\al}_r,\xi_r,\overline{\mu}) \quad &= \qquad
      \parbox{15mm}{
    \begin{fmfgraph*}(15,15)
      %\fmfpen{thick}
      \fmfleft{in}
      \fmflabel{$p$}{in}
      \fmfright{out}
      \fmf{plain}{in,vi}
      \fmf{fermion,tension=0.2}{vi,vo}
      \fmf{boson,left,tension=0.2}{vi,vo}
      \fmf{plain}{vo,out}
      \fmfdot{vi,vo}
    \end{fmfgraph*}
} \quad + \quad
      \parbox{15mm}{
        \begin{fmfgraph*}(15,10)
        \fmfcmd{
                % Please let me know if there?s a more efficient way to do this
                path quadrant, q[], otimes;
                quadrant = (0, 0) -- (0.5, 0) & quartercircle & (0, 0.5) -- (0, 0);
                for i=1 upto 4: q[i] = quadrant rotated (45 + 90*i); endfor
                otimes = q[1] & q[2] & q[3] & q[4] -- cycle;
        }
        \fmfwizard
          %\fmfpen{thick}
          \fmfleft{in}
          \fmfright{out}
          \fmf{fermion,label=$p$}{in,v}
          \fmf{plain}{v,out}
          %\fmfv{decor.shape=circle,decor.filled=empty,decor.size=5thick}{v}
          %\fmfv{decor.shape=cross,decor.filled=empty,decor.size=5thick}{v}
          \fmfv{d.sh=otimes,d.f=empty,d.si=.15w}{v}
        \end{fmfgraph*}
      } \qquad \, ,
\label{chap2:RPT:Sigma1}\\
-\I \Lambda_{1r}^\mu(p,p';\bar{\al}_r,\xi_r,\overline{\mu}) \quad &= \qquad
      \parbox{20mm}{
    \begin{fmfgraph*}(15,15)
      %\fmfpen{thick}
      \fmfleft{in}
      \fmflabel{$\mu$}{in}
      \fmfright{e1,e2}
      \fmflabel{$p$}{e1}
      \fmflabel{$p'$}{e2}
      \fmf{boson}{in,vi}
      \fmf{plain}{e1,v1}
      \fmf{fermion,tension=0.7}{v1,vi}
      \fmf{fermion,tension=0.7}{vi,v2}
      \fmf{plain}{v2,e2}
      \fmffreeze
      \fmf{boson,right,tension=0.7}{v1,v2}
      \fmfdot{vi,v1,v2}
    \end{fmfgraph*}
} \quad + \qquad
      \parbox{20mm}{
        \begin{fmfgraph*}(20,10)
        \fmfcmd{
                % Please let me know if there?s a more efficient way to do this
                path quadrant, q[], otimes;
                quadrant = (0, 0) -- (0.5, 0) & quartercircle & (0, 0.5) -- (0, 0);
                for i=1 upto 4: q[i] = quadrant rotated (45 + 90*i); endfor
                otimes = q[1] & q[2] & q[3] & q[4] -- cycle;
        }
        \fmfwizard
          %\fmfpen{thick}
          \fmfleft{p}
          \fmflabel{$\mu$}{p}
          \fmfright{ei,eo}
          \fmf{boson}{p,v}
          \fmf{vanilla}{ei,v}
          \fmf{fermion}{v,eo}
          \fmfv{d.sh=otimes,d.f=empty,d.si=.15w}{v}
        \end{fmfgraph*}
      } \qquad \, ,
\label{chap2:RPT:Lambda1}
\end{flalign}
\end{subequations}
which can be written in the form:
\begin{subequations}
\label{chap2:RPT:Pi1+Sigma1+Lambda1-gen}
\bea
\Pi_{1r}(q^2;\bar{\al}_r,\overline{\mu}) &=& \Pi_{1}(q^2;\bar{\al}_r) - \delta Z_{1A}(\bar{\al}_r)\, , 
\label{chap2:RPT:Pi1-gen} \\
\Sigma_{1Vr}(p^2;\bar{\al}_r,\xi_r,\overline{\mu}) &=& \Sigma_{1V}(p^2;\bar{\al}_r,\xi_r) - \delta Z_{1\psi}(\bar{\al}_r,\xi_r)\, ,
\label{chap2:RPT:Sigma1-gen}\\ 
\Lambda_{1r}^\mu(p,p';\bar{\al}_r,\xi_r,\overline{\mu}) &=& \Lambda_{1}^\mu(p,p';\bar{\al}_r,\xi_r) + \delta Z_{1\Gamma}(\bar{\al}_r,\xi_r)\,\gamma^\mu\, ,
\label{chap2:RPT:Lambda1-gen}
\eea
\end{subequations}
where, $\Pi_1$, $\Sigma_{1V}$ and $\Lambda_1^\mu$ are the one-loop photon self-energy, fermion self-energy and photon-fermion vertex, respectively. The dependence of these functions on the renormalized parameters is accurate at one-loop 
and Eqs.~(\ref{chap2:RPT:Pi1+Sigma1+Lambda1-gen}) reproduce Eqs.~(\ref{chap2:renormalized-self-energies}) with one-loop accuracy.
In the $\overline{\text{MS}}$ scheme, the one-loop counterterm coefficients $\delta Z_{1A}$, $\delta Z_{1\psi}$ and $\delta Z_{1\Gamma}$ 
are fixed by requiring that they subtract all UV poles from $\Pi_1$, $\Sigma_{1V}$ and $\Lambda_1^\mu$, respectively. They can therefore be defined as:
\begin{subequations}
\label{chap2:RPT:counterterms}
\bea
\delta Z_{1A}(\bar{\al}_r) &=& \mathcal{K}\bigg[ \Pi_{1}(q^2;\bar{\al}_r) \bigg] \quad = \qquad
\mathcal{K}\bigg[~~
      \parbox{15mm}{    
      \begin{fmfgraph*}(15,15)
      %\fmfpen{thick}
      \fmfleft{in}
      \fmfright{out}
      \fmf{boson}{in,ve}
      \fmf{plain,right,tension=0.2}{ve,vw}
      \fmf{plain,right,tension=0.2}{vw,ve}
      \fmf{boson}{vw,out}
      \fmfdot{ve,vw}
    \end{fmfgraph*}
}~~
\bigg] \, , 
\label{chap2:RPT:counterterms:A}\\
\delta Z_{1\psi}(\bar{\al}_r,\xi_r) &=& \mathcal{K}\bigg[ \Sigma_{1V}(p^2;\bar{\al}_r,\xi_r) \bigg] \quad = \qquad
\mathcal{K}\bigg[~~
      \parbox{15mm}{
    \begin{fmfgraph*}(15,15)
      %\fmfpen{thick}
      \fmfleft{in}
      \fmfright{out}
      \fmf{plain}{in,vi}
      \fmf{plain,tension=0.2}{vi,vo}
      \fmf{boson,left,tension=0.2}{vi,vo}
      \fmf{plain}{vo,out}
      \fmfdot{vi,vo}
    \end{fmfgraph*}
}~~
\bigg]\, ,
\label{chap2:RPT:counterterms:S}\\ 
\delta Z_{1\Gamma}(\bar{\al}_r,\xi_r) &=&  - \mathcal{K}\bigg[\Lambda_{1}^\mu(p,p';\bar{\al}_r,\xi_r)/ \gamma^\mu \bigg]\quad = \quad - \quad
\mathcal{K}\bigg[~~
      \parbox{15mm}{
    \begin{fmfgraph*}(15,15)
      %\fmfpen{thick}
      \fmfleft{in}
      \fmfright{e1,e2}
      \fmf{boson}{in,vi}
      \fmf{plain}{e1,v1}
      \fmf{plain,tension=0.7}{v1,vi}
      \fmf{plain,tension=0.7}{vi,v2}
      \fmf{plain}{v2,e2}
      \fmffreeze
      \fmf{boson,right,tension=0.7}{v1,v2}
      \fmfdot{vi,v1,v2}
    \end{fmfgraph*}
}~~
\bigg]\, ,
\label{chap2:RPT:counterterms:G}
\eea
\end{subequations}
where, in the $\overline{\rm{MS}}$ scheme, the $\mathcal{K}$ operator is defined as:
\be
\mathcal{K}~\left( \sum_{n=-\infty}^{+\infty} \frac{c_n}{\veps^n} \right) = \sum_{n=1}^{+\infty} \frac{c_n}{\veps^n}\, .
\label{chap2:def:Sing}
\ee
Notice that the Lorentz structure of the diagrams represented graphically in the argument of  $\mathcal{K}$ in Eqs.~(\ref{chap2:RPT:counterterms:G})
has been projected out. As a consequence, they are all logarithmically divergent (in the case of QED$_4$, see details in the next chapter).
With this graphical convention and at one-loop level, the Ward identity (\ref{chap2:WI:Zpsi}) can be expressed as:
\be
\mathcal{K}\bigg[ ~
      \parbox{15mm}{
    \begin{fmfgraph*}(15,15)
      %\fmfpen{thick}
      \fmfleft{in}
      \fmfright{e1,e2}
      \fmf{boson}{in,vi}
      \fmf{plain}{e1,v1}
      \fmf{plain,tension=0.7}{v1,vi}
      \fmf{plain,tension=0.7}{vi,v2}
      \fmf{plain}{v2,e2}
      \fmffreeze
      \fmf{boson,right,tension=0.7}{v1,v2}
      \fmfdot{vi,v1,v2}
    \end{fmfgraph*}
} ~ \bigg] \quad = \quad - \quad \mathcal{K}\bigg[ ~
      \parbox{15mm}{
    \begin{fmfgraph*}(15,15)
      %\fmfpen{thick}
      \fmfleft{in}
      \fmfright{out}
      \fmf{plain}{in,vi}
      \fmf{plain,tension=0.2}{vi,vo}
      \fmf{boson,left,tension=0.2}{vi,vo}
      \fmf{plain}{vo,out}
      \fmfdot{vi,vo}
    \end{fmfgraph*}
} ~ \bigg]\, .
\label{chap2:ward:graphical}
\ee

The method of counterterms can be extended to higher loops. An equivalent and more systematic (valid at all orders of perturbation theory) way to do so
is the recursive subtraction scheme, the so-called $R$-operation, of Bogoliubov and Parasiuk~\cite{Bogoliubov:1957gp} and
Hepp \cite{Hepp:1966eg}. The power of the BPH recursion is that, for a given graph, it subtracts not only its overall divergence but also its subdivergences.
It is recursive because subdivergences must be subtracted before the overall divergences; and these subdivergences may themselves have subdivergences.
The BPH recursion was solved by Zimmermann \cite{Zimmermann1969}. The solution is known as the {\bf forest formula} and reads:~\footnote{The 
recursive definition of the  $\mathcal{R}'$-operator (solved by Zimmermann's formula) reads:
\be
\mathcal{R}'\,G = G + \sum_{\bar{\Gamma}_d \not= \emptyset} \prod_{\gamma \in \bar{\Gamma}_d} \bigg( - \mathcal{K} \mathcal{R}' \gamma \bigg) \, \star \, G/\bar{\Gamma}_d\, ,
\label{chap2:def:R'-recurs}
\ee
where $\bar{\Gamma}_d$ is the set of all subdivergent graphs which are disjoint (nested ones are not allowed) and there is no constraint on the application of $\mathcal{K}$. In this respect, 
the recursive definition Eq.~(\ref{chap2:def:R'-recurs}) is more self-explanatory than the non-recursive one Eq.~(\ref{chap2:def:R}).  When automated, such as in, \eg, \cite{Batkovich:2014rka}, 
it is the recursive definition which is used. I thank M.\ Kompaniets for discussions on these equations.

From Eqs.~(\ref{chap2:def:R}) and  (\ref{chap2:def:R'-recurs}), we see that the term $\mathcal{K} \mathcal{R}' G$ corresponds to the overall counterterm while
$\mathcal{K} \mathcal{R}' \gamma$ is the counterterm associated with the subgraph $\gamma$ (which may itself contain some subdivergent graphs). 
To see this, let us consider a logarithmically divergent graph $G$ (a graph can be made logarithmic
by proper derivations with respect to mass or external momenta). For such logarithmic graph we identify the counterterm with the renormalization constant.
The renormalized graph is then defined as: $\mathcal{R}\,G = \mathcal{R}'\,G - Z(G)$, where $Z(G)$ is the global counterterm.
Because $\mathcal{K} \mathcal{R}\,G = 0$ and $\mathcal{K} Z(G) = Z(G)$, we have $Z(G) = \mathcal{K} \mathcal{R}'G$ and therefore $Z(\gamma) = \mathcal{K} \mathcal{R}'\gamma$
for any logarithmic subgraph $\gamma$.} \footnote{A nice account on the relation between the BPHZ renormalization prescription
and the Hopf-algebraic approach to renormalization, with applications to QED$_4$, can be found in Ref.~\cite{Kissler:2016gxn}.}
\begin{subequations}
\label{chap2:def:forest}
\bea
\mathcal{R}\,G &=& (1 - \mathcal{K})\,\mathcal{R}'\,G\, ,
\label{chap2:def:R}\\
\mathcal{R}'\,G &=& G + \sum_{\bar{\Gamma}\not= \emptyset} \prod_{\gamma \in \bar{\Gamma}} \bigg( - \mathcal{K} \gamma/\bar{\Gamma} \bigg) \, \star \, G/\bar{\Gamma}\, ,
\label{chap2:def:R'}
\eea
\end{subequations}
where $\mathcal{R}'$ is the so-called incomplete $R$-operation because it subtracts only the subdivergences. 
In Eq.~(\ref{chap2:def:R}), $\mathcal{R}\,G$ corresponds to the finite (renormalized) graph $G$
with all divergences (both subdivergences and overall divergence) subtracted. 
 %The term $\mathcal{K} \mathcal{R}' G$ corresponds to the overall counterterm while
%$\mathcal{K} \mathcal{R}' \gamma$ is the counterterm associated with the subgraph $\gamma$.~\footnote{To see this, let us consider a logarithmically divergent graph $G$ (a graph can be made logarithmic 
%by proper derivations with respect to mass or external momenta). For such logarithmic graph we identify the counterterm with the renormalization constant. 
%The renormalized graph is then defined as: $\mathcal{R}\,G = \mathcal{R}'\,G - Z(G)$, where $Z(G)$ is the global counterterm. 
%Because $\mathcal{K} \mathcal{R}\,G = 0$ and $\mathcal{K} Z(G) = Z(G)$, we have $Z(G) = \mathcal{K} \mathcal{R}'G$ and therefore $Z(\gamma) = \mathcal{K} \mathcal{R}'\gamma$
%for any logarithmic subgraph $\gamma$.} 
 In Eq.~(\ref{chap2:def:R'}), $\bar{\Gamma}$ is the set of all subdivergent graphs which are nested or disjoint, \ie, the so-called normal {\bf forests}.~\footnote{The full forests
include the graph itself.} For nested $\gamma$'s, $\mathcal{K}$ should be applied inside to outside.
 The notation $G/\bar{\Gamma}$ means that the subdiagrams contained in $G$ are shrunk to a vertex.~\footnote{Notice that the shrunk diagram
can be affected by the Lorentz structure of the subdiagram, see \cite{Kissler:2016gxn} and the next chapter for examples.} 
The $\star$ operation amounts to substitute the counterterm in the integrand of the shrunk diagram.
If a graph $G$ has no divergent subgraphs, Eq.~(\ref{chap2:def:R'}) clearly shows that
$\mathcal{R}'\,G = G$. Then, from Eq.~(\ref{chap2:def:R}), we see that $\mathcal{R}\,G = G - \mathcal{K}\,G$. This was the case of the one-loop graphs considered above as we saw from the method of counterterms.

\begin{figure}
  \begin{center}
    %a) \qquad
    \begin{fmfgraph*}(35,30)
      %\fmfpen{thick}
      \fmfleft{i}
      \fmfright{o}
      \fmf{plain}{i,v1}
      \fmf{plain}{v2,o}
      \fmf{phantom,right,tension=0.1,tag=1}{v1,v2}
      \fmf{phantom,right,tension=0.1,tag=2}{v2,v1}
      \fmf{phantom,tension=0.1,tag=3}{v1,v2}
      \fmfdot{v1,v2}
      \fmfposition
      \fmfipath{p[]}
      \fmfiset{p1}{vpath1(__v1,__v2)}
      \fmfiset{p2}{vpath2(__v2,__v1)}
%      \fmfi{plain,label=$k_1+q$}{subpath (0,length(p1)/4) of p1}
      \fmfi{plain}{subpath (0,length(p1)/3) of p1}
%      \fmfi{plain,label=$k_3+q$}{subpath (length(p1)/4,3*length(p1)/4) of p1}
      \fmfi{plain}{subpath (length(p1)/3,2*length(p1)/3) of p1}
%      \fmfi{plain,label=$k_2+q$}{subpath (3*length(p1)/4,length(p1)) of p1}
      \fmfi{plain}{subpath (2*length(p1)/3,length(p1)) of p1}
%      \fmfi{plain,label=$k_2$}{subpath (0,length(p2)/4) of p2}
      \fmfi{plain}{subpath (0,length(p2)/3) of p2}
%      \fmfi{plain,label=$k_3$}{subpath (length(p2)/4,3*length(p2)/4) of p2}
      \fmfi{plain}{subpath (length(p2)/3,2*length(p2)/3) of p2}
%      \fmfi{plain,label=$k_1$}{subpath (3*length(p2)/4,length(p2)) of p2}
      \fmfi{plain}{subpath (2*length(p2)/3,length(p2)) of p2}
%      \fmfi{plain,label=$k_{31}$}{point length(p1)/4 of p1 -- point 3*length(p2)/4 of p2}
      \fmfi{plain}{point length(p1)/3 of p1 -- point 2*length(p2)/3 of p2}
%      \fmfi{plain,label=$k_{32}$}{point 3*length(p1)/4 of p1 -- point length(p2)/4 of p2}
      \fmfi{plain}{point 2*length(p1)/3 of p1 -- point length(p2)/3 of p2}
      \def\vert#1{%
        \fmfiv{decor.shape=circle,decor.filled=full,decor.size=2thick}{#1}}
      \vert{point length(p1)/3 of p1}
      \vert{point 2*length(p1)/3 of p1}
      \vert{point length(p2)/3 of p2}
      \vert{point 2*length(p2)/3 of p2}
    \end{fmfgraph*}
    \qquad
    %\qquad
    \begin{fmfgraph*}(35,30)
      %\fmfpen{thick}
      \fmfleft{i}
      \fmfright{o}
      \fmf{plain}{i,v1}
      \fmf{plain}{v2,o}
      \fmf{phantom,right,tension=0.1,tag=1}{v1,v2}
      \fmf{phantom,right,tension=0.1,tag=2}{v2,v1}
      \fmf{phantom,tension=0.1,tag=3}{v1,v2}
      \fmfdot{v1,v2}
      \fmfposition
      \fmfipath{p[]}
      \fmfiset{p1}{vpath1(__v1,__v2)}
      \fmfiset{p2}{vpath2(__v2,__v1)}
      \fmfi{plain}{subpath (0,length(p1)/4) of p1}
      \fmfi{plain}{subpath (length(p1)/4,3*length(p1)/4) of p1}
      \fmfi{plain}{subpath (3*length(p1)/4,length(p1)) of p1}
      \fmfi{plain}{subpath (0,length(p2)/4) of p2}
      \fmfi{plain}{subpath (length(p2)/4,3*length(p2)/4) of p2}
      \fmfi{plain}{subpath (3*length(p2)/4,length(p2)) of p2}
      \fmfi{plain}{point length(p1)/4 of p1 -- point length(p2)/4 of p2}
      \fmfi{plain}{point 3*length(p1)/4 of p1 -- point 3*length(p2)/4 of p2}
      \def\vert#1{%
        \fmfiv{decor.shape=circle,decor.filled=full,decor.size=2thick}{#1}}
      \vert{point length(p1)/4 of p1}
      \vert{point 3*length(p1)/4 of p1}
      \vert{point length(p2)/4 of p2}
      \vert{point 3*length(p2)/4 of p2}
    \end{fmfgraph*}
    \qquad
    %c) \qquad
    \begin{fmfgraph*}(35,30)
      %\fmfpen{thick}
      \fmfleft{i}
      \fmfright{o}
      \fmf{plain}{i,v1}
      \fmf{plain}{v2,o}
      \fmf{phantom,right,tension=0.1,tag=1}{v1,v2}
      \fmf{phantom,right,tension=0.1,tag=2}{v2,v1}
      \fmf{phantom,tension=0.1,tag=3}{v1,v2}
      \fmfdot{v1,v2}
      \fmfposition
      \fmfipath{p[]}
      \fmfiset{p1}{vpath1(__v1,__v2)}
      \fmfiset{p2}{vpath2(__v2,__v1)}
      \fmfiset{p3}{vpath3(__v2,__v1)}
      \fmfi{plain}{subpath (0,length(p1)/4) of p1}
      \fmfi{plain}{subpath (length(p1)/4,3*length(p1)/4) of p1}
      \fmfi{plain}{subpath (3*length(p1)/4,length(p1)) of p1}
      \fmfi{plain}{subpath (0,length(p2)/2) of p2}
      \fmfi{plain}{subpath (length(p2)/2,length(p2)) of p2}
      \fmfi{plain}{point length(p1)/4 of p1 -- point length(p3)/2 of p3}
      \fmfi{plain}{point 3*length(p1)/4 of p1 -- point length(p3)/2 of p3}
      \fmfi{plain}{point length(p2)/2 of p2 -- point length(p3)/2 of p3}
      \def\vert#1{%
        \fmfiv{decor.shape=circle,decor.filled=full,decor.size=2thick}{#1}}
      \vert{point length(p1)/4 of p1}
      \vert{point 3*length(p1)/4 of p1}
      \vert{point length(p3)/2 of p3}
      \vert{point length(p2)/2 of p2}
    \end{fmfgraph*}
  \caption{\label{chap2:fig:three-loop}
  Three-loop propagator-type diagrams.}
  \end{center}
\end{figure}
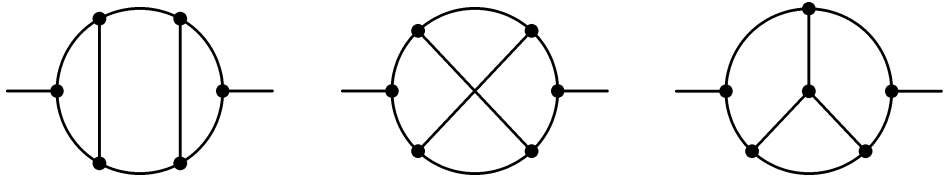

In order to go a little bit further, and see how subdivergences are dealt with,  we apply the forest formula to the scalar two-loop propagator-type diagram of Fig.~\ref{chap2:fig:J}. This graph is characterized by an
overlapping divergence. Its has three normal forests of one element each, consisting of the empty set, the right triangle and the left triangle: 
$\bar{\Gamma} = \{ \{\emptyset \}, \{\gamma_1 \}, \{\gamma_2 \} \}$. In graphical notations, Eq.~(\ref{chap2:def:R'}) for this diagram reads:
\begin{flalign}
\mathcal{R}'\, \bigg[ ~~~~\parbox{16mm}{
    \begin{fmfgraph*}(16,14)
      \fmfleft{i}
      \fmfright{o}
      \fmfleft{ve}
      \fmfright{vo}
      \fmftop{vn}
      \fmftop{vs}
      \fmffreeze
      \fmfforce{(-0.3w,0.5h)}{i}
      \fmfforce{(1.3w,0.5h)}{o}
      \fmfforce{(0w,0.5h)}{ve}
      \fmfforce{(1.0w,0.5h)}{vo}
      \fmfforce{(.5w,0.95h)}{vn}
      \fmfforce{(.5w,0.05h)}{vs}
      \fmffreeze
      \fmf{plain}{i,ve}
      \fmf{plain,left=0.8}{ve,vo}
%      \fmf{phantom,left=0.7,label=$\al_1$,l.d=-0.1w}{ve,vn}
%      \fmf{phantom,right=0.7,label=$\al_2$,l.d=-0.1w}{vo,vn}
      \fmf{plain,left=0.8}{vo,ve}
%      \fmf{phantom,left=0.7,label=$\al_3$,l.d=-0.1w}{vo,vs}
%      \fmf{phantom,right=0.7,label=$\al_4$,l.d=-0.1w}{ve,vs}
      \fmf{plain}{vs,vn}
      \fmf{plain}{vo,o}
      \fmffreeze
      \fmfdot{ve,vn,vo,vs}
    \end{fmfgraph*}
} ~~~~ \bigg] ~~ &= ~~~~
\parbox{16mm}{
    \begin{fmfgraph*}(16,14)
      \fmfleft{i}
      \fmfright{o}
      \fmfleft{ve}
      \fmfright{vo}
      \fmftop{vn}
      \fmftop{vs}
      \fmffreeze
      \fmfforce{(-0.3w,0.5h)}{i}
      \fmfforce{(1.3w,0.5h)}{o}
      \fmfforce{(0w,0.5h)}{ve}
      \fmfforce{(1.0w,0.5h)}{vo}
      \fmfforce{(.5w,0.95h)}{vn}
      \fmfforce{(.5w,0.05h)}{vs}
      \fmffreeze
      \fmf{plain}{i,ve}
      \fmf{plain,left=0.8}{ve,vo}
%      \fmf{phantom,left=0.7,label=$\al_1$,l.d=-0.1w}{ve,vn}
%      \fmf{phantom,right=0.7,label=$\al_2$,l.d=-0.1w}{vo,vn}
      \fmf{plain,left=0.8}{vo,ve}
%      \fmf{phantom,left=0.7,label=$\al_3$,l.d=-0.1w}{vo,vs}
%      \fmf{phantom,right=0.7,label=$\al_4$,l.d=-0.1w}{ve,vs}
      \fmf{plain}{vs,vn}
      \fmf{plain}{vo,o}
      \fmffreeze
      \fmfdot{ve,vn,vo,vs}
    \end{fmfgraph*}
} \qquad - \qquad
\parbox{16mm}{
    \begin{fmfgraph*}(16,14)
      \fmfleft{i}
      \fmfright{o}
      \fmfleft{vv1}
      \fmfright{vo}
      \fmftop{vv2}
      \fmftop{vv3}
      \fmffreeze
      \fmfforce{(-0.3w,0.5h)}{i}
      \fmfforce{(1.3w,0.5h)}{o}
      \fmfforce{(0w,0.5h)}{vv1}
      \fmfforce{(1.0w,0.5h)}{vo}
      \fmfforce{(.5w,0.95h)}{vv2}
      \fmfforce{(.5w,0.05h)}{vv3}
      \fmffreeze
      \fmf{plain}{i,vv1}
      \fmf{plain,left=0.8}{vv1,vo}
%      \fmf{phantom,left=0.7,label=$\al_1$,l.d=-0.1w}{ve,vn}
%      \fmf{phantom,right=0.7,label=$\al_2$,l.d=-0.1w}{vo,vn}
      \fmf{plain,left=0.8}{vo,vv1}
%      \fmf{phantom,left=0.7,label=$\al_3$,l.d=-0.1w}{vo,vs}
%      \fmf{phantom,right=0.7,label=$\al_4$,l.d=-0.1w}{ve,vs}
      \fmf{plain}{vv3,vv2}
      \fmf{plain}{vo,o}
      \fmffreeze
      \fmfdot{vv1,vv2,vo,vv3}
      \fmffreeze
        \fmfcmd{save loc, bmin, bmax;
        forsuffixes $ = 1, 2, 3:
      (loc.$.x, loc.$.y) = vloc __vv.$;
      endfor
      bmax.x = max(loc1x,loc2x,loc3x) + .1w;
      bmax.y = max(loc1y,loc2y,loc3y) + .1h;
      bmin.x = min(loc1x,loc2x,loc3x) - .1w;
      bmin.y = min(loc1y,loc2y,loc3y) - .1h;}
      \fmfi{dashes,width=thin}{(bmin.x,bmin.y)
     -- (bmax.x,bmin.y) -- (bmax.x,bmax.y)
     -- (bmin.x,bmax.y) -- cycle}
    \end{fmfgraph*}
} \qquad - \qquad
\parbox{16mm}{
    \begin{fmfgraph*}(16,14)
      \fmfleft{i}
      \fmfright{o}
      \fmfleft{ve}
      \fmfright{vv1}
      \fmftop{vv2}
      \fmftop{vv3}
      \fmffreeze
      \fmfforce{(-0.3w,0.5h)}{i}
      \fmfforce{(1.3w,0.5h)}{o}
      \fmfforce{(0w,0.5h)}{ve}
      \fmfforce{(1.0w,0.5h)}{vv1}
      \fmfforce{(.5w,0.95h)}{vv2}
      \fmfforce{(.5w,0.05h)}{vv3}
      \fmffreeze
      \fmf{plain}{i,ve}
      \fmf{plain,left=0.8}{ve,vv1}
%      \fmf{phantom,left=0.7,label=$\al_1$,l.d=-0.1w}{ve,vn}
%      \fmf{phantom,right=0.7,label=$\al_2$,l.d=-0.1w}{vo,vn}
      \fmf{plain,left=0.8}{vv1,ve}
%      \fmf{phantom,left=0.7,label=$\al_3$,l.d=-0.1w}{vo,vs}
%      \fmf{phantom,right=0.7,label=$\al_4$,l.d=-0.1w}{ve,vs}
      \fmf{plain}{vv3,vv2}
      \fmf{plain}{vv1,o}
%      \fmffreeze
      \fmfdot{ve,vv2,vv1,vv3}
      \fmffreeze
        \fmfcmd{save loc, bmin, bmax;
        forsuffixes $ = 1, 2, 3:
      (loc.$.x, loc.$.y) = vloc __vv.$;
      endfor
      bmax.x = max(loc1x,loc2x,loc3x) + .1w;
      bmax.y = max(loc1y,loc2y,loc3y) + .1h;
      bmin.x = min(loc1x,loc2x,loc3x) - .1w;
      bmin.y = min(loc1y,loc2y,loc3y) - .1h;}
      \fmfi{dashes,width=thin}{(bmin.x,bmin.y)
     -- (bmax.x,bmin.y) -- (bmax.x,bmax.y)
     -- (bmin.x,bmax.y) -- cycle}
    \end{fmfgraph*}
} 
\nonum \\
&= ~~~~~~
\parbox{16mm}{
    \begin{fmfgraph*}(16,14)
      \fmfleft{i}
      \fmfright{o}
      \fmfleft{ve}
      \fmfright{vo}
      \fmftop{vn}
      \fmftop{vs}
      \fmffreeze
      \fmfforce{(-0.3w,0.5h)}{i}
      \fmfforce{(1.3w,0.5h)}{o}
      \fmfforce{(0w,0.5h)}{ve}
      \fmfforce{(1.0w,0.5h)}{vo}
      \fmfforce{(.5w,0.95h)}{vn}
      \fmfforce{(.5w,0.05h)}{vs}
      \fmffreeze
      \fmf{plain}{i,ve}
      \fmf{plain,left=0.8}{ve,vo}
%      \fmf{phantom,left=0.7,label=$\al_1$,l.d=-0.1w}{ve,vn}
%      \fmf{phantom,right=0.7,label=$\al_2$,l.d=-0.1w}{vo,vn}
      \fmf{plain,left=0.8}{vo,ve}
%      \fmf{phantom,left=0.7,label=$\al_3$,l.d=-0.1w}{vo,vs}
%      \fmf{phantom,right=0.7,label=$\al_4$,l.d=-0.1w}{ve,vs}
      \fmf{plain}{vs,vn}
      \fmf{plain}{vo,o}
      \fmffreeze
      \fmfdot{ve,vn,vo,vs}
    \end{fmfgraph*}
} \qquad - \quad
2\,\mathcal{K}\, \bigg[~ 
   \parbox{15mm}{
    \begin{fmfgraph*}(15,15)
      %\fmfpen{thick}
      \fmfleft{in}
      \fmfright{e1,e2}
      \fmf{plain}{in,vi}
      \fmf{plain}{e1,v1}
      \fmf{plain,tension=0.7}{v1,vi}
      \fmf{plain,tension=0.7}{vi,v2}
      \fmf{plain}{v2,e2}
      \fmffreeze
      \fmf{plain,right,tension=0.7}{v1,v2}
      \fmfdot{vi,v1,v2}
    \end{fmfgraph*} 
}~ \bigg]~ \star ~
\parbox{15mm}{
    \begin{fmfgraph*}(15,15)
      %\fmfpen{thick}
      \fmfleft{in}
      \fmfright{out}
      \fmf{plain}{in,ve}
      \fmf{plain,right,tension=0.2}{ve,vw}
      \fmf{plain,right,tension=0.2}{vw,ve}
      \fmf{plain}{vw,out}
      \fmfdot{ve,vw}
    \end{fmfgraph*}
} \quad \, ,
\label{chap2:forest:G2loop}
\end{flalign}
where, in the first line, the effect of acting with $\mathcal{K}$ on the divergent subdiagram has been represented by enclosing it in a box. 
As we shall see in Chap.~\ref{chap3}, in the cases of QED and reduced QED where diagrams are at most logarithmic,
the $\star$ operation reduces to a simple multiplication. However, as will be shown in Chap.~\ref{chap5}, in the case of graphene the situation is less trivial and the $\star$ operation does not in general reduce to 
simple multiplication; even though the graphical representation is quite convenient, it is much safer in this case to consider directly the integral representation. 
Sticking to the graphical notation for the moment, the overall counterterm for the diagram may be represented as:
\begin{flalign}
\mathcal{K} \mathcal{R}'\, \bigg[ ~~~~\parbox{16mm}{
    \begin{fmfgraph*}(16,14)
      \fmfleft{i}
      \fmfright{o}
      \fmfleft{ve}
      \fmfright{vo}
      \fmftop{vn}
      \fmftop{vs}
      \fmffreeze
      \fmfforce{(-0.3w,0.5h)}{i}
      \fmfforce{(1.3w,0.5h)}{o}
      \fmfforce{(0w,0.5h)}{ve}
      \fmfforce{(1.0w,0.5h)}{vo}
      \fmfforce{(.5w,0.95h)}{vn}
      \fmfforce{(.5w,0.05h)}{vs}
      \fmffreeze
      \fmf{plain}{i,ve}
      \fmf{plain,left=0.8}{ve,vo}
%      \fmf{phantom,left=0.7,label=$\al_1$,l.d=-0.1w}{ve,vn}
%      \fmf{phantom,right=0.7,label=$\al_2$,l.d=-0.1w}{vo,vn}
      \fmf{plain,left=0.8}{vo,ve}
%      \fmf{phantom,left=0.7,label=$\al_3$,l.d=-0.1w}{vo,vs}
%      \fmf{phantom,right=0.7,label=$\al_4$,l.d=-0.1w}{ve,vs}
      \fmf{plain}{vs,vn}
      \fmf{plain}{vo,o}
      \fmffreeze
      \fmfdot{ve,vn,vo,vs}
    \end{fmfgraph*}
} ~~~~ \bigg] ~~ = ~~
\mathcal{K}\, \bigg[~~~~\parbox{16mm}{
    \begin{fmfgraph*}(16,14)
      \fmfleft{i}
      \fmfright{o}
      \fmfleft{ve}
      \fmfright{vo}
      \fmftop{vn}
      \fmftop{vs}
      \fmffreeze
      \fmfforce{(-0.3w,0.5h)}{i}
      \fmfforce{(1.3w,0.5h)}{o}
      \fmfforce{(0w,0.5h)}{ve}
      \fmfforce{(1.0w,0.5h)}{vo}
      \fmfforce{(.5w,0.95h)}{vn}
      \fmfforce{(.5w,0.05h)}{vs}
      \fmffreeze
      \fmf{plain}{i,ve}
      \fmf{plain,left=0.8}{ve,vo}
%      \fmf{phantom,left=0.7,label=$\al_1$,l.d=-0.1w}{ve,vn}
%      \fmf{phantom,right=0.7,label=$\al_2$,l.d=-0.1w}{vo,vn}
      \fmf{plain,left=0.8}{vo,ve}
%      \fmf{phantom,left=0.7,label=$\al_3$,l.d=-0.1w}{vo,vs}
%      \fmf{phantom,right=0.7,label=$\al_4$,l.d=-0.1w}{ve,vs}
      \fmf{plain}{vs,vn}
      \fmf{plain}{vo,o}
      \fmffreeze
      \fmfdot{ve,vn,vo,vs}
    \end{fmfgraph*}
} ~~~~ \bigg] \quad - \quad
2\,\mathcal{K}\, \bigg[~
\mathcal{K}\, \bigg[~
   \parbox{15mm}{
    \begin{fmfgraph*}(15,15)
      %\fmfpen{thick}
      \fmfleft{in}
      \fmfright{e1,e2}
      \fmf{plain}{in,vi}
      \fmf{plain}{e1,v1}
      \fmf{plain,tension=0.7}{v1,vi}
      \fmf{plain,tension=0.7}{vi,v2}
      \fmf{plain}{v2,e2}
      \fmffreeze
      \fmf{plain,right,tension=0.7}{v1,v2}
      \fmfdot{vi,v1,v2}
    \end{fmfgraph*}
}~ \bigg]~ \star ~
\parbox{15mm}{
    \begin{fmfgraph*}(15,15)
      %\fmfpen{thick}
      \fmfleft{in}
      \fmfright{out}
      \fmf{plain}{in,ve}
      \fmf{plain,right,tension=0.2}{ve,vw}
      \fmf{plain,right,tension=0.2}{vw,ve}
      \fmf{plain}{vw,out}
      \fmfdot{ve,vw}
    \end{fmfgraph*}
} ~ \bigg] \quad \, .
\label{chap2:CTc}
\end{flalign}
This example is a simple illustration of the BPHZ renormalization prescription. More interesting examples can be found at three loop level, see Fig.~\ref{chap2:fig:three-loop} 
which represents the three topologies that exist for the three-loop propagator-type diagram. The BPHZ method will be applied at two-loop order to reduced QED and graphene in the next chapters. 

In closing this chapter, let's note that massless propagator-type Feynman diagrams are of extreme practical importance in performing multi-loop calculations beyond two loops.
Some recent results providing the 5 and 6 loop renormalization group functions of various models are based on manipulating such one-scale diagrams, see \cite{Baikov:2016tgj,Batkovich:2016jus}.~\footnote{Another type of one-scale
diagrams, the massive tadpole, was also recently used to reach these high orders, see \cite{Luthe:2017ttc}.} Such breakthrough rest on the following theorem \cite{Chetyrkin:1982nn,Chetyrkin:1984xa}:
\begin{theorem}[Chetyrkin and Tkachov (1982) \& Chetyrkin and Smirnov (1984)]
Any $(L+1)$-loop UV counterterm for any Feynman graph may be expressed in terms of poles and finite parts of some appropriately constructed $L$-loop p-integrals.
\end{theorem}
A key to construct such p-integrals from an arbitrary Feynman graph is that the IR structure of the latter
may be rearranged without affecting its UV structure. This is the method of infra-red rearrangement that we already mentioned.
Such rearrangement may eventually lead to some spurious IR singularities which can be dealt with the help of
a generalization of Bogoliubov's $R$-operation, the so-called $R^*$-operator \cite{Chetyrkin:1982nn,Chetyrkin:1984xa}, see also Ref.~\cite{Grisaru:1986wj} 
and the textbook \cite{kleinert2001critical} for some concrete examples. 
In the following, we shall limit ourselves to two-loop order and we will not need such advanced tools. They may however be of crucial importance in future works extending our results to higher orders.

%\section{Summary of basic formulas in Minkowski space}

%Consider $p$ in Minkowski space and note $p_E$ the (previously used) momentum in Euclidean space.
%We have:
%
%\be
%p^0 = \I p_E^0, \qquad p^2 = - p_E^2\, .
%\ee
%
%Then, in Minkowski space the massless vacuum diagram reads:
%
%\be
%\int \frac{\D^D k}{(-k^2)^\al} \quad = \quad - \pi \, \Omega_D \, \delta(\al-D/2),
%\label{chap2:one-loop-v-int-mink}
%\ee
%
%etc... (list of formulas in Minkowski space)

\end{fmffile}

\cleardoublepage

%% file: Chapter3/rqed.tex
\label{chap3}

\begin{fmffile}{fmf-chap3}

This Chapter presents reduced QED$_{d_\gamma,d_e}$, a relativistic model where the fermion field lives in a space-time of dimension $d_e$ which is lower than the space-time dimension $d_\gamma$ of the gauge field.
In the condensed matter context, this model constitutes a very natural effective relativistic field theory describing (planar) Dirac 
liquids (planar systems with stable Fermi points) that were discussed at length in Chap.~\ref{chap1}, \eg, graphene and graphene-like materials, 
the surface states of some topological insulators and possibly  half-filled fractional quantum Hall 
systems for QED$_{4,3}$ in particular. From the field theory point of view, the model involves an effective (reduced) gauge field 
propagating with a fractional power of the d'Alembertian in marked contrast with usual QEDs. This Chapter initiates a thorough examination of the perturbative structure of such an unconventional model
using the BPHZ renormalization prescription, see Chap.~\ref{chap2}.~\footnote{Let's also note the more recent Hopf
algebraic formulation~\cite{Kreimer:1997dp,Connes:1998qv} of renormalization, see also Ref.~\cite{Panzer:2014kia} for a recent review.
Its application to our model is beyond the scope of our present study.} This is a necessary prerequisite for the study of some of its non-perturbative features which will be investigated in the next chapter.
Most of the results presented in the following be quite general and valid beyond the reduced case (for any $d_\gamma$ and $d_e$).

\section{Lorentz invariant fixed point}
\label{chap3:lorentz}
% only in one file to avoid multiple inclusions
%\nocite{*}

In order to motivate the study of reduced QED, we will here briefly recall the arguments leading to the existence of an infra-red Lorentz invariant fixed point for planar Dirac systems \cite{Gonzalez:1993uz}.
For this, we consider the general model I of Eq.~(\ref{chap1:model-general}) that we reproduce here for clarity:
\bea
S =&& \int \D t\, \D^{D_e} x\, \left[ \bar{\psi}_\sigma \left( \I \gamma^0 \partial_t + \I v \vec{\gamma} \cdot \vec{\nabla}\,\right) \psi^\sigma - e\bar{\psi}_\sigma \,\gamma^0 A_0\, \psi^\sigma
+ e \frac{v}{c}\,\bar{\psi}_\sigma\, \vec{\gamma} \cdot \vec{A}\, \psi^\sigma \right ]
\nonum \\
&&+\, \int \D t\, \D^{D_\gamma} x\,\left[ - \frac{1}{4}\,F^{\mu \nu}\,F_{\mu \nu} - \frac{1}{2\xi}\left(\partial_{\mu}A^{\mu}\right)^2 \right]\, ,
\label{chap3:model-general}
%&&\int \D t\, \D^3 x\, \left[ - \frac{1}{4}\,F^{\mu \nu}\,F_{\mu \nu} -\frac{1}{2a}\,(\partial_\mu A^\mu)^2 \right]\, ,
\eea
where $D_e$ and $D_\gamma$ refer to space dimensionalities while space-time dimensions will be denoted by lower-case letters: $d_e=D_e+1$
and $d_\gamma = D_\gamma + 1$. From dimensional analysis, we see that: $[e] = 2 - d_\gamma/2$, so that the dimensionality of the coupling constant is entirely
determined by the space-time dimension where the gauge field evolves. It is then convenient to parametrize $d_\gamma$ and $d_e$ with the help of two $\veps$-parameters as follows:
\be
d_\gamma = 4 -2\veps_\gamma, \qquad d_e = 4 - 2\veps_e - 2 \veps_\gamma\, .% = 2 + 2 \lambda - 2\veps_\gamma\, ,
\label{chap3:def:de+dg}
\ee
%
%where it is also convenient in some cases to use  $\lambda = 1 -\veps_e$ instead of $\veps_e$. 
The case of graphene corresponds to: $\veps_\gamma \ra 0$ and $\veps_e \ra 1/2$, that is a fermion living in a space of $d_e = 2 +1$-dimensions
interacting with a gauge field in $d_\gamma=3+1$-dimensions.

From model I, Eq.~(\ref{chap3:model-general}), and for arbitrary $d_\gamma$ and $d_e$, we have the following Feynman rules. The fermion propagator reads:
\be
S_0(p) = \frac{\I \Sp}{p^2}\, , \qquad \Sp = \gamma^\mu p_\mu = \gamma^0 p_0 - v \vec{\gamma}\cdot \vec{p}\, .
\label{chap3:gm:fermion-prop0}
\ee
The reduced photon propagator is given by:
\be
\tilde{D}_0^{\mu \nu}(\bar{q}) =
\frac{\I}{(4\pi)^{\veps_e}}\frac{\Gamma(1-\veps_e)}{(-\bar{q}^{\,2})^{1-\veps_e}}\,\left( g^{\mu \nu} - \tilde{\eta}\,\frac{\bar{q}^{\,\mu} \, \bar{q}^{\,\nu}}{\bar{q}^{\,2}} \right)\, ,
\qquad \bar{q}^{\,\mu} = ( q^0/c , \vec{q}\,)\, ,
\label{chap3:gm:gauge-field-prop0}
\ee
where $\tilde{\eta}=1-\tilde{\xi}$ is the gauge fixing parameter related to the reduced gauge field. The later is related to the gauge fixing parameter of the $d_\gamma$-dimensional gauge field, $\eta=1-\xi$,
with the help of  $\tilde{\eta}= (1-\veps_e)\,\eta$. The propagator of Eq.~(\ref{chap3:gm:gauge-field-prop0}) is a reduced one since it has been obtained from the usual $D_\gamma$-dimensional photon propagator
after integrating out all space coordinates perpendicular to the $D_e$-dimensional membrane.
Finally, the vertex function reads:
\be
-\I e \bar{\Gamma}^\mu = \left( -\I e \Gamma^0, -\I e \frac{v}{c}\,\vec \Gamma \,\right), \qquad \Gamma_0^0 = \gamma^0 \, , \qquad \vec{\Gamma}_0 = \vec{\gamma}\, ,
\label{chap3:gm:vertex}
\ee
where a natural distinction between the temporal and space components has been made. As already mentioned in the introduction, 
the coupling of the fermion field to the gauge field is characterized by the dimensionless fine structure constant: 
\be
\al_g = \frac{e^2}{4 \pi \hbar v} \approx 2.2 \, ,
\label{chap2:alg}
\ee
which is of the order of unity due to the fact that $v \approx c / 300$.

Model I, Eq.~(\ref{chap3:model-general}), may be expressed in terms of renormalized parameters and fields by introducing a number of renormalization constants:
\be
\psi = Z_{\psi}^{1/2} \psi_r, \quad A_0 = Z_{A_0}^{1/2} A_{0r}, \quad \vec{A} = Z_{\vec{A}}^{1/2} \vec{A}_r, \quad e = Z_e e_r \mu^{\veps}, \quad v = Z_v v_r\, ,
\label{chap3:gm:QED_d:Z}
\ee
where:
\be
Z_e = \frac{Z_{\Gamma^0}}{Z_\psi Z_{A_0}^{1/2}} = \frac{Z_{\vec \Gamma}}{Z_\psi Z_v Z_{\vec{A}}^{1/2}}\, .
\ee
In the reduced case, $0< \veps_e \leq 1$, the non-integer index characterizing the photon propagator in Eq.~(\ref{chap3:gm:gauge-field-prop0}) corresponds to the fact that the action of the free reduced gauge field
is non-local. As a consequence, see statement below Eq.~(\ref{chap2:model:QED_d:BPH-L}) as well as Refs.~\cite{PhysRevLett.80.5409,PhysRevLett.87.137004} for similar arguments,\footnote{I 
thank Jean-No\"el Fuchs for pointing Refs.~\cite{PhysRevLett.80.5409,PhysRevLett.87.137004} to me.}
the gauge field is not renormalized: $Z_{A_0}=Z_{\vec{A}}=1$. This implies in turn that the gauge fixing parameter is not renormalized either: $Z_\xi=1$. Gauge invariance further
enforces the following Ward identities:
\be
Z_{\Gamma^0} = Z_\psi, \qquad Z_{\vec \Gamma} = Z_\psi Z_v\, .
\label{chap3:gm:WI}
\ee
As a consequence, the charge is not renormalized: $Z_e=1$. The renormalization of the coupling constant is therefore entirely due to the renormalization of the velocity:
\be
\al_g = Z_{\al} \,\al_{gr}, \qquad Z_{\al} = Z_v^{-1}\, .
\label{chap3:gm:alr}
\ee

The renormalization constants $Z_\psi$ and $Z_v$ can be both computed from the fermion self-energy. For the later, we use the following parametrization:
\be
\Sigma(p) = \gamma^0 p_0 \,\Sigma_\om(p^2) - v \vec{\gamma} \cdot \vec{p}\,\,\Sigma_k(p^2)\, ,
\label{chap3:gm:Sigma:param}
\ee
which is such that:
\be
\Sigma_{\om}(k^2) = \frac{\Tr[\gamma^0 k_0\,\Sigma(k)]}{4N_F k_0^2}\, , \quad
\Sigma_{k}(k^2) = \frac{\Tr[\vec{\gamma} \cdot \vec{k}\,\Sigma(k)]}{4N_F v |\vec{k}\,|^2}\, ,
\label{chap3:fsigma-param}
\ee
and from which the (unrenormalized) dressed fermion propagator may be expressed as:
\be
S(p) = \frac{\I \, (1- \Sigma_\om(p^2))^{-1}}{\gamma^0 p_0 - v\,\frac{1-\Sigma_k(p^2)}{1-\Sigma_\om(p^2)}\,\vec{\gamma} \cdot \vec{p}\,}~\, .
\label{chap3:gm:S(p)-gen}
\ee
The renormalization constants are then given by:
\be
\delta Z_\psi = \mathcal{K}\,\bigg[ \frac{1}{1- \Sigma_\om(p^2)} \bigg], \qquad \delta Z_v = \mathcal{K}\,\bigg[ \frac{1- \Sigma_\om(p^2)}{1 - \Sigma_k(p^2)} \bigg]\, ,
\label{chap3:gm:ZpsiZv-gen}
\ee
where $Z_x = 1 + \delta Z_x$ ($x \in \{\psi,v\}$). The constant $Z_\psi$ determines the 
anomalous dimension of the fermion field while the constant $Z_v$ determines the velocity and related coupling constant  $\beta$-functions:
\begin{subequations}
\label{chap3:gm:def:betas}
\bea
&&\beta_v(\al_{gr}) = \mu \frac{\partial v_{r}}{\partial \mu} = \sum_{l=0}^\infty \beta_{v,l} \al_{gr}^{l+1}, \qquad \qquad \qquad \beta_{v,l} = 2 v_r\,(l+1)\,Z_v^{(l+1,1)}\, ,
\label{chap3:gm:def:betav}
\\
&&\beta(\al_{gr}) = \mu \frac{\partial \al_{gr}}{\partial \mu} = -2 \veps_\gamma \al_{gr} + \sum_{l=0}^\infty \beta_{l} \al_{gr}^{l+2}, \qquad \beta_{l} = -2(l+1)\,Z_v^{(l+1,1)}\, .
\label{chap3:gm:def:beta}
\eea
\end{subequations}

Performing all computations in the Feynman gauge, $\xi=1$, the one-loop renormalization constants read \cite{KotikovT-unpublished}:%(!!! check sign in $Z_\psi$ and ??? generalize to an arbitrary gauge ??? !!!):
\begin{subequations}
\label{chap3:gm:res:Zs}
\bea
Z_\psi(\al_{gr}) &=& 1 - \frac{\al_{gr}}{2\pi \veps}\,\frac{x(1-2x^2)}{1-x^2}\,\bigg( 1- \frac{x}{\sqrt{1-x^2}}\,\left( \frac{\pi}{2} - \arcsin(x) \right) \bigg)\, ,
\label{chap3:gm:res:Zpsi}
\\
Z_v(\al_{gr}) &=& 1 + \frac{\al_{gr}}{2\pi \veps}\,\frac{x(1-2x^2)}{1-x^2}\,\bigg( 1- \frac{x}{\sqrt{1-x^2}}\,\left( \frac{\pi}{2} - \arcsin(x) \right) \bigg) 
\nonum \\
&-&  \frac{\al_{gr}}{4\pi \veps}\,\frac{x}{1-x^2}\,\bigg( -1 +  \frac{1}{x\,\sqrt{1-x^2}}\,\left( \frac{\pi}{2} - \arcsin(x) \right) \bigg)\, ,
\label{chap3:gm:res:Zv}
\eea
\end{subequations}
where $0 \leq x=v/c \leq 1$ is a dimensionless parameter measuring the anisotropy between space and time. From $Z_\psi$ we may then deduce the one-loop anomalous dimension of 
the fermion field, Eq.~(\ref{chap2:model:QED_d:anomalousdims-psi}), in the Feynman gauge which reads:
\begin{subequations}
\label{chap3:gm:res:gammapsi+lim}
\bea
\gamma_\psi(\al_{gr}) &=& - \frac{\al_{gr}}{\pi}\,\frac{x(1-2x^2)}{1-x^2}\,\bigg( 1- \frac{x}{\sqrt{1-x^2}}\,\left( \frac{\pi}{2} - \arcsin(x) \right) \bigg)\, ,
\label{chap3:gm:res:gammapsi}
\\
&=& \begin{cases}
-\frac{x \al_{gr}}{\pi}\, \left( 1 - \frac{\pi}{2}\,x + \Ord(x^2) \right) & (\text{case}~x \ra 0)\, , \\
\frac{\al_{gr}}{3\pi}\, \left( 1 - \frac{21}{5}\,(1-x) + \Ord((1-x)^2) \right) & (\text{case}~x \ra 1)\, ,
\end{cases}
\label{chap3:gm:res:gammapsi-lims}
\eea
\end{subequations}
where there is no wave-function renormalization at one-loop in the limit of instantaneous interactions, $x=v/c \ra 0$, and the anomalous dimension changes sign at $x \approx 0.707$.
Similarly, from $Z_v$, we may deduce the one-loop $\beta$-function associated with the velocity:
\begin{subequations}
\label{chap3:gm:res:betav+lim}
\bea
\beta_v(\al_{gr}) &=& -\frac{v_r\,\al_{gr}}{\pi}\,\frac{x(1-2x^2)}{1-x^2}\,\bigg( 1- \frac{x}{\sqrt{1-x^2}}\,\left( \frac{\pi}{2} - \arcsin(x) \right) \bigg)
\nonum \\
&-& \frac{v_r\,\al_{gr}}{2\pi}\,\frac{x}{1-x^2}\,\bigg( -1 +  \frac{1}{x\,\sqrt{1-x^2}}\,\left( \frac{\pi}{2} - \arcsin(x) \right) \bigg)\, ,
\label{chap3:gm:res:betav}\\
&=& \begin{cases}
-\frac{v_r\, \al_{gr}}{4}\, \left( 1 - \frac{1}{2}\,x^2 + \Ord(x^3) \right) & (\text{case}~x \ra 0)\, , \\
-\frac{8 (1-x)\,v_r\,\al_{gr}}{5\pi}\, \left( 1 - \frac{19}{42}\,(1-x) + \Ord((1-x)^2) \right) & (\text{case}~x \ra 1)\, .
\end{cases}
\label{chap3:gm:res:betav-lims}
\eea
\end{subequations}
The function $\beta_v$ has a negative sign implying an increase of the Fermi velocity upon decreasing the energy scale. 
If we restricted ourselves to the case of instantaneous interactions ($x \ra 0$) the Fermi velocity would seem to diverge deep in the infra-red.
With retardation effects included, Eq.~(\ref{chap3:gm:res:betav+lim}) shows that the limit $x = 1$ is a zero of $\beta_v$ (and hence $\beta$). It is therefore an infra-red fixed
point characterized by full Lorentz and scale (possibly even conformal) invariance (see next section for a little more). So, deep in the infra-red, 
the Fermi velocity flows to the velocity of light, $v \ra c$, and the coupling constant to the QED fine structure constant, $\al_g \ra \al = 1/137$. The Coulomb interaction is therefore marginally irrelevant for $x<1$ while
it remains strictly marginal at $x=1$.  The argument was given here at one-loop and for planar systems. It is believed however that it is valid to all orders of perturbation theory and
for all systems having stable Dirac points \cite{Volovik2009book}.

\section{Reduced QED (model)}

The existence of the infra-red Lorentz invariant fixed point implies that condensed matter systems having stable Dirac points and interacting via the long-range Coulomb interaction 
may be described by an effective relativistic field theory where the gauge field lives in a bulk $3+1$-dimensional space-time while the fermion field is restricted to a $2+1$-dimensional one, \eg, pair creation
is restricted to a plane and interaction is fully retarded.  As already stated in Chap.~\ref{chap1}, such an ultra-relativistic model is a peculiar case of the so-called reduced quantum electrodynamics 
(reduced QED or RQED) as refereed to in \cite{Gorbar:2001qt} or pseudo QED as referred to in \cite{Marino:1992xi} or even as mixed-dimension QED as referred to recently in \cite{Hsiao:2017lch}. 
For arbitrary $d_\gamma$ and $d_e$, reduced QED$_{d_\gamma,d_e}$ is a quantum field theory describing the interaction of an abelian $U(1)$ gauge field
living in $d_\gamma$ space-time dimensions with a fermion field living in a reduced space-time of $d_e$ dimensions ($d_e \leqslant d_\gamma$).
In the particular case where gauge and fermion fields live in the same space-time, $d_\gamma=d_e$, reduced QEDs correspond to the usual QEDs such as
QED$_4$, QED$_3$ or QED$_2$, also known as the Schwinger model, which is a celebrated exactly solvable model~\cite{PhysRev.128.2425}. 

Motivations for the study of reduced theories came from interest in branes \cite{Gorbar:2001qt,Dimopoulos:2000iq}, dynamical chiral symmetry breaking in branes which was first studied in \cite{Gorbar:2001qt}, 
conformal field theory \cite{Kaplan:2009kr} (and reference therein), as well as potential applications to condensed matter physics, see Refs.~\cite{Marino:1992xi,Dorey:1991kp,Kovner:1990zz} for early studies.
It is in 2012 that, in \cite{Teber:2012de}, the study of reduced QED$_{d_\gamma,d_e}$ was invoked to describe the infra-red Lorentz invariant fixed point of systems with stable Dirac points with the case
of QED$_{4,3}$ relevant to (intrinsic) graphene and similar planar Dirac materials. 
Since then, that is in the last five years or so, there have been many studies (besides ours) focusing on reduced QED and in particular QED$_{4,3}$, {\it e.g.}, 
transport and optical properties \cite{Herbut:2013kca,Valenzuela:2014uia,Hernandez-Ortiz:2015wua},
quantum Hall effect \cite{Marino:2015uda,Kooi:2017ugi,Son:2015xqa} (in \cite{Son:2015xqa} QED$_{4,3}$ was advocated by Son as a low energy effective QFT describing $1/2$-filled FQHE systems of Dirac composite fermions at the 
self-dual point), dynamical chiral symmetry breaking \cite{Alves:2017fij} in planar systems and interest in long-range interactions \cite{Barros:2017ygp}.
From a more field-theoretic aspect, QED$_{4,3}$ was shown to be unitary by Marino et al.\ \cite{Marino:2014oba}, its properties
under the Landau-Khalatnikov-Frandkin transformation were studied \cite{Ahmad:2016dsb} and it was shown to possess a strong-weak duality mapping the coupling constant $e$ to
$\tilde{e}=8\pi/e$ with a self-dual point at $e^2=8\pi$ (or $\al =2$) \cite{Hsiao:2017lch}. Even more recently, QED$_{4,3}$ has been studied as ``one of the most interesting example'' of an 
interacting boundary conformal field theory \cite{Herzog:2017xha}, see also \cite{Karch:2018uft} and \cite{Dudal:2018pta}, and its supersymmetric extension has been considered in \cite{Herzog:2018lqz}.

Of course, the study of the fixed point may seem to be only of academic interest, {\it a priori}. Our point of view is that the simplicity of the
model, \eg, with respect to (\ref{chap3:model-general}) and even to (\ref{chap1:model-inst}) that will be studied in Chap.~\ref{chap5}, 
allows to obtain very robust results and to set on a firm ground the study of the physics away from the fixed point.
As we shall see in the next Chapters, it turns out that there seems to be a {\it quantitative} agreement between several results obtained so far using this model and known results
in the non-relativistic limit. From a technical point of view, Feynman diagrams with non-integer indices appear which originate from the fact that the reduced theory is non-local.
Such diagrams usually do not appear in $(3+1)$-dimensional theories. Even at the lowest orders of perturbation theory, their exact evaluation requires the use of
advanced multi-loop techniques. 

%In the last four years, many studies focused on reduced QED and in particular QED$_{4,3}$, {\it e.g.}, transport and optical properties \cite{Herbut:2013kca,Valenzuela:2014uia,Hernandez-Ortiz:2015wua}, 
%quantum Hall effect \cite{Marino:2015uda,Kooi:2017ugi,Son:2015xqa} and dynamical chiral symmetry breaking \cite{Alves:2017fij} in planar systems. 
%From a more field-theoretic aspect, QED$_{4,3}$ was shown to be unitary \cite{Marino:2014oba}, its properties 
%under the Landau-Khalatnikov-Frandkin transformation were studied \cite{Ahmad:2016dsb} and it was shown to possess a strong-weak duality mapping the coupling constant $e$ to 
%$\tilde{e}=8\pi/e$ with a self-dual point at $e^2=8\pi$ \cite{Hsiao:2017lch}. 

%\subsection{The model}

The model of massless QED$_{d_\gamma,d_e}$ with $N_F$ species of fermions, model III, simply corresponds to the limit $x=v/c \ra 1$ (vanishing space-time anisotropy) of model I (\ref{chap3:model-general}). The corresponding action was given in Eq.~(\ref{chap1:rqed}) and we repeat it here for clarity:
\begin{flalign}
S = \int \D^{d_e} x\, \bar{\psi}_\sigma \I \gamma^{\mu_e} \big( \partial_{\mu_e} + \I e A_{\mu_e} \big)  \psi^\sigma
- \int \D^{d_\gamma} x\,\left[ \frac{1}{4}\,F^{\mu_\gamma \nu_\gamma}\,F_{\mu_\gamma \nu_\gamma} + \frac{1}{2\xi}\left(\partial_{\mu_\gamma}A^{\mu_\gamma}\right)^2 \right]\, ,
\label{chap3:model:rqed}
\end{flalign}
which is expressed in natural units ($\hbar=c=1$) and where matter indices: $\mu_e\,=\,0,\,1,\,...,\,d_e-1$, are related to the first term in the Lagrangian which is a boundary term describing a fermion field $\psi$ while
gauge indices: $\mu_\gamma \,=\, 0,\,1,\,...,\,d_e-1,\,d_e,\,...,\,d_\gamma-1$, are related to the bulk gauge field $A^\mu$.
In Eq.~(\ref{chap3:model:rqed}), the minimal coupling of the gauge field to the fermion current, $j_\mu A^\mu$, only involves matter indices so that
the conserved current can be defined as:
\be
j^\mu(x) =
\left\{
          \begin{array}{ll}
                \E\, \psibar(x) \gamma^\mu \psi(x) \delta^{(d_\gamma-d_e)}(x) & \,\,\, \mu=\mu_e\, , \\
                0 & \,\,\, \mu = d_e,\,...,\,d_\gamma-1\, , \\
          \end{array}
\right .
\label{chap3:jmu}
\ee
and is localized in a reduced $d_e$-dimensional space ($d_e \leq d_\gamma$). The Feynman rules for model III follow straightforwardly 
from those of model I (\ref{chap3:model-general}) and are summarized in Fig.~\ref{chap3:RQED:fig:FeynmanRules}.
The free massless fermion propagator and fermion-photon vertex are the standard ones:
\be
S_0(p) = \frac{\I}{\Sp}, \qquad \Gamma_0^\mu  = \gamma^\mu,
\label{chap2:RQED:FR:S0+Gamma0}
\ee
%
%where $p=p_0,\,...,\,p_{d_e-1}$ lies in the reduced matter space.
and the reduced gauge field propagator reads (see also Eq.~(\ref{chap1:rqed-int}) where fractional powers appear explicitly at the level of the action):
\be
\tilde{D}_0^{\mu \nu}(q) = \frac{\I}{(4\pi)^{\varepsilon_e}}\frac{\Gamma(1-\varepsilon_e)}{(-q^2)^{1-\varepsilon_e}}\,\left( g^{\mu \nu} - (1-\tilde{\xi})\,\frac{q^{\mu} q^{\nu}}{q^2}\right),
\label{chap3:RQED:FR:Dmunu0}
\ee
where all indices take their values in the $d_e$-dimensional space and the subscript on indices $\mu$ and $\nu$ have been dropped for simplicity. 
The gauge fixing parameter of the reduced gauge field, $\tilde{\xi} = 1 - \tilde{\eta}$, is
related to the gauge fixing parameter of the four-dimensional gauge field, $\xi=1-\eta$, with the help of:
\be
\tilde{\xi} = \veps_e + (1-\veps_e)\, \xi, \qquad \tilde{\eta} = (1-\veps_e)\, \eta\, .
\ee
Following Eq.~(\ref{chap2:QED4:FR:Dperp+Dpara}), the photon propagator can be separated in longitudinal and transverse parts which, in the reduced case, read:
\bea
\tilde{d}_{0 \parallel}(q^2) = \frac{\I \tilde{\xi}}{(4\pi)^{\varepsilon_e}}\frac{\Gamma(1-\varepsilon_e)}{(-q^2)^{1-\varepsilon_e}}\, ,
\qquad
\tilde{d}_{0 \bot}(q^2) = \frac{\I}{(4\pi)^{\varepsilon_e}}\frac{\Gamma(1-\varepsilon_e)}{(-q^2)^{1-\varepsilon_e}}\, .
\label{chap2:RQED:FR:Dperp+Dpara}
\eea
In the case of QED$_{4,3}$: $\varepsilon_e=1/2$ and the reduced propagator has a square root branch-cut whereas for
QED$_{4,2}$: $\varepsilon_e=1$ and the reduced propagator is logarithmic. Notice that reduced QED$_{4,2}$ models a one-dimensional system (a wire) where fermions interact via the long-range (fully retarded)
Coulomb interaction; this case requires additional regularization such as giving a small width to the wire~\cite{Gorbar:2001qt,Kaplan:2009kr}. 
Another case is that of QED$_{4,1}$: $\varepsilon_e=3/2$ which corresponds to a point-like particle in a four-dimensional electromagnetic environment.
In all cases the reduced QFT is non-local. 

\begin{figure}
  \begin{center}
    \bea
      \parbox{20mm}{
        \begin{fmfgraph*}(20,10)
          %\fmfpen{thick}
          \fmfleft{in}
          \fmfright{out}
          \fmf{fermion,label=$p$}{in,out}
        \end{fmfgraph*}
      }
      \quad & = & \frac{\I}{\Sp} \nonum \\
      \parbox{20mm}{
        \begin{fmfgraph*}(20,10)
          %\fmfpen{thick}
          \fmfleft{p}
          \fmflabel{$\mu$}{p}
          \fmfright{ei,eo}
          \fmf{boson}{p,v}
          \fmf{vanilla}{ei,v}
          \fmf{fermion}{v,eo}
          \fmfdot{v}
        \end{fmfgraph*}
      }
      \quad & = & - \I e \gamma^\mu \nonum \\
      \parbox{20mm}{
        \begin{fmfgraph*}(20,10)
          %\fmfpen{thick}
          \fmfleft{in}
          \fmfright{out}
          \fmflabel{$\mu$}{in}
          \fmflabel{$\nu$}{out}
          \fmf{boson,label=$q$}{in,out}
        \end{fmfgraph*}
      }
      \quad & = & \frac{\I}{(4\pi)^{\varepsilon_e}}\frac{\Gamma(1-\varepsilon_e)}{(-q^2)^{1-\varepsilon_e}}\,\left( g^{\mu \nu} - (1-\tilde{\xi})\,\frac{q^\mu q^\nu}{q^2}\right)
      \nonum
    \eea
    \caption{\label{chap3:RQED:fig:FeynmanRules}
    Feynman rules for massless reduced QED$_{d_\gamma,d_e}$ (model III).}
  \end{center}
\end{figure}

Switching on interactions, the dressed fermion propagator and fermion-photon vertex take the same form as in QED$_4$:
\be
S(p) = \frac{\I}{\Sp}\frac{1}{1-\Sigma_V(p^2)}, \qquad \Gamma^\mu (p,p') = \gamma^\mu + \Lambda^\mu (p,p')\, ,
\label{chap3:RQED:Dyson:S+Gamma}
\ee
where $\Sigma(p) = \Sp \Sigma_V(p^2)$ in the massless case.
As for the photon propagator, only its transverse part is affected by interactions as in usual QED:
\be
\tilde{d}_{\bot}(q^2) = \tilde{d}_{0 \, \bot}(q^2) \frac{1}{1- \I q^2 \, \tilde{d}_{0\, \bot}(q^2)\, \Pi(q^2)}, \qquad \tilde{d}_{\parallel}(q^2) = \tilde{d}_{0\,\parallel}(q^2)\, ,
\label{chap3:RQED:Dyson:D}
\ee
where for $\veps_e=0$ we recover Eq.~(\ref{chap2:Dint}) but the form changes for other values of $\veps_e$.

Following similar discussions for QED$_4$ in Sec.~\ref{chap2:subsec:ren} and for the general model (\ref{chap3:model-general}) in the previous section, 
for arbitrary $d_e$ and $d_\gamma$ the dimensions of the fields and coupling are given by:
\bea
[\psi] = \frac{d_e-1}{2} = \frac{3}{2} - \varepsilon_e - \varepsilon_\gamma, \qquad [A^\mu] = \frac{d_\gamma - 2}{2} = 1 - \varepsilon_\gamma, 
\qquad [e] = 2- \frac{d_\gamma}{2} = \varepsilon_\gamma\, ,
\label{chap3:dims}
\eea
where the parameters $\varepsilon_\gamma$ and $\varepsilon_e$ read:
\bea
\varepsilon_\gamma = \frac{4-d_\gamma}{2}, \qquad \varepsilon_e = \frac{d_\gamma - d_e}{2}\, ,
\label{chap3:params}
\eea
in accordance with Eq.~(\ref{chap3:def:de+dg}). As we saw in the previous section, as long as the gauge field is $4$-dimensional ($d_\gamma=4$) the coupling constant is dimensionless whatever space
the fermion field lives in.~\footnote{Notice that, in $p$-space, we have: $[\tilde{A}^\mu (k)] = d_e/2 - [A^\mu] = 1 - \veps_e$, and $\veps_e$ therefore appears as an anomalous dimension for the reduced 
gauge-field in accordance with the form of $p$-space reduced gauge-propagator, see Fig.~\ref{chap3:RQED:fig:FeynmanRules}. Accordingly, gauge fixing is non-local with the usual factor $1/q^4$ appearing in factor of $\xi$ 
replaced by $1/(q^2)^{2-\veps_e}$. Hence, upon performing a gauge transformation: $A_\mu(x) \ra A_\mu(x) + \partial_\mu \varphi(x)$, the correlator of the $\varphi$-field, which is proportional to the longitudinal part
of the photon Green's function, also becomes anomalous. Formally, this amounts to defining a non-local gauge transformation for the reduced gauge-field: 
$\tilde{A}_\mu(x) \ra \tilde{A}_\mu(x) + \partial_\mu^{1+\veps_e} \tilde{\varphi}(x)$ where $\partial_\mu^{1+\veps_e}$ is a fractional derivative, see, \eg, \cite{Shirkov:1989bp} as well
as the next Chapter for more references on non-local gauge-fixing which is also used in the study of QED$_3$. It seems therefore that reduced QED provides a concrete example of the non-local field theories studied in the recent 
Refs.~\cite{LaNave:2017lwf,LaNave:2017nex} and invoked phenomenologically in Ref.~\cite{Limtragool:2016gnl} without knowing the corresponding higher dimensional model.}  
This suggest that all models of reduced QED$_{4,d_e}$ are renormalizable which is indeed the case of standard QED$_4$ ($d_e=d_\gamma=4$) 
but also  of RQED$_{4,3}$ ($d_\gamma=4$ and $d_e=3$) and  RQED$_{4,2}$ ($d_\gamma=4$ and $d_e=2$). This is in agreement with the counting of ultraviolet UV divergences.
Indeed, the SDD of a diagram $G$ in RQED$_{d_\gamma,d_e}$ reads:
\be
\om(G) = d_e - \frac{d_\gamma-2}{2} N_\gamma -\frac{d_e-1}{2}N_e + \frac{d_\gamma - 4}{2} V \, ,
\label{chap3:sdd}
\ee
which generalizes Eq.~(\ref{chap2:QED4:w(G)}) to arbitrary $d_e$ and $d_\gamma$. From Eq.~(\ref{chap3:sdd}) we see that for $d_\gamma=4$ the SDD does not depend on the number of vertices whatever
value $d_e$ takes.  Moreover, we see from Eq.~(\ref{chap3:sdd}) that amongst the most superficially divergent amplitudes, the fermion self-energy ($N_e=2$, $N_\gamma=0$)
and the fermion-gauge vertex ($N_e=2$, $N_\gamma=1$) of all RQED$_{4,d_e}$s have the same SDD as in QED$_4$: $\om(\Sigma) = 1$ and $\om(\Gamma)=0$, respectively.
From the Lorentz structure of the fermion self-energy: $\Sigma(p) = \slashed p\,\Sigma_V(p^2)$, its effective degree of divergence is actually reduced to $D_e=0$ so that both 
fermion self-energy and fermion-gauge vertex are logarithmically divergent in RQED$_{4,d_e}$s.

On the other hand, the SDD of the photon self-energy ($N_e=0$, $N_\gamma=2$) is lowered in RQEDs: $\om(\Pi)=1$ for RQED$_{4,3}$ and $\om(\Pi)=0$ for RQED$_{4,2}$, with respect
to QED$_4$ (where $\om(\Pi)=2$). Taking into account of the tensorial structure of the photon self-energy: $\Pi^{\mu \nu}(q) = \big(\, g^{\mu \nu} q^2 - q^\mu q^\nu\,\big)\,\Pi(q^2)$,
the effective degree of divergence is even lower: $\om(\Pi)=0$ for QED$_4$, $\om(\Pi)=-1$ for RQED$_{4,3}$
and $\om(\Pi)=-2$ for RQED$_{4,2}$. This dimensional analysis therefore suggests that while $\Pi(q^2)$
logarithmically diverges in QED$_4$ it is finite in RQEDs. These elementary facts are summarized in Tab.~\ref{chap3:tab:sdd} 
which displays the degrees of divergence (superficial and effective) of the three most divergent amplitudes in QED$_4$, RQED$_{4,3}$ and
RQED$_{4,2}$.

\begin{center}
\renewcommand{\tabcolsep}{1cm}
\renewcommand{\arraystretch}{1.5}
\begin{table}
    \begin{tabular}{| l || c | c | c |}
      \hline
      \quad & QED$_4$ & RQED$_{4,3}$ & RQED$_{4,2}$ \\
      \hline \hline
      Photon self-energy & 2 (0) & 1 (-1) & 0 (-2)  \\
      \hline
      Fermion self-energy & 1 (0) & 1 (0) & 1 (0) \\
      \hline
      Fermion-photon vertex & 0 (0) & 0 (0) & 0 (0) \\
      \hline
    \end{tabular}
    \caption{Superficial degree of divergence (and effective degree of divergence) of the three most divergent amplitudes in RQED$_{4,d_e}$ for $d_e=4$ (QED$_4$), $d_e=3$ and $d_e=2$.}
    \label{chap3:tab:sdd}
\end{table}
\end{center}

We are now in a position to introduce the renormalization constants associated with a general model of RQED$_{4,d_e}$ along the lines of Eq.~(\ref{chap2:model:QED_d:Z}):
\be
\psi = Z_{\psi}^{1/2} \psi_r, \quad A = Z_A^{1/2} A_r, \quad e = Z_e e_r \mu^{\veps} = \frac{1}{Z_A^{1/2}}\, e_r \mu^{\veps_\gamma}, \quad \xi =  Z_A \xi_r\, ,
\label{chap3:RQED:Z}
\ee
where we have taken into account of the fact that $Z_g=1$, as in QED$_4$ (\ref{chap2:WI:Zxi}), and that the Ward identity 
\be
Z_\psi = Z_\Gamma\, ,
\label{chap3:WI:Zpsi}
\ee
holds for arbitrary $d_e$ which includes the case of QED$_4$, (\ref{chap2:WI:Zpsi}).
From dimensional analysis we expect that $Z_A=1$ in the reduced case ($d_e<4$). This is also in agreement with the argument of the previous section, see also statement below Eq.~(\ref{chap2:model:QED_d:BPH-L}),
that because the free gauge field has a non-local action it does not renormalize. Because the light velocity is a renormalization group invariant, it follows that there
is no renormalization of the coupling constant in the ultra-relativistic limit: $Z_\al = 1$. As a consequence, the $\beta$-function is zero and the coupling remains marginal to
all orders in perturbation theory:
\be
\beta(\al_r) = 0 \qquad d_e <4\, .
\ee
So, contrarily to the case of QED$_4$, quantum corrections do not break scale invariance in RQED$_{4,d_e}$ with $d_e<4$. 
In the condensed matter literature these results are reminiscent of the ($1+1$)-dimensional Tomonaga-Luttinger model (presented in Chap.~\ref{chap1}),
the $\beta$-function of which has been shown to vanish to all orders, see Refs.~\cite{PhysRevLett.67.3852,PhysRevB.47.16107}.

\section{Reduced QED at one loop}

We now proceed in analyzing the model with explicit calculations. Our interest will be mainly in reduced QED$_{4,3}$ but we shall carry out calculations for arbitrary $d_\gamma$ and $d_e$
thereby also recovering the well known cases of QED$_4$ and QED$_3$ (a detailed review of all calculations up to 2-loop in QED$_4$ with conventional renormalization can be found in the book of Grozin \cite{grozin2007lectures}).
From Eqs.~(\ref{chap2:SD:Pi}) and (\ref{chap2:SD:Sigma}), the one-loop fermion self-energy and polarization operators are defined as:
\begin{subequations}
\label{chap3:def:Pi1+Sigma1}
\bea
\I \Pi_1^{\mu \nu}(q) &=&
-\mu^{2\veps_\gamma} \int [\D^{d_e} k] \Tr \bigg[ (-\I e \gamma^\mu)\,S_0(k)\,(-\I e \gamma^\nu)\,S_0(k+q) \bigg]\, ,
\label{chap3:def:Pi1}\\
-\I \Sigma_1(p) &=& \mu^{2\veps_\gamma} \int [\D^{d_e} k] (-\I e \gamma^\mu)\,\tilde{D}_{0,\mu \nu}(p-k)\,S_0(k)\,(-\I e \gamma^\nu)\, ,
\label{chap3:def:Sigma1}
\eea
\end{subequations}
where it is convenient to introduce the renormalization scale from this early stage with the help of Eq.~(\ref{chap2:d4k}); this implies that, in the following, we can take $e=e_r$ 
with one-loop accuracy. Let's give a few details of how the calculations are carried out
in the simple one-loop case starting from the polarization operator. We start by using the fact that:
\be
\Pi(q^2) = \frac{- \Pi^\mu_\mu(q)}{(d_e-1)\,(-q^2)}\, ,
\ee
together with the trace Eq.~(\ref{app:Tr-Pi}), to re-write Eq.~(\ref{chap3:def:Pi1})  as:
\be
\Pi_1(q^2) = 4N_F\,\frac{\I e^2 \mu^{2\veps_\gamma}}{(-q^2)}\,\frac{d_e-2}{d_e-1}\,\int [\D^{d_e} k] \frac{(k,k+q)}{(-k^2)(-(k+q)^2)}\, ,
\label{chap3:Pi1:intermediate}
\ee
where the factor $N_F$ comes from the fermionic loop.
The last integral involves a scalar product and, in Euclidean space, is evaluated as:
\be
\int [\D^{d_e} k] \frac{(k,k+q)}{k^2(k+q)^2} = - q_\mu \int [\D^{d_e} k] \frac{k^\mu}{k^2(k+q)^2} = - \frac{(q^2)^{d_e/2-2}}{(4\pi)^{d_e/2}}\,q^2\,G^{(1,0)}(d_e,1,1)\, ,
\ee
which is a peculiar case of Eq.~(\ref{chap2:oneloop+arrows:res}) with $p=-q$ and $\al=\beta=1$. The function $G^{(n,0)}(d_e,\al,\beta)$ was defined in Eq.~(\ref{chap2:one-loop-Gn}) and, 
in the following, we shall use the fact that $G^{(1,0)}(d_e,1,1) = G(d_e,1,1)/2$. 
Going back to Minkowski space this yields:
\be
\Pi_1(q^2;\bar{\al}) = - 4 N_F \bar{\al} \left( \frac{4\pi}{-q^2} \right)^{\veps_e} \,\left( \frac{\overline{\mu}^{\,2}}{-q^2} \right)^{\veps_\gamma}\, \frac{d_e-2}{2(d_e-1)}\,e^{\gamma_E \veps_\gamma} G(d_e,1,1)\, ,
\label{chap3:res:Pi1}
\ee
where $\bar{\al}=\al/(4\pi)$ and $\overline{\mu}^{\,2}$ was defined in Eq.~(\ref{chap2:muMSbar}). In accordance with power counting, Eq.~(\ref{chap3:res:Pi1}) is UV singular in the limit $\veps_\gamma \ra 0$
only for $\veps_e=0$ as can be seen from the fact that:
\be
G(d_e,1,1) = \frac{\Gamma^2(d_e/2-1) \Gamma(2-d_e/2)}{\Gamma(d_e-2)} = \frac{\Gamma^2(1-\veps_e-\veps_\gamma) \Gamma(\veps_e+\veps_\gamma)}{\Gamma(2-2\veps_e-2\veps_\gamma)}\, .
\ee
The limits $\veps_e \ra 0$ and $\veps_\gamma \ra 0$ do not commute and one has to first set $\veps_e=0$ and then take $\veps_\gamma \ra 0$ in order to recover QED$_4$ 
while the order is unimportant for $\veps_e>0$ because the result is finite.

We may proceed in a similar way for the fermion self-energy. Using:
\be
\Sigma_V(p^2) = \frac{-1}{4N_F\,(-p^2)}\,\Tr[ \Sp \Sigma(p)]\, ,
\ee
together with the trace identity (\ref{app:Tr-Sigma1V}), Eq.~(\ref{chap3:def:Sigma1}) may be rewritten as:~\footnote{Notice that, in Eq.~(\ref{chap3:res:UV+IR}), 
while the first integral is UV divergent, the second one is IR divergent. The appearance of infrared divergences is very common in the process of computing a massless Feynman diagram
in dimensional regularization, in particular when expanding a given diagram in a set of simpler integrals. A way to understand why these IR poles are unavoidable is to recall that
the massless tadpole is zero in dimensional regularization because of a cancellation between IR and UV divergences, see discussion above Eq.~(\ref{chap2:one-loop-v-int2}).
At this point one may wonder if the IR poles generated in such a way have to be differentiated with respect to the UV poles in order not to spoil the whole renormalization process. 
As proved in Ref.~\cite{Chetyrkin:1983wh}, one does not need to care about this issue as long as the quantity being computed is infra-red safe. Fortunately, this is the case for QED$_4$ and RQED$_{4,3}$. One
may then proceed in computing integrals without distinguishing between IR and UV poles at intermediate stages of the calculation as the final poles will be UV ones. In other cases, \eg,
as a result of infrared rearrangement, dangerous IR poles may appear. The latter have to be properly taken into account, \eg, with a mass regulator or proper subtraction using the 
$R^*$-operator \cite{Chetyrkin:1982nn,Chetyrkin:1984xa}, see also Ref.~\cite{Grisaru:1986wj} and the textbook \cite{kleinert2001critical} for some concrete examples. 
I thank M.~Kompaniets for pointing out Ref.~\cite{Chetyrkin:1983wh} to me.}
\begin{flalign}
\Sigma_{1V}(p^2) &= -\frac{\I e^2 \mu^{2\veps_\gamma}}{(-p^2)}\,\frac{\Gamma(1-\veps_e)}{(4\pi)^{\veps_e}}\,(d_e-2)\,\int [\D^{d_e} k] \frac{(p,k)}{(-k^2)(-(p-k)^2)^{1-\veps_e}} 
\nonum \\
&- \I e^2 \mu^{2\veps_\gamma}\,\frac{\Gamma(1-\veps_e)}{(4\pi)^{\veps_e}}\,(1-\tilde{\xi})\,\int [\D^{d_e} k] \frac{(p,k)}{(-k^2)(-(p-k)^2)^{2-\veps_e}}\, .
	\label{chap3:res:UV+IR}
\end{flalign}
In Euclidean space the two integrals can be evaluated using:
\bea
\int [\D^{d_e} k] \frac{(p,k)}{k^2(p-k)^{2\al}} = \frac{(p^2)^{d_e/2-\al}}{(4\pi)^{d_e/2}}\,G^{(1,0)}(d_e,1,\al)\, ,
\eea
and, back in Minkowski space, the fermion self-energy reads:
\be
\Sigma_{1V}(p^2) = \bar{\al} \,\left( \frac{\overline{\mu}^{\,2}}{-p^2} \right)^{\veps_\gamma}\, \Gamma(1-\veps_e)\,\frac{d_e-2}{2}\,
\bigg( \frac{\veps_e}{d_e-2+\veps_e} - \xi\,\bigg)\,e^{\gamma_E \veps_\gamma} G(d_e,1,1-\veps_e)\, ,
\label{chap3:res:Sigma1}
\ee
where we have used the fact that $(1-\tilde{\xi}) = (1-\veps_e)\,(1-\xi)$. Recalling that the photon propagator (internal line in $\Sigma_1$) has
a longitudinal and a transverse part, see (\ref{chap2:RQED:FR:Dperp+Dpara}), a similar decomposition holds for the one-loop fermion self-energy:
\begin{subequations}
\label{chap3:Sigma1:LT}
\bea
\Sigma_{1V}(p^2) &=& \Sigma_{1V}^{(\parallel)}(p^2) + \Sigma_{1V}^{(\perp)}(p^2) \, , 
\label{chap3:Sigma1:L+T}\\
\Sigma_{1V}^{(\parallel)}(p^2) &=& -\frac{\tilde{\xi}}{1-\veps_e} \,\bar{\al} \,\left( \frac{\overline{\mu}^{\,2}}{-p^2} \right)^{\veps_\gamma}\, 
\Gamma(1-\veps_e)\,\frac{d_e-2}{2}\, e^{\gamma_E \veps_\gamma} G(d_e,1,1-\veps_e)\, ,
\label{chap3:Sigma1:L}\\
\Sigma_{1V}^{(\perp)}(p^2) &=& \frac{\veps_e}{1-\veps_e} \, \bar{\al} \,\left( \frac{\overline{\mu}^{\,2}}{-p^2} \right)^{\veps_\gamma}\, 
\Gamma(1-\veps_e)\,\frac{(d_e-2)\,(d_e-1)}{2(d_e-2+\veps_e)}\,e^{\gamma_E \veps_\gamma} G(d_e,1,1-\veps_e)\, .
\label{chap3:Sigma1:T}
\eea
\end{subequations}
in agreement with Eq.~(\ref{chap3:res:Sigma1}).
Moreover, in accordance with power counting, Eq.~(\ref{chap3:res:Sigma1}) is UV singular in the limit $\veps_\gamma \ra 0$ for all $\veps_e$
and the singular part reads:
\be
\delta Z_{1\psi}(\bar{\al}_r,\xi_r) = \mathcal{K}\,\bigg[ \Sigma_{1V}(p^2) \bigg] = \bar{\al}_r \,\bigg( \frac{\veps_e}{2-\veps_e} - \xi_r \bigg)\,\frac{1}{\veps_\gamma}\, .
\ee
Notice that for the one-loop fermion self-energy, the limits $\veps_e \ra 0$ and $\veps_\gamma \ra 0$ commute. In the case of reduced QED$_{4,d_e}$, it is therefore possible to expand
Eq.~(\ref{chap3:res:Sigma1}) in $\veps_\gamma \ra 0$ while keeping $\veps_e$ arbitrary. It's expansion up to $\Ord(\veps_\gamma)$ reads (see Eq.~(\ref{chap3:RQED4,de:exp:G1,1-ee}) for the expansion of the master integral):
\begin{flalign}
&\Sigma_{1V}(p^2) = \bar{\al}\,\Bigg[ \left(\frac{\veps_e}{2-\veps_e} - \xi \right)\,\frac{1}{\veps_\gamma}
- \left(\frac{\veps_e}{2-\veps_e} -\xi \right)\,\bar{L}_p + \frac{2\veps_e}{(2-\veps_e)^2} +
\label{chap3:RQED4,de:Sigma1} \\
&\Bigg . + \biggl( \frac{1}{2}\,\left(\frac{\veps_e}{2-\veps_e} - \xi \right)\,\biggl(\bar{L}_p^2+2\zeta_2-3\Psi_2(2-\veps_e) \biggr) -
\frac{2\veps_e}{(2-\veps_e)^2}\,\left(\bar{L}_p-\frac{2}{2-\veps_e} \right) \biggr)\, \veps_\gamma +  \Ord(\veps_\gamma^2) \Bigg] \, .
\nonum
\end{flalign}
where $\bar{L}_p = L_p - \Psi_1(2-\veps_e)+ \Psi_1(1)$ and $\Psi_1$ and $\Psi_2$ are the digamma and trigamma functions, respectively.

At this point, we may also explicitly check that the Ward identity Eq.~(\ref{chap2:ward:graphical}) holds for reduced QED. From Eq.~(\ref{chap2:SD:Lambda}), the one-loop photon fermion vertex is defined as:
\be
-\I e \Lambda_1^{\mu}(p,p') = \mu^{2\veps_\gamma} \int [\D^{d_e} k] \tilde{D}_0^{\al \beta}(p-k)\,(-\I e \gamma_\al)\,S_0(k)\,(-\I e \gamma^\mu)\,S_0(k+q)\,(-\I e \gamma_\beta)\, ,
\label{chap3:def:Lambda1}
\ee
where $q=p'-p$. It enough to compute this function at $p=p'=0$ in order to extract the singular UV part. 
An infrared singularity then appears which can be regulated by introducing a small mass (as pointed out in Ref.~\cite{grozin2007lectures} this is a simple example of IRR).
%This is a simple example of the so-called infra-red rearrangement (IRR) \cite{Chetyrkin:1981qh,Chetyrkin:1982nn}. 
This leads to:
\be
\Lambda_1^{\mu}(p=p'=0) = \bar{\al}\,\gamma^\mu\,\left( \frac{\overline{\mu}^{\,2}}{m^2} \right)^{\veps_\gamma}\,
\bigg( \frac{(d_e-2)^2}{d_e(1-\veps_e)} - (1 - \xi) \bigg)\,e^{\gamma_E \veps_\gamma}\,\frac{\Gamma(1+\veps_\gamma)}{\veps_\gamma} \, ,
%=
%\xi_r \bar{\al}_r\,\gamma^\mu\,\left[ \frac{1}{\veps} + \Ord(1) \right]\, ,
\label{chap3:Lambda1}
\ee
which is UV singular in the limit $\veps_\gamma \ra 0$ for all $\veps_e$ and its singular part reads:
\be
\delta Z_{1\Gamma}(\bar{\al}_r,\xi_r) = \mathcal{K}\,\bigg[ \Lambda_1^{\mu}(p,p')/ \gamma^\mu \bigg] = -\bar{\al}_r \,\bigg( \frac{\veps_e}{2-\veps_e} - \xi_r \bigg)\,\frac{1}{\veps_\gamma} = - \delta Z_{1\psi}(\bar{\al}_r,\xi_r)\, ,
\label{chap3:ward}
\ee
in agreement with Eq.~(\ref{chap2:ward:graphical}).

\subsection{Case of QED$_4$}

In the case of QED$_4$ we first set $\veps_e=0$ and take the limit $\veps_\gamma \ra 0$. From Eqs.~(\ref{chap3:res:Pi1}) and (\ref{chap3:res:Sigma1}), this leads to:
\begin{subequations}
\label{chap3:QED4:Pi1+Sigma1}
\begin{flalign}
\Pi_1(q^2;\bar{\al}) &= - 4N_F\, \bar{\al} \left( \frac{\overline{\mu}^{\,2}}{-q^2} \right)^{\veps_\gamma}\, \frac{d_\gamma -2}{2(d_\gamma-1)}\,e^{\gamma_E \veps_\gamma} G(4-2\veps_\gamma,1,1)
\nonum \\
&= - 4N_F\, \bar{\al} \left [ \frac{1}{3\veps_\gamma} -\frac{1}{3}\, \left(L_q - \frac{5}{3} \right) + \frac{1}{6} \left( L_q^2 - \frac{10 L_q}{3} - \zeta_2 + \frac{56}{9} \right)\,\veps_\gamma + \Ord(\veps_\gamma^2) \right]\, ,
\label{chap3:QED4:Pi1}\\
\Sigma_{1V}(p^2;\bar{\al},\xi) &= - \xi \bar{\al} \left( \frac{\overline{\mu}^2}{-p^2} \right)^{\veps_\gamma}\, \frac{d_\gamma-2}{2}\,e^{\gamma_E \veps_\gamma} G(4-2\veps_\gamma,1,1)
\nonum \\
&= - \xi \bar{\al} \left [ \frac{1}{\veps_\gamma} -L_p+1 + \frac{1}{2} \left( 4 - \zeta_2 + L_p^2 - 2 L_p \right)\,\veps_\gamma +  \Ord(\veps_\gamma^2) \right]\, ,
\label{chap3:QED4:Sigma1}
\end{flalign}
\end{subequations}
where $L_x = \log(x^2/\overline{\mu}^{\,2})$.
Substituting Eq.~(\ref{chap3:QED4:Pi1+Sigma1}) in Eq.~(\ref{chap2:RPT:counterterms}), the one-loop renormalization constants in the $\overline{\text{MS}}$ scheme read:
\be
\delta Z_{1A}(\bar{\al}_r) = - \frac{4 N_F\,\bar{\al}_r}{3\veps_\gamma}, \qquad \delta Z_{1\psi}(\bar{\al}_r,\xi_r) = - \frac{\xi_r \bar{\al}_r}{\veps_\gamma}\, ,
\label{chap2:QED4:ZA+Zpsi}
\ee
The corresponding anomalous dimensions and $\beta$-function read:
%
%\begin{subequations}
%\label{chap3:QED4:Z+gamma}
\bea
\gamma_{1A}(\bar{\al}_r) = - \frac{8N_F\, \bar{\al}_r}{3}, \qquad \beta(\bar{\al}_r) = \frac{8 N_F\,\bar{\al}_r}{3} + \Ord(\bar{\al}_r^2), \qquad
%\label{chap3:ZA+gammaA}\\
\gamma_{1\psi}(\bar{\al}_r,\xi_r) = - 2\xi_r \bar{\al}_r\, ,
%\label{chap3:Zpsi+gammapsi}
\label{chap3:QED4:Z+gamma}
\eea
%\end{subequations}
%
where we have used the fact that $\xi \bar{\al}=\xi_r \bar{\al}_r$  and $\bar{\al}=\bar{\al}_r$ with one-loop precision. 
The one-loop renormalized self-energies (\ref{chap2:renormalized-self-energies}) then read:
\be
\Pi_{1r}(q^2) = \frac{4N_F\, \bar{\al}_r}{3}\, \left(L_q - \frac{5}{3} \right), \qquad \Sigma_{1Vr}(p^2) = \xi_r \bar{\al}_r (L_p-1)\, .
\label{chap3:QED4:Pi1r+Sigma1r}
\ee
Hence, up to one-loop, the (transverse part of the) renormalized photon propagator and renormalized fermion propagator have the following expansions:
\begin{subequations}
\label{chap2:QED4:Dr+Sr}
\bea
&&\I q^2\,d_{r\perp}(q^2) =  1 + \frac{4 N_F\,\bar{\al}_r}{3}\,\left( L_q - \frac{5}{3} \right) + \Ord(\bar{\al}_r^2) \, ,
\label{chap3:QED4:Dr}\\
&&-\I \Sp S_r(p) = 1 + \xi_r \bar{\al}_r\,(L_p - 1) + \Ord(\bar{\al}_r^2)\, .
\label{chap3:QED4:Sr}
\eea
\end{subequations}

\subsection{Case of reduced QED$_{4,3}$}

In the case of reduced QED$_{4,3}$ we first set $\veps_e=1/2$ and take the limit $\veps_\gamma \ra 0$. From Eqs.~(\ref{chap3:res:Pi1}) and (\ref{chap3:res:Sigma1}), this leads to:
\begin{subequations}
\label{chap3:RQED4,3:Pi1+Sigma1}
\begin{flalign}
\Pi_1(q^2;\bar{\al}) &= - N_F\,\bar{\al}\, \frac{2\pi^2}{\sqrt{-q^2}}\,\left [ 1 - (L_q - \log 4 + 1) \,\veps_\gamma +  \Ord(\veps_\gamma^2) \right]\, ,
\label{chap3:RQED4,3:Pi1}\\
\Sigma_{1V}(p^2;\bar{\al},\xi) &= \bar{\al} \left [ \frac{1-3\xi}{3\veps_\gamma} - \frac{1-3\xi}{3}\,\tilde{L}_p - 2\xi + \frac{10}{9}
\right . 
\nonum \\
&\left . + \left( \frac{1-3\xi}{6}\, ( \tilde{L}_p^2 - 7 \zeta_2 ) + 2\,\left( \xi - \frac{5}{9} \right)\,\tilde{L}_p - 8 \xi +\frac{112}{27} \right)\,\veps_\gamma +  \Ord(\veps_\gamma^2) \right]\, ,
\label{chap3:RQED4,3:Sigma1}
\end{flalign}
\end{subequations}
where $\tilde{L}_x = L_x + \log 4$. For completeness, we also give the expansion of the longitudinal and transverse parts of the fermion self-energy:
\begin{subequations}
\label{chap3:RQED4,3:Sigma1L+T}
\begin{flalign}
	\Sigma_{1V}^{(\parallel)}(p^2) &= -(1+\xi) \,\bar{\al} \, \left [ \frac{1}{\veps_\gamma} + 2 - \tilde{L}_p + 
\frac{\veps_\gamma}{2}\,\left( \tilde{L}_p^2 - 4 \tilde{L}_p - 7\zeta_2 + 16  \right)\,\veps_\gamma +  \Ord(\veps_\gamma^2) \right]\, ,
\label{chap3:RQED4,3:Sigma1L} \\
\Sigma_{1V}^{(\perp)}(p^2) &= \frac{4\,\bar{\al}}{3} \, \left [ \frac{1}{\veps_\gamma} - \tilde{L}_p + \frac{7}{3} +
\frac{\veps_\gamma}{2}\,\left( \tilde{L}_p^2 - \frac{14\,\tilde{L}_p}{3} - 7\zeta_2 + \frac{164}{9}  \right)\,\veps_\gamma +  \Ord(\veps_\gamma^2) \right]\, .
\label{chap3:RQED4,3:Sigma1T} 
\end{flalign}
\end{subequations}

Substituting Eq.~(\ref{chap3:RQED4,3:Pi1+Sigma1}) in Eq.~(\ref{chap2:RPT:counterterms}), the one-loop renormalization constants in the $\overline{\text{MS}}$ scheme read:
\be
\delta Z_{1A}(\bar{\al}_r) = 0, \qquad \delta Z_{1\psi}(\bar{\al}_r,\xi_r) =  \bar{\al}_r\,\frac{1-3\xi_r}{3\veps_\gamma}\, .
\label{chap2:RQED4,3:ZA+Zpsi}
\ee
This yields the following anomalous dimensions:
\bea
\gamma_{1A}(\bar{\al}_r) =  0 , \qquad
%\label{chap3:ZA+gammaA}\\
\gamma_{1\psi}(\bar{\al}_r,\xi_r) =  2\bar{\al}_r\,\frac{1-3\xi_r}{3}\, ,
%\label{chap3:Zpsi+gammapsi}
\label{chap3:QED4,3:Z+gamma}
\eea
and the $\beta$-function vanishes as discussed above.
%\begin{subequations}
%\label{chap2:Z+gamma}
%\bea
%&&\delta Z_{1A}(\bar{\al}_r) = - \frac{4 \bar{\al}_r}{3\veps}, \qquad \gamma_{1A}(\bar{\al}_r) = - \frac{8 \bar{\al}_r}{3}, \qquad \beta(\bar{\al}_r) = \frac{8 \bar{\al}_r}{3} + \Ord(\bar{\al}_r^2)\, ,
%\label{chap2:ZA+gammaA}\\
%&&\delta Z_{1\psi}(\bar{\al}_r,\xi_r) = - \frac{\xi_r \bar{\al}_r}{\veps}, \qquad \gamma_{1\psi}(\bar{\al}_r,\xi_r) = - 2\xi_r \bar{\al}_r\, ,
%\label{chap2:Zpsi+gammapsi}
%\eea
%\end{subequations}
%
The one-loop renormalized self-energies (\ref{chap2:renormalized-self-energies}) then read:
\be
\Pi_{1r}(q^2) = - \frac{N_F e^2}{8\,\sqrt{-q^2}}, \qquad \Sigma_{1Vr}(p^2) = -\bar{\al}_r \left( \frac{1-3\xi_r}{3}\,\tilde{L}_p + 2\xi_r - \frac{10}{9} \right)\, .
\label{chap3:RQED4,3:Pi1r+Sigma1r}
\ee
The expansions of the (transverse part of the) renormalized photon propagator and renormalized fermion propagator read:
\begin{subequations}
\label{chap2:RQED4,3:Dr+Sr}
\bea
&&d_{r\perp}(q^2) =  \frac{\I}{2\sqrt{-q^2}}\, \frac{1}{1 + N_F e^2/16}\, ,
\label{chap3:RQED4,3:Dr}\\
&&-\I \Sp S_r(p) = 1 -\bar{\al}_r \left( \frac{1-3\xi_r}{3}\,\tilde{L}_p + 2\xi_r - \frac{10}{9} \right) + \Ord(\bar{\al}_r^2)\, .
\label{chap3:RQED4,3:Sr}
\eea
\end{subequations}
From Eq.~(\ref{chap3:RQED4,3:Dr}), we see that the transverse part of the renormalized photon propagator essentially remains free.

\subsection{Case of QED$_3$}

In the case of QED$_3$ we first set $\veps_e=0$ and define $\veps_\gamma = 1/2 + \delta_\gamma$ such that $d_\gamma = 3 - 2\delta_\gamma$ where the limit $\delta_\gamma \ra 0$ has to be taken. 
All results are finite and, from Eqs.~(\ref{chap3:res:Pi1}) and (\ref{chap3:res:Sigma1}), they read:
\begin{subequations}
\label{chap3:QED3:Pi1+Sigma1}
\begin{flalign}
\Pi_1(q^2) &= \Pi_{1r}(q^2) = - 4 N_F\,\frac{\al}{\sqrt{4\pi}\sqrt{-q^2}}\, \left( \frac{\overline{\mu}^{\,2}}{-q^2} \right)^{\delta_\gamma}\, \frac{d_\gamma -2}{2(d_\gamma-1)}\,e^{\gamma_E \delta_\gamma} G(3-2\delta_\gamma,1,1)
\nonum \\
&= - \frac{N_F\,e^2}{8\,\sqrt{-q^2}}\,\left [ 1 - (1+L_q - \log 4) \,\delta_\gamma +  \Ord(\delta_\gamma^2) \right]\, ,
\label{chap3:QED3:Pi1}\\
\Sigma_{1V}(p^2) &= \Sigma_{1Vr}(p^2) 
= -\frac{\xi \, \al}{\sqrt{4\pi}\sqrt{-p^2}}\,\left( \frac{\overline{\mu}^{\,2}}{-p^2} \right)^{\delta_\gamma}\, \frac{d_\gamma -2}{2}\,e^{\gamma_E \delta_\gamma} G(3-2\delta_\gamma,1,1)
\nonum \\
&=-\xi\,\frac{e^2}{16\,\sqrt{-p^2}}\,\left [ 1 - (2+L_p - \log 4) \,\delta_\gamma +  \Ord(\delta_\gamma^2) \right]\, , 
\label{chap3:QED3:Sigma1}
\end{flalign}
\end{subequations}
where $e^2$ has dimension of mass. Notice that, in the case of QED$_3$, the transverse part of the renormalized photon propagator has the form:
\bea
d_{r\perp}(q^2) =  \frac{\I}{-q^2}\, \frac{1}{1 + \frac{N_F e^2}{8 \sqrt{-q^2}}} \approx \frac{8}{N_F e^2}\,\frac{\I}{\sqrt{-q^2}}\, ,
\label{chap3:RQED3:Dr}
\eea
where the last equality holds in the limit $Ne^2/8 \gg \sqrt{-q^2}$. This shows that the transverse photon propagator of QED$_3$ is softened in the IR and
has the same ``Coulomb''-like form as the photon propagator of RQED$_{4,3}$, Eq.~(\ref{chap3:RQED4,3:Dr}). This fact will play an important role in the next chapter and we shall come back on it then.

\section{Two-loop polarization operator (general results)}

\begin{figure}
\begin{center}
    a)
    \begin{fmfgraph*}(35,30)
      %\fmfpen{thick}
      \fmfleft{i}
      \fmfright{o}
      \fmf{photon}{i,v1}
      \fmf{photon,label=$q$}{v2,o}
      \fmf{phantom,right,tension=0.1,tag=1}{v1,v2}
      \fmf{phantom,right,tension=0.1,tag=2}{v2,v1}
      \fmf{phantom,tension=0.1,tag=3}{v1,v2}
      \fmfdot{v1,v2}
      \fmfposition
      \fmfipath{p[]}
      \fmfiset{p1}{vpath1(__v1,__v2)}
      \fmfiset{p2}{vpath2(__v2,__v1)}
      \fmfiset{p3}{vpath3(__v1,__v2)}
      \fmfi{fermion,label=$k+q$}{subpath (0,length(p1)) of p1}
      \fmfi{fermion,label=$k$}{subpath (0,length(p2)/4) of p2}
      \fmfi{fermion}{subpath (length(p2)/4,3length(p2)/4) of p2}
      \fmfi{fermion,label=$k$}{subpath (3length(p2)/4,length(p2)) of p2}
      \fmfi{photon}{point length(p2)/4 of p2 .. point length(p3)/2 of p3 .. point 3length(p2)/4 of p2}
      \def\vert#1{%
        \fmfiv{decor.shape=circle,decor.filled=full,decor.size=2thick}{#1}}
      \vert{point length(p2)/4 of p2}
      \vert{point 3length(p2)/4 of p2}
    \end{fmfgraph*}
    \qquad \quad 
    \begin{fmfgraph*}(35,30)
      %\fmfpen{thick}
      \fmfleft{i}
      \fmfright{o}
      \fmf{photon}{i,v1}
      \fmf{photon,label=$q$,label.side=left}{v2,o}
      \fmf{phantom,right,tension=0.1,tag=1}{v1,v2}
      \fmf{phantom,right,tension=0.1,tag=2}{v2,v1}
      \fmf{phantom,tension=0.1,tag=3}{v1,v2}
      \fmfdot{v1,v2}
      \fmfposition
      \fmfipath{p[]}
      \fmfiset{p1}{vpath1(__v1,__v2)}
      \fmfiset{p2}{vpath2(__v2,__v1)}
      \fmfiset{p3}{vpath3(__v1,__v2)}
      \fmfi{fermion,label=$k+q$}{subpath (0,length(p1)/4) of p1}
      \fmfi{fermion}{subpath (length(p1)/4,3length(p1)/4) of p1}
      \fmfi{fermion,label=$k+q$}{subpath (3length(p1)/4,length(p1)) of p1}
      \fmfi{photon}{point length(p1)/4 of p1 .. point length(p3)/2 of p3 .. point 3length(p1)/4 of p1}
      \fmfi{fermion,label=$k$}{subpath (0,length(p2)) of p2}
      \def\vert#1{%
        \fmfiv{decor.shape=circle,decor.filled=full,decor.size=2thick}{#1}}
      \vert{point length(p1)/4 of p1}
      \vert{point 3length(p1)/4 of p1}
    \end{fmfgraph*}
    \qquad \quad
    b) 
    \begin{fmfgraph*}(35,30)
      %\fmfpen{thick}
      \fmfleft{i}
      \fmfright{o}
      \fmf{photon}{i,v1}
      \fmf{photon,label=$q$}{v2,o}
      \fmf{phantom,right,tension=0.1,tag=1}{v1,v2}
      \fmf{phantom,right,tension=0.1,tag=2}{v2,v1}
      \fmf{phantom,tension=0.1,tag=3}{v1,v2}
      \fmfdot{v1,v2}
      \fmfposition
      \fmfipath{p[]}
      \fmfiset{p1}{vpath1(__v1,__v2)}
      \fmfiset{p2}{vpath2(__v2,__v1)}
      \fmfi{fermion,label=$k_1+q$}{subpath (0,length(p1)/2) of p1}
      \fmfi{fermion,label=$k_2+q$}{subpath (length(p1)/2,length(p1)) of p1}
      \fmfi{fermion,label=$k_2$}{subpath (0,length(p2)/2) of p2}
      \fmfi{fermion,label=$k_1$}{subpath (length(p2)/2,length(p2)) of p2}
      \fmfi{photon,label=$k_{12}$}{point length(p1)/2 of p1 -- point length(p2)/2 of p2}
      \def\vert#1{%
        \fmfiv{decor.shape=circle,decor.filled=full,decor.size=2thick}{#1}}
      \vert{point length(p1)/2 of p1}
      \vert{point length(p2)/2 of p2}
    \end{fmfgraph*}
  \caption{\label{chap3:fig:two-loop:Pi}
  Two-loop photon self-energy diagrams ($k_{12} = k_1-k_2$).}
  \end{center}
\end{figure}
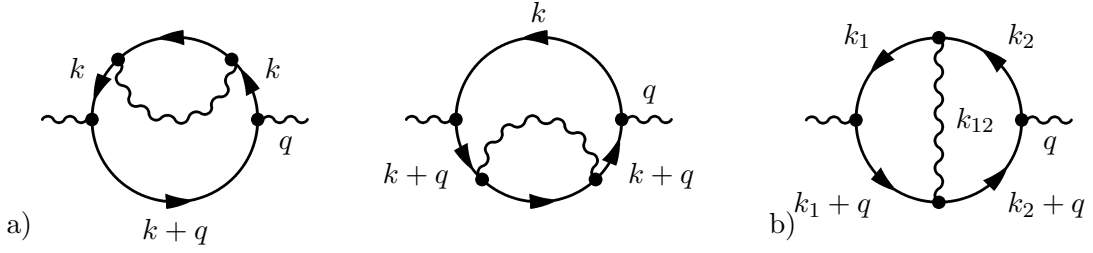

We now go on to two-loop calculations and first focus on the polarization operator. The total two-loop photon self-energy can be decomposed as follows:
\be
\Pi_2^{\mu \nu}(q) = 2\Pi_{2a}^{\mu \nu}(q) + \Pi_{2b}^{\mu \nu}(q)\, ,
\label{chap3:Pi2=2Pi2a+Pi2b}
\ee
where the diagrams are displayed on Fig.~\ref{chap3:fig:two-loop:Pi} and are defined as:
\begin{subequations}
\label{chap3:def:Pi2}
\begin{flalign}
	\I \Pi_{2a}^{\mu \nu}(q) &= -\mu^{2\veps_\gamma} \int [\D^{d_e} k] \Tr \bigg[ (-\I e \gamma^\mu)\,S_0(k+q)\,(-\I e \gamma^\nu)\,S_0(k)\,(-\I \Sk \Sigma_{1V}(k))\,S_0(k) \bigg]\, ,
\label{chap3:def:Pi2a}\\
\I \Pi_{2b}^{\mu \nu}(q) &=
	-\mu^{4\veps_\gamma} \int [\D^{d_e} k_1][\D^{d_e} k_2] \Tr \bigg[ (-\I e \gamma^\nu)\,S_0(k_2+q)\,(-\I e \gamma^\al)\,S_0(k_1+q)\,(-\I e \gamma^\mu)\, \bigg .
\nonum \\
&\times \bigg . S_0(k_1)\,(-\I e \gamma^\beta)\,S_0(k_2)\,\tilde{D}_{0 \al \beta}(k_1-k_2) \bigg]\, ,
\label{chap3:def:Pi2c}
\end{flalign}
\end{subequations}
where $\Sigma_{1V}(k)$ is given by Eq.~(\ref{chap3:res:Sigma1}). Because $\Pi^{\mu \nu}(q)$ is gauge independent, all calculation can be carried out in a specific gauge. 
In the following we shall work in the Feynman gauge, $\xi=1$. The calculation of $\Pi_{2a}(q^2)$, which is recursively one-loop, can then be done along the same lines as the one of $\Pi_1(q^2)$. 
The first steps lead to:
\be
\Pi_{2a}(q^2) = 4N_F\,\frac{\I e^2 \mu^{2\veps_\gamma}}{(-q^2)}\,\frac{d_e-2}{d_e-1}\,\int [\D^{d_e} k] \Sigma_{1V}(k^2)\,\frac{(k,k+q)}{(-k^2)(-(k+q)^2)}\, ,
\label{chap3:Pi2a:intermediate}
\ee
which is identical to Eq.~(\ref{chap3:Pi1:intermediate}) up to the insertion of $\Sigma_{1V}(k^2)$. 
Substituting Eq.~(\ref{chap3:res:Sigma1}), the result reads:
\begin{flalign}
\Pi_{2a}(q^2) =& 4N_F\, \bar{\al}^2 \,\left( \frac{4\pi}{-q^2}\right)^{\veps_e}\, \left( \frac{\overline{\mu}^{\,2}}{-q^2} \right)^{2\veps_\gamma}\,\Gamma(1-\veps_e)\, 
\nonum \\
&\times \, \frac{(d_e-2)^4}{2\,(d_e-1)\,(d_e-2+\veps_e)(d_\gamma-4)}\,e^{2\gamma_E \veps_\gamma} G(d_e,1,1-\veps_e)G(d_e,1,\veps_\gamma)\, .
\label{chap3:res:Pi2a}
\end{flalign}
The calculation of $\Pi_{2b}(q^2)$ is a little more tedious. After some algebra, the result reads:
\begin{flalign}
&\Pi_{2b}(q^2) = -4N_F\, \bar{\al}^2 \,\left( \frac{4\pi}{-q^2}\right)^{\veps_e}\, \left( \frac{\overline{\mu}^{\,2}}{-q^2} \right)^{2\veps_\gamma}\,\Gamma(1-\veps_e)\,\frac{d_e-2}{2(d_e-1)}\,e^{2\gamma_E \veps_\gamma}\,
\bigg [ \,2\, G(d_e,1,1-\veps_e)G(d_e,1,\veps_\gamma)\,\times \bigg .
\nonum \\
&\left . \times \left(d_e-4 + \frac{2(d_e-2)^3}{d_e\,(d_\gamma-4)} - \frac{4(d_e-2)}{d_e+d_\gamma-6} -\frac{4(d_e-2)^2}{(d_e + d_\gamma-4)^2} 
+ \frac{(d_e-2)(d_e^2-8)}{d_e(d_e + d_\gamma-4)} \right) \right .
\nonum \\
&\left . \qquad - G(d_e,1,1,1,1,1-\veps_e)\,\left(d_e - 4 + \frac{4(d_e-2)}{d_e+d_\gamma-6} - \frac{d_e(d_e-2)}{d_e+d_\gamma-4} \right) \right ]\, ,
\label{chap3:res:Pi2c}
\end{flalign}
where the two-loop master integral $G(d_e,1,1,1,1,\al)$ with index $\al=1-\veps_e$ appears. In the case of QED$_4$ and QED$_3$ the photon propagator has a pole, $\veps_e=0$, and the
master integral reduces to $G(d_\gamma,1,1,1,1,1)$ which is well known and actually recursively one-loop, see Eq.~(\ref{chap2:G(1,1,1,1,1)}). 
The case of reduced QED$_{4,3}$ where the photon propagator has a square root branch cut, $\veps_e=1/2$, is more complicated. The function $G(3-2\veps_\gamma,1,1,1,1,1/2)$ relevant to that case
can be obtained from general results due to either Kazakov \cite{Kazakov:1983pk} or Kotikov \cite{Kotikov:1995cw} and that we already advertised in (\ref{chap2:res:I(al):Kazakov}) 
and (\ref{chap2:res:I(al):Kotikov}), respectively. For arbitrary $\veps_e$, the result of Ref.~\cite{Kotikov:1995cw} can be written explicitly in terms of a ${}_3 F_2$ hypergeometric function as:
\begin{flalign}
&G(4-2\veps_e-2\veps_\gamma,1,1,1,1,1-\varepsilon_e) = 2 \Gamma(1- \varepsilon_e - \varepsilon_\gamma)\Gamma(-\varepsilon_\gamma) \Gamma(\varepsilon_e+2\varepsilon_\gamma) \times
\label{chap3:res:I(al):Kotikov}\\
&\times \left [ \frac{- \Gamma(1- \varepsilon_e - \varepsilon_\gamma)}{(1+\varepsilon_\gamma)\Gamma(2-\varepsilon_e)\Gamma(1-2\varepsilon_e-3\varepsilon_\gamma)}\,
\setlength\arraycolsep{1pt}
\pFq{3}{2}{1,\,  2-2\varepsilon_e -2\varepsilon_\gamma,\,  1+\varepsilon_\gamma}{2-\varepsilon_e,\, 2+\varepsilon_\gamma}{1}
+\frac{\pi \cot \pi (\varepsilon_e+2\varepsilon_\gamma)}{\Gamma(2-2\varepsilon_e -2\varepsilon_\gamma)} \right ]\, .
\nonum
\end{flalign}

Eqs.~(\ref{chap3:res:Pi2a}) and (\ref{chap3:res:Pi2c}) provide the general expression for the 2-loop photon self energy in an arbitrary reduced QED$_{d_\gamma, d_e}$. The two-loop diagrams contain 
divergent subgraphs which have to be subtracted in order to renormalize the theory. This can be very easily done using the forest formula Eq.~(\ref{chap2:def:forest}). Graphically,
the renormalization constants associated with the 2-loop diagrams read:
\begin{subequations}
\label{chap3:def:ZA-forest}
\begin{flalign}
2\,\delta Z_{2a\,A}(\bar{\al}_r) =& 2\,\mathcal{K}\, \bigg[~
\parbox{15mm}{
    \begin{fmfgraph*}(15,15)
      %\fmfpen{thick}
      \fmfleft{i}
      \fmfright{o}
      \fmf{photon}{i,v1}
      \fmf{photon}{v2,o}
      \fmf{phantom,right,tension=0.1,tag=1}{v1,v2}
      \fmf{phantom,right,tension=0.1,tag=2}{v2,v1}
      \fmf{phantom,tension=0.1,tag=3}{v1,v2}
      \fmfdot{v1,v2}
      \fmfposition
      \fmfipath{p[]}
      \fmfiset{p1}{vpath1(__v1,__v2)}
      \fmfiset{p2}{vpath2(__v2,__v1)}
      \fmfiset{p3}{vpath3(__v1,__v2)}
      \fmfi{plain}{subpath (0,length(p1)) of p1}
      \fmfi{plain}{subpath (0,length(p2)/4) of p2}
      \fmfi{plain}{subpath (length(p2)/4,3length(p2)/4) of p2}
      \fmfi{plain}{subpath (3length(p2)/4,length(p2)) of p2}
      \fmfi{photon}{point length(p2)/4 of p2 .. point length(p3)/2 of p3 .. point 3length(p2)/4 of p2}
      \def\vert#1{%
        \fmfiv{decor.shape=circle,decor.filled=full,decor.size=2thick}{#1}}
      \vert{point length(p2)/4 of p2}
      \vert{point 3length(p2)/4 of p2}
    \end{fmfgraph*}
} ~\bigg] -
2\,\mathcal{K}\, \bigg[ ~\mathcal{K}\, \bigg[~
   \parbox{15mm}{
    \begin{fmfgraph*}(15,15)
      %\fmfpen{thick}
      \fmfleft{in}
      \fmfright{out}
      \fmf{plain}{in,ve}
      \fmf{plain,right,tension=0.2}{ve,vw}
      \fmf{boson,right,tension=0.2}{vw,ve}
      \fmf{plain}{vw,out}
      \fmfdot{ve,vw}
    \end{fmfgraph*}
}~ \bigg]~\star~
\parbox{15mm}{
    \begin{fmfgraph*}(15,15)
      %\fmfpen{thick}
      \fmfleft{in}
      \fmfright{out}
      \fmf{boson}{in,ve}
      \fmf{plain,right,tension=0.2}{ve,vw}
      \fmf{plain,right,tension=0.2}{vw,ve}
      \fmf{boson}{vw,out}
      \fmfdot{ve,vw}
    \end{fmfgraph*}
} ~ \bigg]\, ,
\label{chap3:deltaZ2Aa}\\
\delta Z_{2b\,A}(\bar{\al}_r) =& \mathcal{K}\, \bigg[~
\parbox{15mm}{
    \begin{fmfgraph*}(15,15)
      %\fmfpen{thick}
      \fmfleft{i}
      \fmfright{o}
      \fmf{photon}{i,v1}
      \fmf{photon}{v2,o}
      \fmf{phantom,right,tension=0.1,tag=1}{v1,v2}
      \fmf{phantom,right,tension=0.1,tag=2}{v2,v1}
      \fmf{phantom,tension=0.1,tag=3}{v1,v2}
      \fmfdot{v1,v2}
      \fmfposition
      \fmfipath{p[]}
      \fmfiset{p1}{vpath1(__v1,__v2)}
      \fmfiset{p2}{vpath2(__v2,__v1)}
      \fmfi{plain}{subpath (0,length(p1)/2) of p1}
      \fmfi{plain}{subpath (length(p1)/2,length(p1)) of p1}
      \fmfi{plain}{subpath (0,length(p2)/2) of p2}
      \fmfi{plain}{subpath (length(p2)/2,length(p2)) of p2}
      \fmfi{photon}{point length(p1)/2 of p1 -- point length(p2)/2 of p2}
      \def\vert#1{%
        \fmfiv{decor.shape=circle,decor.filled=full,decor.size=2thick}{#1}}
      \vert{point length(p1)/2 of p1}
      \vert{point length(p2)/2 of p2}
    \end{fmfgraph*}
} ~\bigg] - 2\,\mathcal{K}\, \bigg[~\mathcal{K}\, \bigg[~
   \parbox{15mm}{
    \begin{fmfgraph*}(15,15)
      %\fmfpen{thick}
      \fmfleft{in}
      \fmfright{e1,e2}
      \fmf{boson}{in,vi}
      \fmf{plain}{e1,v1}
      \fmf{plain,tension=0.7}{v1,vi}
      \fmf{plain,tension=0.7}{vi,v2}
      \fmf{plain}{v2,e2}
      \fmffreeze
      \fmf{boson,right,tension=0.7}{v1,v2}
      \fmfdot{vi,v1,v2}
    \end{fmfgraph*}
}~ \bigg]~\star~
\parbox{15mm}{
    \begin{fmfgraph*}(15,15)
      %\fmfpen{thick}
      \fmfleft{in}
      \fmfright{out}
      \fmf{boson}{in,ve}
      \fmf{plain,right,tension=0.2}{ve,vw}
      \fmf{plain,right,tension=0.2}{vw,ve}
      \fmf{boson}{vw,out}
      \fmfdot{ve,vw}
    \end{fmfgraph*}
} ~\bigg] \, ,
\label{chap3:deltaZ2Ac}
\end{flalign}
\end{subequations}
where, as in Eqs.~(\ref{chap2:RPT:counterterms:G}), it is understood that the Lorentz structure of the diagrams in argument of $\mathcal{K}$ has been projected out so that they are at most logarithmically divergent, \ie, 
the polarization operator is $\Pi(q^2)$ and not $\Pi^{\mu \nu}(q)$ and the fermion self-energy $\Sigma_V(p^2)$ and not $\Sigma(p)$, see, \eg, Eqs.~(\ref{chap3:res2l:QED4:CT:Pi2}) for explicit formulas.
Notice also that the $\star$ operation that was introduced at the level of Eq.~(\ref{chap2:forest:G2loop}) reduces to a simple multiplication in the present case and will be omitted in the following.
Upon computing the total renormalization constant, we see that the last terms in Eqs.~(\ref{chap3:deltaZ2Aa}) and (\ref{chap3:deltaZ2Ac}) cancel eachother because of the Ward identity (\ref{chap3:ward}).
The total two-loop renormalization constant therefore reduces to:
\bea
\delta Z_{2\,A}(\bar{\al}_r) = 2\,\mathcal{K}\, \bigg[~
\parbox{15mm}{
    \begin{fmfgraph*}(15,15)
      %\fmfpen{thick}
      \fmfleft{i}
      \fmfright{o}
      \fmf{photon}{i,v1}
      \fmf{photon}{v2,o}
      \fmf{phantom,right,tension=0.1,tag=1}{v1,v2}
      \fmf{phantom,right,tension=0.1,tag=2}{v2,v1}
      \fmf{phantom,tension=0.1,tag=3}{v1,v2}
      \fmfdot{v1,v2}
      \fmfposition
      \fmfipath{p[]}
      \fmfiset{p1}{vpath1(__v1,__v2)}
      \fmfiset{p2}{vpath2(__v2,__v1)}
      \fmfiset{p3}{vpath3(__v1,__v2)}
      \fmfi{plain}{subpath (0,length(p1)) of p1}
      \fmfi{plain}{subpath (0,length(p2)/4) of p2}
      \fmfi{plain}{subpath (length(p2)/4,3length(p2)/4) of p2}
      \fmfi{plain}{subpath (3length(p2)/4,length(p2)) of p2}
      \fmfi{photon}{point length(p2)/4 of p2 .. point length(p3)/2 of p3 .. point 3length(p2)/4 of p2}
      \def\vert#1{%
        \fmfiv{decor.shape=circle,decor.filled=full,decor.size=2thick}{#1}}
      \vert{point length(p2)/4 of p2}
      \vert{point 3length(p2)/4 of p2}
    \end{fmfgraph*}
} ~\bigg] + \mathcal{K}\, \bigg[~
\parbox{15mm}{
    \begin{fmfgraph*}(15,15)
      %\fmfpen{thick}
      \fmfleft{i}
      \fmfright{o}
      \fmf{photon}{i,v1}
      \fmf{photon}{v2,o}
      \fmf{phantom,right,tension=0.1,tag=1}{v1,v2}
      \fmf{phantom,right,tension=0.1,tag=2}{v2,v1}
      \fmf{phantom,tension=0.1,tag=3}{v1,v2}
      \fmfdot{v1,v2}
      \fmfposition
      \fmfipath{p[]}
      \fmfiset{p1}{vpath1(__v1,__v2)}
      \fmfiset{p2}{vpath2(__v2,__v1)}
      \fmfi{plain}{subpath (0,length(p1)/2) of p1}
      \fmfi{plain}{subpath (length(p1)/2,length(p1)) of p1}
      \fmfi{plain}{subpath (0,length(p2)/2) of p2}
      \fmfi{plain}{subpath (length(p2)/2,length(p2)) of p2}
      \fmfi{photon}{point length(p1)/2 of p1 -- point length(p2)/2 of p2}
      \def\vert#1{%
        \fmfiv{decor.shape=circle,decor.filled=full,decor.size=2thick}{#1}}
      \vert{point length(p1)/2 of p1}
      \vert{point length(p2)/2 of p2}
    \end{fmfgraph*}
} ~\bigg] \, ,
\label{chap3:deltaZ2A}
\eea
where $\delta Z_{2\,A} = 2 \,\delta Z_{2a\,A} + \delta Z_{2b\,A}$.
As known at textbook level \cite{itzykson2012quantum}, this simplification implies that even though the individual diagrams may have subdivergent graphs, 
overall, the subdivergences cancel each other and therefore do not contribute to the final result.~\footnote{As will be discussed in Chap.~\ref{chap5}, this is the essential difference with respect 
to the non-relativistic case where the contribution of the subtracted subdivergent graphs is non-zero due to non-standard Ward identities.}
We now apply all the above formulas to specific cases.

\subsection{Case of QED$_4$}

In the case of QED$_4$ we first set $\veps_e=0$ and take the limit $\veps_\gamma \ra 0$. From Eqs.~(\ref{chap3:res:Pi2a}) and (\ref{chap3:res:Pi2c}), this leads to:
\begin{subequations}
\label{chap3:res2l:QED4:Pi2a+Pi2c}
\begin{flalign}
&\Pi_{2a}(q^2;\bar{\al}) = 4N_F\, \bar{\al}^2 \left( \frac{\overline{\mu}^{\,2}}{-q^2} \right)^{2\veps_\gamma}\, \frac{(d_\gamma-2)^3}{2(d_\gamma-1)(d_\gamma-4)}\,e^{2\gamma_E \veps_\gamma} 
G(4-2\veps_\gamma,1,1)G(4-2\veps_\gamma,1,\veps_\gamma)
\nonum \\
&= 4N_F\, \bar{\al}^2 \left [ \frac{1}{6\veps_\gamma^2} -\frac{1}{3\veps_\gamma}\, \left(L_q - \frac{25}{12} \right) + \frac{1}{6} \left( 2L_q^2 - \frac{25 L_q}{3} - \zeta_2 + \frac{541}{36} \right) + \Ord(\veps_\gamma) \right]\, ,
\label{chap3:res2l:QED4:Pi2a}\\
&\Pi_{2b}(q^2;\bar{\al}) =
-4N_F\, \bar{\al}^2 \left( \frac{\overline{\mu}^{\,2}}{-q^2} \right)^{2\veps_\gamma}\, \frac{d_\gamma-2}{2(d_\gamma-1)(d_\gamma-4)}\, e^{2\gamma_E \veps_\gamma} \, \Bigg
( (d_\gamma^2-7d_\gamma+16)\, G^2(4-2\veps_\gamma,1,1) - \Bigg .
\nonum \\
&\Bigg . -2\frac{d_\gamma^3 -6d_\gamma^2 +20d_\gamma -32}{d_\gamma-4}\,G(4-2\veps_\gamma,1,1)G(4-2\veps_\gamma,1,\veps_\gamma) \Bigg)
\nonum \\
&= -4N_F\, \bar{\al}^2 \left [ \frac{1}{3\veps_\gamma^2} -\frac{2}{3\veps_\gamma}\, \left(L_q - \frac{17}{6} \right) + \frac{1}{3} \left( 2L_q^2 - \frac{34 L_q}{3} - 12\zeta_3 - \zeta_2 + \frac{259}{9} \right) + \Ord(\veps_\gamma) \right]\, ,
\label{chap3:res2l:QED4:Pi2c}
\end{flalign}
\end{subequations}
in agreement with \cite{grozin2007lectures}. The double poles as well as the non-local terms of the type $L_q / \veps$ which appear in Eqs.~(\ref{chap3:res2l:QED4:Pi2a+Pi2c}) are due to the subdivergent graphs.
Upon computing the total two-loop polarization operator, the non-local terms cancel out (together with double poles and zeta function of even argument) and we are left with the simple expression:
\be
\Pi_{2}(q^2;\bar{\al}) = -2N_F\, \bar{\al}^2 \bigg [ \frac{1}{\veps_\gamma} -2 \left( L_q + 4 \zeta_3 - \frac{55}{12} \right) + \Ord(\veps_\gamma)\bigg]\, .
\label{chap3:res2l:QED4:Pi2}
\ee
As we saw above, this simplification results from the Ward identity.  From Eq.~(\ref{chap3:res2l:QED4:Pi2}), the overall counterterm and renormalized two-loop polarization operator read:
%
%\begin{subequations}
%\label{chap2:Z2A+Pi2r}
\bea
\delta Z_{2\,A}(\bar{\al}_r) =  -2N_F\, \frac{\bar{\al}_r^2}{\veps_\gamma}\, , \qquad
%\label{chap2:CT:2Pi2a+Pi2c} \\
\Pi_{2\,r}(q^2;\bar{\al}_r)  = 4N_F\, \bar{\al}_r^2 \, \left( L_q + 4 \zeta_3 - \frac{55}{12} \right)\, .
%\label{chap2:R:2Pi2a+Pi2c}
\label{chap3:res2l:QED4:Z2A+Pi2r}
\eea
%\end{subequations}
%
These results can also be recovered from the more lengthy computation of individual counterterms:
%
%From Eqs.~(\ref{chap2:CTa}) and (\ref{chap2:CTc}), the overall counterterms to $\Pi_{2a}$ and $\Pi_{2b}$, respectively, read:
%
\begin{subequations}
\label{chap3:res2l:QED4:CT:Pi2}
\begin{flalign}
\delta Z_{2a\,A}(\bar{\al}_r) &= \mathcal{K} \mathcal{R}' \bigg[ \Pi_{2a}(q^2;\bar{\al}_r) \bigg] = \mathcal{K} \bigg[ \Pi_{2a}(q^2;\bar{\al}_r) \bigg] 
- \mathcal{K} \bigg[ \mathcal{K} \big[ \Sigma_{1V}(\bar{\al}_r) \big]\,\Pi_{1}(q^2;\bar{\al}_r) \bigg]
\nonum \\
&= - \frac{2 \bar{\al}_r^2}{3 \veps_\gamma^2} + \frac{5 \bar{\al}_r^2}{9 \veps_\gamma}\, ,
\label{chap3:res2l:QED4:CT:Pi2a}\\
\delta Z_{2b\,A}(\bar{\al}_r) &= \mathcal{K} \mathcal{R}' \bigg[\Pi_{2b}(q^2;\bar{\al}_r) \bigg] = \mathcal{K} \bigg[ \Pi_{2b}(q^2;\bar{\al}_r) \bigg]
- 2\,\mathcal{K} \bigg[ \mathcal{K} \big[ \Lambda^{\mu}_{1}(\bar{\al}_r)/\gamma^\mu \big]\,\Pi_{1}(q^2;\bar{\al}_r) \bigg]
\nonum \\
&= + \frac{4 \bar{\al}_r^2}{3 \veps_\gamma^2} - \frac{28 \bar{\al}_r^2}{9 \veps_\gamma}\, ,
\label{chap3:res2l:QED4:CT:Pi2c}
%\\
%\delta Z_{2\,A}(\bar{\al}_r) &= 2 \,\delta Z_{2a\,A}(\bar{\al}_r) + \delta Z_{2b\,A}(\bar{\al}_r) = -2\, \frac{\bar{\al}_r^2}{\veps}\, , 
%\label{chap2:CT:2Pi2a+Pi2c}
\end{flalign}
\end{subequations}
and renormalized diagrams:
\begin{subequations}
\label{chap3:res2l:QED4:R:Pi2}
\begin{flalign}
\Pi_{2a\,r}(q^2;\bar{\al}_r) &= \Pi_{2a}(q^2;\bar{\al}_r) - \mathcal{K} \big[ \Sigma_{1V}(\bar{\al}_r) \big]\,\Pi_{1}(q^2;\bar{\al}_r) - \delta Z_{2a\,A}(\bar{\al}_r)
\nonum \\
&= \frac{2 \bar{\al}_r^2}{3}\,\left(L_q^2 - 5 L_q + \frac{317}{36} \right)\, ,
\label{chap3:res2l:QED4:R:Pi2a}\\
\Pi_{2b\,r}(q^2;\bar{\al}_r) &= \Pi_{2b}(q^2;\bar{\al}_r) - 2\,\mathcal{K} \big[ \Lambda^{\mu}_{1}(\bar{\al}_r)/\gamma^\mu  \big]\,\Pi_{1}(q^2;\bar{\al}_r) - \delta Z_{2b\,A}(\bar{\al}_r)
\nonum \\
&= -\frac{4 \bar{\al}_r^2}{3}\,\left( L_q^2 - 8 L_q -12 \zeta_3 + \frac{203}{9} \right) \, ,
\label{chap3:res2l:QED4:R:Pi2c}
%\\
%\Pi_{2\,r}(q^2;\bar{\al}_r) &= 4 \bar{\al}_r^2 \, \left( L_q + 4 \zeta_3 - \frac{55}{12} \right)\, ,
%\label{chap2:R:2Pi2a+Pi2c}
\end{flalign}
\end{subequations}
where non-local terms as well as zeta functions of even argument are canceled by subtracting the subdivergences. 

From Eq.~(\ref{chap2:Dint}), together with the renormalized one-loop polarization operator of Eq.~(\ref{chap3:QED4:Pi1r+Sigma1r}) and the two-loop one of Eq.~(\ref{chap3:res2l:QED4:Z2A+Pi2r}), 
the expansion of the transverse photon propagator up to two loops reads:
\be
\I q^2\,d_\perp(q^2) =  1 + \frac{4N_F\, \bar{\al}_r}{3}\,\left( L_q - \frac{5}{3} \right) + 4N_F\,\bar{\al}_r^2\,\left( L_q + 4\zeta_3 - \frac{55}{12} + \frac{4N_F}{9}\,\left( L_q - \frac{5}{3} \right)^2  \right) 
+ \Ord(\bar{\al}_r^3) \, .
\label{chap3:res2l:QED4:Tphoton:exp}
\ee
%
%Moreover, from Eqs.~(\ref{chap2:model:QED_d:beta3}) and (\ref{chap2:model:QED_d:anomalousdims-A}), together with Eqs.~(\ref{chap2:ZA+Zpsi}) and (\ref{chap2:Z2A+Pi2r}), 
The two-loop $\beta$-function and anomalous dimension of the gauge field are given by:
\be
\beta(\bar{\al}_r) = - \gamma_{A}(\bar{\al}_r) = \frac{8N_F\, \bar{\al}_r}{3} + 8N_F\, \bar{\al}_r^2 + \Ord(\bar{\al}_r^3)\, .
\label{chap3:res2l:QED4:beta-2loop}
\ee
Of course, all of the above results for QED$_4$ are well known from the literature, see, \eg, \cite{grozin2007lectures}.

\subsection{Case of reduced QED$_{4,3}$}

In the case of reduced QED$_{4,3}$ we first set $\veps_e=1/2$ and take the limit $\veps_\gamma \ra 0$.
From Eqs.~(\ref{chap3:res:Pi2a}) and (\ref{chap3:res:Pi2c}), this leads to:
\begin{subequations}
\label{chap3:res2l:RQED4,3:Pi2a+Pi2c}
\begin{flalign}
&\Pi_{2a}(q^2) =  \frac{N_F\,\al^2}{\sqrt{-q^2}}\,\left [ \frac{1}{12\veps_\gamma} - \frac{L_q}{6} + \frac{1}{9} + \Ord(\veps_\gamma) \right]\, ,
\label{chap3:res2l:RQED4,3:Pi2a}\\
&\Pi_{2b}(q^2) =  \frac{N_F\,\al^2}{\sqrt{-q^2}}\,\left [ -\frac{1}{6\veps_\gamma} + \frac{L_q}{3} + \frac{\pi^2}{4} - \frac{25}{9} + \Ord(\veps_\gamma) \right]\, .
\label{chap3:res2l:RQED4,3:Pi2c}
\end{flalign}
\end{subequations}
While the individual contributions are divergent (with only simple poles arising from divergent subgraphs, see below), the total two-loop polarization operator is finite and reads:
\be
\Pi_{2}(q^2) = \Pi_{2r}(q^2) = -N_F\,\frac{\al_r^2}{\sqrt{-q^2}}\,\frac{92-9\pi^2}{36}, \qquad \delta Z_{2\,A}(\al_r) = 0 \, .
\label{chap3:res2l:RQED4,3:Pi2}
\ee

These results can also be recovered from the computation of individual counterterms in the Feynman gauge ($\xi=1$):
%
%From Eqs.~(\ref{chap2:CTa}) and (\ref{chap2:CTc}), the overall counterterms to $\Pi_{2a}$ and $\Pi_{2b}$, respectively, read:
%
\begin{subequations}
\label{chap3:res2l:RQED4,3:CT:Pi2}
\begin{flalign}
\delta Z_{2a\,A}(\al_r) &= \mathcal{K} \mathcal{R}' \bigg[\Pi_{2a}(q^2) \bigg] = \mathcal{K} \bigg[ \Pi_{2a}(q^2) \bigg]
- \mathcal{K} \bigg[ \mathcal{K} \big[ \Sigma_{1V}(\al_r) \big]\,\Pi_{1}(q^2) \bigg] = 0\, ,
\label{chap3:res2l:RQED4,3:CT:Pi2a}\\
\delta Z_{2b\,A}(\al_r) &= \mathcal{K} \mathcal{R}' \bigg[\Pi_{2b}(q^2;\al_r) \bigg] = \mathcal{K} \bigg[ \Pi_{2b}(q^2;\al_r) \bigg]
- 2\,\mathcal{K} \bigg[ \mathcal{K} \big[ \Lambda^{\mu}_{1}(\al_r)/\gamma^\mu \big]\,\Pi_{1}(q^2;\al_r) \bigg] = 0\, ,
\label{chap3:res2l:RQED4,3:CT:Pi2c}
\end{flalign}
\end{subequations}
which vanish in accordance with the fact that the singularity of each two-loop photon self-energy graph in QED$_{4,3}$ arises solely from its {\it divergent subgraph}. Hence, the renormalized diagrams read (for $\xi=1$):
\begin{subequations}
\label{chap3:res2l:RQED4,3:R:Pi2}
\begin{flalign}
\Pi_{2a\,r}(q^2;\al_r) &= \Pi_{2a}(q^2;\al_r) - \mathcal{K} \big[ \Sigma_{1V}(\al_r) \big]\,\Pi_{1}(q^2;\al_r) - \underbrace{\delta Z_{2a\,A}(\al_r)}_{=0}
\nonum \\
&= \frac{\al^2}{\sqrt{-q^2}}\,\left( -\frac{\tilde{L}_q}{12} + \frac{1}{9} + \frac{1}{12} \right)\, ,
\label{chap3:res2l:RQED4,3:R:Pi2a}\\
\Pi_{2b\,r}(q^2;\al_r) &= \Pi_{2b}(q^2;\al_r) - 2\,\mathcal{K} \big[ \Lambda^{\mu}_{1}(\al_r)/\gamma^\mu \big]\,\Pi_{1}(q^2;\al_r) - \underbrace{\delta Z_{2b\,A}(\al_r)}_{=0}
\nonum \\
&= \frac{\al^2}{\sqrt{-q^2}}\,\left( \frac{\tilde{L}_q}{6} + \frac{\pi^2}{4} - \frac{25}{9} - \frac{1}{6}\right) \, ,
\label{chap3:res2l:RQED4,3:R:Pi2c}
\end{flalign}
\end{subequations}
where again $\tilde{L}_q = L_q + \log(4)$. Upon taking the sum of the individual contributions, Eq.~(\ref{chap3:res2l:RQED4,3:Pi2}) is straightforwardly recovered.

Finally, from the one-loop (\ref{chap3:RQED4,3:Pi1r+Sigma1r}) and two-loop (\ref{chap3:res2l:RQED4,3:Pi2}) results, the total renormalized polarization operator up to two loops can be written as:
\be
\Pi_{r}(q^2) = \Pi_{1r}(q^2)\,\left( 1 + \al_r\,\mathcal{C}^* + \Ord(\al_r^2) \right)\, , \qquad \mathcal{C}^* = \frac{92-9\pi^2}{18\pi}\, ,
\label{chap3:res2l:RQED4,3:Pi-total}
\ee
where $\mathcal{C}^*$ is the so-called {\it interaction correction coefficient}, see Chap.~\ref{chap5} for more. And the transverse photon propagator up to two loops reads:
\be
d_{r\perp}(q^2) =  \frac{\I}{2\sqrt{-q^2}}\, \frac{1}{1 + N_F\frac{\al_r \pi}{4}\,\left( 1 + \al_r \mathcal{C}^* \right)}\, .
\label{chap3:res2l:RQED4,3:Dr:2-loop}
\ee

\subsection{Case of QED$_3$}

In the case of QED$_3$ we first set $\veps_e=0$ and define $\veps_\gamma = 1/2 + \delta_\gamma$ such that $d_\gamma = 3 - 2\delta_\gamma$ where the limit $\delta_\gamma \ra 0$ has to be taken.
From Eqs.~(\ref{chap3:res:Pi2a}) and (\ref{chap3:res:Pi2c}), the individual contributions are again singular:
\begin{subequations}
\label{chap3:res2l:QED3:Pi2a+Pi2c}
\begin{flalign}
&\Pi_{2a}(q^2;\al) = \frac{N_F\, \al^2}{\pi\,(-q^2)}\, \left( \frac{\overline{\mu}^{\,2}}{-q^2} \right)^{2\delta_\gamma}\, \frac{(d_\gamma-2)^3}{2(d_\gamma-1)(d_\gamma-4)}\,e^{2\gamma_E \delta_\gamma} 
G(3-2\delta_\gamma,1,1)G(3-2\delta_\gamma,1,1/2+\delta_\gamma)
\nonum \\
&= \frac{N_F\,\al^2}{2\,(-q^2)}\, \left [ -\frac{1}{2\delta_\gamma} + L_q + \frac{1}{2} + \Ord(\delta_\gamma) \right]\, ,
\label{chap3:res2l:QED3:Pi2a}\\
&\Pi_{2b}(q^2;\bar{\al}) =
-\frac{N_F\,\al^2}{\pi\,(-q^2)}\,\left( \frac{\overline{\mu}^{\,2}}{-q^2} \right)^{2\delta_\gamma}\, \frac{d_\gamma-2}{2(d_\gamma-1)(d_\gamma-4)}\, 
e^{2\gamma_E \delta_\gamma} \, \bigg( (d_\gamma^2-7d_\gamma+16)\, G^2(3-2\delta_\gamma,1,1) - \bigg .
\nonum \\
&\bigg . -2\frac{d_\gamma^3 -6d_\gamma^2 +20d_\gamma -32}{d_\gamma-4}\,G(3-2\delta_\gamma,1,1)G(3-2\delta_\gamma,1,1/2+\delta_\gamma) \bigg)
\nonum \\
&= \frac{N_F\,\al^2}{(-q^2)}\, \left [ \frac{1}{2\delta_\gamma} - L_q +6\zeta_2 - \frac{21}{2} + \Ord(\delta_\gamma) \right]\, , 
\label{chap3:res2l:QED3:Pi2c}
\end{flalign}
\end{subequations}
and, as in the case of reduced QED$_{4,3}$, the singularities are only in terms of simple poles. Contrarily to the case of reduced QED$_{4,3}$ however, 
such singularities correspond to the {\it overall} singularity of each two-loop diagram as photon self-energy subgraphs are all finite in QED$_3$.
Let's pause for a moment and examine the nature (IR or UV) of these singularities. From Eqs.~(\ref{chap3:res2l:QED3:Pi2a+Pi2c}), the singularities look like UV ones. 
Indeed, they originate from the master integral $G(d_\gamma,1,\varepsilon_\gamma) \equiv G(d_\gamma,1,1/2+\delta_\gamma)$ that can be explicited as:
\be
G(3-2\delta_\gamma,1,1/2+\delta_\gamma) =
\frac{\Gamma(1/2-\delta_\gamma) \Gamma(1-2\delta_\gamma)\Gamma(2\delta_\gamma)}{\Gamma(1/2+\delta_\gamma)\Gamma(3/2-3\delta_\gamma)}\, .
\ee
From this equation we see that a pole arises from the last gamma-function and is therefore of UV type, see discussion around Eq.~(\ref{chap2:G:IR+UV}).
As noticed in Ref.~\cite{Kotikov:2013eha}, it turns out, however, that the present example is one in which there is an interchange between
UV and IR types of singularities. 
The above considered UV type of singularity was actually related to our choice of master integral: we took
$G(d_\gamma,1,\varepsilon_\gamma)$. %(or $G(d_e,1,\varepsilon_\gamma)$ in the general case). 
 In our calculations, however, it is the integral
$G(d_\gamma,1,1+\varepsilon_\gamma)$ %(or $G(d_e,1,1+\varepsilon_\gamma) in the general case) 
 that is involved. The two integrals are related:
\be
G(d_\gamma,1,\alpha+1) = - \frac{d_\gamma-2-\alpha}{\alpha}\, G(d_\gamma,1,\alpha) \, .
\ee
And the singularity in $G(d_\gamma,1,1+\varepsilon_\gamma)$ is an IR one as can be seen from:
%for example for $G(1,\varepsilon_\gamma-\varepsilon_e+1)$ we have}
%
\be
G(3-2\delta_\gamma,1,3/2+\delta_\gamma) =
\frac{\Gamma(1/2-\delta_\gamma)
\Gamma(-2\delta_\gamma)\Gamma(1+2\delta_\gamma)}{\Gamma(3/2+\delta_\gamma)\Gamma(1/2-3\delta_\gamma)}\, ,
\ee
where the pole arises from the second gamma-function. Hence, in QED$_3$ the singularities in the individual two-loop photon self-energy diagrams is of IR origin, as it was shown  in the early
Refs.~\cite{Jackiw:1980kv,Guendelman:1982fm,Guendelman:1983dt}, see also Ref.~\cite{King:1985hr} for IR singularities in QED$_3$. This is to be contrasted with the cases of QED$_4$ and RQED$_{4,3}$ where the corresponding singularities are of the UV type.
In the case of QED$_3$, similarly to the case of reduced QED$_{4,3}$, the  singularities cancel out upon summing the individual contributions. 
The total two-loop polarization operator is therefore finite and reads:
\be
\Pi_{2}(q^2) = \Pi_{2r}(q^2) = -\frac{N_F}{8}\,\frac{e_r^4}{(-q^2)}\,\frac{10-\pi^2}{2\pi^2}, \qquad \delta Z_{2\,A}(\al_r) = 0 \, .
\label{chap3:res2l:RQED3:Pi2}
\ee

These results can also be recovered from the computation of individual counterterms:
\begin{subequations}
\label{chap3:res2l:RQED3:CT:Pi2}
\begin{flalign}
\delta Z_{2a\,A}(\al_r) &= \mathcal{K} \mathcal{R}' \bigg[\Pi_{2a}(q^2) \bigg] = \mathcal{K} \bigg[ \Pi_{2a}(q^2) \bigg]
- \mathcal{K} \bigg[ \underbrace{\mathcal{K} \big[ \Sigma_{1V}(\al_r) \big]}_{=0}\,\Pi_{1}(q^2) \bigg]
\nonum \\
&= -\frac{N_F\,\al^2}{4(-q^2)\,\delta_\gamma}\, ,
\label{chap3:res2l:RQED3:CT:Pi2a}\\
\delta Z_{2b\,A}(\al_r) &= \mathcal{K} \mathcal{R}' \bigg[\Pi_{2b}(q^2;\al_r) \bigg] = \mathcal{K} \bigg[ \Pi_{2b}(q^2;\al_r) \bigg]
- 2\,\mathcal{K} \bigg[ \underbrace{\mathcal{K} \big[ \Lambda^{\mu}_{1}(\al_r)/\gamma^\mu \big]}_{=0}\,\Pi_{1}(q^2;\al_r) \bigg]
\nonum \\
&= \frac{N_F\,\al^2}{2(-q^2)\,\delta_\gamma}\, ,
\label{chap3:res2l:RQED3:CT:Pi2c}
\end{flalign}
\end{subequations}
where subgraphs have a vanishing contribution. Hence, the renormalized diagrams read (for $\xi=1$):
\begin{subequations}
\label{chap3:res2l:RQED3:R:Pi2}
\begin{flalign}
\Pi_{2a\,r}(q^2;\al_r) &= \Pi_{2a}(q^2;\al_r) - \underbrace{\mathcal{K} \big[ \Sigma_{1V}(\al_r) \big]}_{=0}\,\Pi_{1}(q^2)  -  \delta Z_{2a\,A}(\al_r) 
\nonum \\
&= \frac{N_F\,\al^2}{2(-q^2)}\,\left( L_q + \frac{1}{2} \right)\, ,
\label{chap3:res2l:RQED3:R:Pi2a}\\
\Pi_{2b\,r}(q^2;\al_r) &= \Pi_{2b}(q^2;\al_r)  - 2\,\underbrace{\mathcal{K} \big[ \Lambda^{\mu}_{1}(\al_r)/\gamma^\mu \big]}_{=0}\,\Pi_{1}(q^2;\al_r)  - \delta Z_{2b\,A}(\al_r) 
\nonum \\
&= \frac{N_F\,\al^2}{(-q^2)}\,\left( -L_q + 6\zeta_2 - \frac{21}{2} \right) \, .
\label{chap3:res2l:RQED3:R:Pi2c}
\end{flalign}
\end{subequations}
Taking the sum of the individual contributions, Eq.~(\ref{chap3:res2l:RQED3:Pi2}) is straightforwardly recovered.

Finally, the total polarization operator up to two loops can be written as:
\be
\Pi_{r}(q^2) = \Pi_{1r}(q^2)\,\left( 1 + \frac{e_r^2}{\sqrt{-q^2}}\,\frac{10-\pi^2}{2\pi^2} + \Ord(e_r^4) \right)\, ,
\label{chap3:res2l:RQED3:Pi-total}
\ee
and the transverse photon propagator up to two loops reads:
\be
d_{r\perp}(q^2) =  \frac{\I}{-q^2}\, \frac{1}{1 + \frac{N_F e_r^2}{8 \sqrt{-q^2}}\,\left( 1 + \frac{e_r^2}{\sqrt{-q^2}}\,\frac{10-\pi^2}{2\pi^2} + \Ord(e_r^4) \right)}\, .
\label{chap3:res2l:RQED3:Dr:2-loop}
\ee

\section{Two-loop polarization operator (uniqueness)}

In the previous section, the two-lop polarization operator was derived for an arbitrary reduced QED$_{d_\gamma,d_e}$, see Eqs.~(\ref{chap3:res:Pi2a}) and (\ref{chap3:res:Pi2c}).
In Eq.~(\ref{chap3:res:Pi2c}), the non-trivial two-loop master integral, $G(d_e,1,1,1,1,1-\veps_e)$, appeared. For arbitrary $\veps_e$, it can be expressed in terms 
of a ${}_3 F_2$ hypergeometric function, see Eqs.~(\ref{chap2:res:I(al):Kazakov}) and (\ref{chap2:res:I(al):Kotikov}) as well as (\ref{chap3:res:I(al):Kotikov}).
In the peculiar case $\veps_e=0$, this function becomes: $G(4-2\veps_\gamma,1,1,1,1,1)$ which is recursively one-loop. Another less trivial limit is the one where $\veps_\gamma =0$ in which
case the function to be computed is: $G(4-2\veps_e,1,1,1,1,1-\veps_e)$.  Defining $\lambda=1-\veps_e$, this function 
can be written as $G(2+2\lambda,1,1,1,1,\lambda)$, that we encountered in Eq.~(\ref{chap2:res:I(lambda)}) and was first computed
by Vasil'ev, Pis'mak and Khonkonen~\cite{Vasiliev:1981dg} using the method of uniqueness in real space (see also discussions in Refs.~\cite{Vasiliev:1992wr,Kivel:1993wq}).
For clarity, we reproduce the result here:
\be
I(\lambda) = G(2\lambda+2,1,1,1,1,\lambda) = 3\,\frac{\Gamma(\lambda)\Gamma(1-\lambda)}{\Gamma(2\lambda)} \,\Big[ \psi'(\lambda) - \psi'(1) \Big] \, ,
\label{chap3:res:I(lambda)}
\ee
where $\psi'(x)$ is the trigamma function. In this section, following Ref.~\cite{Kotikov:2013kcl}, we will present an alternative, hopefully simpler, derivation of this formula using the method of uniqueness
in momentum space, see Sec.~(\ref{chap2:sec:meth:uniqueness}) for a brief presentation of the method. As such, the method will be applicable only to the specific cases
of reduced QED$_{4,d_e}$. Moreover, as we have set $\veps_\gamma=0$ from the start, the results obtained will not be applicable to QED$_4$; so the limit $\lambda \ra 1$ does not make sense. They do however apply
to QED$_{4,3}$ which is finite and for which $\lambda \ra 1/2$. This will allow us to cross check our results for this model and, in particular, recover Eq.~(\ref{chap3:res2l:RQED4,3:Pi2}).

We first start by re-deriving Eq.~(\ref{chap3:res:I(lambda)}). Crucial to the derivation are the previously presented uniqueness relation, Eq.~(\ref{chap2:def:star-triangle}) and
the IBP identity Eq.~(\ref{chap2:def:IBP-5-left}). As a first step, we replace the central line by a 
loop in $I(\lambda)$ in order to make the right triangle unique. The uniqueness relation, Eq.~(\ref{chap2:def:star-triangle}), can then be used. 
In graphical notations this reads:
\be
I(\lambda) \,\, = \quad
\parbox{18mm}{
  \begin{fmfgraph*}(18,16)
    \fmfleft{i}
    \fmfright{o}
    \fmfleft{ve}
    \fmfright{vo}
    \fmftop{vn}
    \fmftop{vs}    
    \fmffreeze
    \fmfforce{(-0.1w,0.5h)}{i}
    \fmfforce{(1.1w,0.5h)}{o}
    \fmfforce{(0w,0.5h)}{ve}
    \fmfforce{(1.0w,0.5h)}{vo}
    \fmfforce{(.5w,0.95h)}{vn}
    \fmfforce{(.5w,0.05h)}{vs}
    \fmffreeze
    \fmf{plain}{i,ve}
    \fmf{plain,left=0.8}{ve,vo}
    \fmf{plain,left=0.8}{vo,ve}
    \fmf{plain,label=$\lambda$,l.d=0.05w}{vs,vn}
    \fmf{plain}{vo,o}
    \fmffreeze
    \fmfdot{ve,vn,vo,vs}
  \end{fmfgraph*}
} \quad  =  \, \frac{1}{\pi^{D/2} G(1,2\lambda)} \quad
\parbox{18mm}{
  \begin{fmfgraph*}(18,16)
    \fmfleft{i}
    \fmfright{o}
    \fmfleft{ve}
    \fmfright{vo}
    \fmftop{vn}
    \fmftop{vs}
    \fmffreeze
    \fmfforce{(-0.1w,0.5h)}{i}
    \fmfforce{(1.1w,0.5h)}{o}
    \fmfforce{(0w,0.5h)}{ve}
    \fmfforce{(1.0w,0.5h)}{vo}
    \fmfforce{(.5w,0.95h)}{vn}
    \fmfforce{(.5w,0.05h)}{vs}
    \fmffreeze
    \fmf{plain}{i,ve}
    \fmf{plain,left=0.8}{ve,vo}
    \fmf{plain,left=0.8}{vo,ve}
    \fmf{plain,left=0.3}{vs,vn}
    \fmf{plain,left=0.3,label=$2\lambda$,l.d=0.05w}{vn,vs}
    \fmf{plain}{vo,o}
    \fmffreeze
    \fmfdot{ve,vn,vo,vs}
  \end{fmfgraph*}
} \quad = \quad
\parbox{16mm}{
  \begin{fmfgraph*}(16,16)
    \fmfleft{i}
    \fmfright{o}
    \fmfleft{ve}
    \fmfright{vo}
    \fmftop{vn}
    \fmftop{vs}
    \fmffreeze
    \fmfforce{(-0.1w,0.5h)}{i}
    \fmfforce{(1.1w,0.5h)}{o}
    \fmfforce{(0w,0.5h)}{ve}
    \fmfforce{(1.0w,0.5h)}{vo}
    \fmfforce{(.5w,0.9h)}{vn}
    \fmfforce{(.5w,0.1h)}{vs}
    \fmffreeze
    \fmf{plain}{i,ve}
    \fmf{plain,left=0.8}{ve,vo}
    \fmf{plain,left=0.8}{vo,ve}
    \fmf{plain}{vs,vn}
    \fmf{phantom,right,label=$\lambda$,l.s=left,l.d=-0.05h}{vo,vn}
    \fmf{phantom,left,label=$\lambda$,l.s=right,l.d=-0.05h}{vo,vs}
    \fmf{plain}{vo,o}
    \fmffreeze
    \fmfdot{ve,vn,vo,vs}
  \end{fmfgraph*}
} \quad \frac{1}{p^{2(1-\lambda)}} \, ,
\label{chap2:first+second-steps}
\ee
where the last equality is actually a peculiar case of the general result already presented in Eq.~(\ref{chap2:GI2}).
Then, using integration by parts, Eq.~(\ref{chap2:def:IBP-5-left}), the last diagram is reduced to sequences of chains and 
simple loops which can immediately be integrated:
\begin{subequations}
\label{chap3:third-step}
\begin{flalign}
&\quad
\nonumber \\
( -2 \delta ) \quad
\parbox{16mm}{
  \begin{fmfgraph*}(16,16)
    \fmfleft{i}
    \fmfright{o}
    \fmfleft{ve}
    \fmfright{vo}
    \fmftop{vn}
    \fmftop{vs}
    \fmffreeze
    \fmfforce{(-0.1w,0.5h)}{i}
    \fmfforce{(1.1w,0.5h)}{o}
    \fmfforce{(0w,0.5h)}{ve}
    \fmfforce{(1.0w,0.5h)}{vo}
    \fmfforce{(.5w,0.9h)}{vn}
    \fmfforce{(.5w,0.1h)}{vs}
    \fmffreeze
    \fmf{plain}{i,ve}
    \fmf{plain,left=0.8}{ve,vo}
    \fmf{plain,left=0.8}{vo,ve}
    \fmf{plain}{vs,vn}
    \fmf{phantom,right=0.1,label=$\lambda+\delta$,l.s=right}{vo,vn}
    \fmf{phantom,left=0.1,label=$\lambda+\delta$,l.s=left}{vo,vs}
    \fmf{plain}{vo,o}
    \fmffreeze
    \fmfdot{ve,vn,vo,vs}
  \end{fmfgraph*}
} \, \, &= \,\, 2 ( \lambda + \delta ) \quad 
\left[ \quad
\parbox{32mm}{
  \begin{fmfgraph*}(32,16)
    \fmfleft{i}
    \fmfright{o}
    \fmfleft{ve}
    \fmfright{vo}
    \fmftop{v}
    \fmffreeze
    \fmfforce{(-0.1w,0.5h)}{i}
    \fmfforce{(1.1w,0.5h)}{o}
    \fmfforce{(0w,0.5h)}{ve}
    \fmfforce{(1.0w,0.5h)}{vo}
    \fmfforce{(.5w,0.5h)}{v}
    \fmffreeze
    \fmf{plain}{i,ve}
    \fmf{plain,left=0.8}{ve,v}
    \fmf{plain,left=0.8}{v,ve}
    \fmf{plain,left=0.8,label=$\lambda+\delta$,l.s=left}{v,vo}
    \fmf{plain,left=0.8,label=$\lambda+\delta+1$,l.s=left}{vo,v}
    \fmf{plain}{vo,o}
    \fmffreeze
    \fmfdot{ve,v,vo}
  \end{fmfgraph*}
} \qquad - \qquad
\parbox{16mm}{
  \begin{fmfgraph*}(16,16)
    \fmfleft{i}
    \fmfright{o}
    \fmfleft{ve}
    \fmfright{vo}
    \fmftop{v}
    \fmffreeze
    \fmfforce{(-0.1w,0.5h)}{i}
    \fmfforce{(1.1w,0.5h)}{o}
    \fmfforce{(0w,0.5h)}{ve}
    \fmfforce{(1.0w,0.5h)}{vo}
    \fmfforce{(.5w,0.9h)}{v}
    \fmffreeze
    \fmf{plain}{i,ve}
    \fmf{plain,left=0.8}{ve,vo}
    \fmf{plain,left=0.8,label=$\lambda+\delta+1$,l.s=left}{vo,ve}
    \fmf{plain,left=0.5}{v,ve}
    \fmf{phantom,right=0.1,label=$\lambda+\delta$,l.s=right}{vo,v}
    \fmf{plain}{vo,o}
    \fmffreeze
    \fmfdot{ve,v,vo}
  \end{fmfgraph*}
} \qquad
\right] \,\, ,
\label{third-step-a}\\
&\quad
\nonumber \\
&\quad
\nonumber \\
& =  \frac{\pi^D 2 ( \lambda + \delta )}{p^{2(1+2\delta)}}\, G(1,1)\, \Big[ G(\lambda+\delta+1,\lambda+\delta) - G(\lambda+\delta+1, 1+\delta) \Big]_1 \, ,
\label{chap3:third-step-b}
\end{flalign}
\end{subequations}
where the parameter $\delta$ has been introduced as a regulator and the bracketed terms in Eq.~(\ref{chap3:third-step-b}) can be explicited as:
%
%\begin{subequations}
%\label{chap3:G(1,1) and first-bracket}
\begin{flalign}
%G(1,1) & = \frac{\Gamma^2(\lambda)\Gamma(1-\lambda)}{\Gamma(2\lambda)} \, ,
%\label{G(1,1)} \\ 
( \lambda + \delta ) \, \Big[ \bullet \Big]_1    =   \frac{\Gamma(-\delta)}{\Gamma(\lambda+\delta)}\, \frac{\Gamma(\lambda-\delta)\Gamma(1+2\delta)}{\Gamma(1+\delta)\Gamma(\lambda-2\delta)}\,
\Bigg[\, \frac{\Gamma(1-\delta) \Gamma(1+\delta) \Gamma(\lambda+2\delta) \Gamma(\lambda-2\delta)}{\Gamma(1-2\delta) \Gamma(1+2\delta)\Gamma(\lambda+\delta) \Gamma(\lambda-\delta)} - 1 \,\Bigg]_2 \, .
\label{chap3:first-bracket}
\end{flalign}
%\end{subequations}
%
At this point it is convenient to use the product expansion of the Gamma function Eq.~(\ref{chap2:gamma-expansion2}) which was expressed in terms of
polygamma functions Eq.~(\ref{chap2:def:polygamma}) and from which the following relation is obtained:
\be
\Gamma(x+\veps) \Gamma(x-\veps) = \Gamma^2(x)\,\exp \Big[\,\,2\, \sum_{m=1}^{\infty} \psi^{(2m-1)}(x) \frac{\veps^{2m}}{(2m)!}\,\, \Big]\,.
\label{chap3:Gamma-expansion-2}
\ee
Making use of Eq.~(\ref{chap3:Gamma-expansion-2}) in the bracket of Eq.~(\ref{chap3:first-bracket}) yields:
\be
\Big[ \bullet \Big]_2 =  \exp \Big[ \,\, 2\,\sum_{m=1}^{\infty} \left( 2^{2m}-1 \right) \left[\psi^{(2m-1)}(\lambda) - \psi^{(2m-1)}(1)\right] \frac{\delta^{2m}}{(2m)!} \,\,\Big]
= 3 \delta^2 \Big[ \psi'(\lambda) - \psi'(1) \Big]_3 + \Ord(\delta^4)\,.
\label{chap3:second-bracket}
\ee
Substituting back Eq.~(\ref{chap3:second-bracket}) in (\ref{chap3:first-bracket}) and performing the remaining $\delta$-expansion yields:
\be
( \lambda + \delta ) \, \Big[ \bullet \Big]_1 = \frac{-3 \delta}{\Gamma(\lambda)}\,\Big[ \bullet \Big]_3 \,\, 
\underset{{\rm Eq.}~(\ref{chap3:third-step})}{\Rightarrow} \quad 
\parbox{16mm}{
  \begin{fmfgraph*}(16,16)
    \fmfleft{i}
    \fmfright{o}
    \fmfleft{ve}
    \fmfright{vo}
    \fmftop{vn}
    \fmftop{vs}
    \fmffreeze
    \fmfforce{(-0.1w,0.5h)}{i}
    \fmfforce{(1.1w,0.5h)}{o}
    \fmfforce{(0w,0.5h)}{ve}
    \fmfforce{(1.0w,0.5h)}{vo}
    \fmfforce{(.5w,0.9h)}{vn}
    \fmfforce{(.5w,0.1h)}{vs}
    \fmffreeze
    \fmf{plain}{i,ve}
    \fmf{plain,left=0.8}{ve,vo}
    \fmf{plain,left=0.8}{vo,ve}
    \fmf{plain}{vs,vn}
    \fmf{phantom,right=0.1,label=$\lambda$,l.s=right}{vo,vn}
    \fmf{phantom,left=0.1,label=$\lambda$,l.s=left}{vo,vs}
    \fmf{plain}{vo,o}
    \fmffreeze
    \fmfdot{ve,vn,vo,vs}
  \end{fmfgraph*}
} \quad = \, \frac{\pi^D}{p^2}\,3\,\frac{\Gamma(\lambda)\Gamma(1-\lambda)}{\Gamma(2\lambda)} \,\Big[ \psi'(\lambda) - \psi'(1) \Big] \,,
\label{chap3:third-bracket}
\ee
where, in the last step, Eq.~(\ref{chap3:third-step}) has been used and $\delta$ sent to zero. 
Extracting the coefficient function from the final result of Eq.~(\ref{chap3:third-bracket}), we obtain the advertised result 
Eq.~(\ref{chap3:res:I(lambda)}) \cite{Vasiliev:1981dg,Vasiliev:1992wr,Kivel:1993wq}.
In the even-dimensional case ($\lambda \ra 1$ or $D \ra 4$) the well-known result: $I(1) = 6 \, \zeta(3)$, is recovered.
On the other hand, in the odd-dimensional case ($\lambda \ra 1/2$ or $D \ra 3$), the result reads: $I(1/2) = 6 \pi \,\zeta(2)$.

We now focus on the computation of the two-loop polarization operator in reduced QED$_{4,d_e}$. From Eqs.~(\ref{chap3:res:Pi2a}) and (\ref{chap3:res:Pi2c}) in the limit $\veps_\gamma \ra 0$, with $\veps_e = 1 -\lambda$ and 
using Eq.~(\ref{chap3:res:I(lambda)}) we obtain the following simpler and more explicit formulas:
\begin{subequations}
\label{chap3:2loopdiagrams-ab}
\begin{flalign}
&2\Pi_{2a}(q^2) = 4N_F\,
\frac{e^4 \, \Gamma(\lambda) \Gamma^2(1+\varepsilon_\gamma)}{
%\Gamma(1-\varepsilon_e) 
(4\pi)^{3+\lambda- 2\varepsilon_\gamma}
\left(q^2\right)^{1-\lambda+2\varepsilon_\gamma}}\, 
\frac{16 \Gamma(1+\lambda)
 \Gamma(1-\lambda)}{ \Gamma(3+2\lambda)}\, \nonumber \\
& \times \biggl\{ \, \lambda^2 \left(\frac{1}{\varepsilon_\gamma} + 
\overline{\psi} + \frac{2}{1+2\lambda}\right)
+ \frac{3\lambda}{2} - 2 + \frac{2}{1+\lambda} \, + \, \Ord(\veps_\gamma) \, \Biggr\} \, ,
\label{chap3:2loop-a} \\
&\Pi_{2b}(q^2) =  4N_F\,
\frac{e^4 \, \Gamma(\lambda) \Gamma^2(1+\varepsilon_\gamma)}{
%\Gamma(1-\varepsilon_e) 
(4\pi)^{3+\lambda- 2\varepsilon_\gamma}
\left(q^2\right)^{1-\lambda+2\varepsilon_\gamma}}\, 
\frac{16 \Gamma(1+\lambda)
 \Gamma(1-\lambda)}{ \Gamma(3+2\lambda)} \nonumber \\
& \times \biggl\{\, -\lambda^2 \left(\frac{1}{\varepsilon_\gamma} + 
\overline{\psi} + \frac{2}{1+2\lambda}\right)
+ \frac{\lambda}{2} - \frac{1}{2} - \frac{3}{2\lambda} - \frac{1}{1+\lambda}
 + \frac{3}{2} \lambda (1+\lambda)
\Bigr[ \psi'(\lambda) - \psi'(1) \Bigl] \, +  \,\Ord(\veps_\gamma) \, \Biggr\} \,,
\label{chap3:2loop-b}
\end{flalign}
\end{subequations}
where $\lambda=1-\veps_e$ and $\overline{\psi} = 3\psi(2\lambda) - 2 \psi(\lambda) + 2\psi(1-\lambda) -3 \psi(1)$.
The 1-loop and total 2-loop contributions therefore read:
\begin{subequations}
\label{chap3:1+2loopdiagrams}
\bea
\Pi_{1}(q^2) =&& - 4N_F\,\frac{e^2 \, \Gamma(\lambda)}{(4\pi)^{1+\lambda} \, \left(q^2\right)^{1-\lambda}}
\frac{\Gamma(1+\lambda) \Gamma(1-\lambda)}{ (1+2\lambda) \Gamma(2\lambda)} \, , 
\label{chap3:1loop}\\
\Pi_{2}(q^2) = && 4N_F\,
\frac{e^4 \, \Gamma(\lambda)}{(4\pi)^{3+\lambda} \, \left(q^2\right)^{1-\lambda}}
\frac{16 \Gamma(1+\lambda) \Gamma(1-\lambda)}{ \Gamma(3+2\lambda)} \, C_1(\lambda) \, , 
\label{chap3:2loop-a+b} \\
C_1(\lambda) = && 2\lambda - \frac{5}{2} - \frac{3}{2\lambda} + \frac{1}{1+\lambda}
+ \frac{3}{2} \lambda (1+\lambda) \Bigl[ \psi'(\lambda) - \psi'(1) \Bigr] \, .
\label{chap3:C1}
\eea
\end{subequations}
From Eqs.~(\ref{chap3:1+2loopdiagrams}) we see that $\Pi_{1}(q^2)$ and $\Pi_{2}(q^2)$ are finite as long as $\lambda \neq 1$.
We can then replace $e^2$ by $4\pi \alpha$ in  (\ref{chap3:1+2loopdiagrams}) which yields:
\be
\label{chap3:2loopdiagramN}
\Pi_{2}(q^2) =  - \frac{\alpha}{\pi\lambda(1+\lambda)} \, C_1(\lambda) \, 
\Pi_{1}(q^2) \, .
\ee
As we noted in the beginning of this section, these results cannot be used for QED$_{4}$
which can be reached from a general RQED$_{d_\gamma,d_e}$,  by first fixing $\veps_e=0$ and then
taking the limit $\varepsilon_\gamma \to 0$. The results (\ref{chap3:1+2loopdiagrams}) 
are singular in the limit $\lambda \to 1$ but this limit corresponds to $\varepsilon_\gamma = 0$ and $\varepsilon_e \to 0$, which does
not lead to QED$_4$.

The total, up to 2 loops, gauge-field self-energy  in RQED$_{4,d_e}$ ($\varepsilon_\gamma = 0$ and arbitrary $\varepsilon_e$)
may then be written as: 
\begin{flalign}
\label{chap3:1+2loopdiagram}
&\Pi(q^2) = \Pi_{1}(q^2) \Bigl(1+\alpha C(\lambda) + \Ord \left( {\bf \alpha}^2\right) 
\Bigr), \nonumber \\
&C(\lambda) = - \frac{1}{\pi\lambda(1+\lambda)} \, C_1(\lambda)
= - \frac{1}{2\pi} \left(3 \Bigl[ \psi'(\lambda+2) - \psi'(1) \Bigr]
+ \frac{4}{1+\lambda} + \frac{1}{(1+\lambda)^2} \right)
 \, .
\end{flalign}
For $\lambda=1/2$, {\it i.e.}, in the case of RQED$_{4,3}$ ($\varepsilon_\gamma = 0$ and 
$\varepsilon_e=1/2$) %which corresponds to an ultra-relativistic model of graphene~\cite{GrapheneRG} (a 2-brane), 
we reproduce the basic result of Eq.~(\ref{chap3:res2l:RQED4,3:Pi2}). Indeed: 
\be
\label{chap3:C+C1}
C_1(1/2)= \frac{9\pi^2-92}{24}, \qquad \mathcal{C}^* = C(1/2)= \frac{92-9\pi^2}{18\pi} \,.
\ee
At this point we note that a coefficient similar to $C_1(1/2)$ has been obtained earlier in Ref.~\cite{Gusynin:2000zb}
in the framework of the $1/N_F$-expansion of QED$_{3}$ with $N_F$ fermions. The similarity of the coefficients
comes from the corresponding similarity of the form of the effective photon propagator
in RQED$_{4,3}$ and in the $1/N_F$-expansion of QED$_{3}$ in the limit of large $N_F$, Eq.~(\ref{chap3:RQED3:Dr}). We will come back on this fact in Chap.~\ref{chap4}.

\section{Two-loop fermion self-energy}

\begin{figure}
    a)
    \begin{fmfgraph*}(45,40)
      %\fmfpen{thick}
      \fmfleft{i1}
      \fmfright{o1}
      \fmf{phantom}{i1,i2}
      \fmf{plain,tension=1.4}{i2,i3}
      \fmf{plain}{i3,i4}
      \fmf{fermion,tension=0.4,label=$p+k$}{i4,o4}
      \fmf{plain}{o4,o3}
      \fmf{plain,tension=1.4}{o3,o2}
      \fmf{phantom}{o2,o1}
      \fmffreeze
      %\fmf{plain,left,tension=0.1,tag=1}{i1,o1}
      \fmf{phantom,left,tension=0.1,tag=2}{i2,o2}
      \fmf{phantom,left,tension=0.1,tag=3}{i3,o3}
      \fmf{phantom,left,tension=0.1,tag=4}{i4,o4}
      \fmfdot{i3,o3}
      \fmfposition
      \fmfipath{p[]}
      %\fmfiset{p1}{vpath1(__i1,__o1)}
      \fmfiset{p2}{vpath2(__i2,__o2)}
      \fmfiset{p3}{vpath3(__i3,__o3)}
      \fmfiset{p4}{vpath4(__i4,__o4)}
      \fmfi{photon}{subpath (0,3length(p3)/8) of p3}
      \fmfi{photon,label=$k$}{subpath (5length(p3)/8,length(p3)) of p3}
      \fmfi{fermion}{point 3length(p3)/8 of p3 .. point length(p2)/2 of p2 .. point 5length(p3)/8 of p3}
      \fmfi{fermion}{point 5length(p3)/8 of p3 .. point length(p4)/2 of p4 .. point 3length(p3)/8 of p3}
      \def\vert#1{%
        \fmfiv{decor.shape=circle,decor.filled=full,decor.size=2thick}{#1}}
      \vert{point 3length(p3)/8 of p3}
      \vert{point 5length(p3)/8 of p3}
   \end{fmfgraph*}
 \quad   b)
    \begin{fmfgraph*}(45,40)
      %\fmfpen{thick}
      \fmfleft{i1}
      \fmfright{o1}
      \fmf{plain,tension=1.7}{i1,i2}
      \fmf{fermion,label=$p+k$}{i2,i3}
      \fmf{plain}{i3,o3}
      \fmf{fermion,label=$p+k$}{o3,o2}
      \fmf{plain,tension=1.7}{o2,o1}
      \fmffreeze
      \fmf{photon,left,tension=0.1,label=$k$,label.side=right}{i2,o2}
      \fmf{photon,left,tension=0.1}{i3,o3}
      \fmfdot{i2,i3,o2,o3}
   \end{fmfgraph*}
 \quad   c)
    \begin{fmfgraph*}(45,40)
      %\fmfpen{thick}
      \fmfleft{i1}
      \fmfright{o1}
      \fmf{plain,tension=1.7}{i1,i2}
      \fmf{fermion,label=$p+k_1$}{i2,i3}
      \fmf{fermion,label=$k_{12}$,label.side=left}{i3,o3}
      \fmf{fermion,label=$-k_2$}{o3,o2}
      \fmf{plain,tension=1.7}{o2,o1}
      \fmffreeze
      \fmf{photon,left,tension=0.1,label=$k_1$,label.side=left}{i2,o3}
      \fmf{photon,right,tension=0.1,label=$p+k_2$,label.side=right}{i3,o2}
      \fmfdot{i2,i3,o2,o3}
   \end{fmfgraph*}
  \caption{\label{chap3:fig:two-loop:Sigma}
  Two-loop fermion self-energy diagrams ($k_{12} = k_1-k_2$).}
\end{figure}
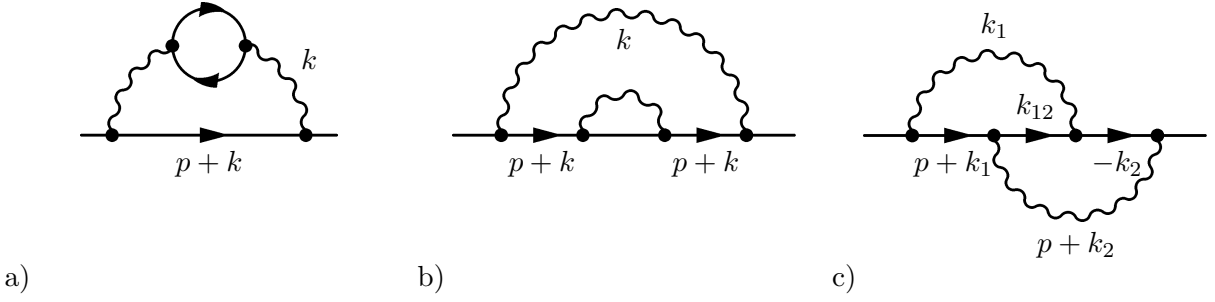

We may proceed in a similar way for the two-loop fermion self-energy:
\be
\Sigma_2(p) = \Sigma_{2a}(p) + \Sigma_{2b}(p) + \Sigma_{2c}(p) \, ,
\ee
where the diagrams are represented on Fig.~\ref{chap3:fig:two-loop:Sigma}. The latter are defined as:
\begin{subequations}
\label{chap3:def:Sigma2}
\begin{flalign}
-\I \Sigma_{2a}(p) &= \mu^{2\veps_\gamma}\int [\D^{d_e} k] (-\I e \gamma^\al)\,S_0(p+k)\,(-\I e \gamma^\beta)\,\tilde{D}_{0\,\al \mu}(k)\,\I \Pi_1^{\mu \nu}(k)\,\tilde{D}_{0\,\nu \beta}(k)\, ,
\label{chap3:def:Sigma2a} \\
-\I \Sigma_{2b}(p) &= \mu^{2\veps_\gamma}\int [\D^{d_e} k] (-\I e \gamma^\mu)\,S_0(p+k)\,(-\I e \Sigma_1(p+k))\,S_0(p+k)\,(-\I e \gamma^\nu)\,\tilde{D}_{0\,\mu \nu}(k)\, ,
\label{chap3:def:Sigma2b} \\
-\I \Sigma_{2c}(p) &= \mu^{2\veps_\gamma}\int [\D^{d_e} k_1] [\D^{d_e} k_2] (-\I e \gamma^\mu)\,S_0(k_2)\,\tilde{D}_{0\,\beta \mu}(k_2+p)\,(-\I e \gamma^\al)\,S_0(k_1-k_2)
\nonum \\
&\qquad \times \tilde{D}_{0\,\al \nu}(k_1)\,(-\I e \gamma^\beta)\,\,S_0(k_1+p)\,(-\I e \gamma^\nu)\, ,
\label{chap3:def:Sigma2c}
\end{flalign}
\end{subequations}
where $\Pi_1^{\mu \nu}(k) = (g^{\mu \nu} - k^{\mu} k^{\nu}/k^2)\,\Pi_1(k^2)$ the one-loop photon self-energy insertion with $\Pi_1(k^2)$ given by Eq.~(\ref{chap3:res:Pi1})
and $\Sigma_1(k) = \Sk \Sigma_{V1}(k^2)$ is a one-loop fermion self-energy insertion with $\Sigma_{V1}(k^2)$ given by Eq.~(\ref{chap3:res:Sigma1}).

The first diagram, Fig.~\ref{chap3:fig:two-loop:Sigma}a, is the so-called bubble diagram. It is gauge invariant thanks to the Ward identity $k_\mu\Pi^{\mu \nu}=0$ 
which implies that all terms proportional to the gauge fixing term vanish. The diagram is recursively one-loop and can be straightforwardly evaluated from Eq.~(\ref{chap3:def:Sigma2a}) after some algebra. 
The final result reads:
\begin{flalign}
\Sigma_{V2a}(p^2) = 4N_F\,\bar{\al}^2 \, \left( \frac{\overline{\mu}^{\,2}}{-p^2} \right)^{2\veps_\gamma}\,\Gamma^2(1-\veps_e)\,\frac{(d_e-2)^2}{2(2d_\gamma-d_e-6)}\,
e^{2\gamma_E \veps_\gamma}\,G(d_e,1,1)G(d_e,1,\varepsilon_\gamma-\varepsilon_e),
\label{chap3:res:Sigma2a}
\end{flalign}
where the factor of $N_F$ is due to the fermion loop.  The second diagram, Fig.~\ref{chap3:fig:two-loop:Sigma}b, is the so-called rainbow diagram. It is also recursively one-loop but it is not 
gauge invariant. Proceeding along the same lines as for the bubble diagram, Eq.~(\ref{chap3:def:Sigma2b}) yields:
\bea
\Sigma_{V2b}(p^2) &=& \bar{\al}^2 \, \left( \frac{\overline{\mu}^{\,2}}{-p^2} \right)^{2\veps_\gamma}\,\Gamma^2(1-\veps_e)\,\frac{(d_e-2)(d_\gamma-3)(d_\gamma+d_e-4)}{2(d_\gamma-4)}\,
\left( \xi - \frac{d_\gamma - d_e}{d_\gamma+d_e-4} \right)^2\,
\nonum \\
&\times&e^{2\gamma_E \veps_\gamma}\,G(d_e,1,1-\varepsilon_e)G(d_e,1-\varepsilon_e,\varepsilon_\gamma)\, ,
\label{chap3:res:Sigma2b}
\eea
where $\xi$ is the gauge fixing parameter. Finally, the third diagram, Fig.~\ref{chap3:fig:two-loop:Sigma}c, which is the so-called crossed photon diagram
is a truly two-loop diagram. Proceeding along the same lines as for the two previous diagrams, Eq.~(\ref{chap3:def:Sigma2c}) yields:
\begin{flalign}
&\Sigma_{V2c}(p^2) = - \bar{\al}^2 \, \left( \frac{\overline{\mu}^{\,2}}{-p^2} \right)^{2\veps_\gamma}\,\Gamma^2(1-\veps_e)\,\frac{d_e-2}{2}\,e^{2\gamma_E \veps_\gamma}\,
\label{chap3:res:Sigma2c} \\
&\times \Bigg \{ \left( \frac{(d_e-6)(3d_\gamma-d_e-4)}{2(d_\gamma+d_e-4)} -\xi\,\frac{(d_e-2)(d_\gamma-d_e)}{d_\gamma+d_e-4} + \xi^2\,\frac{d_e-2}{2} \right)\,G^2(d_e,1,1-\varepsilon_e) + \Bigg .
\nonum \\
& \Bigg . + \left( 5d_e -d_\gamma - 1 + \frac{16}{d_\gamma-4} - \frac{4(d_e-3)(d_e-2)}{d_\gamma+d_e-4} +2\xi\, \frac{(d_\gamma-3)(d_\gamma-d_e)}{d_\gamma+d_e-4} -\xi^2\,(d_\gamma-3) \right)\,\times \Bigg .
\nonum \\
&\qquad \times G(d_e,1,1-\varepsilon_e)G(d_e,1-\varepsilon_e,\varepsilon_\gamma) + 
\nonum \\
& \Bigg . +4G(d_e,1-\varepsilon_e,1,-1-\varepsilon_e,1,1) - 4 G(d_e,1-\varepsilon_e,1,-\varepsilon_e,1,1)+(8-d_e)G(d_e,-\varepsilon_e,1,-\varepsilon_e,1,1) \Bigg \}\, ,
\nonum
\end{flalign}
where, besides the primitively one-loop master integrals, $G^2(d_e,1,1-\varepsilon_e)$ and $G(d_e,1,1-\varepsilon_e)G(d_e,1-\varepsilon_e,\varepsilon_\gamma)$,
complicated two-loop propagator diagrams of the type of Eq.~(\ref{chap2:def:I(al,beta)}) enter the expression of the self-energy. It is interesting to notice that
none of the complicated terms depend on the gauge parameter, $\xi$. That is, working in the Feynman gauge, $\xi=1$, (or any other gauge for that matter) simplifies the calculations but does
not reduce the complexity of the diagrams entering the expression of the fermion self-energy.

For arbitrary $\veps_e$ and $\veps_\gamma$, the complicated 2-loop massless propagator-type diagrams appearing in Eq.~(\ref{chap3:res:Sigma2c}) can be evaluated with the help of
the general expression for $G(D,\al,1,\beta,1,1)$, see Eq.~(\ref{chap2:res:Gab}) and Ref.~\cite{Kotikov:2013eha} for other representations. As noticed in Ref.~\cite{Kotikov:2013eha}, it turns out that in the special case 
where $\veps_\gamma \ra 0$, that is in the case of reduced QED$_{4,d_e}$ which is of main interest to us for applications, the complicated diagrams can be computed exactly using IBP 
for arbitrary $\veps_e$. The reason is that the (UV) divergent diagrams, $G(d_e,1-\varepsilon_e,1,-1-\varepsilon_e,1,1)$, $G(d_e,1-\varepsilon_e,1,-\varepsilon_e,1,1)$
and $G(d_e,-\varepsilon_e,1,-\varepsilon_e,1,1)$ can all be expressed, with the help of well chosen IBP identities such as Eq.~(\ref{chap2:sec:IBP:2loop-homogc-graph}), in terms of the (UV) convergent one, 
$G(d_e,1-\varepsilon_e,1,1-\varepsilon_e,1,1)$. It turns out that the coefficient of this complicated convergent diagram is $\veps_\gamma$ which then does not contribute to the final answer
in the limit $\veps_\gamma \ra 0$. So the self-energy is obtained exactly for arbitrary $\veps_e$ to $\Ord(\veps_\gamma^0)$ which is enough for renormalization at two loop level. We shall not
proceed here on repeating the proofs which can be found in Ref.~\cite{Kotikov:2013eha}; instead, we simply quote from Ref.~\cite{Kotikov:2013eha} the final expression for $\Sigma_{V2c}$ after IBP relations are used. 
This expression reads:
\begin{flalign}
&\Sigma_{V2c}(p^2) = - \bar{\al}^2 \, \left( \frac{\overline{\mu}^{\,2}}{-p^2} \right)^{2\veps_\gamma}\,\Gamma^2(1-\veps_e)\,\frac{d_e-2}{2}\,e^{2\gamma_E \veps_\gamma}\,
\nonum \\
&\times \Bigg \{ \left[ d_e-4 + \frac{(d_e-2)(d_\gamma-3d_e+4)}{2(d_\gamma+d_e-4)}  -\frac{(d_\gamma+d_e-6)( d_\gamma(d_e-4)+8)}{(2d_\gamma+d_e-10)(2d_\gamma+d_e-8)}
\right . \Bigg .
\nonum \\
&\qquad -\frac{4(d_\gamma-d_e)}{d_\gamma+d_e-4} -  \frac{d_\gamma-d_e}{2d_\gamma+d_e-8}\, \left( d_e-8 - 4 \, \frac{d_\gamma+d_e-6}{d_\gamma+d_e-4} \right) 
\nonum \\
&\qquad \left . -\xi\,\frac{(d_e-2)(d_\gamma-d_e)}{d_\gamma+d_e-4} +\xi^2\,\frac{d_e-2}{2} \right]\,G^2(d_e,1,1-\varepsilon_e)
%\Bigg .
\nonum \\
&+ \left[ 2d_e - d_\gamma -1  + \frac{4(d_e-2)(d_\gamma-1)}{d_\gamma+d_e-4}+\frac{8(d_\gamma-1)}{d_\gamma-4} + \frac{2(d_e-8)(d_\gamma-d_e)}{d_\gamma+d_e-6}-
\frac{4(d_\gamma-2)(d_\gamma-d_e)}{(d_\gamma-4)(d_\gamma+d_e-4)} \right . 
\nonum \\
&\qquad \left . + 2\xi\,\frac{(d_\gamma-3)(d_\gamma-d_e)}{d_\gamma+d_e-4} -\xi^2 (d_\gamma-3) \right ]\,G(d_e,1,1-\varepsilon_e)G(1-\varepsilon_e,\varepsilon_\gamma)  
\nonum \\
&\Bigg . - \frac{(d_\gamma-4)(d_\gamma(d_e-4)+8)}{(2d_\gamma+d_e-8)(2d_\gamma+d_e-10)}\,G(d_e,1-\varepsilon_e,1,1-\varepsilon_e,1,1) \Bigg \}\, ,
\label{chap3:res:Sigma2c-2}
\end{flalign}
and it is clearly seen that the last term, which is the only one depending on a complicated diagram, comes with a factor $d_\gamma -4$.

Summing up all individual self-energy contributions, Eqs.~(\ref{chap3:res:Sigma2a}), (\ref{chap3:res:Sigma2b}) and Eq.~(\ref{chap3:res:Sigma2c-2}), yields the total two-loop fermion self-energy in an arbitrary gauge:
\begin{flalign}
&\Sigma_{V2}(p^2) = \bar{\al}^2 \, \left( \frac{\overline{\mu}^{\,2}}{-p^2} \right)^{2\veps_\gamma}\,\Gamma^2(1-\veps_e)\,\frac{d_e-2}{2}\,e^{2\gamma_E \veps_\gamma}\,
\Bigg \{ 4N_F\,\frac{d_e-2}{2d_\gamma-d_e-6}\,G(d_e,1,1)G(d_e,1,\varepsilon_\gamma-\varepsilon_e) \Bigg .
\nonum \\
&\quad + \left[ 4-d_e - \frac{(d_e-2)(d_\gamma-3d_e+4)}{2(d_\gamma+d_e-4)} +\frac{(d_\gamma+d_e-6)( d_\gamma(d_e-4)+8)}{(2d_\gamma+d_e-10)(2d_\gamma+d_e-8)}
\right . \Bigg .
\nonum \\
&\qquad + \frac{4(d_\gamma-d_e)}{d_\gamma+d_e-4} + \frac{d_\gamma-d_e}{2d_\gamma+d_e-8}\, \left( d_e-8 - 4 \, \frac{d_\gamma+d_e-6}{d_\gamma+d_e-4} \right) 
\nonum \\
&\qquad \left . +\xi\,\frac{(d_e-2)(d_\gamma-d_e)}{d_\gamma+d_e-4} -\xi^2\,\frac{d_e-2}{2} \right]\,G^2(d_e,1,1-\varepsilon_e)
%\Bigg .
\nonum \\
&\Bigg . + \left[ d_\gamma - 2d_e + 1  - \frac{4(d_e-2)(d_\gamma-1)}{d_\gamma+d_e-4} - \frac{8(d_\gamma-1)}{d_\gamma-4} - \frac{2(d_e-8)(d_\gamma-d_e)}{d_\gamma+d_e-6} +
\frac{4(d_\gamma-2)(d_\gamma-d_e)}{(d_\gamma-4)(d_\gamma+d_e-4)} \right .
\nonum \\
&\qquad \left . - 2\xi\,\frac{(d_\gamma-3)(d_\gamma-d_e)}{d_\gamma+d_e-4} + \xi^2 (d_\gamma-3) 
+ \frac{(d_\gamma-3)(d_\gamma+d_e-4)}{d_\gamma-4}\,\left( \xi - \frac{d_\gamma - d_e}{d_\gamma+d_e-4} \right)^2
\right ]\,
\nonum \\
&\qquad \times G(d_e,1,1-\varepsilon_e)G(1-\varepsilon_e,\varepsilon_\gamma)
\nonum \\
& \Bigg . + \frac{(d_\gamma-4)(d_\gamma(d_e-4)+8)}{(2d_\gamma+d_e-8)(2d_\gamma+d_e-10)}\,G(d_e,1-\varepsilon_e,1,1-\varepsilon_e,1,1) \Bigg \}\, .
\label{chap3:res:Sigma2}
\end{flalign}
Again, the last line of Eq.~(\ref{chap3:res:Sigma2}) contains the UV-convergent diagram $G(d_e,1-\varepsilon_e,1,1-\varepsilon_e,1,1)$
with a prefactor proportional to $d_\gamma -4$. It therefore vanishes for all RQED$_{4,d_e}$ and in particular in RQED$_{4,3}$ which enables us to avoid computing the complicated diagram $G(3-2\veps_\gamma,1/2,1,1/2,1,1)$
and obtain an expansion valid up to $\Ord(\varepsilon_\gamma^0)$ for the two-loop fermion self-energy.

Eq.~(\ref{chap3:res:Sigma2}) simplifies for usual QEDs where $d_\gamma=d_e$ corresponding to $\varepsilon_e=0$. In this case, the complicated diagram reduces to the well known $G(d_\gamma,1,1,1,1,1)$ 
and Eq.~(\ref{chap3:res:Sigma2}) becomes:
\begin{flalign}
&\Sigma_{V2}(p^2) = \bar{\al}^2 \, \left( \frac{\overline{\mu}^{\,2}}{-p^2} \right)^{2\veps_\gamma}\,(d_\gamma-2)\,e^{2\gamma_E \veps_\gamma}\,
\Bigg \{ 2N_F\,\frac{d_\gamma-2}{d_\gamma-6}\,G(d_\gamma,1,1)G(d_\gamma,1,\varepsilon_\gamma) \Bigg . 
\nonum \\
&\Bigg . -\frac{1}{4}\,\bigg(d_\gamma-6+(d_\gamma-2)\xi^2\bigg)G^2(d_\gamma,1,1) 
- \frac{1}{2}\,\frac{d_\gamma-3}{d_\gamma-4}\,\bigg(d_\gamma + 4 -  \xi^2(3d_\gamma-8) \bigg)\,G(d_\gamma,1,1)G(d_\gamma,1,\varepsilon_\gamma)\quad \Bigg \}\, ,
\label{chap3:res:QED4:Sigma2}
\end{flalign}
in agreement with well known results in the literature, see {\it e.g.}, Ref.~\cite{grozin2007lectures}.

On the basis of these results, the computation of renormalization constants and renormalized fermion self-energies can be conveniently carried out using the forest formula Eq.~(\ref{chap2:def:forest}). Graphically,
the renormalization constants associated with the individual 2-loop fermion self-energy diagrams read:
\begin{subequations}
\label{chap3:def:Zpsi-forest}
\begin{flalign}
&\delta Z_{2a\,\psi}(\bar{\al}_r) = \mathcal{K}\, \bigg[~
\parbox{20mm}{
    \begin{fmfgraph*}(20,20)
      %\fmfpen{thick}
      \fmfleft{i1}
      \fmfright{o1}
      \fmf{phantom}{i1,i2}
      \fmf{plain,tension=1.4}{i2,i3}
      \fmf{plain}{i3,i4}
      \fmf{plain}{i4,o4}
      \fmf{plain}{o4,o3}
      \fmf{plain,tension=1.4}{o3,o2}
      \fmf{phantom}{o2,o1}
      \fmffreeze
      %\fmf{plain,left,tension=0.1,tag=1}{i1,o1}
      \fmf{phantom,left,tension=0.1,tag=2}{i2,o2}
      \fmf{phantom,left,tension=0.1,tag=3}{i3,o3}
      \fmf{phantom,left,tension=0.1,tag=4}{i4,o4}
      \fmfdot{i3,o3}
      \fmfposition
      \fmfipath{p[]}
      %\fmfiset{p1}{vpath1(__i1,__o1)}
      \fmfiset{p2}{vpath2(__i2,__o2)}
      \fmfiset{p3}{vpath3(__i3,__o3)}
      \fmfiset{p4}{vpath4(__i4,__o4)}
      \fmfi{photon}{subpath (0,3length(p3)/8) of p3}
      \fmfi{photon}{subpath (5length(p3)/8,length(p3)) of p3}
      \fmfi{plain}{point 3length(p3)/8 of p3 .. point length(p2)/2 of p2 .. point 5length(p3)/8 of p3}
      \fmfi{plain}{point 5length(p3)/8 of p3 .. point length(p4)/2 of p4 .. point 3length(p3)/8 of p3}
      \def\vert#1{%
        \fmfiv{decor.shape=circle,decor.filled=full,decor.size=2thick}{#1}}
      \vert{point 3length(p3)/8 of p3}
      \vert{point 5length(p3)/8 of p3}
   \end{fmfgraph*}
}~ \bigg] - \mathcal{K}\, \bigg[~\mathcal{K}\, \bigg[~
   \parbox{15mm}{
    \begin{fmfgraph*}(15,15)
      %\fmfpen{thick}
      \fmfleft{in}
      \fmfright{out}
      \fmf{boson}{in,ve}
      \fmf{plain,right,tension=0.2}{ve,vw}
      \fmf{plain,right,tension=0.2}{vw,ve}
      \fmf{boson}{vw,out}
      \fmfdot{ve,vw}
    \end{fmfgraph*}
} ~ \bigg] ~ \star ~
\parbox{15mm}{
    \begin{fmfgraph*}(15,15)
      %\fmfpen{thick}
      \fmfleft{in}
      \fmfright{out}
      \fmf{plain}{in,ve}
      \fmf{plain,right,tension=0.2}{ve,vw}
      \fmf{boson,right,tension=0.2,label=$\perp$,l.s=left}{vw,ve}
      \fmf{plain}{vw,out}
      \fmfdot{ve,vw}
    \end{fmfgraph*}
}~ \bigg]\, ,
\label{chap3:deltaZ2apsi}
\\
&\delta Z_{2b\,\psi}(\bar{\al}_r,\xi_r) = \mathcal{K}\, \bigg[~
\parbox{20mm}{
    \begin{fmfgraph*}(20,20)
      %\fmfpen{thick}
      \fmfleft{i1}
      \fmfright{o1}
      \fmf{plain,tension=1.7}{i1,i2}
      \fmf{plain}{i2,i3}
      \fmf{plain}{i3,o3}
      \fmf{plain}{o3,o2}
      \fmf{plain,tension=1.7}{o2,o1}
      \fmffreeze
      \fmf{photon,left,tension=0.1}{i2,o2}
      \fmf{photon,left,tension=0.1}{i3,o3}
      \fmfdot{i2,i3,o2,o3}
   \end{fmfgraph*}
} ~\bigg] -  \mathcal{K}\, \bigg[~\mathcal{K}\, \bigg[~
   \parbox{15mm}{
    \begin{fmfgraph*}(15,15)
      %\fmfpen{thick}
      \fmfleft{in}
      \fmfright{out}
      \fmf{plain}{in,ve}
      \fmf{plain,right,tension=0.2}{ve,vw}
      \fmf{boson,right,tension=0.2}{vw,ve}
      \fmf{plain}{vw,out}
      \fmfdot{ve,vw}
    \end{fmfgraph*}
}~ \bigg] ~ \star ~
   \parbox{15mm}{
    \begin{fmfgraph*}(15,15)
      %\fmfpen{thick}
      \fmfleft{in}
      \fmfright{out}
      \fmf{plain}{in,ve}
      \fmf{plain,right,tension=0.2}{ve,vw}
      \fmf{boson,right,tension=0.2}{vw,ve}
      \fmf{plain}{vw,out}
      \fmfdot{ve,vw}
    \end{fmfgraph*}
}~ \bigg]\, ,
\label{chap3:deltaZ2bpsi}
\\
&\delta Z_{2c\,\psi}(\bar{\al}_r,\xi_r) = \mathcal{K}\, \bigg[~
\parbox{20mm}{
    \begin{fmfgraph*}(20,20)
      %\fmfpen{thick}
      \fmfleft{i1}
      \fmfright{o1}
      \fmf{plain,tension=1.7}{i1,i2}
      \fmf{plain}{i2,i3}
      \fmf{plain}{i3,o3}
      \fmf{plain}{o3,o2}
      \fmf{plain,tension=1.7}{o2,o1}
      \fmffreeze
      \fmf{photon,left,tension=0.1}{i2,o3}
      \fmf{photon,right,tension=0.1}{i3,o2}
      \fmfdot{i2,i3,o2,o3}
   \end{fmfgraph*}
} \bigg] + 2 \, \mathcal{K}\, \bigg[~\mathcal{K}\, \bigg[~
   \parbox{15mm}{
    \begin{fmfgraph*}(15,15)
      %\fmfpen{thick}
      \fmfleft{in}
      \fmfright{out}
      \fmf{plain}{in,ve}
      \fmf{plain,right,tension=0.2}{ve,vw}
      \fmf{boson,right,tension=0.2}{vw,ve}
      \fmf{plain}{vw,out}
      \fmfdot{ve,vw}
    \end{fmfgraph*}
}~ \bigg] ~ \star ~
   \parbox{15mm}{
    \begin{fmfgraph*}(15,15)
      %\fmfpen{thick}
      \fmfleft{in}
      \fmfright{out}
      \fmf{plain}{in,ve}
      \fmf{plain,right,tension=0.2}{ve,vw}
      \fmf{boson,right,tension=0.2}{vw,ve}
      \fmf{plain}{vw,out}
      \fmfdot{ve,vw}
    \end{fmfgraph*}
}~ \bigg]\, ,
\label{chap3:deltaZ2cpsi}
\end{flalign}
\end{subequations}
where it is again understood that the graphs in argument of $\mathcal{K}$ represent the logarithmic part of the corresponding diagrams.
Notice that in Eq.~(\ref{chap3:deltaZ2apsi}) the contraction of the one-loop polarization operator subgraph (which is transverse due to current conservation)
resulted in the appearance of the transverse part of the one-loop fermion self-energy Eq.~(\ref{chap3:Sigma1:T}). This is an example well known in the literature, see, \eg, Ref.~\cite{Kissler:2016gxn}, of the sensitivity 
of the contraction procedure to the Lorentz structure of subdiagrams (once the contraction done, the $\star$ operator reduces here too to simple multiplication). 
Interestingly, the transverse part is non-zero only in the reduced case ($\veps_e >0$). 
Moreover, in Eq.~(\ref{chap3:deltaZ2cpsi}), we have used the Ward identity relating the divergent one-loop vertex part to the divergent
one-loop fermion self-energy part, Eq.~(\ref{chap3:ward}) or (\ref{chap2:ward:graphical}) in graphical notations. Summing all the individual contributions in Eqs.~(\ref{chap3:def:Zpsi-forest})
yields the total 2-loop fermion renormalization constant:
\begin{flalign}
&\delta Z_{2\,\psi}(\bar{\al}_r,\xi_r) = \mathcal{K}\, \bigg[~
\parbox{20mm}{
    \begin{fmfgraph*}(20,20)
      %\fmfpen{thick}
      \fmfleft{i1}
      \fmfright{o1}
      \fmf{phantom}{i1,i2}
      \fmf{plain,tension=1.4}{i2,i3}
      \fmf{plain}{i3,i4}
      \fmf{plain}{i4,o4}
      \fmf{plain}{o4,o3}
      \fmf{plain,tension=1.4}{o3,o2}
      \fmf{phantom}{o2,o1}
      \fmffreeze
      %\fmf{plain,left,tension=0.1,tag=1}{i1,o1}
      \fmf{phantom,left,tension=0.1,tag=2}{i2,o2}
      \fmf{phantom,left,tension=0.1,tag=3}{i3,o3}
      \fmf{phantom,left,tension=0.1,tag=4}{i4,o4}
      \fmfdot{i3,o3}
      \fmfposition
      \fmfipath{p[]}
      %\fmfiset{p1}{vpath1(__i1,__o1)}
      \fmfiset{p2}{vpath2(__i2,__o2)}
      \fmfiset{p3}{vpath3(__i3,__o3)}
      \fmfiset{p4}{vpath4(__i4,__o4)}
      \fmfi{photon}{subpath (0,3length(p3)/8) of p3}
      \fmfi{photon}{subpath (5length(p3)/8,length(p3)) of p3}
      \fmfi{plain}{point 3length(p3)/8 of p3 .. point length(p2)/2 of p2 .. point 5length(p3)/8 of p3}
      \fmfi{plain}{point 5length(p3)/8 of p3 .. point length(p4)/2 of p4 .. point 3length(p3)/8 of p3}
      \def\vert#1{%
        \fmfiv{decor.shape=circle,decor.filled=full,decor.size=2thick}{#1}}
      \vert{point 3length(p3)/8 of p3}
      \vert{point 5length(p3)/8 of p3}
   \end{fmfgraph*}
}~ \bigg] + 
\mathcal{K}\, \bigg[~
\parbox{20mm}{
    \begin{fmfgraph*}(20,20)
      %\fmfpen{thick}
      \fmfleft{i1}
      \fmfright{o1}
      \fmf{plain,tension=1.7}{i1,i2}
      \fmf{plain}{i2,i3}
      \fmf{plain}{i3,o3}
      \fmf{plain}{o3,o2}
      \fmf{plain,tension=1.7}{o2,o1}
      \fmffreeze
      \fmf{photon,left,tension=0.1}{i2,o2}
      \fmf{photon,left,tension=0.1}{i3,o3}
      \fmfdot{i2,i3,o2,o3}
   \end{fmfgraph*}
} ~\bigg] +
\mathcal{K}\, \bigg[~
\parbox{20mm}{
    \begin{fmfgraph*}(20,20)
      %\fmfpen{thick}
      \fmfleft{i1}
      \fmfright{o1}
      \fmf{plain,tension=1.7}{i1,i2}
      \fmf{plain}{i2,i3}
      \fmf{plain}{i3,o3}
      \fmf{plain}{o3,o2}
      \fmf{plain,tension=1.7}{o2,o1}
      \fmffreeze
      \fmf{photon,left,tension=0.1}{i2,o3}
      \fmf{photon,right,tension=0.1}{i3,o2}
      \fmfdot{i2,i3,o2,o3}
   \end{fmfgraph*}
} \bigg] 
\nonum \\
& \qquad - \mathcal{K}\, \bigg[~\mathcal{K}\, \bigg[~
   \parbox{15mm}{
    \begin{fmfgraph*}(15,15)
      %\fmfpen{thick}
      \fmfleft{in}
      \fmfright{out}
      \fmf{boson}{in,ve}
      \fmf{plain,right,tension=0.2}{ve,vw}
      \fmf{plain,right,tension=0.2}{vw,ve}
      \fmf{boson}{vw,out}
      \fmfdot{ve,vw}
    \end{fmfgraph*}
} ~ \bigg] ~ \star ~
\parbox{15mm}{
    \begin{fmfgraph*}(15,15)
      %\fmfpen{thick}
      \fmfleft{in}
      \fmfright{out}
      \fmf{plain}{in,ve}
      \fmf{plain,right,tension=0.2}{ve,vw}
      \fmf{boson,right,tension=0.2,label=$\perp$,l.s=left}{vw,ve}
      \fmf{plain}{vw,out}
      \fmfdot{ve,vw}
    \end{fmfgraph*}
}~ \bigg] 
+ \mathcal{K}\, \bigg[~\mathcal{K}\, \bigg[~
   \parbox{15mm}{
    \begin{fmfgraph*}(15,15)
      %\fmfpen{thick}
      \fmfleft{in}
      \fmfright{out}
      \fmf{plain}{in,ve}
      \fmf{plain,right,tension=0.2}{ve,vw}
      \fmf{boson,right,tension=0.2}{vw,ve}
      \fmf{plain}{vw,out}
      \fmfdot{ve,vw}
    \end{fmfgraph*}
}~ \bigg] ~ \star ~
   \parbox{15mm}{
    \begin{fmfgraph*}(15,15)
      %\fmfpen{thick}
      \fmfleft{in}
      \fmfright{out}
      \fmf{plain}{in,ve}
      \fmf{plain,right,tension=0.2}{ve,vw}
      \fmf{boson,right,tension=0.2}{vw,ve}
      \fmf{plain}{vw,out}
      \fmfdot{ve,vw}
    \end{fmfgraph*}
}~ \bigg]\, ,
\label{chap3:deltaZ2psi-forest}
\end{flalign}
where, contrarily to the case of the polarization operator Eq.~(\ref{chap3:deltaZ2A}), subdivergent graphs do contribute.

\subsection{Case of QED$_4$}

In the case of QED$_4$, we first set $\veps_e=0$ and take the limit $\veps_\gamma \ra 0$. From Eqs.~(\ref{chap3:res:Sigma2a}), (\ref{chap3:res:Sigma2b}) and Eq.~(\ref{chap3:res:Sigma2c-2}), this leads to:
\begin{subequations}
\label{chap3:res:QED4:Sigma2a+b+c}
\begin{flalign}
\Sigma_{2aV}(p^2;\bar{\al},\xi) &= 4\,N_F\, \bar{\al}^2 \left( \frac{\overline{\mu}^2}{-p^2} \right)^{2\veps_\gamma}\, \frac{(d_\gamma-2)^2}{2(d_\gamma-6)}\,e^{2 \gamma_E \veps_\gamma} 
G(4-2\veps_\gamma,1,1)G(4-2\veps_\gamma,1,2-d_\gamma/2)
\nonum \\
&= N_F\,\bar{\al}^2 \left [ \frac{1}{\veps_\gamma} - 2 L_p + \frac{7}{2} +  \Ord(\veps_\gamma) \right]\, ,
\label{chap3:res:QED4:Sigma2a} \\
\Sigma_{2bV}(p^2;\bar{\al},\xi) &= \xi^2 \bar{\al}^2 \left( \frac{\overline{\mu}^2}{-p^2} \right)^{2\veps_\gamma}\, \frac{(d_\gamma-2)^2(d_\gamma-3)}{d_\gamma-4}\,e^{2 \gamma_E \veps_\gamma} 
G(4-2\veps_\gamma,1,1)G(4-2\veps_\gamma,1,2-d_\gamma/2)
\nonum \\
&= \frac{\xi^2 \bar{\al}^2}{2} \left [ \frac{1}{\veps_\gamma^2} - \frac{2}{\veps_\gamma}\,\left( L_p - \frac{5}{4} \right)  + 2 \left( L_p^2 - \frac{5L_p}{2} - \frac{\zeta_2}{2} + \frac{31}{8} \right) +  \Ord(\veps_\gamma) \right]\, ,
\label{chap3:res:QED4:Sigma2b} \\
\Sigma_{2cV}(p^2;\bar{\al},\xi) &= - \bar{\al}^2 \left( \frac{\overline{\mu}^2}{-p^2} \right)^{2\veps_\gamma}\, \frac{d_\gamma-2}{4}\,e^{2 \gamma_E \veps_\gamma} \,
\bigg[ \bigg( d_\gamma -6 + (d_\gamma-2)\xi^2 \bigg)\,G^2(4-2\veps_\gamma,1,1) \bigg .
\nonum \\
&\bigg . \qquad + 2\,\frac{d_\gamma-3}{d_\gamma-4}\,\bigg( d_\gamma + 4 - (d_\gamma-4)\xi^2 \bigg)\, G(4-2\veps_\gamma,1,1)G(4-2\veps_\gamma,1,2-d_\gamma/2) \bigg ]\,
\nonum \\
&= \frac{3 \bar{\al}^2}{4} \left [ \frac{1}{\veps_\gamma} - 2L_p + \frac{5}{6}\right] -
\xi^2 \bar{\al}^2 \left [ \frac{1}{\veps_\gamma^2} - \frac{2}{\veps_\gamma}\,\left( L_p - \frac{9}{8} \right) + 2 \left( L_p^2 - \frac{9L_p}{4} - \frac{\zeta_2}{2} + \frac{47}{16} \right) \right]  +  \Ord(\veps_\gamma)\, ,
\label{chap3:res:QED4:Sigma2c}
\end{flalign}
\end{subequations}
which is valid for an arbitrary gauge fixing parameter $\xi$. Combining Eqs.~(\ref{chap3:res:QED4:Sigma2a+b+c}) with Eqs.~(\ref{chap3:def:Zpsi-forest}), 
the individual counter-terms read:
\begin{subequations}
\label{chap3:res:QED4:Z2a+b+c-psi}
\begin{flalign}
\delta Z_{2a\,\psi}(\bar{\al}_r) &= \mathcal{K}\, \bigg[ \Sigma_{2aV}(p^2;\bar{\al}_r) \bigg] - 
\mathcal{K}\, \bigg[\mathcal{K}\, \bigg[\Pi_{1}(q^2;\bar{\al}_r,\xi_r) \bigg]\,\underbrace{\Sigma_{1V}^{(\perp)}(p^2;\bar{\al}_r,\xi_r)}_{=0} \bigg]
\nonum \\
&= \frac{N_F\,\bar{\al}_r^2}{\veps_\gamma}\, ,
\label{chap3:res:QED4:Z2apsi} \\
\delta Z_{2b\,\psi}(\bar{\al}_r,\xi_r) &= \mathcal{K}\, \bigg[ \Sigma_{2bV}(p^2;\bar{\al}_r,\xi_r) \bigg] - \mathcal{K}\, \bigg[\mathcal{K}\, \bigg[\Sigma_{1V}(p^2;\bar{\al}_r,\xi_r) \bigg]\,\Sigma_{1V}(p^2;\bar{\al}_r,\xi_r) \bigg]
\nonum \\
&= - \frac{\xi_r^2 \bar{\al}_r^2}{2}\,\bigg( \frac{1}{\veps_\gamma^2} - \frac{1}{2\veps_\gamma} \bigg)\, ,
\label{chap3:res:QED4:Z2bpsi} \\
\delta Z_{2c\,\psi}(\bar{\al}_r,\xi_r) &= \mathcal{K}\, \bigg[ \Sigma_{2cV}(p^2;\bar{\al}_r,\xi_r) \bigg] +2\, \mathcal{K}\, \bigg[\mathcal{K}\, \bigg[\Sigma_{1V}(p^2;\bar{\al}_r,\xi_r) \bigg]\,\Sigma_{1V}(p^2;\bar{\al}_r,\xi_r) \bigg]
\nonum \\
&= \frac{3 \bar{\al}_r^2}{4\veps_\gamma} + \xi_r^2 \bar{\al}_r^2\,\bigg( \frac{1}{\veps_\gamma^2} - \frac{1}{4\veps_\gamma} \bigg)\, ,
\label{chap3:res:QED4:Z2cpsi} 
\end{flalign}
\end{subequations}
where, in the first line, we have used the fact that $\Sigma_{1V}^{(\perp)}=0$ in QED$_4$, see Eq.~(\ref{chap3:Sigma1:T}). The sum of Eqs.~(\ref{chap3:res:QED4:Z2a+b+c-psi}) yields the total counterterm at two-loop order:
\be
\delta Z_{2\,\psi} =  \frac{\bar{\al}_r^2}{\veps_\gamma} \left( N_F + \frac{3}{4} \right ) + \frac{\xi_r^2 \bar{\al}_r^2}{2\veps^2} \, .
\label{chap3:res2l:QED4:Z2psi}
\ee

The individual renormalized diagrams are also straightforward to compute and read:
\begin{subequations}
\label{chap3:res:QED4:Sigma2ra+b+c}
\begin{flalign}
\Sigma_{2aVr}(p^2;\bar{\al}_r) &= \Sigma_{2aV}(p^2;\bar{\al}_r) - \mathcal{K}\, \bigg[\Pi_{1}(q^2;\bar{\al}_r,\xi_r) \bigg]\,\underbrace{\Sigma_{1V}^{(\perp)}(p^2;\bar{\al}_r,\xi_r)}_{=0} - \delta Z_{2a\,\psi}(\bar{\al}_r)
\nonum \\
&= -2 N_F \bar{\al}_r^2 \left [ L_p - \frac{7}{4} \right]\, ,
\label{chap3:res:QED4:Sigma2ar} \\
\Sigma_{2bVr}(p^2;\bar{\al}_r,\xi_r) &= \Sigma_{2bV}(p^2;\bar{\al}_r,\xi_r) - \mathcal{K} \bigg[ \Sigma_{1V}(p^2;\bar{\al}_r,\xi_r) \bigg]~\Sigma_{1V}(p^2;\bar{\al}_r,\xi_r) - \delta Z_{2b\,\psi}(\bar{\al}_r,\xi_r)
\nonum \\
&= \frac{\xi_r^2 \bar{\al}_r^2}{2} \left ( L_p^2 - 3L_p  + \frac{15}{4} \right)\, ,
\label{chap3:res:QED4:Sigma2br} \\
\Sigma_{2cVr}(p^2;\bar{\al}_r,\xi_r) &= \Sigma_{2cVr}(p^2;\bar{\al}_r,\xi_r) + 2\mathcal{K} \bigg[ \Sigma_{1V}(p^2;\bar{\al}_r,\xi_r) \bigg]~\Sigma_{1V}(p^2;\bar{\al}_r,\xi_r) - \delta Z_{2c\,\psi}(\bar{\al}_r,\xi_r)
\nonum \\
&= -\frac{3 \bar{\al}_r^2}{2} \left( L_p - \frac{5}{12} \right ) - \xi_r^2 \bar{\al}_r^2 \left ( L_p^2 - \frac{5L_p}{2} + \frac{15}{8} \right)\, .
\label{chap3:res:QED4:Sigma2cr}
\end{flalign}
\end{subequations}
The sum of Eqs.~(\ref{chap3:res:QED4:Sigma2ra+b+c}) yields the total two-loop renormalized fermion self-energy:
\be
\Sigma_{2Vr}(p^2;\bar{\al}_r,\xi_r) = -\bar{\al}_r^2 \bigg( 2 L_p \left( N_F + \frac{3}{4} \right) - \frac{7}{2} N_F - \frac{5}{8} \bigg) - \frac{\xi_r^2 \bar{\al}_r^2}{2}\,L_p(L_p-2)\, .
\label{chap3:res2l:QED4:Sigma2r}
\ee

Combining the above two-loop results with the one-loop ones derived in the previous paragraphs, we deduce the anomalous dimension up to two loops:
\be
\gamma_{\psi}(\bar{\al}_r,\xi_r) = - 2\xi_r \bar{\al}_r + 4\,\left( N_F + \frac{3}{4} \right)\,\bar{\al}_r^2 + \Ord(\bar{\al}_r^3)\, ,
\label{chap3:res2l:QED4:gammapsi}
\ee
where the two-loop contribution is gauge-independent, as well as the expansion of the renormalized fermion propagator up to two loops:
\be
-\I \Sp S_r(p) = 1 + \xi_r \bar{\al}_r\,(L_p - 1) + \bar{\al}_r^2\,\bigg( \xi_r^2 \left( 1 + \frac{L_p}{2}\,(L_p - 2)\right) - 2 L_p \left( N_F + \frac{3}{4} \right) + \frac{7}{2} N_F + \frac{5}{8}  \bigg) \, . 
\label{chap3:res2l:QED4:Sr}
\ee
All these results are of course well known from the literature, see, \eg, \cite{grozin2007lectures}, and recovering them constitutes a basic check of our general formulas. 
Notice for example, that the result of Eq.~(\ref{chap3:res2l:QED4:gammapsi})
suggests that the gauge for which $\gamma_{\psi}=0$ is given by: $\xi_r (\bar{\al}_r) = 2 (N_F+3/4) \bar{\al}_r + \Ord(\bar{\al}_r^2)$. 
In the case of $N_F=0$ (the so-called quenched approximation), we recover the early result of Johnson et al.\ \cite{Johnson:1964da}: 
$\xi_r (\bar{\al}_r, N_F=0) = (3/2) \bar{\al}_r + \Ord(\bar{\al}_r^2)$.~\footnote{I thank Valery Gusynin from pointing this to me.}

\subsection{Case of reduced QED$_{4,3}$}

In the case of reduced QED$_{4,3}$, we first set $\veps_e=1/2$ and take the limit $\veps_\gamma \ra 0$. From Eqs.~(\ref{chap3:res:Sigma2a}), (\ref{chap3:res:Sigma2b}) and Eq.~(\ref{chap3:res:Sigma2c-2}), this leads to:
\begin{subequations}
\label{chap3:res:RQED4,3:Sigma2a+b+c}
\begin{flalign}
&\Sigma_{2aV}(p^2;\bar{\al},\xi) = -\frac{2\pi^2\,N_F\, \bar{\al}^2}{3}\, \left [ \frac{1}{\veps_\gamma} - 2 L_p  + \Ord(\veps_\gamma) \right]\, ,
\label{chap3:res:RQED4,3:Sigma2a} \\
&\Sigma_{2bV}(p^2;\bar{\al},\xi) = \bar{\al}^2\, \left [ \frac{(1-3\xi)^2}{18\veps_\gamma^2} + \frac{1-3\xi}{\veps_\gamma}\,\left( \frac{11}{27} - \frac{7\xi}{9} - \frac{1-3\xi}{9}\,\tilde{L}_p \right)  + 
\frac{(1-3\xi)^2}{9}\,\left( \tilde{L}^2_p - \frac{9}{2}\,\zeta_2 \right) - \right .
\nonum \\
&\left . - \frac{2\,(11+9\xi(7\xi-6))}{27}\,\tilde{L}_p  + \frac{2\,(103 + 81\xi(7\xi-6))}{81} + \Ord(\veps_\gamma) \right]\, ,
\label{chap3:res:RQED4,3:Sigma2b} \\
&\Sigma_{2cV}(p^2;\bar{\al},\xi) = \bar{\al}^2 \left [ -\frac{(1-3\xi)^2}{9\veps_\gamma^2} + \frac{1}{\veps_\gamma}\,\left( \frac{2(1-3\xi)^2}{9}\,\tilde{L}_p  + \frac{(34-39\xi)\xi}{9} - \frac{37}{27} \right)
-\frac{2(1-3\xi)^2}{9}\,\tilde{L}_p^2 + \right .
\nonum \\
&\left . + \frac{2\,(37+3\xi(39\xi-34))}{27}\,\tilde{L}_p + \frac{71 + 21\xi(3\xi -2 )}{9}\,\zeta_2 - \frac{2(695-798\xi + 891\xi^2)}{81} + \Ord(\veps_\gamma) \right]\, ,
\label{chap3:res:RQED4,3:Sigma2c}
\end{flalign}
\end{subequations}
which is valid for an arbitrary gauge fixing parameter $\xi$. Combining Eqs.~(\ref{chap3:res:RQED4,3:Sigma2a+b+c}) with Eqs.~(\ref{chap3:def:Zpsi-forest}),
the individual counter-terms read:
\begin{subequations}
\label{chap3:res:RQED4,3:Z2a+b+c-psi}
\begin{flalign}
\delta Z_{2a\,\psi}(\bar{\al}_r) &= \mathcal{K}\, \bigg[ \Sigma_{2aV}(p^2;\bar{\al}_r) \bigg] -
\mathcal{K}\, \bigg[\underbrace{\mathcal{K}\, \bigg[\Pi_{1}(q^2;\bar{\al}_r,\xi_r) \bigg]}_{=0}\,\Sigma_{1V}^{(\perp)}(p^2;\bar{\al}_r,\xi_r) \bigg]
\nonum \\
&= -\frac{2\pi^2\,N_F\,\bar{\al}_r^2}{3\veps_\gamma}\, ,
\label{chap3:res:RQED4,3:Z2apsi} \\
\delta Z_{2b\,\psi}(\bar{\al}_r,\xi_r) &= \mathcal{K}\, \bigg[ \Sigma_{2bV}(p^2;\bar{\al}_r,\xi_r) \bigg] - \mathcal{K}\, \bigg[\mathcal{K}\, \bigg(\Sigma_{1V}(p^2;\bar{\al}_r,\xi_r) \bigg)\,\Sigma_{1V}(p^2;\bar{\al}_r,\xi_r) \bigg]
\nonum \\
&= - \frac{(1-3\xi_r)^2 \bar{\al}_r^2}{18}\,\bigg( \frac{1}{\veps_\gamma^2} - \frac{2}{3\veps_\gamma} \bigg)\, ,
\label{chap3:res:RQED4,3:Z2bpsi} \\
\delta Z_{2c\,\psi}(\bar{\al}_r,\xi_r) &= \mathcal{K}\, \bigg[ \Sigma_{2cV}(p^2;\bar{\al}_r,\xi_r) \bigg] +2\, \mathcal{K}\, \bigg[\mathcal{K}\, \bigg(\Sigma_{1V}(p^2;\bar{\al}_r,\xi_r) \bigg)\,\Sigma_{1V}(p^2;\bar{\al}_r,\xi_r) \bigg]
\nonum \\
&= \bar{\al}_r^2\,\bigg( \frac{(1-3\xi_r)^2}{9\veps_\gamma^2} - \frac{17-6\xi_r + 9 \xi_r^2}{27\veps_\gamma} \bigg)\, ,
\label{chap3:res:RQED4,3:Z2cpsi}
\end{flalign}
\end{subequations}
where, in the first line, we have used the fact that $\Pi_{1}(q^2)$ is finite in QED$_{4,3}$. The sum of Eqs.~(\ref{chap3:res:RQED4,3:Z2a+b+c-psi}) yields the total counterterm at two-loop order:
\be
\delta Z_{2\,\psi} =  -\frac{4\,\bar{\al}_r^2}{\veps_\gamma}\, \left( \zeta_2 N_F + \frac{4}{27} \right ) + \frac{(1-3\xi_r)^2\, \bar{\al}_r^2}{18\veps^2} \, .
\label{chap3:res2l:RQED4,3:Z2psi}
\ee

The individual renormalized diagrams are also straightforward to compute and read:
\begin{subequations}
\label{chap3:res:RQED4,3:Sigma2ra+b+c}
\begin{flalign}
&\Sigma_{2aVr}(p^2;\bar{\al}_r) = \Sigma_{2aV}(p^2;\bar{\al}_r) - \underbrace{\mathcal{K}\, \bigg[\Pi_{1}(q^2;\bar{\al}_r,\xi_r) \bigg]}_{=0}\,\Sigma_{1V}^{(\perp)}(p^2;\bar{\al}_r,\xi_r) 
- \delta Z_{2a\,\psi}(\bar{\al}_r)
\nonum \\
&= 8 N_F \zeta_2\,\bar{\al}_r^2\, L_p\, ,
\label{chap3:res:RQED4,3:Sigma2ar} \\
&\Sigma_{2bVr}(p^2;\bar{\al}_r,\xi_r) = \Sigma_{2bV}(p^2;\bar{\al}_r,\xi_r) - \mathcal{K} \bigg[ \Sigma_{1V}(p^2;\bar{\al}_r,\xi_r) \bigg]~\Sigma_{1V}(p^2;\bar{\al}_r,\xi_r) - \delta Z_{2b\,\psi}(\bar{\al}_r,\xi_r)
\nonum \\
&= \bar{\al}^2\, \left [ \frac{(1-3\xi_r)^2}{18}\,\left( \tilde{L}^2_p - 2\,\zeta_2\right) - \frac{4\,(1+\xi_r(6\xi_r-5))}{9}\,\tilde{L}_p  + \frac{2\,(47 -210\xi_r +243\xi_r^2)}{81} \right]\, ,
\label{chap3:res:RQED4,3:Sigma2br} \\
&\Sigma_{2cVr}(p^2;\bar{\al}_r,\xi_r) = \Sigma_{2cVr}(p^2;\bar{\al}_r,\xi_r) + 2\mathcal{K} \bigg[ \Sigma_{1V}(p^2;\bar{\al}_r,\xi_r) \bigg]~\Sigma_{1V}(p^2;\bar{\al}_r,\xi_r) - \delta Z_{2c\,\psi}(\bar{\al}_r,\xi_r)
\nonum \\
&=\bar{\al}^2\, \left [-\frac{(1-3\xi_r)^2}{9}\,\tilde{L}_p^2 + \frac{64}{9}\,\zeta_2 + \frac{2\,(3+\xi_r(7\xi_r-6))}{3}\,\tilde{L}_p + \frac{2\xi_r(82-81\xi_r)}{27} - \frac{1166}{81}  \right]\, ,
\label{chap3:res:RQED4,3:Sigma2cr}
\end{flalign}
\end{subequations}
where $\tilde{L}_p=L_p+\log 4$. The sum of Eqs.~(\ref{chap3:res:RQED4,3:Sigma2ra+b+c}) yields the total two-loop renormalized fermion self-energy:
\bea
\Sigma_{2Vr}(p^2;\bar{\al}_r,\xi_r) =&& \bar{\al}_r^2 \bigg[ 8 N_F \zeta_2\,L_p - \frac{(1-3\xi_r)^2}{18}\,\tilde{L}_p^2 + \frac{64-(1-3\xi_r)^2}{9}\,\zeta_2 +
\nonum \\
&&\qquad + \frac{2\,(7+\xi_r(9\xi_r-8))}{9}\,\tilde{L}_p - \frac{8(134 - 9\xi_r)}{81} \bigg]\, .
\label{chap3:res2l:RQED4,3:Sigma2r}
\eea

Combining the above two-loop results with the one-loop ones derived in the previous paragraphs, we deduce the anomalous dimension up to two loops:
\be
\gamma_{\psi}(\bar{\al}_r,\xi_r) = 2\bar{\al}_r\,\frac{1-3\xi_r}{3} -16\,\left( \zeta_2 N_F + \frac{4}{27} \right)\, \bar{\al}_r^2 + \Ord(\bar{\al}_r^3)\, ,
\label{chap3:res2l:RQED4,3:gammapsi}
\ee
where the two-loop contribution is gauge-independent, as well as the expansion of the renormalized fermion propagator up to two loops:
\begin{flalign}
&-\I \Sp S_r(p) = 1 +\bar{\al}_r \left( \frac{10}{9} - 2\xi_r - \frac{1-3\xi_r}{3}\,\tilde{L}_p \right) + 
\bar{\al}_r^2\,\bigg( 8 N_F \zeta_2\,L_p + \frac{(1-3\xi_r)^2}{18}\,\tilde{L}_p^2 + 
\nonum \\
&\qquad + \frac{2\,(11+3\xi_r(8-9\xi_r))}{27}\,\tilde{L}_p - \frac{4(27+\xi_r(8 - 9\xi_r))}{9} + \zeta_2\,\left(7 + \frac{2\xi_r}{3} - \xi_r^2 \right)  \bigg)
+\Ord(\bar{\al}_r^3)\, .
\label{chap3:res2l:RQED4,3:Sr}
\end{flalign}

\subsection{Case of QED$_3$}

In the case of QED$_3$ we first set $\veps_e=0$ and define $\veps_\gamma = 1/2 + \delta_\gamma$ such that $d_\gamma = 3 - 2\delta_\gamma$ where the limit $\delta_\gamma \ra 0$ has to be taken.
From Eqs.~(\ref{chap3:res:Sigma2a}), (\ref{chap3:res:Sigma2b}) and Eq.~(\ref{chap3:res:Sigma2c-2}), this leads to:
\begin{subequations}
\label{chap3:res:QED3:Sigma2a+b+c}
\begin{flalign}
\Sigma_{2aV}(p^2;\bar{\al},\xi) &= \frac{N_F\, \al^2}{\pi\,(-p^2)}\,\left( \frac{\overline{\mu}^2}{-p^2} \right)^{2\delta_\gamma}\, \frac{(d_\gamma-2)^2}{2(d_\gamma-6)}\,e^{2 \gamma_E \delta_\gamma}\, 
G(3-2\delta_\gamma,1,1)G(3- 2\delta_\gamma,1,1/2+\delta_\gamma)
\nonum \\
&= \frac{N_F\, \al^2}{6\,(-p^2)}\, \left [ -\frac{1}{\delta_\gamma} + 2 L_p - \frac{4}{3} +  \Ord(\delta_\gamma) \right]\, ,
\label{chap3:res:QED3:Sigma2a} \\
\Sigma_{2bV}(p^2;\bar{\al},\xi) &= \frac{\xi^2 \al^2}{4\pi\,(-p^2)}\, \left( \frac{\overline{\mu}^2}{-p^2} \right)^{2\delta_\gamma}\, \frac{(d_\gamma-2)^2(d_\gamma-3)}{d_\gamma-4}\,
e^{2 \gamma_E \delta_\gamma} G(3-\delta_\gamma,1,1)G(3-\delta_\gamma,1,1/2+\delta_\gamma)
\nonum \\
&= \frac{\xi^2 \al^2}{2\,(-p^2)} \left [ 1 +  \Ord(\delta_\gamma) \right]\, ,
\label{chap3:res:QED3:Sigma2b} \\
\Sigma_{2cV}(p^2;\bar{\al},\xi) &= - \frac{\al^2}{4\pi\,(-p^2)}\, \left( \frac{\overline{\mu}^2}{-p^2} \right)^{2\delta_\gamma}\, \frac{d_\gamma-2}{4}\,e^{2 \gamma_E \delta_\gamma} \,
\bigg[ \bigg( d_\gamma -6 + (d_\gamma-2)\xi^2 \bigg)\,G^2(3-\delta_\gamma,1,1) \bigg .
\nonum \\
&\bigg . \qquad + 2\,\frac{d_\gamma-3}{d_\gamma-4}\,\bigg( d_\gamma + 4 - (d_\gamma-4)\xi^2 \bigg)\, G(3-\delta_\gamma,1,1)G(3-\delta_\gamma,1,1/2+\delta_\gamma) \bigg ]\,
\nonum \\
&= \frac{\al^2}{4\,(-p^2)} \left [ \frac{9}{2}\zeta_2 - 7 - \xi^2\,\left(1 + \frac{3}{2}\zeta_2 \right)  +  \Ord(\delta_\gamma) \right]\, ,
\label{chap3:res:QED3:Sigma2c}
\end{flalign}
\end{subequations}
which is valid for an arbitrary gauge fixing parameter $\xi$ and where only the first diagram is divergent. Recall from the discussion below Eqs.~(\ref{chap3:res2l:QED3:Pi2a+Pi2c}) 
that this singularity is of IR nature and arises from the master integral $G(3- 2\delta_\gamma,1,1/2+\delta_\gamma)$. It does not affect diagrams b and c because of the factors $d_\gamma -3$.
Combining Eqs.~(\ref{chap3:res:QED3:Sigma2a+b+c}) with Eqs.~(\ref{chap3:def:Zpsi-forest}),
the individual counter-terms read:
\begin{subequations}
\label{chap3:res:QED3:Z2a+b+c-psi}
\begin{flalign}
\delta Z_{2a\,\psi}(\bar{\al}_r) &= \mathcal{K}\, \bigg[ \Sigma_{2aV}(p^2;\bar{\al}_r) \bigg] -
\mathcal{K}\, \bigg[\underbrace{\mathcal{K}\, \bigg[\Pi_{1}(q^2;\bar{\al}_r,\xi_r) \bigg]}_{=0}\,\underbrace{\Sigma_{1V}^{(\perp)}(p^2;\bar{\al}_r,\xi_r)}_{=0} \bigg]
\nonum \\
&= -\frac{N_F\,\al_r^2}{6 (-p^2)\,\delta_\gamma}\, ,
\label{chap3:res:QED3:Z2apsi} \\
\delta Z_{2b\,\psi}(\bar{\al}_r,\xi_r) &= \underbrace{\mathcal{K}\, \bigg[ \Sigma_{2bV}(p^2;\bar{\al}_r,\xi_r) \bigg]}_{=0} - 
\mathcal{K}\, \bigg[ \underbrace{\mathcal{K}\, \bigg[\Sigma_{1V}(p^2;\bar{\al}_r,\xi_r) \bigg]}_{=0}\,\Sigma_{1V}(p^2;\bar{\al}_r,\xi_r) \bigg]
\nonum \\
&= 0\, ,
\label{chap3:res:QED3:Z2bpsi} \\
\delta Z_{2c\,\psi}(\bar{\al}_r,\xi_r) &= \underbrace{\mathcal{K}\, \bigg[ \Sigma_{2cV}(p^2;\bar{\al}_r,\xi_r) \bigg]}_{=0} + 
2\, \mathcal{K}\, \bigg[ \underbrace{\mathcal{K}\, \bigg[\Sigma_{1V}(p^2;\bar{\al}_r,\xi_r) \bigg]}_{=0}\,\Sigma_{1V}(p^2;\bar{\al}_r,\xi_r) \bigg]
\nonum \\
&= 0\, ,
\label{chap3:res:QED3:Z2cpsi}
\end{flalign}
\end{subequations}
where all one-loop diagrams are finite in QED$_3$. The sum of Eqs.~(\ref{chap3:res:QED3:Z2a+b+c-psi}) yields the total counterterm at two-loop order:
\be
\delta Z_{2\,\psi} =  -\frac{N_F\, \al_r^2}{6(-p^2)\,\delta_\gamma} \, .
\label{chap3:res2l:QED3:Z2psi}
\ee
Notice that, because the theory is super-renormalizable, the counterterm depends on momentum through the dimensionless combination $\al_r / \sqrt{-p^2}$.

The individual renormalized diagrams are also straightforward to compute and read:
\begin{subequations}
\label{chap3:res:QED3:Sigma2ra+b+c}
\begin{flalign}
\Sigma_{2aVr}(p^2;\bar{\al}_r) &= \Sigma_{2aV}(p^2;\bar{\al}_r) - \underbrace{\mathcal{K}\, \bigg[\Pi_{1}(q^2;\bar{\al}_r,\xi_r) \bigg]}_{=0}\,\underbrace{\Sigma_{1V}^{(\perp)}(p^2;\bar{\al}_r,\xi_r)}_{=0}
- \delta Z_{2a\,\psi}(\bar{\al}_r)
\nonum \\
&= \frac{N_F\,\al_r^2}{3(-p^2)}\,\left [ L_p - \frac{2}{3} \right]\, ,
\label{chap3:res:QED3:Sigma2ar} \\
\Sigma_{2bVr}(p^2;\bar{\al}_r,\xi_r) &= \Sigma_{2bV}(p^2;\bar{\al}_r,\xi_r) - \underbrace{\mathcal{K} \bigg[ \Sigma_{1V}(p^2;\bar{\al}_r,\xi_r) \bigg]}_{=0}~\Sigma_{1V}(p^2;\bar{\al}_r,\xi_r) - 
\underbrace{\delta Z_{2b\,\psi}(\bar{\al}_r,\xi_r)}_{=0}
\nonum \\
&= \frac{\xi_r^2\,\al_r^2}{2(-p^2)}\, ,
\label{chap3:res:QED3:Sigma2br} \\
\Sigma_{2cVr}(p^2;\bar{\al}_r,\xi_r) &= \Sigma_{2cVr}(p^2;\bar{\al}_r,\xi_r) + 2 \underbrace{\mathcal{K} \bigg[ \Sigma_{1V}(p^2;\bar{\al}_r,\xi_r) \bigg]}_{=0}~\Sigma_{1V}(p^2;\bar{\al}_r,\xi_r) 
- \underbrace{\delta Z_{2c\,\psi}(\bar{\al}_r,\xi_r)}_{=0}
\nonum \\
&= \frac{\al^2}{4\,(-p^2)} \left [ \frac{9}{2}\zeta_2 - 7 - \xi^2\,\left(1 + \frac{3}{2}\zeta_2 \right) \right]\, .
\label{chap3:res:QED3:Sigma2cr}
\end{flalign}
\end{subequations}
The sum of Eqs.~(\ref{chap3:res:QED3:Sigma2ra+b+c}) yields the total two-loop renormalized fermion self-energy:
\bea
\Sigma_{2Vr}(p^2;\bar{\al}_r,\xi_r) = \frac{\al_r^2}{4\,(-p^2)}\, \bigg[ \frac{4N_F}{9}\, \left ( 3L_p - 2 \right) + \frac{9}{2}\,\zeta_2 - 7 + \xi_r^2\,\left(1 - \frac{3}{2}\zeta_2 \right) \bigg]\, .
\label{chap3:res2l:QED3:Sigma2r}
\eea

Combining the above two-loop results with the one-loop ones derived in the previous paragraphs, we deduce the anomalous dimension up to two loops:~\footnote{The reader should be warned
that, in the case of QED$_3$, all poles are infra-red ones, \eg, the theory is badly IR divergent as was discussed previously above Eq.~(\ref{chap3:res2l:RQED3:Pi2}). 
The notion of an anomalous dimension shoud be then considered with great care in this case.}
\be
\gamma_{\psi}(\bar{\al}_r,\xi_r) = - \frac{2 N_F\,\al_r^2}{3\,(-p^2)} + \Ord(\al_r^3) = - \frac{N_F\,e_r^4}{24\pi^2\,(-p^2)} + \Ord(e^6)\, ,
\label{chap3:res2l:QED3:gammapsi}
\ee
where only the two-loop contribution is non zero and it is also gauge-independent, as well as the expansion of the renormalized fermion propagator up to two loops:
\begin{flalign}
-\I \Sp S_r(p) &= 1 - \frac{\pi \xi_r \al_r}{4 \sqrt{-p^2}} + \frac{\al_r^2}{4 (-p^2)}\,\bigg[ \frac{4N_F}{9}\, \left ( 3L_p - 2 \right) + \frac{9}{2}\,\zeta_2 - 7 + \xi_r^2 \bigg]
+\Ord(\al_r^3)
\nonum \\
&= 1 - \frac{\xi_r e^2}{16 \sqrt{-p^2}} + \frac{e^4}{64 \pi^2 (-p^2)}\,\bigg[ \frac{4N_F}{9}\, \left ( 3L_p - 2 \right) + \frac{9}{2}\,\zeta_2 - 7 + \xi_r^2 \bigg]
+\Ord(e^6)\, .
\label{chap3:res2l:QED3:Sr}
\end{flalign}

\subsection{Case of reduced QED$_{4,d_e}$}

For the fermion self-energy, it is possible to derive general formulas for reduced QED$_{4,d_e}$. These formulas are very instructive as they reveal the general perturbative structure of these models.
There are however some limitations in there applications. We shall therefore not proceed in a detailed (diagram by diagram) analysis but rather focus on the final formulas derived in Ref.~\cite{Kotikov:2013eha}  
and discuss them. For calculational purposes, let's note that all formulas can be obtained with the help of the following general expansions (in the limit $\veps_\gamma \ra 0$) of the master integrals \cite{Kotikov:2013eha}: 
\begin{subequations}
\label{chap3:RQED4,de:exp:Gfuncs}
\begin{flalign}
&G(4-2\veps_e-2\veps_\gamma,1,1-\veps_e)=\frac{\exp[-\gamma_E\varepsilon_\gamma]}{\varepsilon_\gamma(1-\veps_e-2\varepsilon_\gamma)}
\frac{1}{\Gamma(1-\veps_e)}
\biggl(1+\overline{\Psi}_1\varepsilon_\gamma + \frac{\varepsilon_\gamma^2}{2}
\Bigl(\overline{\Psi}_1^2+2\zeta_2 -3\Psi_2(1-\veps_e)\Bigr)\biggr)\, , 
\label{chap3:RQED4,de:exp:G1,1-ee}  \\
&G(4-2\veps_e-2\veps_\gamma,1,\varepsilon_\gamma-\veps_e)=
-\frac{\exp[-\gamma_E\varepsilon_\gamma]}{2\varepsilon_\gamma}
\frac{\varepsilon_\gamma-\veps_e}{(1-\veps_e-3\varepsilon_\gamma)(2-\veps_e-3\varepsilon_\gamma)}
\frac{1}{\Gamma(1-\veps_e)}
\times
\nonum \\
&\qquad \times \biggl(1+\overline{\Psi}_1\varepsilon_\gamma + \frac{\varepsilon_\gamma^2}{2}
\Bigl(\overline{\Psi}_1^2+8\zeta_2 -9\Psi_2(1-\veps_e)\Bigr)\biggr) \, ,
\label{chap3:RQED4,de:exp:G1,eg-ee}  \\
&G(4-2\veps_e-2\veps_\gamma,1-\veps_e,\varepsilon_\gamma)=
-\frac{\exp[-\gamma_E\varepsilon_\gamma]}{2(1-2\varepsilon_\gamma)}
\frac{1-\veps_e-2\varepsilon_\gamma}{(1-\veps_e-3\varepsilon_\gamma)(2-\veps_e-3\varepsilon_\gamma)}
\frac{1}{\Gamma(1-\veps_e)}
\times
\nonum \\
&\qquad \times \biggl(1+\overline{\Psi}_1\varepsilon_\gamma + \frac{\varepsilon_\gamma^2}{2}
\Bigl(\overline{\Psi}_1^2+4\zeta_2 -5\Psi_2(1-\veps_e)\Bigr)\biggr)\, , 
\label{chap3:RQED4,de:exp:G1-ee,eg}  \\
&G(4-2\veps_e-2\veps_\gamma,1,1)=\exp[-\gamma_E\varepsilon_\gamma]
\frac{K_1}{\Gamma(1-\veps_e)}\times
\nonum \\
&\qquad \times \biggl(1+\overline{\Psi}_2\varepsilon_\gamma + \frac{\varepsilon_\gamma^2}{2}
\Bigl(\overline{\Psi}_2^2+\Psi_2(\veps_e)+2\Psi_2(1-\veps_e)-4\Psi_2(2-2\veps_e)\Bigr) \biggr),
%\nonum
\label{chap3:RQED4,de:exp:G11}
\end{flalign}
\end{subequations}
where 
\begin{subequations}
\label{chap3:RQED4,de:params}
\begin{flalign}
&\overline{\Psi}_1=\Psi_1(1-\veps_e)-\Psi_1(1), \qquad
\overline{\Psi}_2=\Psi_1(\veps_e)-2\Psi_1(1-\veps_e)+2\Psi_1(2-2\veps_e) -\Psi_1(1)\, ,
\label{chap3:RQED4,de:params:Psis} \\
&K_1 = \frac{\Gamma^3(1-\veps_e)\Gamma(\veps_e)}{\Gamma(2-2\veps_e)}, \qquad \bar{L}_p = L_p - \Psi_1(2-\veps_e)+ \Psi_1(1)\, ,
\label{chap3:RQED4,de:params:Lp+K1}
\end{flalign}
\end{subequations}
and, anticipating the formulas below, the expression of $\bar{L}_p$ is repeated for clarity (it already appeared in Eq.~(\ref{chap3:RQED4,de:Sigma1})). 

Using the above expansions in combination with Eqs.~(\ref{chap3:res:Sigma2a}), (\ref{chap3:res:Sigma2b}) and (\ref{chap3:res:Sigma2c-2}) and after tedious algebra, 
the total counterterm at two-loop order reads:
\be
\delta Z_{2\,\psi} =  \frac{\bar{\al}_r^2}{2\veps_\gamma^2}\,\left(\xi_r-\frac{\veps_e}{2-\veps_e}\right)^2 
- \frac{2 \bar{\al}_r^2}{\veps_\gamma}\,\left( N_F K_1\,\frac{\veps_e}{2-\veps_e} -\frac{(3-2\veps_e)\,[1-\veps_e\,(3-\veps_e)]}{(1-\veps_e)(2-\veps_e)^3} \right) \, .
\label{chap3:res2l:RQED4,de:Z2psi}
\ee
Combining this result with the corresponding one-loop counterterm, we deduce the anomalous dimension up to two loops:
\be
\gamma_{\psi}(\bar{\al}_r,\xi_r) =  2\left(\frac{\veps_e}{2-\veps_e} - \xi_r \right)\,\bar{\al}_r
-8\,\left( N_F K_1 \frac{\veps_e}{2-\veps_e}
- \frac{(3-2\veps_e)\,[1-\veps_e\,(3-\veps_e)]}{(1-\veps_e)(2-\veps_e)^3} \right)\,\bar{\al}_r^2 + \Ord(\bar{\al}_r^3)\, ,
\label{chap3:res2l:RQED4,de:gammapsi}
\ee
where the two-loop contribution is gauge-independent. This anomalous dimension does not depend on the external momentum and digamma functions.
Such a property is an extension of the standard rule \cite{Vladimirov:1979zm} (discussed in Chap.~\ref{chap2}) for results related to anomalous dimensions in standard
QFT ($\veps_e=0$ and $\veps_\gamma \to 0$): they should not depend on the external momenta and (in $\overline{MS}$-scheme) on $\gamma_E$ and $\zeta_2$. 
Notice that for RQED$_{4,3}$ there is a contribution $\zeta_2$ (see Eq.~(\ref{chap3:res2l:RQED4,3:gammapsi}) above) but, as is now clear from the general case, this contribution appears 
not from the expansion of $\Gamma$-functions in  $\veps_\gamma$ but from the factor $K_1$ in the expression $G(d_e,1,1)$, see Eq.~(\ref{chap3:RQED4,de:exp:G11}).

%The individual renormalized diagrams are also straightforward to compute and read:
%%
%\begin{subequations}
%\label{chap3:res:RQED4,de:Sigma2ra+b+c}
%\begin{flalign}
%&\Sigma_{2aVr}(p^2;\bar{\al}_r) = \Sigma_{2aV}(p^2;\bar{\al}_r) - \mathcal{K}\, \bigg(\Pi_{1}(q^2;\bar{\al}_r,\xi_r) \bigg)\,\Sigma_{1V}^{(\perp)}(p^2;\bar{\al}_r,\xi_r)
%- \delta Z_{2a\,\psi}(\bar{\al}_r)
%\nonum \\
%&= ...\, ,
%\label{chap3:res:RQED4,de:Sigma2ar} \\
%&\Sigma_{2bVr}(p^2;\bar{\al}_r,\xi_r) = \Sigma_{2bV}(p^2;\bar{\al}_r,\xi_r) - \mathcal{K} \bigg( \Sigma_{1V}(p^2;\bar{\al}_r,\xi_r) \bigg)~\Sigma_{1V}(p^2;\bar{\al}_r,\xi_r) - \delta Z_{2b\,\psi}(\bar{\al}_r,\xi_r)
%\nonum \\
%&= ...\, , 
%\label{chap3:res:RQED4,de:Sigma2br} \\
%&\Sigma_{2cVr}(p^2;\bar{\al}_r,\xi_r) = \Sigma_{2cVr}(p^2;\bar{\al}_r,\xi_r) + 2\mathcal{K} \bigg( \Sigma_{1V}(p^2;\bar{\al}_r,\xi_r) \bigg)~\Sigma_{1V}(p^2;\bar{\al}_r,\xi_r) - \delta Z_{2c\,\psi}(\bar{\al}_r,\xi_r)
%\nonum \\
%&= ... \, .
%\label{chap3:res:RQED4,de:Sigma2cr}
%\end{flalign}
%\end{subequations}
%%
%The sum of Eqs.~(\ref{chap3:res:RQED4,de:Sigma2ra+b+c}) yields the total two-loop renormalized fermion self-energy:
%%
%\bea
%\Sigma_{2Vr}(p^2;\bar{\al}_r,\xi_r) =&& \bar{\al}_r^2 \bigg[ ... \bigg]\, .
%\label{chap3:res2l:RQED4,de:Sigma2r}
%\eea
%%

The total two-loop renormalized fermion self-energy can be obtained in a similar way. Combining it with the one-loop 
result derived in the previous paragraphs, yields the expansion of the renormalized fermion propagator up to two loops:
\begin{flalign}
&-\I \Sp S_r(p) = 1 
+\bar{\al}_r \Biggl( \left(\xi_r-\frac{\veps_e}{2-\veps_e}\right)\,\bar{L}_p + \frac{2\veps_e}{(2-\veps_e)^2} \Biggr) 
+\bar{\al}_r^2\,\Biggl[2\xi_r\,\frac{\veps_e}{(2-\veps_e)^2}\,\bar{L}_p + \Biggr .
\nonum \\
&+\frac{1}{2}\,\left(\xi_r-\frac{\veps_e}{2-\veps_e}\right)^2\,\biggl( \bar{L}_p^2+\zeta_2-\Psi_2(2-\veps_e) \biggr)
+ 4\,\frac{3-2\veps_e}{(2-\veps_e)^2}\,\bigg(\Psi_2(1-\veps_e) - \zeta_2 \bigg) - \Biggr .
\nonum \\
&- 2\bar{L}_p\,\frac{6-22\veps_e+19\veps_e^2-5\veps_e^3}{(1-\veps_e)(2-\veps_e)^3}
-2\,\frac{7+23\veps_e-30\veps_e^2+8\veps_e^3}{(1-\veps_e)(2-\veps_e)^4}-2\,\frac{\veps_e^2}{(1-\veps_e)^2(2-\veps_e)^2} + 
\nonumber \\
&\Biggl. + 4N_F K_1\,\frac{\veps_e}{2-\veps_e}\,\left( \bar{L}_p+\frac{1}{2}\,(\overline{\Psi}_1 - \overline{\Psi}_2)
+\frac{1}{\veps_e(1-\veps_e)(2-\veps_e)}-\frac{1}{2-\veps_e} \right) \Biggr ]
+\Ord(\bar{\al}_r^3)\, .
\label{chap3:res2l:RQED4,de:Sr}
\end{flalign}

As can be seen from the above expressions, the Laurent series in $\veps_\gamma$ is plagued by singularities for integer values of $\veps_e$. The case $\veps_e=2$ is unphysical ($d_e=0$). 
The case $\veps_e=3/2$ (case of a $0$-brane or RQED$_{4,1}$) corresponds to the quantum mechanics of a point particle coupled to a  $3+1$-dimensional electromagnetic environment.
The singularity at $\veps_e=1$ (case of a $1$-brane or RQED$_{4,2}$) appears starting from two-loop and requires some additional regularization that we shall not pursue here.
Finally, at two-loop, a singularity appears for $\veps_e=0$ (case of QED$_4$) as can be seen
in particular from the last term in Eq.~(\ref{chap3:res2l:RQED4,de:Sr}).
This singularity is gauge invariant and can be traced back to the UV behaviour of the one-loop polarization operator entering the bubble diagram
which is contained in the master integral $G(d_e,1,1)$, see Eq.~(\ref{chap3:RQED4,de:exp:G11}) and in particular the expression of $K_1$ which contains a $1/\veps_e$ pole, Eq.~(\ref{chap3:RQED4,de:params:Lp+K1}). 
The above results are therefore valid essentially in application to RQED$_{4,3}$ (case of a $2$-brane) for which $\veps_e=1/2$ and which is of main interest to us. 
From the general formulas of this paragraph one may indeed recover all the results previously derived in the case of QED$_{4,3}$.

\subsection{Case of reduced QED$_{3,2}$}

In all the cases considered up to now, the complicated master integral $G(d_e,1-\veps_e,1,1-\veps_e,1,1)$ did not contribute. As discussed above, in the case $\veps_e=0$ (QED$_4$ and QED$_3$), this integral
reduces to the well known $G(d_\gamma,1,1,1,1,1)$ which is easy to evaluate. On the other hand, in the case of QED$_{4,3}$, the non-trivial convergent integral $G(3-2\veps_\gamma,1/2,1,1/2,1,1)$ appears
with a factor $\veps_\gamma$ and therefore does not contribute up to $\Ord(1)$ which is enough for two-loop renormalization. At two-loop level, there are other cases which may not be so attractive from the physics point of view 
but are interesting to consider from the field theory point of view as they require computing the non-trivial integral. An example is given by reduced 
QED$_{3,2}$ ($\veps_\gamma = 1/2 +\delta_\gamma$ with $\delta_\gamma \ra 0$ and $\veps_e = 1/2$, so that $d_\gamma = 3 -2\delta_\gamma$ and $d_e=2-2\delta_\gamma$)  
which corresponds to a fermion in $1+1$-dimensions embedded in a $2+1$-dimensional electromagnetic environment. This example has been considered in Ref.~\cite{Teber:2014hna} and we shall summarize the results below. 
Notice that similarly to QED$_3$ ($\veps_e = 0$ and $\delta_\gamma \ra 0$), RQED$_{3,2}$ is super-renormalizable and therefore asymptotically free.
However, contrarily to QED$_3$ where infra-red (IR) divergences yield an anomalous dimension to the fermion field at two loop, RQED$_{3,2}$
is finite at two-loop and the corrections to the fermion propagator take a very simple form, as will be shown below.

The computation of the fermion self-energy is standard and follows from applying Eqs. (\ref{chap3:res:Sigma2a}), (\ref{chap3:res:Sigma2b}) and (\ref{chap3:res:Sigma2c-2}).
In the last formula, the difficulty is in evaluating $G(2-2\delta_\gamma,1/2,1,1/2,1,1)$. This can be done using the general formulas derived in Ref.~\cite{Kotikov:2013eha} and in particular
Eq.~(\ref{chap2:res:Gab}) that was advertised in Chap.~\ref{chap2}. In expanded form, the result reads:
\begin{flalign}
&G(2-2\delta_\gamma,1/2,1,1/2,1,1) =
e^{-2\gamma_E \delta_\gamma}\,\bigg(\frac{1}{\delta_\gamma^2} + \frac{8(1-\ln 2)}{\delta_\gamma} - 56 - 5 \zeta_2 - 64 \ln 2 + 32 \ln^2 2  \bigg .
\nonum \\
& \quad \bigg . + \delta_\gamma\, \Bigl( 240 - 40 \zeta_2 - \frac{110}{3}\,\zeta_3 - 384 \,G - 128 \ln 2 + 40\, \zeta_2 \ln 2 + 256 \ln^2 2 - \frac{256}{3} \ln^3 2 \Bigr)
+ O(\delta_\gamma^2) \bigg)\, ,
\label{chap3:G-RQED32}
\end{flalign}
where $G$ is Catalan's constant. The total two-loop self-energy then reduces to:
\begin{flalign}
&\Sigma_{2V}(p^2) = -\frac{\al^2}{(-p^2)}\,\Bigg( N_F + \frac{(1-\xi)^2}{4} - 
%- \frac{\delta_\gamma}{2}\,\Bigl( 2N_F(1 + 2\bar{L}_p - 8 \ln 2) + 25 + 10 a -3\xi^2 -32\,G + (1-\xi)^2 \bar{L}_p \Bigr)
\Bigg .
\\
&\Bigg . \qquad - \frac{\delta_\gamma}{2}\,\Bigl( 2N_F(1 + 2\bar{L}_p - 8 \ln 2) + 25 + 10 a -3\xi^2 -32\,G + (1-\xi)^2 \bar{L}_p \Bigr) +  O(\delta_\gamma^2)\Bigg)\, ,
\nonum
\end{flalign}
where now $\bar{L}_p = L_p + 4 \ln 2$. As anticipated, the self-energy is finite and there is therefore no wave-function renormalization ($Z_\psi = 1$).
Adding the one-loop contribution, the expansion of the renormalized fermion propagator up to two loops reads:
\begin{flalign}
&-i {\slashed p}\,S_r(p) = 1 + \frac{\al}{2\,\sqrt{-p^2}}\,\bigg( 1-\xi +\delta_\gamma \big(4 - (1-\xi)\bar{L}_p \big) + O(\delta_\gamma^2) \bigg)
\\
&\quad +  \frac{\al^2}{(-p^2)} \,\bigg(-N_F + \delta_\gamma\,\Bigl( N_F(1 + 2 \bar{L}_p - 8 \ln 2 ) + 16(1-G)-\frac{3}{2}\,(1-\xi)^2  \Bigr) + O(\delta_\gamma^2) \bigg) + O(\tilde{\al}^3)\, .
\nonum
\end{flalign}
The $O(1)$ two-loop correction is again gauge-invariant and reduces to a very simple form proportional to $N_F$ while the $O(\delta_\gamma)$
correction involves $\pi$, $\ln 2$ as well as the Clausen function $\text{Cl}_2(\pi/2)=G$.

\section{Conclusion}

To conclude this Chapter, we have provided a detailed overview of the analysis of the perturbative structure of reduced QED$_{d_\gamma,d_e}$ up to two loops for the polarisation operator
and the fermion self-energy. This provides a detailed analysis of the general model Eq.~(\ref{chap1:model-general}) in the ultra-relativistic limit $x=v/c \ra 1$.
The main focus was on reduced QED$_{4,3}$ (graphene at the IR Lorentz invariant fixed point) which is somehow intermediate between QED$_4$ and QED$_3$ but our general expressions may be applied to any model. 
They include Eqs.~(\ref{chap3:res:Pi1}) and (\ref{chap3:res:Sigma1}) for the one-loop polarization operator and fermion self-energy, respectively, Eqs.~(\ref{chap3:res:Pi2a}) and (\ref{chap3:res:Pi2c})
for the two-loop polarization operator and Eqs.~(\ref{chap3:res:Sigma2a}), (\ref{chap3:res:Sigma2b}) and (\ref{chap3:res:Sigma2c-2}) for the two-loop fermion self-energy.
Some results related to this model  will be used in the next chapters. The main two are the following:
\begin{itemize}
\item the fermion anomalous dimension of reduced QED$_{4,3}$ up to two loops, Eq.~(\ref{chap3:res2l:RQED4,3:gammapsi}), that we reproduce here for clarity:
\be
\gamma_{\psi}(\bar{\al}_r,\xi_r) = 2 \bar{\al}_r\,\frac{1-3\xi_r}{3} -16\,\left( \zeta_2 N_F + \frac{4}{27} \right)\, \bar{\al}_r^2 + \Ord(\bar{\al}_r^3)\, ,
\label{chap3:res2l:RQED4,3:gammapsi-2}
\ee
will play an important role in studying the critical properties of reduced QED$_{4,3}$ and in particular the dynamical generation of a mass in Chap.~\ref{chap4}.

\item the renormalized polarization operator of reduced QED$_{4,3}$ up to two loops, Eq.~(\ref{chap3:res2l:RQED4,3:Pi-total}) that we also reproduce for clarity:
\be
\Pi_{r}(q^2) = \Pi_{1r}(q^2)\,\left( 1 + \al_r\,\mathcal{C}^* + \Ord(\al_r^2) \right)\, , \qquad \mathcal{C}^* = \frac{92-9\pi^2}{18\pi}\, .
\label{chap3:res2l:RQED4,3:Pi-total-2}
\ee
As will be discussed in Chap.~\ref{chap5}, this operator is related to the optical conductivity of the system and  the term proportional to the interaction correction coefficient $\mathcal{C}^*$
describes (small) deviations of this conductivity with respect to the case of free fermions. This result will also play an important role in the study of dynamical symmetry breaking in QED$_3$
in Chap.~\ref{chap4}. 
%Anticipating the next chapter, a rather surprising fact is that a coefficient similar to $\mathcal{C}^*$ actually appeared previously in a different context: the work of Gusynin et al.\ \cite{Gusynin:2000zb}
%(and, although not explicitely, in even earlier works of Gracey \cite{Gracey:1993sn} and Kotikov \cite{Kotikov:1993wr}) on $1/N$ expansion of QED$_3$. 
\end{itemize}

Anticipating the next Chapter, a rather surprising fact is that a coefficient similar to $\mathcal{C}^*$ actually appeared previously in a different context: the work of Gusynin et al.\ \cite{Gusynin:2000zb}
(and, although not explicitly, in even earlier works of Gracey \cite{Gracey:1993sn} and Kotikov \cite{Kotikov:1993wr}) on $1/N$ expansion of QED$_3$.
As will be explained in more details in the next Chapter, this similarity comes from the corresponding similarity of the infra-red effective photon propagators of QED$_{4,3}$ and QED$_3$. 
We will then also prove that there is a correspondence between our result for the 2-loop fermion anomalous dimension of reduced QED$_{4,3}$, Eq.~(\ref{chap3:res2l:RQED4,3:gammapsi-2}), 
and the NLO fermion anomalous dimension of QED$_{3}$ derived by Gracey \cite{Gracey:1993sn}.

%Despite the appearence of non-trivial master integrals (with respect to usual QFTs), all results so far have been derived analytically ``by hand''. 
%As we will be clear after the next two chapters, there is interest in going beyond two loop order for some specific problems. 
%In this case, one will clearly need the help of computer algebra systems. We shall come back on this in the main conclusion.

\end{fmffile}

\cleardoublepage

%% file: Chapter4/dchisb.tex
\label{chap4}

\begin{fmffile}{fmf-chap4}

\fmfcmd{%
vardef cross_bar (expr p, len, ang) =
((-len/2,0)--(len/2,0))
rotated (ang + angle direction length(p)/2 of p)
shifted point length(p)/2 of p
enddef;
style_def crossed expr p =
cdraw p;
ccutdraw cross_bar (p, 5mm,  45);
ccutdraw cross_bar (p, 5mm, -45)
enddef;}

In the previous Chapter, we have focused on perturbative calculations exploring the weak-coupling structure of various types of QEDs with a special focus on reduced QED$_{4,3}$.
On the basis of these results, we move on in this Chapter to the study of strong coupling phenomena which require going beyond finite orders. 
The focus will be on dynamical chiral symmetry breaking and associated dynamical mass (or gap in the condensed matter context) generation in QED$_3$ and QED$_{4,3}$. 
As a starting point, we will consider QED$_3$ and solve the Schwinger-Dyson equations for the fermion propagator at next-to-leading order in the $1/N$-expansion. Then, we will present a mapping from 
large-$N$ QED$_3$ to reduced QED$_{4,3}$ from which results for QED$_{4,3}$ will be rather straightforwardly derived.

\section{Critical properties of QED$_3$} 

Quantum Electrodynamics in $2+1$ dimensions (QED$_3$) has been extensively studied during more than three decades now.
As reviewed in the Introduction, originally, the interest in QED$_3$ came from its similarities to ($3+1$)-dimensional QCD and the fact that phenomena such as dynamical
chiral symmetry breaking (D$\chi$SB) and mass generation may be studied systematically in such a toy model \cite{Pisarski:1984dj,Appelquist:1985vf,Appelquist:1986fd,Appelquist:1986qw,Appelquist:1988sr}.
Later, a strong interest in QED$_3$ arose in connection with planar condensed matter physics systems
having relativistic-like low-energy excitations \cite{Marston:1989zz,Ioffe:1989zz,PhysRev.71.622,Semenoff:1984dq}. 
The study of a dynamically generated gap in the fermion spectrum of, \eg, planar Dirac liquids such as graphene,
has now become an active area of research and we shall come back on this in Sec.~\ref{chap4:sec:rqed}.
In all cases, the understanding of the phase structure of QED$_3$ is a crucial pre-requisite to understand
non-perturbative dynamic phenomena in more realistic particle and condensed matter physics models.

Three-dimensional QED is described by model III, Eq.~(\ref{chap1:rqed}), in the case where $d_\gamma = d_e = d = 3$. The corresponding Lagrangian density reads:
\bea
L_{\text{QED}_3} = \bar{\psi}_\sigma \left (\I \slashed \partial  - e \slashed A \right )\psi^\sigma - \frac{1}{4}\,F_{\mu \nu}^2 \qquad (\sigma =1,\cdots,N)\, ,
\label{chap4:qed3}
\eea
where, as before, $\psi_\sigma$ is a four-component spinor and the implicit summation over the flavour index $\sigma$ runs from $1$ to $N$ (equivalently this corresponds to 
$2N$ two-component spinors $\chi_i$, $i=1,...,2N$).~\footnote{In order to simplify notations, the fermion flavour number that was denoted by $N_F$ in previous chapters will be noted $N$ here.} 
Due to the reduced dimensionality of the system ($d=3$) the model of Eq.~(\ref{chap4:qed3}) has a global global $U(2N)$ flavour symmetry
sometimes refereed to as a ``chiral'' symmetry.  A (parity-invariant) fermion mass term, $m \bar{\psi} \psi$, breaks this symmetry to $U(N) \times U(N)$ (the case of a parity
non-invariant mass will not be considered here). As we saw in Chap.~\ref{chap3}, in the massless case, loop expansions are plagued by infrared divergences (starting at two loops).
The latter soften upon analyzing the model in a $1/N$-expansion \cite{Appelquist:1981vg,Appelquist:1981sf,Jackiw:1980kv}.
Since the theory is super-renormalizable, the mass scale is then given by the dimensionful coupling constant: $a = N e^2/8$, which is kept fixed as $N \rightarrow \infty$. 
Early studies of this model \cite{Pisarski:1984dj,Appelquist:1988sr} suggested that the physics is rapidly damped at momentum scales $p \gg a$ 
and that a (parity-invariant) fermion mass term breaking the flavour symmetry is dynamically generated at scales which are orders of magnitude smaller than 
the intrinsic scale $a$. Since then, D$\chi$SB in QED$_3$ and the dependence of the dynamical fermion mass on $N$ 
have been the subject of extensive studies, see, {\it e.g.}, 
\cite{Nash:1989xx,Pennington:1990bx,Curtis:1992gm,Pisarski:1991kg,Atkinson:1989fp,Karthik:2015sgq,Karthik:2016ppr,Dagotto:1988id,Dagotto:1989td,Hands:2004bh,%
Strouthos:2008kc,Azcoiti:1993fb,Azcoiti:1995mi,Appelquist:2004ib,Raya:2013ina,Appelquist:1999hr,Giombi:2015haa,Giombi:2016fct,Braun:2014wja,Janssen:2016nrm,Kotikov:1993wr,Kotikov:2011kg,Gusynin:1995bb,Bashir:2005wt,%
Bashir:2009fv,Kubota:2001kk,DiPietro:2015taa,Herbut:2016ide,Gusynin:2016som}.

A central issue is related to the value of the critical fermion number, $N_c$, which is such that D$\chi$SB takes place only for $N<N_c$. 
It turns out that the values that can be found in the literature vary from $N_c \ra \infty$ 
\cite{Pisarski:1984dj,Pennington:1990bx,Curtis:1992gm,Pisarski:1991kg,Azcoiti:1993fb,Azcoiti:1995mi} 
corresponding to D$\chi$SB for all values of $N$, all the way to $N_c \ra 0$ in the case where no sign of D$\chi$SB is found \cite{Atkinson:1989fp,Karthik:2015sgq,Karthik:2016ppr}, see Tab.~\ref{chap4:tab:Nc-values} for a summary.
Recent works based on conformal field theory techniques tend to narrow this range but the upper bound found for $N_c$ still varies: $N_c <3/2$ \cite{Appelquist:1999hr} and recently
$N_c < 4.4$ \cite{Giombi:2015haa}. It seems that the most recent and accurate estimate obtained with these techniques is: $N_c =1 + \sqrt{2} \approx 2.41$ \cite{Giombi:2016fct}. 
Other works attempt to localize the transition with the criterion that parity- and $U(2N)$-invariant
four-fermion interactions should become relevant at $N_c$. This led to $3<N_c<4$ in the early study of \cite{Kubota:2001kk}, to $N_c=9/4<2.25$ \cite{DiPietro:2015taa} and $N_c \approx 2.89$ \cite{Herbut:2016ide}
very recently. Let's note also that some recent works suggest that there might be two different critical flavor
numbers \cite{Braun:2014wja,Janssen:2016nrm}: $N_c$ and $N_c^{\text{conf}}$ which are such that $N_c < N_c^{\text{conf}}$. The third intermediate phase, if it exists and for which the flavour number is in the range:
$N_c < N < N_c^{\text{conf}}$, would be characterized by spontaneous breaking of Lorentz symmetry. The predicted critical numbers read: $N_c \approx 4$ \cite{Braun:2014wja}, $N_c \leq 4.422$ \cite{Janssen:2016nrm} 
and $4.1<N_c^{\text{conf}}<10.0$ \cite{Braun:2014wja}, $N_c^{\text{conf}} \approx 6.24$ \cite{Janssen:2016nrm}. As far as we understand, all these
works are restricted to a leading order (LO) computation of $N_c$ and the question of the stability of all these approaches upon going to higher orders remains open.

\begin{center}
\renewcommand{\tabcolsep}{0.25cm}
\renewcommand{\arraystretch}{1.5}
\begin{table}
    \begin{tabular}{  c | c | c }
      \hline
             $N_c$                                         &       {\bf Method}                                         &       {\bf Year} \\
      \hline \hline
          $\infty$                       		   &     Schwinger-Dyson (LO)                            	&       1984 \cite{Pisarski:1984dj}    \\
      \hline
          $\infty$                                         &     Schwinger-Dyson (non-perturbative, Landau gauge)       &       1990, 1992 \cite{Pennington:1990bx,Curtis:1992gm} \\
      \hline
	  $\infty$					   &     RG study						&	1991 \cite{Pisarski:1991kg}	\\
      \hline
          $\infty$                     			   &     lattice simulations		                        &       1993, 1996 \cite{Azcoiti:1993fb,Azcoiti:1995mi}    \\
      \hline
	  $< 4.4$					   &	 F-theorem						&	2015 \cite{Giombi:2015haa}	 \\	
      \hline
          $(4/3)(32/\pi^2)=4.32$                           &     Schwinger-Dyson (LO, resummation)              	&       1989 \cite{Nash:1989xx}    \\
      \hline
          $4.422$ 				 	   &     RG study (one-loop) $\quad$ ($N_c^{\text{conf}} \approx 6.24$)    &       2016 \cite{Janssen:2016nrm}       \\
      \hline
          $4$ 				                   &     functional RG $\quad$ ($4.1<N_c^{\text{conf}}<10.0$)   &       2014 \cite{Braun:2014wja}       \\
      \hline
          $3<N_c< 4$                                       &     RG study                                               &       2001 \cite{Kubota:2001kk}       \\
      \hline
          $3.5 \pm 0.5$  				   &     lattice simulations		                        &       1988, 1989  \cite{Dagotto:1988id,Dagotto:1989td} \\
%      \hline
% typo in abstract it seems + Proceedings associated with the 93 paper         $<3.78$                                          &     Schwinger-Dyson (NLO, Landau gauge)                    &       2011 \cite{Kotikov:2011kg} \\
      \hline
          $3.31$                                          &     Schwinger-Dyson (NLO, Landau gauge)                    &       1993 \cite{Kotikov:1993wr} \\
      \hline
\textcolor{red}{$3.29$} 				   &     \textcolor{red}{Schwinger-Dyson (NLO, Landau gauge)}	&       \textcolor{red}{2016} \cite{Kotikov:2016wrb} \\
      \hline
          $32/ \pi^2 \approx 3.24$                         &     Schwinger-Dyson (LO, Landau gauge)              	&       1988 \cite{Appelquist:1988sr}   \\
      \hline
\textcolor{red}{$3.0084 - 3.0844$}                         &     \textcolor{red}{Schwinger-Dyson (NLO, resummation)}    &       \textcolor{red}{2016} \cite{Kotikov:2016prf} \\
      \hline
          $2.89$   	                                   &     RG study (one-loop)                                    &       2016 \cite{Herbut:2016ide}     \\
      \hline
\textcolor{red}{$2.85$}                                    &     \textcolor{red}{Schwinger-Dyson (NLO, resummation, $\forall \xi$)}  &       \textcolor{red}{2016} \cite{Gusynin:2016som,Kotikov:2016prf}     \\
      \hline
          $1 + \sqrt{2} = 2.41$                            &     F-theorem	                                        &       2016 \cite{Giombi:2016fct}     \\
      \hline
          $<9/4=2.25$                                      &     RG study (one-loop)	                                &       2015 \cite{DiPietro:2015taa}     \\
      \hline
          $<3/2$                                           &     Free energy constraint					&       1999 \cite{Appelquist:1999hr}     \\
      \hline
          $1< N_c<4$                                       &     lattice simulations				        &       2004, 2008 \cite{Hands:2004bh,Strouthos:2008kc}     \\
      \hline
          0  					           &     Schwinger-Dyson (non-perturbative, Landau gauge)       &       1990 \cite{Atkinson:1989fp}     \\
      \hline
          0                                                &     lattice simulations                                    &       2015, 2016 \cite{Karthik:2015sgq,Karthik:2016ppr}     \\
      \hline
    \end{tabular}
    \caption{D$\chi$SB in QED$_3$: some values of $N_c$ obtained over the years with different methods. Values presented in this manuscript are in red. The value to be retained is the gauge-invariant $N_c=2.85$.}
    \label{chap4:tab:Nc-values}
\end{table}
\end{center}

Of importance to us in the following, is the approach of Appelquist et al.\ \cite{Appelquist:1988sr} who found that $N_c = 32/ \pi^2 \approx 3.24$ by solving
the Schwinger-Dyson (SD) gap equation in the Landau gauge using a LO $1/N$-expansion, see, \eg, \cite{Roberts:1994dr} for an (early) review on SD equations and some of their applications. 
Lattice simulations in agreement with a finite non-zero value of $N_c$ can be found in \cite{Dagotto:1988id,Dagotto:1989td}.
Soon after the analysis of \cite{Appelquist:1988sr}, Nash approximately included next-to-leading order (NLO) corrections and performed a clever partial resummation
of the fermion anomalous dimension at the level of the gap equation; this led to the suppression of the gauge dependence of $N_c$ at LO with the result:
$N_c=(4/3)(32/\pi^2)=4.32$ \cite{Nash:1989xx}.~\footnote{Including NLO corrections computed in an approximate way, Nash obtained: $N_c \approx 3.28$. We have been informed by V.~Gusynin, see also \cite{Gusynin:2016som}, that
Eq.~(15) in Ref.~\cite{Nash:1989xx} contains an error: ``341'' should be replaced by ``277'' which then leads to $N_c=3.52$. So the reliable result of Nash is his gauge-independent LO one:
$N_c=4.32$.} It was Kotikov who, in the beginning of the 90s, realized that, within the approach of Appelquist et al., powerful methods of multi-loop massless Feynman diagrams
(described in Chap.~\ref{chap2}) could be applied to the computation of $N_c$. He attempted an exact computation of NLO corrections \cite{Kotikov:1993wr,Kotikov:2011kg} but difficulties were encountered
with the strong gauge-dependence of the result at NLO. While the well-known results of Nash \cite{Nash:1989xx} were in favour of a strong stability
of the $1/N$ expansion, the ones of Kotikov \cite{Kotikov:1993wr,Kotikov:2011kg} showed that a similar property apparently holds only in the Landau gauge. In this gauge, he obtained: $N_c = 3.31$
though the numerical estimate was based on an approximate evaluation of the most complicated diagrams.
It turns out that the strong gauge dependence found in the early work \cite{Kotikov:1993wr,Kotikov:2011kg} is in agreement with the more recent studies of \cite{Bashir:2005wt} in the so-called rainbow approximation.
Interestingly, the last years witnessed a strong progress in the study of the gauge dependence of D$\chi$SB
in various models, see Refs.~\cite{Ahmad:2016dsb,Bashir:2009fv} as well as references and
discussions therein. The progress is related to the use of the Landau-Khalatnikov-Fradkin
transformation~\cite{Landau:1955zz,Fradkin:1955jr,Johnson:1959zz,Zumino:1959wt,Sonoda:2000kn}. It turns out that
the application of this transformation~\cite{Bashir:2009fv} to large-$N$ QED$_3$ has revealed the almost complete lack of gauge dependence
for $N_c$ when using the Ball-Chiu vertex \cite{Ball:1980ay}.

These facts have motivated in \cite{Kotikov:2016wrb} a refined study of \cite{Kotikov:1993wr,Kotikov:2011kg} leading to an exact computation of
all NLO corrections in the Landau gauge. As we will review in the following, this allowed to extract a high precision estimate: $N_c \approx 3.29$, a value which is surprisingly close to the (erroneous) one of Nash.
Soon after, the work \cite{Kotikov:2016prf} extended these results in two very non-trivial ways. First, all (exact) calculations were carried out for an arbitrary non-local gauge. Second, a Nash-like resummation was performed.
This confirmed the absence of gauge dependence at LO and allowed also to explicitly prove the strong suppression of the gauge dependence of $N_c$ at NLO. Moreover, as noticed in \cite{Kotikov:2016prf} and will be described below, 
the weak gauge-dependence of the gap equation is entirely due to a ``rest'' given by the most complicated master integrals entering the problem; neglecting this rest the gap equation becomes completely gauge-invariant
and leads to: $N_c=2.85$. It turns out that exactly the same value of $N_c$ has been found by Gusynin and Pyatkovskiy \cite{Gusynin:2016som} which, at the time of writing, is probably the only other recent work which also 
undertook an NLO computation of $N_c$ though with a different approach than \cite{Kotikov:2016wrb,Kotikov:2016prf}. The prescription of Gusynin and Pyatkovskiy \cite{Gusynin:2016som} is an NLO expansion
in terms of the parameter $\alpha$ which is the index parametrising the mass-function (see text for more) rather than the mass function itself. 
Once this prescription is implemented at the level of the gap equation found in \cite{Kotikov:2016prf}, the ``rest'' has to be neglected
in order to achieve NLO accuracy.~\footnote{I thank Valery Gusynin for illuminating discussions on this crucial point.} 
With this prescription, both works \cite{Gusynin:2016som} and \cite{Kotikov:2016prf} are in perfect agreement and yield order by order fully gauge-invariant methods to compute $N_c$.
Thirty years after the seminal work of Nash, the papers \cite{Gusynin:2016som} and \cite{Kotikov:2016wrb,Kotikov:2016prf} therefore bring a definite and complete solution to NLO computations  of $N_c$ in 
QED$_3$ within a SD gap equation approach. In the following, we will review the derivations leading to this solution. %Let us just mention at this point that the prescription of 
%Gusynin and Pyatkovskiy \cite{Gusynin:2016som}, when combined with the technical approach of \cite{Kotikov:2016wrb,Kotikov:2016prf} based on the multi-loop techniques of Chap.~\ref{chap2}, allows
%for a systematic method to compute $N_c$ in QED$_3$, \eg, results can be extended to NNLO.

\subsection{Some properties of the model}

It is instructive to first review some of the properties of (\ref{chap4:qed3}). 
 Our presentation will follow the original one due to Appelquist et al. \cite{Appelquist:1986fd} but similar 
 arguments can be found in almost any paper dealing with dynamical chiral symmetry breaking in QED$_3$, see, \eg, Ref.~\cite{Kubota:2001kk}. 

In three dimensions, a possible spinorial representation of the Lorentz group, $SO(2,1)$, is provided by $2\times2$ Dirac $\gamma$-matrices which can be taken as the Pauli matrices:
\be
\gamma^0 = \sigma_2, \qquad \gamma^1 = \I \sigma_3, \qquad \gamma^2 = \I \sigma_1 \, .
\label{chap4:gamma:2x2}
\ee
There are no other $2\times2$ matrix anticommuting with these $\gamma^\mu$.~\footnote{This is clear from the fact that the above $\gamma$-matrices together with the identity form a complete set
of four $2\times2$ matrices. Any arbitrary $2\times2$ matrix $\Gamma$ can be decomposed on this set as:
\be
\Gamma = \frac{1}{2}\,\Tr \bigl[ \Gamma \bigr] + \frac{1}{2}\,\gamma^\mu\,\Tr \bigl[ \gamma_\mu \Gamma \bigr]\, .
\label{chap4:gamma:2x2:decomp}
\ee
}
 There is therefore no available generator for a chiral symmetry that would be broken by a mass term.

One may consider on the other hand a (reducible) representation of the Lorentz group in terms of four-component spinors (as we did in all previous chapters). The corresponding $4\times4$ Dirac $\gamma$-matrices can be taken as:
\be
\gamma^0 \,=\, \left( \begin{array}{cc} \sigma_3       &  0               \\
                                        0              &  -\sigma_3       \end{array}\right),\quad
\gamma^1 \,=\, \I\, \left( \begin{array}{cc} \sigma_1  &  0 \\
                                        0              &  -\sigma_1    \end{array}\right),\quad
\gamma^2 \,=\, \I\,\left( \begin{array}{cc} \sigma_2   &  0   \\
                                        0              &  -\sigma_2    \end{array}\right).
\label{chap4:gamma:4x4}
\ee
In this case there are two other $4\times4$ matrices which anti-commute with these matrices:
\be
\gamma^3 \,=\, 							
			\I\, \left( \begin{array}{cc}    0     &  -\I \\
                                          		\I     &  0    \end{array}\right),\quad
\gamma^5 \,=\, \gamma_5 \,=\, \I \gamma^0 \gamma^1 \gamma^2 \gamma^3\, = \, 	
			    \left( \begin{array}{cc}     0     &  1   \\
                                                	 1     &  0     \end{array}\right)\, .
\label{chap4:gamma:4x4:3+5}
\ee
One may also introduce:
\be
\gamma^{35} = \gamma^3 \gamma^5 = \frac{1}{2}\,[\gamma^3,\gamma^5] = \,
			  \left( \begin{array}{cc}     1     &  0   \\
                                                       0     &  -1     \end{array}\right)\, .
\ee
For each four-component spinor, the massless model of Eq.~(\ref{chap4:qed3}) has a global $U(2)$ symmetry with generators ${\bf 1}, \gamma^3, \gamma^5$ and $\gamma^{35}$, \eg, it is invariant
under the ``chiral'' transformation $\psi \ra e^{\I \al \gamma^3}$ and $\psi \ra e^{\I \beta \gamma^5}$. For $N$ flavours, this global symmetry is raised to $U(2N)$.\footnote{Notice that in three dimensions, we have a complete set of 16 matrices in the space 
of $4\times4$ matrices: 
\be
\{\Gamma^A \}_{d=3} = \{ {\bf 1}, \gamma^\mu, \gamma^3, \gamma^5, \gamma^3 \gamma^\mu, \gamma^5 \gamma^\mu, \gamma^{35}, \sigma^{\mu \nu}\}\, ,
\ee
where ${\bf 1}$ is the $4\times4$ unit matrix, $\sigma^{\mu \nu} = (\I/2)[\gamma^\mu,\gamma^\nu]$ and only the components $\mu<\nu$ of this tensor are included ($\mu,\nu=0,1,2$).
Any arbitrary $4\times4$ matrix $\Gamma$ can be decomposed on this set as:
\bea
\Gamma & = & \frac{1}{4}\, {\bf 1}\,\Tr \bigl[ \Gamma \bigr] + \frac{1}{4}\,\gamma^\mu\,\Tr \bigl[ \gamma_\mu \Gamma \bigr] + \frac{1}{4}\,\gamma^3\,\Tr \bigl[ \gamma_3 \Gamma \bigr]
+ \frac{1}{4}\,\gamma_5\,\Tr \bigl[ \gamma_5 \Gamma \bigr]
\nonum \\
& - &\frac{1}{4}\,\gamma^3 \gamma^\mu\,\Tr \bigl[ \gamma_3 \gamma_\mu \Gamma \bigr] -\frac{1}{4}\,\gamma_5 \gamma^\mu\,\Tr \bigl[ \gamma_5 \gamma_\mu \Gamma \bigr]
+ \frac{1}{8}\,\sigma^{\mu \nu}\,\Tr \bigl[ \sigma_{\mu \nu} \Gamma \bigr]\, .
\eea
Notice that both in three and four dimensions, we use the fact that the trace of $\gamma_5$ with an even number of $\gamma^\mu$ vanishes
(as well as the fact that the trace of any odd number of $\gamma$-matrices vanishes). Notice also that in dimensional regularization, \eg, $D=3-2\veps$, we depart from $D=3$ and the above complete set of matrices can be modified.
This is the origin of the appearance of ``evanescent'' operators in some cases, see, \eg, \cite{Gracey:2016mio}.
}
{\!\!$^{,}$}\footnote{ For a usual four-dimensional model the complete set
also includes 16 matrices:
\be
\{\Gamma^A \}_{d=4} = \{ {\bf 1}, \gamma^\mu, \gamma^5, \gamma^5 \gamma^\mu, \sigma^{\mu \nu}\}\, ,
\ee
where $\mu,\nu=0,1,2,3$ and only the components $\mu<\nu$ of the $\sigma^{\mu \nu}$  tensor are included. Any arbitrary $4\times4$ matrix $\Gamma$ can be decomposed on this set as:
\be
\Gamma = \frac{1}{4}\, {\bf 1}\,\Tr \bigl[ \Gamma \bigr] + \frac{1}{4}\,\gamma^\mu\,\Tr \bigl[ \gamma_\mu \Gamma \bigr] + \frac{1}{4}\,\gamma_5\,\Tr \bigl[ \gamma_5 \Gamma \bigr]
-\frac{1}{4}\,\gamma_5 \gamma^\mu\,\Tr \bigl[ \gamma_5 \gamma_\mu \Gamma \bigr] + \frac{1}{8}\,\sigma^{\mu \nu}\,\Tr \bigl[ \sigma_{\mu \nu} \Gamma \bigr]\, .
\ee
}

In order to better appreciate the $U(2N)$ flavour symmetry (we will also discuss the effect of a parity transformation) it is instructive to go to a two-component fermion representation, $\chi^i$ ($i=1,\cdots,2N$).
The later  are related to the four-components ones with the help of:
\be
\psi_\sigma = \left( \begin{array}{c}     \chi_{\sigma}   \\
                                          \chi_{_{N+\sigma}}   \end{array}\right)\, ,
\qquad 
\bar{\psi}_\sigma = \left( \bar{\chi}_\sigma\, , \bar{\chi}_{_{N+\sigma}} \right)\,\gamma^{35}\, , \qquad \bar{\chi}_\sigma = \chi^\dagger \sigma_3 \qquad (\sigma =1,\cdots,N)\, .
\ee
In terms of these two-component spinors the massless Lagrangian (\ref{chap4:qed3}) becomes:
\be
L_{\text{QED}_3} = \bar{\chi}_i \,\sigma^\mu \left(\I \partial_\mu  - e A_\mu \right )\chi^i - \frac{1}{4}\,F_{\mu \nu}^2 \qquad (i=1,\cdots,2N)\, .
\ee
Similarly, the following fermion bilinears can be explicited:
\begin{subequations}
\label{chap4:fermion-bilinear}
\bea
&&\bar{\psi}_\sigma \,\gamma^\mu \, \psi^\sigma =  \bar{\chi}_i \,\sigma^\mu \, \chi^i \qquad \qquad \qquad (\text{parity even},~~U(2N)~\text{invariant})\, ,
\label{chap4:fermion-bilinear:kinetic}
\\
&&\bar{\psi}_\sigma \, \psi^\sigma =  \bar{\chi}_\sigma\, \chi^\sigma - \bar{\chi}_{_{N+\sigma}}\, \chi^{N+\sigma} \qquad (\text{parity even},~~\text{breaks}~U(2N))\, ,
\label{chap4:fermion-bilinear:massPE}
\\
&&\bar{\psi}_\sigma \, \gamma^{35}\, \psi^\sigma = \bar{\chi}_i \, \chi^i ~~~\qquad \qquad \qquad (\text{parity odd},~~U(2N)~\text{invariant})\, ,
\label{chap4:fermion-bilinear:massPO}
\eea
\end{subequations}
where the parity transformation~\footnote{For completeness, with our choice of $\gamma$ matrices, the time reversal transformation is given by: $\psi'(x_T') = \I \gamma^3 \gamma^2 K \psi(x)$ with $x_T'=(-t,x,y)$ 
where $K$ is the anti-linear (complex conjugation) operator. The mass $\bar{\psi}_\sigma \, \psi^\sigma$ is even under time reversal while the mass $\bar{\psi}_\sigma \, \gamma^{35}\, \psi^\sigma$ is odd.
Moreover, the charge conjugation transformation is given by: $\psi'(x) = \I \gamma^1 \gamma^0 (\overline{\psi}(x))^{T}$.} is given by 
$\psi'(x_P') = \I \gamma^3 \gamma^1 \psi(x)$ with $x_P'=(t,-x,y)$ and the $U(2N)$ transformation by $\chi_i'=U_i^j\chi_j$ where $U$ is a $U(2N)$ matrix.
The first fermion bilinear, (\ref{chap4:fermion-bilinear:kinetic}), is invariant under both transformations which translates the invariance of the fermion kinetic energy in (\ref{chap4:qed3}).
The second fermion bilinear, (\ref{chap4:fermion-bilinear:massPE}), is the parity invariant mass term which, as anticipated in the previous section, clearly breaks $U(2N)$ into $U(N) \times U(N)$. 
It is on the dynamical generation of this (parity-invariant) mass term that we will focus on in the following.
The parity odd mass term is represented by the third fermion bilinear, (\ref{chap4:fermion-bilinear:massPO}), but we shall not consider it any further.

At this point, we shall anticipate the discussions of next sections and recall from Chap.~\ref{chap2} that the main peculiarity of QED$_3$ in the large-$N$ limit comes from the form of its photon propagator
(see Eq.~(\ref{chap3:RQED3:Dr})):
\bea
d_{r\perp}(q^2) =  \frac{\I}{-q^2}\, \frac{1}{1 + \frac{a}{\sqrt{-q^2}}} \approx \frac{1}{a}\,\frac{\I}{\sqrt{-q^2}}\, ,
\label{chap4:QED3:Dr}
\eea
where the last equality holds in the IR limit: $a \gg \sqrt{-q^2}$. This shows that the transverse photon propagator of QED$_3$ is softened in the IR and
has the same ``Coulomb''-like form as the photon propagator of RQED$_{4,3}$, Eq.~(\ref{chap3:RQED4,3:Dr}). The Coulomb form implies that the theory is not confining.
But, most importantly, it implies that quantum corrections will not break scale invariance order by order in the $1/N$ expansions. Indeed, just as in the case
of RQED$_{4,3}$, large-$N$ QED$_3$ becomes renormalizable in the IR and power counting shows that the photon self-energy becomes finite (neither UV not IR poles are generated).
The gauge field is therefore not renormalized and the dimensionful coupling constant $a=Ne^2/8$ therefore does not run.
Either in QED$_3$ or in reduced QED$_{4,3}$, if scale (actually conformal) invariance is to be broken, it has to be 
through dynamical symmetry breaking. For QED$_3$, these arguments show that such breaking will take place at energy scales much smaller than $a$. In order to better appreciate this fact, let's continue following 
Appelquist et al.\ \cite{Appelquist:1986fd} and consider the following dimensionless effective running coupling constant:
\be
\bar{a}(p) = \frac{a}{p\left( 1 + a/p \right)}\, .
\ee
In the UV limit: $a \ll p$, we have $\bar{a}(p) \ra 0$ which corresponds to a free stable UV fixed point in accordance with the fact that QED$_3$ is asymptotically free. 
On the other hand, in the IR limit: $a \gg p$, we have $\bar{a}(p) \ra 1$ which corresponds to a stable non-trivial IR fixed point. Notice that, because each photon line comes with two vertices, at each order
in perturbation theory the mass scale $e^2$ drops out in the combination $e^2/a=8/N$ and we are left with a dimensionless coupling constant $\sim 1/\sqrt{N}$.
This implies that D$\chi$SB will never take place at any finite order of the $1/N$-expansion (or loop expansion for RQED$_{4,3}$). So, in order to understand whether or D$\chi$SB will effectively take place at 
scales $p \ll a$, we have to go beyond perturbation theory. For this purpose, we will therefore consider solving the Schwinger-Dyson equations for the fermion propagator.

\subsection{Schwinger-Dyson equations} 
\label{chap4:sec:LO-QED3}

We follow the conventions set in Chaps.~\ref{chap2} and \ref{chap3}. The fermion propagator and self-energy read:
\be
S^{-1}(p) = \frac{1}{\I}\,(\Sp -\Sigma(p)), \qquad \Sigma(p) = \Sp\,\Sigma_V(p^2) + \Sigma_S(p^2)\, ,
\ee
where the vector part of the self-energy, $\Sigma_V$, is related to wave function renormalization and the newly added scalar part, $\Sigma_S$, 
is related to the parity-invariant dynamical mass term (\ref{chap4:fermion-bilinear:massPE}). The vector and scalar parts can be obtained from $\Sigma(p)$ with the help of the following formulas:
\be
\Sigma_V(p^2) = - \frac{ \Tr [ \Sp\,\Sigma(p)]}{4\,(-p^2)}, \qquad \Sigma_S(p^2) = \frac{1}{4}\,\Tr [\Sigma(p)]\, .
\ee
It is then convenient to define the (gauge-invariant) combination:
\be
\Sigma_S'(p^2) = \frac{\Sigma_S(p^2)}{1 - \Sigma_V(p^2)}\, ,
\label{chap4:Sigma'-def}
\ee
so that:
\be
S(p) = \frac{-\I}{1-\Sigma_V(p^2)}\,\frac{\Sp + \Sigma_S'(p^2)}{-p^2 + \Sigma_S'^2(p^2)}\, .
\ee
The following graphical notations will be used for fermionic lines:
\begin{subequations}
\bea
      \parbox{20mm}{
        \begin{fmfgraph*}(20,10)
          %\fmfpen{thick}
          \fmfleft{in}
          \fmfright{out}
          \fmf{crossed,label=$p$}{in,out}
        \end{fmfgraph*}
      } \quad & = & \quad \frac{-\I}{1-\Sigma_V(p^2)}\,\frac{\Sigma_S'(p^2)}{-p^2 + \Sigma_S'^2(p^2)}\, , \\
      \parbox{20mm}{
        \begin{fmfgraph*}(20,10)
          %\fmfpen{thick}
          \fmfleft{in}
          \fmfright{out}
          \fmf{plain,label=$p$}{in,out}
        \end{fmfgraph*}
      } \quad & = & \quad \frac{-\I}{1-\Sigma_V(p^2)}\,\frac{\Sp}{-p^2 + \Sigma_S'^2(p^2)}\, ,
\eea
\end{subequations}
where the cross denotes a mass insertion.

With these conventions, the SD equation for the fermion propagator, Eq.~(\ref{chap2:SD:Sigma}), may be decomposed into scalar and vector components
as follows:
\begin{subequations}
\label{chap4:SD-sigma+A}
\bea
&&\Sigma_S(p^2) = -\frac{2a}{N}\,\int [\D^D k] \frac{\Tr[ \gamma^\mu\,D_{\mu \nu}(p-k)\,(\Sk + \Sigma_S'(k^2))\,\Gamma^\nu(p,k)]}{[1-\Sigma_V(k^2)]\,[-k^2 + \Sigma_S'^2(k^2)]}\, ,
\label{chap4:SD-sigma}\\
&&\Sigma_V(p^2) = \frac{2 a}{N\,(-p^2)}\,\int [\D^D k] \frac{\Tr[\Sp\, \gamma^\mu\,D_{\mu \nu}(p-k)\,(\Sk + \Sigma_S'(k^2))\,\Gamma^\nu(p,k)]}{[1-\Sigma_V(k^2)]\,[-k^2 + \Sigma_S'^2(k^2)]}\, ,
\label{chap4:SD-A}
\eea
\end{subequations}
where $\Gamma^{\nu}(p,k)$ is the vertex function,  $D_{\mu \nu}(p)$ is the photon propagator in a non-local $\xi$-gauge \cite{Nash:1989xx}:
\be
D^{\mu \nu}(q) =  \frac{\I P_{\mu \nu}(q;1-\xi)}{(-q^2)\,[1 - \Pi(q^2)]}, \qquad P_{\mu \nu}(q;\eta) = g^{\mu \nu} - \eta\,\frac{q^\mu q^\nu}{q^2}\, ,
\label{chap4:photon}
\ee
and $\Pi(q^2)$ is the polarization part. Notice that the non-local gauge is identical to the usual gauge fixing in the special case of the Landau gauge, $\xi=0$. 
The use of a non-local gauge insures that the theory remains scale invariant order by order in the $1/N$-expansion for all values of $\xi$ as the mass scale 
$e^2$ drops out in the combination $e^2/a=8/N$ upon using Eq.~(\ref{chap4:photon}). As will be seen in the following, this will allow us to work in dimensional regularization 
with the IR photon-propagator for arbitrary values of $\xi$.

\subsection{Leading order}

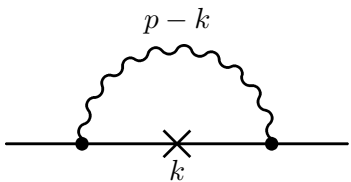
\begin{figure}
  \begin{center}
    \begin{fmfgraph*}(45,35)
      \fmfleft{in}
      \fmfright{out}
      \fmf{plain}{in,vi}
      \fmf{crossed,tension=0.2,label=$k$}{vi,vo}
      \fmf{wiggly,left,tension=0.2,label=$p-k$}{vi,vo}
      \fmf{plain}{vo,out}
      \fmfdot{vi,vo}
    \end{fmfgraph*}
    \caption{\label{chap4:fig:diags-LO}
        LO diagram to the dynamically generated mass $\Sigma(p)$. The crossed line denotes mass insertion.}
  \end{center}
\end{figure}

To start with, we focus on the gap equation (\ref{chap4:SD-sigma}) and consider the LO approximations in the $1/N$ expansion. The latter are given by:
\be
\Sigma_V(p) = 0, \quad \Pi(q^2)=\Pi_1(q^2) = -\frac{a}{(-q^2)^{1/2}}, \quad \Gamma^{\nu}(p,k) = \gamma^{\nu}\, ,
\label{chap4:LO0}
\ee
where the fermion mass has been neglected~\footnote{A study of the fermion mass contribution to $\Pi(q^2)$ can be
found, for example, in \cite{Gusynin:1995bb}.} in the calculation of $\Pi_1(q^2)$ which was computed in the last chapter. 
A single diagram contributes to the gap equation (\ref{chap4:SD-sigma}) at LO, see Fig.~\ref{chap4:fig:diags-LO}, and the latter reads (after Wick rotation):
\be
  \Sigma_S (p^2) =\frac{8(2+\xi)a}{N}
  %\frac{8a(2 + \tilde{\xi})}{N}
   \int \frac{d^3 k}{(2 \pi )^3}
\frac{ \Sigma_S (k^2) }{ \left( k^2 + \Sigma_S^2(k^2)
\right)
\bigl[ (p-k)^2 + a \,|p-k| \bigr]} \, . 
\label{chap4:SD-sigma-LO1} 
\ee

Let's first follow the original analysis of Appelquist et al.\ \cite{Appelquist:1988sr}. Performing the angular integration in Eq.~(\ref{chap4:SD-sigma-LO1}) yields:
\be
  \Sigma_S (p^2) =\frac{2(2+\xi)a}{\pi^2 N |p|}
  %\frac{8a(2 + \tilde{\xi})}{N}
   \int_0^{\infty} \! \! \! d |k|\,
\frac{ |k| \Sigma_S (|k|^2) }{k^2 + \Sigma_S^2(|k|^2)}
\ln \left(\frac{|k|+|p|+a}{|k-p|+a}\right)
% (p-k)^2 + a \mid p-k \mid\bigr]}
\, .
\label{chap4:SD-sigma-LO2}
\ee
The study of Eq.~(\ref{chap4:SD-sigma-LO2}) in Ref.~\cite{Appelquist:1988sr} revealed the existence
of a critical number of fermion flavours $N_c$, such that for $N>N_c$, $\Sigma (p) =0$.
As it was argued in this reference, QED$_3$ is strongly damped for $|p|>a$, {\it i.e.},
all relevant physics occur at $|p|/a <1$. Hence, only the lowest order terms in $|p|/a$ have to be kept on the r.h.s.\ of Eq.~(\ref{chap4:SD-sigma-LO2})
with a hard cut-off at $|p|=a$. Moreover, considering $N$
close to $N_c$, the value of $\Sigma (|k|)$ can be made arbitrarily small. Thus,
$k^2 + \Sigma^2(|k|)$ can be replaced by $k^2$ on the r.h.s.\ of Eq.~(\ref{chap4:SD-sigma-LO2})
which then further simplifies as:
\be
\Sigma_S (p^2) =\frac{4(2+\xi)}{\pi^2 N}
   \int_0^{a} \! \! \! d |k| \,
\frac{\Sigma_S (|k|^2) }{ \mbox{Max}({|k|,|p|})}\, .
\label{chap4:SD-sigma-LO3}
\ee
The mass function may then be parametrized as \cite{Appelquist:1988sr}:
\be
\Sigma_S' (k^2) = B \, (-k^2)^{ -\alpha} \, ,
\label{chap4:sigma-parametrization}
\ee
% Thus, we have for large $a$
(with an arbitrary $B$ value) where the index $\al$ has to be self-consistently determined. Substituting (\ref{chap4:sigma-parametrization}) in Eq.~(\ref{chap4:SD-sigma-LO3}), the gap equation reads:
\bea
1 = \frac{(2+\xi)\beta}{L} \quad \text{where} \quad \beta = \frac{1}{\alpha \left( 1/2 - \alpha \right)} \quad \text{and} \quad  L \equiv \pi^2 N\, .
\label{chap4:gap-eqn-LO}
\eea
Solving the gap equation, the following values of $\al$ are obtained:
\begin{eqnarray}
\alpha_{\pm} = \frac{1}{4}\,\left( 1 \pm \sqrt{1 - \frac{16(2+\xi)}{L}} \right) \, ,
\label{chap4:al-LO}
\end{eqnarray}
which reproduces the solution given by Appelquist et al.\ in Ref.~\cite{Appelquist:1988sr}.
Their analysis yields a strongly gauge-dependent critical number of fermions:
$N_c \equiv N_c(\xi) = 16(2+\xi)/ \pi^2$, which is such that $\Sigma_S(p^2) = 0$ for $N>N_c$ and:
\be
\Sigma_S(0) \simeq a\,\exp \Bigl[ - \frac{2 \pi}{(N_c/N - 1)^{1/2}} \Bigr]\, ,
\label{chap4:dynm-qed3}
\ee
for $N<N_c$. Restricting ourselves to the Landau gauge, we recover the result that was put forward by Appelquist et al.:
$N_c(\xi=0) = 32/ \pi^2 \approx 3.24$ ({\it i.e.}, $L_c = 32$). Essential to this analysis is the fact that 
D$\chi$SB occurs when $\alpha$ becomes complex, that is for $N<N_c$.

As it was shown in Refs.~\cite{Kotikov:1993wr,Kotikov:2011kg}, the same result for $\Sigma_S(p^2)$ can be obtained
in another much more straightforward way. Taking the limit of large $a$, the linearized version of Eq.~(\ref{chap4:SD-sigma-LO1})
has the following form:~\footnote{Notice that the derivation of Eq.~(\ref{chap4:dynm-qed3}) requires a finite UV cut-off $a$
and the use of $\Sigma(0)$ as an IR cut-off, see, \eg, Ref.~\cite{Appelquist:1988sr}, the recent \cite{Gusynin:2016som} as well as early references related to four-dimensional theories 
\cite{Fomin:1978rk,Fukuda:1976zb,Bardeen:1985sm}. The peculiar scaling in Eq.~(\ref{chap4:dynm-qed3}) is sometimes referred to in the literature as a Miransky scaling, see \cite{Fomin:1978rk} where such type
of scaling appeared for the first time within the study of the instability of massless QED$_4$ and \cite{Fomin:1984tv} for a review.
On the other hand, the critical regime on which we will focus 
is reached in the limit $p \ll a \ra \infty$ (and for $\Sigma_S(p^2) \ll p$ in the linearized case). Such an approach to criticality was used in early works on four-dimensional theories, see, \eg, 
Refs.~\cite{Fomin:1978rk,Bardeen:1985sm} and also \cite{Atkinson:1993mz} for a bifurcation analysis. It is in the critical regime that the form of the mass function Eq.~(\ref{chap4:sigma-parametrization}) holds.
This form is particularly well suited to apply dimensional regularization and techniques from massless Feynman diagram calculations as first noticed in \cite{Kotikov:1993wr}. Notice 
that, in some case, working with dimensional regularization to solve SD equations was found to be more difficult than using a cut-off, see, \eg, \cite{Gusynin:1998se}.  
I thank Valery Gusynin for discussions related to all of these points.}
\be
  \Sigma_{1S} (p^2) =
%\frac{16}{N}
 \frac{8(2 + \xi)}{N}
   \int \frac{d^3 k}{(2 \pi )^3} \frac{ \Sigma_{1S}(k^2) }{k^2 \, |p-k| } \, .
\label{chap4:SD-sigma-LO4}
\ee
With the help of the ansatz (\ref{chap4:sigma-parametrization}), one can then see that
the r.h.s.\ of Eq.~(\ref{chap4:SD-sigma-LO4}) may be calculated with the help of the standard rules of perturbation theory for massless Feynman diagrams, see Chap.~\ref{chap2}.
Indeed, given these rules, the computation of Eq.~(\ref{chap4:SD-sigma-LO4}) is straightforward and reads:
\be
\Sigma_{1S}(p) = \frac{4(2+\xi)B}{N}\,\frac{(-p^2)^{-\al}}{(4\pi)^{3/2}}\, \frac{2\beta}{\pi^{1/2}}\, .
\label{chap4:sigma-LO-res}
\ee
This  immediately yields the gap equation (\ref{chap4:gap-eqn-LO}) and, hence, the results
of Eq.~(\ref{chap4:al-LO}) together with the critical value $N_c(\xi) =  16(2+\xi)/ \pi^2 $ at which the index $\alpha$
becomes complex.

The gauge-dependent fermion wave function may be computed in a similar way. At LO, Eq.~(\ref{chap4:SD-A}) simplifies as:
\be
\Sigma_{1V}(p^2) = \I\,\frac{2\mu^{2\veps}}{N\,(-p^2)}\,\int [\D^D k] \frac{\Tr[\Sp\,P_{\mu \nu}(p-k;1-\xi) \,\gamma^\mu\,\Sk\,\gamma^\nu]}{[-(p-k)^2]^{1/2}\,[-k^2]}\, ,
\ee
where the integral has been dimensionally regularized with $D=3 -2 \veps$. 
Taking the trace, performing a Wick rotation and computing the integral on the r.h.s.\ yields:
\be
  \Sigma_{1V}(p^2) =
\frac{\Gamma(1+\veps)(4\pi)^{\ep} \mu^{2\ep}}{p^{2\ep}} \, C_1(\xi) = 
  \frac{\overline{\mu}^{2\ep}}{p^{2\ep}} \, C_1(\xi) \, + \Ord(\ep) \, ,
\label{chap4:SD-A2}
\ee
where the $\overline{MS}$ parameter $\overline{\mu}$ was defined in (\ref{chap2:muMSbar}) and
\be
C_1(\xi)=
+\frac{2}{3\pi^2N} \left((2-3\xi)\left[\frac{1}{\ep} - 2\ln 2\right]+ \frac{14}{3} - 6\xi\right)\, .
\label{chap4:C1}
\ee
We note that in the $\xi=2/3$-gauge, the value of $\Sigma_{1V}$ is finite and
$C_1(\xi=2/3)= +4/(9\pi^2N)$.
From Eqs.~(\ref{chap4:SD-A2}) and (\ref{chap4:C1}), the LO fermion anomalous dimension constant may be extracted:
%$\gamma_\psi = \mu (d/d\mu) A(p) = 4(2 - 3\tilde{\xi})/(3\pi^{2}N)$ %
\be
\gamma_\psi = \mu \frac{\D \Sigma_{1V}}{\D \mu} = \frac{4(2 - 3\xi)}{3\pi^{2}N}\, , 
\label{chap4:lambdaA-LO}
\ee
a result which coincides with the one of \cite{Nash:1989xx,Gracey:1993iu,Fischer:2004nq}. 

%\begin{widetext}

%\begin{figure}[t]
%    \includegraphics[width=0.96\textwidth]{fig2}
%    \caption{\label{chap4:fig:diags-NLO}
%        NLO diagrams to the dynamically generated mass $\Sigma(p)$. The shaded blob is defined in Eq.~(\ref{polar-2loops}).}
%%  \end{center}
%\end{figure}
%
%\end{widetext}

\subsection{Next-to-leading order: self-energy contributions} 
\label{chap4:sec:NLO}

%\subsection{Self-energy contributions}

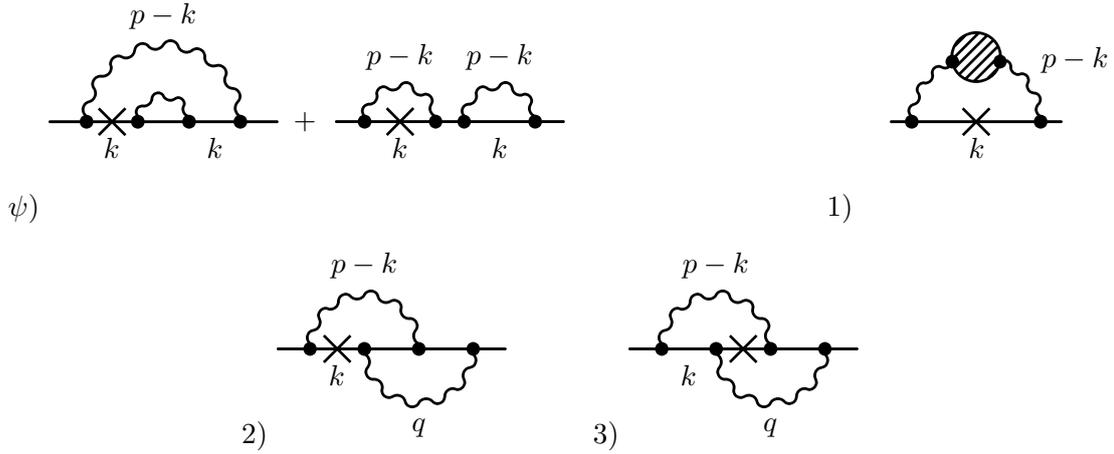
\begin{figure}
  \begin{center}
    $\psi$)
    \begin{fmfgraph*}(30,25)
      %\fmfpen{thick}
      \fmfleft{i1}
      \fmfright{o1}
      \fmflabel{$+$}{o1}
      \fmf{plain,tension=1.4}{i1,i2}
      \fmf{crossed,label=$k$}{i2,i3}
      \fmf{plain}{i3,o3}
      \fmf{plain,label=$k$}{o3,o2}
      \fmf{plain,tension=1.4}{o2,o1}
      \fmffreeze
      \fmf{photon,left,tension=0.1,label=$p-k$,label.side=left}{i2,o2}
      \fmf{photon,left,tension=0.1}{i3,o3}
      \fmfdot{i2,i3,o2,o3}
   \end{fmfgraph*}
$\quad ~$
    \begin{fmfgraph*}(30,25)
      %\fmfpen{thick}
      \fmfleft{i1}
      \fmfright{o1}
      \fmf{plain}{i1,i2}
      \fmf{crossed,tension=0.2,label=$k$}{i2,i3}
      \fmf{wiggly,left,tension=0.2,label=$p-k$}{i2,i3}
      \fmf{plain}{i3,o3}
%      \fmf{plain}{o3,o2}
      \fmf{plain,tension=0.2,label=$k$}{o3,o2}
      \fmf{wiggly,left,tension=0.2,label=$p-k$}{o3,o2}
      \fmf{plain}{o2,o1}
      \fmfdot{i2,i3,o2,o3}
   \end{fmfgraph*}
$\qquad \qquad \qquad \qquad ~$   1)
    \begin{fmfgraph*}(30,25)
      %\fmfpen{thick}
% lower line from left to right: i1, i2, i3 (vertex), i4, o4, o3 (vertex), o2, o1
% paths are arcs from one vertex to another
      \fmfleft{i1}
      \fmfright{o1}
      \fmf{phantom}{i1,i2}
      \fmf{plain,tension=1.4}{i2,i3}
      \fmf{plain}{i3,i4}
      \fmf{crossed,tension=0.4,label=$k$}{i4,o4}
      \fmf{plain}{o4,o3}
      \fmf{plain,tension=1.4}{o3,o2}
      \fmf{phantom}{o2,o1}
      \fmffreeze
      %\fmf{plain,left,tension=0.1,tag=1}{i1,o1}
      \fmf{phantom,left,tension=0.1,tag=2}{i2,o2}
      \fmf{phantom,left,tension=0.1,tag=3}{i3,o3}
      \fmf{phantom,left,tension=0.1,tag=4}{i4,o4}
%      \fmf{phantom,left,tag=5}{i1,o1}
      \fmfdot{i3,o3}
      \fmfposition
      \fmfipath{p[]}
      %\fmfiset{p1}{vpath1(__i1,__o1)}
      \fmfiset{p2}{vpath2(__i2,__o2)}
      \fmfiset{p3}{vpath3(__i3,__o3)}
      \fmfiset{p4}{vpath4(__i4,__o4)}
%      \fmfiset{p5}{vpath5(__i1,__o1)}
      \fmfi{photon}{subpath (0,3length(p3)/8) of p3}
      \fmfi{photon,label=$p-k$}{subpath (5length(p3)/8,length(p3)) of p3}
%      \fmfi{plain}{point 3length(p3)/8 of p3 .. point length(p2)/2 of p2 .. point 5length(p3)/8 of p3}
%      \fmfi{plain}{point 5length(p3)/8 of p3 .. point length(p4)/2 of p4 .. point 3length(p3)/8 of p3}
      \def\vert#1{%
        \fmfiv{decor.shape=circle,decor.filled=full,decor.size=2thick}{#1}}
      \vert{point 3length(p3)/8 of p3}
      \vert{point 5length(p3)/8 of p3}
      \def\blob#1{%
        \fmfiv{decor.shape=circle,decor.filled=shaded,decor.size=9thick}{#1}}
      \blob{point length(p3)/2 of p3}
   \end{fmfgraph*}
\\
$\qquad$
\\
%$\quad ~$ 
2)
    \begin{fmfgraph*}(30,25)
      %\fmfpen{thick}
      \fmfleft{i1}
      \fmfright{o1}
      \fmf{plain,tension=1.7}{i1,i2}
      \fmf{crossed,label=$k$}{i2,i3}
      \fmf{plain}{i3,o3}
      \fmf{plain}{o3,o2}
      \fmf{plain,tension=1.7}{o2,o1}
      \fmffreeze
      \fmf{photon,left,tension=0.1,label=$p-k$}{i2,o3}
      \fmf{photon,right,tension=0.1,label=$q$}{i3,o2}
      \fmfdot{i2,i3,o2,o3}
   \end{fmfgraph*}
$\qquad ~$ 3)
    \begin{fmfgraph*}(30,25)
      %\fmfpen{thick}
      \fmfleft{i1}
      \fmfright{o1}
      \fmf{plain,tension=1.7}{i1,i2}
      \fmf{plain,label=$k$}{i2,i3}
      \fmf{crossed}{i3,o3}
      \fmf{plain}{o3,o2}
      \fmf{plain,tension=1.7}{o2,o1}
      \fmffreeze
      \fmf{photon,left,tension=0.1,label=$p-k$}{i2,o3}
      \fmf{photon,right,tension=0.1,label=$q$}{i3,o2}
      \fmfdot{i2,i3,o2,o3}
   \end{fmfgraph*}
    \caption{\label{chap4:fig:diags-NLO}
        NLO diagrams to the dynamically generated mass $\Sigma(p)$. The shaded blob is defined in Eq.~(\ref{chap4:polar-2loops}).}
  \end{center}
%  \end{center}
\end{figure}

We now consider the NLO contributions and parametrize them as:
\be
\Sigma_{2S}(p^2) = \left(\frac{8}{N}\right)^2 B\,\frac{(-p^2)^{-\al}}{(4\pi)^{3}}\,
\left( \Sigma_{\psi} + \Sigma_1 + 2\,\Sigma_2 + \Sigma_3 \right) \, ,
\ee
where each contribution to the linearized gap equation is represented graphically in Fig.~\ref{chap4:fig:diags-NLO}.
Adding these contributions to the LO result, Eq.~(\ref{chap4:sigma-LO-res}), 
the gap equation has the following general form:
\begin{eqnarray}
1 = \frac{(2+\xi) \beta}{L} +
\frac{\overline{\Sigma}_\psi (\xi) + \overline{\Sigma}_1(\xi) + 2\,\overline{\Sigma}_2(\xi) + \overline{\Sigma}_3(\xi)}{L^2} \, ,
\label{chap4:gap-eqn-NLO}
\end{eqnarray}
where $\overline{\Sigma}_i = \pi \Sigma_i$, $(i=1,2,3,\psi)$.
In \cite{Kotikov:2016wrb}, these contributions were computed in the Landau gauge, $\xi=0$. After very tedious and lengthy calculations, 
these computations could be extended to an arbitrary non-local gauge \cite{Kotikov:2016prf}. We now summarize these results.

The contribution  $\Sigma_\psi$, see Fig.~\ref{chap4:fig:diags-NLO} $\psi$), originates from the LO value of $\Sigma_V(p^2)$ and is singular.
Using dimensional regularization, and for an arbitrary parameter $\xi$, it reads:
\bea
%&&
\overline{\Sigma}_\psi(\xi) = 4 \,
%+\frac{16}{3}\, 
\frac{\overline{\mu}^{2\ep}}{p^{2\ep}}\, \beta 
\left[ \left(\frac{4}{3}(1-\xi)-\xi^2\right)\left[
\frac{1}{\ep} + \Psi_1 - \frac{\beta}{4} \right] 
%\right .
%\nonum \\
%&&\left . 
+ \left(\frac{16}{9}-\frac{4}{9}\xi-2\xi^2\right)
\right] \, ,
\label{chap4:sigma-NLO-A.1}
\eea
where %$\overline{\Sigma}_i = \pi \Sigma_i$, $(i=1,2,3.A)$ and 
\be
\Psi_1 = \Psi(\alpha)+  \Psi(1/2-\alpha)-2\Psi(1) + \frac{3}{1/2-\alpha} %+ \ln \left(\frac{\overline{\mu}^2}{p^2}\right)
-2 \ln 2\, ,
\label{chap4:psi1}
\ee
and $\Psi$ is the digamma function. 

The contribution of diagram 1) in Fig.~\ref{chap4:fig:diags-NLO} is finite and reads:
\be
\overline{\Sigma}_1(\xi) = -2(2+\xi)\, \beta\, \hat{\Pi}, \qquad \hat{\Pi} = \frac{92}{9} - \pi^2\, ,
\label{chap4:sigma-NLO-1}
\ee
where the gauge dependence comes from the fact that we work in a non-local gauge
and $\hat{\Pi}$ arises from the two-loop polarization operator in dimension $D=3$~\cite{Gracey:1993sn,Gusynin:2000zb,Teber:2012de,Kotikov:2013kcl} which may be graphically represented as:
\be
\parbox{8mm}{
    \begin{fmfgraph*}(8,7)
      \fmfleft{i}
      \fmfright{o}
      \fmfleft{ve}
      \fmfright{vo}
      \fmffreeze
      \fmfforce{(-0.3w,0.5h)}{i}
      \fmfforce{(1.3w,0.5h)}{o}
      \fmfforce{(0w,0.5h)}{ve}
      \fmfforce{(1.0w,0.5h)}{vo}
      \fmffreeze
      \fmf{photon}{i,ve}
      \fmf{photon}{vo,o}
      \fmffreeze
      \fmfdot{ve,vo}
      \fmf{phantom,tag=1}{ve,vo}
      \fmfposition
      \fmfipath{p[]}
      \fmfiset{p1}{vpath1(__ve,__vo)}
      \def\blob#1{%
        \fmfiv{decor.shape=circle,decor.filled=shaded,decor.size=1w}{#1}}
      \blob{point length(p1)/2 of p1}
    \end{fmfgraph*}
} \qquad = \quad 2 \times ~
\parbox{8mm}{
    \begin{fmfgraph*}(14,8)
      %\fmfpen{thick}
      \fmfleft{i}
      \fmfright{o}
      \fmf{photon}{i,v1}
      \fmf{photon}{v2,o}
      \fmf{phantom,right,tension=0.1,tag=1}{v1,v2}
      \fmf{phantom,right,tension=0.1,tag=2}{v2,v1}
      \fmf{phantom,tension=0.1,tag=3}{v1,v2}
      \fmfdot{v1,v2}
      \fmfposition
      \fmfipath{p[]}
      \fmfiset{p1}{vpath1(__v1,__v2)}
      \fmfiset{p2}{vpath2(__v2,__v1)}
      \fmfiset{p3}{vpath3(__v1,__v2)}
      \fmfi{plain}{subpath (0,length(p1)) of p1}
      \fmfi{plain}{subpath (0,length(p2)/4) of p2}
      \fmfi{plain}{subpath (length(p2)/4,3length(p2)/4) of p2}
      \fmfi{plain}{subpath (3length(p2)/4,length(p2)) of p2}
      \fmfi{photon}{point length(p2)/4 of p2 .. point length(p3)/2 of p3 .. point 3length(p2)/4 of p2}
      \def\vert#1{%
        \fmfiv{decor.shape=circle,decor.filled=full,decor.size=2thick}{#1}}
      \vert{point length(p2)/4 of p2}
      \vert{point 3length(p2)/4 of p2}
%      \vert{point length(p1)/2 of p1}
%      \vert{point length(p2)/2 of p2}
%      \vert{point length(p1)/2 of p1}
%      \vert{point length(p2)/2 of p2}
    \end{fmfgraph*}
} \qquad + \quad
\parbox{8mm}{
    \begin{fmfgraph*}(14,8)
      %\fmfpen{thick}
      \fmfleft{i}
      \fmfright{o}
      \fmf{photon}{i,v1}
      \fmf{photon}{v2,o}
      \fmf{phantom,right,tension=0.1,tag=1}{v1,v2}
      \fmf{phantom,right,tension=0.1,tag=2}{v2,v1}
      \fmf{phantom,tension=0.1,tag=3}{v1,v2}
      \fmfdot{v1,v2}
      \fmfposition
      \fmfipath{p[]}
      \fmfiset{p1}{vpath1(__v1,__v2)}
      \fmfiset{p2}{vpath2(__v2,__v1)}
      \fmfi{plain}{subpath (0,length(p1)/2) of p1}
      \fmfi{plain}{subpath (length(p1)/2,length(p1)) of p1}
      \fmfi{plain}{subpath (0,length(p2)/2) of p2}
      \fmfi{plain}{subpath (length(p2)/2,length(p2)) of p2}
      \fmfi{photon}{point length(p1)/2 of p1 -- point length(p2)/2 of p2}
      \def\vert#1{%
        \fmfiv{decor.shape=circle,decor.filled=full,decor.size=2thick}{#1}}
      \vert{point length(p1)/2 of p1}
      \vert{point length(p2)/2 of p2}
    \end{fmfgraph*}
} \qquad ,
\label{chap4:polar-2loops}
\ee
and was already encountered in Chap.~\ref{chap3} when studying reduced QED$_{4,3}$.

\ms

The contribution of diagram 2) in Fig.~\ref{chap4:fig:diags-NLO} is again singular. Dimensionally regularizing it yields:
\begin{subequations}
\begin{flalign}
&\overline{\Sigma}_2 (\xi)= -2\,\frac{\overline{\mu}^{2\ep}}{p^{2\ep}}\, \beta \, 
\bigg[ \frac{(2+\xi)(2-3\xi)}{3}\,\left(\frac{1}{\veps} + \Psi_1 - \frac{\beta}{4} \right) 
%\bigg .
%\label{chap4:sigma-NLO-22} \\
%&&\bigg . 
+ \frac{\beta}{4}\,\left(\frac{14}{3}\,(1-\xi) + \xi^2 \right) 
+ \frac{28}{9} + \frac{8}{9}\,\xi - 4 \xi^2 \bigg] 
\nonum\\
&\qquad \qquad + (1-\xi)\,\hat{\Sigma}_{2} \, ,
\label{chap4:sigma-NLO-22} \\
%\nonum\\
&\hat{\Sigma}_2(\alpha) = (4\alpha-1) \beta \Bigl[\Psi'(\alpha) - \Psi'(1/2-\alpha)\Bigl] 
%\nonum \\
%&&
+\frac{\pi}{2\alpha}\,\tilde{I}_1(\alpha) + \frac{\pi}{2(1/2-\alpha)} \tilde{I}_1(\alpha+1) \, ,
\label{chap4:sigma-NLO-22_L} 
\end{flalign}
\end{subequations}
where $\Psi'$ is the trigamma function and $\tilde{I}_1(\alpha)$ is a dimensionless integral that is defined as:
\begin{flalign}
I_1(\alpha) \equiv \frac{(p^2)^{-\al}}{(4\pi)^3}\, \tilde{I}_1(\alpha)
%\label{chap4:I1-def} \\
= \int \frac{d^3k_1}{(2\pi)^3} \frac{d^3k_2}{(2\pi)^3}
   \frac{1}{|p-k_1|k_1^{2\alpha} (k_1-k_2)^2 (p-k_2)^2|k_2|}\, .
\label{chap4:I1-def}
\end{flalign}
This integral obeys the following relation (it can be obtained by analogy with the ones in Ref.~\cite{Kazakov:1984bw}):
\bea
\tilde{I}_1(\alpha+1) =  \frac{(\alpha-1/2)^2}{\alpha^2}
\tilde{I}_1(\alpha) - \frac{1}{\pi \alpha^2}
\Bigl[ \Psi'(\alpha) - \Psi'(1/2-\alpha) \Bigr]\, .
%  \frac{\Gamma^2(\alpha)}{\pi \Gamma^2(\alpha+1/2)}\, .
\label{chap4:I1-relation}
\eea
Using the results of Ref.~\cite{Kotikov:1995cw}, the integral $\tilde{I}_1(\alpha)$ can be represented
in the form of a two-fold series
\begin{flalign}
&\tilde{I}_1(\alpha) = \sum_{n=0}^{\infty} \sum_{l=0}^{\infty}\,\frac{B(l,n,1,1/2)}{(n+1/2)\,\Gamma(1/2)}
\times \Biggl[  \frac{2}{n+1/2} \left(  \frac{1}{l+n+\alpha} +  \frac{1}{l+n+3/2-\alpha} \right) \Biggr .
\nonum \\ 
&\Biggl . \qquad \qquad \qquad +\frac{1}{(l+n+\alpha)^2} + \frac{1}{(l+n+3/2-\alpha)^2} \Biggr], \,
\label{chap4:I1-series}
\end{flalign}
where
\bea
B(m,n,\alpha,1/2) = \frac{\Gamma(m+n+\alpha) \Gamma(m+\alpha-1/2)}{m!\Gamma(m+n+3/2)
\Gamma(\alpha)\Gamma(\alpha-1/2)} \, .
\label{chap4:B}
\eea
Notice that the singularities in $\overline{\Sigma}_\psi(\xi)$ and $\overline{\Sigma}_2(\xi)$ cancel each other and their sum is therefore finite.
This cancellation corresponds to the one of the logarithms, $\ln(p/\al)$, in Ref.~\cite{Nash:1989xx}; the importance of such cancellations
was discussed before, in Ref.~\cite{Appelquist:1988sr}. Defining: $\overline{\Sigma}_{2A} (\xi)= \overline{\Sigma}_{A} (\xi) + 2\overline{\Sigma}_{2} (\xi)$, the latter reads:
\bea
\overline{\Sigma}_{2A}(\xi) =  2(1-\xi) \hat{\Sigma}_2(\alpha) - \left(\frac{14}{3}(1-\xi)+\xi^2\right) \beta^2 - 8\beta\left(\frac{2}{3}(1+\xi)-\xi^2\right) \, .
\eea

Finally, the contribution of diagram 3) in Fig.~\ref{chap4:fig:diags-NLO} is finite and reads:
\begin{subequations}
\begin{flalign}
&\overline{\Sigma}_3(\xi) = \hat{\Sigma}_3(\alpha,\xi) + \Bigl(3+4\xi-2\xi^2 \Bigr) \beta^2\, , 
 \\
&\hat{\Sigma}_3(\alpha,\xi) = \frac{1}{4}\bigl(1+8\xi+\xi^2+2\alpha (1-\xi^2)\bigr)
%\left(\frac{1}{2}+\xi+\xi^2+2\alpha \xi(1-\xi)\right)
\pi\tilde{I}_2(\alpha)
%\nonum \\
+ \frac{1}{2}\bigl(1+4\xi-\alpha (1-\xi^2)\bigr)
%\left(\frac{1}{2}+2\xi-\alpha (1-\xi)^2\right)
\pi\tilde{I}_2(1+\alpha) 
\nonum \\
&\qquad \qquad + \frac{1}{4}\bigl(-7-16\xi+3\xi^2\bigr)
%\Bigl(2+3\xi+\xi^2-\alpha (1-3\xi)(1-\xi)\Bigr)
\pi\tilde{I}_3(\alpha)\, .
\label{chap4:sigma-NLO-3AG}
\end{flalign}
\end{subequations}
The dimensionless integrals  in Eq.~(\ref{chap4:sigma-NLO-3AG}) are defined as: $\tilde{I}_2(\alpha)= \tilde{I}(\gamma=1/2,\alpha)$ and
$\tilde{I}_3(\alpha)= \tilde{I}(\gamma=-1/2,1+\alpha)$, where:
\be
  I(\gamma,\alpha)  \equiv \frac{(p^2)^{-\al -\gamma+1/2}}{(4\pi)^3}\, \tilde{I}(\gamma,\alpha)
  =\int \frac{d^3k_1}{(2\pi)^3} \frac{d^3k_2}{(2\pi)^3}
   \frac{1}{(p-k_1)^{2\gamma}k_1^{2} (k_1-k_2)^{2\alpha} (p-k_2)^2|k_2|} \, .
   \label{chap4:I-def}
\ee
They satisfy the following relations:
\begin{flalign}
 \tilde{I}_2(\alpha) =  \tilde{I}_2(3/2-\alpha),~~~
 \tilde{I}_3(\alpha) = \frac{2}{4\alpha -1} \Bigl(\alpha  \tilde{I}_2(1+\alpha) - (1/2-\alpha)
  \tilde{I}_2(\alpha) \Bigr)
- \frac{\beta^2}{\pi}\, ,
\label{chap4:I2-I3-relations}
\end{flalign}
and, thus, only one of them is independent.
Using the results of Ref.~\cite{Kotikov:1995cw}, the integral $\tilde{I}_2(\alpha)$ can be represented in the form of a 
three-fold series:
\begin{subequations}
\label{chap4:I2}
\begin{flalign}
&\tilde{I}_2(\alpha)
%\frac{ \tilde{I}_2(\alpha)}{N}
= \sum_{n=0}^{\infty} \sum_{m=0}^{\infty}
B(m,n,\beta,1/2) \sum_{l=0}^{\infty}
B(l,n,1,1/2)
\times  C(n,m,l,\alpha) \, ,
\label{chap4:I2-series}\\
&C(n,m,l,\alpha) =
%\Biggl[
\frac{1}{(m+n+\alpha)(l+n+\alpha)}
%\nonumber \\ &&
+
 \frac{1}{(m+n+\alpha)(l+m+n+1)}
%\nonumber \\ &&
\nonum \\
&+  \frac{1}{(m+n+1/2)(l+m+n+\alpha)}
+ \frac{1}{(m+n+1/2)(l+n+3/2-\alpha)}
\nonumber \\ 
&+  \frac{1}{(n+l+\alpha)(l+m+n+\alpha)}
%\nonumber \\ &&
+ \frac{1}{(l+n+3/2-\alpha)(l+n+m+\alpha)}\, .
%  \Biggr]
\label{chap4:I2-C}
%\nonumber
\end{flalign}
\end{subequations}

\subsection{Next-to-leading order: gap equation (1)} 

Combining all of the above results, the gap equation (\ref{chap4:gap-eqn-NLO}) may be written in an explicit form as:
\begin{flalign}
1 = \frac{(2+\xi)\beta}{L} + \frac{1}{L^2}\,
\Bigl[8 S(\al,\xi) - 2(2+\xi) \hat{\Pi} \beta \Bigl . 
%\nonum \\
%&&\Bigr . 
+ \left(-\frac{5}{3}+ \frac{26}{3}\xi -3\xi^2\right) \beta^2 - 8\beta\left(\frac{2}{3}(1-\xi)-\xi^2\right) \Bigr]\, , \qquad
\label{chap4:gap-eqn-NLO-explicit}
\end{flalign}
where %$S(\al,\xi) = \Bigl(\hat{\Sigma}_3(\al,\xi)+2(1-\xi) \hat{\Sigma}_2(\alpha)\Bigr)/8$.
\be
S(\al,\xi) = \Bigl(\hat{\Sigma}_3(\al,\xi)+2(1-\xi) \hat{\Sigma}_2(\alpha)\Bigr)/8 \, .
\label{chap4:delta}
\ee
At this point, we consider Eq.~(\ref{chap4:gap-eqn-NLO-explicit}) directly at the critical point $\alpha=1/4$, {\it i.e.}, at $\beta=16$. 
This yields:
\bea
L_c^2 -16(2+\xi) L_c - 8\left [ S(\xi) - 4(2+\xi) \hat{\Pi} %\right .
%\nonum \\
%&& \left . 
- 16 \left(4- 50 \xi / 3 + 5\xi^2\right) \right ] = 0 \, ,
 \label{chap4:Lc-eqn}
\eea
where $S(\xi)= S(\al=1/4,\xi)= (\hat{\Sigma}_3(\xi)+2(1-\xi) \hat{\Sigma}_2)/8$ with $\hat{\Sigma}_2 = \hat{\Sigma}_2(\al=1/4)$ and 
$\hat{\Sigma}_3(\xi)= \hat{\Sigma}_3(\al=1/4,\xi)$. Solving Eq.~(\ref{chap4:Lc-eqn}), we have two standard solutions:
%
%\begin{subequations}
\bea
%\begin{flalign}
%&&
L_{c,\pm} = 8(2+\xi) \pm \sqrt{d_1(\xi)}, \quad %&&
%\label{chap4:Lc-solutions} \\ 
%&&
d_1(\xi)= 8\left [ S(\xi)-8\left(4-\frac{112}{3}\xi+9\xi^2 + \frac{2+\xi}{2} \hat{\Pi} \right) \right ] \, .%&&
\label{chap4:Lc-solutions}
%\nonum %\label{chap4:d1}
%\end{flalign}
\eea
%\end{subequations}
%
In order to provide a numerical estimate for $N_c$, we have used the series representations (\ref{chap4:I1-series}) to evaluate the integrals: 
$\pi \tilde{I}_1(\alpha=1/4) \equiv R_1$ and $\pi \tilde{I}_2(\alpha=1/4 + i\delta) \equiv R_2 - iP_2  \delta + O(\delta^2)$ where $\delta \ra 0$ regulates an artificial singularity in $\pi \tilde{I}_3(\alpha=1/4)=R_2 + P_2/4$.
With 10000 iterations for each series, the following numerical estimates are obtained: %$R_1=163.7428$, $R_2=209.175$, $P_2=1260.720$, 
\be
R_1=163.7428, \quad R_2=209.175, \quad P_2=1260.720 \, ,
\label{chap4:Is-numerics}
\ee
from which the complicated part of the self-energies can be evaluated:
\be
\hat{\Sigma}_2= 4 R_1, \quad \hat{\Sigma}_3(\xi)=(\xi^2-1) R_2 - (7+16 \xi -3 \xi^2)\,P_2 / 16\, .
\label{chap4:sigma2.3}
\ee
Combining these values with the one of $\hat{\Pi}$, yields: 
\be
N_c(\xi=0)=3.29, \quad N_c(\xi=2/3)=3.09\, ,
\label{chap4:Nc-no-resum}
\ee
where ``$-$'' solutions are unphysical and there is no solution in the Feynman gauge.
The range of $\xi$-values for which there is a solution corresponds to $\xi_- \leq  \xi \leq  \xi_+$, where $\xi_{+}=0.88$ and  $\xi_{-}=-2.36$. %The Feynman gauge is excluded from this range.

\subsection{Next-to-leading order: gap equation (2)} 

Following Ref.~\cite{Nash:1989xx}, we would like to resum the LO term together with part of the NLO corrections containing terms $\sim \beta^2$.
In order to do so, we will now rewrite the gap equation (\ref{chap4:gap-eqn-NLO-explicit}) in a form which is suitable for resummation.
This amounts to extract the terms  $\sim \beta$ and $\sim \beta^2$ from the complicated parts of the fermion self-energy, Eqs.~(\ref{chap4:sigma-NLO-22_L}) and (\ref{chap4:sigma-NLO-3AG}).
All calculations done this yields:
\begin{subequations}
\label{chap4:tSigma2}
\bea
&&\hat{\Sigma}_2(\alpha) = \beta \bigl(3\beta -8 \bigr) + \tilde{\Sigma}_2(\alpha)\, , 
\label{chap4:Sigma2} \\
&&\tilde{\Sigma}_2 = \tilde{\Sigma}_2(\alpha=1/4)=4 \tilde{R}_1, \qquad \qquad \tilde{R}_1=3.7428\, ,
\label{chap4:tSigma2-1/4} 
\eea
\end{subequations}
where the rest, $\tilde{\Sigma}_2 = \tilde{\Sigma}_2(\alpha=1/4)$, was determined by imposing Eq.~(\ref{chap4:sigma2.3}). Similarly:
%we have $\overline{\Sigma}_3(\xi) = -4 \xi (4+\xi) \beta + \tilde{\Sigma}_3(\alpha,\xi)$ where:
%
\begin{subequations}
\label{chap4:tSigma3}
\begin{flalign}
&\overline{\Sigma}_3(\xi) = -4 \xi (4+\xi) \beta + \tilde{\Sigma}_3(\alpha,\xi)\, ,  
\label{chap4:tsigma-NLO-3AG0} \\
&\tilde{\Sigma}_3(\alpha,\xi) = \frac{1}{4}\bigl(1+8\xi+\xi^2+2\alpha (1-\xi^2)\bigr)
%\left(\frac{1}{2}+\xi+\xi^2+2\alpha \xi(1-\xi)\right)
\pi\tilde{J}_2(\alpha)
%\nonum \\
+ \frac{1}{2}\bigl(1+4\xi-\alpha (1-\xi^2)\bigr)
%\left(\frac{1}{2}+2\xi-\alpha (1-\xi)^2\right)
\pi\tilde{J}_2(1+\alpha) 
\nonum \\
& \qquad \qquad- \frac{1}{4}\bigl(-7-16\xi+3\xi^2\bigr)
%\Bigl(2+3\xi+\xi^2-\alpha (1-3\xi)(1-\xi)\Bigr)
\pi\tilde{J}_3(\alpha)\, ,
\label{chap4:tsigma-NLO-3AG} \\
&\tilde{\Sigma}_3(\xi) = \bigl(\xi^2-1\bigr)  \tilde{R}_2 -  \bigl(7+16\xi -3\xi^2 \bigr)\,\frac{\tilde{P}_2}{16}\, ,
\label{chap4:tR2}
%\\
%&&\pi \tilde{J}_2(\alpha=1/4) = \pi \tilde{J}_2(\alpha=5/4) = \tilde{R}_2^N = 17.175,~~ \pi \tilde{J}_3(\alpha=1/4) = \tilde{R}_2^N
%+ \frac{\tilde{P}_2^N}{4},~~ \tilde{P}_2^N = -19.28\, ,
%\label{chap4:tsigma-NLO-3AG.1} 
\end{flalign}
\end{subequations}
where the form of the rest, $\tilde{\Sigma}_3(\xi)=\tilde{\Sigma}_3(\alpha=1/4,\xi)$, is imposed by Eq.~(\ref{chap4:sigma2.3}). 
Equating Eqs.~(\ref{chap4:tR2}) with (\ref{chap4:tsigma-NLO-3AG}) for $\al=1/4$ together with using the values:
\begin{subequations}
\label{chap4:tJ}
\begin{flalign}
&\pi \tilde{J}_2(\alpha=1/4) = \pi \tilde{J}_2(\alpha=5/4) = \tilde{R}_2^N = 17.175\, , 
\\
&\pi \tilde{J}_3(\alpha=1/4) = \tilde{R}_2^N + \tilde{P}_2^N/4\, ,
\\
&\tilde{P}_2^N = -19.28\, , 
\end{flalign}
\end{subequations}
yields:
%
%\be
%\tilde{\Sigma}_3(\xi) = \bigl(\xi^2-1\bigr)  \tilde{R}_2 -  \bigl(7+16\xi -3\xi^2 \bigr) \frac{\tilde{P}_2}{16}
%\label{chap4:tR2}
%\ee
%
%Considering the results (\ref{sigma2.3}) and (\ref{tsigma-NLO-3AG0})-(\ref{tsigma-NLO-3AG.1}), we see that
%
\be
\tilde{R}_2 = \tilde{R}_2^N -16 = 1.175, \qquad \qquad \tilde{P}_2 = \tilde{P}_2^N = -19.28\, .
\label{chap4:til}
\ee
With the help of the results (\ref{chap4:tSigma2}) and (\ref{chap4:tSigma3}), the gap equation (\ref{chap4:gap-eqn-NLO-explicit}) may be written as:
\begin{flalign}
&1 = \frac{(2+\xi)\beta}{L} + \frac{1}{L^2}
%\Bigl[8 \tilde{S}(\al,\xi) +\left(\frac{4}{3}(1-\xi) -\xi^2\right) \beta^2 + 
\Bigl[8 \tilde{S}(\al,\xi) -2(2+\xi) \hat{\Pi} \beta  %\Bigr .
%\nonum \\
%&&\Bigl . 
+ \left(\frac{2}{3}-\xi\right)\bigl(2+\xi\bigr)\, \beta^2 +
4\beta\left(\xi^2-\frac{4}{3}\xi - \frac{16}{3}\right) \Bigr]\, , \quad
 \label{chap4:gap-eqn-NLO-explicit_1} 
\end{flalign}
%
%\end{widetext}
where %$\tilde{S}(\al,\xi) = \Bigl(\tilde{\Sigma}_3(\al,\xi)+2(1-\xi) \tilde{\Sigma}_2(\al) \Bigr)/8$.
\be
\tilde{S}(\al,\xi) = \Bigl(\tilde{\Sigma}_3(\al,\xi)+2(1-\xi) \tilde{\Sigma}_2(\al) \Bigr)/8 \, .
\label{chap4:tdelta} 
\ee
At this point Eqs.~(\ref{chap4:gap-eqn-NLO-explicit}) and (\ref{chap4:gap-eqn-NLO-explicit_1}) are strictly equivalent to each other and yield the same values for $N_c(\xi)$.

\subsection{Next-to-leading order: resummation} 

Eq.~(\ref{chap4:gap-eqn-NLO-explicit_1}) is the convenient starting point to perform a resummation of the fermion anomalous dimension.
Up to second order, the expansion of the latter reads:
\be
\gamma_\psi = \frac{\lambda^{(1)}}{L} + \frac{\lambda^{(2)}}{L^2} + \cdots, \quad \lambda^{(1)} = 4 \left(\frac{2}{3}-\xi \right)\, ,
\label{chap4:lambdaA}
\ee
where $\lambda^{(1)}$ is the LO part and $\lambda^{(2)}$ the NLO one. The latter can be obtained
from Gracey's calculations~\cite{Gracey:1993sn} and reads:
\bea
\lambda^{(2)} = - 8 \left(\frac{8}{27} + \left(\frac{2}{3}-\xi \right) \hat{\Pi} \right)\, .
\label{chap4:lambdaA-NLO}
\eea

As can be seen from Eq.~(\ref{chap4:gap-eqn-NLO-explicit_1}), the NLO term $\sim \beta^2$ is proportional to the LO fermion anomalous dimension. This term, together
with the LO term in the gap equation, can be thought of the first and zeroth order terms, respectively, of an expansion in $\gamma_\psi$.
Following Nash, it is possible to resum the full expansion of $\gamma_\psi$ at the level of the gap equation. 
In order to implement this resummation, let us first consider the integral:
\be
%1=\frac{1}{\Sigma (p)} \frac{4(2+\xi)}{\pi^2 N}
   \int_0^{a} \! \! \! d |k| \,
\frac{\Sigma (|k|) }{ \mbox{Max}({|k|,|p|})} {\left[\frac{\mbox{Max}({|k|,|p|})}{\mbox{Min}({|k|,|p|})}\right]}^{\lambda} ~~ ,
\label{app:SD-sigma-LO3-l}
\ee
with some arbitrary $\lambda$. Using the fact that: $\Sigma (p) = B (p^2)^{ -\alpha}$, we have:
\begin{flalign}
1=\frac{1}{\Sigma (p)}
%\frac{4(2+\xi)}{\pi^2 N}
   \int_0^{a} \! \! \! d |k| \,
\frac{\Sigma (|k|) }{ \mbox{Max}({|k|,|p|})} {\left[\frac{ \mbox{Max}({|k|,|p|})}{ \mbox{Min}({|k|,|p|})}\right]}^{\lambda}
&= \left(\frac{1}{2\alpha-\lambda} + \frac{1}{1-2\alpha-\lambda} \right) 
\label{app:SD-sigma-LO3-la} \\
&=  \frac{1-2\lambda}{(2\alpha-\lambda)(1-2\alpha-\lambda)}
\, ,
\nonum
\end{flalign}
which is linear in $\beta$ in the limit $\lambda \ra 0$.
Taking the derivative of $\lambda$ and putting $\lambda=0$, we have another important integral:
\begin{flalign}
1=\frac{1}{\Sigma (p)}
%\frac{4(2+\xi)}{\pi^2 N}
   \int_0^{a} \! \! \! d |k| \,
\frac{\Sigma (|k|) }{ \mbox{Max}({|k|,|p|})} \ln \left[\frac{ \mbox{Max}({|k|,|p|})}{ \mbox{Min}({|k|,|p|})}\right]
&= \left(\frac{1}{(2\alpha)^2} + \frac{1}{(1-2\alpha)^2} \right) 
\label{app:SD-sigma-LO3-lo} \\
&=  \frac{\beta}{16}\bigl(\beta -8 \bigr)\, ,
\nonum
%\label{app:SD-sigma-LO3-lo}
\end{flalign}
which now contains terms linear and quadratic in $\beta$.
So, now, we can represent the gap equation (\ref{chap4:gap-eqn-NLO-explicit_1}) in the form:
\be
1 = \frac{4(2+\xi)}{L\Sigma (p)}  \int_0^{a} \! \! \! d |k| \, \frac{\Sigma (|k|) }{ \mbox{Max}({|k|,|p|})}
\left\{1 + \frac{4(2-3\xi)}{3L} \ln \left[\frac{ \mbox{Max}({|k|,|p|})}{ \mbox{Min}({|k|,|p|})}\right] \right\}
+ \frac{\Delta(\al,\xi)}{L^2} \, ,
\label{app:gap-eqn-NLO-explicit_2}
\ee
where
\bea
\Delta(\al,\xi)  =   8 \tilde{S}(\al,\xi) - 4\beta\,(\xi^2 +4\xi + 8/3)  - 2\beta\,(2+\xi)\,\hat{\Pi} \, .
\label{chap4:Delta}
\eea
Following Nash \cite{Nash:1989xx}, the integral (\ref{app:gap-eqn-NLO-explicit_2}) may be viewed as the first two orders of the expansion of the integral (\ref{app:SD-sigma-LO3-l})
with the anomalous dimension $\lambda$ corresponding fermion anomalous dimension:
\be
\gamma_\psi = \frac{\lambda^{(1)}}{L} + \frac{\lambda^{(2)}}{L^2} + ... , \qquad \lambda^{(1)}= 4 \left(\frac{2}{3}-\xi \right)\, ,
\label{app:lambda}
\ee
where $\lambda^{(1)}$ is the LO part and $\lambda^{(2)}$ the NLO one. In order to resum this contribution, we perform the following replacement:
\be
 \int_0^{a} \! \! \! d |k| \, \frac{\Sigma (|k|) }{ \mbox{Max}({|k|,|p|})}
\left\{1 + \frac{4(2-3\xi)}{3L} \ln \left[\frac{ \mbox{Max}({|k|,|p|})}{ \mbox{Min}({|k|,|p|})}\right] \right\} \to
\int_0^{a} \! \! \! d |k| \, \frac{\Sigma (|k|) }{ \mbox{Max}({|k|,|p|})}
 {\left[\frac{ \mbox{Max}({|k|,|p|})}{ \mbox{Min}({|k|,|p|})}\right]}^{\gamma_\psi}\, .
\ee
After this replacement, the gap equation  (\ref{app:gap-eqn-NLO-explicit_2}) takes the form:
\be
1 = \frac{4(2+\xi)}{L}  \frac{1-2\gamma_\psi}{(2\alpha-\gamma_\psi)(1-2\alpha-\gamma_\psi)}  + \frac{\Delta(\al,\xi)}{L^2} \, .
\label{app:gap-eqn-NLO-explicit_3}
\ee
It is convenient to multiply Eq.~(\ref{app:gap-eqn-NLO-explicit_3}) by the factor $(2\alpha-\gamma_\psi)(1-2\alpha-\gamma_\psi)$. This yields:
\be
(2\alpha-\gamma_\psi)(1-2\alpha-\gamma_\psi) = \frac{4(2+\xi)}{L} \bigl(1-2\gamma_\psi\bigr) + (2\alpha-\gamma_\psi)(1-2\alpha-\gamma_\psi)
\frac{\Delta(\al,\xi)}{L^2} \, .
\ee
Note that the l.h.s. can be represented as $2\alpha(1-2\alpha) - \gamma_\psi(1-\gamma_\psi)$ which leads to:
\be
2\alpha(1-2\alpha) = \gamma_\psi(1-\gamma_\psi) + \frac{4(2+\xi)}{L} \bigl(1-2\gamma_\psi\bigr) + (2\alpha-\gamma_\psi)(1-2\alpha-\gamma_\psi)
\frac{\Delta(\al,\xi)}{L^2} \, .
\label{app:gap-eqn-NLO-explicit_4}
\ee
From Eq.~(\ref{app:gap-eqn-NLO-explicit_4}), we see that, after resummation, $\gamma_\psi$, Eq.~(\ref{app:lambda}), will contribute to the gap equation up to NLO. 
With the help of the expression of $\lambda^{(2)}$ in Eq.~(\ref{chap4:lambdaA-NLO}), we have:
\bea
\gamma_\psi(1-\gamma_\psi) + \frac{4(2+\xi)}{L} \bigl(1-2\gamma_\psi\bigr) &=& \frac{4(2+\xi) + \lambda^{(1)}}{L} +  \frac{1}{L^2} \left(
\lambda^{(2)} - {\bigl(\lambda^{(1)}\bigr)}^2 - 8(2+\xi) \lambda^{(1)}\right) \nonumber \\
&=& \frac{32}{3L} +  \frac{1}{L^2} \left(\lambda^{(2)} - 4\lambda^{(1)}
\left(\frac{14}{3}+\xi \right) \right)\, ,
\eea
which shows the complete cancellation of the $\xi$-dependence at LO. Now it is convenient to return to the standard form for the gap equation
by multiplying Eq.~(\ref{app:gap-eqn-NLO-explicit_4}) by the factor $1/[ 2\alpha(1-2\alpha)]$. This yields:
\be
1 = \frac{8\beta}{3L} +   \frac{\beta}{4L^2} \left(\lambda^{(2)} - 4\lambda^{(1)}
\left(\frac{14}{3}+\xi \right) \right) + \frac{\Delta(\al,\xi)}{L^2} \, ,
\label{app:gap-eqn-NLO-explicit_5a}
\ee
or more explicitly:
\be
1 = \frac{8\beta}{3L}  + \frac{1}{L^2}\, \Bigl [ 8 \tilde{S}(\al,\xi) - \frac{16}{3} \, \beta \, \left(\frac{40}{9} + \hat{\Pi} \right) \Bigr] \, ,
\label{chap4:gap-eqn-NLO-explicit_5b}
\ee
where $\Delta(\al,\xi)$ was given in Eq.~(\ref{chap4:Delta}).
Interestingly, the LO term in Eq.~(\ref{chap4:gap-eqn-NLO-explicit_5b}) is now gauge independent. 
Remarkably, there is also a strong suppression of the gauge dependence at NLO as $\xi$-dependent terms do exist but they enter the gap equation only through the rest, $\tilde{S}$, which is very small numerically.

We now consider Eq.~(\ref{chap4:gap-eqn-NLO-explicit_5b}) at the critical point, $\al=1/4$ ($\beta=16$), which yields:
\bea
%L_c^2 -\frac{128}{3} L_c - \tilde{\Delta}(\xi) = 0 \, ,
L_c^2 -\frac{128}{3} L_c - \Bigl[ 8 \tilde{S}(\xi) - \frac{256}{3}\,\left(\frac{40}{9} + \hat{\Pi} \right) \Bigr] = 0 \, .
\label{chap4:Lc-eqn_1}
\eea
%
%where $ \tilde{\Delta}(\xi)= \tilde{\Delta}(\al=1/4,\xi)$. 
Solving Eq.~(\ref{chap4:Lc-eqn_1}), we have two standard solutions:
%
%\begin{subequations}
\begin{flalign}
L_{c,\pm} = \frac{64}{3} \pm \sqrt{d_2(\xi)}, \qquad d_2(\xi)= {\left(\frac{64}{3}\right)}^2 + \Bigl[ 8 \tilde{S}(\xi) - \frac{256}{3}\,\left(\frac{40}{9} + \hat{\Pi} \right) \Bigr]\, . %\, ,
\label{chap4:Lc-solutions_1} %\\
%&&d_2(\xi)= {\left(\frac{64}{3}\right)}^2 + \Bigl[ 8 \tilde{S}(\xi) - \frac{256}{3}\,\left(\frac{40}{9} + \hat{\Pi} \right) \Bigr]\, . \quad
%8\left(S(\xi)-8\left(4-\frac{112}{3}\xi+3\xi^2\right)- 4(2+\xi) \hat{\Pi}\right) 
%\label{chap4:d2}
\end{flalign}
%\end{subequations}
%
In order to provide a numerical estimate for $N_c$,  we have used the values of $\tilde{R}_1$, $\tilde{R}_2$ and $\tilde{P}_2$ of Eqs.~(\ref{chap4:tSigma2}) and (\ref{chap4:til}).
Combining these values with: %$\tilde{S}(\xi=0)=\tilde{R}_1 - \tilde{R}_2/8 - 7\tilde{P}_2/128$, 
%$\tilde{S}(\xi=1)= -5\tilde{P}_2/ 32$ and $\tilde{S}(\xi=2/3)=\tilde{R}_1 / 3 - 5\tilde{R}_2 / 72 - 49 \tilde{P}_2/384$,
%
\begin{subequations}
\bea
&&\tilde{S}(\xi=0)=\tilde{R}_1-\frac{\tilde{R}_2}{8}-\frac{7\tilde{P}_2}{128}\, , 
\\ 
&&\tilde{S}(\xi=1)= -\frac{5\tilde{P}_2}{32}\, ,
\\
&&\tilde{S}(\xi=2/3)=\frac{\tilde{R}_1}{3}-\frac{5\tilde{R}_2}{72} - \frac{49 \tilde{P}_2}{384}\, ,
\eea
\end{subequations}
together with the value of $\hat{\Pi}$, yields, for $L_c(\xi)$ and $N_c(\xi)$ (``$-$'' solutions being unphysical): 
\begin{subequations}
\bea
&&L_c(0)=30.44, \qquad L_c(2/3)=29.98, \qquad L_c(1)=29.69\, , %\qquad
\label{chap4:Lc-res} \\
&&N_c(0)=3.08,  \qquad N_c(2/3)=3.04,  \qquad N_c(1)=3.01\, . %\qquad
\label{chap4:Nc-values}
\eea
\end{subequations}
Actually, solutions exist for a broad range of values of $\xi$: $\xi_- \leq  \xi \leq  \xi_+$,
where $\xi_{+}=4.042$ and  $\xi_{-}=-8.412$; this is consistent with the weak $\xi$-dependence of the gap equation. 
Moreover, following \cite{Gorbar:2001qt}, it seems that a %the ``right(est)'' 
good gauge choice is one close to $\xi=2/3$ where the LO fermion wave function is finite (there is also strong support for the Landau gauge, see the last section for conclusion and a more detailed discussion of the results).
 %Indeed, as can be seen by comparing Eqs.~(\ref{Nc-no-resum}) and (\ref{Nc-values}), with (without) resummation, the value of $N_c(\xi)$ increases (decreases) with increasing $\xi$. 
Indeed, as can be seen by comparing Eqs.~(\ref{chap4:Nc-no-resum}) and (\ref{chap4:Nc-values}), upon resumming the theory, the value  of $N_c(\xi)$ increases (decreases) for small (large)
values of $\xi$. For $\xi=2/3$, the value of $N_c$ is very stable, decreasing
only by $1$-$2\%$ during resummation. Finally, if we neglect the rest, {\it i.e.}, $\tilde{S}(\xi)=0$ in Eq.~(\ref{chap4:Lc-eqn_1}), the gap equation becomes $\xi$-independent and we have:
\be
\overline{L}_c=28.0981, \qquad \overline{N}_c=2.85\, .
\label{chap4:overlN}
\ee
The results of Eq.~(\ref{chap4:overlN}) are in full agreement with the recent results of \cite{Gusynin:2016som}. We shall come back on this important (gauge-invariant) result in the Conclusion of this Chapter.

%Actually, the prescription of \cite{Gusynin:2016som} consists
%in an NLO expansion for the index $\al$ rather than for the mass function. corrections
%have been analysed in an approximation corresponding to $\tilde{S}(\xi)=0$, {\it i.e.}, taking into account
%only the NLO terms $\sim \beta$ and $\sim \beta^2$. 

\section{Critical properties of reduced QED$_{4,3}$} 
\label{chap4:sec:rqed}

As discussed at length in the Introduction %and in Chap.~\ref{chap3}, reduced QED$_{4,3}$ is a relativistic low-energy effective action for graphene and other Dirac systems 
 %at their Lorentz invariant fixed point.  
a commonly raised theoretical issue in the physics of Dirac materials, and in particular graphene, is to know whether or not the strong and long-ranged Coulomb interaction may drive the semi-metallic 
system into an insulating phase thereby dynamically generating a gap in the single particle fermionic spectrum, see Chap.~\ref{chap1} and references therein.
Let's recall that this excitonic instability should take place for values of the coupling constant larger than a critical one, $\al_c$, {\it i.e.}, a gap is dynamically generated for
$\al > \al_c$ for the physical number of fermion flavours (spin): $N=2$. Alternatively, the instability would also manifest at $N <N_c$ where 
$N_c$ is a critical fermion flavour number for which $\al \to \infty$. A quantitative analysis of such an instability is important given the fact that
the possibility to generate in a controlled way a fermion gap in graphene and graphene-like materials is crucial for, {\it e.g.}, the development of graphene-based transistors  \cite{nevius2015}.
Such an instability is the condensed matter physics analogue of the dynamical mass generation and associated dynamical
chiral symmetry breaking (D$\chi$SB) in QED$_3$ that we have studied in the previous sections. Of theoretical importance, is to
clarify the precise relation between QED$_3$ and effective field theories describing planar Dirac liquids.

A puzzling fact is that, despite the strength of the interaction, there is no experimental evidence for the existence of a gap of more than $0.1$meV
in clean suspended graphene \cite{elias2011cones}. This may be the indication that $\al$ is not large enough in actual samples and/or that the latter may be subject to additional sources of screening,
{\it e.g.}, from electrons in $\sigma$ bands \cite{Ulybyshev:2013swa}. Neglecting the latter and focusing on clean graphene at zero temperature, there have been many attempts 
to compute $\al_c$ on the basis of an elementary model 
of massless Dirac fermions interacting via the instantaneous Coulomb interaction, {\it i.e.}, the limit $v/c \ra 0$ which is indeed quite realistic. 
Various methods were used such as, {\it e.g.}, analytical or numerical solutions of LO SD equations 
\cite{Khveshchenko:2001zz,Gorbar:2002iw,Leal:2003sg,Khveshchenko:2008ye,Liu09.PhysRevB.79.205429,Gamayun:2009em,Wang2012,Popovic13.PhysRevB.88.205429,Gonzalez15.PhysRevB.92.125115,Carrington:2017hlc}, renormalization group studies 
\cite{Son07.PhysRevB.75.235423,Gonzalez12.PhysRevB.85.085420}, 
lattice simulations \cite{Drut:2008rg,Buividovich12.PhysRevB.86.245117,Ulybyshev:2013swa} and a combination of Bethe-Salpeter and functional renormalization group approaches \cite{Katanin16.PhysRevB.93.035132}.
All computations agree on the fact that $\al_c \sim \Ord(1)$ but there is still
no general agreement on the precise value of $\al_c$ 
even though recent results \cite{Wang2012,Popovic13.PhysRevB.88.205429,Gonzalez15.PhysRevB.92.125115,Katanin16.PhysRevB.93.035132} 
seem to indicate that it is indeed larger than the bare value of $\al$ ($\al_g =2.2$) in agreement with the experimentally observed semi-metallic behaviour of graphene.

Following \cite{Kotikov:2016yrn}, we will address the problem of dynamical gap generation in planar Dirac liquids from an original point of view with respect to the literature on the subject.
We shall focus on the deep infra-red Lorentz invariant fixed point, {\it i.e.}, the limit $v/c \ra 1$, where these systems
may be effectively described by reduced QED$_{4,3}$ that has been studied in Chap.~\ref{chap3}.
The results that we will present below provide an exact analytical solution to the SD equations of RQED$_{4,3}$ up to NLO including a full resummation of the fermion anomalous dimension and leading to 
gauge-invariant results for $\al_c$ and $N_c$. They are based on an important correspondence between RQED$_{4,3}$ and large-$N$ QED$_3$ which will allow us
to transcribe the exact NLO gap equation found for QED$_3$ in the last sections to the case of RQED$_{4,3}$ (as we will see, it is crucial for that matter that large-$N$ QED$_3$ was solved for an arbitrary gauge-fixing term). 
Solving the gap equation will yield high precision estimates of $\al_c$ and $N_c$. 

\subsection{Correspondence between reduced QED$_{4,3}$ and QED$_3$}
\label{chap4:sec:cor}

Our goal is to study dynamical gap generation in model III, Eq.~(\ref{chap1:rqed}), in the case $d_e =3$ and $d_\gamma=4$. The perturbative structure of the model has been extensively studied in Chap.~\ref{chap3}.
Similarly to QED$_3$, this model has a $U(2N)$ flavour symmetry and dynamical mass generation
leads to a spontaneous breakdown of this symmetry to $U(N) \times U(N)$.~\footnote{Notice that the $U(4)$ (for $N=2$) symmetry appears only for the continuum model (relativistic or not). A lattice description of graphene
based on a tight-biding model still features some elements of this continuous symmetry: sublattice symmetry (invariance upon exchanging the two triangular sublattices that form the honeycomb lattice). 
The breaking of the $U(4)$ symmetry may indeed be seen as breaking this sublattice symmetry. The corresponding gap is analogous to the (topologically trivial) one found, \eg, in boron nitride 
systems due to the asymmetry between the two sublattices.} Notice, however, that while QED$_3$ is super-renormalizable
and has an intrinsic mass scale fixed by the dimensionful coupling constant $a =N e^2/8$, RQED$_{4,3}$ is renormalizable
and has a dimensionless coupling $\al = e^2/(4\pi)$. As noticed in \cite{Gorbar:2001qt}, D$\chi$SB
in RQED$_{4,3}$ corresponds to the so-called conformal phase transition (CPT) whereas it is only a pseudo-CPT in QED$_3$ \cite{Miransky:1996pd}.
Nevertheless, in both cases, the dynamical mass satisfies the Miransky scaling albeit with different coefficients in front of the exponential, see \cite{Gorbar:2001qt}:~\footnote{As recalled previously: 
see \cite{Fomin:1978rk} where such type of scaling appeared for the first time within the study of the instability of massless QED$_4$ and \cite{Fomin:1984tv} for a review.}
\be
\Sigma_S(0) \simeq \mu\,\exp \Bigl[ - \frac{2 \pi}{(\al/\al_c - 1)^{1/2}} \Bigr]\, ,
\label{chap4:dynm-rqed3}
\ee
where $\mu$ is the renormalization scale (and $N$ is kept fixed), {\it c.f.}, compare with Eq.~(\ref{chap4:dynm-qed3}) for QED$_3$.
Just as we did in the case of large-$N$ QED$_3$, in the following, we shall not be interested by the full scaling but rather focus on the critical behaviour of RQED$_{4,3}$, {\it i.e.}, 
consider the transition point $\al=\al_c$ or $N = N_c$ where the dynamical mass first appears.

In order to study the dynamical generation of such a mass we need to solve the SD equations for the fermion propagator. 
This is possible to do on the basis of the analysis of Chap.~\ref{chap3}. An alternative way to derive the same results is based on a %there is no need to repeat these calculations for RQED$_{4,3}$. A 
simple correspondence between RQED$_{4,3}$ and large-$N$ QED$_3$. As we shall demonstrate in the following, the latter will allow us to straightforwardly study the critical properties of RQED$_{4,3}$ 
on the basis of the results obtained for large-$N$ QED$_3$ without any  further complicated calculation.

We first recall the photon propagator  of RQED$_{4,3}$ which was given in Eq.~(\ref{chap3:RQED:FR:Dmunu0}) (for $\veps_e = 1/2$) and that we reproduce here for clarity: %(in Euclidean space):
\begin{flalign}
D_{\text{RQED}}^{\mu \nu}(p)
= \frac{\I}{2\,[-p^2]^{1/2}} P^{\mu \nu}\left(p\,;\frac{\eta}{2}\right),~~~
P^{\mu \nu}(p\,;\tilde{\eta}) = g^{\mu \nu} - \tilde{\eta} \frac{p^{\mu}p^{\nu}}{p^2}\, . 
\label{chap4:D-RQED}
\end{flalign}
From Eq.~(\ref{chap4:D-RQED}) we see that we may define an effective gauge fixing parameter, $\tilde{\eta} = 1 - \tilde{\xi}$, 
for the reduced gauge field which is related to the gauge fixing parameter, $\eta = 1-\xi$, 
of the corresponding $4$-dimensional gauge field as follows:
\be
%\tilde{\eta}=1-\tilde{\xi},~~ \bar{\tilde{\eta}}=1-\bar{\tilde{\xi}},~~
\tilde{\eta}= \frac{\eta}{2}, \qquad \tilde{\xi} = \frac{1+\xi}{2}\, .
\label{chap4:defxi}
\ee

Next, we consider the photon propagator %$D^{\mu \nu}_{\text{QED}3}(p)$ 
of QED$_{3}$ %in the $1/N$-expansion \cite{Pisarski84,AppelquistNW88} 
in a non-local $\tilde{\xi}$-gauge:
\be
D^{\mu \nu}_{\text{QED}3}(p) = %\frac{g_{\mu \nu} - (1-\tilde{\xi})p_{ \mu} p_{ \nu} / p^2}
%-i {d^{\mu \nu}(\tilde{\eta})}{p^2 \left[1 + \Pi (p) \right]} =i 
\frac{\I\,P^{\mu \nu}(\tilde{\eta})}{[-p^2]\, \left[1 - \Pi (p^2) \right]} \, ,
\label{chap4:D-QED3}
\ee
where $\Pi(p^2)$ is the polarization operator. Notice that, because in QED$_3$ the gauge field is $2+1$-dimensional, it is actually the reduced gauge-fixing parameter, $\tilde{\xi}$, which enters its expression.
% and $ \Gamma ^{ \nu}(p,k)$ is the vertex function.
As we saw in Chap.~\ref{chap2}, and as displayed also in Eq.~(\ref{chap4:LO0}), at the LO of the $1/N$-expansion, the polarization operator reads:
\be
\Pi_1(p^2) = -\frac{a}{[-p^2]^{1/2}}\, .% a = \frac{Ne^2}{8}
\label{chap4:Pi1}
\ee
The behaviour of $D^{\mu \nu}_{\text{QED}3}(p)$ is softened in the infra-red
because: %changes its infrared behavior \cite{Kotikov93+12,KoShiTe} because
\be
D^{\mu \nu}_{\text{QED}3}(p) = \frac{8\I}{e^2 N [-p^2]^{1/2}}\, P^{\mu \nu}(p\,;\tilde{\eta}) \, .
\label{chap4:D-QED3-N}
\ee
As we commented on before, following \cite{Teber:2012de,Kotikov:2013kcl,Kotikov:2013eha}, the photon propagators of QED$_3$ in the IR limit, Eq.~(\ref{chap4:D-QED3-N}), and the
one of RQED$_{4,3}$, Eq.~(\ref{chap4:D-RQED}) have the same form. One may easily pass from one form to the other with the help of the following transformation:
\be
\frac{1}{L} \to \frac{e^2}{16\pi^2} = \frac{\al}{4\pi} \equiv g, \quad \tilde{\eta} \to \frac{\eta}{2} \quad \left( \tilde{\xi} \to \frac{1+\xi}{2} \right) \, ,
\label{chap4:transform.1}
\ee
where $L=\pi^2 N$ as in previous sections.
The transformation (\ref{chap4:transform.1}) will allow us to transcribe the solution
of dynamic mass generation in QED$_3$ using the $1/N$-expansion presented in the previous paragraphs \cite{Kotikov:2016wrb,Gusynin:2016som,Kotikov:2016prf}
to the case of RQED$_{4,3}$ using the loop expansion. In the following, LO will either refer to LO 
in the $1/N$-expansion for QED$_3$ or to the one-loop order for RQED$_{4,3}$. Similarly, NLO will either refer
to NLO in the $1/N$-expansion for QED$_3$ or to the two-loop order for RQED$_{4,3}$. Of course, the solution of SD
equations, combined with the various resummations we shall perform in the following, is non-perturbative in nature
and beyond the reach of a simple $1/N$ or loop expansion.

\subsection{Leading order} 
\label{chap4:sec:LO}

In order to illustrate how the correspondence works, we first compute the critical coupling at LO.
Combining the LO QED$_3$ result of Eq.~(\ref{chap4:gap-eqn-LO})  with the transformation (\ref{chap4:transform.1}), the LO gap equation for the critical coupling of RQED$_{4,3}$ reads:
%At LO the critical value for $L$ has the form
\be
%L_c = 16(2+\tilde{\xi})  =  8(5+\xi)\, ,
1 = \frac{16(2+\tilde{\xi})}{L_c} \quad \to \quad 1 = 16(2+\tilde{\xi})g_c = 8(5+ \xi) g_c \, ,
\label{chap4:GE-LO}
\ee
where we find it convenient in this chapter to introduce the notation:
\be
g_c \equiv \bar{\al}_c = \frac{\al_c}{4\pi}\, . 
\ee
This yields:
\be
\al_c(\xi) = \frac{\pi}{2(5 + \xi)}\, .
\label{chap4:alc-LO}
\ee
The LO critical coupling is seen to be strongly gauge dependent but does not depend on the fermion flavour number, $N$. 
Its value in various gauges, including Landau ($\xi=0,\tilde{\xi}=1/2$) and Feynman ($\xi=\tilde{\xi}=1$) gauges, 
reads:
\begin{subequations}
\label{chap4:alc-LO-values}
%\bea
\begin{flalign}
&\alpha_c (\xi=0)=0.3142, \quad ~~ \alpha_c(\xi=1) =0.2618\, ,
%L_c = 40,~~ g_c=\frac{1}{40},~~ e^2_c=\frac{2\pi^2}{5}=3.9478,~~e_c=1.9869, ~~\alpha_c=0.3142,
%\label{alc-LO-LG} \\
%&&\xi=1: \qquad \alpha_c=0.2618 \, .
% ~~ L_c = 48,~~ g_c=\frac{1}{49},~~ e^2_c=\frac{\pi^2}{3}=3.2899,~~e_c=1.8138, ~~\alpha_c=0.2618 \, ,
% \label{alc-LO-FG}
\label{chap4:alc-LO-values1}\\
&\alpha_c(\xi=-1) =0.3927, \quad \alpha_c(\xi=1/3) = 0.2945\, .
\label{chap4:alc-LO-values2}
\end{flalign}
%\eea
\end{subequations}
The gauge ($\xi=-1,\tilde{\xi}=0$) corresponds to the Landau gauge for the reduced gauge field while 
the gauge ($\xi=1/3,\tilde{\xi}=2/3$) will be discussed later.
 
Following, {\it e.g.}, \cite{Gorbar:2002iw,Khveshchenko:2008ye,Liu09.PhysRevB.79.205429}
the dynamical screening of the interaction may be included, in the so-called random-phase approximation (RPA), 
by resumming the one-loop polarization operator, Eq.~(\ref{chap4:Pi1}); notice that the LO polarization operator in QED$_3$ and the one-loop
polarization operator in RQED$_{4,3}$ are equal.
Contrarily to the case of QED$_{3}$, however, in RQED$_{4,3}$ such a resummation does not change the infrared property of the corresponding photon propagator
\cite{Teber:2012de} (see Eq.~(\ref{chap3:RQED4,3:Dr})) which then reads:
\be
D_{RQED}^{\mu \nu}(p) =
\frac{\I}{2\,[-p^2]^{1/2}\,(1+N e^2/16)} P^{\mu \nu}\left(p\,;\frac{\eta}{2}\right) \, .
\label{chap4:D-RQED-RPA}
\ee
This RPA resummation may be taken into account with the help of a simple redefinition of the coupling constant which becomes:
\be
\tilde{\al} = \frac{\al}{1+Ne^2/16}\, .
\label{chap4:al-RPA}
\ee
Accordingly, the transformation (\ref{chap4:transform.1}) has to be replaced by:
\be
\frac{1}{L} \to \frac{\tilde{e}^2}{16\pi^2} = \frac{\tilde{\al}}{4\pi} \equiv \tilde{g}, \quad \tilde{\eta} \to \frac{\eta}{2} \quad \left( \tilde{\xi} \to \frac{1+\xi}{2} \right) \, .
\label{chap4:transform.2}
\ee
The gap equation (\ref{chap4:GE-LO}) then immediately yields the LO RPA critical coupling constant:
\be
\al_c(\xi) = \frac{\pi}{2(5+\xi) - N \pi^2/4}\, ,
\label{chap4:alc-LO+RPA}
\ee
which now is not only gauge dependent but also depends on the number of fermion flavours, $N$.
For $N=2$, Eq.~(\ref{chap4:alc-LO+RPA}) yields the following values in various gauges:
\begin{subequations}
\label{chap4:alc-LO+RPA-values}
%\bea
\begin{flalign}
&\alpha_c (\xi=0) = 0.6202, \quad ~~ \alpha_c(\xi=1) = 0.4447\, ,
\label{chap4:alc-LO+RPA-values1}\\
&\alpha_c(\xi=-1) = 1.0249, \quad \alpha_c(\xi=1/3) = 0.5481\, .
\label{chap4:alc-LO+RPA-values2}
\end{flalign}
%\eea
\end{subequations}
Dynamical screening therefore increases the value of the critical coupling, compare (\ref{chap4:alc-LO+RPA-values}) with (\ref{chap4:alc-LO-values}).
It is also convenient to find the critical fermion flavour number $N_c$ for which $\al_c \to \infty$.
Eq.~(\ref{chap4:alc-LO+RPA}) yields:
\be
N_c(\xi) = \frac{8(5 + \xi)}{\pi^2}\, .
\label{chap4:Nc-LO}
\ee
This number coincides 
%exactly 
 with the critical flavour number $N_c$ for D$\chi$SB in QED$_3$ which is defined as $N_c=L_c/\pi^2$.
In various gauges, its value reads:
\begin{subequations}
\label{chap4:Nc-LO+RPA-values}
%\bea
\begin{flalign}
&N_c(\xi=0) = 4.0529, \quad ~~ N_c(\xi=1) = 4.8634\, ,
\label{chap4:Nc-LO+RPA-values1}\\
&N_c(\xi=-1) = 3.2423, \quad N_c(\xi=1/3) = 4.3230\, .
\label{chap4:Nc-LO+RPA-values2}
\end{flalign}
%\eea
\end{subequations}
Notice that the value $N_c(1/3) = 128/(3\pi^2)$ has already been obtained in \cite{Gorbar:2001qt} where the importance of the 
$\xi=1/3$ gauge has been emphasized following the seminal work of Nash \cite{Nash:1989xx}, see also discussions in \cite{Kotikov:2016prf}. 
As shown by Nash \cite{Nash:1989xx} (see previous sections), the fermion anomalous dimension may be resummed at the level of the gap equation.
The peculiar $\xi=1/3$ gauge is the one where the fermion anomalous dimension vanishes at LO and for which Nash's 
resummation does not affect much the results. We shall confirm this below.

\subsection{Next-to-Leading order: transformations} 
\label{sec:NLO}

From our NLO results for QED$_3$ \cite{Kotikov:2016prf}, combined with the transformations (\ref{chap4:transform.1}) 
or (\ref{chap4:transform.2}), we may compute the NLO coupling constant of RQED$_{4,3}$. However, in order to properly do so, some additional
replacements are necessary.

The first additional replacement is related to the fact that, in QED$_3$, the NLO polarization operator $\hat{\Pi}_2$ \cite{Kotikov:2016prf} (that was simply noted $\hat{\Pi}$ in Eq.~(\ref{chap4:sigma-NLO-1})) 
contributed to the NLO result within the framework of the $1/N$ expansion:
\be
\Pi_2^{(\text{QED$_3$})}(p) = \frac{2a}{N\pi^2}\,\frac{\hat{\Pi}_2}{|p|}, \quad \hat{\Pi}_2 = \frac{92}{9} - \pi^2 \, .
\label{chap4:Pi2-QED3}
\ee
On the other hand, within the framework of standard perturbation theory which applies to RQED$_{4,3}$, it is only the LO (one-loop) polarization operator, Eq.~(\ref{chap4:Pi1}), which contributes at NLO (two-loop).
From the ratio of Eqs.~(\ref{chap4:Pi2-QED3}) and (\ref{chap4:Pi1}) we define $\hat{\Pi}_1$ which is such that:
\be
\frac{\Pi_2^{(\text{QED$_3$})}(p)}{\Pi_1(p)} = \frac{\hat{\Pi}_2}{\hat{\Pi}_1}, \quad \hat{\Pi}_1 = \frac{N \pi^2}{2} \, .
\label{chap4:hatPi1}
\ee
So, in going from large-$N$ QED$_3$ to RQED$_{4,3}$, the first additional replacement reads:
\be
\hat{\Pi}_2  \quad  \to \quad \hat{\Pi}_1 \, .
\label{chap4:transform.3}
\ee
As we shall see below, of importance will be the fact that the values of $\hat{\Pi}_1$ are large in comparison with those of $\hat{\Pi}_2$. 

The second additional replacement is related to the use of the non-local $\tilde{\xi}$-gauge in the case of QED$_{3}$ which comes from the IR property of $\Pi(q^2)$. Transforming back to the non-local
$\tilde{\xi}$-gauge of RQED$_{4,3}$ (which originates from integrating the gauge-field in the third spacial dimension) amounts to apply the following simple condition: 
\be
\tilde{\xi} \hat{\Pi}_1 =  0\, .
\label{chap4:transform.4}
\ee

With these additional transformations we are now in a position to transcribe the NLO results of large-$N$ QED$_3$ to the case of RQED$_{4,3}$.
In the following, we shall first solve the NLO gap equation without Nash's resummation and then with Nash's resummation.~\footnote{Let's just remark that these 
transformations allow to recover the result for the two-loop fermion anomalous dimension in RQED$_{4,3}$ derived in the previous
section, see Eq.~(\ref{chap3:res2l:RQED4,3:gammapsi-2}), from Gracey's result \cite{Gracey:1993sn} for the NLO fermion anomalous dimension in QED$_{3}$, see Eqs.~(\ref{chap4:lambdaA}) and (\ref{chap4:lambdaA-NLO}).
Indeed, we have:
\bea
\gamma_\psi = \frac{4}{L}\,\left(\frac{2}{3}-\tilde{\xi} \right) - \frac{8}{L^2}\,\left(\frac{8}{27} + \left(\frac{2}{3} - \tilde{\xi} \right) \hat{\Pi}_2 \right) + \Ord(1/L^3)
\, \, \ra \, \, \gamma_{\psi} = 2 g\,\frac{1-3\xi}{3} -16\,\left( \zeta_2 N + \frac{4}{27} \right)\,g^2 + \Ord(g^3)\, ,
\nonum
\eea
where the final expression corresponds to Eq.~(\ref{chap3:res2l:RQED4,3:gammapsi-2}) up to minor differences in notations. In principle, it is also possible to do the reverse operation: 
from RQED$_{4,3}$ to large-$N$ QED$_3$, but this requires a little care because of the non-local nature of the gauge-fixing in QED$_3$, see (\ref{chap4:transform.4}).}

\subsection{Next-to-leading order: without Nash-like resummation} 

Combining the QED$_3$ NLO result of  Eq.~(\ref{chap4:Lc-eqn}) with the transformations (\ref{chap4:transform.1}), (\ref{chap4:transform.3}) and (\ref{chap4:transform.4}), the NLO gap equation 
for the critical coupling of RQED$_{4,3}$ without RPA resummation reads: 
\begin{flalign}
1 = 16(2+\tilde{\xi}) g_c - 8\left(S(\tilde{\xi})- 16 \left(4-\frac{50}{3}\tilde{\xi} + 5\tilde{\xi}^2\right) - 8 \hat{\Pi}_1 \right)\,g_c^2 \, ,
%4(2+\tilde{\xi}) \hat{\Pi}\right) = 0 \, ,
%&&L_c^2 -16(2+\tilde{\xi}) L_c 
%\nonum \\
%&&- 8\left(S(\tilde{\xi})- 16 \left(4-\frac{50}{3}\tilde{\xi} + 5\tilde{\xi}^2\right) - 8 \hat{\Pi}_2 \right) = 0 \, ,
%4(2+\tilde{\xi}) \hat{\Pi}\right) = 0 \, ,
\label{chap4:GE-NLO1}
\end{flalign}
where we have kept $\tilde{\xi} = (1+\xi)/2$ to facilitate comparison with the results of \cite{Kotikov:2016prf}.
In Eq.~(\ref{chap4:GE-NLO1}), $S(\tilde{\xi})$ contains the contribution of the complicated diagrams $I_1$, $I_2$ and $I_3$
having representations \cite{Kotikov:2016wrb} in the form of two-fold series (for $I_1$) and three-fold ones (for $I_2$ and $I_3$),
respectively. Solving Eq.~(\ref{chap4:GE-NLO1}), we have two standard solutions:
%
%\begin{subequations}
%\label{Lc1-NLO-QED3}
\begin{flalign}
\al_{c,\pm}(\tilde{\xi}) = \frac{4\pi}{8(2+\tilde{\xi}) \pm \sqrt{d_1(\tilde{\xi})}}\, ,
\label{chap4:alc-NLO1}
%
%&g_{c,\pm}^{-1} = 8(2+\tilde{\xi}) \pm \sqrt{d_1(\tilde{\xi})}\, ,
%\label{Lc1-QED3}\\
%&d_1(\tilde{\xi})= 8\left(S(\tilde{\xi})-8\left(4-\frac{112}{3}\tilde{\xi}+9\tilde{\xi}^2\right)- 8
%%4(2+\tilde{\xi})
%\hat{\Pi}_1 \right) \, .
% \label{d1-QED3}
%&L_{c,\pm} = 8(2+\tilde{\xi}) \pm \sqrt{d_1(\tilde{\xi})}\, ,
%\label{Lc1-QED3}\\ 
%&d_1(\tilde{\xi})= 8\left(S(\tilde{\xi})-8\left(4-\frac{112}{3}\tilde{\xi}+9\tilde{\xi}^2\right)- 8
%%4(2+\tilde{\xi})
%\hat{\Pi}_2 \right) \, .
% \label{d1-QED3}
\end{flalign}
%\end{subequations}
%
where %$d_1(\tilde{\xi})= 8\left(S(\tilde{\xi})-8\left(4-112 \tilde{\xi} / 3 + 9\tilde{\xi}^2\right)- 8 \hat{\Pi}_1 \right)$.
\begin{flalign}
&d_1(\tilde{\xi})= 8\left(S(\tilde{\xi})-8\left(4-\frac{112}{3}\tilde{\xi}+9\tilde{\xi}^2\right)- 8
%%%4(2+\tilde{\xi})
\hat{\Pi}_1 \right) \, .
\label{chap4:d1}
\end{flalign}
It turns out that the ``$-$'' solution is unphysical and has to be rejected because $\al_{c,-}<0$. So, the physical solution is unique and corresponds to
$\al_{c} = \al_{c,+}$. For numerical applications, we use the numerical estimates provided in Eq.~(\ref{chap4:Is-numerics}) and that we reproduce for clarity:
%
%\begin{subequations}
\begin{flalign}
R_1=163.7428, ~~ R_2=209.175, ~~ P_2=1260.720 \, ,
\label{chap4:R1.R2.P2-values} % \\
\end{flalign}
%\end{subequations}
%
which enter the expression of $S(\tilde{\xi})$:
\begin{flalign}
S(\tilde{\xi}) =
(1-\tilde{\xi})R_1
- (1-\tilde{\xi}^2) \frac{R_2}{8} - (7+16 \tilde{\xi} -3 \tilde{\xi}^2) \frac{P_2}{128}\, .
\label{chap4:S-expr}
\end{flalign}
%
%In the absence of RPA resummations, these results can be immediately transcribed to the case of RQED$_{4,3}$ with the help of the
%transformation (\ref{transform.1}) together with (\ref{transform.3}). 
In various $\xi$-gauges: $\xi=0,1,-1,1/3$ that respectively correspond to $\tilde{\xi}=1/2,1,0,2/3$, 
Eq.~(\ref{chap4:R1.R2.P2-values}) allows us to obtain the numerical value of: %$S(\xi=0)=R_1/2-3 R_2/32 - 57 P_2/512$,
%$S(\xi=1)= -5P_2/32$, $S(\xi=-1)=R_1-R_2/8-7P_2/128$ and
%$S(\xi=1/3)=R_1/3-5 R_2/72 - 49 P_2/384$.
%
\begin{subequations}
\label{chap4:S-xi-values}
\begin{flalign}
&S(\xi=0)=\frac{R_1}{2}-\frac{3 R_2}{32} - \frac{57 P_2}{512}\, ,
\label{chap4:S-oxi=0} \\
&S(\xi=1)= -\frac{5P_2}{32}\, ,
\label{chap4:S-oxi=1} \\
&S(\xi=-1)=R_1-\frac{R_2}{8}-\frac{7P_2}{128}\, ,
\label{chap4:S-oxi=-1} \\ 
&S(\xi=1/3)=\frac{R_1}{3}-\frac{5 R_2}{72} - \frac{49 P_2}{384}\, .
\label{chap4:S-oxi=1/3}
\end{flalign}
\end{subequations}
Notice that the solutions of Eq.~(\ref{chap4:alc-NLO1}) are physical provided that the following inequality is satisfied: %$d_1(\tilde{\xi}) \geq 0$.
\be
d_1(\tilde{\xi}) \geq 0 \, .
\label{chap4:inequality.1}
\ee
In the absence of RPA resummation, Eq.~(\ref{chap4:inequality.1}) %this inequality 
is satisfied only in the unphysical case $N=0$ for which $\hat{\Pi}_1=0$.
For $N>0$, the large value of $\hat{\Pi}_1$ makes $d_1(\tilde{\xi})$ negative which in turn implies that Eq.~(\ref{chap4:GE-NLO1}) has no physical solutions.

Fortunately, the situation strongly improves upon performing the RPA resummation.
Indeed, in this case, the inequality Eq.~(\ref{chap4:inequality.1}) 
is satisfied for all the values of $\tilde{\xi}$ we consider, excepting the case $\xi=\tilde{\xi}=1$ which corresponds to the Feynman gauge.
To see this, consider Eq.~(\ref{chap4:al-RPA}) which can be re-written as:
\be
\al = \frac{\tilde{\al}}{1 - \hat{\Pi}_1 \tilde{\al}/(2\pi)}, \quad g = \frac{\tilde{g}}{1 - 2\hat{\Pi}_1 \tilde{g}}\, ,
\label{chap4:al.vs.al-RPA}
\ee
where $\hat{\Pi}_1$ defined in Eq.~(\ref{chap4:hatPi1}) was made explicit. Substituting Eq.~(\ref{chap4:al.vs.al-RPA}) in (\ref{chap4:GE-NLO1}), 
we see that the contribution of $\hat{\Pi}_1$ cancels out from the new gap equation which reads:
\begin{flalign}
1 = 16(2+\tilde{\xi}) \tilde{g}_c
- 8\left(S(\tilde{\xi})- 16 \left(4-\frac{50}{3}\tilde{\xi} + 5\tilde{\xi}^2\right) \right)\,\tilde{g}_c^2 \, .
\label{chap4:GE-NLO2}
\end{flalign}
Hence, Eq.~(\ref{chap4:GE-NLO2}) has a broader range of solutions than Eq.~(\ref{chap4:GE-NLO1}) including
all $\xi$-values such as: $-5.695 < \xi < 0.915$ or, equivalently: 
$-2.348 < \tilde{\xi} < 0.957$; the Feynman gauge is still excluded because $S(\xi=1)$ is large and negative. 
Numerical applications for $\tilde{\al}_c = \tilde{\al}_{c,+}$ then yield:
\be
\tilde{\al}_c(0)= 0.3804,~\tilde{\al}_c(-1)= 0.3794,~ \tilde{\al}_c(1/3)= 0.3924\, .
\label{chap4:talc-NLO2}
\ee
Substituting the values (\ref{chap4:talc-NLO2}) in Eq.~(\ref{chap4:al.vs.al-RPA}) yields, for $N=2$:
%
%\begin{flalign}
\be
\al_c(0)= 0.9451, ~\al_c(-1)= 0.9389, ~\al_c(1/3)= 1.0227\, .
\ee
%\end{flalign}
%

The critical number, $N_c$, for which $\al_c \to \infty$ and is such that a finite critical
coupling exists for $N<N_c$ takes the following values:
%
%\begin{flalign}
\be
N_c(0)= 3.3472, ~N_c(-1)= 3.3561, ~ N_c(1/3)= 3.2450\, .
\label{chap4:Nc-NLO2}
\ee
%\end{flalign}
%
The numerical values in (\ref{chap4:Nc-NLO2}) are a little larger than in QED$_3$ where $N_c$ is defined as $N_c=L_c/\pi^2$ because of the additional
(small) factor $\hat{\Pi}_2$ coming with a negative sign, see \cite{Kotikov:2016prf}.

\subsection{Next-to-leading order: Nash's resummation}  

Following the derivations made in the case of QED$_3$, we now resum the ``basic'' part of the NLO corrections corresponding to the fermion anomalous dimension.
Combining the QED$_3$ resummed gap equation of Eq.~(\ref{chap4:Lc-eqn_1}) with the transformations (\ref{chap4:transform.1}), (\ref{chap4:transform.3}) and (\ref{chap4:transform.4}), the NLO resummed gap equation
for the critical coupling of RQED$_{4,3}$ without RPA resummation reads:
\be
1 = \frac{128}{3} g_c + 8 \left( \tilde{S}(\tilde{\xi}) - \frac{1280}{27} - \frac{32}{3} \hat{\Pi}_1 \right)g_c^2 \, ,
\label{chap4:GE-NLO3} 
\ee
where $\tilde{S}(\tilde{\xi})$ contains the rest of $S(\tilde{\xi})$, see \cite{Kotikov:2016prf}, after the extraction of the ``most important''
contributions.
Similarly to the case of QED$_3$ \cite{Kotikov:2016prf}, the striking feature of Eq.~(\ref{chap4:GE-NLO3}) is the absence of $\tilde{\xi}$-dependence at LO
and it's strong suppression at NLO: the $\tilde{\xi}$-dependence does exist at NLO but only via $\tilde{S}(\tilde{\xi})$ which, as we shall see shortly, is small numerically.
Solving Eq.~(\ref{chap4:GE-NLO3}), we have the two standard solutions:
\be
\al_{c,\pm} = \frac{4\pi}{64/3 \pm \sqrt{d_2(\tilde{\xi})}}\, ,
\label{chap4:alc-NLO3}
\ee
where %$d_2(\tilde{\xi})= 8 \left( \tilde{S}(\tilde{\xi}) + 256/27 - 32\hat{\Pi}_1/3 \right)$.
\be
d_2(\tilde{\xi})= 8 \left( \tilde{S}(\tilde{\xi}) + \frac{256}{27} - \frac{32}{3} \hat{\Pi}_1 \right)\, .
\label{chap4:d2}
\ee
As before, the ``$-$'' solution is unphysical and has to be rejected because $\al_{c,-}<0$. So, the physical solution is unique and corresponds to
$\al_{c} = \al_{c,+}$. In order to provide  numerical estimates, we use the values $\tilde{R}_1$, $\tilde{R}_2$ and $\tilde{P}_2$, from Eqs.~(\ref{chap4:tSigma2}) and (\ref{chap4:til}) that we reproduce for clarity:
\be
\tilde{R}_1=3.7428, \quad \tilde{R}_2=1.175, \quad \tilde{P}_2=-19.280 \, .
\label{ichap4:tIs-numerics}
\ee
They enter the expression of $\tilde{S}(\tilde{\xi})$:
\be
\tilde{S}(\tilde{\xi}) =  (1-\tilde{\xi})\tilde{R}_1 - (1-\tilde{\xi}^2) \frac{\tilde{R}_2}{8} - (7+16 \tilde{\xi} -3 \tilde{\xi}^2) \frac{\tilde{P}_1}{128}\, .
\label{chap4:sigma2.3b} 
\ee
For the gauge choices: $\xi=0,1,-1,1/3$ that respectively correspond to $\tilde{\xi}=1/2,1,0,2/3$,
we may then obtain numerical values of: %$S(\xi=0)= \tilde{R}_1/2-3 \tilde{R}_2/32 - 57 \tilde{P}_2/512$,
%$S(\xi=1)= -5\tilde{P}_2/32$, $S(\xi=-1)= \tilde{R}_1-\tilde{R}_2/8-7\tilde{P}_2/128$ and
%$S(\xi=1/3)= \tilde{R}_1/3- 5 \tilde{R}_2/72 - 49 \tilde{P}_2/384$.
%
\begin{subequations}
\label{ichap4:tS-xi-values}
\begin{flalign}
&S(\xi=0)= \frac{\tilde{R}_1}{2}-\frac{3 \tilde{R}_2}{32} - \frac{57 \tilde{P}_2}{512}\, ,
\label{chap4:tS-oxi=0} \\
&S(\xi=1)= -\frac{5\tilde{P}_2}{32}\, ,
\label{chap4:tS-oxi=1} \\
&S(\xi=-1)= \tilde{R}_1-\frac{\tilde{R}_2}{8}-\frac{7\tilde{P}_2}{128}\, ,
\label{chap4:tS-oxi=-1} \\
&S(\xi=1/3)= \frac{\tilde{R}_1}{3}-\frac{5 \tilde{R}_2}{72} - \frac{49 \tilde{P}_2}{384}\, .
\label{chap4:tS-oxi=1/3}
\end{flalign}
\end{subequations}
Notice that the solutions of Eq.~(\ref{chap4:alc-NLO3}) are physical provided that: %$d_2(\tilde{\xi}) \geq 0$.
%. It is easy to see that 
\be
d_2(\tilde{\xi}) \geq 0
\label{inequality.2}
\ee
%For any $\tilde{\xi}$-values, the 
As in the previous case, this inequality is satisfied only for the nonphysical case $N=0$.

Fortunately, the situation once again strongly improves upon the additional implementation
of the RPA resummation. Substituting Eq.~(\ref{chap4:al.vs.al-RPA}) in (\ref{chap4:GE-NLO3}),
we see that the contribution of $\hat{\Pi}_1$ cancels out from the new gap equation which reads:
\be
1 = \frac{128}{3} \tilde{g}_c + 8 \left( \tilde{S}(\tilde{\xi}) - \frac{1280}{27} \right) \tilde{g}_c^2 \, .
\label{chap4:GE-NLO4}
\ee
Hence, Eq.~(\ref{chap4:GE-NLO4}) has a broader range of solutions than Eq.~(\ref{chap4:GE-NLO3}) including
all $\xi$-values such as: $-19.668 < \xi < 8.928$ or, equivalently:
$-9.334 < \tilde{\xi} < 4.964$. This shows that the improvement is even better that in the absence of Nash's resummation because,
in the present case, a physical solution also exists in the Feynman gauge $\xi=\tilde{\xi}=1$. 
Numerical applications for $\tilde{\al}_c = \tilde{\al}_{c,+}$ yield:
\begin{subequations}
\label{chap4:talc-NLO4}
\begin{flalign}
&\tilde{\al}_c(\xi=0)= 0.3966, \quad ~~ \tilde{\al}_c(\xi=1)= 0.4011\, ,
\\
&\tilde{\al}_c(\xi=-1)= 0.3931, \quad \tilde{\al}_c(\xi=1/3)= 0.3980\, .
\end{flalign}
\end{subequations}
Substituting the values (\ref{chap4:talc-NLO4}) in Eq.~(\ref{chap4:al.vs.al-RPA}) yields, for $N=2$:
\begin{subequations}
\label{chap4:alc-NLO4-N=2}
\begin{flalign}
&\al_c(\xi=0)= 1.0521, \quad ~~ \al_c(\xi=1)= 1.0841\, ,
\\
&\al_c(\xi=-1)= 1.0278, \quad \al_c(\xi=1/3)= 1.0619\, .
\end{flalign}
\end{subequations}
These values are very close to each other proving the very weak gauge variance of our results.
For the sake of completeness, we give the value of the critical coupling constant in the case $N=1$:
%where the critical coupling is reduced:
%
\begin{subequations}
\label{chap4:alc-NLO4-N=1}
\begin{flalign}
&\al_c(\xi=0)= 0.5761, \quad ~~ \al_c(\xi=1)= 0.5855\, ,
\\
&\al_c(\xi=-1)= 0.5687, \quad \al_c(\xi=1/3)= 0.5790\, ,
\end{flalign}
\end{subequations}
and in the case $N=3$:
%On the other hand, in the case $N=3$ the critical coupling is increased:
%
\begin{subequations}
\label{chap4:alc-NLO4-N=3}
\begin{flalign}
&\al_c(\xi=0)= 6.0588, \quad ~~ \al_c(\xi=1)= 7.2971\, ,
\\
&\al_c(\xi=-1)= 5.3310, \quad \al_c(\xi=1/3)= 6.3960\, .
\end{flalign}
\end{subequations}
For higher (integer) values of $N$, there is no instability. This can be seen by computing the value
of $N$, $N_c$, for which $\al_c \to \infty$. As in the previous case, see Eq.~(\ref{chap4:Nc-NLO2}),
this value 
%exactly
coincides with the critical value $N_c$ for D$\chi$SB in QED$_3$:
\begin{subequations}
\label{chap4:Nc-NLO4}
\begin{flalign}
&N_c(\xi=0)= 3.2102, \quad ~~ N_c(\xi=1)= 3.1745\, ,
\\
&N_c(\xi=-1)= 3.2388, \quad N_c(\xi=1/3)= 3.1991\, .
\end{flalign}
\end{subequations}
As anticipated above, we see that the ``right(est)'' gauge choice \cite{Gorbar:2001qt} is the one close to $(\xi=1/3,\tilde{\xi}=2/3)$ where
the results are more or less the same before and after Nash's resummation. 

At this point, we would like to remark that the weakness of the gauge dependence 
of our results makes it unimportant from the point of view of physical applications. Moreover, such gauge dependence is not specific to reduced QED; actually, as the
mapping we have used suggests, it originates from a similar feature taking place in QED$_3$, see \cite{Kotikov:2016prf}. 
Nevertheless, the existence of such a gauge dependence, even though very weak, may call into question the applicability of our approach.
It is indeed well known that gauge dependence does not affect the critical value of $N_c$ in the case of QED$_3$, see \cite{Bashir:2009fv},
a statement which is based on an application of the Landau-Khalatnikov-Fradkin (LKF) transformation \cite{Landau:1955zz,Fradkin:1955jr} to QED$_3$. 
Recently, an application of the LKF transformation has been carried out in the case of reduced QED
\cite{Ahmad:2016dsb}. Note, however, that in our study of QED$_3$, \cite{Kotikov:2016prf}, we have worked in a non-local gauge
which is quite popular now in the $3$-dimensional case. In this case, a direct application of the LKF
transformation is quite problematic.  We hope to return to this problem in our future studies. 

We finally consider the case where $\tilde{S}(\tilde{\xi})=0$ in (\ref{chap4:GE-NLO4}).
In this case, there no gauge dependence at all and we have: $\tilde{\al}_c = 0.41828$.
Hence: %$\alpha_c(N=2)=1.2196$, $\alpha_c(N=1)=0.6229$ and $\quad \alpha_c(N=3)=28.9670$.
\begin{subequations}
\begin{flalign}
&\alpha_c(N=2)=1.2196 \, ,
\\
&\alpha_c(N=1)=0.6229, \quad \alpha_c(N=3)=28.9670\, .
\end{flalign}
\end{subequations}
The number $N_c$ coincides with $N_c=L_c/\pi^2$ in QED$_3$ and has the following value:
\be
\overline{N}_c=3.0440 \, ,
\ee
which is a little less than the ones in (\ref{chap4:Nc-NLO4}). We shall come back on this important (gauge-invariant) result in the Conclusion of this Chapter.

\subsection{Comparison with other results}
\label{chap4:sec:compare}

Our results for $\alpha_c$ (%$0.94 < \alpha_c < 1.02$ without Nash's resummation, $1.03 < \alpha_c < 1.08$
%with Nash's resummation and 
in particular the fully gauge-invariant ones: $\al_c=1.22$) are in good agreement with $\alpha_c=0.92$ \cite{Gamayun:2009em} and $\alpha_c=1.13$ 
\cite{Khveshchenko:2008ye}. These last results were obtained as improvements of previous studies: \cite{Gamayun:2009em} took into account of the dynamical screening of
the interactions with respect to \cite{Gorbar:2002iw} where the value $\alpha_c=1.62$ was found in the static
approximation, {\it i.e.}, RPA with polarization operator at zero frequency; \cite{Khveshchenko:2008ye} took into account Fermi velocity renormalization
with respect to the earlier work in \cite{Khveshchenko:2001zz}; see also discussions in Refs.~\cite{Gamayun:2009em}
and \cite{Katanin16.PhysRevB.93.035132} as well as a detailed summary of these results in the review \cite{RevModPhys.84.1067} and in \cite{Wang2012}. 
Our results are also in good agreement with lattice Monte-Carlo simulations where  $\al_c = 1.11 \pm 0.06$ was obtained in \cite{Drut:2008rg} 
and $\al_c = 0.9 \pm 0.2$ in \cite{Buividovich12.PhysRevB.86.245117}.
%
%with $\alpha_c=1.08$ in
% \cite{Drut:2008rg} (as it was shown in \cite{Katanin:2015asa}).
%
Moreover, in the strong coupling regime, $\alpha_c \to \infty$, our critical values for $N_c$ (in particular, the gauge invariant one: $\overline{N}_c=3.04$) 
are close to $N_c=7.2/2=3.6$ obtained in  \cite{Khveshchenko:2008ye} and $N_c=3.52$ \cite{Liu09.PhysRevB.79.205429}.
% !!! OK ref found !!!
%{\it (perhaps, I have found a wrong reference \cite{Liu:2009jt})} {\bf (see then recent review
%\cite{KotovUPGC12}).
%
%Note that our results were done for the nonphysical case $v_F=c$, where $v_f$ is Fermi-velocity and the results 
%\cite{Khveshchenko:2001zz,Gorbar:2002iw} have been obtained in the statical approximation $v_F=0$ or with some including of the
%$v_F$ contributions in \cite{Khveshchenko:2008ye,Gamayun:2009em}.
These results for $\al_c$ would not be compatible with the semimetallic behaviour of graphene observed experimentally \cite{elias2011cones} if we were to compare
them with the bare coupling constant $\al \approx 2.2$ in clean suspended graphene. We may however argue that the
renormalization of the Fermi velocity observed in \cite{elias2011cones} would rather be compatible with a coupling constant of about $0.73$ which is indeed smaller than all 
of the above values theoretically obtained for $\al_c$. 

Nevertheless, the results \cite{Khveshchenko:2001zz,Gorbar:2002iw,Khveshchenko:2008ye,Gamayun:2009em} were then
criticized for not properly taking into account dynamical screening of interactions and/or wave function and/or velocity renormalizations.
Attempts to better take into account (some or all of) these effects at the level of SD equations led to larger values: $3.2 < \al_c < 3.3$ \cite{Wang2012}, 
$\al_c = 7.65$ \cite{Popovic13.PhysRevB.88.205429} and $\al_c = 3.1$ \cite{Gonzalez15.PhysRevB.92.125115}. As discussed in \cite{Katanin16.PhysRevB.93.035132}, where the value $\al_c = 3.7$ was obtained using different methods,
the result $\alpha_c=3.1$ seems to be the most reliable within the Schwinger-Dyson approach.  
Such large values are well above the bare value of $\al$ and therefore compatible with the semi-metallic ground state observed experimentally. Notice however that, very recently, the value $2.06$, which is closer to our result, 
was obtained in \cite{Carrington:2017hlc} within an SD approach carefully taking into account retardation effects and the running of the Fermi velocity.

\begin{center}
\renewcommand{\tabcolsep}{0.25cm}
\renewcommand{\arraystretch}{1.5}
\begin{table}
    \begin{tabular}{ c| c | c | c }
      \hline
             $\al_c$ 				&	$N_c$				&       {\bf Method}                                         &       {\bf Year} \\
      \hline \hline
          $7.65$                                &                       		&     SD (LO, dynamic RPA, running $v$)                      &       2013 \cite{Popovic13.PhysRevB.88.205429}     \\
      \hline
          $3.7$                                 &                       		&     FRG, Bethe-Salpeter                                    &       2016 \cite{Katanin16.PhysRevB.93.035132}     \\
      \hline
           $3.2 < \al_c < 3.3$                  &                       		&     SD (LO, dynamic RPA, running $v$)                      &       2012 \cite{Wang2012}     \\
      \hline
           $3.1$  	                        &                       		&     SD (LO, bare vertex approximation)                     &       2015 \cite{Gonzalez15.PhysRevB.92.125115}     \\
      \hline
      \hline
          $2.06$                                &     					&     SD (LO, dynamic RPA, running $v$)                      &       2017 \cite{Carrington:2017hlc}     \\
      \hline	
          $1.62$                        	&                       		&     SD (LO, static RPA)	                             &       2002 \cite{Gorbar:2002iw}     \\
      \hline
                                                &          $3.52$                       &     SD (LO)                                                &       2009 \cite{Liu09.PhysRevB.79.205429}     \\
      \hline
\textcolor{red}{$1.22$}				&    \textcolor{red}{$3.04$}            &  \textcolor{red}{SD (NLO, RPA, resummation, $v/c \ra 1$, $\forall \xi$)} & \textcolor{red}{2016} \cite{Kotikov:2016yrn}    \\    
     \hline
          $1.13$                         	&          $3.6$             		&     SD (LO, static RPA, running $v$)                       &       2008 \cite{Khveshchenko:2008ye}        \\
      \hline
          $1.11 \pm 0.06$               	&                       		&     Lattice simulations                                    &       2008 \cite{Drut:2008rg}       \\
      \hline
\textcolor{red}{$1.03 < \alpha_c < 1.08$}	& \textcolor{red}{$3.17 < N_c <  3.24$}	&  \textcolor{red}{SD (NLO, RPA, resummation, $v/c \ra 1$)}  &       \textcolor{red}{2016} \cite{Kotikov:2016yrn}    \\
      \hline
          $1.02$                                &                       		&   SD (LO, dynamic RPA, running $v$)                        &       2011 \cite{WangLiu11b}     \\
      \hline
\textcolor{red}{$0.94 < \alpha_c < 1.02$}	& \textcolor{red}{$3.24 < N_c <  3.36$}	&  \textcolor{red}{SD (NLO, RPA, $v/c \ra 1$)} 		     &       \textcolor{red}{2016}  \cite{Kotikov:2016yrn} \\
      \hline
          $0.99$                                &                      			&    RG study                                                &       2012 \cite{Gonzalez12.PhysRevB.85.085420}     \\
      \hline
          $0.92$				&					&    SD (LO, dynamic RPA)                                    &       2009 \cite{Gamayun:2009em}    \\
      \hline
          $0.9 \pm 0.2$                         &                       		&    Lattice simulations                                     &       2012 \cite{Buividovich12.PhysRevB.86.245117}    \\
      \hline
          $0.833$ 	                        &                       		&    RG study                                                &       2008 \cite{VafekCase08}    \\
      \hline
    \end{tabular}
    \caption{D$\chi$SB in graphene: some values of $\al_c(N=2)$ and $N_c$ obtained over the years together with elements of the  different methods used. The double line corresponds to: $\al_g =2.2$. 
Values presented in this manuscript are in red. The values to be retained are the gauge-invariant ones: $\al_c=1.22$ and $N_c=3.04$.}
    \label{chap4:tab:alphac-values}
\end{table}
\end{center}

\section{Conclusion}
\label{chap4:sec:conclusions}

\begin{figure}
\begin{center}
\includegraphics[scale=0.75]{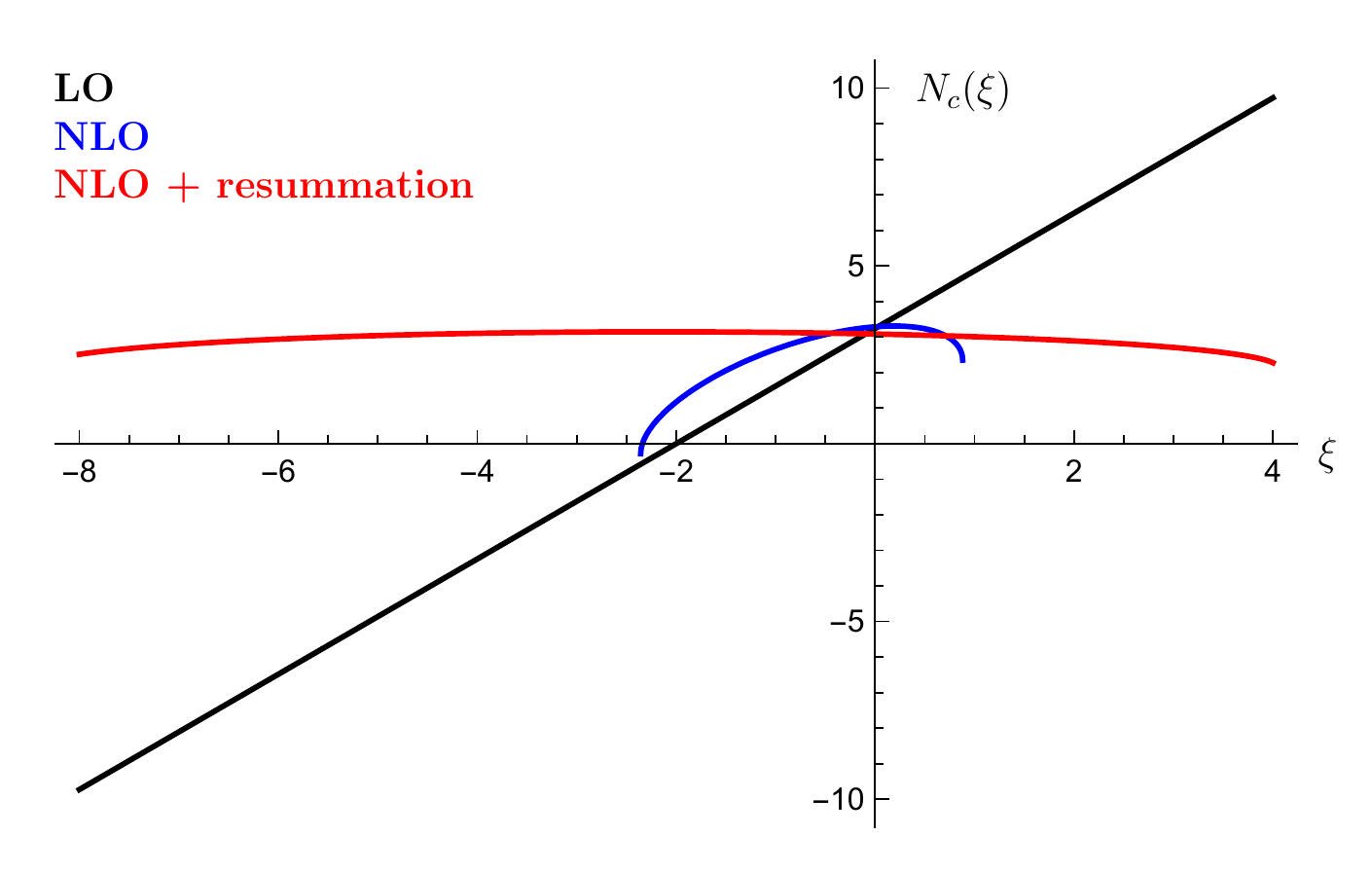}
\caption{\label{chap4:fig:qed3:Nc-vs-xi-curves}
        D$\chi$SB in QED$_3$: plots of $N_c(\xi)$ versus $\xi$ in different approximations.}
\end{center}
\end{figure}

In this Chapter we have performed an accurate study of D$\chi$SB in QED$_3$ by including $1/N^2$ corrections to the SD equation exactly and taking into account the full
$\xi$-dependence of the gap equation. Following Nash, the fermion anomalous dimension has been resummed at the level of the gap equation
leading to a very weak gauge-variance of the critical fermion number $N_c$.~\footnote{Notice that, in the study of SD equations for QED$_4$, a weakly gauge variant critical coupling was found in \cite{Atkinson:1993mz}. This work made use of the
Curtis-Pennington (CP) ansatz for the vertex function. I was informed by Valery Gusynin that, while the CP vertex satifies Ward-Takahashi identities, it does not transform correctly under the LKF transformation; this
may explain the weak gauge variance of the result found in \cite{Atkinson:1993mz}.} 
 %Thirty years after the seminal work of Nash, we bring a definite and complete
%solution to NLO computations in QED$_3$ within the SD approach initiated by Appelquist et al. 
 Our results before resummation read:
\begin{subequations}
\begin{flalign}
&L_c(0)=32.45, 			\qquad ~~  L_c(0.1903)=32.66 \qquad ~~ L_c(2/3)=30.51\, ,
\label{chap4:qed3:Lc-no-resum} \\
&N_c(0)=3.2880, 		\qquad N_c(0.1903) = 3.3095 \qquad N_c(2/3)=3.0915\, ,
\label{chap4:qed3:Nc-no-resum}
\end{flalign}
\end{subequations}
where solutions exist for the range of gauge-fixing parameters:  $-2.36 \leq  \xi \leq  0.88$ (no solution in the Feynman gauge) and for $\xi_0=0.1903$ we have $\D N_c(\xi_0)/\D \xi = 0$ so that $N_c<N_c(0.1903) = 3.3095$. 
After Nash's resummation they become:
\begin{subequations}
\begin{flalign}
&L_c(1)=29.69, 	\qquad ~~ L_c(2/3)=29.98, \qquad ~~L_c(0)=30.44\, , %\quad L_c(-2.1849)=31.06\, , %\qquad
\label{chap4:qed3:Lc-res+resum} \\
&N_c(1)=3.0084,	\qquad  N_c(2/3)=3.0377,  \qquad   N_c(0)=3.0844\, ,%\quad N_c(-2.1849)= 3.1471\, , %\qquad
\label{chap4:qed3:Nc-values+resum}
\end{flalign}
\end{subequations}
with a very weak gauge-dependence and where now solutions exist in the range: $-8.412 \leq  \xi \leq  4.042$; for $\xi_0=-2.1849$, we have $\D N_c(\xi_0)/\D \xi = 0$ so that 
with resummation $N_c<N_c(-2.1849)= 3.1471$. Fig.~\ref{chap4:fig:qed3:Nc-vs-xi-curves} summarizes these results with plots of $N_c(\xi)$ in various approximations; the 
displayed weakening of the gauge-dependence upon going from NLO to NLO with resummation is quite impressive. %This strong suppression 
%of the gauge dependence of $N_c$ at NLO that gives increasing support to the stability of the critical point. Our exact NLO result indicates that 
 %%the ``rightest'' gauge seems to be the one corresponding to the gauge fixing parameter 
%for $\xi=2/3$, where the fermion wave function is finite at LO, the value of $N_c=3.0377$ is very stable with or without resummation. A closer look at Fig.~\ref{chap4:fig:qed3:Nc-vs-xi-curves} shows
%that there are actually two values of $\xi$ for which NLO results with and without resummation coincide: $N_c(\xi_1=-0.4367) = 3.1074$ and $N_c(\xi_2=0.7092) = 3.0342$ where $\xi_2$ is very close to $2/3$.
Moreover, Fig.~\ref{chap4:fig:qed3:Nc-vs-xi-curves} gives strong support for working in the Landau gauge because all results cross in a close vicinity of $\xi=0$.
 %Our work therefore gives increasing support to the stability of the critical point. 
However, at this point, let us recall that we also obtained a fully gauge-invariant result:
\be
\boxed{\overline{L}_c=28.0981, \qquad \overline{N}_c=2.85\, ,}
\label{chap4:overlN-2}
\ee
by neglecting the rest, {\it i.e.}, $\tilde{S}(\xi)=0$ in Eq.~(\ref{chap4:Lc-eqn_1}). As we noticed earlier, the corresponding gap equation together with the
results of Eq.~(\ref{chap4:overlN-2}) are in full agreement with the recent results of \cite{Gusynin:2016som} who used a different method to solve the SD equations as well as a different regulator (hard cut-off). 
It turns out that the prescription of Gusynin and Pyatkovskiy \cite{Gusynin:2016som} is an NLO expansion
in terms of the parameter $\alpha$ rather than the mass function itself, see \cite{Gusynin:2016som} for more details. Applying this prescription to our gap equation justifies the neglect of 
$\tilde{S}(\xi)$ in order to achieve NLO accuracy. Hence, this proves the {\it complete} gauge invariance of the gap equation
order by order in the $1/N$-expansion.  So the final value of $N_c$ that we shall retain is the fully gauge-invariant one of Eq.~(\ref{chap4:overlN-2}). 
Thirty years after the seminal work of Nash, the works \cite{Gusynin:2016som} and \cite{Kotikov:2016prf} bring a definite and complete solution to NLO computations in QED$_3$ within the SD approach.
We conclude this part by further adding that the complete agreement between the results of \cite{Gusynin:2016som} and \cite{Kotikov:2016prf} proves that the use of dimensional regularization 
within a (non-perturbative) SD type approach (which is sometimes controversial) is 
completely equivalent to using a standard hard cut-off regulator. Moreover, the approach based on sophisticated multi-loop techniques is systematic in nature; combined with the prescription of \cite{Gusynin:2016som}, it
provides a powerful and gauge-invariant method to compute $N_c$ at higher orders.

\begin{figure}
\begin{center}
\includegraphics[scale=0.75]{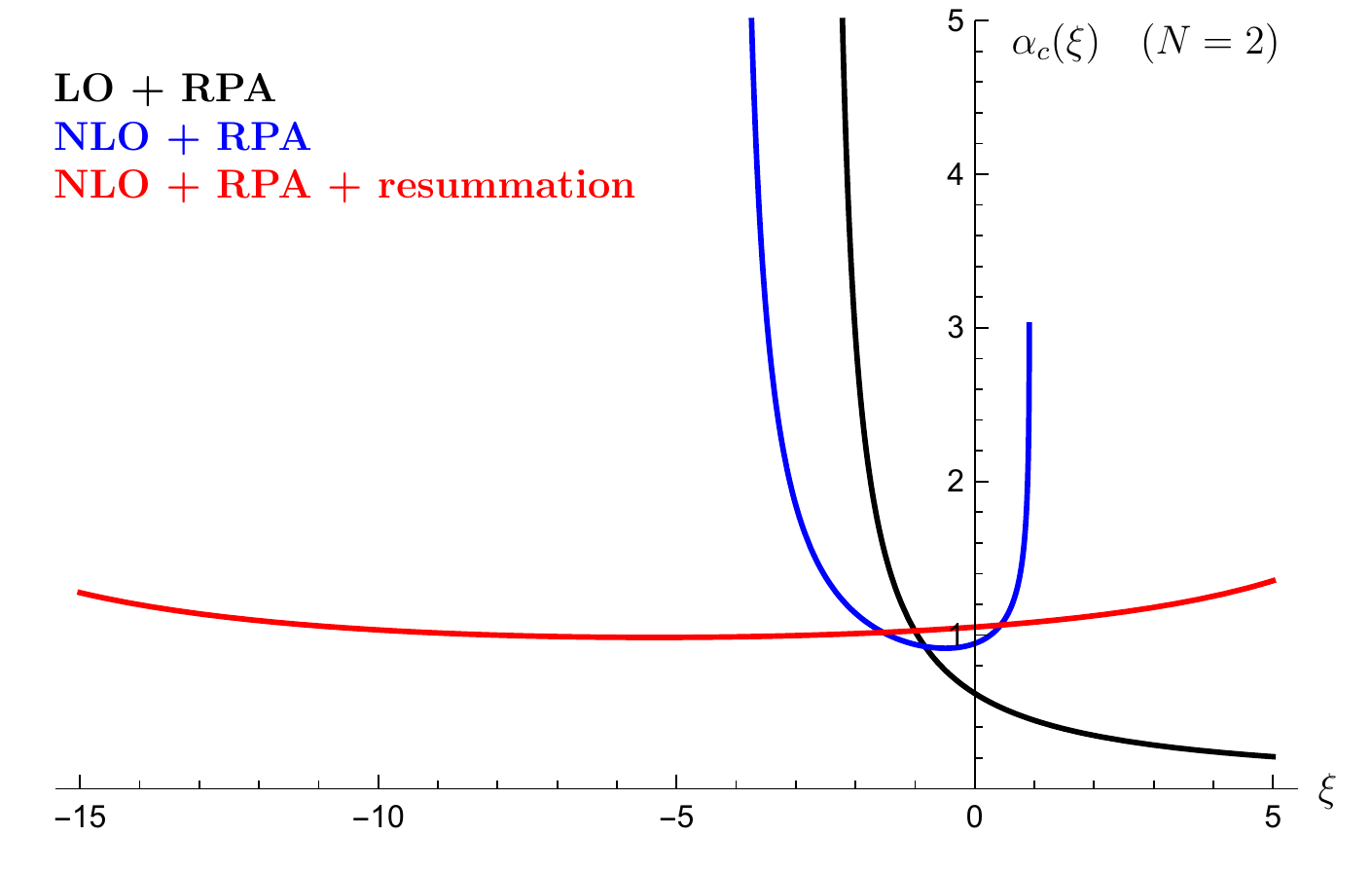}
\caption{\label{chap4:fig:rqed43:alc-vs-xi-curves}
        D$\chi$SB in reduced QED$_{4,3}$: plots of $\al_c(\xi)$ versus $\xi$ for $N=2$ in different approximations.}
\end{center}
\end{figure}

We have also studied D$\chi$SB in reduced QED$_{4,3}$ which describes planar Dirac materials at the IR Lorentz-invariant fixed point. 
We have discovered a very nice mapping between large-$N$ QED$_3$ and reduced QED which originates from the fact that the photon propagators in both models have the same form.
Using this mapping we were able to transcribe the exact NLO gap equation found for QED$_3$ to the case of RQED$_{4,3}$. Similarly to QED$_3$, a Nash-like resummation was performed at the
level of the gap equation for RQED$_{4,3}$. An additional RPA (or Khveshchenko-like) resummation was crucial for solutions to exist in the case of RQED$_{4,3}$. Our study provides
high precision estimates of the critical coupling constant, $\al_c$, and critical fermion flavour number, $N_c$. 
Dynamical screening and wave function renormalization were fully taken into account (there is no velocity renormalization at the fixed point).
We found that $\al_c \sim \Ord(1)$, so that $\al_c \gg \al_{\text{QED}}$ and, at the IR fixed point, the system is deep in the semimetallic phase in qualitative
agreement with experiments in actual samples. More precisely, our results before Nash's resummation read (case $N=2$ for $\al_c$):
\begin{subequations}
\begin{flalign}
&\al_c(\xi=-1)= 0.9389, \qquad ~\al_c(\xi=0)= 0.9451, \qquad ~\al_c(\xi=1/3)= 1.0227\, ,
\label{chap4:rqed4,3:alc-no-resum} \\
&N_c(\xi=-1)= 3.3561, 	\qquad N_c(\xi=0)= 3.3472,   \qquad N_c(\xi=1/3)= 3.2450\, ,
\label{chap4:rqed4,3:Nc-no-resum}
\end{flalign}
\end{subequations}
where ($\xi=-1,\tilde{\xi}=0$) corresponds to the Landau gauge for the reduced gauge field, ($\xi=0,\tilde{\xi}=1/2$) and ($\xi=1/3,\tilde{\xi}=2/3$).
Notice that for $\xi_0=-0.4939$, we have $\D \al_c(\xi_0)/\D \xi = 0$ so that $\al_c>\al_c(-0.4939) = 0.9143$ ($N=2$) and $N_c < N_c(-0.4939) = 3.3925$.
Solutions exist in the range:  $-5.695 < \xi < 0.915$ or, equivalently: $-2.348 < \tilde{\xi} < 0.957$ and the Feynman gauge ($\xi=1,\tilde{\xi}=1$) is excluded.
After Nash's resummation, we obtain (case $N=2$ for $\al_c$):
\begin{subequations}
\begin{flalign}
&\al_c(-1)= 1.0278, \qquad ~\al_c(0)= 1.0521,  \qquad ~\al_c(1/3)= 1.0619, \qquad  ~\al_c(1)= 1.0841\, ,
\label{chap4:rqed4,3:alc+resum} \\
&N_c(-1)= 3.2388,   \qquad   N_c(0)= 3.2102, \qquad N_c(1/3)= 3.1991,  \qquad  N_c(1)= 3.1745\, ,
\label{chap4:rqed4,3:Nc+resum}
\end{flalign}
\end{subequations}
where the gauge-dependence is very weak and solutions exist for $-19.668 < \xi < 8.928$ or, equivalently: $-9.334 < \tilde{\xi} < 4.964$ (including Feynman gauge). 
for $\xi_0=-5.3698$, we have $\D \al_c(\xi_0)/\D \xi = 0$ so that $\al_c>\al_c(\xi=-5.3698) = 0.9847$ ($N=2$) and $N_c < N_c(\xi=-5.3698)= 3.2930$.
Fig.~\ref{chap4:fig:rqed43:alc-vs-xi-curves} summarizes all these results and display the strong reduction in the gauge-dependence of $\al_c(\xi)$ upon performing a Nash resummation at NLO.
This figure also shows that all curves cross in the vicinity of $\xi=-1$ giving support for  working in the Landau gauge for the reduced gauge field.
However, at this point, let us recall that we also obtained fully gauge-invariant results:
\begin{subequations}
\label{chap4:rqed4,3:alc-Nc-gi}
\begin{empheq}[box=\fbox]{align}
%\begin{subequations}
%\label{chap4:rqed4,3:alc-Nc-gi}
%\begin{flalign}
&\alpha_c(N=2)=1.2196, \quad \alpha_c(N=1)=0.6229, \quad \alpha_c(N=3)=28.9670\, ,
\label{chap4:rqed4,3:alc-gi}
\\
&\overline{N}_c=3.0440\, ,
\label{chap4:rqed4,3:Nc-gi}
%\end{flalign}
%\end{subequations}
\end{empheq}
\end{subequations}
by neglecting the rest: $\tilde{S}(\tilde{\xi})=0$ in (\ref{chap4:GE-NLO4}). The results of Eqs.~(\ref{chap4:rqed4,3:alc-Nc-gi}) are equivalent to those of
Eq.~(\ref{chap4:overlN-2}) for QED$_3$ and satisfy the prescription of Gusynin and Pyatkovskiy \cite{Gusynin:2016som}. Hence, among all the above results we shall retain the fully gauge-invariant ones of 
Eqs.~(\ref{chap4:rqed4,3:alc-Nc-gi}). We conclude by adding that the striking feature of these results is that our values of $\al_c$ and $N_c$ are 
in good {\it quantitative} agreement with results obtained in the non-relativistic limit \cite{Leal:2003sg,Khveshchenko:2008ye,Drut:2008rg,Liu09.PhysRevB.79.205429,Gamayun:2009em,Buividovich12.PhysRevB.86.245117}
 including lattice simulations \cite{Drut:2008rg,Buividovich12.PhysRevB.86.245117}. Such an agreement between the two extreme limits, $v/c \ra 0$ and $v/c \ra 1$,
seems to suggest that the study of the fixed point is not only of academic interest. As we shall see in the next Chapter there is another quantity
which is in good quantitative agreement in the two extreme limits: the interaction correction coefficient to the optical conductivity.

\end{fmffile}

\cleardoublepage

%% file: Chapter5/graphene.tex
\label{chap5}

\begin{fmffile}{fmf-chap5}

In this final Chapter we depart from the IR Lorentz invariant fixed point and 
focus on the non-relativistic limit with instantaneous Coulomb interaction (limit $v/c \ra 0$). This is quite a realistic limit
for planar Dirac systems given the fact that the bare Fermi velocity is much smaller than the velocity of light, $v \approx c / 300$.
At a technical level, the broken Lorentz invariance significantly complicates all calculations. Nevertheless, we will show that the powerful 
multi-loop techniques presented in Chap.~\ref{chap2}, augmented with rules to deal with massive Feynman diagrams, are quite efficient to address this limit. 
The focus will be in particular on the computation of the two-loop interaction correction to the minimal conductivity of disorder-free intrinsic graphene, a quantity which
is of experimental interest. 

%\begin{fmffile}{fmfcond}
%%%%%%%%%%%%%%%%%%%%%%%%%%%%%%%%%%%%%%%

\section{Introduction}

As we discussed in details in Chap.~\ref{chap1}, a remarkable feature of disorder-free intrinsic graphene and related planar Dirac liquids is that,
despite the fact that they have a vanishing density of states at the Fermi points, the chiral nature of the charge carriers yields
a minimal ac conductivity in the collisionless regime ($\om \gg \Gamma,\, T$):~\footnote{As recalled in Chap.~\ref{chap1}, the conductivity, $\sigma(\om,\vec q\,,T,\mu,\Gamma, \Delta)$, 
is in general a complicated function of frequency ($\om$), momentum ($\vec q$\,), temperature ($T$), chemical potential ($\mu$), scattering rates ($\Gamma$), the gap ($\Delta$) in case the system is an insulator, .... 
see, \eg, \cite{Gusynin:2006ym,Gusynin:2009-ac} and \cite{Peres:2010mx} for a review. 
The collisionful regime ($\om \ll \Gamma,\, T$) is relevant to the study of the dc limit; it is complicated 
by the fact that one has to take into account temperature and disorder. This case has led to numerous works, see previous references. The collisionless limit is simpler and is relevant
to the study of the conductivity in the optical regime, \eg, at $\om \sim 1$eV (visible range of the spectrum) where experiments are carried out \cite{Nair:2008zz,PhysRevLett.101.196405}. It is on this regime 
that we will focus in this Chapter and we shall simply set: $\mu=\Delta=\vec q\,=\Gamma=T=0$. Certainly, once the effect of electron-electron interactions is understood in this limit, more realistic cases may be considered. Notice that, in this Chapter, we keep on setting $\hbar =1$.}
\be
\sigma_0 = \frac{e^2}{4}\, ,
\label{chap5:sigma0}
\ee
which is universal. This result, which was predicted long ago to hold for free Dirac fermions~\cite{Fradkin86.PhysRevB.33.3263,Lee93.PhysRevLett.71.1887,Ludwig94.PhysRevB.50.7526}, 
agrees to within $1$-$2\%$ with optical experiments \cite{Nair:2008zz,PhysRevLett.101.196405}.
This is rather surprising because the long-range Coulomb interaction among charge carriers is not only unscreened but also supposed to be strong as witnessed by 
the fine structure constant of suspended graphene:
\be
\al_g = \frac{e^2}{4 \pi v} \approx 2.2 \, ,
\label{chap5:alg}
\ee
which is of the order of unity due to the fact that $v \approx c / 300$. 
Moreover, as recalled in the Introduction, it is well known that Kohn's theorem~\cite{Kohn61PhysRev.123.1242} does 
not apply to pseudo-relativistic systems thereby allowing electron-electron interactions to affect Eq.~(\ref{chap5:sigma0}).
There has therefore been extensive theoretical attempts to understand the effect of electron-electron interactions on the
optical conductivity of graphene in the collisionless limit, see, {\it e.g.}, 
Refs.~\cite{PhysRevB.83.195401,Herbut08.PhysRevLett.100.046403,Mishchenko2008,Juricic:2010dm,Sheehy09.PhysRevB.80.193411,Abedinpour11.PhysRevB.84.045429,Sodemann12.PhysRevB.86.115408,%
Gazzola13.0295-5075-104-2-27002,Rosenstein13.PhysRevLett.110.066602,Link16.PhysRevB.93.235447,Boyda:2016emg,Stauber17.PhysRevLett.118.266801}.
The latter can be defined via a density-density correlation function:
\be
\sigma(q_0) = - \lim_{\vec{q} \ra 0} \, \frac{\I q_0}{|\vec{q}\,|^2}\,\Pi^{00}(q_0,\vec{q}\,)\, ,
\label{chap5:sigma-dd}
\ee
where, in real time, $\Pi^{0 0} (t,\vec{q}\,) = \B T \rho(t,\vec{q}\,) \rho(0,-\vec{q}\,) \K$, $\rho$ is the charge density and $T$ the time-ordering operator. 
Equivalently, from current conservation, it can also be defined 
via a current-current correlation function (Kubo formula):
\be
\tilde{\sigma}(q_0) = \frac{1}{\I q_0}\,\frac{K^{11}(q_0,\vec{q}=0\,) + K^{22}(q_0,\vec{q}=0\,)}{2}\, ,
\label{chap5:sigma-cc}
\ee
where in real time, $K^{i j} (t,\vec{q}\,) = \B T j^i(t,\vec{q}\,) j^j(0,-\vec{q}\,) \K$ and $\vec{j}$ is the charge current. Despite the strength of the interactions, 
one may first focus on the lowest order interaction corrections to $\Pi^{00}(q)$ and $K^{ij}(q)$:
\be
\sigma(q_0) = \sigma_0\, \bigg( 1 + \mathcal{C} \al_{g} + \Ord(\al_{g}^2) \bigg), \quad \tilde{\sigma}(q_0) = \sigma_0\, \bigg( 1 + \tilde{\mathcal{C}} \al_{g} + \Ord(\al_{g}^2) \bigg)\, , 
\label{chap5:pert-exp}
\ee
in order to extract the numerical value of the first order interaction-correction coefficients, $\mathcal{C}$ and $\tilde{\mathcal{C}}$, respectively.
On physical grounds, one expects that $\mathcal{C} = \tilde{\mathcal{C}}$, independent on the method used. The diagrams that have to be computed are those of Fig.~\ref{chap5:fig:2loop-polarization-cc}
for the Kubo formula and Fig.~\ref{chap5:fig:2loop-polarization-dd} for the density-density correlation function (conventions used for the vertices in these figures will be explained in the following).

It turns out, however, that different theoretical results can be found in the literature so that the value of the first order coefficient is controversial. 
As quoted by Ref.~\cite{Rosenstein13.PhysRevLett.110.066602}, these results read:
\begin{subequations}
\label{chap5:c123}
\bea
\mathcal{C}^{(1)}  = \frac{25-6\pi}{12} \approx 0.512\, ,
\label{chap5:c1}\\
\mathcal{C}^{(2)}  = \frac{19-6\pi}{12} \approx 0.013\, ,
\label{chap5:c2}\\
\mathcal{C}^{(3)}  = \frac{11-3\pi}{6} \approx 0.263\, .
\label{chap5:c3}
\eea
\end{subequations}
Starting from the Kubo formula, Herbut et al.\ obtained~\cite{Herbut08.PhysRevLett.100.046403}:
$\tilde{\mathcal{C}}^{(\Lambda)}  = \mathcal{C}^{(1)}$.
%
%\be
%\tilde{\mathcal{C}}^{(\Lambda)}  = \mathcal{C}^{(1)}\, .
%\frac{25-6\pi}{12} \approx 0.512\, .
%\label{chap5:tc}
%\ee
%
They used a hard cut-off in order to regularize the UV-divergences arising from individual two-loop diagrams
and which ultimately cancel out in their sum. On the other hand, starting from Eq.~(\ref{chap5:sigma-dd}), Mishchenko obtained~\cite{Mishchenko2008}:
\be
\mathcal{C}^{(\Lambda)} = \mathcal{C}_{a}^{(\Lambda)} + \mathcal{C}_{b}^{(\Lambda)} = \mathcal{C}^{(2)},
%\frac{19-6\pi}{12} \approx 0.013\, 
\qquad \mathcal{C}_{a}^{(\Lambda)}  = \frac{1}{4}, \qquad \mathcal{C}_{b}^{(\Lambda)}  = \frac{8-3\pi}{6}\, ,
\label{chap5:c}
\ee
where individual two-loop diagrams are finite in this case and the hard cut-off only regularizes the divergent self-energy subgraph
of the two diagrams in Fig.~\ref{chap5:fig:2loop-polarization-dd}a (the latter contribute to the value $\mathcal{C}_{a}$ while 
$\mathcal{C}_{b}$ comes from the diagram in Fig.~\ref{chap5:fig:2loop-polarization-dd}b). 
Still using a hard cut-off, a third result, $\mathcal{C}_{kin}^{(\Lambda)} = \mathcal{C}^{(3)}$, was even obtained with the help of a kinetic equation approach~\cite{Mishchenko2008}.
%
%\be
%\tilde{\tilde{\mathcal{C}}}^{(\Lambda)} = \mathcal{C}^{(3)}
%\frac{11-3\pi}{6} \approx 0.263\, .
%\ee
%
According to Mishchenko's analysis, these discrepancies are due to the long-range nature of the Coulomb interaction. He explains that ``the polarization
function method ... gives results (in the form of convergent integrals) {\it independent} of the cut-off procedure'' so that calculations with Eq.~(\ref{chap5:sigma-dd}) are reliable. 
On the other hand, both the Kubo, Eq.~(\ref{chap5:sigma-cc}), and kinetic equation approaches involve (singular) current vertices and thus ``{\it fail} if the hard cut-off is implemented without a proper 
modification of the current vertices''. He then advocated the use of a soft cut-off in order to properly regularize
the UV-divergent integrals finding, in all cases, a single result given by $\mathcal{C}^{(2)}$ \cite{Mishchenko2008}. 

In Ref.~\cite{Juricic:2010dm} (henceforth referred to as JVH) the coefficient was recomputed with the help of dimensional regularization (DR). We have repeatedly used this technique
in this manuscript. Certainly, it is incomparably more efficient that the cut-off approach as it preserves the symmetries of the model. Starting from Eq.~(\ref{chap5:sigma-dd}), JVH obtained:
\be
\mathcal{C}^{(D)} =  \mathcal{C}_{a}^{(D)} + \mathcal{C}_{b}^{(D)} = \mathcal{C}^{(3)}\, \qquad \mathcal{C}_{a}^{(D)}  = \frac{1}{2}, \qquad \mathcal{C}_{b}^{(D)}  = \frac{8-3\pi}{6}\, ,
\label{chap5:c-D}
\ee
in disagreement with Eq.~(\ref{chap5:c}). They obtain the same result via the Kubo formula: $\tilde{\mathcal{C}}^{(D)} = \mathcal{C}^{(3)}$, in disagreement with both
Eq.~(\ref{chap5:c}) and their previous result~\cite{Herbut08.PhysRevLett.100.046403}: $\tilde{\mathcal{C}}^{(\Lambda)}  = \mathcal{C}^{(1)}$. 
%Eq.~(\ref{chap5:tc}). 
Recently, support in favor of JVH's result, $\mathcal{C}^{(3)}$, came from yet another approach based on a full tight-binding computation \cite{Rosenstein13.PhysRevLett.110.066602}.
However, the most commonly accepted result is, up to date, the value $\mathcal{C}^{(2)}$ of Mishchenko since it has been recovered by a majority of groups, see, {\it e.g.}, 
Refs.~\cite{Sheehy09.PhysRevB.80.193411,Abedinpour11.PhysRevB.84.045429,Sodemann12.PhysRevB.86.115408,Gazzola13.0295-5075-104-2-27002,Link16.PhysRevB.93.235447,Boyda:2016emg,Stauber17.PhysRevLett.118.266801}. 
Incidentally, this is also the only result, among those of Eqs.~(\ref{chap5:c123}), which is consistent with the experimental uncertainties~\cite{Nair:2008zz,PhysRevLett.101.196405} as
$\mathcal{C}^{(2)} \al_g \approx 2\%$. On the numerical side, let's note that very recent tight-binding computation \cite{Link16.PhysRevB.93.235447} and Quantum Monte Carlo simulations \cite{Boyda:2016emg} are both in favour of $\mathcal{C}^{(2)}$.
But the latest result of Stauber et al.\ \cite{Stauber17.PhysRevLett.118.266801}, using a self-consistent Hartree-Fock approach, found $\mathcal{C}^{(3)}$ in case of unscreened interactions while 
self-screening (equivalent to RPA) has to be included in order to reduce the interaction correction coefficient to a value compatible with $\mathcal{C}^{(2)}$, see Tab.~\ref{chap5:tab:C-values} for a summary of 
some results.

Let us note at this point that there is a limit where the result for the interaction correction coefficient does not raise any doubt: this is the deep IR limit corresponding to the Lorentz-invariant fixed point
that was described at length in Chap.~\ref{chap3}. In this limit, the result (\ref{chap3:res2l:RQED4,3:Pi-total-2}) was obtained for the (renormalized) polarization operator up to two loops.
It is the expression of $\Pi(q^2)$ which is given in this formula, which is related to the current-current correlator $\Pi^{\mu \nu}(q)$ with the help of: 
$\Pi^{\mu \nu}(q) = (g^{\mu \nu}q^2 - q^\mu q^\nu)\,\Pi(q^2)$. The corresponding conductivity reads: 
\be
\tilde{\sigma}(q_0) = \sigma_0\, \bigg( 1 + \mathcal{C}^* \al + \Ord(\al^2) \bigg)\, , \qquad \mathcal{C}^* = \frac{92-9\pi^2}{18\pi}\, .
\label{chap5:sigma-cc*}
\ee
Of course, at the fixed point $\al = 1/137$ and the product $\mathcal{C}^* \al \approx 10^{-4}$ is very small leading to almost unobservable effects. However, it is interesting to note that 
$\mathcal{C}^{*} = 0.056$, a value which is of the same order of magnitude as $\mathcal{C}^{(2)}$ (and, surprisingly, in very good agreement with the result of the recent 
self-consistent Hartree-Fock calculation with self-screened interactions).\footnote{The interest in comparing $\mathcal{C}^{(2)}$ to $\mathcal{C}^{*}$
(rather than $\mathcal{C}^{(2)}\alpha_g$ to $\mathcal{C}^{*}\alpha$) comes from the existence of the more general model I which is valid for arbitrary $v/c$, see Chaps.~\ref{chap1} and \ref{chap2}. 
So there is actually a non-trivial interaction correction {\it function}
$\mathcal{C}(v/c)$ which encodes relativistic corrections (the dependence of the fine structure constant on $v$ is trivial). Presently, only the limiting
values $\mathcal{C}(v/c \rightarrow 0) = \mathcal{C}^{(2)}$ (see the proof in the following pages) and $\mathcal{C}(v/c \rightarrow 1) = \mathcal{C}^{*}$ are
known. It is surprising that these values are of the same order as if $\mathcal{C}(v/c)$ was only weakly dependent on $v/c$. A study of $\mathcal{C}(v/c)$ for arbitrary
$v/c$ is beyond the scope of the present manuscript and we leave it for our future investigations.}

At this point, let's recall from previous chapters that a self-dual point was found in \cite{Hsiao:2017lch} for reduced QED$_{4,3}$ at $e_{\text{sd}}^2=8\pi$. 
This leads to $\al_{\text{sd}} =2$ which is surprisingly close to the bare coupling constant of graphene. 
At this value of the coupling, the optical conductivity could be computed exactly with the result reading: 
\be
\sigma_{\text{sd}}(q_0) = \frac{4}{\pi}\,\sigma_0 \qquad (\text{at $\al_{\text{sd}}=2$})\, .
\ee
This result leads to a $27\%$ deviation of the conductivity with respect to the free fermion result ($4/\pi = 1.273$). As noticed in \cite{Hsiao:2017lch}, the perturbative result (\ref{chap5:sigma-cc*})
leads only to a $\mathcal{C}^* \al_{\text{sd}} = 11\%$ deviation with respect to free fermions clearly showing that at large couplings higher orders cannot be neglected. Of course, at the fixed point $\al \ll 1$ and
the result (\ref{chap5:sigma-cc*}) is reliable. But clearly, for graphene, results based on the loop expansion (\ref{chap5:pert-exp}) may also receive strong corrections from higher orders and the perturbative approach
is rightly questionable. A way to overcome this difficulty may be to reorder the perturbative series in the form of a $1/N$-expansion or, in other words, to preform an RPA-like resummation, see, {\eg}, 
\cite{PhysRevLett.113.105502} for an attempt to carry out NLO computations. This is a very interesting suggestion which certainly requires further study. In the following, we shall pursue a more modest goal
and restrict ourselves to the study of Eq.~(\ref{chap5:pert-exp}) without any additional resummation. The reason is that, in our opinion, the problem of an {\it accurate} evaluation of NLO 
corrections cannot reasonably be addressed before a {\it full} understanding of the first few orders of the loop expansion is achieved.  

\begin{center}
\renewcommand{\tabcolsep}{0.25cm}
\renewcommand{\arraystretch}{1.5}
\begin{table}
\begin{tabular}{|c|c|c|}
\hline
             $\mathcal{C}$     		                        &            {\bf Method}                                          &       {\bf Year} \\
\hline
\hline
$\mathcal{C}^{(1)}  = (25-6\pi)/12 \approx 0.512$       	& Eq.~(\ref{chap5:sigma-cc}) hard cut-off			   &    2008 \cite{Herbut08.PhysRevLett.100.046403}    \\
\hline
\hline
\textcolor{red}{$\mathcal{C}^{(2)}  = (19-6\pi)/12 \approx 0.013$}		& Eq.~(\ref{chap5:sigma-dd}) hard cut-off	   &    2008 \cite{Mishchenko2008} \\
\cline{2-3}
%%\multirow{10}{*}{$\quad$} 
			     					& Eq.~(\ref{chap5:sigma-cc}) and kinetic equations, soft cut-off   &    2008 \cite{Mishchenko2008}           \\
\cline{2-3}
        							& Eqs.~(\ref{chap5:sigma-cc}) hard cut-off			   &    2009 \cite{Sheehy09.PhysRevB.80.193411}            \\
\cline{2-3}
        							& Eq.~(\ref{chap5:sigma-dd}) hard cut-off			   &    2011 \cite{Abedinpour11.PhysRevB.84.045429}        \\
\cline{2-3}
        							& Eq.~(\ref{chap5:sigma-dd}) hard cut-off			   &    2012 \cite{Sodemann12.PhysRevB.86.115408}           \\
\cline{2-3}
                                                                & Eqs.~(\ref{chap5:sigma-cc}) hard cut-off, implicit regularization  &  2013 \cite{Gazzola13.0295-5075-104-2-27002}           \\
\cline{2-3}
                                                                & \textcolor{red}{Eqs.~(\ref{chap5:sigma-cc}), (\ref{chap5:sigma-dd}),  DR + CR}  &    \textcolor{red}{2014} \cite{Teber:2014ita}            \\
\cline{2-3}
                                                                & lattice (tight-binding) simulations 				   &    2016  \cite{Link16.PhysRevB.93.235447}           \\
\cline{2-3}
                                                                & Quantum Monte Carlo calculations				   &    2016  \cite{Boyda:2016emg}           \\
\cline{2-3}
($0.05$)     	                                                & Hartree-Fock simulations (self-screened)                         &    2017 \cite{Stauber17.PhysRevLett.118.266801}\\
\cline{2-3}
								& \textcolor{red}{Eqs.~(\ref{chap5:sigma-cc}), (\ref{chap5:sigma-dd}),  DR + BPHZ}  &    \textcolor{red}{2018} \cite{Teber:2018qcn}           \\
\hline
\hline
$\mathcal{C}^{(3)}  = (11-3\pi)/6 \approx 0.263$		& kinetic equations, hard cut-off				   &    2008 \cite{Mishchenko2008}                  \\
\cline{2-3}
%%\multirow{10}{*}{$\quad$}
  			                                        & Eqs.~(\ref{chap5:sigma-cc}), (\ref{chap5:sigma-dd}), DR          &  2010 \cite{Juricic:2010dm}          \\
\cline{2-3}
								& lattice (tight-binding) simulations				   &    2013 \cite{Rosenstein13.PhysRevLett.110.066602}\\
\cline{2-3}
($1/4=0.25$)                                                    & Hartree-Fock simulations (unscreened)                            &    2017 \cite{Stauber17.PhysRevLett.118.266801}\\
\hline
\hline
\textcolor{red}{$\mathcal{C}^{*}  = (92-9\pi^2)/(18\pi) \approx 0.056$}          & \textcolor{red}{DR + CR ($v/c \ra 1$)}          &  \textcolor{red}{2012} \cite{Teber:2012de}                  \\
\hline
\end{tabular}
    \caption{Some values of $\mathcal{C}$ obtained over the years together with elements of the different methods used. In case of numerical simulations, we cite the numerical value obtained whenever available and 
when it slightly differs from the main $3$ results found in the literature, $\mathcal{C}^{(i)}$ ($i=1,2,3$). For the sake of completeness, the value at the IR fixed point, $\mathcal{C}^{*}$, has been also added. 
DR is for dimensional regularization. CR is for conventional renormalization. BPHZ stands for the use of the forest formula.}
    \label{chap5:tab:C-values}
\end{table}
\end{center}

In the following we will focus on the computation of the first order interaction correction to the minimal conductivity of graphene in the non-relativistic limit ($v/c \ra 0$) with 
the help of dimensional regularization \cite{Teber:2014ita}. Our approach will make use of the multi-loop techniques introduced in Chap.~\ref{chap2} and that were used to compute the 
result presented above, Eq.~(\ref{chap5:sigma-cc*}), in the ultra-relativistic limit ($v/c \ra 1$).
As will be seen in the following the non-relativistic case is a little more subtle than the ultra-relativistic one. Anticipating the conclusion, we will show in the following 
that our approach is in favour of the result first derived by Mishchenko. Our final result reads: 
\be
\mathcal{C}^{(\rm{DR})} = \tilde{\mathcal{C}}^{(\rm{DR})} = \mathcal{C}^{(D)} + \mathcal{C'}^{(D)} = \mathcal{C}^{(2)}\, \qquad \mathcal{C'}^{(D)} = -\frac{1}{4}\, ,
\label{chap5:c-DR}
\ee
where $\mathcal{C'}^{(D)}$ originates from one-loop counterterms. As far as dimensional regularization is concerned, our analysis reveals that the origin of the discrepancy between the different results
found in the literature does not lie in the regularization method but, more simply, in the renormalization procedure itself. 
This has been proven in \cite{Teber:2014ita} using conventional renormalization (CR). In the following we will give a stronger proof
based on the BPHZ (Bogoliubov-Parasiuk-Hepp-Zimmermann) renormalization prescription, see Chap.~\ref{chap2}.~\footnote{Let's also note the more recent Hopf 
algebraic formulation~\cite{Kreimer:1997dp,Connes:1998qv} of renormalization, see also Ref.~\cite{Panzer:2014kia} for a recent review. 
Its application to our model is beyond the scope of our present study.} The later will provide a clear explanation for why radiative corrections 
to the optical conductivity of graphene are finite and perfectly well determined.\footnote{We are therefore
not in the field theoretic situation reviewed by Jackiw in \cite{Jackiw:1999qq}, see also \cite{Gazzola13.0295-5075-104-2-27002} for a related nice paper on the optical conductivity of graphene. 
According to Ref.~\cite{Jackiw:1999qq}, it seems that there are indeed peculiar cases where corrections may take arbitrary values; they cannot be determined by the theory and require experiments to be fixed.
As reviewed in Ref.~\cite{Jackiw:1999qq}, one such case is when an anomaly is present. Our opinion is that no such anomaly exists in the problem under consideration (which is $(2+1)$-dimensional)
and that the interaction correction has a unique value independent of the method used to evaluate it (hence it is well determined).}

\section{Master integrals} 

%In the following, we shall work with in a space of dimension, $D = 2 -2\veps$.
Contrary to JVH, who introduced Feynman parameters, we shall compute the multi-loop dimensionally regularized integrals using algebraic methods, {\it c.f.}, Chap.~\ref{chap2}, therefore
providing an independent check of the calculations. The implementation of these methods requires the knowledge of some basic integrals
such as the massless one-loop propagator-type integral with $n\leq 2$ \cite{Kazakov:1986mu}
%a traceless product:
%
\begin{subequations}
\label{chap5:massless-p-integral}
\begin{flalign}
&\int [\D^{D} q]\,\frac{q^{\mu_1}\dotsc q^{\mu_n}
%q^{\mu_1 \mu_2 \dotsc \mu_n}
}{[q^2]^{\al}[(q-k)^2]^\beta}  =
\frac{(k^2)^{D/2-\al-\beta}}{(4\pi)^D}\,\times \,\left[k^{\mu_1} \dotsc k^{\mu_n}
%k^{\mu_1 \mu_2 \dotsc \mu_n}
\, G^{(n,0)}_0(\al,\beta)\, + \delta^2_n \frac{g^{\mu_1 \mu_2}}{D} 
\, G^{(1,1)}_1(\al,\beta)\right], ~~~
\label{chap5:massless} \\
&G^{(n,m)}_i(\al,\beta) = \frac{a_n(\al)a_m(\beta)}{a_{n+m-i}
%(\beta-\tilde{\alpha}-i)},~~
(\al+\beta-D/2-i)}, 
\qquad a_n(\al) = \frac{\Gamma(D/2-\al + n)}{\Gamma(\al)} \, ,
\end{flalign}
\end{subequations}
where $[\D^D q] = \D^{D} q / (2\pi)^{D}$ and  $\delta^2_n$ is the Kronecker
symbol.
%, $\tilde{\alpha}= D/2-\al$ 
%and
%
%\begin{subequations}
%\bea 
%&&G^{(n,m)}_i(\al,\beta) = \frac{a_n(\al)a_m(\beta)}{a_{n+m-i}
%%(\beta-\tilde{\alpha}-i)},~~
%(\al+\beta-D/2-i)}, \\
%&&a_n(\al) = \frac{\Gamma
%(\tilde{\alpha}+n)
%(D/2-\al + n)
%}{\Gamma(\al)} \, .
%\eea
%\end{subequations}
%
The simplified notation: $G(\al,\beta)=G^{(0,0)}_0(\al,\beta)$, will also be used.
As we shall see in the following, the computation of the optical conductivity involves semi-massive tadpole diagrams. 
In particular, the one-loop semi-massive tadpole diagram reads:
\begin{flalign}
%\label{chap5:tadpole}
\int \frac{[\D^D k]}{[k^2]^{\al}[k^2+m^2]^\beta} 
%\int \frac{\D^{D} k}{(2\pi)^{D}} \frac{1}{[k^2]^{\al}[k^2+m^2]^\beta} 
=
\frac{(m^2)^{D/2-\al-\beta}}{(4\pi)^{D/2}}\,B(\beta,\al)\, ,
\qquad B(\beta,\al) = \frac{\Gamma(D/2-\al)\,\Gamma(\al + \beta - D/2)}{\Gamma(D/2)\,\Gamma(\beta)}\, .
\label{chap5:tadpole}
\end{flalign}
%
%\bea
%B(\beta,\al) = \frac{\Gamma(D/2-\al)\,\Gamma(\al + \beta - D/2)}{\Gamma(D/2)\,\Gamma(\beta)}\, .
%\label{chap5:tadpole-B}
%\eea
%
These formulas can be used to compute all required 2-loop semi-massive tadpole diagrams. As will be seen in the following,
the latter are of the form:
\begin{flalign}
I_n(\al) =  \int  \frac{[\D^{D_e}k_1][\D^{D_e}k_2]\,(\vec{k}_1 \cdot \vec{k}_2\,)^n
[|\vec{k}_1 - \vec{k}_2\,|^{2}]^{-1/2}}
{[|\vec{k}_1\,|^2]^\al\,[|\vec{k}_1\,|^2+m^2]\,[|\vec{k}_2\,|^2]^{\al}\,[|\vec{k}_2\,|^2+m^2]
%[|\vec{k}_1 - \vec{k}_2\,|^{2}]^{1/2}
} =  \frac{(m^2)^{D_e+n-2\al-5/2}}{(4\pi)^{D_e}} \, \tilde{I}_n(\al)
\, .
%\nonum
\label{chap5:I(n)}
\end{flalign}
The diagrams with different $\al$ values are related to each other by the relation:
\be
\frac{1}{k^{2\al}\,[k^2+m^2]} =
\frac{1}{m^2} \left[ \frac{1}{k^{2\al}} -
\frac{1}{k^{2(\al-1)}\,[k^2+m^2]} \right] \, .
\ee
The master integrals $I_n(\al)$ with some particular $\al$ values can be calculated by 
%with help of 
Eqs.~(\ref{chap5:massless-p-integral})-(\ref{chap5:tadpole}) when one of the massive propagators
can be replaced by a massless one with the help of a Mellin-Barnes transformation \cite{Boos:1990rg} (see also Eq.~(\ref{chap2:Mellin-Barnes})):
\be
\frac{1}{k^2 + m^2} = \frac{1}{2\I \pi}\,\int_{-\I \infty}^{+\I \infty} \D s \,\Gamma(-s) \Gamma(1+s) \frac{(m^2)^s}{(k^2)^{1+s}}\, . 
\ee
The results have the following form for $\veps_\gamma \ra 0$:
\be
\tilde{I}_0(1/2)= -\tilde{I}_1(3/2)= \tilde{I}_2(5/2)= 
\,\pi^2\, .
\label{chap5:masters}
\ee
We note that really only one of the diagrams, for example $I_0(1/2)$, is
independent. The two others can be expressed through $I_0(1/2)$ using
integration by parts identities. However, this procedure is quite long
and we will not prove it here. 
The diagrams with other  $\al$ values can be expressed as combinations
of the ones in Eq.~(\ref{chap5:masters}) and of simpler diagrams, which can be calculated  
directly with help of Eqs.~(\ref{chap5:massless-p-integral})-(\ref{chap5:tadpole}).
So, we have, for the diagrams contributing to $\Pi^{00}(q)$ at two loops (see 
Eq.(\ref{chap5:pi2b00-inter}) below) and $\veps_\gamma \ra 0$:
\be
\tilde{I}_1(1/2)= \,\pi(4-\pi)\, ,~~ \tilde{I}_2(3/2)= \,\pi\left(\pi -
\frac{4}{3}\right) \, .
\label{chap5:master-dd}
\ee
On the other hand, the diagrams contributing to $\Pi(q^2)$ (see Eq.~(\ref{chap5:pi2bmumu-inter})) are UV-singular and read:
\bea
&&\tilde{I}_1(-1/2)=  
\pi^2 - 2 G\left(1/2,1/2\right)
\, B\left(1,(3- D_e)/2\right)
\, , \label{chap5:master-cc} \\
&&\tilde{I}_2(1/2)=  
\pi^2 - \frac{4\pi}{3} 
- (3-D_e)\,G\left(1/2,1/2\right)
\,B\left(1,(3- D_e)/2\right)\, ,
\nonumber
\eea
with the accuracy $ O(\veps_\gamma)$.

\section{Feynman rules and renormalization} 

The effective low-energy action that we wish to consider is the one of model II, Eq.~(\ref{chap1:model-inst}), with $D_e=2$ and $D_\gamma=3$ that we reproduce here for clarity:
\be
S = \int \D t\, \D^{2} x\, \bar{\psi}_\sigma \left[ \gamma^0 \left( \I \partial_t -eA_0 \right) + \I v \vec{\gamma} \cdot \vec{\nabla}\,\right] \psi^\sigma
+\, \frac{1}{2}\,\int \D t\, \D^{3} x\, \left( \vec{\nabla} A_0 \right)^2 \, ,
\label{chap5:model-inst}
\ee
where, as before, $v$ and $e$ are the bare Fermi velocity and charge, respectively,
$\psi_\sigma$ is a four-component spinor field describing a fermion of specie $\sigma$ ($\sigma=1,\cdots,N_F$ and for graphene $N_F=2$) and
$A_0$ is the gauge field mediating the instantaneous Coulomb interaction.
The Dirac matrices, $\gamma^\mu = (\gamma^0,\vec{\gamma}\,)$ satisfy the usual algebra $\{\gamma^\mu,\gamma^\nu \} = 2g^{\mu \nu}$,
with metric tensor $g^{\mu \nu} = \text{diag}(+,-,-)$. 

From Eq.~(\ref{chap5:model-inst}), the bare momentum space fermion propagator reads (we use pseudo-relativistic notations with
$p^\mu=(p^0,v \vec{p}\,)$, see App.~\ref{app:pr} for our conventions):
\be
S_0(p) = \frac{i\Sp}{p^2}\, , \qquad \Sp = \gamma^\mu p_\mu = \gamma^0 p_0 - v \vec{\gamma}\cdot \vec{p}\, .
\label{chap5:fermion-prop0}
\ee
The effective photon propagator reduces to the instantaneous Coulomb interaction and reads:
\be
V_0(\vec{q}\,) = \frac{\I}{2 (|\vec{q}\,|^2)^{1/2}}\, .
\label{chap5:gauge-field-prop0}
\ee
The bare vertex reduces to the temporal part: 
\be
-\I e \Gamma_0^0 = -\I e \gamma^0 \, .
\label{chap5:vertex0}
\ee
These rules can be compared with the ones of the more general model I (\ref{chap1:model-general}), see Eqs.~(\ref{chap3:gm:fermion-prop0}), (\ref{chap3:gm:gauge-field-prop0}) and (\ref{chap3:gm:vertex}).
In particular, in the limit $v/c \ra 0$ vector photons decouple, see Eq.~(\ref{chap3:gm:vertex}). 
 %vector photons decouple as witnessed from the fact that: $\vec{\Gamma}_0 = -\frac{v}{c} \vec{\gamma} \ra 0$, see Eq.~(\ref{chap3:gm:vertex}).
 %Nevertheless, for our future purposes, it will be convenient to define $\Gamma_{\text{eff}}^\mu = (\Gamma^0, \vec{\Gamma}_{\text{eff}})$ which is such that $\Gamma_{\text{eff}\,0}^\mu = \gamma^\mu = (\gamma^0, \vec{\gamma}\,)$
Nevertheless, for our future purposes, it will be convenient to define $\Gamma^\mu = (\Gamma^0, \vec{\Gamma}\,)$ such that $\Gamma_{0}^\mu = \gamma^\mu = (\gamma^0, \vec{\gamma}\,)$
and employ the following graphical notations:
\be
-\I e \gamma^0 = ~~ 
      \parbox{20mm}{
        \begin{fmfgraph*}(20,10)
          %\fmfpen{thick}
          \fmfleft{p}
          \fmfright{ei,eo}
          \fmf{boson}{p,v}
          \fmf{vanilla}{ei,v}
          \fmf{fermion}{v,eo}
          \fmfv{decor.shape=circle,decor.filled=empty,decor.size=2thick}{v}
        \end{fmfgraph*}
      } ~~, \qquad
-\I e \vec{\gamma} = ~~
      \parbox{20mm}{
        \begin{fmfgraph*}(20,10)
          %\fmfpen{thick}
          \fmfleft{p}
          \fmfright{ei,eo}
          \fmf{boson}{p,v}
          \fmf{vanilla}{ei,v}
          \fmf{fermion}{v,eo}
%          \fmfdot{v}
	\fmfv{decor.shape=circle,decor.filled=shaded,decor.size=2thick}{v}
          %\fmfiv{decor.shape=cross,decor.filled=empty,decor.size=5thick}{#1}}
          %\vertc{v}
        \end{fmfgraph*}
      } ~~, \qquad
-\I e \gamma^\mu = ~~
      \parbox{20mm}{
        \begin{fmfgraph*}(20,10)
          %\fmfpen{thick}
          \fmfleft{p}
          \fmfright{ei,eo}
          \fmf{boson}{p,v}
          \fmf{vanilla}{ei,v}
          \fmf{fermion}{v,eo}
          \fmfdot{v}
        \end{fmfgraph*}
      } ~~.
\ee

As in the relativistic case, in conventional DR, these Feynman rules stay the same but momenta, Dirac matrices and metric tensor are extended to span a $D_e$-dimensional space
(keeping $\Tr \left[ {\bf 1} \right] = 4 N_F$) with 
\be
D_e = 2 - 2\veps_\gamma\, ,
\label{chap5:def:De}
\ee
and the renormalization scale $\mu$ has now dimension of momentum. All bare parameters and fields are then related to renormalized ones via renormalization constants in a standard way: 
%$\psi_n = Z_{\psi}^{1/2}\psi_{nr}$, $A_0 = Z_A^{1/2} A_{0r}$ and $v_0 = Z_v v$.
%
\be
\psi = Z_{\psi}^{1/2} \psi_r, \qquad A_0 = Z_{A_0}^{1/2} A_{0r}, \qquad e = Z_e e_r \mu^{\veps}, \qquad v = Z_v v_r\, ,
\label{chap5:renormZ}
\ee
where:
\be
Z_e = \frac{Z_{\Gamma^0}}{Z_\psi Z_{A_0}^{1/2}}\, ,
\ee
and such that the basic correlation functions of the model are renormalized as follows:
\be
S(k) = Z_\psi S_r(k), \qquad V(\vec{q}\,) = Z_{A_0} V_{r}(\vec{q}\,), \qquad \Gamma^\mu(k,q) = Z_{\Gamma^\mu}^{-1} \Gamma_r^\mu(k,q)\, ,
\ee
where the notation implicitly assumes that renormalized quantities depend on $\mu$.
The action of the free reduced gauge field is non-local so the gauge field is not renormalized: $Z_{A_0}=1$, see Chap.~\ref{chap3:lorentz} for similar arguments at the level of model I. 
Moreover, as we will check explicitly below, the Ward identity: $Z_{\Gamma^0} = Z_\psi$ still holds. 
As a consequence, the charge is not renormalized: $Z_e=1$. The renormalization of the coupling constant is therefore entirely due to the renormalization of the velocity which is the only 
running parameter of the model:
\be
\al_g = Z_{\al} \,\al_{gr}, \qquad Z_{\al} = Z_v^{-1}\, .
\label{chap5:alr}
\ee
The velocity $\beta$-function is defined by the usual expression, see Eq.~(\ref{chap3:gm:def:betav}), that we reproduce here for clarity:
\bea
\beta_v(\al_{gr}) = \mu \frac{\partial v_{r}}{\partial \mu} = \sum_{l=0}^\infty \beta_{v,l} \al_{gr}^{l+1}, \qquad \qquad \qquad \beta_{v,l} = 2 v_r\,(l+1)\,Z_v^{(l+1,1)}\, .
\label{chap5:gm:def:betav}
\eea

\section{Graphene at one-loop}

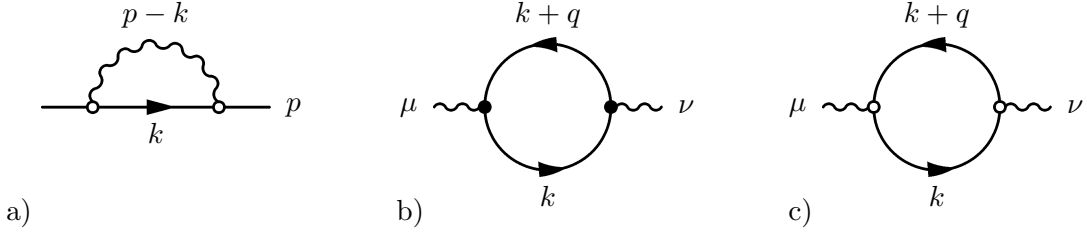
\begin{figure}
\begin{center}
  a)
    \begin{fmfgraph*}(30,30)
      %\fmfpen{thick}
      \fmfleft{in}
      \fmflabel{$p$}{out}
      \fmfright{out}
      \fmflabel{$p$}{out}
      \fmf{plain}{in,vi}
      \fmf{fermion,tension=0.2,label=$k$}{vi,vo}
      \fmf{boson,left,tension=0.2,label=$p-k$}{vi,vo}
      \fmf{plain}{vo,out}
      \fmfv{decor.shape=circle,decor.filled=empty,decor.size=2thick}{vi,vo}
    \end{fmfgraph*}
  \qquad \qquad
  b)
    \begin{fmfgraph*}(30,30)
      %\fmfpen{thick}
      \fmfleft{in}
      \fmflabel{$\mu$}{in}
      \fmfright{out}
      \fmflabel{$\nu$}{out}
      \fmf{boson}{in,ve}
      \fmf{fermion,right,tension=0.2,label=$k$}{ve,vw}
      \fmf{fermion,right,tension=0.2,label=$k+q$}{vw,ve}
      \fmf{boson}{vw,out}
%      \fmfv{decor.shape=circle,decor.filled=empty,decor.size=2thick}{ve,vw}
      \fmfdot{ve,vw}
    \end{fmfgraph*}
  \qquad \qquad
c)
    \begin{fmfgraph*}(30,30)
      %\fmfpen{thick}
      \fmfleft{in}
      \fmflabel{$\mu$}{in}
      \fmfright{out}
      \fmflabel{$\nu$}{out}
      \fmf{boson}{in,ve}
      \fmf{fermion,right,tension=0.2,label=$k$}{ve,vw}
      \fmf{fermion,right,tension=0.2,label=$k+q$}{vw,ve}
      \fmf{boson}{vw,out}
      \fmfv{decor.shape=circle,decor.filled=empty,decor.size=2thick}{ve,vw}
    \end{fmfgraph*}
  \caption{\label{chap5:fig:one-loop}
  One-loop: a) fermion self-energy and photon self-energies: b) $\Pi^{\mu \nu}$, c) $\Pi^{00}$.}
\end{center}
\end{figure}

We start by analysing model II, Eq.~(\ref{chap5:model-inst}), at one-loop. 

\subsection{One-loop fermion self-energy}

The one-loop fermion self-energy, Fig.~\ref{chap5:fig:one-loop}a, is defined as:
\be
-i\Sigma_1(k) = \mu^{2\veps_\gamma}\,\int [\D^{d_e} q]\, (-\I e \gamma^0) \,S_0(k+q)\, (-\I e \gamma^0)\,V_0(q)\, ,
\label{chap5:def:fermion-self}
\ee
where $d_e = 1 + D_e$ is the space-time dimension. The following parametrization will be used (see Eq.~(\ref{chap3:gm:Sigma:param})):
\be
\Sigma(k) = \gamma^0 k_0 \,\Sigma_\om(k^2) - v \vec{\gamma} \cdot \vec{k}\,\,\Sigma_k(k^2)\, ,
\label{chap5:gm:Sigma:param}
\ee
which is such that:
\be
\Sigma_{\om}(k^2) = \frac{\Tr[\gamma^0 k_0\,\Sigma(k)]}{4N_F k_0^2}\, , \quad
\Sigma_{k}(k^2) = \frac{\Tr[\vec{\gamma} \cdot \vec{k}\,\Sigma(k)]}{4N_F v |\vec{k}\,|^2}\, .
\label{chap5:fsigma-param}
\ee
Using this parametrization together with the standard rules for integrating massless Feynman diagrams straightforwardly yields:
\be
\Sigma_{1k}(|\vec{k}\,|^2) =  -\frac{\al_g}{8}\,\left(\frac{\overline{\mu}^{\,2}}{|\vec{k}\,|^2}\right)^{\veps_\gamma}\,e^{\gamma_E\veps_\gamma}\,G(1/2,1/2)\,  ,
\label{chap5:fsigma2}
\ee
and $\Sigma_{1\om}(k^2) =0$. The fact that the one-loop fermion self-energy depends only on momentum is due to 
the instantaneous nature of the interaction. At one-loop, there is therefore no wave-function renormalization: 
\be
Z_{1\psi} = 1 + \delta Z_{1\psi}, \qquad \delta Z_{1\psi} = 0\, .
\label{chap5:Zpsi}
\ee
Performing the $\veps_\gamma$-expansion in Eq.~(\ref{chap5:fsigma2}), yields, with one-loop accuracy:
\be
\Sigma_{k1}(|\vec{k}\,|^{2}) = -\frac{\al_g}{8}\, \left( \frac{1}{\veps_\gamma} - L_k + 4 \log 2 + \Ord(\veps_\gamma) \right)\, ,
\label{chap5:fsigma2-exp}
\ee
where $L_k = \log(|\vec{k}\,|^2/\overline{\mu}^{\,2})$. 
The UV-divergent self-energy leads to a renormalization of the Fermi velocity~\cite{Gonzalez:1993uz}:
\be
Z_{1v} = 1 + \delta Z_{1v}, \qquad \delta Z_{1v}= -\frac{\al_{gr}}{8 \veps_\gamma}\, .
\label{chap5:Zv}
\ee
From Eq.~(\ref{chap5:gm:def:betav}) we see that the corresponding beta-function is negative: 
\be
\beta_v = - v_r \al_{gr}/4 + \Ord(\al_{gr}^2)\, ,
\ee
a result which agrees with (\ref{chap3:gm:res:betav-lims}) in the limit $x=v/c \ra 0$ and implies that Fermi velocity grows in the infrared~\cite{Gonzalez:1993uz}. 
Graphically, these results can be summarized as follows:
\be
\delta Z_{1v} = \mathcal{K}\bigg[ \Sigma_{1k}(k^2) \bigg] ~ = ~~
\mathcal{K}\bigg[~~
      \parbox{15mm}{
    \begin{fmfgraph*}(15,15)
      %\fmfpen{thick}
      \fmfleft{in}
      \fmfright{out}
      \fmf{plain}{in,vi}
      \fmf{plain,tension=0.2}{vi,vo}
      \fmf{boson,left,tension=0.2}{vi,vo}
      \fmf{plain}{vo,out}
      \fmfv{decor.shape=circle,decor.filled=empty,decor.size=2thick}{vi,vo}
    \end{fmfgraph*}
}~~
\bigg] ~ = ~ - \frac{\al_{gr}}{8 \veps_\gamma}\, .
\label{chap5:dZ1v}
\ee

\subsection{One-loop photon self-energy (free fermion conductivity)}

We may proceed in a similar way with the one-loop photon self-energy, Fig.~\ref{chap5:fig:one-loop}b, defined in the usual way as:
\be
\I \Pi_1^{\mu \nu}(q) = - \mu^{2\veps_\gamma}\,\int [\D^{d_e}k]\, \Tr \left[ (-\I e \gamma^\mu)\,S_0(k+q)\,(-\I e \gamma^\nu)\,S_0(k) \right] \, .
\label{chap5:pi1munu}
\ee
Focusing on $\Pi^{00}$, performing the trace, going to euclidean space ($q_0 = \I q_{E0}$),
integrating over frequencies and taking the $\vec{q} \ra 0$ limit, yields:
\be
\Pi_1^{00}(m,\vec{q} \ra 0) = \frac{N_F e^2}{2v}\, \mu^{2\veps_\gamma}\,|\vec{q}\,|^2\,\frac{D_e-1}{D_e}\, 
\int \frac{[\D^{D_e}k]}{|\vec{k}\,|\, [|\vec{k}\,|^2 + m^2]}\, ,
\ee
which is of the form Eq.~(\ref{chap5:tadpole}) with $m = q_{E0}/2 v$. This is immediately integrated to give:
\be
\Pi_1^{00}(m,\vec{q} \ra 0) = \frac{N_F e^2}{8\pi v m}\,|\vec{q}\,|^2\,\left(\frac{\overline{\mu}^{\,2}}{m^2}\right)^{\veps_\gamma}\,\frac{D_e-1}{D_e}\,e^{\gamma_E\veps_\gamma}\,B(1,1/2)\, .
\ee
Performing the $\veps_\gamma$-expansion yields:
\be
\Pi_1^{00}(q_{0},\vec{q} \ra 0) = -\frac{N_F e^2}{8}\,\frac{|\vec{q}\,|^2}{\I q_0}\,\biggl( 1 - (1 + L_{q_0})\,\veps_\gamma + \Ord(\veps_\gamma^2) \biggr)\, ,
\label{chap5:pi100}
\ee
which is finite as expected and where $L_{q_0}=\log(-q_0^2/(4v^2 \overline{\mu}^{\,2}))$.
 %We note that Fermi velocity renormalization plays a crucial role in Eq.~(\ref{chap5:pi100}) as it brings the factor $(Z_v)^{2\veps_\gamma} = 1-\al/4$ to $\Ord(\veps_\gamma)$ accuracy.
Combining Eqs.~(\ref{chap5:sigma-dd}) and (\ref{chap5:pi100}), we arrive at the (renormalized) one-loop conductivity: 
\be
\sigma_{1r}(q_0) = \frac{N_F e_r^2}{8}\, , %\biggl( 1 - (1 + L_{q_0})\,\veps_\gamma + \Ord(\veps_\gamma^2) \biggr)\, ,
 %\sigma_0 \,\biggl( 1 - (1 + L_{q_0})\,\veps_\gamma + \Ord(\veps_\gamma^2) \biggr)\, ,
\label{chap5:sigma1}
\ee
which, for $N_F=2$, corresponds to the well known universal minimal conductivity $\sigma_0$, Eq.~(\ref{chap5:sigma0}).
%$\sigma_1(q_0) = \sigma_0$.%\,(1 + \mathcal{C}'^{(D)} \al + \Ord(\al^2))$, with $\mathcal{C}'^{(D)} = -1/4$.

We may now proceed in a similar way with the help of the Kubo formula Eq.~(\ref{chap5:sigma-cc}). 
In order to better exploit the $O(2)$ space rotational symmetry of the system, we shall derive an alternate formula based on Eq.~(\ref{chap5:pi1munu}).
As before, the transversality of $\Pi^{\mu \nu}$ allows to parametrize it as follows:
\be
\Pi^{\mu \nu}(q) = (g^{\mu \nu}q^2 - q^\mu q^\nu)\,\Pi(q^2), \quad
\Pi(q^2) = \frac{- \Pi^\mu_\mu(q)}{(d_e-1)(-q^2)}\, .
\label{chap5:def:pi2}
\ee
Then:
\be
\tilde{\sigma}(q_0) = \I q_0 \, K(q_0), \qquad K(q_0) = v^2 \Pi(q_0^2,|\vec{q}\,|^2 \ra 0) \, ,
\label{chap5:sigma-cc2}
\ee
and should be properly renormalized in the course of the computation. Following the steps of the $\Pi^{00}$ computation, at one-loop, we have:
\be
K_1(m) =  \frac{N_F e^2}{8 \pi v m} \,\left(\frac{\overline{\mu}^{\,2}}{|\vec{k}\,|^2}\right)^{\veps_\gamma}\,\frac{D_e-1}{D_e}\,e^{\gamma_E\veps_\gamma}\,B(1,-1/2)\, ,
\ee
and the $\veps_\gamma$-expansion yields: %with two-loop accuracy:
\bea
K_1(q_{0}) = \,\frac{N_F \, e^2}{8 \,\I q_0}\,\biggl( 1 - (1 + L_{q_0})\,\veps_\gamma + \Ord(\veps_\gamma^2) \biggr)\, . %\left( 1 - \frac{\al}{4} \right)\, .
\label{chap5:K1-res}
\eea
Combining Eqs.~(\ref{chap5:sigma-cc2}) and (\ref{chap5:K1-res}), we arrive once again at: 
\be
\tilde{\sigma}_{1r}(q_0) = \frac{N_F e_r^2}{8}\, , %\biggl( 1 - (1 + L_{q_0})\,\veps_\gamma + \Ord(\veps_\gamma^2) \biggr)\, ,
\label{chap5:sigmat1}
\ee
in perfect agreement with Eq.~(\ref{chap5:sigma1}).

%$\tilde{\sigma}_1(q_0) = \sigma_0$.%\,(1 + \tilde{\mathcal{C}}'^{(D)} \al + \Ord(\al^2))$, with $\tilde{\mathcal{C}}'^{(D)} = -1/4$.

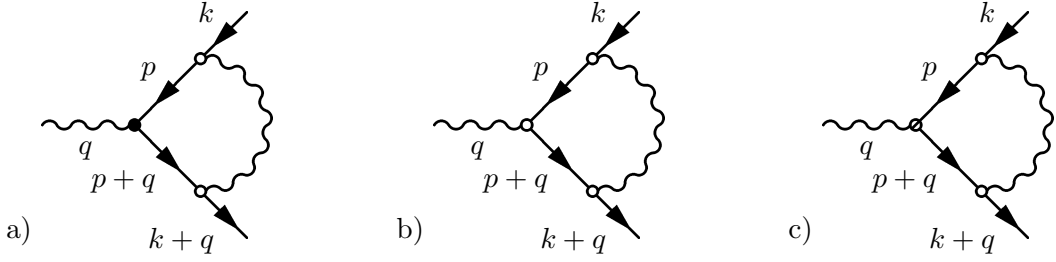
\begin{figure}
\begin{center}
  a)
    \begin{fmfgraph*}(30,30)
      %\fmfpen{thick}
      \fmfleft{in}
      \fmfright{e1,e2}
      \fmf{boson,label=$q$}{in,vi}
      \fmf{fermion,label=$k+q$,l.s=right}{v1,e1}
      \fmf{fermion,tension=0.7,label=$p+q$,l.s=right}{vi,v1}
      \fmf{fermion,tension=0.7,label=$p$,l.s=right}{v2,vi}
      \fmf{fermion,label=$k$,l.s=right}{e2,v2}
      \fmffreeze
      \fmf{boson,right,tension=0.7}{v1,v2}
      \fmfdot{vi}
      \fmfv{decor.shape=circle,decor.filled=empty,decor.size=2thick}{v1,v2}
   \end{fmfgraph*}
 \qquad \qquad
b)
    \begin{fmfgraph*}(30,30)
      %\fmfpen{thick}
      \fmfleft{in}
      \fmfright{e1,e2}
      \fmf{boson,label=$q$}{in,vi}
      \fmf{fermion,label=$k+q$,l.s=right}{v1,e1}
      \fmf{fermion,tension=0.7,label=$p+q$,l.s=right}{vi,v1}
      \fmf{fermion,tension=0.7,label=$p$,l.s=right}{v2,vi}
      \fmf{fermion,label=$k$,l.s=right}{e2,v2}
      \fmffreeze
      \fmf{boson,right,tension=0.7}{v1,v2}
      \fmfv{decor.shape=circle,decor.filled=empty,decor.size=2thick}{vi,v1,v2}
    \end{fmfgraph*}
 \qquad \qquad
c)
    \begin{fmfgraph*}(30,30)
      %\fmfpen{thick}
      \fmfleft{in}
      \fmfright{e1,e2}
      \fmf{boson,label=$q$}{in,vi}
      \fmf{fermion,label=$k+q$,l.s=right}{v1,e1}
      \fmf{fermion,tension=0.7,label=$p+q$,l.s=right}{vi,v1}
      \fmf{fermion,tension=0.7,label=$p$,l.s=right}{v2,vi}
      \fmf{fermion,label=$k$,l.s=right}{e2,v2}
      \fmffreeze
      \fmf{boson,right,tension=0.7}{v1,v2}
      \fmfv{decor.shape=circle,decor.filled=shaded,decor.size=2thick}{vi}
      \fmfv{decor.shape=circle,decor.filled=empty,decor.size=2thick}{v1,v2}
    \end{fmfgraph*}
  \caption{\label{chap5:fig:one-loop-vertex}
  One-loop vertex parts: a) $\Lambda_1^\mu$, b) $\Lambda_1^0$ and c) $\vec{\Lambda}_1$.}
\end{center}
\end{figure}

\subsection{One-loop fermion-photon vertex and Ward identities}

At this point it is also instructive to look at the vertex part: $\Gamma^\mu = \gamma^\mu + \Lambda_1^\mu + \Ord(\al_g^2)$ where $\Lambda_1^\mu$ is the one-loop correction.
The latter is defined as, see Fig.~\ref{chap5:fig:one-loop-vertex}a:
\be
-\I e \Lambda_1^\mu(k,q) = \mu^{2\veps_\gamma}\,\int [\D^{d_e}p]\, V_0(p-k)\,(-\I e\gamma^0)\,S_0(p+q)(-ie\gamma^\mu)\,S_0(p)(-\I e \gamma^0)\, .
\label{chap5:def:vertex}
\ee
%
%At one loop, the temporal part is defined as:
%%
%\be
%-\I e \Lambda_1^0(k,k') = \mu^{2\veps_\gamma}\,\int [\D^{d_e}q]\,(-\I e \gamma^0)\,\frac{\I (\Sq + \Sk)}{(q+k)^2}\,(-\I e \gamma^0)\,\frac{\I (\Sq + \Sk')}{(q+k')^2}\,(-\I e \gamma^0)\,\frac{\I}{2 |\vec{q}\,|}\, .
%\ee
%%
We first consider the temporal part, Fig.~\ref{chap5:fig:one-loop-vertex}b. In order to single out it's UV divergent part we will evaluate it for $k=q=0$. Performing the trace and going to euclidean space yields:
\be
\Lambda_1^0(0,0) = \frac{\mu^{2\veps_\gamma}\,e^2}{2}\,\gamma^0\,\int [\D^{D_e}p] [\D p_{E0}]\,\frac{p_{E0}^2 - v^2 |\vec{p}\,|^2}{[p_{E0}^2 + v^2 |\vec{p}\,|^2]^2\,|\vec{p}\,|} = 0\, ,
\ee
where the frequency integral vanishes identically. The temporal part of the vertex is therefore not renormalized at one-loop. 
Graphically, this result can be summarized as follows:
\be
\delta Z_{1\Gamma^0} = - \mathcal{K}\bigg[ \Lambda_1^0/\gamma^0 \bigg] ~ = ~ -
\mathcal{K}\bigg[~~
      \parbox{15mm}{
    \begin{fmfgraph*}(15,15)
      %\fmfpen{thick}
      \fmfleft{in}
      \fmfright{e1,e2}
      \fmf{boson}{in,vi}
      \fmf{plain}{e1,v1}
      \fmf{plain,tension=0.7}{v1,vi}
      \fmf{plain,tension=0.7}{vi,v2}
      \fmf{plain}{v2,e2}
      \fmffreeze
      \fmf{boson,right,tension=0.7}{v1,v2}
      \fmfv{decor.shape=circle,decor.filled=empty,decor.size=2thick}{vi,v1,v2}
    \end{fmfgraph*}
}~~
\bigg] ~  = 0 \, .
\label{chap5:WI1}
\ee
Together with Eq.~(\ref{chap5:Zpsi}), this implies that: $Z_{1\psi} = Z_{1\Gamma^0} = 1$
and the Ward identity, $Z_{\psi} = Z_{\Gamma^0}$, is (trivially) satisfied at one-loop.

Let's now turn to the vector part of the vertex, see Fig.~\ref{chap5:fig:one-loop-vertex}c, focusing again of the case where $k=q=0$.
%%
%\be
%-\I e \vec{\Lambda}_1(k,k') = \mu^{2\veps_\gamma}\,\int [\D^{d_e}q]\,(-\I e \gamma^0)\,\frac{\I (\Sq + \Sk)}{(q+k)^2}\,(-\I e \vec{\gamma}\,)\,\frac{\I (\Sq + \Sk')}{(q+k')^2}\,(-\I e \gamma^0)\,\frac{\I}{2 |\vec{q}\,|}\, .
%\ee
%
%We will also evaluate it only for $k=k'=0$. 
 Performing the trace and going to euclidean space yields:
\be
\vec{\Lambda}_1(0,0) = \frac{\mu^{2\veps_\gamma}\,e^2}{2}\,\vec{\gamma}\,\int [\D^{D_e}p] [\D p_{E0}]\,\frac{p_{E0}^2 + v^2 |\vec{p}\,|^2 (D_e-2)/De}{[p_{E0}^2 + v^2 |\vec{p}\,|^2]^2\,|\vec{p}\,|}\, ,
\ee
where now the frequency integral is non-zero. Performing the remaining integrations, we arrive at:
\be
\vec{\Lambda}_1(0,0) = \vec{\gamma}\,\,\frac{\al_g}{8}\,\left(\frac{\overline{\mu}^{\,2}}{m^2}\right)^{\veps_\gamma}\,\frac{e^{\gamma_E\veps_\gamma}\,\Gamma(1+\veps_\gamma)}{\veps_\gamma}\, ,
\ee
which shows that the vector part of the vertex is UV singular ($m$ is just an arbitrary IR regulator). Extracting the pole part, the corresponding renormalization constant together with its graphical representation
are given by:
\be
\delta Z_{1 \vec{\Gamma}\,} = - \mathcal{K}\bigg[ \vec{\Lambda}_1/\vec{\gamma}\, \bigg] ~ = ~ -
\mathcal{K}\bigg[~~
      \parbox{15mm}{
    \begin{fmfgraph*}(15,15)
      %\fmfpen{thick}
      \fmfleft{in}
      \fmfright{e1,e2}
      \fmf{boson}{in,vi}
      \fmf{plain}{e1,v1}
      \fmf{plain,tension=0.7}{v1,vi}
      \fmf{plain,tension=0.7}{vi,v2}
      \fmf{plain}{v2,e2}
      \fmffreeze
      \fmf{boson,right,tension=0.7}{v1,v2}
      \fmfv{decor.shape=circle,decor.filled=shaded,decor.size=2thick}{vi}
      \fmfv{decor.shape=circle,decor.filled=empty,decor.size=2thick}{v1,v2}
    \end{fmfgraph*}
}~~
\bigg] ~  = - \frac{\al_{gr}}{8 \veps_\gamma} \, .
\label{chap5:vecLambda1}
\ee
At one-loop, this result is consistent with the Ward identity: $Z_{\vec{\Gamma}\,} = Z_{v} Z_{\psi}$ which may be graphically represented as:
\be
\mathcal{K}\bigg[~~
      \parbox{15mm}{
    \begin{fmfgraph*}(15,15)
      %\fmfpen{thick}
      \fmfleft{in}
      \fmfright{out}
      \fmf{plain}{in,vi}
      \fmf{plain,tension=0.2}{vi,vo}
      \fmf{boson,left,tension=0.2}{vi,vo}
      \fmf{plain}{vo,out}
      \fmfv{decor.shape=circle,decor.filled=empty,decor.size=2thick}{vi,vo}
    \end{fmfgraph*}
}~~
\bigg] ~ = ~ -
\mathcal{K}\bigg[~~
      \parbox{15mm}{
    \begin{fmfgraph*}(15,15)
      %\fmfpen{thick}
      \fmfleft{in}
      \fmfright{e1,e2}
      \fmf{boson}{in,vi}
      \fmf{plain}{e1,v1}
      \fmf{plain,tension=0.7}{v1,vi}
      \fmf{plain,tension=0.7}{vi,v2}
      \fmf{plain}{v2,e2}
      \fmffreeze
      \fmf{boson,right,tension=0.7}{v1,v2}
      \fmfv{decor.shape=circle,decor.filled=shaded,decor.size=2thick}{vi}
      \fmfv{decor.shape=circle,decor.filled=empty,decor.size=2thick}{v1,v2}
    \end{fmfgraph*}
}~~
\bigg] ~\, .
\label{chap5:WI2}
\ee
Notice that the peculiar Ward identities Eqs.~(\ref{chap5:WI1}) and (\ref{chap5:WI2}) are rather unusual with respect to those which can be found in the case of usual (relativistic or Lorentz-invariant) QEDs.
This will play a crucial role in deriving the correct interaction correction to the optical conductivity in the non-relativistic limit.

\section{Optical conductivity at two loops}

\begin{figure}
    a)
    \begin{fmfgraph*}(35,30)
      %\fmfpen{thick}
      \fmfleft{i}
      \fmfright{o}
      \fmf{photon}{i,v1}
      \fmf{photon,label=$q$}{v2,o}
      \fmf{phantom,right,tension=0.1,tag=1}{v1,v2}
      \fmf{phantom,right,tension=0.1,tag=2}{v2,v1}
      \fmf{phantom,tension=0.1,tag=3}{v1,v2}
      \fmfv{decor.shape=circle,decor.filled=full,decor.size=2thick}{v1,v2}
      \fmfposition
      \fmfipath{p[]}
      \fmfiset{p1}{vpath1(__v1,__v2)}
      \fmfiset{p2}{vpath2(__v2,__v1)}
      \fmfiset{p3}{vpath3(__v1,__v2)}
      \fmfi{fermion,label=$k+q$}{subpath (0,length(p1)) of p1}
      \fmfi{fermion,label=$k$}{subpath (0,length(p2)/4) of p2}
      \fmfi{fermion}{subpath (length(p2)/4,3length(p2)/4) of p2}
      \fmfi{fermion,label=$k$}{subpath (3length(p2)/4,length(p2)) of p2}
      \fmfi{photon}{point length(p2)/4 of p2 .. point length(p3)/2 of p3 .. point 3length(p2)/4 of p2}
      \def\vert#1{%
        \fmfiv{decor.shape=circle,decor.filled=empty,decor.size=2thick}{#1}}
      \vert{point length(p2)/4 of p2}
      \vert{point 3length(p2)/4 of p2}
    \end{fmfgraph*}
    \qquad \quad
    \begin{fmfgraph*}(35,30)
      %\fmfpen{thick}
      \fmfleft{i}
      \fmfright{o}
      \fmf{photon}{i,v1}
      \fmf{photon,label=$q$,label.side=left}{v2,o}
      \fmf{phantom,right,tension=0.1,tag=1}{v1,v2}
      \fmf{phantom,right,tension=0.1,tag=2}{v2,v1}
      \fmf{phantom,tension=0.1,tag=3}{v1,v2}
      \fmfv{decor.shape=circle,decor.filled=full,decor.size=2thick}{v1,v2}
      \fmfposition
      \fmfipath{p[]}
      \fmfiset{p1}{vpath1(__v1,__v2)}
      \fmfiset{p2}{vpath2(__v2,__v1)}
      \fmfiset{p3}{vpath3(__v1,__v2)}
      \fmfi{fermion,label=$k+q$}{subpath (0,length(p1)/4) of p1}
      \fmfi{fermion}{subpath (length(p1)/4,3length(p1)/4) of p1}
      \fmfi{fermion,label=$k+q$}{subpath (3length(p1)/4,length(p1)) of p1}
      \fmfi{photon}{point length(p1)/4 of p1 .. point length(p3)/2 of p3 .. point 3length(p1)/4 of p1}
      \fmfi{fermion,label=$k$}{subpath (0,length(p2)) of p2}
      \def\vert#1{%
        \fmfiv{decor.shape=circle,decor.filled=empty,decor.size=2thick}{#1}}
      \vert{point length(p1)/4 of p1}
      \vert{point 3length(p1)/4 of p1}
    \end{fmfgraph*}
    \qquad \quad
    b) 
    \begin{fmfgraph*}(35,30)
      %\fmfpen{thick}
      \fmfleft{i}
      \fmfright{o}
      \fmf{photon}{i,v1}
      \fmf{photon,label=$q$}{v2,o}
      \fmf{phantom,right,tension=0.1,tag=1}{v1,v2}
      \fmf{phantom,right,tension=0.1,tag=2}{v2,v1}
      \fmf{phantom,tension=0.1,tag=3}{v1,v2}
      \fmfv{decor.shape=circle,decor.filled=full,decor.size=2thick}{v1,v2}
      \fmfposition
      \fmfipath{p[]}
      \fmfiset{p1}{vpath1(__v1,__v2)}
      \fmfiset{p2}{vpath2(__v2,__v1)}
      \fmfi{fermion}{subpath (0,length(p1)/2) of p1}
      \fmfi{fermion,label=$k+q$}{subpath (length(p1)/2,length(p1)) of p1}
      \fmfi{fermion,label=$k$}{subpath (0,length(p2)/2) of p2}
      \fmfi{fermion}{subpath (length(p2)/2,length(p2)) of p2}
      \fmfi{photon}{point length(p1)/2 of p1 -- point length(p2)/2 of p2}
      \def\vert#1{%
        \fmfiv{decor.shape=circle,decor.filled=empty,decor.size=2thick}{#1}}
      \vert{point length(p1)/2 of p1}
      \vert{point length(p2)/2 of p2}
    \end{fmfgraph*}
    \caption{\label{chap5:fig:2loop-polarization-cc}
     Two-loop vacuum polarization, $\Pi_2^{\mu \nu}$, diagrams.}
\end{figure}
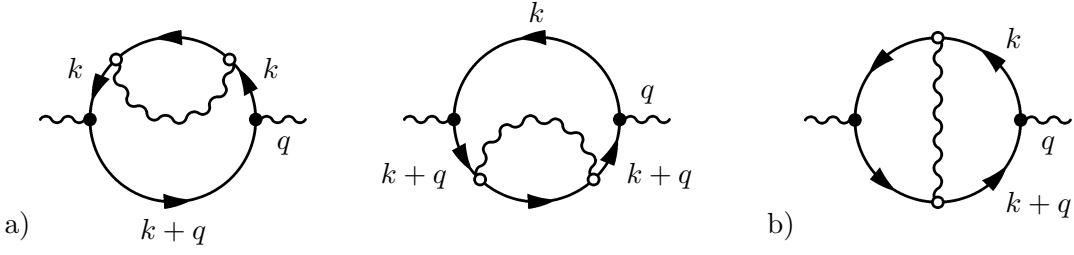
%\FloatBarrier
%

\begin{figure}
    a)
    \begin{fmfgraph*}(35,30)
      %\fmfpen{thick}
      \fmfleft{i}
      \fmfright{o}
      \fmf{photon}{i,v1}
      \fmf{photon,label=$q$}{v2,o}
      \fmf{phantom,right,tension=0.1,tag=1}{v1,v2}
      \fmf{phantom,right,tension=0.1,tag=2}{v2,v1}
      \fmf{phantom,tension=0.1,tag=3}{v1,v2}
      \fmfv{decor.shape=circle,decor.filled=empty,decor.size=2thick}{v1,v2}
      \fmfposition
      \fmfipath{p[]}
      \fmfiset{p1}{vpath1(__v1,__v2)}
      \fmfiset{p2}{vpath2(__v2,__v1)}
      \fmfiset{p3}{vpath3(__v1,__v2)}
      \fmfi{fermion,label=$k+q$}{subpath (0,length(p1)) of p1}
      \fmfi{fermion,label=$k$}{subpath (0,length(p2)/4) of p2}
      \fmfi{fermion}{subpath (length(p2)/4,3length(p2)/4) of p2}
      \fmfi{fermion,label=$k$}{subpath (3length(p2)/4,length(p2)) of p2}
      \fmfi{photon}{point length(p2)/4 of p2 .. point length(p3)/2 of p3 .. point 3length(p2)/4 of p2}
      \def\vert#1{%
        \fmfiv{decor.shape=circle,decor.filled=empty,decor.size=2thick}{#1}}
      \vert{point length(p2)/4 of p2}
      \vert{point 3length(p2)/4 of p2}
    \end{fmfgraph*}
    \qquad \quad
    \begin{fmfgraph*}(35,30)
      %\fmfpen{thick}
      \fmfleft{i}
      \fmfright{o}
      \fmf{photon}{i,v1}
      \fmf{photon,label=$q$,label.side=left}{v2,o}
      \fmf{phantom,right,tension=0.1,tag=1}{v1,v2}
      \fmf{phantom,right,tension=0.1,tag=2}{v2,v1}
      \fmf{phantom,tension=0.1,tag=3}{v1,v2}
      \fmfv{decor.shape=circle,decor.filled=empty,decor.size=2thick}{v1,v2}
      \fmfposition
      \fmfipath{p[]}
      \fmfiset{p1}{vpath1(__v1,__v2)}
      \fmfiset{p2}{vpath2(__v2,__v1)}
      \fmfiset{p3}{vpath3(__v1,__v2)}
      \fmfi{fermion,label=$k+q$}{subpath (0,length(p1)/4) of p1}
      \fmfi{fermion}{subpath (length(p1)/4,3length(p1)/4) of p1}
      \fmfi{fermion,label=$k+q$}{subpath (3length(p1)/4,length(p1)) of p1}
      \fmfi{photon}{point length(p1)/4 of p1 .. point length(p3)/2 of p3 .. point 3length(p1)/4 of p1}
      \fmfi{fermion,label=$k$}{subpath (0,length(p2)) of p2}
      \def\vert#1{%
        \fmfiv{decor.shape=circle,decor.filled=empty,decor.size=2thick}{#1}}
      \vert{point length(p1)/4 of p1}
      \vert{point 3length(p1)/4 of p1}
    \end{fmfgraph*}
    \qquad \quad
    b) 
    \begin{fmfgraph*}(35,30)
      %\fmfpen{thick}
      \fmfleft{i}
      \fmfright{o}
      \fmf{photon}{i,v1}
      \fmf{photon,label=$q$}{v2,o}
      \fmf{phantom,right,tension=0.1,tag=1}{v1,v2}
      \fmf{phantom,right,tension=0.1,tag=2}{v2,v1}
      \fmf{phantom,tension=0.1,tag=3}{v1,v2}
      \fmfv{decor.shape=circle,decor.filled=empty,decor.size=2thick}{v1,v2}
      \fmfposition
      \fmfipath{p[]}
      \fmfiset{p1}{vpath1(__v1,__v2)}
      \fmfiset{p2}{vpath2(__v2,__v1)}
      \fmfi{fermion}{subpath (0,length(p1)/2) of p1}
      \fmfi{fermion,label=$k+q$}{subpath (length(p1)/2,length(p1)) of p1}
      \fmfi{fermion,label=$k$}{subpath (0,length(p2)/2) of p2}
      \fmfi{fermion}{subpath (length(p2)/2,length(p2)) of p2}
      \fmfi{photon}{point length(p1)/2 of p1 -- point length(p2)/2 of p2}
      \def\vert#1{%
        \fmfiv{decor.shape=circle,decor.filled=empty,decor.size=2thick}{#1}}
      \vert{point length(p1)/2 of p1}
      \vert{point length(p2)/2 of p2}
    \end{fmfgraph*}
    \caption{\label{chap5:fig:2loop-polarization-dd}
     Two-loop vacuum polarization, $\Pi_2^{0 0}$, diagrams.}
\end{figure}
%\FloatBarrier
%

We now proceed on computing the 2-loop corrections displayed on Fig.~\ref{chap5:fig:2loop-polarization-cc}:
$\Pi_2^{\mu \nu}(q) = 2 \Pi_{2a}^{\mu \nu}(q) + \Pi_{2b}^{\mu \nu}(q)$
where $\Pi_{2a}$ is the so-called self-energy correction and $\Pi_{2b}$ is the so-called vertex correction. The latter are defined in the usual way as:
\begin{subequations}
\label{chap5:def:pi2loop}
\begin{flalign}
&\I \Pi_{2a}^{\mu \nu}(q) =
- \mu^{2\veps_\gamma}\, \int [\D^{d_e}k ]\, \Tr \left[ (-\I e\gamma^\nu) \,S_0(k+q)\,(-\I e\gamma^\mu)\, S_0(k)\left( -\I \Sigma_1(k) \right)\,S_0(k) \right]\, ,
\label{chap5:def:pi2a} \\
&\I \Pi_{2b}^{\mu \nu}(q) = - \mu^{2\veps_\gamma}\, \int [\D^{d_e}k]\,\Tr \left[ (-\I e\gamma^\nu)\, S_0(k+q)\, (-\I e \Lambda_1^\mu(k,q))\, S_0(k) \right]\, ,
\label{chap5:def:pi2b}
\end{flalign}
\end{subequations}
where the fermion self-energy was defined in Eq.~(\ref{chap5:def:fermion-self}) and the fermion-photon vertex in Eq.~(\ref{chap5:def:vertex}).
For completeness, the diagrams associated to $\Pi^{00}(a)$ are displayed on Fig.~\ref{chap5:fig:2loop-polarization-dd}. Our convention for the conductivity will be that:
\be
\sigma_2 = \sigma_{2a} + \sigma_{2b},\, \quad
\sigma_{2a} = - \lim_{\vec{q} \ra 0} \, \frac{\I q_0}{|\vec{q}\,|^2}\,2 \,\Pi_{2a}^{00}(q_0,\vec{q}\,), \quad 
\sigma_{2b} = - \lim_{\vec{q} \ra 0} \, \frac{\I q_0}{|\vec{q}\,|^2}\,\Pi_{2b}^{00}(q_0,\vec{q}\,)\, ,
\label{chap5:sigma-dd2}
\ee
and similarly for $\tilde{\sigma}_2$.

\subsection{Density-density correlation function approach}

Let's first focus on $\Pi_{2a}^{0 0}$, Fig.~\ref{chap5:fig:2loop-polarization-dd}a. Performing the trace, going to euclidean space, integrating over frequencies and taking the $\vec{q} \ra 0$ limit 
 %and substituting the expression of the fermion self-energy Eq.~(\ref{chap5:fsigma2}), 
leads to:
\begin{flalign}
\Pi_{2a}^{00}(m,\vec{q} \ra 0) =  \frac{N_F \, e^2}{4v}\,\mu^{2\veps_\gamma}\,|\vec{q}\,|^2\,\frac{D_e-1}{D_e}\, \int [\D^{D_e} k]\,\Sigma_{1k}(|\vec{k}^{\,2}|)\,
\frac{|\vec{k}\,|^2 - m^2}{|\vec{k}\,|\, [|\vec{k}\,|^2 + m^2]^2}\, .
%\Pi_{2a}^{00}(q_{E0},\vec{q} \ra 0) = - \frac{N_F}{32}\,|\vec{q}\,|^2\,\frac{e_0^4}{v_0^2\,(4\pi)^{D_e/2}}\,\frac{D_e-1}{D_e}\,G(1/2,1/2)\,
%\int \frac{[\D^{D_e}k ]\, (|\vec{k}\,|^2 - m_0^2)}{[|\vec{k}\,|^2]^{1/2+\veps_\gamma}\, [|\vec{k}\,|^2 + m_0^2]^2}\, .
\label{chap5:pi2a-inter}
\end{flalign}
Substituting the expression of the fermion self-energy Eq.~(\ref{chap5:fsigma2}), the integral is again of the semi-massive one-loop tadpole type and is straightforwardly evaluated as:
\be
\Pi_{2a}^{00}(m,\vec{q} \ra 0) = - \frac{N_F e^2}{16\pi v}\,\frac{|\vec{q}\,|^2}{m}\,\frac{\al_g}{8}\,\left(\frac{\overline{\mu}^{\,2}}{m^2}\right)^{2\veps_\gamma}\,
\frac{(D_e-1)\,(D_e-2-2\veps_\gamma)}{D_e}\,e^{2\gamma_E\veps_\gamma}\,G(1/2,1/2)\,B(1,1/2+\veps_\gamma)\, . 
\ee
Performing the $\veps_\gamma$-expansion yields:
\be
2\,\Pi_{2a}^{00}(q_0,\vec{q} \ra 0) = - \frac{N_F\, e^2}{8}\,\frac{|\vec{q}\,|^2}{\I q_0}\,\frac{\al_g}{2}\, .
\label{chap5:pi2a00}
\ee
Combining Eqs.~(\ref{chap5:sigma-dd}) and (\ref{chap5:pi2a00}), we arrive at the (bare) conductivity: 
\be
\sigma_{2a}(q_0) = \frac{N_F\, e^2}{8}\,\frac{\al_g}{2}\, .
\label{chap5:sigma2a}
\ee
This shows that $\sigma_{2a}(q_0) = \sigma_0\,\mathcal{C}_a^{(D)} \al_g$,
and we recover the result of JVH: $\mathcal{C}_{a}^{(D)} =1/2$. Proceeding in a similar way for the vertex correction, Fig.~\ref{chap5:fig:2loop-polarization-dd}b, 
the latter can be written in terms of master integrals as follows:
\begin{flalign}
\Pi_{2b}^{0 0}(m,\vec{q} \ra 0\,) = \frac{N_F\,e^4}{8\,v^2}\,\frac{|\vec{q}\,|^2}{D_e}\, 
\times
\biggl\{ (D_e-1)\, I_1(1/2) - m^2 \, I_2(3/2)-m^2 (D_e-2)\, I_0(1/2) \biggr\}\, .
\label{chap5:pi2b00-inter}
\end{flalign}
Using the results of Eqs.~(\ref{chap5:master-dd}), yields:
\be
\Pi_{2b}^{00}(q_{0},\vec{q} \ra 0) = -\frac{N_F \,e^2}{8}\,\frac{|\vec{q}\,|^2}{\I q_0}\,\al_g\,\frac{8-3 \pi}{6}\, .
\label{chap5:pi2b00}
\ee
Combining Eqs.~(\ref{chap5:sigma-dd}) and (\ref{chap5:pi2b00}), we arrive at: 
\be
\sigma_{2b}(q_0) = \frac{N_F\, e^2}{8}\,\al_g\,\frac{8-3 \pi}{6}\, ,
\label{chap5:sigma2b}
\ee
which shows that $\sigma_{2b}(q_0) = \sigma_0\,\mathcal{C}_b^{(D)} \al_g$ allowing to recover the result: $\mathcal{C}_b^{(D)} =  (8-3\pi)/6$. Adding the two-loop contributions, 
the total (bare) optical conductivity reads:
\be
\sigma_{2}(q_0) = \frac{N_F\, e^2}{8}\,\al_g\,\frac{11-3 \pi}{6}\, ,
\ee
and we recover the result of JVH, Eq.~(\ref{chap5:c-D}): $\mathcal{C}^{(D)} = \mathcal{C}_{a}^{(D)} + \mathcal{C}_{b}^{(D)} = (11-3\pi)/6$. 
%Adding moreover the contribution of the one loop counter-term, $\mathcal{C}'^{(D)}=-1/4$, the total conductivity, with two loop accuracy, reads: 
%$\sigma(q_0) = \sigma_{1}(q_0) + \sigma_{2}(q_0) = \sigma_0\,(1+\mathcal{C}^{(DR)} \al + \Ord(\al^2))$, and we finally arrive at the advertised result, Eq.~(\ref{chap5:c-DR}).
%: $\mathcal{C}_b^{(DR)} = \mathcal{C}^{(D)} + \mathcal{C}'^{(D)} = (19-6\pi)/12$.

We now proceed on computing the renormalized conductivity using Zimmermann's forest formula Eq.~(\ref{chap2:def:forest}). 
Using this formula, the renormalized diagrams contributing to the density-density correlation function, $\Pi_{2\,\al\,r}^{00}(q)$ ($\al = a,b$), are related to the bare ones as follows:
\bea
\Pi_{2\,\al\,r}^{00} = \mathcal{R}\,\Pi_{2\,\al}^{00} = (1 - \mathcal{K})\,\mathcal{R}'\,\Pi_{2\,\al}^{00}, \qquad \mathcal{R}'\,\Pi_{2\,\al}^{00} = \Pi_{2\al}^{00} + \Pi_{2\al'}^{00}  \qquad \qquad (\al=a,\,b)\, ,
\label{chap5:forest:dd}
\eea
where $\Pi_{2\al'}^{00}$ may be graphically represented as:
%
%\begin{subequations}
\begin{flalign}
\Pi_{2a'}^{00} = -
\mathcal{K}\, \bigg[~
   \parbox{15mm}{
    \begin{fmfgraph*}(15,15)
      \fmfleft{in}
      \fmfright{out}
      \fmf{plain}{in,vi}
      \fmf{plain,tension=0.2}{vi,vo}
      \fmf{boson,left,tension=0.2}{vi,vo}
      \fmf{plain}{vo,out}
      \fmfv{decor.shape=circle,decor.filled=empty,decor.size=2thick}{vi,vo}
    \end{fmfgraph*}
}~ \bigg]~\star~
\parbox{15mm}{
    \begin{fmfgraph*}(15,15)
      \fmfleft{in}
      \fmfright{out}
      \fmf{boson}{in,ve}
      \fmf{plain,right,tension=0.2}{ve,vw}
      \fmf{plain,right,tension=0.2}{vw,ve}
      \fmf{boson}{vw,out}
      \fmfv{decor.shape=circle,decor.filled=empty,decor.size=2thick}{ve,vw}
    \end{fmfgraph*}
}\, , \qquad
\Pi_{2b'}^{00} = -2
\mathcal{K}\, \bigg[~
   \parbox{15mm}{
    \begin{fmfgraph*}(15,15)
      \fmfleft{in}
      \fmfright{e1,e2}
      \fmf{boson}{in,vi}
      \fmf{plain}{e1,v1}
      \fmf{plain,tension=0.7}{v1,vi}
      \fmf{plain,tension=0.7}{vi,v2}
      \fmf{plain}{v2,e2}
      \fmffreeze
      \fmf{boson,right,tension=0.7}{v1,v2}
      \fmfv{decor.shape=circle,decor.filled=empty,decor.size=2thick}{vi,v1,v2}
    \end{fmfgraph*}
}~ \bigg]~\star~
\parbox{15mm}{
    \begin{fmfgraph*}(15,15)
      \fmfleft{in}
      \fmfright{out}
      \fmf{boson}{in,ve}
      \fmf{plain,right,tension=0.2}{ve,vw}
      \fmf{plain,right,tension=0.2}{vw,ve}
      \fmf{boson}{vw,out}
      \fmfv{decor.shape=circle,decor.filled=empty,decor.size=2thick}{ve,vw}
    \end{fmfgraph*}
}\, ~~.
\label{chap5:R':a+b}
\end{flalign}
%\end{subequations}
%
The peculiarity of the present non-relativistic theory is that the one-loop fermion self-energy and fermion-photon vertex subdiagrams appearing in Eq.~(\ref{chap5:R':a+b}) are not related by a Ward identity
and therefore do not cancel each other (contrary to what happens in usual QED). The case of $\Pi_{2b}^{00}$ is trivial: this diagram is finite overall so 
$\mathcal{K} \Pi_{2b}^{00}=0$ and, from Eq.~(\ref{chap5:WI1}), its subdiagram is also finite so: $\Pi_{2b'}^{00}=0$. This diagram is therefore absolutely convergent in Weinberg's sense so that:
\be
\sigma_{2br}(q_0) =  \sigma_{2b} = \frac{N_F\, e^2}{8}\,\al_g\,\frac{8-3 \pi}{6}\, .
\label{chap5:sigma2br}
\ee
The case of $\Pi_{2a}^{00}$ is more interesting: this diagram is also finite overall so $\mathcal{K} \Pi_{2a}^{00}=0$. However, its subdiagram is divergent, see Eq.~(\ref{chap5:dZ1v})
and needs to be subtracted. In order to compute $\Pi_{2a'}^{00}$ we go to the integral representation of Eq.~(\ref{chap5:R':a+b}) as the $\star$ operation
does not reduce to a simple multiplication (the diagram is not logarithmic and care must be taken in projecting out its pseudo-Lorentz structure). 
This can be straightforwardly done with the help of Eq.~(\ref{chap5:pi2a-inter}) and leads to:
\begin{flalign}
\Pi_{2a'}^{00}(m,\vec{q} \ra 0) =  -\frac{N_F \, e^2}{4v}\,\mu^{2\veps_\gamma}\,|\vec{q}\,|^2\,\frac{D_e-1}{D_e}\, \int [\D^{D_e} k]\,\mathcal{K} \bigg[ \Sigma_{1k}(|\vec{k}^{\,2}|) \bigg]\,
\frac{|\vec{k}\,|^2 - m^2}{|\vec{k}\,|\, [|\vec{k}\,|^2 + m^2]^2}\, .
%\Pi_{2a}^{00}(q_{E0},\vec{q} \ra 0) = - \frac{N_F}{32}\,|\vec{q}\,|^2\,\frac{e_0^4}{v_0^2\,(4\pi)^{D_e/2}}\,\frac{D_e-1}{D_e}\,G(1/2,1/2)\,
%\int \frac{[\D^{D_e}k ]\, (|\vec{k}\,|^2 - m_0^2)}{[|\vec{k}\,|^2]^{1/2+\veps_\gamma}\, [|\vec{k}\,|^2 + m_0^2]^2}\, .
\label{chap5:pi2a'-inter}
\end{flalign}
Substituting Eq.~(\ref{chap5:dZ1v}) and computing the integral yields:
\be
\Pi_{2a'}^{00}(m,\vec{q} \ra 0) = \frac{N_F e^2}{16\pi v}\,\frac{|\vec{q}\,|^2}{m}\,\frac{\al_g}{8\veps_\gamma}\,\left(\frac{\overline{\mu}^{\,2}}{m^2}\right)^{\veps_\gamma}\,
\frac{(D_e-1)\,(D_e-2)}{D_e}\,e^{2\gamma_E\veps_\gamma}\,G(1/2,1/2)\,B(1,1/2)\, ,
\ee
which is finite in the limit $\veps_\gamma \ra 0$ and reads:
\be
\Pi_{2a'}^{00}(q_0,\vec{q} \ra 0) = \frac{N_F e^2}{8}\,\frac{|\vec{q}\,|^2}{\I q_0}\,\frac{\al_g}{8}\, .
\ee
The fact that both $\Pi_{2a'}^{00}$ and $\Pi_{2a}^{00}$ are finite implies that $\mathcal{K}\,\mathcal{R}'\,\Pi_{2\,a}^{00}=0$ which was expected since there is no global divergence to subtract.
However, the subtraction of the subdivergence brings a {\it finite} contribution to the renormalized function:
\be
2\Pi_{2ar}^{00}(q_0,\vec{q} \ra 0) = 2\Pi_{2a}^{00} + 2\Pi_{2a'}^{00} = -\frac{N_F e^2}{8}\,\frac{|\vec{q}\,|^2}{\I q_0}\,\al_g\,\biggl( \frac{1}{2} - \frac{1}{4} \biggr)\, .
\ee
The corresponding renormalized conductivity reads:
\be
\sigma_{2ar}(q_0) = \frac{N_F\, e^2}{8}\,\al_g\,\biggl( \frac{1}{2} - \frac{1}{4} \biggr)\, ,
\label{chap5:sigma2ar}
\ee
and is decreased with respect to its bare value Eq.~(\ref{chap5:sigma2a}) in accordance with the fact that:  $\mathcal{C}'^{(D)}=-1/4$. 
Hence,  the total two-loop renormalized conductivity reads:
\be 
\sigma_{2r}(q_0) = \sigma_{2ar} + \sigma_{2br} = \frac{N_F\, e^2}{8}\, \al_g\,\biggl( \frac{11-3 \pi}{6} - \frac{1}{4} \biggr) = \frac{N_F\, e^2}{8}\, \al_g\,\frac{19-6\pi}{12} \, , 
\label{chap5:sigma2r:final}
\ee
in accordance with the advertised result, Eq.~(\ref{chap5:c-DR}).

\subsection{Current-current correlation function approach}

We now proceed from Eq.~(\ref{chap5:sigma-cc2}) with $K_2(q_0) = 2K_{2a}(q_0) + K_{2b}(q_0)$. %Calculations of trace and frequency integral are done along the same lines as for $\Pi_2^{00}$.
For the self-energy correction $K_{2a}$, Fig.~\ref{chap5:fig:2loop-polarization-cc}a, the computation of the trace and the frequency integral lead to:
\begin{flalign}
K_{2a}(m) = -\frac{N_F e^2}{4v m^2}\,\frac{D_e-1}{D_e}\,\mu^{2\veps_\gamma}\,\int [\D^{D_e}k]\,\Sigma_{1k}(|\vec{k}^{\,2}|)\,
\frac{|\vec{k}\,|\,(|\vec{k}\,|^2 - m^2)}{[|\vec{k}\,|^2 + m^2]^2}\, ,
\label{chap5:pi2amumu-inter}
\end{flalign}
where the integral is again of the semi-massive tadpole type. Performing the integration yields:
\begin{flalign}
K_{2a}(m) = - \frac{N_F e^2}{16\pi v m}\,\frac{\al_g}{4}\,\left(\frac{\overline{\mu}^{\,2}}{m^2}\right)^{2\veps_\gamma}\, \frac{(D_e-1)(D_e-2\veps_\gamma)}{D_e}\,
e^{2\gamma_E\veps_\gamma}\,G(1/2,1/2)\,B(1,-1/2+\veps_\gamma)\, ,
\end{flalign}
which, contrarily to the case of $\Pi^{00}$, is now divergent. The $\veps_\gamma$-expansion reads:
\be
2K_{2a}(q_{0}) = \frac{N_F\,e^2}{8\I q_0}\,\frac{\al_g}{4}\,\left( -\frac{1}{\veps_\gamma} + 3 + 2 L_{q_0} - 4 \log 2 + \Ord(\veps_\gamma) \right)\, ,
\label{chap5:pi2amumu-expr}
\ee
where again $L_{q_0}=\log(-q_0^2/(4v^2 \overline{\mu}^{\,2}))$.
 Similarly, after lengthy calculations, %after performing the trace, going to euclidean space, integrating over frequencies and taking the $\vec{q} \ra \vec{0}$ limit, 
the vertex correction, Fig.~\ref{chap5:fig:2loop-polarization-cc}b, can be expressed in terms of master integrals as follows:
\begin{flalign}
K_{2b}(m) = \frac{N_F e^4}{8\,m^2 v^2}\,\frac{1}{D_e} \times \biggl\{-(D_e-1)\,m^2\,I_1(1/2) + I_2(1/2) + (D_e-2)\,I_0(-1/2) \biggr\}\, .
\label{chap5:pi2bmumu-inter}
\end{flalign}
Using Eqs.~(\ref{chap5:master-cc}), this can be explicited as:
\begin{flalign}
K_{2b}(q_{E0}) = \frac{N_Fe^2}{16\pi v m}\,\frac{\al_g}{2}\,\left(\frac{\overline{\mu}^{\,2}}{m^2}\right)^{2\veps_\gamma}\,e^{2\gamma_E\veps_\gamma}\,
\times \biggl [ \pi \,\left( \pi - \frac{2}{3} \right)  - \frac{3-D_e}{D_e}\,G(1/2,1/2)\,B(1,1/2+\veps_\gamma) \biggr]\, .
\end{flalign}
Performing the $\veps_\gamma$-expansion, the final result can be put in the form:
\be
K_{2b}(q_0) = -2K_{2a}(q_0)
%-\frac{N_F\,e^2}{8\,\I q_{0}}\,\frac{\al}{4}\,\left( - \frac{1}{\veps_\gamma} + 2 L_q + 3 - 4 \log 2 + \Ord(\veps_\gamma) \right) 
+\frac{N_F\,e^2}{8\,\I q_{0}}\,\al_g\,\frac{11 - 3\pi}{6}\, .
\label{chap5:pi2bmumu-expr}
\ee
Singular terms cancel from the sum of Eqs.~(\ref{chap5:pi2amumu-expr}) and (\ref{chap5:pi2bmumu-expr}). Combining these equations with Eq.~(\ref{chap5:sigma-cc2})
yields the total (bare) optical conductivity at two-loops: 
\be
\tilde{\sigma}_{2}(q_0) = \frac{N_Fe^2}{8}\,\al_g\,\frac{11-3\pi}{6}\, ,
\ee
 and we recover the result of JVH: $\tilde{\mathcal{C}}^{(D)} = (11-3\pi)/6$. 

We now proceed on computing the renormalized conductivity using Zimmermann's forest formula Eq.~(\ref{chap2:def:forest}).
Similarly to the density-density case, the renormalized diagrams contributing to the current-current correlation function, $K_{2\,\al\,r}(q_0)$ ($\al = a,b$), are related to the bare ones as follows:
\bea
K_{2\,\al\,r} = \mathcal{R}\,K_{2\,\al} = (1 - \mathcal{K})\,\mathcal{R}'\,K_{2\,\al}, \qquad \mathcal{R}'\,K_{2\,\al} = K_{2\al} + K_{2\al'}  \qquad \qquad (\al=a,\,b)\, ,
\label{chap5:forest:cc}
\eea
where $K_{2\al'}$ may be graphically represented as:
%
%\begin{subequations}
\begin{flalign}
K_{2a'} = -
\mathcal{K}\, \bigg[~
   \parbox{15mm}{
    \begin{fmfgraph*}(15,15)
      \fmfleft{in}
      \fmfright{out}
      \fmf{plain}{in,vi}
      \fmf{plain,tension=0.2}{vi,vo}
      \fmf{boson,left,tension=0.2}{vi,vo}
      \fmf{plain}{vo,out}
      \fmfv{decor.shape=circle,decor.filled=empty,decor.size=2thick}{vi,vo}
    \end{fmfgraph*}
}~ \bigg]~\star~
\parbox{15mm}{
    \begin{fmfgraph*}(15,15)
      \fmfleft{in}
      \fmfright{out}
      \fmf{boson}{in,ve}
      \fmf{plain,right,tension=0.2}{ve,vw}
      \fmf{plain,right,tension=0.2}{vw,ve}
      \fmf{boson}{vw,out}
      \fmfv{decor.shape=circle,decor.filled=full,decor.size=2thick}{ve,vw}
    \end{fmfgraph*}
}\, , \qquad
K_{2b'} = -2
\mathcal{K}\, \bigg[~
   \parbox{15mm}{
    \begin{fmfgraph*}(15,15)
      \fmfleft{in}
      \fmfright{e1,e2}
      \fmf{boson}{in,vi}
      \fmf{plain}{e1,v1}
      \fmf{plain,tension=0.7}{v1,vi}
      \fmf{plain,tension=0.7}{vi,v2}
      \fmf{plain}{v2,e2}
      \fmffreeze
      \fmf{boson,right,tension=0.7}{v1,v2}
      \fmfv{decor.shape=circle,decor.filled=full,decor.size=2thick}{vi}
      \fmfv{decor.shape=circle,decor.filled=empty,decor.size=2thick}{v1,v2}
    \end{fmfgraph*}
}~ \bigg]~\star~
\parbox{15mm}{
    \begin{fmfgraph*}(15,15)
      \fmfleft{in}
      \fmfright{out}
      \fmf{boson}{in,ve}
      \fmf{plain,right,tension=0.2}{ve,vw}
      \fmf{plain,right,tension=0.2}{vw,ve}
      \fmf{boson}{vw,out}
      \fmfv{decor.shape=circle,decor.filled=full,decor.size=2thick}{ve,vw}
    \end{fmfgraph*}
}\, ~~.
\label{chap5:R':Ka+b}
\end{flalign}
%\end{subequations}
%
Contrarily to the density-density case, both $K_{2a}$ and $K_{2b}$ need to be properly renormalized. Indeed, in the present case, both one-loop fermion self-energy and fermion-photon vertex 
subgraphs appearing in Eqs.~(\ref{chap5:R':Ka+b}) are singular. One may wonder at this point if these contributions will cancel each other due to the Ward identity Eq.~(\ref{chap5:WI2}).
As will be shown in the following, this is not the case. The proof requires some care as the $\star$ operation does not (necessarily) reduce to a simple multiplication. 
Notice for example that only the vector part of the fermion-photon vertex subdiagram is singular; but the full vertex appears in Eq.~(\ref{chap5:R':Ka+b}).
So, one needs to be careful in projecting out only the vector component. In order to do this we will use the integral representation of Eqs.~(\ref{chap5:R':Ka+b})

We first consider $K_{2a}$. With the help of Eq.~(\ref{chap5:pi2amumu-inter}) the contribution of $K_{2a'}$ is defined as:
\begin{flalign}
K_{2a'}(m) = \frac{N_F e^2}{4v m^2}\,\frac{D_e-1}{D_e}\,\mu^{2\veps_\gamma}\,\int [\D^{D_e}k]\,\mathcal{K} \biggl[ \Sigma_{1k}(|\vec{k}^{\,2}|) \biggr]\,
\frac{|\vec{k}\,|\,(|\vec{k}\,|^2 - m^2)}{[|\vec{k}\,|^2 + m^2]^2}\, .
\label{chap5:pi2a'mumu-inter}
\end{flalign}
Substituting Eq.~(\ref{chap5:dZ1v}) and computing the integral yields:
\begin{flalign}
K_{2a'}(m) = \frac{N_F e^2}{16\pi v m}\,\frac{\al_g}{8 \veps_\gamma}\,\left(\frac{\overline{\mu}^{\,2}}{m^2}\right)^{2\veps_\gamma}\, (D_e-1)\,e^{\gamma_E\veps_\gamma}\,B(1,-1/2)\, .
\end{flalign}
The $\veps_\gamma$-expansion reads:
\be
2K_{2a'}(q_{0}) = \frac{N_F\,e^2}{8\I q_0}\,\frac{\al_g}{4}\,\left( \frac{1}{\veps_\gamma} - 2 - L_{q_0} + \Ord(\veps_\gamma) \right)\, .
\label{chap5:pi2a'mumu-expr}
\ee
Adding the contribution of $2K_{2a}$ the pole terms cancel each other and we find that:
\be
2 K_{2\,ar} = 2 \mathcal{R}'\,K_{2\,a} = \frac{N_F\,e^2}{8\I q_0}\,\frac{\al_g}{4}\,\biggl(1 + L_{q_0} - 4 \log 2 + \Ord(\veps_\gamma) \biggr)\, .
\label{chap5:K2ar}
\ee
The fact that $\mathcal{K} \mathcal{R}'\,K_{2\,a} = 0$ means that the overall counter-term is zero in accordance with the fact that the only singularity which needs to be subtracted is the
one associated with the fermion self-energy subgraph. From Eq.~(\ref{chap5:K2ar}), we may now derive the expression for the renormalized conductivity associated with 
Fig.~\ref{chap5:fig:2loop-polarization-cc}a:
\be
\tilde{\sigma}_{2ar}(q_0) = \frac{N_Fe^2}{8}\,\frac{\al_g}{4}\,\biggl(1 + L_{q_0} - 4 \log 2 + \Ord(\veps_\gamma) \biggr)\, .
\label{chap5:tsigma2ar}
\ee

We now consider the case of  $K_{2b}$. In order to compute $K_{2b'}$, we find it convenient to go back to the initial definition Eq.~(\ref{chap5:def:pi2b}) replacing
$\Lambda_1^\mu$ by $\vec{\gamma}\,\mathcal{K}\bigg[ \vec{\Lambda}_1/\vec{\gamma}\, \bigg]$ with the appropriate $\gamma$-matrix contractions. Performing the trace, going to euclidean space, 
integrating over frequencies and taking the $\vec{q} \ra 0$ limit leads to:
\be
K_{2b'}(m) = - \frac{N_F e^2}{v m^2}\,\mathcal{K}\bigg[ \vec{\Lambda}_1/\vec{\gamma}\, \bigg]\,\frac{D_e-1}{D_e}\,\mu^{2\veps_\gamma}\,
\int [\D^{D_e}k]\,\frac{|\vec{k}\,|}{|\vec{k}\,|^2 + m^2}\, .
\label{chap5:def:pi2b'}
\ee
The integral is easily computed with the rules for evaluation semi-massive tadpoles and the result reads:
\be
K_{2b'}(m) = - \frac{N_F e^2}{4\pi v m}\,\mathcal{K}\bigg[ \vec{\Lambda}_1/\vec{\gamma}\, \bigg]\,\left(\frac{\overline{\mu}^{\,2}}{m^2}\right)^{\veps_\gamma}\,\frac{D_e-1}{D_e}\,
e^{\gamma_E\veps_\gamma}\,B(1,-1/2)\, .
\label{chap5:expr:pi2b'}
\ee
Performing the $\veps_\gamma$-expansion then yields:
\be
K_{2b'}(q_0) = \frac{N_F e^2}{8 \I q_0}\,\frac{\al_g}{4}\,\left( -\frac{1}{\veps_\gamma} + 1 + L_{q_0} + \Ord(\veps_\gamma) \right)\, .
\label{chap5:expand:pi2b'}
\ee
Adding the contribution of $K_{2b}$, the pole terms cancel each other and we find that:
\be
K_{2\,br} = 2 \mathcal{R}'\,K_{2\,b} = \frac{N_F\,e^2}{8\I q_0}\,\frac{\al_g}{4}\,\biggl(-2 - L_{q_0} + 4 \log 2 + \Ord(\veps_\gamma) \biggr) + \frac{N_F\,e^2}{8\,\I q_{0}}\,\al\,\frac{11 - 3\pi}{6}\, \, .
\label{chap5:K2br}
\ee
The fact that $\mathcal{K} \mathcal{R}'\,K_{2\,b} = 0$ means that the overall counter-term is zero in accordance with the fact that the only singularity which needs to be subtracted is the
one associated with the vector photon-fermion vertex subgraph. From Eq.~(\ref{chap5:K2br}), we may now derive the expression for the renormalized conductivity associated with
Fig.~\ref{chap5:fig:2loop-polarization-cc}b:
\be
\tilde{\sigma}_{2br}(q_0) = \frac{N_F\,e^2}{8}\,\al_g\,\frac{11 - 3\pi}{6} +  \frac{N_Fe^2}{8}\,\frac{\al_g}{4}\,\biggl(-2 - L_{q_0} + 4 \log 2 + \Ord(\veps_\gamma) \biggr)\, .
\label{chap5:tsigma2br}
\ee

Hence, adding Eqs.~(\ref{chap5:tsigma2ar}) and (\ref{chap5:tsigma2br}), the total two-loop renormalized conductivity reads:
\be
\tilde{\sigma}_{2r}(q_0) = \tilde{\sigma}_{2ar} + \tilde{\sigma}_{2br} = \frac{N_F\, e^2}{8}\, \al_g\,\biggl( \frac{11-3 \pi}{6} - \frac{1}{4} \biggr) = \frac{N_F\, e^2}{8}\, \al_g\,\frac{19-6\pi}{12} \, ,
\label{chap5:tsigma2r:final}
\ee
in accordance with the advertised result, Eq.~(\ref{chap5:c-DR}) and therefore with the result obtained from the density-density correlation function, Eq.~(\ref{chap5:sigma2r:final}).

\section{Conclusion} 

In this Chapter, we have elaborated a field theoretic renormalization approach to the computation of electron-electron interactions in graphene and related planar Dirac liquids in the non-relativistic limit ($v/c \ra 0$).
As we have seen, the broken Lorentz invariance brings some peculiarities with respect to the relativistic case (non-standard Ward identities) as well as some complications (semi-massive Feynman diagrams instead of the massless
ones). Focusing on the optical conductivity of these materials, we have provided a clear proof that radiative corrections affecting this experimentally relevant observable are {\it finite} and {\it well determined}.
Our proof makes use of the powerful BPHZ prescription. It confirms the validity of our previous approach based on conventional renormalization \cite{Teber:2012de} but is considerably more robust and allowed us 
to perform a refined diagram by diagram analysis.  Both the density (where individual diagrams are overall finite) and Kubo formula (where individual diagrams are explicitly singular) 
approaches were shown to yield a single well-defined result for the two-loop interaction correction to the minimal conductivity: $\mathcal{C}^{(2)} = (19-6\pi)/12 \approx 0.013$, 
a result first found by Mishchenko \cite{Mishchenko2008} and which is  compatible with experimental uncertainties \cite{Nair:2008zz,PhysRevLett.101.196405}. 
At this point, let's recall that Mishchenko's analysis warns against the use of the Kubo formula with a hard cut-off regulator. Reassuringly, in dimensional regularization no such problem is encountered and
both approaches can be safely used. Our formalism, can be straightforwardly extended to higher orders,~\footnote{As we have discussed in the Introduction, 
for coupling constant $\al_g \approx 1$, perturbation theory is highly questionable. For the polarization operator, the appearance of a small numerical constant 
$\mathcal{C}^{(2)}$ in factor of $\alpha_g$ brings a small parameter $\mathcal{C}^{(2)}\alpha_g$.  This may restore the validity of perturbation theory for the 
optical conductivity provided that higher order terms may be neglected as well. However, beyond first order, the value of $\mathcal{C}_n$
at order $n$ is unknown and it is therefore an open question as to whether or not $\mathcal{C}_n\alpha_g^n$ is small.}
other quantities and/or systems of other dimensionality, {\eg}, the optical conductivity of 3D Dirac materials.
It also constitutes a solid base on which strong-coupling approaches may be developed, {\eg}, dynamical mass generation in the non-relativistic limit.~\footnote{The computation of NLO 
corrections in Ref.~\cite{PhysRevLett.113.105502} was approximate as an NLO diagram with two-loop polarization insertion was neglected (see lines below
Eq.~(9) in that paper). It turns out that this important diagram is the one proportional to $\mathcal{C}\alpha_g$. Recall from Chap.~\ref{chap4}, that this kind of diagram was computed
exactly: see diagram 1 in Fig.~\ref{chap4:fig:diags-NLO} together with Eqs.~\ref{chap4:sigma-NLO-1} and \ref{chap4:polar-2loops} as well as Sec.~\ref{sec:NLO} 
where $\hat{\Pi}_1$ and $\hat{\Pi}_2$ are numbers analogous to $\mathcal{C}$. The fact that this number is small (as for $\hat{\Pi}_2$ in the case of QED$_3$ 
where it is actually proportional to $\mathcal{C}^*$) or large (as for $\hat{\Pi}_1$ in the case of reduced QED$_{4,3}$) strongly impacts the results. Though this would require a more careful analysis,
our proof that $\mathcal{C} = \mathcal{C}^{(2)}$ is indeed small for graphene may further justify
the neglect of the corresponding diagram in Ref.~\cite{PhysRevLett.113.105502} which may in turn insure the good convergence of the RPA observed in Ref.~\cite{PhysRevLett.113.105502}.}
 We leave these issues for our future investigations.

\end{fmffile}

\cleardoublepage

%% file: Conclusions/conclusions.tex
\label{chap6}

This manuscript has focused on the field-theoretic study of strongly interacting low dimensional quantum systems with relativistic-like low-energy excitations nowadays often referred to as 
planar Dirac liquids. These include, \eg, graphene and graphene-like materials, the surface states of some topological insulators and possibly half-filled fractional quantum Hall systems. 
As we have discussed at length, these systems admit an infra-red Lorentz-invariant fixed point at which their low-energy properties are described by effective relativistic quantum field theories such as the so-called reduced 
QED and QED$_3$. With the help of advanced multi-loop techniques, our goal was to achieve an accurate and systematic computation of radiative corrections 
in these models both at the perturbative and non-perturbative levels. We have also considered the more subtle non-relativistic counterparts of these models (away from the IR fixed point) 
which are closer to the experimental reality. A wide range of challenging theoretical problems at the interface between low- and high-energy physics and relevant to condensed matter physics experiments have been addressed: 
from the study of spectral and transport properties of these systems, to dynamical chiral symmetry breaking phenomena and associated dynamical mass or gap generation. 
Some of the most remarkable achievements presented include: 1) in the relativistic limit: the complete exact solution to NLO computation in QED$_3$ and reduced QED$_{4,3}$ with high precision estimates for the critical flavour number
and critical coupling constant separating massive and massless phases; 2) in the non-relativistic limit: the exact computation of the two-loop interaction correction coefficient to the optical conductivity of graphene; 3) from
the point of multi-loop calculations: the computation of new master integrals and a better general understanding of the structure of odd-dimensional quantum field theories.

It should be noted that the use of the multi-loop techniques as done in the present manuscript and related research papers is rather non-standard. From the point of view of condensed matter physics, these 
techniques are mostly unknown. Their use in this context represents a rather original way to investigate many-body effects in Dirac liquids. We are looking forward to new experiments further
unveiling the subtle role of electron-electron interactions in these systems and challenging our theoretical understanding. 
From the field theoretic point of view, these techniques are of course very well known to some experts around the world mostly coming from high-energy physics and mathematical physics. 
As reviewed in the Introduction, the focus there is on high order calculations essentially in even-dimensional 
models with applications to statistical mechanics and particle physics (collider physics within and beyond the Standard Model). In this respect, the odd-dimensional field theories which are relevant to our
study, not even to mention non-relativistic ones, remain largely unexplored in comparison with their even-dimensional counter-parts. Moreover, the solution of Schwinger-Dyson type equations using dimensional regularization
is also non-standard. The fact that we have managed with our approach (which is systematic in nature) to achieve such a task represents a significant progress with respect to the study of the non-perturbative
structure of quantum field theories.

The success of the use of the field theoretic renormalization approach combined with sophisticated multi-loop techniques demonstrated by the present manuscript opens the way for promissing new developments. Some of them include 
straightforward extensions, \eg, to 3D Dirac liquids, by enriching the present models and/or considering other (mixed dimensional) models. Others, much less trivial, require going to higher orders (in perturbative and 
non-perturbative, \eg, Schwinger-Dyson type, frameworks) in order to reach a better precision and gain a deeper understanding 
on the phase-structure of the models and phase diagram of the related experimental systems. Certainly, an automation of the present (``hand-made'') calculations would be highly desirable to pursue such tasks.
Of course, this also opens more formal new perspectives such as a better understanding of the number theoretic content of odd-dimensional (eventually non-relativistic) field theories, a difficult subject which has been extensively 
worked out for some four-dimensional relativistic models but remains largely unexplored beyond these. We leave these issues for our future investigations.

\cleardoublepage

%% file: Appendix/conventions.tex
\label{app:conv}

\begin{fmffile}{fmf-app}

\section{Basic notations and conventions}

Throughout the manuscript, the space-time dimension is denoted by a lower case $d$ and the space dimension by an upper case $D$.
The notations $d_e$ and $D_e$ refer to the space-time and space, respectively, dimensions of the fermion field;
we have: $d_e=D_e+1$. Similarly, $d_\gamma=D_\gamma +1$ for the gauge field. 
Example: in graphene $D_e=2$ and $D_\gamma=3$ and therefore $d_e=2+1$ and $d_\gamma = 3+1$.

The QED fine structure constant is defined as:
\be
\al = \frac{e^2}{4\pi \hbar c} = \frac{1}{137}\, .
\ee
It is sometimes, \eg, in the case of QED$_4$, convenient to use the coupling constant:
\be
\bar{\al} = \frac{\al}{4\pi}\, .
\ee
Otherwise explicitly mentioned, we work with units such that:
\be
\hbar=1.
\label{app0:units}
\ee
When relativistic invariance is insured, we shall also set $c=1$.

\section{(Pseudo-)Relativistic notations}
\label{app:pr}

In an arbitrary $d$-dimensional space-time, the Minkowski metric $g_{\mu \nu}$ and its inverse $g^{\mu \nu}$, are given by:
\be
g_{\mu \nu} = g^{\mu \nu}= (1,-1,\cdots,-1)\, ,
\label{app:metric}
\ee
where Greek indices run over all components: $\mu =0,1,2,...,D$ ($\mu=0$ is the time component). Roman indices will be used to denote the spatial components.

General four-vectors are defined as:
\bea
&&C^\mu = (C^0,\vec C\,), \quad C_\mu = g_{\mu \nu} C^\nu = (C^0, - \vec C\,),
\\
&&C \cdot D = g_{\mu \nu} C^\mu D^\nu = C^0 D^0 - \vec C \cdot \vec D\, .
\eea
For example, the four-position vector reads: 
\be
x^\mu = (x^0,\vec x\,) = (ct,\vec x\,), \qquad x_\mu = (x^0, -\vec x\,) = (ct,-\vec x\,)\, .
\ee
Similarly, the four-potential vector reads: $A^\mu = (A^0,\vec A\,)$ where $A_0=\varphi$ is the scalar potential. 
We work in a system of (Gaussian) units where electric and magnetic fields are related to the components of the four-potential with the help of:
\be
\vec E = - \frac{1}{c}\,\frac{\partial \vec A}{\partial t} - \vec \nabla \varphi, \qquad \vec B = \vec \nabla \times \vec A\, .
\ee
%

%The four-momentum vector reads: $p^\mu = (p^0,\vec p\,)=(E,\vec p\,)$ and $p_\mu = (p^0, -\vec p\,)=(E,-\vec p\,)$.

The four-gradient is defined as:
\be
\partial^\mu = \frac{\partial}{\partial x_\mu} = (\frac{\partial}{\partial x_0},-\vec \nabla\,)=(\frac{1}{c}\,\frac{\partial}{\partial t},-\vec \nabla\,), \qquad
\partial_\mu = \frac{\partial}{\partial x^\mu} = (\frac{\partial}{\partial x^0},\vec \nabla\,) = (\frac{1}{c}\,\frac{\partial}{\partial t},\vec \nabla\,)\, .
\ee
For example, the Lagrangian density of the electromagnetic field is explicited as:
\be
\mathcal{L}_\gamma = -\frac{1}{4}\,F^{\mu \nu} F_{\mu \nu} = \frac{|\vec E\,|^2 - |\vec B\,|^2}{2}\, .
\ee
On the other hand, the Lorentz gauge condition, $\partial_\mu A^\mu=0$ is explicited as:
\be
\partial_\mu A^\mu= \frac{1}{c}\,\frac{\partial \varphi}{\partial t} + \vec \nabla \cdot \vec A = 0\, .
\ee
For $c \ra \infty$ (instantaneous interactions) the above condition reduces to the Coulomb gauge.
%Formally, the case of instantaneous interactions corresponds to taking the limit $c \ra 0$

As a second example, the four-momentum reads:
\be
p^\mu = \I \partial^\mu = (E/c,\vec p\,), \qquad E= c \,\I \,\frac{\partial}{\partial x^0} = \I \,\frac{\partial}{\partial t}, \quad \vec p = - \I \,\vec \nabla\, .
\ee

In the case of graphene, we need to switch to pseudo-relativistic notations in order to describe in a compact way a massless fermion propagating at the velocity $v$. 
This may be done by defining a four-pseudo-gradient:
\be
\tilde{\partial}^\mu = (\frac{\partial}{\partial t},-v \vec \nabla\,), \qquad \tilde{\partial}_\mu = (\frac{\partial}{\partial t},v \vec \nabla\,)\, .
\ee
The corresponding four-pseudo momentum is then defined as:
\be
\tilde{p}^\mu = (\tilde{p}^0,v \vec p\,) = (E, v \vec p\,), \qquad \tilde{p}^2 = E^2 + v^2 |\vec p\,|^2\, ,
\ee
where $\vec p$ is the momentum vector.
In the main text, the tildes are omitted. In the case $v=c$, the natural unit $v=c=1$ is the most convenient one.
%This may be done by defining pseudo-gamma matrices $\tilde{\gamma}^\mu$ the components of which are related to the usual
%gamma matrices $\gamma^\mu=(\gamma^0,\vec \gamma\,)$ with the help of: $\tilde{\gamma}^\mu = (c\,\gamma^0, v\,\vec \gamma\,)$. Then:
%%
%\be
%\hat{\partial}_\mu = \tilde{\gamma}^\mu \partial_\mu = \gamma^0 \frac{\partial}{\partial t} + v\,\vec \nabla, \qquad \hat{p}_\mu = \tilde{\gamma}^\mu p_\mu = \gamma^0 E - v \vec \gamma \cdot \vec p\, .
%\ee
%%
%In this case, $p^2 = E^2 + v^2 |\vec p\,|^2$. This is equivalent to defining a four-pseudo-momentum vector: $\tilde{p}^\mu = (\tilde{p}^0,v \vec p\,) = (E,v \vec p\,)$. 

\section{Dirac gamma matrices}
\label{app:Dirac}

In $d$-dimensional space-time, we consider a representation where the $d$ gamma matrices are $4\times 4$ matrices satisfying:\footnote{We assume this holds in odd dimensions.} 
\be
\{\gamma^\mu,\gamma^\nu \} = 2\,g^{\mu \nu}, \qquad \Tr[ {\bf 1} ] = 4N_F\, ,
\ee
where $\Tr$ is the trace operator (in gamma-matrix space) and the factor of $N_F$ comes from the fact that we have $N_F$ fermion species. 
In conventional dimensional regularization (when $d$ is non-integer) these relations are preserved.
The slashed notation defines:
\be
\Sk \equiv \gamma^\mu k_\mu\, .
\ee
Notice that, if $p$ is the momentum in pseudo-relativistic notations, then: $\hat{p} = \gamma^0 E - v \vec \gamma \cdot \vec p$.

%Conventions for conventional dimensional regularization:
%%
%\begin{subequations}
%\bea
%g^{\mu \nu} = \text{diag}(1,-1,-1, \dotsc,-1), \qquad \{\gamma^\mu,\gamma^\nu \} = 2\,g^{\mu \nu}, \qquad \Tr[ \mathbb{1} ] = 4\, ,
%\eea
%\end{subequations}
%%

Contraction identities:
\begin{subequations}
\begin{flalign}
&\gamma^\mu \gamma_\mu = g^{\mu}_{\mu} = d, \quad \gamma^\mu \gamma^\al \gamma_\mu = -(d-2)\,\gamma^\al, \quad
\gamma^\mu \gamma^\al \gamma^\beta \gamma_\mu = 4 g^{\al \beta} + (d-4)\,\gamma^\al \gamma^\beta\, ,
\\
& \gamma^\mu \gamma^\al \gamma^\beta \gamma^\delta \gamma_\mu = -2 \gamma^\delta \gamma^\beta \gamma^\al -(d-4)\, \gamma^\al \gamma^\beta \gamma^\delta\, ,
\end{flalign}
\end{subequations}
This implies, for example: $\gamma^\mu \Sk \gamma^\beta \gamma_\mu = 4k^\beta + (d-4)\,\Sk\, \gamma^\beta$.

Trace identities:
\begin{subequations}
\begin{flalign}
&\Tr[\gamma^\mu \gamma^\nu] = 4N_F g^{\mu \nu}, \quad \Tr[\gamma^\mu \gamma^\al \gamma^\beta \gamma^\nu] = 4N_F\,( g^{\mu \al} g^{\beta \nu} - g^{\mu \beta} g^{\al \nu} + g^{\mu \nu} g^{\al \beta})\, ,
\\
&\Tr[\gamma^\mu \gamma^\al\gamma_\mu \gamma^\beta] = -4N_F\,(d-2)\,g^{\al \beta}\, ,
\end{flalign}
\end{subequations}
This implies, for example: $\Tr[\Sk \Sp ] = 4N_F\,(p,k)$.

Some simple examples of traces useful in calculations:
\begin{subequations}
\label{app:traces}
\begin{flalign}
&
\frac{1}{4N_F}\,\Tr[ \gamma^\mu\,\Sk\,\gamma^\nu\,(\Sk + \Sq)] = -(d-2)\,(k,k+q)\, ,
\label{app:Tr-Pi}
\\
%\frac{1}{4}\,\Tr[ \gamma^\mu\,\left( g_{\mu \nu} - \xi\,\frac{(p-k)_\mu (p-k)_\nu}{(p-k)^2} \right)\,\gamma^\nu] = \frac{1}{4}\,\Tr[ \gamma^\mu\gamma_\mu - \xi] = D - \xi\, .
&\frac{1}{4N_F}\,\Tr[ \gamma^\mu\,P_{\mu \nu}(p-k;\eta)\,\gamma^\nu] = \frac{1}{4}\,\Tr[ \gamma^\mu\gamma_\mu - 1 + \xi] = D - \eta\, ,
\label{app:Tr-Sigma1S}
\\
&\frac{1}{4N_F}\,\Tr[\Sp\, \gamma^\mu\,P_{\mu \nu}(p-k;\eta)\,\Sk\,\gamma^\nu] = -(D-2)\,(p,k) - \eta\,\frac{[-p^2]}{[-(k-p)^2]}\,(p,k)\, ,
\label{app:Tr-Sigma1V}
\end{flalign}
\end{subequations}
where, in the last trace, all terms proportional to $k^2$ have been neglected as they lead to the (vanishing in dimensional regularization) massless tadpole diagram 
and $P_{\mu \nu}(q;\eta) = g_{\mu \nu} - \eta\,q_\mu q_\nu/q^2$.

The Euclidean $\gamma$-matrices satisfy (see \ref{app:sec:m-e}):
\be
(\gamma_{_E}^\mu)^\dagger = \gamma_{_E}^\mu, \qquad \qquad \{ \gamma_{_E}^\mu, \gamma_{_E}^\nu \} = 2 \delta^{\mu \nu}\, .
\ee
So, for example, the Euclidean $\gamma$-matrices corresponding to Eqs.~(\ref{chap4:gamma:4x4})  and (\ref{chap4:gamma:4x4:3+5}) read:
\be
\gamma_{_E}^0 \,=\, \left( \begin{array}{cc} \sigma_3       &  0               \\
                                        0              &  -\sigma_3       \end{array}\right),\quad
\gamma_{_E}^1 \,=\, \left( \begin{array}{cc} \sigma_1  &  0 \\
                                        0              &  -\sigma_1    \end{array}\right),\quad
\gamma_{_E}^2 \,=\, \left( \begin{array}{cc} \sigma_2   &  0   \\
                                        0              &  -\sigma_2    \end{array}\right), \quad
\gamma_{_E}^3 \,=\,
                        \left( \begin{array}{cc}    0     &  -\I \\
                                                        \I     &  0    \end{array}\right)\, .
\ee
The two additional matrices read:
\be
\gamma^5 \,=\, \gamma_{_E}^0 \gamma_{_E}^1 \gamma_{_E}^2 \gamma_{_E}^3\, = \,
                            \left( \begin{array}{cc}     0     &  1   \\
                                                         1     &  0     \end{array}\right), \qquad
\gamma^{35} = \I \gamma_{_E}^3 \gamma^5 = \,
                          \left( \begin{array}{cc}     1     &  0   \\
                                                       0     &  -1     \end{array}\right)\, .
\ee

\section{Minkowski vs Euclidean}
\label{app:sec:m-e}

Table \ref{tab:eucl-mink} summarizes the conventions on how various quantities transform upon going from Minkowski to Euclidean space and vice versa.

In Euclidean space, the action of Eq.~(\ref{chap1:rqed}) reads:
\bea
S_E = \int \D^{d_e} x_{_{E}}\, \bar{\psi}_{_{E}\,\sigma} \left( \slashed \partial_{_E} - \I e \slashed A_{_E} \right) \psi_{_{E}}^\sigma + \int \D^{d_\gamma} x_{_{E}}\,\left[ \frac{1}{4}\,F_{_E}^{\mu \nu}\,F_{_E}^{\mu \nu} 
+ \frac{1}{2\xi}\left(\partial_{_E}^{\mu}A_{_E}^{\mu}\right)^2 \right]\, .
\label{app:rqed-Eucl}
\eea
This yields the following Euclidean space Feynman rules (we drop the subscript $E$ for simplicity):
\begin{subequations}
\label{app:FR-Eucl}
\bea
&&S_0(p) = \frac{1}{\I \gamma \cdot p} \,
\\
&&\tilde{D}_0^{\mu \nu}(q) = \frac{\Gamma(1-\varepsilon_e)}{(4\pi)^{\varepsilon_e}\,(q^2)^{1-\varepsilon_e}}\,\left( \delta^{\mu \nu} - (1-\tilde{\xi})\,\frac{q^{\mu} q^{\nu}}{q^2}\right)\, ,
\\
&&\I e \Gamma_0^\mu = \I e \gamma^\mu\, .
\eea
\end{subequations}
The photon propagator is conveniently separated in a longitudinal and a transverse part:
\be
\tilde{D}^{\mu \nu}(q) = \tilde{d}_\perp(q^2)\,\left( g^{\mu \nu} - \frac{q^{\mu} q^{\nu}}{q^2} \right) + \tilde{d}_\parallel(q^2)\frac{q^{\mu} q^{\nu}}{q^2}\, .
\ee
In the free case:
\bea
\tilde{d}_{0 \parallel}(q^2) = \tilde{\xi}\,\frac{\Gamma(1-\varepsilon_e)}{(4\pi)^{\varepsilon_e}\,(q^2)^{1-\varepsilon_e}}\, ,
\qquad
\tilde{d}_{0 \bot}(q^2) = \frac{\Gamma(1-\varepsilon_e)}{(4\pi)^{\varepsilon_e}\,(q^2)^{1-\varepsilon_e}}\, .
\eea
With interactions, the Euclidean fermion and photon propagators become:
\begin{subequations}
\label{app:S+D}
\begin{flalign}
&S(p) = \frac{1}{\I \gamma \cdot p + \Sigma(p)} \,
\\
&\tilde{d}_{\bot}(q^2) = \tilde{d}_{0 \, \bot}(q^2) \frac{1}{1+ q^2 \, \tilde{d}_{0\, \bot}(q^2)\, \Pi(q^2)}, \qquad \tilde{d}_{\parallel}(q^2) = \tilde{d}_{0\,\parallel}(q^2)\, ,
\end{flalign}
\end{subequations}
with the corresponding Euclidean Schwinger-Dyson equations reading:
\vspace{0.5cm}
\begin{subequations}
\label{app:SDE}
\begin{flalign}
&\Sigma(p) \quad =
\quad
      \parbox{20mm}{
    \begin{fmfgraph*}(20,15)
      \fmfleft{vi}
      \fmfright{vo}
%      \fmf{double}{in,vi}
      \fmf{heavy,tension=0.2,label=$k$}{vi,vo}
      \fmf{dbl_wiggly,left,tension=0.2,label=$p-k$}{vi,vo}
%      \fmf{double}{vo,out}
      \fmfdot{vi}
      \fmfv{decor.shape=circle,decor.filled=shaded,decor.size=7thick}{vo}
    \end{fmfgraph*}
      }
 \quad = \quad
\int [\D^4 k] \,(e \gamma^\mu)\,D_{\mu \nu}(p-k)\,S(k)\,(e \Gamma^\nu(p,k))\, ,
\label{app:SDE:Sigma}
\\
\nonum
\\
&\Pi^{\mu \nu}(q) \quad =
\quad
  \parbox{20mm}{
  \begin{fmfgraph*}(20,18)
      \fmfleft{vi}
      \fmfright{vo}
      \fmf{heavy,right=0.6,label=$k$,l.d=0.1h}{vi,vo}
      \fmf{heavy,right=0.6,label=$k+q$,l.d=0.05h}{vo,vi}
      \fmfdot{vi}
      \fmfv{decor.shape=circle,decor.filled=shaded,decor.size=7thick}{vo}
  \end{fmfgraph*}
}
 \quad = \quad
-\int [\D^4 k]\, \Tr \bigg[ (e \gamma^\mu)\,S(k)\,(e \Gamma^\nu(p,k))\,S(k+q) \bigg]\, .
\label{app:SDE:Pi}
\end{flalign}
\end{subequations}

\begin{center}
\renewcommand{\tabcolsep}{0.25cm}
\renewcommand{\arraystretch}{1.5}
\begin{table}
    \begin{tabular}{  l || c | c | c }
      \hline
        ~~                  &       {\bf Minkowski}                                         &       {\bf Euclidean}                                         &       {\bf Relation} \\
      \hline \hline
      position $x^\mu$      &    $x_M^\mu = (x_M^0,\vec x\,)$                       &     $x_E^\mu = (x_E^0,\vec x\,)$                      &       $x_M^0 = -\I x_E^0$    \\
      \hline
      momentum $p^\mu$      &    $p_M^\mu = (p_M^0,\vec p_M\,)$                     &     $p_E^\mu = (p_E^0,\vec p_E\,)$                  &       $p_M^0 = \I p_E^0$ and $\vec p_M=-\vec p_E$     \\
      \hline
      4-vector $A^\mu$      &    $A_M^\mu = (A_M^0,\vec A\,)$                       &     $A_E^\mu = (A_E^0,\vec A\,)$                      &       $A_M^0 = \I A_E^0$     \\
      \hline
      $\gamma$-matrix       &    $\gamma_M^\mu = (\gamma^0,\vec \gamma_M\,)$        &     $\gamma_E^\mu = (\gamma^0,\vec \gamma_E\,)$       &       $\vec \gamma_M = \I \vec \gamma_E$     \\
      \hline
      $x \cdot p$           &    $x_M \cdot p_M = x_M^0 p_M^0 - \vec x \cdot \vec p$    &     $x_E \cdot p_E = x_E^0 p_E^0 + \vec x \cdot \vec p_E$   &       $x_M \cdot p_M = x_E \cdot p_E$     \\
      \hline
      $A \cdot B$           &    $A_M \cdot B_M = A_M^0 B_M^0 - \vec A \cdot \vec B$    &     $A_E \cdot B_E = A_E^0 B_E^0 + \vec A \cdot \vec B$   &       $A_M \cdot B_M = - A_E \cdot B_E$     \\
      \hline
      $\gamma \cdot \gamma$ &    $\gamma_M \cdot \gamma_M = \gamma^0 \gamma^0 - \vec \gamma_M \cdot \vec \gamma_M$    &     $\gamma_E \cdot \gamma_E = \gamma^0 \gamma^0 +\vec \gamma_E \cdot \vec \gamma_E$   &       
$\gamma_M \cdot \gamma_M = + \gamma_E \cdot \gamma_E$     \\
      \hline
      $\slashed x = \gamma^\mu x_\mu$           &    $\slashed x_M = \gamma^0 x_{M0} - \vec \gamma_M \cdot \vec x$    &     $\slashed \partial_E  = \gamma^0 x_{E0} + \vec \gamma_E \cdot \vec x$   &
$\slashed x_M = -\I \slashed x_E$     \\
      \hline
      $\slashed p = \gamma^\mu p_\mu$           &    $\slashed p_M = \gamma^0 p_{M0} - \vec \gamma_M \cdot \vec p_M$    &     $\slashed p_E  = \gamma^0 p_{E0} + \vec \gamma_E \cdot \vec p_E$   &
$\slashed p_M = \I \slashed p_E$     \\
      \hline
      $\slashed \partial = \gamma^\mu \partial_\mu$           &    $\slashed \partial_M = \gamma^0 \partial_{M0} + \vec \gamma_M \cdot \vec \nabla$    &     $\slashed \partial_E  = \gamma^0 \partial_{E0} + \vec \gamma_E \cdot \vec \nabla$   &       
$\slashed \partial_M = \I \slashed \partial_E$     \\
      \hline
      $\slashed A = \gamma^\mu A_\mu$           &    $\slashed \partial_M = \gamma^0 A_{M0} - \vec \gamma_M \cdot \vec A$    &     $\slashed \partial_E  = \gamma^0 A_{E 0} + \vec \gamma_E \cdot \vec A$   &     
$\slashed A_M = -\I \slashed A_E$     \\
      \hline
      $\slashed D$           &    $\slashed D_M = \slashed \partial_M +\I e \slashed A_M $    &     $\slashed D_E  = \slashed \partial_E - \I e \slashed A_E$   &
$\slashed D_M = \I \slashed D_E$     \\
      \hline
    \end{tabular}
    \caption{Euclidean and Minkowski space conventions.}
    \label{tab:eucl-mink}
\end{table}
\end{center}

\end{fmffile}

\cleardoublepage

%% file: Appendix/gegenbauer.tex
\label{app:gegen}

In this Appendix, we provide a rather detailed analysis of the formulas which are at the basis of the Gegenbauer polynomial technique.
Some of them can be found in the classic monograph on Orthogonal Polynomials by Gabor Szeg\"o \cite{szego1939orthogonal}. 
Other useful references are given. An attempt is made to provide proofs whenever possible.

\section{Orthogonal polynomials}

Orthogonal polynomials are a class of polynomials $\{ p_n(x) \}_{n=0}^\infty$ defined over a range $[a,b]$ that obey an orthogonality relation:
\be
\int_a^b\,p_n(x)p_m(x)w(x)\,\D x=\delta_{n,m} c_n,
\label{app:gegen:ortho}
\ee
where $w(x)$ is a weighting function and $n$ is the degree of the polynomial. 
In the case where $c_n=1$ the polynomials are orthonormal. 
A given class of polynomials is entierely defined by $a$, $b$ and $w(x)$.
Such polynomials provide a convenient basis to expand solutions of various differential equations. 

\subsection{Jacobi polynomials}

Following Szeg\"o \cite{szego1939orthogonal} let's consider the Jacobi polynomials (also known as hypergeometric polynomials). They are
noted $P_n^{(\al,\beta)}(x)$ where $\al$ and $\beta$ are two indices. They satify the following orthogonality relation:
\be
\int_{-1}^1\,P_n^{(\al,\beta)}(x)P_m^{(\al,\beta)}(x)\,(1-x)^\al(1+x)^\beta(x)\,\D x=\delta_{n,m}\,\frac{2^{\al+\beta+1}}{2n+\al+\beta+1}\,\frac{\Gamma(n+\al+1)\Gamma(n+\beta+1)}{n!\,\Gamma(n+\al+\beta+1)}\, ,
\label{app:gegen:jacobi:orth}
\ee
where $\al>-1$ and $\beta>-1$ to insure the integrability of $w(x)$.
The normalization condition reads:
\be
P_n^{(\al,\beta)}(1)= \binom{n+\al}{n}\, ,
\label{app:gegen:jacobi:norm}
\ee
%
%where $\binom{n}{k}=\frac{n!}{k!(n-k)!}$ is a binomial coefficient which may be generalized as follows:
where $\binom{z}{k}$ for $z$ non-integer is a generalized binomial coefficient:

\be
\binom{z}{k} = \frac{\Gamma(z+1)}{\Gamma(k+1)\Gamma(z-k+1)} \quad \text{and} \quad \binom{z}{k} = 0 \quad  \text{for} \quad k<0 \, .
\ee
The Jacobi polynomials also satisfy the following symmetry relation:
\be
P_n^{(\al,\beta)}(-x) = (-1)^n\,P_n^{(\beta,\al)}(x)\, .
\label{app:gegen:jacobi:symm}
\ee
Eq.~(\ref{app:gegen:jacobi:symm}) together with Eq.~(\ref{app:gegen:jacobi:norm}) yields:
\be
P_n^{(\al,\beta)}(-1)= (-1)^n\,\binom{n+\beta}{n}\, .
\label{app:gegen:jacobi:norm2}
\ee
The Jacobi polynomials are a polynomial solution of the Jacobi differential equation:
\be
(1-x^2)y''+[\beta-\alpha-(\alpha+\beta+2)x]y'+n(n+\alpha+\beta+1)y=0\, . 
\label{app:gegen:jacobi:diffeq}
\ee
They satisfy the following Rodrigues-type formula:
\be
P_n^{(\al,\beta)}(x) = \frac{(-1)^n}{2^n\,n!}\,(1-x)^{-\al}(1+x)^{-\beta}\,\frac{\D^n}{\D\,x^n}\,\left[ (1-x)^{\al+n}(1+x)^{\beta+n} \right].
\label{app:gegen:jacobi:rodrigues}
\ee
From Eq.~(\ref{app:gegen:jacobi:rodrigues}) the following representations are obtained:
\begin{subequations}
\label{app:gegen:jacobi:repres}
\begin{flalign}
P_n^{(\al,\beta)}(x) & =  \sum_{k=0}^n\, \binom{n+\al}{n-k}\,\binom{n+\beta}{k}\,\left( \frac{x-1}{2} \right)^k\,\left( \frac{x+1}{2} \right)^{n-k}\, ,
\label{app:gegen:jacobi:repres-a}
\\
& =  \binom{n+\al}{n}\,\left( \frac{x+1}{2} \right)^n\,\sum_{k=0}^n\,\frac{n(n-1) \cdots (n-k+1)}{(\al+1)(\al+2) \cdots (\al+k)}\,\binom{n+\beta}{k}\,\left( \frac{x-1}{x+1} \right)^k \, ,
\label{app:gegen:jacobi:repres-b}
\\
& =  \binom{n+\al}{n}\,\left( \frac{x+1}{2} \right)^n\,
\pFq{2}{1}{-n,-n-\beta}{\al+1}{\frac{x-1}{x+1}} \, .
\label{app:gegen:jacobi:repres-c}
%F \left(-n,-n-\beta;\al+1;\frac{x-1}{x+1} \right).
\end{flalign}
\end{subequations}
In Eq.~(\ref{app:gegen:jacobi:repres-c}) the Gaussian hypergeometric function has been introduced as defined in the second line above or as:
\be
\pFq{2}{1}{a,b}{c}{z} = \sum_{k=0}^n\,\frac{(a)_k\,(b)_k}{(c)_k}\,\frac{z^n}{n!},\qquad |z|<1,\qquad c \not= 0,-1,-2,\cdots,
\label{app:gegen:2F1:def}
\ee
where $(z)_n$ is the Pochhammer symbol (or rising factorial) which is defined as:
\be
(z)_n = \frac{\Gamma(z+n)}{\Gamma(z)} = 
\left\{
          \begin{array}{l r}
                z(z+1) \cdots (z+n-1), & n>0 \\
                1, & ~ n=0.
          \end{array}
\right.
\ee
The hypergeometric function also converges on the unit circle $|z|=1$ if $\Re[c-a-b]>0$.

Other basic properties of the Jacobi polynomials include the derivatives:
\be
\frac{\D^k}{\D\,x^k}\,P_n^{(\al,\beta)}(x) = \frac{\Gamma(\al+\beta+n+1+k)}{2^k\Gamma(\al+\beta+n+1)}\,P_{n-k}^{(\al+k,\beta+k)}(x),
\ee
a recurrence relation:
\bea
&& 2n(n+\al+\beta)(2n+\al+\beta-2)\,P_n^{(\al,\beta)}(x) =  
\label{app:gegen:jacobi:recurr} \\
&& = (2n+\al+\beta-1)\,\left\{ (2n+\al+\beta)(2n+\al+\beta-2)x +\al^2-\beta^2\right\}\,P_{n-1}^{(\al,\beta)}(x)
\nonum \\
&&- 2(n+\al-1)(n+\beta-1)(2n+\al+\beta)P_{n-2}^{(\al,\beta)}(x),
\eea
for $n=2,3,\cdots$ and the generating function:
\be
\sum_{n=0}^{\infty}\,P_n^{(\al,\beta)}(x)\,w^n = 2^{\al+\beta}\,R^{-1}\,(1-w+R)^{-\al}\,(1+w+R)^{-\beta}, \quad
R = R(x,w) = \sqrt{1-2xw+w^2}.
\label{app:gegen:jacobi:generatfunc} 
\ee

Among the Jacobi polynomials of particular importance are the following polynomials for which all the above formulas apply:
\begin{itemize}
\item The Legendre polynomials correspond to the case:
\be
\al=\beta=0,\qquad w(x)=1, \qquad P_n(x) = P_n^{(0,0)}(x).
\label{app:gegen:legendre:def}
\ee
These polynomials are well known for their relation to the solution of the Laplace equation in spherical coordinates (see below for more)
and their applications to multipole expansions. 
This expansion is readily obtained from the generating function for the Legendre polynomials (Eq.~(\ref{app:gegen:jacobi:generatfunc}) with $\al=\beta=0$)
which reads:
\be
\frac{1}{\sqrt{1-2xw+w^2}} = \sum_{n=0}^\infty\,P_n(x) w^n \, .
\label{app:gegen:legendre:generat}
\ee
Using Eq.~(\ref{app:gegen:legendre:generat}) a three-dimensional Newtonian potential can be expanded as follows (Legendre 1782):
\bea
\frac{1}{|x_1 - x_2|} & = & \frac{1}{\sqrt{x_1^2-2 x_1 \cdot x_2 + x_2^2}} 
\nonum \\
& = & \sum_{l=0}^\infty\,P_l(\hat{x}_1 \cdot \hat{x}_2)\,\Bigg [\frac{(x_2)^{l/2}}{(x_1^2)^{l/2+1/2}}\, \Theta(x_1^2-x_2^2) + (x_1 \longleftrightarrow x_2) \Bigg]\, ,
\label{app:gegen:legendre:multipole}
\eea
where $\hat{x}= x/\sqrt{x^2}$, $\Theta(z) = 0 $ for $z<0$ and $\Theta(z) = 1 $ for $z \geq 0$.

\item The Chebychev polynomials correspond to the case:~\footnote{The normalization of the Chebychev polynomials of the first kind requires some care,
see \cite{szego1939orthogonal}(4.7.2) and (4.7.8). From the definition of the Gegenbauer polynomial it is given by: 
\be
\lim_{\lambda \ra 0} \lambda^{-1}\,C_n^{\lambda}(x) = (2/n)T_n(x).
\nonum
\ee
}
\be
\al=\beta=\pm\frac{1}{2},\quad w(x)=(1-x^2)^{\pm \frac{1}{2}}, \quad T_n(x) = \frac{(1)_n}{(1/2)_n}\,P_n^{(-\frac{1}{2},-\frac{1}{2})}(x), 
\quad U_n(x) = \frac{(2)_n}{(3/2)_n}\,P_n^{(\frac{1}{2},\frac{1}{2})}(x),
\label{app:gegen:chebychev:def}
\ee
where standard normalization conventions were used to define $T_n(x)$ and $U_n(x)$ the Chebychev polynomials of the first and second kind, respectively. 
The generating function for the Chebychev polynomials of the second kind can be written as:~\footnote{We will prove this for the more general case of Gegenbauer polynomials.}
\be
\frac{1}{1-2xw+w^2} = \sum_{n=0}^\infty\,U_n(x) w^n.
\label{app:gegen:chebychev:generat}
\ee
The Chebychev polynomials together with Eq.~(\ref{app:gegen:chebychev:generat}) were used to expand propagator in four-dimensional QED in, \eg,
the early \cite{Rosner:1967zz}. In position space the expansion reads:
\begin{flalign}
\frac{1}{(x_1 - x_2)^2} & = & \sum_{n=0}^\infty\,U_n(\hat{x}_1 \cdot \hat{x}_2)\,\Bigg[ \frac{(x_2^2)^{n/2}}{(x_1^2)^{n/2+1}}\,\Theta(x_1^2-x_2^2) + (x_1^2 \longleftrightarrow x_2^2) \Bigg], 
\quad \hat{x} = \frac{x}{\sqrt{x^2}}\, .
\label{app:gegen:chebychev:multipole}
\end{flalign}
The orthogonality of these polynomials on the unit sphere in $\R^4$ makes it possible to calculate some Feynman diagrams. With the advent of dimensional regularization
a generalization to arbitrary dimension was needed. This is done by the Gegenbauer polynomials, see \cite{Chetyrkin:1980pr} and references therein.
\end{itemize}

\subsection{Legendre polynomials, spherical harmonics and products of symmetric traceless tensors}

As mentionned above, Legendre polynomials are very well known in relation with solutions to the Laplace equation in spherical coordinates.
Let $F(x)$ be a general solution of the Laplace equation:
\be
\frac{1}{r^2}\,\frac{\partial^2}{\partial r^2}\,( r F ) + \Delta_\theta F = 0, \qquad 
\Delta_\theta = \frac{1}{\sin \theta}\,\frac{\partial}{\partial \theta} \,\left( \sin \theta \, \frac{\partial}{\partial \theta} \right)
+ \frac{1}{\sin^2 \theta}\,\frac{\partial^2}{\partial \phi^2}\, ,
\label{app:gegen:legendre:LaplaceEq}
\ee
where $\Delta_\theta$ is the angular part of the Laplacian which is related to the orbital momentum by the relation:
$L^2 = - \hbar^2 \Delta_\theta$. Separating variables, the most general solution can be written as: 
\be
F(x) = \sum_{l=0}^\infty \sum_{m=-l}^l\, c_{l,m} F_l^m(x), \quad F_l^m(x) =  R_l(r) Y_l^m(\theta,\phi), \quad R_l(r) = a_l r^l + b_l r^{-l-1}\, ,
\label{app:gegen:legendre:generalsol1}
\ee
where $r = \sqrt{x^2}$ and $Y_l^m(\theta,\phi)$ is the spherical harmonic which is such that:
\be
\Delta_\theta Y_l^m = -l(l+1) Y_l^m\, .
\label{app:gegen:legendre:DeltaThetaYlm}
\ee
The latter are related to associated Legendre polynomials, $P_l^m$, with the help of the relation:
\be
Y_l^m(\theta,\phi) = \sqrt{\frac{2l+1}{4\pi}\,\frac{(l-m)!}{(l+m)!}}\,P_l^m(\cos \theta)\,e^{\I m \phi}, \qquad P_l^m(x) = (-1)^m\,(1-x^2)^{m/2}\,\frac{\D^m}{\D x^m} \,P_l(x)\, ,
\label{app:gegen:legendre:YlmPlm}
\ee
where $P_l(x)$ are the usual Legendre polynomials presented above. In case of azimutal symmetry, that is, for a function $F$ invariant under rotations around the $z$-axis, only the $m=0$ harmonic contributes
for every $l$. In this case we recover the above mutlipole-type expansion Eq.~(\ref{app:gegen:legendre:multipole}):
\be
F(r,\theta) = \sum_{l=0}^\infty c_{l} R_l(r) P_l(\cos \theta), \qquad  Y_l^0(\theta) = \sqrt{\frac{2l+1}{4\pi}}\,P_l(\cos \theta)\, ,
\label{app:gegen:legendre:generalsol-z1}
\ee
where $c_l$ is a constant and $\cos \theta$ is the angle between $x$ and the $z$-axis.

It turns out that there is another way to write the solutions of the Laplace equation: in terms of traceless symmetric tensors.
Let's consider the function:
\be
F_l(x) = r^lC_{\mu_1 \mu_2 \cdots \mu_l}^{(l)} \,\hat{x}^{\mu_1} \hat{x}^{\mu_2} \cdots \hat{x}^{\mu_l} =
C_{\mu_1 \mu_2 \cdots \mu_l}^{(l)} \,x^{\mu_1} x^{\mu_2} \cdots x^{\mu_l}\, ,
\label{app:gegen:legendre:TST1}
\ee
where radial and angular parts are separated and the angular part is expressed in terms of 
$C_{\mu_1 \mu_2 \cdots \mu_l}^{(l)}$ which is a traceless symmetric tensor (TST) of rank $l$ with indices $\mu$ running over the 3 components of $x$.
Due to the tracelessness of the tensor, that is: $g^{\mu_i \mu_j}\,C_{\mu_1 \mu_2 \cdots \mu_i \cdots \mu_j \cdots \mu_l}^{(l)}=0$, this function turns out to be harmonic for every $l$. 
This can be checked easily at the level of the rank 2 tensor ($l=2$):
\begin{flalign}
\Delta F_2(x) = \Delta C_{\mu_1 \mu_2}^{(2)}\,x^{\mu_1} x^{\mu_2} =
C_{\mu_1 \mu_2}^{(2)}\,\frac{\partial^2}{\partial x^\nu \partial x^\nu}\,x^{\mu_1} x^{\mu_2}
= 2 C_{\mu_1 \mu_2}^{(2)}\,g^{\mu_1 \nu}\,g^{\mu_2 \nu} = 2 C_{\mu_1 \mu_2}^{(2)}\,g^{\mu_1 \mu_2} = 0\, ,
\end{flalign}
and similarly for higher rank tensors.
Moreover, using the separated form of $F_l(r)$ we can see that:
\be
\Delta_\theta F_l(\hat{x}) = - l(l+1) F_l(\hat{x})\, ,
\label{app:gegen:legendre:DeltaFl}
\ee
where $\hat{x}$ is the unit vector in the $x$-direction and therefore: $F_l(\hat{x}) = C_{\mu_1 \mu_2 \cdots \mu_l}^{(l)} \,\hat{x}^{\mu_1} \hat{x}^{\mu_2} \cdots \hat{x}^{\mu_l}$
depends only on the angular variables. The similarity between Eqs.~(\ref{app:gegen:legendre:DeltaFl}) and (\ref{app:gegen:legendre:DeltaThetaYlm}) 
suggests that there is a relation between traceless symmetric tensors and spherical harmonics.
Combinatorial arguments indicate that, for a given $l$, there are $2l+1$ linearly independent TST. This is consistent with the correspondence between TSTs
and $Y_l^m$ as, for a fixed $l$, $m$ takes $2l+1$ values. 

Taking into account the radial part, the most general solution of the Laplace equation in terms of TST reads:
\be
F(x) = \sum_{l=0}^\infty\,R_l(r)\,C_{\mu_1 \mu_2 \cdots \mu_l}^{(l)} \,\hat{x}^{\mu_1} \hat{x}^{\mu_2} \cdots \hat{x}^{\mu_l}\, ,
\label{app:gegen:legendre:generalsol2}
\ee
and should be equivalent to Eq.~(\ref{app:gegen:legendre:generalsol1}).
When azimutal symmetry is present the angular part of the function $F$ depends only on $\theta$ (and not on $\phi$). The TST can therefore be built from the $\hat{z}^\mu$ (which
are such that: $\hat{x} \cdot \hat{z} = \cos \theta$) as follows:
\be
C_{\mu_1 \mu_2 \cdots \mu_l}^{(l)} = \hat{z}^{\mu_1 \mu_2 \cdots \mu_l}\, ,
\label{app:gegen:legendre:Cvsz}
\ee
where the notation $x^{\mu_1 \mu_2 \cdots \mu_n}$ refers to the traceless symmetric tensor of rank $n$ made out of $x^{\mu_1}$, $x^{\mu_2}$, ..., $x^{\mu_n}$
and the indices $\mu_i$ run over the 3 components of $x$. For example, the rank zero TST is a scalar of value $1$ and the rank one TST is the vector of components $x^\mu$.
Nontrivial examples start from rank two as can be see from the following examples:
\be
x^{\mu_1 \mu_2} = x^{\mu_1} x^{\mu_2} - \frac{x^2}{3}\,g^{\mu_1 \mu_2}, \quad 
x^{\mu_1 \mu_2 \mu_3} = x^{\mu_1} x^{\mu_2} x^{\mu_3} - \frac{x^2}{5}\,\left( g^{\mu_1 \mu_2}\,x^{\mu_3} + g^{\mu_2 \mu_3}\,x^{\mu_1} + g^{\mu_3 \mu_1}\,x^{\mu_2} \right)\, .
\ee
The inverse relations:
\be
x^{\mu_1} x^{\mu_2} = \frac{x^2}{3}\,g^{\mu_1 \mu_2} + x^{\mu_1 \mu_2}, \qquad 
x^{\mu_1} x^{\mu_2} x^{\mu_3} = \frac{x^2}{5}\,\left( g^{\mu_1 \mu_2}\,x^{\mu_3} + g^{\mu_2 \mu_3}\,x^{\mu_1} + g^{\mu_3 \mu_1}\,x^{\mu_2} \right) + x^{\mu_1 \mu_2 \mu_3}\, ,
\ee
correspond to the decomposition of the dyadic product of rank one tensors into irreducible 
ones.~\footnote{The dyadic product of two rank 1 tensors can indeed be decomposed as:
\be
x^\mu y^\nu = \frac{x \cdot y}{3}\,g^{\mu \nu} 
+ \left( \frac{x^\mu y^\nu - x^\nu y^\mu}{2}\right) 
+ \left( \frac{x^\mu y^\nu + x^\nu y^\mu}{2} - \frac{x \cdot y}{3}\,g^{\mu \nu}  \right)\, ,
\ee
where the number of components of the objects on the right hand side matches the multiplicities of the spherical harmonics with $l=0$, $l=1$ and $l=2$. 
The later are referred to as irreducible, or spherical, tensors as they transform in a well defined way under the action of SO($3$).} 
These formulas can be generalized to:
\begin{subequations}
\label{app:gegen:legendre:decomp:tensors}
\begin{flalign}
x^{\mu_1 \mu_2 \cdots \mu_l} &= \hat{S}\,\sum_{k=0}^{[l/2]}\,\frac{(-1)^k}{(k)!!}\,\frac{l!}{(l-2k)!}\,\frac{(2l-2k-1)!!}{(2l-1)!!}\,g^{\mu_1 \mu_2} 
\cdots g^{\mu_{2k-1} \mu_{2k}} \, x^{2k}\, x^{\mu_{2k+1}} \cdots x^{\mu_l}\, ,
\label{app:gegen:legendre:decomp:tensors:TST2utens} \\
x^{\mu_1} x^{\mu_2} \cdots x^{\mu_l} &= \hat{S}\,\sum_{k=0}^{[l/2]}\,\frac{1}{(2k)!!}\,\frac{l!}{(l-2k)!}\,\frac{(2l-4k+1)!!}{(2l-2k+1)!!}\,g^{\mu_1 \mu_2}
\cdots g^{\mu_{2k-1} \mu_{2k}} \, x^{2k}\, x^{\mu_{2k+1} \cdots \mu_l}\, ,
\label{app:gegen:legendre:decomp:tensors:utens2TST}
\end{flalign}
\end{subequations}
where $\hat{S}$ is a symmetrizer with respect to the indices $\mu_i$. Eqs.~(\ref{app:gegen:legendre:decomp:tensors}) correspond to the decomposition of a TST 
of rank $l$ on usual tensors and vice-versa. These formulas are a peculiar case of more general formulas that we shall demonstrate
in the section concerning Gegenbauer polynomials, see Eqs.~(\ref{app:gegen:gegenbauer:decomp:tensors}). We mention also that, from the properties of TSTs, the following identities must hold:
\be
x^{\mu_1 \mu_2 \cdots \mu_n}\,z^{\mu_1 \mu_2 \cdots \mu_n} = x^{\mu_1 \mu_2 \cdots \mu_n}\,z^{\mu_1}z^{\mu_2} \cdots z^{\mu_n} = x^{\mu_1}x^{\mu_2} \cdots x^{\mu_n}\, z^{\mu_1 \mu_2 \cdots \mu_n}\, .
\label{app:gegen:legendre:TSTidentities}
\ee
The most general solution of the Laplace equation with rotational symmetry around the $z$-axis then reads:
\be
F(r,\theta) = \sum_{l=0}^\infty c_{l}' R_l(r) \hat{z}^{\mu_1 \mu_2 \cdots \mu_l}\,\hat{x}^{\mu_1 \mu_2 \cdots \mu_l}\, ,
\label{app:gegen:legendre:generalsol-z2}
\ee
where $c_l'$ is a constant, Eq.~(\ref{app:gegen:legendre:TSTidentities}) has been used and $\cos \theta = \hat{x} \cdot \hat{z}$. Both solutions Eq.~(\ref{app:gegen:legendre:generalsol-z2})
and Eq.~(\ref{app:gegen:legendre:generalsol-z1}) should be identical at every order $l$. This implies that every Legendre polynomial can be expressed as a product of traceless
symmetric tensors:
\be
P_l(\cos \theta) = \text{constant}\,\hat{z}^{\mu_1 \mu_2 \cdots \mu_l}\,\hat{x}^{\mu_1 \mu_2 \cdots \mu_l}\, ,
\label{app:gegen:legendre:LegendreTST1}
\ee
where the constant has to be determined. From the lowest rank tensors we have the following products:
\be
\hat{z}^0 \, \hat{x}^0 = 1, \qquad \hat{z}^{\mu_1} \, \hat{x}^{\mu_1} = \cos \theta, \qquad \hat{z}^{\mu_1 \mu_2} \, \hat{x}^{\mu_1 \mu_2} = \cos^2 \theta - \frac{1}{3}, \cdots
\ee
which are polynomials in $\cos \theta$ and indeed proportional to the Legendre polynomials of the same order:
\be
P_0(x) = 1, \qquad P_1(x) = x, \qquad P_2(x) = \frac{3}{2}\,\left(x^2 - \frac{1}{3} \right)\, \cdots
\ee
In the general case the constant can be determined by using the Rodrigues formula for Legendre polynomials and computing the coefficient of the highest power in $x=\cos \theta$.
From Eq.~(\ref{app:gegen:jacobi:rodrigues}) in the case $\al=\beta=0$ this yields:
\be
P_l(x) = \frac{1}{2^l\,l!} \, \frac{\D^2}{\D x^l}\,\left[ (x^2-1)^l\right] = \frac{(2l)!}{2^l\,(l!)^2}\,x^l + (\text{lower order terms})\, .
\label{app:gegen:legendre:LegendreTST2}
\ee
Comparing Eqs.~(\ref{app:gegen:legendre:LegendreTST2}) and (\ref{app:gegen:legendre:LegendreTST1}) fixes the constant and yields:
\be
P_l(\hat{x}\cdot\hat{z}) = \frac{(2l)!}{2^l\,(l!)^2}\,\frac{x^{\mu_1 \mu_2 \cdots \mu_l}\,z^{\mu_1 \mu_2 \cdots \mu_l}}{(x^2\,z^2)^{l/2}}\, .
\label{app:gegen:legendre:LegendreTST3}
\ee
Inversely:
\be
x^{\mu_1 \mu_2 \cdots \mu_l}\,z^{\mu_1 \mu_2 \cdots \mu_l} = \frac{2^l\,(l!)^2}{(2l)!}\,P_l(\hat{x}\cdot \hat{z})\,(x^2\,z^2)^{l/2}\, .
\label{app:gegen:legendre:TSTLegendre}
\ee

We are now in a position to express the multipole expansion in terms of products of traceless symmetric tensors. Indeed, combining Eqs.~(\ref{app:gegen:legendre:LegendreTST3}) and 
(\ref{app:gegen:legendre:multipole}) yields:
\bea
\frac{1}{|x_1 - x_2|} = 
\sum_{l=0}^\infty\,\frac{(2l)!}{2^l\,(l!)^2}\,x_1^{\mu_1 \mu_2 \cdots \mu_l}\,x_2^{\mu_1 \mu_2 \cdots \mu_l}\,\Bigg [\frac{\Theta(x_1^2-x_2^2)}{(x_1^2)^{l+1/2}} 
+\frac{\Theta(x_2^2-x_1^2)}{(x_2^2)^{l+1/2}} \Bigg]\, .
\label{app:gegen:legendre:multipoleTST}
\eea

These results can be generalized to the case where there is no symmetry around the $z$-axis (associated Legendre polynomials and products of TST). We shall not need such results
which can be found in Guth's lectures \cite{GuthLecturesEM}. Moreover, these results can be generalized to the four-dimensional case by relating Chebychev polynomials to products of TST
and applying it to the multipole expansion Eq.~(\ref{app:gegen:chebychev:multipole}). We shall not do this either. Instead, we go over to Gegenbauer polynomials which allow us to generalize
the above results to any dimension.

\section{Gegenbauer polynomials}

The Gegenbauer polynomials (or ultraspherical polynomials) are noted $C_n^\lambda$ where $n$ is the degree of the polynomial and $\lambda$ an index that
will later be related to the dimensionality of space-time ($\lambda = D/2-1$ for the Gegenbauer polynomials to be orthogonal on the unit sphere in $\R^D$).
They are a special case of Jacobi polynomials corresponding to:
\begin{flalign}
& \al=\beta=\lambda-\frac{1}{2},\qquad w(x)=(1-x^2)^{\lambda-\frac{1}{2}}, \quad C_n^\lambda(x) = \frac{(2\lambda)_n}{(\lambda+\frac{1}{2})_n}\,P_n^{(\lambda-\frac{1}{2},\lambda-\frac{1}{2})}(x),
\label{app:gegen:gegenbauer:def} \\
& C_0^\lambda (x) = 1, \quad C_1^\lambda(x) = 2\lambda x, \quad C_2^\lambda(x) = 2\lambda(\lambda+1)x^2-\lambda, \quad C_n^\lambda(1) = \frac{\Gamma(n+2\lambda)}{\Gamma(2\lambda)\,n!},
\label{app:gegen:gegenbauer:cases}
\end{flalign}
with standard normalization~\cite{szego1939orthogonal}. From the relation between $\lambda$ and $D$ we see that
the Gegenbauer polynomials reduce to Legendre polynomials for $\lambda=1/2$ (that is $D=3$) and Chebychev polynomials for $\lambda=1$ (that is $D=4$).
Conversly, Gegenbauer polynomials generalize Legendre and Chebychev polynomials to a space-time of arbitrary dimension $D=2\lambda +2$.

From the properties of the Jacobi polynomials we can immediately list some basic properties of the Gegenbauer polynomials:
\begin{itemize}
\item \underline{Orthogonality relation:}
\be
\int_{-1}^1\,C_n^\lambda(x)C_m^\lambda(x)(1-x^2)^{\lambda-1/2}\,\D x=\delta_{n,m}\,\frac{2^{1-2\lambda}\pi}{(n+\lambda)\,n!}\,\frac{\Gamma(n+2\lambda)}{\Gamma^2(\lambda)},
\label{app:gegen:gegenbauer:orth}
\ee
\item \underline{Recurrence relation:}
\be
C_n^\lambda(x) = \frac{1}{n}\,\left[ 2x(n+\lambda-1)C_{n-1}^\lambda(x) - (n+2\lambda-2)C_{n-2}^\lambda(x) \right],
\label{app:gegen:gegenbauer:recurr}
\ee
where $n=2,3,\cdots$.
\item \underline{Generating function:}
\be
\frac{1}{(1-2xw+w^2)^\lambda} = \sum_{k=0}^\infty\,C_k^\lambda(x)\,w^k.
\label{app:gegen:gegenbauer:generatfunc}
\ee
Notice that Eq.~(\ref{app:gegen:gegenbauer:generatfunc}) is different from Eq.~(\ref{app:gegen:jacobi:generatfunc}) but is actually simpler 
and more useful.
It can be derived starting from: $h(w) = \sum_{k=0}^\infty\,C_k^\lambda(x)\,w^k$ and using the recurrence relation 
Eq.~(\ref{app:gegen:gegenbauer:recurr}). This yields:
\be
\frac{h'(w)}{h(w)} = \frac{-2\lambda(w-x)}{1-2xw+w^2}, \qquad h(0) = C_0^\lambda(x)=1.
\nonum
\ee
Solving the differential equation then yields Eq.~(\ref{app:gegen:gegenbauer:generatfunc}).

\item \underline{Representations:}
\begin{subequations}
\label{app:gegen:gegenbauer:repres}
\bea
C_n^\lambda(x) & = & \sum_{k=0}^{[n/2]}\,(-1)^k\,\frac{\Gamma(n-k+\lambda)}{\Gamma(\lambda)\,k!\,(n-2k)!}\,(2x)^{n-2k}\, ,
\label{app:gegen:gegenbauer:repres1} \\
C_n^\lambda(x) & = & \frac{(2\lambda)_n}{n!}\,\pFq{2}{1}{-n,n+2\lambda}{\lambda+\frac{1}{2}}{\frac{1-x}{2}}\, .
\label{app:gegen:gegenbauer:repres2}
\eea
\end{subequations}
Eq.~(\ref{app:gegen:gegenbauer:repres1}) can be derived by expanding the left-hand side of Eq.~(\ref{app:gegen:gegenbauer:generatfunc}):
\bea
\frac{1}{(1-2xw+w^2)^\lambda} & = & \sum_{m=0}^\infty\, \binom{-\lambda}{m}\,(w^2-2xw)^m
\nonum \\
& = & \sum_{m=0}^\infty\, (-1)^m\,\binom{m-1+\lambda}{m}\,(w^2-2xw)^m
\nonum \\
& = & \sum_{m=0}^\infty\, \sum_{k=0}^m \, (-1)^k \,\binom{m-1+\lambda}{m}\,\binom{m}{k}\,(2x)^{m-k}\,w^{m+k}
\nonum \\
& = & \sum_{m=0}^\infty\, \sum_{k=0}^m \, (-1)^k \,\frac{\Gamma(m+\lambda)}{\Gamma(\lambda)\,k!\,(m-k)!}\,(2x)^{m-k}\,w^{m+k}\, .
\nonum
\eea
In the last equation we can make the change $m=n-k$ which implies that $2k \leq n$. This yields Eq.~(\ref{app:gegen:gegenbauer:repres1}).

\item \underline{Rodrigues formula:}
\be
C_n^{\lambda}(x) = \frac{(-2)^n}{n!}\,\frac{\Gamma(n+\lambda)\Gamma(n+2\lambda)}{\Gamma(\lambda)\Gamma(2n+2\lambda)}\,
(1-x^2)^{-\lambda+1/2}\,\frac{\D^n}{\D\,x^n}\,\left[ (1-x^2)^{n+\lambda-1/2} \right] \, .
\label{app:gegen:gegenbauer:rodrigues}
\ee
This formula is obtained from Eq.~(\ref{app:gegen:jacobi:rodrigues}) in the case $\al=\beta=1/2$.

\item \underline{Relation between Gegenbauer polynomials and products of traceless symmetric tensors:}
\be
x^{\mu_1 \mu_2 \cdots \mu_n}\,z^{\mu_1 \mu_2 \cdots \mu_n} = \frac{n!\,\Gamma(\lambda)}{2^n\,\Gamma(n+\lambda)}\,C_n^\lambda(\hat{x}\cdot \hat{z})\,(x^2\,z^2)^{n/2}\, .
\label{app:gegen:gegenbauer:TSTGegen}
\ee
This equation generalizes Eq.~(\ref{app:gegen:legendre:TSTLegendre}) to arbitrary dimensions.
Indeed, Gegenbauer polynomials of order $n$ are, up to a constant, proportional to the product of two traceless symmetric tensors of rank $n$.
The constant of proportionality can be derived along the same lines as Eq.~(\ref{app:gegen:legendre:TSTLegendre})
by using Rodrigues formula Eq.~(\ref{app:gegen:gegenbauer:rodrigues}) for Gegenbauer polynomials and computing the coefficient of the highest order term  $x^n$. This yields:
\be
C_n^\lambda(x) = \frac{2^n\,\Gamma(n+\lambda)}{n!\,\Gamma(\lambda)}\,x^n + (\text{lower order terms})\, .
\ee
As a byproduct of Eq.~(\ref{app:gegen:gegenbauer:TSTGegen}), we have:
\be
x^{\mu_1 \mu_2 \cdots \mu_n}\,x^{\mu_1 \mu_2 \cdots \mu_n} = \frac{\Gamma(\lambda) \Gamma(n+2\lambda)}{2^n\,\Gamma(2\lambda)\,\Gamma(n+\lambda)}\,x^{2n}\, ,
\label{app:gegen:gegenbauer:TST-prod}
\ee
where the expression of $C_n^\lambda(1)$, Eq.~(\ref{app:gegen:gegenbauer:cases}), was used.

\item \underline{Canonical decomposition:} 
\be
\frac{(2x)^n}{n!} = \sum_{k=0}^{[n/2]}\,C_{n-2k}^\lambda(x)\,\frac{(n-2k+\lambda)\Gamma(\lambda)}{k!\,\Gamma(n-k+\lambda+1)}\, .
\label{app:gegen:gegenbauer:candec}
\ee
In order to prove this formula, we follow \cite{Chetyrkin:1980pr} and consider a general 
homogeneous polynomial $\mathcal{P}_n(x)=(2 x \cdot z)^n$. Such a polynomial
can be decomposed into harmonic polynomials, or traceless symmetric tensors, noted $\mathcal{P}^{(n)}(x)=(2 x \cdot z)^{(n)}$ and inversely.
The general decomposition can be written as:
\begin{subequations}
\bea
(2 x \cdot z)^{(n)} &=& \sum_{k=0}^{[n/2]}\,c_k^{(n)}\,(x^2z^2)^k\,(2 x \cdot z)^{n-2k}\, \qquad (c_0^{(n)}=1)\, ,
\label{app:gegen:gegenbauer:decomp:TST2utens-gen} \\
(2 x \cdot z)^n &=& \sum_{k=0}^{[n/2]}\,d_k^{(n)}\,(x^2z^2)^k\,(2 x \cdot z)^{(n-2k)}\, \qquad (d_0^{(n)}=1)\, .
\label{app:gegen:gegenbauer:decomp:utens2TST-gen}
\eea
\end{subequations}
In order to find the coefficients $c_k^{(n)}$ and $d_k^{(n)}$ we use the fact that TSTs are harmonic:
\be
\Box_x \,(2 x \cdot z)^{(n)} = 0, \qquad \Box_x = \frac{\D^2}{\D x^\mu \D x_\mu}\, ,
\ee
where $\Box_x$ is the d'Alembert operator. Other useful formulas include:
\bea
\frac{\partial x^\mu}{\partial x^\nu} = \delta_\nu^{\mu} \quad (\delta_\mu^{\mu} = D = 2\lambda +2), \qquad \frac{\partial \,x^{2\al}}{\partial x^\mu} = 2\al\, x_\mu\, (x^2)^{\al-1} \, 
\nonum \\
\frac{\partial \,(2 x \cdot z)^{\beta}}{\partial x^\mu} = 2 \beta \, z_\mu \, (2 x \cdot z)^{\beta-1}, \qquad x^\mu\,\frac{\partial \,(2 x \cdot z)^{(n)}}{\partial x^\mu} = n \, (2 x \cdot z)^{(n)}\, ,
\nonum \\
\Box_x \,x^{2\al} = 4\al\,(\al +\lambda)\,(x^2)^{\al-1}, \qquad \Box_x \,(2 x \cdot z)^{\beta} = 4 \beta\,(\beta-1)\,z^2\,(2 x \cdot z)^{\beta-2}\, .
\eea
Then, imposing the constraint of harmonicity on Eq.~(\ref{app:gegen:gegenbauer:decomp:TST2utens-gen}) and identifying terms of same power yields a recurrence equation which can easily be solved:
\bea
&&c_k^{(n)} = \frac{(2k-2-n)(2k-1-n)}{k(k-n-\lambda)}\,c_{k-1}^{(n)}
\nonum \\
&&\Rightarrow c_k^{(n)} = \frac{(2k-2-n)\cdots (2-2-n)(2k-1-n)\cdots (2-1-n)}{k!\,(k-n-\lambda)\cdots (1-n-\lambda)}\, .
\nonum
\eea
The solution on the rhs can be written in a more compact way by using the reflection formula: $\Gamma(z)\Gamma(1-z) = \pi/\sin(\pi z)$. As a matter of fact:
\bea
&&\frac{1}{(k-n-\lambda)\cdots (1-n-\lambda)} = \frac{\Gamma(1-n-\lambda)}{\Gamma(1-n-\lambda+k)} = (-1)^k\,\frac{\Gamma(n+\lambda-k)}{\Gamma(n+\lambda)}, 
\nonum \\
&&(2k-2-n)\cdots (2-2-n)(2k-1-n)\cdots (2-1-n) = \frac{\Gamma(2k-1-n+1)}{\Gamma(-n)}= \frac{n!}{(n-2k)!}\, .
\nonum
\eea
This yields:
\be
c_k^{(n)} = \frac{(-1)^k}{k!}\,\frac{n!\,\Gamma(n+\lambda-k)}{(n-2k)!\,\Gamma(n+\lambda)}\, .
\ee
Similarly, we impose the constraint of harmonicity on Eq.~(\ref{app:gegen:gegenbauer:decomp:utens2TST-gen}) and expand the resulting lhs in harmonic polynomials.
In this case, after identifying terms of equal power, a double recurrence (in $n$ and $k$) is found which can also be easily solved:
\be
d_k^{(n)} = \frac{n(n-1)}{k(n+\lambda-k)}\,d_{k-1}^{(n-2)} ~~ \Longrightarrow ~~ d_k^{(n)} = \frac{n(n-1)\cdots (n-2k+2)(n-2k+1)}{k!\,(n+\lambda-k)\cdots (n+\lambda-2k+1)}\, .
\nonum
\ee
The solution can be straightforwardly written in a more compact way using gamma functions:
\be
d_k^{(n)} = \frac{n!\, \Gamma(n+\lambda-2k+1)}{k!\,(n-2k)!\,\Gamma(n+\lambda-k+1)}\, .
\ee
Hence, we arrive at the important formulas:
\begin{subequations}
\label{app:gegen:gegenbauer:decomp:galpolyn}
\begin{flalign}
(2 x \cdot z)^{(n)} &= \sum_{k=0}^{[n/2]}\, \frac{(-1)^k}{k!}\,\frac{n!\,\Gamma(n+\lambda-k)}{(n-2k)!\,\Gamma(n+\lambda)}\,(x^2z^2)^k\,(2 x \cdot z)^{n-2k}\, ,
\label{app:gegen:gegenbauer:decomp:TST2utens} \\
(2 x \cdot z)^n &= \sum_{k=0}^{[n/2]}\,\frac{n!\, \Gamma(n+\lambda-2k+1)}{k!\,(n-2k)!\,\Gamma(n+\lambda-k+1)}\,(x^2z^2)^k\,(2 x \cdot z)^{(n-2k)}\, .
\label{app:gegen:gegenbauer:decomp:utens2TST}
\end{flalign}
\end{subequations}
Using the relation, Eq.~(\ref{app:gegen:gegenbauer:TSTGegen}), between TST and Gegenbauer polynomials,
Eq.~(\ref{app:gegen:gegenbauer:decomp:utens2TST}), yields:
\be
\frac{(2 x \cdot z)^2}{(x^2 z^2)^{n/2}} = \sum_{k=0}^{[n/2]}\,C_{n-2k}^{\lambda}(\hat{x} \cdot \hat{z})\,\frac{n!\, (n+\lambda-2k)\Gamma(\lambda)}{k!\,\Gamma(n+\lambda-k+1)}\, .
\label{app:gegen:gegenbauer:candec-2}
\ee
Defining $x \equiv x\cdot z / (x^2 z^2)^{1/2}$ yields the advertized result Eq.~(\ref{app:gegen:gegenbauer:candec}).

As a by product, Eqs.~(\ref{app:gegen:gegenbauer:decomp:galpolyn}) yield the coefficients of the decomposition of TSTs over usual tensors and vice versa:
\begin{subequations}
\label{app:gegen:gegenbauer:decomp:tensors}
\begin{flalign}
x^{\mu_1}\,x^{\mu_2}\, \cdots\, x^{\mu_n} &= \hat{S}\,\sum_{k=0}^{[n/2]}\,\frac{n!\,\Gamma(n-2k+\lambda+1)}{2^{2k}\,k!\,(n-2k)!\,\Gamma(n-k+\lambda+1)}\,
g^{\mu_1 \mu_2} \dotsc g^{\mu_{2k-1}\mu_{2k}}\,x^{2k}\,x^{\mu_{2k+1} \dotsc \mu_n}\, ,
\label{app:gegen:gegenbauer:decomp:tensors:uten2TST} \\
x^{\mu_1 \mu_2 \cdots \mu_n} &= \hat{S}\,\sum_{k=0}^{[n/2]}\,\frac{n!\,(-1)^k\,\Gamma(n-k+\lambda)}{2^{2k}\,k!\,(n-2k)!\,\Gamma(n+\lambda)}\,
g^{\mu_1 \mu_2} \dotsc g^{\mu_{2k-1}\mu_{2k}}\,x^{2k}\,x^{\mu_{2k+1}} \dotsc x^{\mu_n}\, ,
\label{app:gegen:gegenbauer:decomp:tensors:TST2uten}
\end{flalign}
\end{subequations}
where $\hat{S}$ is a symmetrizer with respect to the indices $\mu_i$.
In the three-dimensional case these equations reduce to Eqs.~(\ref{app:gegen:legendre:decomp:tensors}). Peculiar cases in arbitrary dimension include:
\begin{subequations}
\begin{flalign}
&x^{\mu_1 \mu_2} = x^{\mu_1} x^{\mu_2} - \frac{x^2}{D}\,g^{\mu_1 \mu_2}\, ,
\\
&x^{\mu_1 \mu_2 \mu_3} = x^{\mu_1} x^{\mu_2} x^{\mu_3} - \frac{x^2}{D+2}\,\left( g^{\mu_1 \mu_2}\,x^{\mu_3} + g^{\mu_2 \mu_3}\,x^{\mu_1} + g^{\mu_3 \mu_1}\,x^{\mu_2} \right)\, ,
\\
&x^{\mu_1 \mu_2 \mu_3 \mu_4} = x^{\mu_1} x^{\mu_2} x^{\mu_3} x^{\mu_4} - \frac{x^2}{D+4}\,\left( g^{\mu_1 \mu_2}\,x^{\mu_3}x^{\mu_4} + g^{\mu_2 \mu_3}\,x^{\mu_4} x^{\mu_1} + \right .
\nonum \\
&\left . + g^{\mu_3 \mu_4}\,x^{\mu_1} x^{\mu_2} + g^{\mu_4 \mu_1}\,x^{\mu_2} x^{\mu_3} + g^{\mu_1 \mu_3}\,x^{\mu_2} x^{\mu_4} + g^{\mu_4 \mu_2}\,x^{\mu_3} x^{\mu_1} \right) +
\nonum \\
&+\frac{x^4}{(D+2)(D+4)}\,\left( g^{\mu_1 \mu_2}\,g^{\mu_3 \mu_4} + g^{\mu_2 \mu_3}\,g^{\mu_4 \mu_1} + g^{\mu_1 \mu_3}\,g^{\mu_2 \mu_4} \right)\, .
\end{flalign}
\end{subequations}

\item \underline{Relation between Gegenbauer polynomials of different indices:}
\be
C_n^\delta(x) =  \sum_{k=0}^{[n/2]}\,C_{n-2k}^\lambda(x)\,\frac{(n-2k+\lambda)\Gamma(\lambda)}{k!\,\Gamma(\delta)}\,
\frac{\Gamma(n+\delta-k)\Gamma(k+\delta-\lambda)}{\Gamma(n-k+\lambda+1) \Gamma(\delta-\lambda)}\, .
\label{app:gegen:gegenbauer:diffind}
\ee
This formula is obtained by combining Eqs.~(\ref{app:gegen:gegenbauer:repres1}) and (\ref{app:gegen:gegenbauer:candec}):
\begin{flalign}
C_n^\delta(x) =  \sum_{p=0}^{[n/2]}\,\sum_{k=p}^{[n/2]}\,C_{n-2k}^\lambda(x)\,\frac{(-1)^p(n-2k+\lambda)\Gamma(\lambda)}{p!\,\Gamma(\delta)}\,
\frac{\Gamma(n+\delta-p)}{(k-p)!\Gamma(n-p-k+\lambda+1)}\, .
\end{flalign}
The two sums can be decoupled by noticing that in the sum over $k$ all terms corresponding to $0\leq k \leq p$ vanish due to the presence of $(k-p)!$
in the denominator. 
%Similarly, due to the same factorial, the sum over $p$ can be extended to infinity without affecting the final result. 
Hence:
\begin{flalign}
C_n^\delta(x) =  \sum_{k=0}^{[n/2]}\,C_{n-2k}^\lambda(x)\,\frac{(n-2k+\lambda)\Gamma(\lambda)}{\Gamma(\delta)}\,\sum_{p=0}^{\infty}\,
\frac{(-1)^p\,\Gamma(n+\delta-p)}{p!\,\Gamma(k+1-p)\,\Gamma(n-p-k+\lambda+1)}\, .
\end{flalign}
The last sum has been extended to infinity because it corresponds to a Gaussian hypergeometric function of argument unity with a negative upper index: $-k$ (it
therefore reduces to a polynomial of degree $k$ all higher order terms being zero). To see this, we use the reflection formula which yields:
\bea
&&\sum_{p=0}^{\infty}\, \frac{(-1)^p\,\Gamma(n+\delta-p)}{p!\,\Gamma(k+1-p)\,\Gamma(n-p-k+\lambda+1)} 
\nonum \\
&&\qquad = \frac{\Gamma(n+\delta)}{\Gamma(1+k) \Gamma(n-k+\lambda+1)}\,{}_2F_1(-k,k-n-\lambda;1-\delta-n;1)\, .
\eea
The hypergeometric function can then be evaluated using the Gauss summation formula. Hence, the two-fold series is brought to a one-fold series and the advertised
result, Eq.~(\ref{app:gegen:gegenbauer:diffind}), is obtained.

\item \underline{Addition theorem:}
\begin{flalign}
& C_n^\lambda(\cos \phi_1\, \cos \theta_1 + \sin \phi_1\,\sin \theta_1\,\cos \theta_2) = \frac{\Gamma(2\lambda-1)}{\Gamma^2(\lambda)}\,
 \sum_{m=0}^n\,4^m\,\frac{\Gamma(n-m+1) \Gamma^2(m+\lambda)}{\Gamma(n+m+2\lambda)}
\nonum \\
& \times (2m+2\lambda-1)(\sin \phi_1\,\sin \theta_1)^m\,C_{n-m}^{\lambda+m}(\cos \phi_1)\,C_{n-m}^{\lambda+m}(\cos \theta_1)\,C_m^{\lambda-1/2}(\cos \theta_2) \, .
\label{app:gegen:gegenbauer:additionth}
\end{flalign}
The proof of this formula can be found in \cite{Carslon:1971}.

\item \underline{Orthogonality relation on the unit sphere in $\R^D$:}

\be
\frac{1}{\Omega_D}\, \int\,\D_D\, \hat{x}\,C_n^\lambda(\hat{z} \cdot \hat{x})C_m^\lambda(\hat{x} \cdot \hat{z})=\delta_{n,m}\,\frac{\lambda\,\Gamma(n+2\lambda)}{\Gamma(2\lambda)\,(n+\lambda)\,n!}, 
\qquad \lambda = \frac{D}{2} -1\, .
\label{app:gegen:gegenbauer:orth3}
\ee
where $\D_D\,\hat{x}$ is the surface element of the unit $D$-dimensional sphere~\footnote{Let's recall that the hyperspherical volume element is given by:
\bea
\D\, V_D & = & x^{D-1} \D\, x\,(\sin \theta_1)^{D-2}\,(\sin \theta_2)^{D-3}\,\cdots\,\sin \theta_{D-2}\,\D\, \theta_1\,\D\, \theta_2\,\cdots \, \D \,\theta_{D-2}\, \D \,\theta_{D-1},
\nonum \\
& = & x^{D-1} \D\, x\,\D_D\, \hat{x}, \qquad \theta_1,\,\theta_2,\,\cdots,\,\theta_{D-2} \in [0,\pi], \quad \theta_{D-1} \in [0,2\pi]\, .
\label{app:gegen:hyperspher:volume}
\eea
}
\begin{subequations}
\label{app:gegen:hyperspher:surface}
\bea
\int\,\D_D\, \hat{x} &=& \int\,\D_{D-1}\, \hat{x}\, \int_0^\pi\,\D \,\theta_1\,(\sin \theta_1)^{D-2}\, , 
\label{app:gegen:hyperspher:surface-a} \\
&=& \int\,\D_{D-2}\, \hat{x}\, \int_0^\pi\,\D \,\theta_1\,(\sin \theta_1)^{D-2}\,\int_0^\pi\,\D \,\theta_2\,(\sin \theta_2)^{D-3}\, , 
\label{app:gegen:hyperspher:surface-b} \\
\Omega_D &=& \int\,\D_D\, \hat{x} = \frac{2 \pi^{D/2}}{\Gamma(D/2)}\, ,
\label{app:gegen:hyperspher:surface-c}
\eea
\end{subequations}
and $\theta_1$ is the angle between the unit vectors $\hat{x}$ and $\hat{z}$.

Eq.~(\ref{app:gegen:gegenbauer:orth3}) can be derived by first re-writing the orthogonality relation Eq.~(\ref{app:gegen:gegenbauer:orth}) as:
\be
\int_{0}^{\pi}\,C_n^\lambda(\cos \theta_1)C_m^\lambda(\cos \theta_1)(\sin \theta_1)^{2\lambda}\,\D \theta_1
=\delta_{n,m}\,\frac{\pi \,2^{1-2\lambda}}{(n+\lambda)\,n!}\,\frac{\Gamma(n+2\lambda)}{\Gamma^2(\lambda)} \, .
\label{app:gegen:gegenbauer:orth2}
\ee
Both sides of this equation are then integrated over $\D_{D-1}\, \hat{x}$. It turns out that, for $\lambda=D/2-1$, the term 
$\D_{D-1} \hat{x}\, \D \theta_1 (\sin \theta_1)^{2\lambda} = \D_{D-1} \,\hat{x}\, \D \theta_1 \,(\sin \theta_1)^{D-2}$
corresponds to the surface element of the unit $D$-dimensional sphere, see Eq.~(\ref{app:gegen:hyperspher:surface-a}).
Choosing $\theta_1$ as the angle between the unit vectors $\hat{x}$ and $\hat{z}$ and after simple calculations, Eq.~(\ref{app:gegen:gegenbauer:orth3}) is derived.

\item \underline{Convolution formula:}
\be
\frac{1}{\Omega_D}\, \int\,\D_D\,\hat{x}\,C_n^\lambda(\hat{x}_1 \cdot \hat{x})C_m^\lambda(\hat{x} \cdot \hat{x}_2)=\delta_{n,m}\,\frac{\lambda}{n+\lambda}\,C_n^\lambda(\hat{x}_1 \cdot \hat{x}_2).
\label{app:gegen:gegenbauer:conv}
\ee
Following Ref.~\cite{Celmaster:1980ji} Eq.~(\ref{app:gegen:gegenbauer:conv}) can be derived by choosing a coordinate system such that $\hat{x}_1=(1,0_{D-1})$, $\hat{x}_2=(\cos \phi_1,\sin \phi_1, 0_{D-2})$,
$\hat{x} = (\cos \theta_1, \sin \theta_1 \cos \theta_2,\hat{k}_{D-2})$ and using Eq.~(\ref{app:gegen:hyperspher:surface-b}). Then, the integral on the left hand side of Eq.~(\ref{app:gegen:gegenbauer:conv})
can be written as:
\begin{flalign}
& \int\,\D_D\,\hat{x}\,C_n^\lambda(\hat{x}_1 \cdot \hat{x})C_m^\lambda(\hat{x} \cdot \hat{x}_2) 
\nonum \\
&=
\int\,\D_{D-2}\,\hat{x}\,\int_0^\pi\,\D \,\theta_1\,(\sin \theta_1)^{D-2}\,\int_0^\pi\,\D \,\theta_2\,(\sin \theta_2)^{D-3}\,
C_n^\lambda(\cos \theta_1)C_m^\lambda(\cos \theta_1\cos \phi_1+\sin \theta_1 \cos \theta_2\sin \phi_1)\, .
\nonum \\
&\stackrel{\text{Eq.~(\ref{app:gegen:gegenbauer:additionth})}}{=}
\frac{\Gamma(2\lambda-1)}{\Gamma^2(\lambda)}\,
 \sum_{k=0}^m\,4^k\,\frac{\Gamma(m-k+1)\Gamma^2(k+\lambda)}{\Gamma(m+k+2\lambda)}
\, (2k+2\lambda-1)\, C_{m-k}^{\lambda+k}(\cos \phi_1)
\nonum \\
&  \int\,\D_{D-2}\,\hat{x}\,\int_0^\pi\,\D \,\theta_1\,(\sin \theta_1)^{k+D-2}\,C_n^\lambda(\cos \theta_1)\,C_{m-k}^{\lambda+k}(\cos \theta_1)\,
\int_0^\pi\,\D \,\theta_2\,(\sin \theta_2)^{k+D-3}\,C_k^{\lambda-1/2}(\cos \theta_2) \, .
\end{flalign}
The last integral can be evaluated using the orthogonality relation Eq.~(\ref{app:gegen:gegenbauer:orth2}) with $\lambda \ra \lambda -1/2$ and using the fact that $C_0^{\lambda-1/2}(\cos \theta_2)=1$.
This yields, for the last integral:
\be
\int_0^\pi\,\D \,\theta_2\,(\sin \theta_2)^{k+D-3}\,C_k^{\lambda-1/2}(\cos \theta_2)C_0^{\lambda-1/2}(\cos \theta_2) = 
\delta_{k,0}\,\frac{\pi \,2^{2-2\lambda}}{\lambda-1/2}\,\frac{\Gamma(2\lambda-1)}{\Gamma^2(\lambda-1/2)}\, .
\ee
With $k=0$ the integral over $\theta_1$ can also be evaluated straightforwardly by using Eq.~(\ref{app:gegen:gegenbauer:orth2}). Moreover, $\cos \phi_1=\hat{x}_1 \cdot \hat{x}_2$.
All together, and after some simple calculations, the left hand side of Eq.~(\ref{app:gegen:gegenbauer:conv}) simplifies as:
\be
\frac{1}{\Omega_D}\, \int\,\D_D\,\hat{x}\,C_n^\lambda(\hat{x}_1 \cdot \hat{x})C_m^\lambda(\hat{x} \cdot \hat{x}_2)=
\delta_{n,m}\,\frac{\lambda}{n+\lambda}\,\left( \frac{\sqrt{\pi} 2^{1-2\lambda} \Gamma(2\lambda)}{\Gamma(\lambda) \Gamma(\lambda+1/2)} \right)^2\,C_n^\lambda(\hat{x}_1 \cdot \hat{x}_2)\, .
\ee
Using the duplication formula: $\Gamma(\lambda) \Gamma(\lambda+1/2) = \sqrt{\pi} 2^{1-2\lambda} \Gamma(2\lambda)$ the squared term reduces to $1$ which proves Eq.~(\ref{app:gegen:gegenbauer:conv}).

\item \underline{Propagator expansion:} the above formulas allow us to generalize the propagator expansions of Eqs.~(\ref{app:gegen:legendre:multipole}) and (\ref{app:gegen:chebychev:multipole}) 
to arbitrary dimensions
\begin{flalign}
\frac{1}{(x_1 - x_2)^{2\lambda}}  = 
\sum_{n=0}^\infty\,C_n^\lambda(\hat{x}_1 \cdot \hat{x}_2)\,\Bigg[ \frac{(x_1^2)^{n/2}}{(x_2^2)^{n/2+\lambda}}\,\Theta(x_2^2-x_1^2) + (x_1^2 \longleftrightarrow x_2^2) \Bigg],
\quad \hat{x} = \frac{x}{\sqrt{x^2}}\, .
\label{app:gegen:gegenbauer:multipole}
\end{flalign}
where $\lambda = D/2-1$ is the ordinary index of a $D$-dimensional propagator in $x$-space. For a propagator with arbitrary index:
\bea
\frac{1}{(x_1 - x_2)^{2\beta}}  = 
\sum_{n=0}^\infty\,C_n^\beta(\hat{x}_1 \cdot \hat{x}_2)\,\Bigg[ \frac{(x_1^2)^{n/2}}{(x_2^2)^{n/2+\beta}}\,\Theta(x_2^2-x_1^2) + (x_1^2 \longleftrightarrow x_2^2) \Bigg]\, ,
\label{app:gegen:gegenbauer:multipole2}
\eea
where $C_n^\beta(x)$ can then be related to $C_{n-2k}^\lambda(x)$ ($0 \leq k \leq [n/2]$) with the help of Eq.~(\ref{app:gegen:gegenbauer:diffind}).

The expansion can also be written in terms of traceless products by 
using Eq.~(\ref{app:gegen:gegenbauer:TSTGegen}) which relates Gegenbauer polynomials of index $\lambda=D/2-1$ to traceless products. 
The expansion has the form:
\be
\frac{1}{(x_1 - x_2)^{2\lambda}}  = 
\sum_{n=0}^\infty\,\frac{2^n \Gamma(n+\lambda)}{n!\,\Gamma(\lambda)}\,x_1^{\mu_1 \cdots \mu_n}\,x_2^{\mu_1 \cdots \mu_n}\,
\Bigg[ \frac{1}{(x_2^2)^{n+\lambda}}\,\Theta(x_2^2-x_1^2) + (x_1^2 \longleftrightarrow x_2^2) \Bigg]\, .
\label{app:gegen:gegenbauer:multipole3}
\ee
For a propagator of arbitrary index one has first to use Eq.~(\ref{app:gegen:gegenbauer:diffind}) and then Eq.~(\ref{app:gegen:gegenbauer:TSTGegen}).
Expansions in terms of traceless products are the most convenient ones in computing dimensionally regularized propagator-type massless 2-loop Feynman diagrams.

\end{itemize}

%% file: hdr.bbl
\providecommand{\href}[2]{#2}\begingroup\raggedright\begin{thebibliography}{100}

\bibitem{Anderson1984basic}
P.~W.~Anderson, ``Basic Notions Of Condensed Matter Physics,''
\newblock The Benjamin/Cummings Publishing Company, 1984.

\bibitem{tHooft:1995wad}
  G.~'t Hooft, ``Under the spell of the gauge principle,''
  Adv.\ Ser.\ Math.\ Phys.\  {\bf 19} (1994) 1.

\bibitem{Berestetskii82}
V.\ B.~Berestetskii, E.\ M.~Lifshitz and L.\ P.~Pitaevski{\u\i}, ``Quantum Electrodynamics,''
\newblock Course of theoretical physics, Volume 4, Lev D.~Landau, Butterworth-Heinemann, 1982.

\bibitem{itzykson2012quantum}
C.~Itzykson and J.\ B.~Zuber, ``Quantum Field Theory,''
\newblock Dover Books on Physics, Dover Publications, 2012.

\bibitem{peskin1995introduction}
M.\ E.~Peskin and D.\ V.~Schroeder, ``An Introduction to Quantum Field Theory,''
\newblock Advanced book classics, Addison-Wesley Publishing Company, 1995.

\bibitem{Weinberg:1995mt}
S.~Weinberg, ``The Quantum theory of fields. Vol. 1: Foundations,''
\newblock Cambridge University Press, 2005.
%%CITATION = INSPIRE-406190;%%

\bibitem{Tomonaga:1946zz}
S.-I.~Tomonaga, ``On a relativistically invariant formulation of the quantum theory of wave fields,''
\href{http://dx.doi.org/10.1143/PTP.1.27}{Prog.\ Theor.\ Phys.\  {\bf 1} (1946) 27};
Z.~Koba, T.~Tati, and S.-I.~Tomonaga, ``On a Relativistically Invariant Formulation of the Quantum Theory of Wave Fields.
II: Case of Interacting Electromagnetic and Electron Fields''
\href{http://dx.doi.org/10.1143/ptp/2.3.101}{Prog. Theor. Phys. (1947) 2};
 %\bibitem{Tomonaga:1948zz}
S.~I.~Tomonaga and J.~R.~Oppenheimer, ``On Infinite Field Reactions in Quantum Field Theory,''
\href{http://dx.doi.org/10.1103/PhysRev.74.224}{Phys.\ Rev.\  {\bf 74} (1948) 224}.

\bibitem{Schwinger:1948yk+Schwinger:1948yj+Schwinger:1949ra}
J.~S.~Schwinger, ``Quantum electrodynamics. I: A covariant formulation,''
\href{http://dx.doi.org/10.1103/PhysRev.74.1439}{Phys.\ Rev.\  {\bf 74} (1948) 1439};
 %\bibitem{Schwinger:1948yj}
%  J.~S.~Schwinger,
``Quantum electrodynamics. II: Vacuum polarization and selfenergy,''
\href{http://dx.doi.org/10.1103/PhysRev.75.651}{Phys.\ Rev.\  {\bf 75} (1948) 651};
 %\bibitem{Schwinger:1949ra}
%  J.~S.~Schwinger,
``Quantum electrodynamics. III: The electromagnetic properties of the electron: Radiative corrections to scattering,''
\href{http://dx.doi.org/10.1103/PhysRev.76.790}{Phys.\ Rev.\  {\bf 76} (1949) 790}.
%  doi:10.1103/PhysRev.76.790

\bibitem{Feynman:1948fi+Feynman:1949zx+Feynman:1950ir}
R.~P.~Feynman, ``Relativistic cutoff for quantum electrodynamics,''
\href{http://dx.doi.org/10.1103/PhysRev.74.1430}{Phys.\ Rev.\  {\bf 74} (1948) 1430};
 %\bibitem{Feynman:1949zx}
%  R.~P.~Feynman,
``Space-time approach to quantum electrodynamics,''
\href{http://dx.doi.org/10.1103/PhysRev.76.769}{Phys.\ Rev.\  {\bf 76} (1949) 769};
 %%CITATION = doi:10.1103/PhysRev.76.769;%%
%\bibitem{Feynman:1950ir}
%  R.~P.~Feynman,
``Mathematical formulation of the quantum theory of electromagnetic interaction,''
\href{http://dx.doi.org/10.1103/PhysRev.80.440}{Phys.\ Rev.\  {\bf 80} (1950) 440}.

\bibitem{PhysRev.75.486}
F.\ J.~Dyson, ``The Radiation Theories of Tomonaga, Schwinger, and Feynman,''
\href{http://link.aps.org/doi/10.1103/PhysRev.75.486}{Phys.\ Rev.\ {\bf 75} (1949) 486}.

\bibitem{Stueckelberg:1953dz}
E.~C.~G.~Stueckelberg and A.~Petermann, ``La normalisation des constantes dans la th\'eorie des quanta. Normalization of constants in the quanta theory,''
\href{http://dx.doi.org/10.5169/seals-112426}{Helv.\ Phys.\ Acta {\bf 26} (1953) 499}.
 %%CITATION = doi:10.5169/seals-112426;%%

\bibitem{GellMann:1954fq}
M.~Gell-Mann and F.~E.~Low, ``Quantum electrodynamics at small distances,''
\href{http://dx.doi.org/10.1103/PhysRev.95.1300}{Phys.\ Rev.\  {\bf 95} (1954) 1300}.
 %%CITATION = doi:10.1103/PhysRev.95.1300;%%

\bibitem{Bogolyubov:1956gh}
  N.~N.~Bogolyubov and D.~V.~Shirkov, ``Charge renormalization group in quantum field theory,''
\href{http://dx.doi.org/10.1007/BF02823486}{Nuovo Cim.\  {\bf 3} (1956) 845.}
  %%CITATION = doi:10.1007/BF02823486;%%

\bibitem{Bogoliubov:1957gp}
  N.~N.~Bogoliubov and O.~S.~Parasiuk, ``On the Multiplication of the causal function in the quantum theory of fields,''
\href{http://dx.doi.org/10.1007/BF02392399}{Acta Math.\  {\bf 97} (1957) 227}.
  %%CITATION = doi:10.1007/BF02392399;%%

\bibitem{Hepp:1966eg}
  K.~Hepp, ``Proof of the Bogolyubov-Parasiuk theorem on renormalization,''
\href{http://dx.doi.org/10.1007/BF01773358}{Commun.\ Math.\ Phys.\  {\bf 2} (1966) 301}.
 % doi:10.1007/BF01773358
  %%CITATION = doi:10.1007/BF01773358;%%

\bibitem{Zimmermann1969}
W.~Zimmermann, `Convergence of Bogoliubov's method of renormalization in momentum space,''
\href{http://dx.doi.org/10.1007/BF01645676}{Communications in Mathematical Physics {\bf 15} (1969) 208};
W.~Zimmermann, in ``Lectures on elementary particle and quantum field theory'', 1970. Brandies University Summer Institute in Theoretical Physics (MIT Press Cambridge, Massachusetts). 

\bibitem{Bogolyubov:1980nc}
N.~N.~Bogolyubov and D.~V.~Shirkov, ``Introduction To The Theory Of Quantized Fields,''
\newblock  Intersci.\ Monogr.\ Phys.\ Astron.\  {\bf 3} (1959) 1.
%%CITATION = IMTPA,3,1;%%

%\bibitem{Bogolyubov:1983gp}
%N.~N.~Bogolyubov and D.~V.~Shirkov, ``Quantum Fields,''
%\newblock Reading, Usa: Benjamin/Cummings, 1983.

\bibitem{Yang:1954ek}
C.~N.~Yang and R.~L.~Mills, ``Conservation of Isotopic Spin and Isotopic Gauge Invariance,''
\href{http://dx.doi.org/10.1103/PhysRev.96.191}{Phys.\ Rev.\ {\bf 96} (1954) 191}.
%%CITATION = doi:10.1103/PhysRev.96.191;%%

\bibitem{Glashow61+Salam:1964ry+Weinberg:1967tq}
S.~Glashow, ``{Partial Symmetries of Weak Interactions},''
\href{http://dx.doi.org/10.1016/0029-5582(61)90469-2}{Nucl.\ Phys.\ {\bf 22}  (1961)  579};
 %\bibitem{Salam:1964ry}
A.~Salam and J.~C. Ward, ``{Electromagnetic and Weak Interactions},''
\href{http://dx.doi.org/10.1016/0031-9163(64)90711-5}{Phys.\ Lett.\ {\bf 13}  (1964)  168};
 %%CITATION = PHLTA,13,168;%%.
%
%\bibitem{Weinberg:1967tq}
S.~Weinberg, ``{A Model of Leptons},''
\href{http://dx.doi.org/10.1103/PhysRevLett.19.1264}{Phys.\ Rev.\ Lett.\ {\bf 19} (1967)  1264}.
%%CITATION = PRLTA,19,1264;%%.

\bibitem{'tHooft:1971rn}
G.~'t Hooft, ``Renormalizable Lagrangians for Massive Yang-Mills Fields,''
\href{http://dx.doi.org/10.1016/0550-3213(71)90139-8}{Nucl.\ Phys.\ B {\bf 35} (1971) 167}.

\bibitem{'tHooft:1972fi}
G.~'t Hooft and M.~J.~G.~Veltman, ``Regularization and Renormalization of Gauge Fields,''
\href{http://dx.doi.org/10.1016/0550-3213(72)90279-9}{Nucl.\ Phys.\ B {\bf 44} (1972) 189}.

\bibitem{Bollini:1972ui}
C.~G.~Bollini and J.~J.~Giambiagi, ``Dimensional Renormalization: The Number of Dimensions as a Regularizing Parameter,''
\href{http://dx.doi.org/10.1007/BF02895558}{Nuovo Cim.\ B {\bf 12} (1972) 20}.
%%CITATION = doi:10.1007/BF02895558;%%

\bibitem{Cicuta:1972jf}
G.~M.~Cicuta and E.~Montaldi, ``Analytic renormalization via continuous space dimension,''
\href{http://dx.doi.org/10.1007/BF02756527}{Lett.\ Nuovo Cim.\  {\bf 4} (1972) 329}.
  %%CITATION = doi:10.1007/BF02756527;%%

\bibitem{Ashmore:1972uj}
J.~F.~Ashmore, ``A Method of Gauge Invariant Regularization,''
\href{http://dx.doi.org/10.1007/BF02824407}{Lett.\ Nuovo Cim.\  {\bf 4} (1972) 289}.
  %%CITATION = doi:10.1007/BF02824407;%%

\bibitem{'tHooft:1973mm}
G.~'t Hooft, ``Dimensional regularization and the renormalization group,''
\href{http://dx.doi.org/10.1016/0550-3213(73)90376-3}{Nucl.\ Phys.\ B {\bf 61} (1973) 455}.
  %%CITATION = doi:10.1016/0550-3213(73)90376-3;%%

\bibitem{Politzer:1973fx}
H.~D.~Politzer, ``Reliable Perturbative Results for Strong Interactions?,''
\href{http://dx.doi.org/10.1103/PhysRevLett.30.1346}{Phys.\ Rev.\ Lett.\  {\bf 30} (1973) 1346}.
%%CITATION = doi:10.1103/PhysRevLett.30.1346;%%

\bibitem{Gross:1973id}
D.~J.~Gross and F.~Wilczek, ``Ultraviolet Behavior of Nonabelian Gauge Theories,''
\href{http://dx.doi.org/10.1103/PhysRevLett.30.1343}{Phys.\ Rev.\ Lett.\  {\bf 30} (1973) 1343}.
%%CITATION = doi:10.1103/PhysRevLett.30.1343;%%

\bibitem{Dimopoulos:1981zb}
  S.~Dimopoulos and H.~Georgi, ``Softly Broken Supersymmetry and SU(5),''
  \href{http://dx.doi.org/10.1016/0550-3213(81)90522-8}{Nucl.\ Phys.\ B {\bf 193} (1981) 150}.

\bibitem{'tHooft:1973jz}
G.~'t Hooft, ``A Planar Diagram Theory for Strong Interactions,''
\href{http://dx.doi.org/0.1016/0550-3213(74)90154-0}{Nucl.\ Phys.\ B {\bf 72} (1974) 461}.
%%CITATION = doi:10.1016/0550-3213(74)90154-0;%%

\bibitem{Brezin:1994eb}
E.~Br\'ezin and S.~R.~Wadia, ``The Large N expansion in quantum field theory and statistical physics: From spin systems to two-dimensional gravity,''
\newblock  Singapore, Singapore: World Scientific (1993).
  %2 citations counted in INSPIRE as of 15 Jul 2016

\bibitem{Maldacena:1997re+Gubser:1998bc+Witten:1998qj}
%\bibitem{Maldacena:1997re}
J.~M.~Maldacena, ``The Large N limit of superconformal field theories and supergravity,''
\href{http://dx.doi.org/10.1023/A:1026654312961}{Int.\ J.\ Theor.\ Phys.\  {\bf 38} (1999) 1113  [Adv.\ Theor.\ Math.\ Phys.\  {\bf 2} (1998) 231]} [\href{https://arxiv.org/abs/hep-th/9711200}{arXiv:hep-th/9711200}];
 %%CITATION = doi:10.1023/A:1026654312961;%%
S.~S.~Gubser, I.~R.~Klebanov and A.~M.~Polyakov, ``Gauge theory correlators from noncritical string theory,''
\href{http://dx.doi.org/10.1016/S0370-2693(98)00377-3}{Phys.\ Lett.\ B {\bf 428} (1998) 105} [\href{https://arxiv.org/abs/hep-th/9802109}{arXiv:hep-th/9802109}];
 %%CITATION = doi:10.1016/S0370-2693(98)00377-3;%%
E.~Witten, ``Anti-de Sitter space and holography,''
\newblock  Adv.\ Theor.\ Math.\ Phys.\  {\bf 2} (1998) 253 [\href{https://arxiv.org/abs/hep-th/9802150}{arXiv:hep-th/9802150}].
%%CITATION = HEP-TH/9802150;%%

\bibitem{Nicolai:2007zza}
  H.~Nicolai, ``String theory: Back to basics,''
\href{http://dx.doi.org/10.1038/449797a}{Nature {\bf 449} (2007) 797}.
  %%CITATION = doi:10.1038/449797a;%%

%\bibitem{Aharony:1999ti}
%O.~Aharony, S.~S.~Gubser, J.~M.~Maldacena, H.~Ooguri and Y.~Oz, ``Large N field theories, string theory and gravity,''
%\href{http://dx.doi.org/10.1016/S0370-1573(99)00083-6}{Phys.\ Rept.\  {\bf 323} (2000) 183}.
%%%CITATION = doi:10.1016/S0370-1573(99)00083-6;%%

%\bibitem{Anderson:1975nw}
%P.~W.~Anderson, ``Uses of Solid State Analogies in Elementary Particle Theory,'' 
%\newblock Conf.\ Proc.\ C {\bf 750926} (1975) 311.
%%CITATION = CONFP,C750926,311;%%

\bibitem{Wilson:1971bg}
K.~G.~Wilson, ``Renormalization group and critical phenomena. 1. Renormalization group and the Kadanoff scaling picture,''
\href{http://dx.doi.org/10.1103/PhysRevB.4.3174}{Phys.\ Rev.\ B {\bf 4} (1971) 3174}.
%%CITATION = doi:10.1103/PhysRevB.4.3174;%%

\bibitem{Wilson:1971dh}
K.~G.~Wilson, ``Renormalization group and critical phenomena. 2. Phase space cell analysis of critical behavior,''
\href{http://dx.doi.org/10.1103/PhysRevB.4.3184}{Phys.\ Rev.\ B {\bf 4} (1971) 3184}.
%%CITATION = doi:10.1103/PhysRevB.4.3184;%%

\bibitem{Wilson:1971dc}
K.~G.~Wilson and M.~E.~Fisher, ``Critical exponents in 3.99 dimensions,''
\href{http://dx.doi.org/10.1103/PhysRevLett.28.240}{Phys.\ Rev.\ Lett.\  {\bf 28} (1972) 240}.
%%CITATION = doi:10.1103/PhysRevLett.28.240;%%

\bibitem{Brezin:1974eb}
E.~Br\'ezin, J.~C.~Le Guillou and J.~Zinn-Justin, ``Wilson's theory of critical phenomena and Callan-Symanzik equations in 4-epsilon dimensions,''
\href{http://dx.doi.org/10.1103/PhysRevD.8.434}{Phys.\ Rev.\ D {\bf 8} (1973) 434};
``Addendum to Wilson's theory of critical phenomena and Callan-Symanzik equations in 4-epsilon dimensions,''
\href{http://dx.doi.org/10.1103/PhysRevD.9.1121}{Phys.\ Rev.\ D {\bf 9} (1974) 1121}.
  %%CITATION = doi:10.1103/PhysRevD.9.1121, 10.1103/PhysRevD.10.2046;%%

\bibitem{zinn2002quantum}
J.~Zinn-Justin, ``Quantum Field Theory and Critical Phenomena,''
\newblock Clarendon Press, 2002.

%\bibitem{Wilson:1973jj}
%K.~G.~Wilson and J.~B.~Kogut, ``The Renormalization group and the epsilon expansion,''
%\href{http://dx.doi.org/10.1016/0370-1573(74)90023-4}{Phys.\ Rept.\  {\bf 12} (1974) 75}.
%%CITATION = doi:10.1016/0370-1573(74)90023-4;%%

\bibitem{Stanley:1968gx}
H.~E.~Stanley, ``Spherical model as the limit of infinite spin dimensionality,''
\href{http://dx.doi.org/0.1103/PhysRev.176.718}{Phys.\ Rev.\  {\bf 176} (1968) 718}.
%%CITATION = doi:10.1103/PhysRev.176.718;%%

\bibitem{Vladimirov:1979zm}
A.~A.~Vladimirov, ``Method for Computing Renormalization Group Functions in Dimensional Renormalization Scheme,''
\href{http://dx.doi.org/10.1007/BF01018394}{Theor.\ Math.\ Phys.\  {\bf 43} (1980) 417 [Teor.\ Mat.\ Fiz.\  {\bf 43} (1980) 210]}.
  %%CITATION = doi:10.1007/BF01018394;%%

\bibitem{Chetyrkin:1982nn}
K.~G.~Chetyrkin and F.~V.~Tkachov, ``Infrared R Operation And Ultraviolet Counterterms In The Ms Scheme,''
\href{http://dx.doi.org/10.1016/0370-2693(82)90358-6}{Phys.\ Lett.\ B {\bf 114} (1982) 340}.
  %%CITATION = doi:10.1016/0370-2693(82)90358-6;%%

\bibitem{Chetyrkin:1983wh}
  K.~G.~Chetyrkin and V.~A.~Smirnov, ``Dimensional Regularization And Infrared Divergences,''
\href{https://link.springer.com/article/10.1007/BF01016818}{Theor.\ Math.\ Phys.\  {\bf 56} (1984) 770}
   [Teor.\ Mat.\ Fiz.\  {\bf 56} (1983) 206].

\bibitem{Chetyrkin:1984xa}
K.~G.~Chetyrkin and V.~A.~Smirnov, ``R* Operation Corrected,''
\href{http://dx.doi.org/10.1016/0370-2693(84)91291-7}{Phys.\ Lett.\ B {\bf 144} (1984) 419}.
  %%CITATION = doi:10.1016/0370-2693(84)91291-7;%%

\bibitem{Smirnov:1986me}
V.~A.~Smirnov and K.~G.~Chetyrkin, ``R* Operation in the Minimal Subtraction Scheme,''
\href{http://dx.doi.org/10.1007/BF01017902}{Theor.\ Math.\ Phys.\  {\bf 63} (1985) 462  [Teor.\ Mat.\ Fiz.\  {\bf 63} (1985) 208]}.
  %%CITATION = doi:10.1007/BF01017902;%%

\bibitem{Chetyrkin:2017ppe}
  K.~G.~Chetyrkin, ``Combinatorics of $\mathbf{R}$-, $\mathbf{R^{-1}}$-, and $\mathbf{R^*}$-operations and asymptotic expansions of feynman integrals in the limit of large momenta and masses,''
\href{https://arxiv.org/abs/1701.08627}{arXiv:1701.08627 [hep-th]}.

\bibitem{D'Eramo:1971zz}
M.~D'Eramo, G.~Parisi and L.~Peliti, ``Theoretical Predictions For Critical Exponents At The Lambda Point Of Bose Liquids,''
\href{http://dx.doi.org/0.1007/BF02774121}{Lett.\ Nuovo Cim.\  {\bf 2} (1971)  878}.

\bibitem{Vasiliev:1981dg}
A.~N.~Vasiliev, Y.~M.~Pismak and J.~R.~Honkonen, ``1/$N$ Expansion: Calculation of the Exponents $\eta$ and Nu in the Order 1/$N^2$ for Arbitrary Number of Dimensions,''
\href{http://dx.doi.org/10.1007/BF01019296}{Theor.\ Math.\ Phys.\  {\bf 47} (1981) 465 [Teor.\ Mat.\ Fiz.\  {\bf 47} (1981) 291]}.

\bibitem{Usyukina:1983gj}
N.~I.~Usyukina, ``Calculation Of Many Loop Diagrams Of Perturbation Theory,''
\href{http://dx.doi.org/10.1007/BF01017127}{Theor.\ Math.\ Phys.\  {\bf 54} (1983) 78 [Teor.\ Mat.\ Fiz.\  {\bf 54} (1983) 124]}.
%%CITATION = doi:10.1007/BF01017127;%%

% method of uniqueness + expansion O(veps^3)
\bibitem{Kazakov:1983ns}
D.~I.~Kazakov, ``Calculation Of Feynman Integrals By The Method Of 'uniqueness',''
\href{http://dx.doi.org/10.1007/BF01018044}{ Theor.\ Math.\ Phys.\  {\bf 58} (1984) 223 [Teor.\ Mat.\ Fiz.\  {\bf 58} (1984) 343]} [\href{https://inspirehep.net/record/197610?ln=en}{inspires}].
%%CITATION = doi:10.1007/BF01018044;%%

\bibitem{Kazakov:1986mu}
  D.~I.~Kazakov and A.~V.~Kotikov, ``The Method of Uniqueness: Multiloop Calculations in {QCD},''
\href{http://dx.doi.org/10.1007/BF01041909}{Theor.\ Math.\ Phys.\  {\bf 73} (1988) 1264    [Teor.\ Mat.\ Fiz.\  {\bf 73} (1987) 348]}.

%conformal bootstrap
\bibitem{Vasiliev:1982dc}
A.~N.~Vasiliev, Y.~M.~Pismak and J.~R.~Honkonen, ``1/n Expansion: Calculation Of The Exponent Eta In The Order 1/n**3 By The Conformal Bootstrap Method,''
\href{http://dx.doi.org/10.1007/BF01015292}{Theor.\ Math.\ Phys.\  {\bf 50} (1982) 127 [Teor.\ Mat.\ Fiz.\  {\bf 50} (1982) 195]}.
  %%CITATION = doi:10.1007/BF01015292;%%

\bibitem{Tkachov:1981wb}
F.~V.~Tkachov, ``A Theorem on Analytical Calculability of Four Loop Renormalization Group Functions,''
\href{http://dx.doi.org/10.1016/0370-2693(81)90288-4}{Phys.\ Lett.\ B {\bf 100} (1981) 65}.
%%CITATION = doi:10.1016/0370-2693(81)90288-4;%%

\bibitem{Chetyrkin:1981qh}
K.~G.~Chetyrkin and F.~V.~Tkachov, ``Integration by Parts: The Algorithm to Calculate beta Functions in 4 Loops,''
 %\href{http://dx.doi.org/10.1016/0550-3213(81)90199-1}{Nucl.\ Phys.\ B {\bf 192} (1981) 159}.
\href{http://inspirehep.net/record/171845}{Nucl.\ Phys.\ B {\bf 192} (1981) 159}.
%%CITATION = doi:10.1016/0550-3213(81)90199-1;%%

\bibitem{Chetyrkin:1980pr}
K.~G.~Chetyrkin, A.~L.~Kataev and F.~V.~Tkachov, ``New Approach to Evaluation of Multiloop Feynman Integrals: The Gegenbauer Polynomial x Space Technique,''
 %\href{http://dx.doi.org/10.1016/0550-3213(80)90289-8}{Nucl.\ Phys.\ B {\bf 174} (1980) 345}.
\href{http://inspirehep.net/record/159610}{Nucl.\ Phys.\ B {\bf 174} (1980) 345}.
%%CITATION = doi:10.1016/0550-3213(80)90289-8;%%

% expansion O(veps^5)
\bibitem{Broadhurst:1986bx}
D.~J.~Broadhurst, ``Exploiting the 1.440 Fold Symmetry of the Master Two Loop Diagram,''
\href{http://dx.doi.org/10.1007/BF01552503}{Z.\ Phys.\ C {\bf 32} (1986) 249}.
%%CITATION = doi:10.1007/BF01552503;%%

% expansion O(veps^6)
\bibitem{Barfoot:1987kg}
D.~T.~Barfoot and D.~J.~Broadhurst, ``$Z$(2) X S(6) Symmetry of the Two Loop Diagram,''
\href{http://dx.doi.org/10.1007/BF01412581}{Z.\ Phys.\ C {\bf 41} (1988) 81}.
%%CITATION = doi:10.1007/BF01412581;%%

\bibitem{Collins:1984xc}
J.~C.~Collins, ``Renormalization : An Introduction to Renormalization, The Renormalization Group, and the Operator Product Expansion,''
\newblock Cambridge Monographs on Mathematical Physics, 1986.
%%CITATION = INSPIRE-209810;%%

\bibitem{kleinert2001critical}
H.~Kleinert and V.~Schulte-Frohlinde, ``Critical Properties of $\Phi^4$-theories,''
\newblock World Scientific, 2001.

\bibitem{Vasil'evbook}
A.N.Vasil'ev, ``The Field Theoretic Renormalization Group in Critical Behavior theory and Stochastic Dynamics,''
\newblock Chapman and Hall/CRC, London, 2004.

\bibitem{grozin2007lectures}
A.\ G.~Grozin, ``Lectures on QED and QCD: Practical Calculation and Renormalization of One- and Multi-loop Feynman Diagrams,''
\newblock World Scientific, 2007 [\href{https://arxiv.org/abs/hep-ph/0508242}{arXiv:hep-ph/0508242}].

\bibitem{smirnov2013analytic}
V.\ A.~Smirnov, ``Analytic Tools for Feynman Integrals,''
\newblock Springer Tracts in Modern Physics, Springer Berlin Heidelberg, 2013.

\bibitem{Tarasov:1980au}
O.~V.~Tarasov, A.~A.~Vladimirov and A.~Y.~Zharkov, ``The Gell-Mann-Low Function of QCD in the Three Loop Approximation,''
\href{http://dx.doi.org/10.1016/0370-2693(80)90358-5}{Phys.\ Lett.\ B {\bf 93} (1980) 429}.
  %%CITATION = doi:10.1016/0370-2693(80)90358-5;%%

% method of uniqueness + expansion O(veps^4)
\bibitem{Kazakov:1984km}
D.~I.~Kazakov, ``The Method Of Uniqueness, A New Powerful Technique For Multiloop Calculations,''
\href{http://dx.doi.org/10.1016/0370-2693(83)90816-X}{Phys.\ Lett.\ B {\bf 133} (1983) 406}.
  %%CITATION = doi:10.1016/0370-2693(83)90816-X;%%

\bibitem{Gorishnii:1983gp}
  S.~G.~Gorishnii, S.~A.~Larin, F.~V.~Tkachov and K.~G.~Chetyrkin,
  ``Five Loop Renormalization Group Calculations in the $g \phi^4$ in Four-dimensions Theory,''
\href{http://dx.doi.org/10.1016/0370-2693(83)90324-6}{Phys.\ Lett.\  {\bf 132B} (1983) 351}.

% method of uniqueness + expansion O(veps^4)
\bibitem{Kazakov:1983pk}
D.~I.~Kazakov, ``Multiloop Calculations: Method of Uniqueness and Functional Equations,''
\href{http://dx.doi.org/10.1007/BF01034829}{ Theor.\ Math.\ Phys.\  {\bf 62} (1985) 84 [Teor.\ Mat.\ Fiz.\  {\bf 62} (1984) 127]} [\href{https://inspirehep.net/record/196741?ln=en}{inspires}].
%%CITATION = doi:10.1007/BF01034829;%%

% alternative way to compute ChT(alpha,beta) of Vasil'ev et al 1981
\bibitem{Gracey:1992ew}
J.~A.~Gracey, ``On the evaluation of massless Feynman diagrams by the method of uniqueness,''
\href{http://dx.doi.org/10.1016/0370-2693(92)91812-N}{ Phys.\ Lett.\ B {\bf 277} (1992) 469}.
%%CITATION = doi:10.1016/0370-2693(92)91812-N;%%

\bibitem{Kotikov:1995cw}
A.~V.~Kotikov, ``The Gegenbauer polynomial technique: The Evaluation of a class of Feynman diagrams,''
\href{http://dx.doi.org/10.1016/0370-2693(96)00226-2}{Phys.\ Lett.\ B {\bf 375} (1996) 240} [\href{https://arxiv.org/abs/hep-ph/9512270}{arXiv:hep-ph/9512270}].
%%CITATION = doi:10.1016/0370-2693(96)00226-2;%%

% IBP for 3 arbitrary indices + expansion O(veps^7 and veps^8)
\bibitem{Broadhurst:1996ur}
D.~J.~Broadhurst, J.~A.~Gracey and D.~Kreimer, ``Beyond the triangle and uniqueness relations: Nonzeta counterterms at large N from positive knots,''
\href{http://dx.doi.org/10.1007/s002880050500}{Z.\ Phys.\ C {\bf 75} (1997) 559} [\href{https://arxiv.org/abs/hep-th/9607174}{arXiv:hep-th/9607174}].
%%CITATION = doi:10.1007/s002880050500;%%

\bibitem{Panzer:2013cha}
  E.~Panzer, ``On the analytic computation of massless propagators in dimensional
regularization,''
\href{http://dx.doi.org/10.1016/j.nuclphysb.2013.05.025}{Nucl.\ Phys.\ B {\bf 874}, 567 (2013)} [\href{https://arxiv.org/abs/1305.2161}{arXiv:1305.2161 [hep-th]}].

\bibitem{Kotikov:1990kg}
A.~V.~Kotikov, ``Differential equations method: New technique for massive Feynman diagrams calculation,''
\href{http://dx.doi.org/10.1016/0370-2693(91)90413-K}{Phys.\ Lett.\ B {\bf 254} (1991) 158}.
  %%CITATION = doi:10.1016/0370-2693(91)90413-K;%%

\bibitem{Boos:1990rg}
E.~E.~Boos and A.~I.~Davydychev, ``A Method of evaluating massive Feynman integrals,''
\href{http://dx.doi.org/10.1007/BF01016805}{Theor.\ Math.\ Phys.\  {\bf 89} (1991) 1052  [Teor.\ Mat.\ Fiz.\  {\bf 89} (1991) 56]}.
  %%CITATION = doi:10.1007/BF01016805;%%

\bibitem{Kotikov:1991hm}
A.~V.~Kotikov, ``Differential equations method: The Calculation of vertex type Feynman diagrams,''
\href{http://dx.doi.org/10.1016/0370-2693(91)90834-D}{Phys.\ Lett.\ B {\bf 259} (1991) 314}.
  %%CITATION = doi:10.1016/0370-2693(91)90834-D;%%

\bibitem{Kotikov:1991pm}
A.~V.~Kotikov, ``Differential equation method: The Calculation of N point Feynman diagrams,''
\href{http://dx.doi.org/10.1016/0370-2693(91)90536-Y}{Phys.\ Lett.\ B {\bf 267} (1991) 123};
Erratum: \href{http://dx.doi.org/10.1016/0370-2693(92)91582-T}{[Phys.\ Lett.\ B {\bf 295} (1992) 409]}.
  %%CITATION = doi:10.1016/0370-2693(91)90536-Y, 10.1016/0370-2693(92)91582-T;%%

\bibitem{Tarasov:1996br}
O.~V.~Tarasov,``Connection between Feynman integrals having different values of the
space-time dimension,''
\href{http://dx.doi.org/10.1103/PhysRevD.54.6479}{Phys.\ Rev.\ D {\bf 54} (1996) 6479} [\href{https://arxiv.org/abs/hep-th/9606018}{hep-th/9606018}];
%\cite{Tarasov:1997kx}
%\bibitem{Tarasov:1997kx}
%  O.~V.~Tarasov,
  ``Generalized recurrence relations for two loop propagator integrals
with arbitrary masses,''
\href{http://dx.doi.org/10.1016/S0550-3213(97)00376-3}{Nucl.\ Phys.\ B {\bf 502} (1997) 455} [\href{https://arxiv.org/abs/hep-ph/9703319}{hep-ph/9703319}].
%  doi:10.1016/S0550-3213(97)00376-3
%  [hep-ph/9703319].
  %%CITATION = doi:10.1016/S0550-3213(97)00376-3;%%

%\cite{Lee:2009dh}
\bibitem{Lee:2009dh}
  R.~N.~Lee,
  ``Space-time dimensionality D as complex variable: Calculating loop
integrals using dimensional recurrence relation and analytical
properties with respect to D,''
\href{http://dx.doi.org/10.1016/j.nuclphysb.2009.12.025}{Nucl.\ Phys.\ B {\bf 830} (2010) 474} [\href{https://arxiv.org/abs/0911.0252}{arXiv:0911.0252 [hep-ph]}].

% example where I(\lambda) appears in some form
\bibitem{Broadhurst:1996yc}
D.~J.~Broadhurst and A.~V.~Kotikov, ``Compact analytical form for nonzeta terms in critical exponents at order 1/N**3,''
\href{http://dx.doi.org/10.1016/S0370-2693(98)01146-0}{Phys.\ Lett.\ B {\bf 441} (1998) 345} [\href{https://arxiv.org/abs/hep-th/9612013}{arXiv:hep-th/9612013}].
%%CITATION = doi:10.1016/S0370-2693(98)01146-0;%%

% expansion O(veps^9)
\bibitem{Broadhurst:2002gb}
D.~J.~Broadhurst, ``Where do the tedious products of zeta's come from?,''
\href{http://dx.doi.org/10.1016/S0920-5632(03)80214-1}{Nucl.\ Phys.\ Proc.\ Suppl.\  {\bf 116} (2003) 432} [\href{https://arxiv.org/abs/hep-ph/0211194}{arXiv:hep-ph/0211194}].
%%CITATION = doi:10.1016/S0920-5632(03)80214-1;%%

% automated expansion
\bibitem{Bierenbaum:2003ud}
I.~Bierenbaum and S.~Weinzierl, ``The Massless two loop two point function,''
\href{http://dx.doi.org/10.1140/epjc/s2003-01389-7}{Eur.\ Phys.\ J.\ C {\bf 32} (2003) 67} [\href{https://arxiv.org/abs/hep-ph/0308311}{arXiv:hep-ph/0308311}].
%%CITATION = doi:10.1140/epjc/s2003-01389-7;%%

\bibitem{Brown:2008um}
  F.~Brown, ``The Massless higher-loop two-point function,''
\href{http://dx.doi.org/10.1007/s00220-009-0740-5}{Commun.\ Math.\ Phys.\  {\bf 287} (2009) 925} [\href{https://arxiv.org/abs/0804.1660}{arXiv:0804.1660 [math.AG]}].

\bibitem{Brown:2009ta}
  F.~C.~S.~Brown, ``On the periods of some Feynman integrals,''
  \href{http://inspirehep.net/record/832791}{arXiv:0910.0114 [math.AG]}.

\bibitem{Kleinert:1991rg}
H.~Kleinert, J.~Neu, V.~Schulte-Frohlinde, K.~G.~Chetyrkin and S.~A.~Larin, ``Five loop renormalization group functions of O(n) symmetric phi**4 theory and epsilon expansions of critical exponents up to epsilon**5,''
\href{http://dx.doi.org/10.1016/0370-2693(91)91009-K}{Phys.\ Lett.\ B {\bf 272} (1991) 39} Erratum: [\href{http://dx.doi.org/10.1016/0370-2693(93)91768-I}{Phys.\ Lett.\ B {\bf 319} (1993) 545}].
  %%CITATION = doi:10.1016/0370-2693(91)91009-K, 10.1016/0370-2693(93)91768-I;%%

\bibitem{Gracey:1993sn}
J.~A.~Gracey, ``Electron mass anomalous dimension at O(1/(Nf(2)) in quantum electrodynamics,''
\href{http://dx.doi.org/10.1016/0370-2693(93)91017-H}{Phys.\ Lett.\ B {\bf 317} (1993) 415} [\href{https://arxiv.org/abs/hep-th/9312055}{arXiv:hep-th/9312055}].
  %%CITATION = doi:10.1016/0370-2693(93)91017-H;%%

\bibitem{Vasiliev:1992wr}
A.~N.~Vasiliev, S.~E.~Derkachov, N.~A.~Kivel and A.~S.~Stepanenko, ``The 1/n expansion in the Gross-Neveu model: Conformal bootstrap calculation of the index eta in order 1/n**3,''
\href{http://dx.doi.org/10.1007/BF01019324}{Theor.\ Math.\ Phys.\  {\bf 94}, 127 (1993) [Teor.\ Mat.\ Fiz.\  {\bf 94}, 179 (1993)]}.
  %%CITATION = doi:10.1007/BF01019324;%%

\bibitem{Gracey:1993kx}
J.~A.~Gracey, ``The Conformal bootstrap equations for the four Fermi interaction in arbitrary dimensions,''
\href{http://dx.doi.org/10.1007/BF01566688}{Z.\ Phys.\ C {\bf 59} (1993) 243}.
  %%CITATION = doi:10.1007/BF01566688;%%

% examples where I(\lambda) appears in some form
\bibitem{Kivel:1993wq}
N.~A.~Kivel, A.~S.~Stepanenko and A.~N.~Vasiliev, ``On calculation of (2+epsilon) RG functions in the Gross-Neveu model from large N expansions of critical exponents,''
\href{http://dx.doi.org/10.1016/0550-3213(94)90411-1}{Nucl.\ Phys.\ B {\bf 424} (1994) 619} [\href{https://arxiv.org/abs/hep-th/9308073}{arXiv:hep-th/9308073}].
%%CITATION = doi:10.1016/0550-3213(94)90411-1;%%

\bibitem{Gracey:1993iu}
J.~A.~Gracey, ``Computation of critical exponent eta at O(1/N(f)**2) in quantum electrodynamics in arbitrary dimensions,''
\href{http://dx.doi.org/10.1016/0550-3213(94)90257-7}{Nucl.\ Phys.\ B {\bf 414} (1994) 614} [\href{https://arxiv.org/abs/hep-th/9312055}{arXiv:hep-th/9312055}].
  %%CITATION = doi:10.1016/0550-3213(94)90257-7;%%

\bibitem{Kreimer:1997dp}
D.~Kreimer, ``On the Hopf algebra structure of perturbative quantum field theories,''
\href{http://inspirehep.net/record/446989?ln=en}{Adv.\ Theor.\ Math.\ Phys.\  {\bf 2} (1998) 303} [\href{https://arxiv.org/abs/q-alg/9707029}{arXiv:q-alg/9707029}].
  %%CITATION = Q-ALG/9707029;%%

\bibitem{Connes:1998qv}
A.~Connes and D.~Kreimer, ``Hopf algebras, renormalization and noncommutative geometry,''
\href{http://dx.doi.org/10.1007/s002200050499}{Commun.\ Math.\ Phys.\  {\bf 199} (1998) 203} [\href{https://arxiv.org/abs/hep-th/9808042}{arXiv:hep-th/9808042}].
  %%CITATION = doi:10.1007/s002200050499;%%

\bibitem{Cartier:2001}
        P.~Cartier, ``A mad day's work: from Grothendieck to Connes and Kontsevich The evolution of concepts of space and symmetry,''
\href{http://www.ams.org/journals/bull/2001-38-04/S0273-0979-01-00913-2/S0273-0979-01-00913-2.pdf}{Bull.\ Amer.\ Math.\ Soc.\ {\bf 38} (2001), 389}.

\bibitem{Kontsevich:1999}
M.~Kontsevich, ``Operads and Motives in Deformation Quantization,''
\href{http://dx.doi.org/10.1023/A:1007555725247}{Lett.\ Math.\ Phys.\ {\bf 48} (1999)35} [\href{https://arxiv.org/abs/math/9904055}{arXiv:math/9904055 [math.QA]}].

\bibitem{Connes:2004zi}
  A.~Connes and M.~Marcolli, ``Renormalization and motivic galois theory,''
\href{https://arxiv.org/abs/math/0409306}{math/0409306 [math-nt]}.

\bibitem{Broadhurst:1987ei}
  D.~J.~Broadhurst, ``The Master Two Loop Diagram With Masses,''
\href{http://dx.doi.org/10.1007/BF01551921}{Z.\ Phys.\ C {\bf 47} (1990) 115}.

\bibitem{Veltman}
M.\ J.\ G.~Veltman, SCHOONSHIP (1963).

\bibitem{Hearn}
A.\ C.~Hearn, \href{http://www.reduce-algebra.com}{REDUCE}.

\bibitem{Vermaseren:2000nd}
  J.~A.~M.~Vermaseren, ``New features of FORM,''
  \href{http://inspirehep.net/record/541001}{math-ph/0010025}; \href{http://www.nikhef.nl/~form/}{FORM}.

\bibitem{Bauer:2000cp}
  C.~W.~Bauer, A.~Frink and R.~Kreckel, ``Introduction to the GiNaC framework for symbolic computation within the C++ programming language,''
  \href{http://inspirehep.net/record/526688}{J.\ Symb.\ Comput.\  {\bf 33} (2000) 1   [cs/0004015 [cs-sc]].}

\bibitem{wolfram1991mathematica}
 S.~Wolfram, ``Mathematica: a system for doing mathematics by computer,''
\newblock (1991) Addison-Wesley Pub. Co., Advanced Book Program

\bibitem{Weinzierl:2002cg}
  S.~Weinzierl, ``Computer algebra in particle physics,''
  \href{http://inspirehep.net/record/595845}{hep-ph/0209234}.

\bibitem{Nogueira:1991ex}
  P.~Nogueira, ``Automatic Feynman graph generation,''
\href{http://dx.doi.org/10.1006/jcph.1993.1074}{J.\ Comput.\ Phys.\  {\bf 105} (1993) 279} [\href{http://inspirehep.net/record/315611}{inspirehep/315611}].

\bibitem{Batkovich:2014bla}
  D.~Batkovich, Y.~Kirienko, M.~Kompaniets and S.~Novikov,
``GraphState - a tool for graph identification and labelling,''
\href{https://arxiv.org/abs/1409.8227}{arXiv:1409.8227 [hep-ph]}.

\bibitem{Seidensticker:1999bb}
  T.~Seidensticker, ``Automatic application of successive asymptotic expansions of Feynman diagrams,''
\href{https://arxiv.org/abs/hep-ph/9905298}{hep-ph/9905298}.

% Laporta's algorithm
\bibitem{Laporta:2001dd}
  S.~Laporta, ``High precision calculation of multiloop Feynman integrals by difference equations,''
\href{ http://dx.doi.org/10.1142/S0217751X00002159}{Int.\ J.\ Mod.\ Phys.\ A {\bf 15} (2000) 5087} [\href{https://arxiv.org/abs/hep-ph/0102033}{hep-ph/0102033}].

\bibitem{Baikov:1996rk}
  P.~A.~Baikov, ``Explicit solutions of the three loop vacuum integral recurrence relations,''
\href{http://dx.doi.org/10.1016/0370-2693(96)00835-0}{Phys.\ Lett.\ B {\bf 385} (1996) 404} [\href{https://arxiv.org/abs/hep-ph/9603267}{hep-ph/9603267}].

\bibitem{Studerus:2009ye}
  C.~Studerus, ``Reduze-Feynman Integral Reduction in C++,''
\href{http://dx.doi.org/10.1016/j.cpc.2010.03.012}{Comput.\ Phys.\ Commun.\  {\bf 181} (2010) 1293} \href{https://arxiv.org/abs/0912.2546}{[arXiv:0912.2546 [physics.comp-ph]]}.

\bibitem{vonManteuffel:2012np}
  A.~von Manteuffel and C.~Studerus, ``Reduze 2 - Distributed Feynman Integral Reduction,''
\href{https://arxiv.org/abs/1201.4330}{arXiv:1201.4330 [hep-ph]}.

\bibitem{Smirnov:2008iw}
  A.~V.~Smirnov, ``Algorithm FIRE -- Feynman Integral REduction,''
\href{http://dx.doi.org/10.1088/1126-6708/2008/10/107}{JHEP {\bf 0810} (2008) 107} [\href{https://arxiv.org/abs/0807.3243}{arXiv:0807.3243 [hep-ph]}].

\bibitem{Maierhoefer:2017hyi}
  P.~Maierhoefer, J.~Usovitsch and P.~Uwer, ``Kira - A Feynman Integral Reduction Program,''
\href{http://dx.doi.org/10.1016/j.cpc.2018.04.012}{Comput.\ Phys.\ Commun.\  {\bf 230} (2018) 99}
[\href{https://arxiv.org/abs/1705.05610}{arXiv:1705.05610 [hep-ph]}].

\bibitem{Lee:2013mka}
  R.~N.~Lee, ``LiteRed 1.4: a powerful tool for reduction of multiloop integrals,''
\href{http://dx.doi.org/10.1088/1742-6596/523/1/012059}{J.\ Phys.\ Conf.\ Ser.\  {\bf 523} (2014) 012059} [\href{https://arxiv.org/abs/1310.1145}{arXiv:1310.1145 [hep-ph]}].

\bibitem{Binoth:2000ps}
  T.~Binoth and G.~Heinrich, ``An automatized algorithm to compute infrared divergent multiloop integrals,''
  \href{http://dx.doi.org/10.1016/S0550-3213(00)00429-6}{Nucl.\ Phys.\ B {\bf 585} (2000) 741} 
  [\href{http://inspirehep.net/record/525717?ln=en}{hep-ph/0004013}.]

\bibitem{Bogner:2007cr}
  C.~Bogner and S.~Weinzierl, ``Resolution of singularities for multi-loop integrals,''
  \href{http://dx.doi.org/10.1016/j.cpc.2007.11.012}{Comput.\ Phys.\ Commun.\  {\bf 178} (2008) 596}
  [\href{http://inspirehep.net/record/761982}{arXiv:0709.4092 [hep-ph]}].

\bibitem{Bogner2009}
C.~Bogner, ``Mathematical aspects of Feynman integrals,'' 
 \href{http://ubm.opus.hbz-nrw.de/volltexte/2010/2215/}{PhD, Mainz (2009)}.

\bibitem{Panzer:2014caa}
  E.~Panzer, ``Algorithms for the symbolic integration of hyperlogarithms with applications to Feynman integrals,''
  \href{http://dx.doi.org/10.1016/j.cpc.2014.10.019}{Comput.\ Phys.\ Commun.\  {\bf 188} (2015) 148}
  [\href{http://inspirehep.net/record/1285754}{arXiv:1403.3385 [hep-th]}].

\bibitem{Panzer:2015ida}
  E.~Panzer, ``Feynman integrals and hyperlogarithms,''
  \href{http://inspirehep.net/record/1377774}{PhD, Humboldt-Universit\"at zu Berlin [arXiv:1506.07243 [math-ph]].}

\bibitem{Schnetz:2013hqa}
  O.~Schnetz, ``Graphical functions and single-valued multiple polylogarithms,''
  \href{http://dx.doi.org/10.4310/CNTP.2014.v8.n4.a1}{Commun.\ Num.\ Theor.\ Phys.\  {\bf 08} (2014) 589}
  %doi:10.4310/CNTP.2014.v8.n4.a1
  [\href{http://inspirehep.net/record/1221279http://inspirehep.net/record/1221279}{arXiv:1302.6445 [math.NT]}].

\bibitem{Golz:2015rea}
  M.~Golz, E.~Panzer and O.~Schnetz, ``Graphical functions in parametric space,''
  \href{http://dx.doi.org/10.1007/s11005-016-0935-6}{Lett.\ Math.\ Phys.\ {\bf 107} (2017) no.6, 1177}
  [\href{https://arxiv.org/abs/1509.07296}{arXiv:1509.07296 [math-ph]}].

\bibitem{Batkovich:2014rka}
  D.~V.~Batkovich and M.~Kompaniets, ``Toolbox for multiloop Feynman diagrams calculations using $R^{*}$ operation,''
  \href{http://dx.doi.org/10.1088/1742-6596/608/1/012068}{J.\ Phys.\ Conf.\ Ser.\  {\bf 608} (2015) no.1,  012068}
  [\href{http://inspirehep.net/record/1327265}{arXiv:1411.2618 [hep-th]}].

\bibitem{Herzog:2017bjx}
  F.~Herzog and B.~Ruijl, ``The R$^{*}$-operation for Feynman graphs with generic numerators,''
\href{http://dx.doi.org/10.1007/JHEP05(2017)037}{JHEP {\bf 1705} (2017) 037} [\href{https://arxiv.org/abs/1703.03776}{arXiv:1703.03776 [hep-th]}].

\bibitem{Gorishnii:1989gt}
  S.~G.~Gorishnii, S.~A.~Larin, L.~R.~Surguladze and F.~V.~Tkachov,
``Mincer: Program for Multiloop Calculations in Quantum Field Theory for the Schoonschip System,''
\href{http://dx.doi.org/10.1016/0010-4655(89)90134-3}{Comput.\ Phys.\ Commun.\  {\bf 55} (1989) 381};
%\bibitem{Larin:1991fz}
  S.~A.~Larin, F.~V.~Tkachov and J.~A.~M.~Vermaseren, ``The FORM version of MINCER,'' \href{http://inspirehep.net/record/30575}{NIKHEF-H-91-18 [KEK scan]}.

\bibitem{Ueda:2016yjm}
  T.~Ueda, B.~Ruijl and J.~A.~M.~Vermaseren,
``Forcer: a FORM program for 4-loop massless propagators,''
\href{http://dx.doi.org/10.22323/1.260.0070}{PoS LL {\bf 2016} (2016) 070}  [\href{https://arxiv.org/abs/1607.07318}{arXiv:1607.07318 [hep-ph]}].

\bibitem{Ruijl:2017cxj}
  B.~Ruijl, T.~Ueda and J.~A.~M.~Vermaseren,
``Forcer, a FORM program for the parametric reduction of four-loop massless propagator diagrams,''
\href{https://arxiv.org/abs/1704.06650}{arXiv:1704.06650 [hep-ph]}.

\bibitem{Gracey:2016mio}
  J.~A.~Gracey, T.~Luthe and Y.~Schroder, ``Four loop renormalization of the Gross-Neveu model,''
\href{http://dx.doi.org/10.1103/PhysRevD.94.125028}{Phys.\ Rev.\ D {\bf 94} (2016) no.12,  125028} [\href{https://arxiv.org/abs/1609.05071}{arXiv:1609.05071 [hep-th]}].

\bibitem{Gracey:2018qba}
  J.~A.~Gracey,
``Large $N$ critical exponents for the chiral Heisenberg Gross-Neveu universality class,''
\href{http://dx.doi.org/10.1103/PhysRevD.97.105009}{Phys.\ Rev.\ D {\bf 97} (2018) no.10,  105009}
[\href{https://arxiv.org/abs/1801.01320}{arXiv:1801.01320 [hep-th]}].

\bibitem{Gracey:2018fwq}
  J.~A.~Gracey,
``Fermion bilinear operator critical exponents at $O(1/N^2)$ in the QED-Gross-Neveu universality class,''
\href{https://arxiv.org/abs/1808.07697}{arXiv:1808.07697 [hep-th]}.

\bibitem{Mihaila:2017ble}
  L.~N.~Mihaila, N.~Zerf, B.~Ihrig, I.~F.~Herbut and M.~M.~Scherer,
 ``Gross-Neveu-Yukawa model at three loops and Ising critical behavior of Dirac systems,''
\href{http://dx.doi.org/10.1103/PhysRevB.96.165133}{Phys.\ Rev.\ B {\bf 96} (2017) no.16,  165133}
[\href{https://arxiv.org/abs/1703.08801}{arXiv:1703.08801 [cond-mat.str-el]}].

\bibitem{Zerf:2017zqi}
  N.~Zerf, L.~N.~Mihaila, P.~Marquard, I.~F.~Herbut and M.~M.~Scherer,
 ``Four-loop critical exponents for the Gross-Neveu-Yukawa models,''
 \href{http://dx.doi.org/10.1103/PhysRevD.96.096010}{Phys.\ Rev.\ D {\bf 96} (2017) no.9,  096010}
 [\href{https://arxiv.org/abs/1709.05057}{arXiv:1709.05057 [hep-th]}].

\bibitem{Baikov:2016tgj}
P.~A.~Baikov, K.~G.~Chetyrkin and J.~H.~K\"uhn, ``Five-Loop Running of the QCD coupling constant,''
\href{http://dx.doi.org/10.1103/PhysRevLett.118.082002}{Phys.\ Rev.\ Lett.\  {\bf 118} (2017) no.8,  082002}
[\href{http://arxiv.org/abs/1606.08659}{arXiv:1606.08659 [hep-ph]}].
  %%CITATION = ARXIV:1606.08659;

\bibitem{Chetyrkin:2017bjc}
  K.~G.~Chetyrkin, G.~Falcioni, F.~Herzog and J.~A.~M.~Vermaseren,
  ``Five-loop renormalisation of QCD in covariant gauges,''
  \href{http://dx.doi.org/10.1007/JHEP12(2017)006}{JHEP {\bf 1710} (2017) 179};
  Addendum: [\href{http://dx.doi.org/10.1007/JHEP10(2017)179}{JHEP {\bf 1712} (2017) 006}]
 [\href{https://arxiv.org/abs/1709.08541}{arXiv:1709.08541 [hep-ph]}].

\bibitem{Luthe:2016xec}
  T.~Luthe, A.~Maier, P.~Marquard and Y.~Schr\"oder, ``Five-loop quark mass and field anomalous dimensions for a general gauge group,''
  \href{http://dx.doi.org/10.1007/JHEP01(2017)081}{JHEP {\bf 1701} (2017) 081}
  [\href{http://inspirehep.net/record/1504230}{[arXiv:1612.05512 [hep-ph]}].

\bibitem{Luthe:2017ttc}
  T.~Luthe, A.~Maier, P.~Marquard and Y.~Schr\"oder, ``Complete renormalization of QCD at five loops,''
  \href{http://dx.doi.org/10.1007/JHEP03(2017)020}{JHEP {\bf 1703} (2017) 020}
  [\href{http://inspirehep.net/record/1510578}{arXiv:1701.07068 [hep-ph]}].

\bibitem{Luthe:2017ttg}
  T.~Luthe, A.~Maier, P.~Marquard and Y.~Schroder,
 ``The five-loop Beta function for a general gauge group and anomalous dimensions beyond Feynman gauge,''
 \href{http://dx.doi.org/10.1007/JHEP10(2017)166}{JHEP {\bf 1710} (2017) 166}
 [\href{https://arxiv.org/abs/1709.07718}{arXiv:1709.07718 [hep-ph]}].

\bibitem{Herzog:2017ohr}
  F.~Herzog, B.~Ruijl, T.~Ueda, J.~A.~M.~Vermaseren and A.~Vogt, ``The five-loop beta function of Yang-Mills theory with fermions,''
  \href{http://dx.doi.org/10.1007/JHEP02(2017)090}{JHEP {\bf 1702} (2017) 090}
  [\href{http://inspirehep.net/record/1507902}{[arXiv:1701.01404 [hep-ph]}].

\bibitem{Batkovich:2016jus}
D.~V.~Batkovich, K.~G.~Chetyrkin and M.~V.~Kompaniets, ``Six loop analytical calculation of the field anomalous dimension and the critical exponent $\eta$ in $O(n)$-symmetric $\varphi^4$ model,''
\href{http://dx.doi.org/10.1016/j.nuclphysb.2016.03.009}{Nucl.\ Phys.\ B {\bf 906} (2016) 147} [\href{https://arxiv.org/abs/1601.01960}{arXiv:1601.01960 [hep-th]}].
  %%CITATION = doi:10.1016/j.nuclphysb.2016.03.009;%%

\bibitem{Kompaniets:2016hct}
M.~Kompaniets and E.~Panzer, ``Renormalization group functions of $\phi^4$ theory in the MS-scheme to six loops,''
  PoS LL {\bf 2016} (2016) 038 [\href{http://arxiv.org/abs/1606.09210}{arXiv:1606.09210 [hep-th]}].
  %%CITATION = ARXIV:1606.09210;%%

\bibitem{Kompaniets:2017yct}
  M.~V.~Kompaniets and E.~Panzer, ``Minimally subtracted six loop renormalization of $O(n)$-symmetric $\phi^4$ theory and critical exponents,''
\href{http://dx.doi.org/10.1103/PhysRevD.96.036016}{Phys.\ Rev.\ D {\bf 96} (2017) no.3,  036016} [\href{https://arxiv.org/abs/1705.06483}{arXiv:1705.06483 [hep-th]}].

\bibitem{Schnetz:2016fhy}
  O.~Schnetz, ``Numbers and Functions in Quantum Field Theory,''
\href{http://dx.doi.org/10.1103/PhysRevD.97.085018}{Phys.\ Rev.\ D {\bf 97} (2018) no.8,  085018}
[\href{http://inspirehep.net/record/1472848}{arXiv:1606.08598 [hep-th]}].

\bibitem{Marboe:2016igj}
  C.~Marboe and V.~Velizhanin, ``Twist-2 at seven loops in planar $ \mathcal{N} $ = 4 SYM theory: full result and analytic properties,''
  \href{http://dx.doi.org/10.1007/JHEP11(2016)013}{JHEP {\bf 1611} (2016) 013}
  [\href{http://inspirehep.net/record/1477226}{[arXiv:1607.06047 [hep-th]}].

\bibitem{abrikosov1975methods}
A.\ A.~Abrikosov, L.\ P.~Gorkov and I.\ E.~Dzyaloshinski, ``Methods of Quantum Field Theory in Statistical Physics,''
\newblock Dover Books on Physics Series, 1975.

\bibitem{landau1956}
L.\ D.~Landau, ``The Theory of a Fermi Liquid,''
\href{http://www.jetp.ac.ru/cgi-bin/e/index/e/3/6/p920?a=list}{Sov.\ Phys.\ JETP {\bf 3} (1956) 920 [Zh.\ Eksp.\ Teor.\ Fiz.\ {\bf 30} (1956) 1058]}.

\bibitem{landau1957}
L.\ D.~Landau, ``Oscillations in a Fermi Liquid,''
\href{http://www.jetp.ac.ru/cgi-bin/e/index/e/5/1/p101?a=list}{Sov.\ Phys.\ JETP {\bf 5} (1957) 101 [Zh.\ Eksp.\ Teor.\ Fiz.\ {\bf 32} (1957) 59]}.

\bibitem{landau1958}
L.\ D.~Landau, ``On the Theory of the Fermi Liquid,''
\href{http://www.jetp.ac.ru/cgi-bin/e/index/e/8/1/p70?a=list}{Sov.\ Phys.\ JETP {\bf 8} (1959) 70 [Zh.\ Eksp.\ Teor.\ Fiz.\ {\bf 35} (1958) 97]}.

\bibitem{PinesN66}
D.~Pines and  P.~Nozi\`eres, ``The Theory of Quantum Liquids: Normal Fermi Liquids,''
\newblock Addison-Wesley, New York, 1966.

\bibitem{Shankar:1993pf}
R.~Shankar, ``Renormalization group approach to interacting fermions,''
\href{http://dx.doi.org/10.1103/RevModPhys.66.129}{Rev.\ Mod.\ Phys.\  {\bf 66} (1994) 129} [\href{https://arxiv.org/abs/cond-mat/9307009}{arXiv:cond-mat/9307009}].
%%CITATION = doi:10.1103/RevModPhys.66.129;%%

\bibitem{Polchinski:1992ed}
J.~Polchinski, ``Effective field theory and the Fermi surface,''
\newblock In *Boulder 1992, Proceedings, Recent directions in particle theory* 235-274, and Calif. Univ. Santa Barbara - NSF-ITP-92-132 (92,rec.Nov.) 39 p. (220633) Texas Univ. Austin - UTTG-92-20 (92,rec.Nov.) 39 p
[\href{https://arxiv.org/abs/hep-th/9210046}{hep-th/9210046}].
%%CITATION = HEP-TH/9210046;%%

\bibitem{Bardeen:1957mv}
J.~Bardeen, L.~N.~Cooper and J.~R.~Schrieffer, ``Theory of superconductivity,''
\href{http://dx.doi.org/10.1103/PhysRev.108.1175}{Phys.\ Rev.\  {\bf 108} (1957) 1175}.
%%CITATION = doi:10.1103/PhysRev.108.1175;%%

\bibitem{bogoliubov1958}
N.~N.~Bogoliubov, V.~V.~Tolmachov and D.~V.~Shirkov, ``A New Method in the Theory of Superconductivity,''
\href{http://dx.doi.org/10.1002/prop.19580061102}{Fortschr. Phys. {\bf 6} (1958) 605}.

\bibitem{Nambu60.PhysRev.117.648}
Y.~Nambu, ``Quasi-Particles and Gauge Invariance in the Theory of Superconductivity,''
\href{http://link.aps.org/doi/10.1103/PhysRev.117.648}{Phys.\ Rev.\ {\bf 117} (1960) 648}.

\bibitem{eliashberg1959}
G.\ M.~Eliashberg, ``Interactions between Electrons and Lattice Vibrations in a Superconductor,''
\href{http://www.jetp.ac.ru/cgi-bin/e/index/e/11/3/p696?a=list}{Sov.\ Phys.\ JETP {\bf 11} (1960) 696 [Zh.\ Eksp.\ Teor.\ Fiz.\ {\bf 38} (1960) 966]}.

\bibitem{Miransky:1994vk}
V.~A.~Miransky, ``Dynamical symmetry breaking in quantum field theories,''
\newblock  Singapore, Singapore: World Scientific (1993).

\bibitem{Nambu:1960xd}
Y.~Nambu, ``Axial vector current conservation in weak interactions,''
\href{http://dx.doi.org/10.1103/PhysRevLett.4.380}{Phys.\ Rev.\ Lett.\  {\bf 4} (1960) 380}.
%%CITATION = doi:10.1103/PhysRevLett.4.380;%%

\bibitem{Nambu:1961tp}
Y.~Nambu and G.~Jona-Lasinio, ``Dynamical Model of Elementary Particles Based on an Analogy with Superconductivity. 1.,''
\href{http://dx.doi.org/10.1103/PhysRev.122.345}{Phys.\ Rev.\  {\bf 122} (1961) 345}.
  %%CITATION = doi:10.1103/PhysRev.122.345;%%

\bibitem{VaksL61}
V.~G.~Vaks and A.~I.~Larkin, ``On the Application of the Methods of Superconductivity Theory to the Problem of the Masses of Elementary Particles,''  
\href{http://www.jetp.ac.ru/cgi-bin/e/index/e/13/1/p192?a=list}{Sov.\ Phys.\ JETP {\bf 13} (1961) 192 [ZhETF {\bf 40} (1961) 282]}.

\bibitem{Cohen:1988sq}
  A.~G.~Cohen and H.~Georgi, ``Walking Beyond the Rainbow,''
\href{http://dx.doi.org/10.1016/0550-3213(89)90109-0}{Nucl.\ Phys.\ B {\bf 314} (1989) 7} [\href{http://inspirehep.net/record/261200}{inspires}].

%excitonic insulator
\bibitem{KeldyshK64+HalperinR68}
L.~V.~Keldysh and Yu.~V.~Kopaev, 
\newblock Fiz. Tverd. Tela 6, 2791 (1964) [English transl.: Soviet Phys.\ Solid State 6, 2219 (1965)];
B.~I.~Halperin and T.~M.~Rice, 
\newblock Solid State Physics, F.~Seitz, D.~Turnbull and H.~Ehrenreich, Eds.,  Academic Press Inc., New York, 1968,  {\bf 21}.

\bibitem{RevModPhys.40.755}
B.\ I.~Halperin and T.\ M.~Rice, ``Possible Anomalies at a Semimetal-Semiconductor Transition,''
\href{http://link.aps.org/doi/10.1103/RevModPhys.40.755}{Rev.\ Mod.\ Phys.\ {\bf 40} (1968) 755}.

\bibitem{khomskii2010basic}
D.\ I.~Khomskii, ``Basic Aspects of the Quantum Theory of Solids: Order and Elementary Excitations,''
\newblock Cambridge University Press, 2010.

\bibitem{abrikosov1971}
A.\ A.~Abrikosov and S.\ D.~Beneslavskii, ``Possible Existence of Substances Intermediate Between Metals and Dielectrics,''
\href{http://www.jetp.ac.ru/cgi-bin/e/index/e/32/4/p699?a=list}{Sov.\ Phys.\ JETP {\bf 32} (1971) 699 [ZhETF {\bf 59} (1971) 1280]}.

\bibitem{HTc-1986}
J.\ G.~Bednorz and K.\ A.~M\"uller, ``Possible high $T_c$ superconductivity in the Ba-La-Cu-O system,''
\href{http://dx.doi.org/10.1007/BF01303701}{Z. Physik B - Condensed Matter (1986) 64}.

\bibitem{PhysRevLett.48.1559}
D.\ C.~Tsui, H.\ L.~Stormer and A.\ C.~Gossard, ``Two-Dimensional Magnetotransport in the Extreme Quantum Limit,''
\href{http://link.aps.org/doi/10.1103/PhysRevLett.48.1559}{Phys.\ Rev.\ Lett.\ {\bf 48} (1982) 1559}.

\bibitem{H2S-2015}
A.\ P.~Drozdov, M.\ I.~Eremets,  I.\ A.~Troyan, V.~Ksenofontov  and S.\ I.~Shylin,
``Conventional superconductivity at 203 kelvin at high pressures in the sulfur hydride system,''
\href{http://link.aps.org/10.1038/nature14964}{Nature {\bf 525} (2015) 73} [\href{https://arxiv.org/abs/1506.08190}{arXiv:1506.08190 [cond-mat.supr-con]}].

\bibitem{RevModPhys.78.17}
P.\ A.~Lee, N.~Nagaosa and X.-G.~Wen, ``Doping a Mott insulator: Physics of high-temperature superconductivity,''
\href{http://link.aps.org/doi/10.1103/RevModPhys.78.17}{Rev.\ Mod.\ Phys.\ {\bf 78} (2006) 17} [\href{https://arxiv.org/abs/cond-mat/0410445}{arXiv:cond-mat/0410445 [cond-mat.str-el]}].

\bibitem{PhysRevB.37.580}
G.~Baskaran and P.~W.~Anderson, ``Gauge theory of high-temperature superconductors and strongly correlated Fermi systems,''
\href{http://link.aps.org/doi/10.1103/PhysRevB.37.580}{Phys.\ Rev.\ B {\bf 37} (1988) 580}.

\bibitem{PhysRevLett.63.1996}
C.\ M.~Varma, P.\ B.~Littlewood, S.~Schmitt-Rink, E.~Abrahams and A.\ E.~Ruckenstein, ``Phenomenology of the normal state of Cu-O high-temperature superconductors,''
\href{http://link.aps.org/doi/10.1103/PhysRevLett.63.1996}{Phys.\ Rev.\ Lett.\ {\bf 63} (1989) 1996}.

%\bibitem{PhysRevB.65.165113}
%X.-G.~Wen, ``Quantum orders and symmetric spin liquids,''
%\href{http://link.aps.org/doi/10.1103/PhysRevB.65.165113}{Phys.\ Rev.\ B {\bf 65} (2002) 165113} [\href{https://arxiv.org/abs/cond-mat/0107071}{arXiv:cond-mat/0107071}].

\bibitem{PhysRevLett.86.3871}
W.~Rantner and X.-G.~Wen, ``Electron Spectral Function and Algebraic Spin Liquid for the Normal State of Underdoped High ${T}_{c}$ Superconductors,''
\href{http://link.aps.org/doi/10.1103/PhysRevLett.86.3871}{Phys.\ Rev.\ Lett.\ {\bf 86} (2001) 3871} [\href{https://arxiv.org/abs/cond-mat/0010378}{arXiv:cond-mat/0010378}].

\bibitem{Rantner2002}
W.~Rantner, ``The algebraic spin liquid of a possible model description for the normal state of underdoped high temperature superconductors,''
\href{http://hdl.handle.net/1721.1/29303}{Massachusetts Institute of Technology, 2002}.

\bibitem{Marston:1989zz}
J.~B.~Marston and I.~Affleck, ``Large-n limit of the Hubbard-Heisenberg model,''
\href{http://dx.doi.org/10.1103/PhysRevB.39.11538}{Phys.\ Rev.\ B {\bf 39} (1989) 11538}.
%%CITATION = doi:10.1103/PhysRevB.39.11538;%%

\bibitem{Ioffe:1989zz}
L.~B.~Ioffe and A.~I.~Larkin, ``Gapless fermions and gauge fields in dielectrics,''
\href{http://link.aps.org/10.1103/PhysRevB.39.8988}{Phys.\ Rev.\ B {\bf 39} (1989) 8988}.
%%CITATION = doi:10.1103/PhysRevB.39.8988;%%

\bibitem{PhysRevLett.64.2450}
N.~Nagaosa and P.\ A.~Lee, ``Normal-state properties of the uniform resonating-valence-bond state,''
\href{http://link.aps.org/doi/10.1103/PhysRevLett.64.2450}{Phys.\ Rev.\ Lett.\ {\bf 64} (1990) 2450}.

\bibitem{PhysRevB.46.5621}
P.\ A.~Lee and N.~Nagaosa, ``Gauge theory of the normal state of high-${\mathit{T}}_{\mathit{c}}$ superconductors,''
\href{http://link.aps.org/doi/10.1103/PhysRevB.46.5621}{Phys.\ Rev.\ B {\bf 46} (1992) 5621}.

\bibitem{PhysRevLett.65.653}
L.\ B.~Ioffe and P.\ B.~Wiegmann, ``Linear temperature dependence of resistivity as evidence of gauge interaction,''
\href{http://link.aps.org/doi/10.1103/PhysRevLett.65.653}{Phys.\ Rev.\ Lett.\ {\bf 65} (1990) 653}.

\bibitem{PhysRevLett.87.257003}
M.~Franz and Z.~Te\v{s}anovi\'c, ``Algebraic Fermi Liquid from Phase Fluctuations: ``Topological'' Fermions, Vortex ``Berryons,'' and ${\mathrm{QED}}_{3}$ Theory of Cuprate Superconductors,''
\href{http://link.aps.org/doi/10.1103/PhysRevLett.87.257003}{Phys.\ Rev.\ Lett.\ {\bf 87} (2001) 257003} [\href{https://arxiv.org/abs/cond-mat/0012445}{arXiv:cond-mat/0012445 [cond-mat.supr-con]}].

\bibitem{PhysRevB.66.054535}
M.~Franz, Z.~Te\v{s}anovi\'c and O.~Vafek, ``${\mathrm{QED}}_{3}$ theory of pairing pseudogap in cuprates: From $d$-wave superconductor to antiferromagnet via an algebraic Fermi liquid,''
\href{http://link.aps.org/doi/10.1103/PhysRevB.66.054535}{Phys.\ Rev.\ B {\bf 66} (2002) 054535} [\href{http://arxiv.org/abs/cond-mat/0203333}{cond-mat/0203047}].

\bibitem{Herbut:2002yq}
I.~F.~Herbut, ``QED(3) theory of underdoped high temperature superconductors,''
\href{http://dx.doi.org/10.1103/PhysRevB.66.094504}{Phys.\ Rev.\ B {\bf 66} (2002) 094504} [\href{http://arxiv.org/abs/cond-mat/0202491}{cond-mat/0202491}].
  %%CITATION = doi:10.1103/PhysRevB.66.094504;%%

\bibitem{PhysRevB.65.165113}
X.-G.~Wen, ``Quantum orders and symmetric spin liquids,''
\href{http://link.aps.org/doi/10.1103/PhysRevB.65.165113}{Phys.\ Rev.\ B {\bf 65} (2002) 165113} [\href{https://arxiv.org/abs/cond-mat/0107071}{arXiv:cond-mat/0107071}].

%\bibitem{PhysRevB.78.054432}
%C.~Xu, ``Renormalization group studies on four-fermion interaction instabilities on algebraic spin liquids,''
%\href{http://link.aps.org/doi/10.1103/PhysRevB.78.054432}{Phys.\ Rev.\ B {\bf 78} (2008) 054432} [\href{https://arxiv.org/abs/0803.0794}{arXiv:0803.0794 [cond-mat.str-el]}].

\bibitem{Appelquist:1981vg}
  T.~Appelquist and R.~D.~Pisarski, ``High-Temperature Yang-Mills Theories and Three-Dimensional Quantum Chromodynamics,''
\href{http://dx.doi.org/10.1103/PhysRevD.23.2305}{Phys.\ Rev.\ D {\bf 23} (1981) 2305}.

\bibitem{Appelquist:1981sf}
  T.~Appelquist and U.~W.~Heinz, ``Three-dimensional O(n) Theories At Large Distances,''
\href{http://dx.doi.org/10.1103/PhysRevD.24.2169}{Phys.\ Rev.\ D {\bf 24} (1981) 2169.}

\bibitem{Pisarski:1984dj}
R.~D.~Pisarski, ``Chiral Symmetry Breaking in Three-Dimensional Electrodynamics,''
\href{http://dx.doi.org/10.1103/PhysRevD.29.2423}{Phys.\ Rev.\ D {\bf 29} (1984) 2423}.
%\href{http://inspirehep.net/record/14971?ln=en}{Phys.\ Rev.\ D {\bf 29} (1984) 2423}.
  %%CITATION = doi:10.1103/PhysRevD.29.2423;%%

\bibitem{Appelquist:1985vf}
  T.~Appelquist, M.~J.~Bowick, E.~Cohler and L.~C.~R.~Wijewardhana,  ``Chiral Symmetry Breaking in (2+1)-dimensions,''
\href{http://dx.doi.org/10.1103/PhysRevLett.55.1715}{Phys.\ Rev.\ Lett.\  {\bf 55} (1985) 1715} [\href{http://inspirehep.net/record/214293}{inspirehep.net/record/214293}].

\bibitem{Appelquist:1986fd}
  T.~W.~Appelquist, M.~J.~Bowick, D.~Karabali and L.~C.~R.~Wijewardhana, ``Spontaneous Chiral Symmetry Breaking in Three-Dimensional QED,''
\href{http://dx.doi.org/10.1103/PhysRevD.33.3704}{Phys.\ Rev.\ D {\bf 33} (1986) 3704} [\href{http://inspirehep.net/record/226980}{inspirehep.net/record/226980}].
  %doi:10.1103/PhysRevD.33.3704

\bibitem{Appelquist:1986qw}
T.~Appelquist, M.~J.~Bowick, D.~Karabali and L.~C.~R.~Wijewardhana, ``Spontaneous Breaking of Parity in (2+1)-dimensional {QED},''
\href{http://dx.doi.org/10.1103/PhysRevD.33.3774}{Phys.\ Rev.\ D {\bf 33} (1986) 3774} [\href{https://inspirehep.net/record/17809}{inspirehep.net/record/17809}].
  %%CITATION = doi:10.1103/PhysRevD.33.3774;%%

\bibitem{Appelquist:1988sr}
T.~Appelquist, D.~Nash and L.~C.~R.~Wijewardhana, ``Critical Behavior in (2+1)-Dimensional QED,''
\href{http://dx.doi.org/10.1103/PhysRevLett.60.2575}{Phys.\ Rev.\ Lett.\  {\bf 60} (1988) 2575}
[\href{http://inspirehep.net/record/261201?ln=en}{inspirehep.net/record/261201}].
  %%CITATION = doi:10.1103/PhysRevLett.60.2575;%%

\bibitem{Nash:1989xx}
D.~Nash, ``Higher Order Corrections in (2+1)-Dimensional QED,''
\href{http://dx.doi.org/10.1103/PhysRevLett.62.3024}{Phys.\ Rev.\ Lett.\  {\bf 62} (1989) 3024} [\href{http://inspirehep.net/record/277524?ln=en}{inspirehep.net/record/277524}].
  %%CITATION = doi:10.1103/PhysRevLett.62.3024;%%

\bibitem{PhysRevLett.64.721}
T.~Appelquist and D.~Nash, ``Critical behavior in (2+1)-dimensional QCD,''
\href{http://link.aps.org/doi/10.1103/PhysRevLett.64.721}{Phys.\ Rev.\ Lett.\  {\bf 64} (1990) 721}.

\bibitem{Pennington:1990bx}
M.~R.~Pennington and D.~Walsh, ``Masses from nothing: A Nonperturbative study of QED in three-dimensions,''
\href{http://dx.doi.org/10.1016/0370-2693(91)91392-9}{Phys.\ Lett.\ B {\bf 253} (1991) 246}.
%\href{http://inspirehep.net/record/299468?ln=en}{Phys.\ Lett.\ B {\bf 253} (1991) 246}.
  %%CITATION = doi:10.1016/0370-2693(91)91392-9;%%

\bibitem{Curtis:1992gm}
D.~C.~Curtis, M.~R.~Pennington and D.~Walsh, ``Dynamical mass generation in QED in three-dimensions and the 1/N expansion,''
\href{http://dx.doi.org/10.1016/0370-2693(92)91572-Q}{Phys.\ Lett.\ B {\bf 295} (1992) 313}.
  %%CITATION = doi:10.1016/0370-2693(92)91572-Q;%%

\bibitem{Pisarski:1991kg}
R.~D.~Pisarski, ``Fermion mass in three-dimensions and the renormalization group,''
\href{http://dx.doi.org/10.1103/PhysRevD.44.1866}{Phys.\ Rev.\ D {\bf 44} (1991) 1866}.
  %%CITATION = doi:10.1103/PhysRevD.44.1866;%%

\bibitem{Atkinson:1989fp}
D.~Atkinson, P.~W.~Johnson and P.~Maris, ``Dynamical Mass Generation in {QED} in Three-dimensions: Improved Vertex Function,''
\href{http://dx.doi.org/10.1103/PhysRevD.42.602}{Phys.\ Rev.\ D {\bf 42} (1990) 602}.
  %%CITATION = doi:10.1103/PhysRevD.42.602;%%

\bibitem{Karthik:2015sgq}
N.~Karthik and R.~Narayanan, ``No evidence for bilinear condensate in parity-invariant three-dimensional QED with massless fermions,''
\href{http://dx.doi.org/10.1103/PhysRevD.93.045020}{ Phys.\ Rev.\ D {\bf 93} (2016)  045020} [\href{https://arxiv.org/abs/1512.02993}{arXiv:1512.02993 [hep-lat]}].
  %%CITATION = doi:10.1103/PhysRevD.93.045020;%%

\bibitem{Karthik:2016ppr}
N.~Karthik and R.~Narayanan, ``Scale-invariance of parity-invariant three-dimensional QED,''
\href{http://dx.doi.org/10.1103/PhysRevD.94.065026}{Phys.\ Rev.\ D {\bf 94} (2016) no.6,  065026} [\href{https://arxiv.org/abs/1606.04109}{arXiv:1606.04109 [hep-th]}].
  %%CITATION = ARXIV:1606.04109;%%

\bibitem{Dagotto:1988id}
E.~Dagotto, J.~B.~Kogut and A.~Kocic, ``A Computer Simulation of Chiral Symmetry Breaking in (2+1)-Dimensional QED with N Flavors,''
\href{http://dx.doi.org/10.1103/PhysRevLett.62.1083}{Phys.\ Rev.\ Lett.\  {\bf 62} (1989) 1083}.
  %%CITATION = doi:10.1103/PhysRevLett.62.1083;%%

\bibitem{Dagotto:1989td}
E.~Dagotto, A.~Kocic and J.~B.~Kogut, ``Chiral Symmetry Breaking in Three-dimensional {QED} With $N$(f) Flavors,''
\href{http://dx.doi.org/10.1016/0550-3213(90)90665-Z}{Nucl.\ Phys.\ B {\bf 334} (1990) 279}.
  %%CITATION = doi:10.1016/0550-3213(90)90665-Z;%%

\bibitem{Hands:2004bh}
S.~J.~Hands, J.~B.~Kogut, L.~Scorzato and C.~G.~Strouthos, ``Non-compact QED(3) with N(f) = 1 and N(f) = 4,''
\href{http://dx.doi.org/10.1103/PhysRevB.70.104501}{Phys.\ Rev.\ B {\bf 70} (2004) 104501} [\href{https://arxiv.org/abs/hep-lat/0404013}{arXiv:hep-lat/0404013}].
  %%CITATION = doi:10.1103/PhysRevB.70.104501;%%

\bibitem{Strouthos:2008kc}
C.~Strouthos and J.~B.~Kogut, ``The Phases of Non-Compact QED(3),''
\newblock  PoS LAT {\bf 2007} (2007) 278 [\href{https://arxiv.org/abs/0804.0300}{arXiv:0804.0300 [hep-lat]}].
  %%CITATION = ARXIV:0804.0300;%%

\bibitem{Azcoiti:1993fb}
V.~Azcoiti and X.~Q.~Luo, ``(2+1)-dimensional compact QED with dynamical Fermions,''
\href{http://dx.doi.org/10.1016/0920-5632(93)90315-W}{Nucl.\ Phys.\ Proc.\ Suppl.\  {\bf 30} (1993) 741}.
  %%CITATION = doi:10.1016/0920-5632(93)90315-W;%%

\bibitem{Azcoiti:1995mi}
  V.~Azcoiti, V.~Laliena and X.~Q.~Luo, ``Investigation of spontaneous symmetry breaking from a nonstandard approach,''
\href{http://dx.doi.org/10.1016/0920-5632(96)00123-5}{Nucl.\ Phys.\ Proc.\ Suppl.\  {\bf 47} (1996) 565} [\href{https://arxiv.org/abs/hep-lat/9605022}{hep-lat/9605022}].

\bibitem{Appelquist:2004ib}
T.~Appelquist and L.~C.~R.~Wijewardhana, ``Phase structure of noncompact QED3 and the Abelian Higgs model,''
\href{https://arxiv.org/abs/hep-ph/0403250}{hep-ph/0403250}.
  %%CITATION = HEP-PH/0403250;%%

\bibitem{Raya:2013ina}
K.~Raya, A.~Bashir, S.~Hern\'andez-Ortiz, A.~Raya and C.~D.~Roberts, ``Multiple solutions for the fermion mass function in QED3,''
\href{http://dx.doi.org/10.1103/PhysRevD.88.096003}{Phys.\ Rev.\ D {\bf 88} (2013) 096003} [\href{https://arxiv.org/abs/1305.2955}{arXiv:1305.2955 [nucl-th]}].
  %%CITATION = doi:10.1103/PhysRevD.88.096003;%%

\bibitem{Appelquist:1999hr}
T.~Appelquist, A.~G.~Cohen and M.~Schmaltz, ``A New constraint on strongly coupled gauge theories,''
\href{http://dx.doi.org/10.1103/PhysRevD.60.045003}{Phys.\ Rev.\ D {\bf 60} (1999) 045003} [\href{https://arxiv.org/abs/hep-th/9901109}{arXiv:hep-th/9901109}].
  %%CITATION = doi:10.1103/PhysRevD.60.045003;%%

\bibitem{Giombi:2015haa}
S.~Giombi, I.~R.~Klebanov and G.~Tarnopolsky, ``Conformal QED$_d$, $F$-Theorem and the $\epsilon$ Expansion,''
\href{http://dx.doi.org/10.1088/1751-8113/49/13/135403}{J.\ Phys.\ A {\bf 49} (2016)  135403} [\href{https://arxiv.org/abs/1508.06354}{arXiv:1508.06354 [hep-th]}].
%\href{http://inspirehep.net/record/1389863?ln=en}{J.\ Phys.\ A {\bf 49} (2016)  135403}.
  %%CITATION = doi:10.1088/1751-8113/49/13/135403;%%

\bibitem{Braun:2014wja}
  J.~Braun, H.~Gies, L.~Janssen and D.~Roscher,
  ``Phase structure of many-flavor QED$_3$,''
\href{http://dx.doi.org/10.1103/PhysRevD.90.036002}{Phys.\ Rev.\ D {\bf 90} (2014) no.3,  036002} [\href{https://arxiv.org/abs/1404.1362}{arXiv:1404.1362 [hep-ph]}].

\bibitem{Janssen:2016nrm}
  L.~Janssen, ``Spontaneous breaking of Lorentz symmetry in $(2+\epsilon)$-dimensional QED,''
\href{http://dx.doi.org/10.1103/PhysRevD.94.094013}{Phys.\ Rev.\ D {\bf 94} (2016) no.9,  094013} [\href{https://arxiv.org/abs/1604.06354}{arXiv:1604.06354 [hep-th]}].

\bibitem{Kotikov:1993wr}
A.~V.~Kotikov, ``Critical behavior of 3D electrodynamics,''
\href{http://www.jetpletters.ac.ru/ps/1193/article_18004.shtml}{Sov.\ Phys.\ JETP {\bf 58} (1993) 731 [Zh.\ Eksp.\ Teor.\ Fiz.\ {\bf 58} (1993) 785]}.

\bibitem{Kotikov:2011kg}
A.~V.~Kotikov, ``On the Critical Behavior of (2+1)-Dimensional QED,''
\href{http://dx.doi.org/10.1134/S1063778812070058}{Phys.\ Atom.\ Nucl.\  {\bf 75} (2012) 890} [\href{https://arxiv.org/abs/1104.3888}{arXiv:1104.3888 [hep-ph]}].
  %%CITATION = doi:10.1134/S1063778812070058;%%

\bibitem{Gusynin:1995bb}
V.~P.~Gusynin, A.~H.~Hams and M.~Reenders, ``(2+1)-dimensional QED with dynamically massive fermions in the vacuum polarization,''
\href{http://dx.doi.org/10.1103/PhysRevD.53.2227}{Phys.\ Rev.\ D {\bf 53} (1996) 2227} [\href{https://arxiv.org/abs/hep-ph/9509380}{arXiv:hep-ph/9509380}].
  %%CITATION = doi:10.1103/PhysRevD.53.2227;%%

\bibitem{Bashir:2005wt}
A.~Bashir and A.~Raya, ``Truncated Schwinger-Dyson equations and gauge covariance in QED3,''
\href{http://dx.doi.org/10.1007/s00601-007-0177-3}{Few Body Syst.\  {\bf 41} (2007) 185} [\href{https://arxiv.org/abs/hep-ph/0511291}{arXiv:hep-ph/0511291}].
  %%CITATION = doi:10.1007/s00601-007-0177-3;%%

\bibitem{Bashir:2009fv}
A.~Bashir, A.~Raya, S.~Sanchez-Madrigal and C.~D.~Roberts, ``Gauge invariance of a critical number of flavours in QED3,''
\href{http://dx.doi.org/10.1007/s00601-009-0069-9}{Few Body Syst.\  {\bf 46} (2009) 229} [\href{https://arxiv.org/abs/0905.1337}{arXiv:0905.1337 [hep-ph]}].
  %%CITATION = doi:10.1007/s00601-009-0069-9;%%

\bibitem{Giombi:2016fct}
  S.~Giombi, G.~Tarnopolsky and I.~R.~Klebanov, ``On $C_{J}$ and $C_{T}$ in Conformal QED,'' 
\href{http://dx.doi.org/10.1007/JHEP08(2016)156}{JHEP {\bf 1608} (2016) 156} [\href{https://arxiv.org/abs/1602.01076}{arXiv:1602.01076 [hep-th]}].
%\href{http://inspirehep.net/record/1419101?ln=en}{JHEP {\bf 1608} (2016) 156}.

\bibitem{Kubota:2001kk}
  K.~I.~Kubota and H.~Terao, ``Dynamical symmetry breaking in QED(3) from the Wilson RG point of view,''
\href{http://dx.doi.org/10.1143/PTP.105.809}{Prog.\ Theor.\ Phys.\  {\bf 105} (2001) 809} [\href{https://arxiv.org/abs/hep-ph/0101073}{hep-ph/0101073}].

\bibitem{DiPietro:2015taa}
L.~Di Pietro, Z.~Komargodski, I.~Shamir and E.~Stamou, ``Quantum Electrodynamics in d=3 from the epsilon-Expansion,''
\href{http://dx.doi.org/10.1103/PhysRevLett.116.131601}{Phys.\ Rev.\ Lett.\  {\bf 116} (2016) 131601} [\href{https://arxiv.org/abs/1508.06278}{arXiv:1508.06278 [hep-th]}].
%\href{http://inspirehep.net/record/1389875?ln=en}{Phys.\ Rev.\ Lett.\  {\bf 116} (2016) 131601}.
  %%CITATION = doi:10.1103/PhysRevLett.116.131601;%%

\bibitem{Herbut:2016ide}
I.~F.~Herbut, ``Chiral symmetry breaking in three-dimensional quantum electrodynamics as fixed point annihilation,''
\href{http://dx.doi.org/10.1103/PhysRevD.94.025036}{Phys.\ Rev.\ D {\bf 94} (2016) no.2,  025036} [\href{https://arxiv.org/abs/1605.09482}{arXiv:1605.09482 [hep-th]}].
%\href{http://inspirehep.net/record/1466462?ln=en}{Phys.\ Rev.\ D {\bf 94} (2016) 025036}. %arXiv:1605.09482 [hep-th].
  %%CITATION = ARXIV:1605.09482;%%

\bibitem{Gusynin:2016som}
V.~P.~Gusynin and P.~K.~Pyatkovskiy, ``Critical number of fermions in three-dimensional QED,''
\href{http://dx.doi.org/10.1103/PhysRevD.94.125009}{Phys.\ Rev.\ D {\bf 94} (2016) no.12,  125009} [\href{https://arxiv.org/abs/1607.08582}{arXiv:1607.08582 [hep-ph]}].
%\href{http://inspirehep.net/record/1478381?ln=en}{arXiv:1607.08582 [hep-ph]}.
  %%CITATION = ARXIV:1607.08582;%%

\bibitem{Polyakov:1975rs}
  A.~M.~Polyakov, ``Compact Gauge Fields and the Infrared Catastrophe,''
\href{http://dx.doi.org/10.1016/0370-2693(75)90162-8}{Phys.\ Lett.\  {\bf 59B} (1975) 82}.

\bibitem{Polyakov:1976fu}
  A.~M.~Polyakov, ``Quark Confinement and Topology of Gauge Groups,''
\href{http://dx.doi.org/10.1016/0550-3213(77)90086-4}{Nucl.\ Phys.\ B {\bf 120} (1977) 429}.

\bibitem{PhysRevB.70.214437}
M.~Hermele, T.~Senthil, M.\ P.~A.~Fisher, P.\ A.~Lee, N.~Nagaosa and X.-G.~Wen, ``Stability of $U(1)$ spin liquids in two dimensions,''
\href{http://dx.doi.org/10.1103/PhysRevB.70.214437}{Phys.\ Rev.\ B {\bf 70} (2004) 214437} [\href{https://arxiv.org/abs/cond-mat/0404751}{arXiv:cond-mat/0404751 [cond-mat.str-el]}].

\bibitem{Pufu:2013vpa}
  S.~S.~Pufu, ``Anomalous dimensions of monopole operators in three-dimensional quantum electrodynamics,''
\href{http://dx.doi.org/10.1103/PhysRevD.89.065016}{Phys.\ Rev.\ D {\bf 89} (2014) no.6,  065016} [\href{https://arxiv.org/abs/1303.6125}{arXiv:1303.6125 [hep-th]}].

\bibitem{Chester:2015wao}
  S.~M.~Chester, M.~Mezei, S.~S.~Pufu and I.~Yaakov, ``Monopole operators from the $4-\epsilon$ expansion,''
\href{http://dx.doi.org/10.1007/JHEP12(2016)015}{JHEP {\bf 1612} (2016) 015} [\href{https://arxiv.org/abs/1511.07108}{arXiv:1511.07108 [hep-th]}].

\bibitem{Jain:1989tx}
  J.~K.~Jain, ``Composite fermion approach for the fractional quantum Hall effect,''
\href{http://dx.doi.org/10.1103/PhysRevLett.63.199}{Phys.\ Rev.\ Lett.\  {\bf 63} (1989) 199.}

\bibitem{Jain:book}
  J.~K.~Jain, ``Composite Fermions,''
\newblock Cambridge University Press, 2007.

\bibitem{Son:2015xqa}
  D.~T.~Son, ``Is the Composite Fermion a Dirac Particle?,''
\href{http://dx.doi.org/10.1103/PhysRevX.5.031027}{Phys.\ Rev.\ X {\bf 5} (2015) no.3,  031027} [\href{https://arxiv.org/abs/1502.03446}{arXiv:1502.03446 [cond-mat.mes-hall]}].

\bibitem{PfeiferCF:1993}
R.\ L.~Willett, R.\ R.~Ruel, K.\ W.~West and L.\ N.~Pfeiffer, ``Experimental demonstration of a Fermi surface at one-half filling of the lowest Landau level, ''
\href{https://doi.org/10.1103/PhysRevLett.71.3846}{Phys.\ Rev.\ Lett.\ {\bf 71} (1993) 3846}. 

\bibitem{JainCF:1994}
V.\ J.~Goldman, B.~Su and J.\ K.~Jain, ``Detection of composite fermions by magnetic focusing,''
\href{https://doi.org/10.1103/PhysRevLett.72.2065}{Phys.\ Rev.\ Lett.\ {\bf 72} (1994) 2065}.

\bibitem{PfeifferCF:1996}
R.\ L.~Willett and L.\ N.~Pfeiffer, ``Composite fermions examined with surface acoustic waves,''
\href{https://doi.org/10.1016/0039-6028(96)00325-1}{Surface Science {\bf 361} (1996) 38}.

\bibitem{RezayiCF:1994}
E.~Rezayi and N.~Read, ``Fermi-liquid-like state in a half-filled Landau level,''
\href{https://doi.org/10.1103/PhysRevLett.72.900}{Phys.\ Rev.\ Lett.\ {\bf 72} (1994) 900}. 

\bibitem{PanCF:2017}
W.~Pan, W.~Kang, K.\ W.~Baldwin, K.\ W.~West, L.\ N.~Pfeiffer and D.\ C.~Tsui, ``Berry phase and anomalous transport of the composite fermions at the half-filled Landau level,''
\href{https://doi.org/10.1038/nphys4231}{Nature Physics {\bf 13} (2017) 1168} [\href{https://arxiv.org/abs/1702.07307}{arXiv:1702.07307 [cond-mat.mes-hall]}].

\bibitem{Hsiao:2017lch}
  W.~H.~Hsiao and D.~T.~Son, ``Duality and Universal Transport in a Mixed-Dimension Electrodynamics,''
\href{http://dx.doi.org/10.1103/PhysRevB.96.075127}{Phys.\ Rev.\ B {\bf 96} (2017) 075127} [\href{https://arxiv.org/abs/1705.01102}{arXiv:1705.01102 [cond-mat.mes-hall]}].

\bibitem{Xu:2015lxa}
  C.~Xu and Y.~Z.~You, ``Self-dual Quantum Electrodynamics as Boundary State of the three dimensional Bosonic Topological Insulator,''
\href{http://dx.doi.org/10.1103/PhysRevB.92.220416}{Phys.\ Rev.\ B {\bf 92} (2015) no.22,  220416} [\href{https://arxiv.org/abs/1510.06032}{arXiv:1510.06032 [cond-mat.str-el]}].

\bibitem{Cheng:2016pdn}
  M.~Cheng and C.~Xu, ``Series of (2+1)-dimensional stable self-dual interacting conformal field theories,''
\href{http://dx.doi.org/10.1103/PhysRevB.94.214415}{Phys.\ Rev.\ B {\bf 94} (2016) no.21,  214415} [\href{https://arxiv.org/abs/1609.02560}{arXiv:1609.02560 [cond-mat.str-el]}].

\bibitem{Seiberg:2016gmd}
  N.~Seiberg, T.~Senthil, C.~Wang and E.~Witten, ``A Duality Web in 2+1 Dimensions and Condensed Matter Physics,''
\href{http://dx.doi.org/10.1016/j.aop.2016.08.007}{Annals Phys.\  {\bf 374} (2016) 395} [\href{https://arxiv.org/abs/1606.01989}{arXiv:1606.01989 [hep-th]}].

\bibitem{Zhang:1988wy}
S.~C.~Zhang, T.~H.~Hansson and S.~Kivelson, ``An effective field theory model for the fractional quantum hall effect,''
\href{http://dx.doi.org/10.1103/PhysRevLett.62.82}{Phys.\ Rev.\ Lett.\  {\bf 62} (1988) 82}.
  %%CITATION = doi:10.1103/PhysRevLett.62.82;%%

\bibitem{Fradkin:1991wy}
E.~Fradkin and A.~Lopez, ``Fractional Quantum Hall effect and Chern-Simons gauge theories,''
\href{http://dx.doi.org/10.1103/PhysRevB.44.5246}{Phys.\ Rev.\ B {\bf 44} (1991) 5246}.
  %%CITATION = doi:10.1103/PhysRevB.44.5246;%%

%\bibitem{wilczek1990fractional}
%F.~Wilczek, ``Fractional Statistics and Anyon Superconductivity,''
%\newblock International journal of modern physics, 1990.

\bibitem{Witten:1988hf}
E.~Witten, ``Quantum Field Theory and the Jones Polynomial,''
\href{http://dx.doi.org/10.1007/BF01217730}{Commun.\ Math.\ Phys.\  {\bf 121} (1989) 351}.
  %%CITATION = doi:10.1007/BF01217730;%%

\bibitem{Moore:1991ks}
G.~W.~Moore and N.~Read, ``Nonabelions in the fractional quantum Hall effect,''
\href{http://dx.doi.org/10.1016/0550-3213(91)90407-O}{Nucl.\ Phys.\ B {\bf 360} (1991) 362}.
  %%CITATION = doi:10.1016/0550-3213(91)90407-O;%%

\bibitem{PhysRevLett.79.2526}
L.~Saminadayar, D.\ C.~Glattli, Y.~Jin and B.~Etienne, ``Observation of the $\mathit{e}\mathit{/}3$ Fractionally Charged Laughlin Quasiparticle,''
\href{http://link.aps.org/doi/10.1103/PhysRevLett.79.2526}{Phys.\ Rev.\ Lett.\ {\bf 79} (1997) 2526} [\href{https://arxiv.org/abs/cond-mat/9706307}{arXiv:cond-mat/9706307}].

\bibitem{picciotto1997fqhe}
R.\ de-Picciotto, M.~Reznikov, M.~Heiblum, V.~Umansky, G.~Bunin and D.~Mahalu, ``Direct observation of a fractional charge,''
\href{http://dx.doi.org/10.1038/38241}{Nature {\bf 389} (1997) 162} [\href{https://arxiv.org/abs/cond-mat/9707289}{arXiv:cond-mat/9707289 [cond-mat.mes-hall]}].

%\bibitem{martin2004fqhe}
%J.~Martin, S.~Ilani, B.~Verdene, J.~Smet, V.~Umansky, D.~Mahalu, D.~Schuh, G.~Abstreiter and A.~Yacoby, ``Localization of Fractionally Charged Quasi-Particles,''
%\href{http://dx.doi.org/10.1126/science.1099950}{Science {\bf 305} (2004) 980}.

\bibitem{PhysRevD.13.3398}
R.~Jackiw and C.~Rebbi, ``Solitons with fermion number $1/2$,''
\href{http://link.aps.org/doi/10.1103/PhysRevD.13.3398}{Phys.\ Rev.\ Lett.\ {\bf 13} (1976) 3398}.

\bibitem{0038-5670-30-5-R02}
I.\ V.~Krive and A.\ S.~Rozhavskii, ``Fractional charge in quantum field theory and solid-state physics,''
\href{http://stacks.iop.org/0038-5670/30/i=5/a=R02}{Soviet Physics Uspekhi {\bf 30} (1987) 370}.

\bibitem{PhysRevLett.36.432}
M.\ J.~Rice, A.\ R.~Bishop and D.\ K.~Campbell, ``Weakly Pinned Fr\"ohlich Charge-Density-Wave Condensates: A New, Nonlinear, Current-Carrying Elementary Excitation,''
\href{http://link.aps.org/doi/10.1103/PhysRevLett.36.432}{Phys.\ Rev.\ Lett.\ {\bf 36} (1976) 432}.

\bibitem{brazovskii78}
S.\ A.~Brazovskii, ``Electronic excitations in Peierls-Fr\"ohlich state,''
\href{http://www.jetpletters.ac.ru/ps/1581/article_24247.shtml}{JETP Letters {\bf 28} (1978) 606 [Pis'ma Zh.\ Eksp.\ Teor.\ Fiz.\ {\bf 28} (1978) 656]}.

\bibitem{brazovskii1980}
S.\ A.~Brazovskii, ``Self-localized excitations in the Peierls-Fr\"ohlich state,''
\href{http://www.jetp.ac.ru/cgi-bin/e/index/e/51/2/p342?a=list}{Sov.\ Phys.\ JETP {\bf 51} (1980) 342 [Zh.\ Eksp.\ Teor.\ Fiz.\ {\bf 78} (1980) 677]}.

\bibitem{PhysRevLett.42.1698}
W.\ P.~Su, J.\ R.~Schrieffer and A.\ J.~Heeger, ``Solitons in Polyacetylene,''
\href{http://link.aps.org/doi/10.1103/PhysRevLett.42.1698}{Phys.\ Rev.\ Lett.\ {\bf 42} (1979) 1698}.

\bibitem{rice1979}
M.\ J.~Rice, ``Charged $\pi$-phase kinks in lightly doped polyacetylene,''
\href{http://dx.doi.org/10.1016/0375-9601(79)90905-8}{Phys.\ Lett.\ A {\bf 71} (1979) 152}.

\bibitem{RevModPhys.60.781}
A.\ J.~Heeger, S.~Kivelson and J.\ R.~Schrieffer, ``Solitons in conducting polymers,''
\href{http://link.aps.org/doi/10.1103/RevModPhys.60.781}{Rev.\ Mod.\ Phys.\ {\bf 60} (1988) 781}.

\bibitem{PhysRevLett.104.227602}
F.~Kagawa, S.~Horiuchi, H.~Matsui, R.~Kumai, Y.~Onose, T.~Hasegawa and Y.~Tokura, 
``Electric-Field Control of Solitons in a Ferroelectric Organic Charge-Transfer Salt,''
\href{http://link.aps.org/doi/10.1103/PhysRevLett.104.227602}{Phys.\ Rev.\ Lett.\ {\bf 22} (2010) 227602}.

\bibitem{0953-8984-13-18-311}
S.~Teber, B.\ P.~Stojkovic, S.\ A.~Brazovskii and A.\ R.~Bishop, ``Statistics of charged solitons and formation of stripes,''
\href{http://stacks.iop.org/0953-8984/13/i=18/a=311}{J.\ Phys.: Condens.~Matter {\bf 13} (2001) 4015} [\href{https://arxiv.org/abs/cond-mat/0112081}{arXiv:cond-mat/0112081 [cond-mat.stat-mech]}]. 

\bibitem{Karpov16}
P.~Karpov and S.~Brazovskii, ``Phase transitions in ensembles of solitons induced by an optical pumping or a strong electric field,''
\href{http://dx.doi.org/10.1103/PhysRevB.94.125108}{Phys.\ Rev.\ B {\bf 94} (2016) 125108} [\href{https://arxiv.org/abs/1605.04395}{arXiv:1605.04395 [cond-mat.str-el]}].

\bibitem{PhysRev.134.A1416}
W.\ A.~Little, ``Possibility of Synthesizing an Organic Superconductor,''
\href{http://link.aps.org/doi/10.1103/PhysRev.134.A1416}{Phys.\ Rev.\ {\bf 134} (1964) A1416}.

\bibitem{saito1998physical}
R.~Saito, G.~Dresselhaus and M.\ S.~Dresselhaus, ``Physical Properties of Carbon Nanotubes,''
\newblock Imperial College Press, 1998.

\bibitem{giamarchi2003quantum}
T.~Giamarchi, ``Quantum Physics in One Dimension,''
\newblock International Series of Monographs on Physics, Clarendon Press, 2003.

\bibitem{gogolin2004bosonization}
A.\ O.~Gogolin, A.\ A.~Nersesyan and A.\ M.~Tsvelik, ``Bosonization and Strongly Correlated Systems,''
\newblock Cambridge University Press, 2004. 

\bibitem{nanotubestomonaga1998}
M.~Bockrath, D.\ H.~Cobden, J.~Lu, A.\ G.~Rinzler, R.\ E.~Smalley, L.~Balents and P.\ L.~McEuen, ``Luttinger-liquid behaviour in carbon nanotubes,''
\href{http://dx.doi.org/10.1038/17569}{Nature {\bf 397} (1999) 598} [\href{https://arxiv.org/abs/cond-mat/9812233}{arXiv:cond-mat/9812233 [cond-mat.mes-hall]}].

\bibitem{nanotubestomonaga2003}
H.~Ishii, H.~Kataura, H.~Shiozawa, H.~Yoshioka, H.~Otsubo, Y.~Takayama, T.~Miyahara, S.~Suzuki, 
Y.~Achiba, M.~Nakatake, T.~Narimura, M.~Higashiguchi, K.~Shimada, H.~Namatame and M.~Taniguchi, ``Direct observation of Tomonaga-Luttinger-liquid state in carbon nanotubes at low temperatures,''
\href{http://dx.doi.org/10.1038/nature02074}{Nature {\bf 426} (2003) 540}. 

\bibitem{egger2001}
R.~Egger, A.~Bachtold, M.\ S.~Fuhrer, M.~Bockrath, D.\ H.~Cobden and P.\ L.~McEuen, ``Luttinger Liquid Behavior in Metallic Carbon Nanotubes,''
\newblock Interacting Electrons in Nanostructures, Edited by R.\ Haug, H.\ Schoeller, Lecture Notes in Physics {\bf 579} (2001) 125 [\href{https://arxiv.org/abs/cond-mat/0008008}{arXiv:cond-mat/0008008 [cond-mat.mes-hall]}].

\bibitem{Tomonaga:1950zz}
S.-I.~Tomonaga, ``Remarks on Bloch's Method of Sound Waves applied to Many-Fermion Problems,''
\href{http://dx.doi.org/10.1143/PTP.5.544}{Prog.\ Theor.\ Phys.\  {\bf 5} (1950) 544}.
  %%CITATION = doi:10.1143/PTP.5.544;%%

\bibitem{Bloch1933}
F.~Bloch, ``Bremsverm{\"o}gen von Atomen mit mehreren Elektronen,''
\href{http://dx.doi.org/10.1007/BF01344553}{ZS.  Phys. {\bf 81} (1933) 5}.

\bibitem{Bloch1934}
F.~Bloch, ``Inkoh\"arente R\"ontgenstreuung und Dichteschwankungen eines entarteten Fermigases,''
\href{http://www.e-periodica.ch/digbib/view?pid=hpa-001:1934:7#387}{Helv.\ Phys.\ Acta {\bf 7} (1934) 385}.

\bibitem{Luttinger:1963zz}
J.~M.~Luttinger, ``An Exactly Soluble Model of a Many-Fermion System,''
\href{http://dx.doi.org/10.1063/1.1704046}{J.\ Math.\ Phys.\  {\bf 4} (1963) 1154}.
  %%CITATION = doi:10.1063/1.1704046;%%

\bibitem{Mattis:1964wp}
D.~C.~Mattis and E.~H.~Lieb, ``Exact solution of a many fermion system and its associated boson field,''
\href{http://dx.doi.org/10.1063/1.1704281}{J.\ Math.\ Phys.\  {\bf 6} (1965) 304}.
  %%CITATION = doi:10.1063/1.1704281;%%

\bibitem{bychkov1966}
Yu.\ A.~Bychkov, L.\ P.~Gor'kov, I.\ E.~Dzyaloshinskii, ``Possibility of Superconductivity Type Phenomena in a One-dimensional System,''
\href{http://www.jetp.ac.ru/cgi-bin/e/index/e/23/3/p489?a=list}{Sov.\ Phys.\ JETP {\bf 23} (1966) 489 [Zh.\ Eksp.\ Teor.\ Fiz.\ {\bf 23} (1966) 738]}.

\bibitem{dzyaloshinskii1972}
I.\ E.~Dzyaloshinskii and A.\ I.~Larkin, ``Possible States of Quasi-unidimensional Systems,''
\href{http://www.jetp.ac.ru/cgi-bin/e/index/e/34/2/p422?a=list}{Sov.\ Phys.\ JETP {\bf 34} (1972) 422 [Zh.\ Eksp.\ Teor.\ Fiz.\ {\bf 61} (1972) 791]}.

\bibitem{solyom1979}
J.~S\'olyom, ``The Fermi gas model of one-dimensional conductors,''
\href{http://dx.doi.org/10.1080/00018737900101375}{Adv.\ Phys.\ {\bf 28} (1979) 1979}.

\bibitem{diatlov1957}
I.\ T.~Diatlov, V.\ V.~Sudakov and K.\ A.~Ter-Martirosian, ``Asymptotic Meson-Meson Scattering Theory,''
\href{http://www.jetp.ac.ru/cgi-bin/e/index/e/5/4/p631?a=list}{Sov.\ Phys.\ JETP {\bf 5} (1957) 631 [Zh.\ Eksp.\ Teor.\ Fiz.\ {\bf 32} (1957) 767]}.

%\bibitem{sudakov1957}
%V.\ V.~Sudakov, ``???,''
%\href{???}{Sov.\ Phys.\ Doklady {\bf 1} (1957) 662.}.

\bibitem{PhysRev.178.1072}
B.~Roulet, J.~Gavoret and P.~Nozi\`eres, ``Singularities in the X-Ray Absorption and Emission of Metals. I. First-Order Parquet Calculation,''
\href{http://link.aps.org/doi/10.1103/PhysRev.178.1072}{Phys.\ Rev.\ {\bf 178} (1969) 1072}.

\bibitem{PhysRev.178.1084}
B.~Roulet, J.~Gavoret and P.~Nozi\`eres, ``Singularities in the X-Ray Absorption and Emission of Metals. II. Self-Consistent Treatment of Divergences,''
\href{http://link.aps.org/doi/10.1103/PhysRev.178.1084}{Phys.\ Rev.\ {\bf 178} (1969) 1084}.

\bibitem{PhysRevB.55.3200}
A.\ T.~Zheleznyak, V.\ M.~Yakovenko and I.\ E.~Dzyaloshinskii, ``Parquet solution for a flat Fermi surface,''
\href{http://link.aps.org/doi/10.1103/PhysRevB.55.3200}{Phys.\ Rev.\ B {\bf 55} (1997) 3200} [\href{https://arxiv.org/abs/cond-mat/9609118}{arXiv:cond-mat/9609118}].

\bibitem{dzyaloshinskii1973}
I.\ E.~Dzyaloshinskii and A.\ I.~Larkin, ``Correlation functions for a one-dimensional Fermi system with long-range interaction (Tomonaga model),''
\href{http://www.jetp.ac.ru/cgi-bin/e/index/e/38/1/p202?a=list}{Sov.\ Phys.\ JETP {\bf 38} (1974) 201 [Zh.\ Eksp.\ Teor.\ Fiz.\ {\bf 65} (1974) 411]}.

\bibitem{efetov1975}
K.\ B.~Efetov and A.\ I.~Larkin, ``Correlation functions in one-dimensional systems with a strong interaction,''
\href{http://www.jetp.ac.ru/cgi-bin/e/index/e/42/2/p390?a=list}{Sov.\ Phys.\ JETP {\bf 42} (1975) 390 [Zh.\ Eksp.\ Teor.\ Fiz.\ {\bf 69} (1975) 764]}.

\bibitem{Coleman:1974bu}
S.~R.~Coleman, ``The Quantum Sine-Gordon Equation as the Massive Thirring Model,''
\href{http://dx.doi.org/10.1103/PhysRevD.11.2088}{Phys.\ Rev.\ D {\bf 11} (1975) 2088}.
  %%CITATION = doi:10.1103/PhysRevD.11.2088;%%

\bibitem{PhysRevD.11.3026}
S.~Mandelstam, ``Soliton operators for the quantized sine-Gordon equation,''
\href{http://link.aps.org/doi/10.1103/PhysRevD.11.3026}{Phys.\ Rev.\ D {\bf 11} (1975) 3026.}

\bibitem{PhysRevB.9.2911}
A.~Luther and I.~Peschel, ``Single-particle states, Kohn anomaly, and pairing fluctuations in one dimension,''
\href{http://link.aps.org/doi/10.1103/PhysRevB.9.2911}{Phys.\ Rev.\ B {\bf 9} (1974) 2911}.

%\bibitem{PhysRevLett.33.589}
%A.~Luther and V.\ J.~Emery, ``Backward Scattering in the One-Dimensional Electron Gas,''
%\href{http://link.aps.org/doi/10.1103/PhysRevLett.33.589}{Phys.\ Rev.\ Lett.\ {\bf 33} (1974) 589}.

\bibitem{Belavin:1984vu}
A.~A.~Belavin, A.~M.~Polyakov and A.~B.~Zamolodchikov, ``Infinite Conformal Symmetry in Two-Dimensional Quantum Field Theory,''
\href{http://link.aps.org/10.1016/0550-3213(84)90052-X}{Nucl.\ Phys.\ B {\bf 241} (1984) 333}.
  %%CITATION = doi:10.1016/0550-3213(84)90052-X;%%

\bibitem{Dotsenko:1984nm}
Vl.~S.~Dotsenko and V.~A.~Fateev, ``Conformal Algebra and Multipoint Correlation Functions in Two-Dimensional Statistical Models,''
\href{http://link.aps.org/10.1016/0550-3213(84)90269-4}{Nucl.\ Phys.\ B {\bf 240} (1984) 312}.
  %%CITATION = doi:10.1016/0550-3213(84)90269-4;%%

\bibitem{korepin1997quantum}
V.\ E.~Korepin, N.\ M.~Bogoliubov and A.\ G.~Izegrin, ``Quantum Inverse Scattering Method and Correlation Functions,''
\newblock Cambridge Monographs on Mathematical Physics, 1997.

\bibitem{Haldane:1981zza}
F.~D.~M.~Haldane, ``Luttinger liquid theory of one-dimensional quantum fluids. I. Properties of the Luttinger model and their extension to the general 1D interacting spinless Fermi gas,''
\href{http://stacks.iop.org/0022-3719/14/i=19/a=010}{J.\ Phys.\ C {\bf 14} (1981) 2585}.
  %%CITATION = JPAGA,C14,2585;%%

\bibitem{jompol2009}
Y.~Jompol, C.\ J.\ B.~Ford, J.\ P.~Griffiths, I.~Farrer, G.\ A.\ C.~Jones, D.~Anderson, D.\ A.~Ritchie, T.\ W.~Silk and A.\ J.~Schofield, 
``Probing Spin-Charge Separation in a Tomonaga-Luttinger Liquid,''
\href{http://dx.doi.org/10.1126/science.1171769}{Science {\bf 325} (2009) 597} [\href{https://arxiv.org/abs/1002.2782}{arXiv:1002.2782 [cond-mat.str-el]}].

\bibitem{steinberg2008frac}
H.~Steinberg, G.~Barak, A.~Yacoby, L.\ N.~Pfeiffer, K.\ W.~West, B.\ I.~Halperin and K.\ Le Hur, ``Charge fractionalization in quantum wires,''
\href{http://dx.doi.org/10.1038/nphys810}{Nature Physics {\bf 4} (2008) 116}.

\bibitem{Delamotte:2007pf}
  B.~Delamotte, ``An Introduction to the nonperturbative renormalization group,''
\href{http://dx.doi.org/10.1007/978-3-642-27320-9_2}{Lect.\ Notes Phys.\  {\bf 852} (2012) 49} [\href{https://arxiv.org/abs/cond-mat/0702365}{cond-mat/0702365 [cond-mat.stat-mech]}].

\bibitem{Wetterich:1992yh}
  C.~Wetterich, ``Exact evolution equation for the effective potential,''
\href{http://dx.doi.org/10.1016/0370-2693(93)90726-X}{Phys.\ Lett.\ B {\bf 301} (1993) 90}.

\bibitem{RevModPhys.84.299}
W.~Metzner, M.~Salmhofer, C.~Honerkamp, V.~Meden and K.~Sch\"onhammer, ``Functional renormalization group approach to correlated fermion systems,''
\href{http://link.aps.org/doi/10.1103/RevModPhys.84.299}{Rev.\ Mod.\ Phys.\ {\bf 84} (2012) 299} [\href{https://arxiv.org/abs/1105.5289}{arXiv:1105.5289 [cond-mat.str-el]}].

\bibitem{Gies:2006wv}
  H.~Gies, ``Introduction to the functional RG and applications to gauge theories,''
\href{http://dx.doi.org/10.1007/978-3-642-27320-9_6}{Lect.\ Notes Phys.\  {\bf 852} (2012) 287} [\href{https://arxiv.org/abs/hep-ph/0611146}{hep-ph/0611146}].

\bibitem{Janssen:2012oca}
  L.~Janssen, ``Critical phenomena in (2+1)-dimensional relativistic fermion systems,'' 
\href{http://inspirehep.net/record/1128310}{PhD Jena U.\ (2012)}.
  %%CITATION = INSPIRE-1128310;%%

\bibitem{novoselov2004}
K.\ S.~Novoselov, A.\ K.~Geim, S.\ V.~Morozov, D.~Jiang, Y.~Zhang, S.\ V.~Dubonos, I.\ V.~Grigorieva and A.\ A.~Firsov, ``Electric Field Effect in Atomically Thin Carbon Films,''
\href{http://dx.doi.org/10.1126/science.1102896}{Science  {\bf 306} (2004) 666} [\href{https://arxiv.org/abs/cond-mat/0410550}{arXiv:cond-mat/0410550 [cond-mat.mtrl-sci]}].

\bibitem{novoselov2005}
K.\ S.~Novoselov, D.~Jiang, F.~Schedin, T.\ J.~Booth, V.\ V.~Khotkevich, S.\ V.~Morozov and A.\ K.~Geim, ``Two-dimensional atomic crystals,''
\href{http://dx.doi.org/10.1073/pnas.0502848102}{PNAS {\bf 102} (2005) 10451} [\href{https://arxiv.org/abs/cond-mat/0503533}{arXiv:cond-mat/0503533 [cond-mat.mtrl-sci]}].

\bibitem{Novoselov:2005kj}
K.~S.~Novoselov, A.~K.~Geim, S.~V.~Morozov, D.~Jiang, M.~I.~Katsnelson, I.~V.~Grigorieva, S.~V.~Dubonos and A.~A.~Firsov, ``Two-dimensional gas of massless Dirac fermions in graphene,''
\href{http://dx.doi.org/10.1038/nature04233}{Nature {\bf 438} (2005) 197} [\href{https://arxiv.org/abs/cond-mat/0509330}{arXiv:cond-mat/0509330 [cond-mat.mes-hall]}].

\bibitem{Schakel:1991af}
  A.~M.~J.~Schakel, ``Relativistic quantum Hall effect,''
\href{http://dx.doi.org/10.1103/PhysRevD.43.1428}{Phys.\ Rev.\ D {\bf 43} (1991) 1428}.

\bibitem{Gusynin:2005pk}
  V.~P.~Gusynin and S.~G.~Sharapov, ``Unconventional integer quantum Hall effect in graphene,''
\href{http://dx.doi.org/10.1103/PhysRevLett.95.146801}{Phys.\ Rev.\ Lett.\  {\bf 95} (2005) 146801} [\href{https://arxiv.org/abs/cond-mat/0506575}{cond-mat/0506575}].

\bibitem{Gusynin:2007ix}
V.~P.~Gusynin, S.~G.~Sharapov and J.~P.~Carbotte, ``AC conductivity of graphene: from tight-binding model to 2+1-dimensional quantum electrodynamics,''
\href{http://dx.doi.org/10.1142/S0217979207038022}{Int.\ J.\ Mod.\ Phys.\ B {\bf 21} (2007) 4611} [\href{http://inspirehep.net/record/769598}{arXiv:0706.3016 [cond-mat.mes-hall]}].

\bibitem{CastroNeto:2009zz}
A.~H.~Castro Neto, F.~Guinea, N.~M.~R.~Peres, K.~S.~Novoselov and A.~K.~Geim, ``The electronic properties of graphene,''
\href{http://dx.doi.org/10.1103/RevModPhys.81.109}{Rev.\ Mod.\ Phys.\  {\bf 81} (2009) 109} [\href{https://arxiv.org/abs/0709.1163}{arXiv:0709.1163 [cond-mat.other]}].
  %%CITATION = doi:10.1103/RevModPhys.81.109;%%

\bibitem{katsnelson2012graphene}
M.\ I.~Katsnelson, ``Graphene: Carbon in Two Dimensions,''
\newblock Cambridge University Press, 2012.

\bibitem{Nair:2008zz}
R.~R.~Nair, P.~Blake, A.~N.~Grigorenko, K.~S.~Novoselov, T.~J.~Booth, T.~Stauber, N.~M.~R.~Peres and A.~K.~Geim, ``Fine structure constant defines visual transparency of graphene,''
\href{http://dx.doi.org/10.1126/science.1156965}{Science {\bf 320} (2008) 1308}.
  %%CITATION = doi:10.1126/science.1156965;%%

\bibitem{PhysRevLett.101.196405}
K.\ F.~Mak,  M.\ Y.~Sfeir, Y.~Wu, C.\ H.~Lui, J.\ A.~Misewich and T.\ F.~Heinz, ``Measurement of the Optical Conductivity of Graphene,''
\href{http://link.aps.org/doi/10.1103/PhysRevLett.101.196405}{Phys.\ Rev.\ Lett.\ {\bf 101} (2008) 196405} [\href{https://arxiv.org/abs/0810.1269}{arXiv:0810.1269 [cond-mat.mtrl-sci]}].

\bibitem{PhysRev.71.622}
P.\ R.~Wallace, ``The Band Theory of Graphite,''
\href{http://link.aps.org/doi/10.1103/PhysRev.71.622}{Phys.\ Rev.\ {\bf 71} (1947) 622}.

\bibitem{Semenoff:1984dq}
G.~W.~Semenoff, ``Condensed Matter Simulation of a Three-dimensional Anomaly,''
\href{http://dx.doi.org/10.1103/PhysRevLett.53.2449}{Phys.\ Rev.\ Lett.\  {\bf 53} (1984) 2449}.
  %%CITATION = doi:10.1103/PhysRevLett.53.2449;%%

\bibitem{Volovik2009book}
G.\ E.~Volovik, ``The Universe in a Helium Droplet,''
\newblock Oxford University Press, Oxford, 2009.

\bibitem{meyer2007ripples}
J.\ C.~Meyer, A.\ K.~Geim, M.\ I.~Katsnelson, K.\ S.~Novoselov, T.\ J.~Booth and S.~Roth, ``The structure of suspended graphene sheets,''
\href{http://dx.doi.org/10.1038/nature05545}{Nature {\bf 446} (2007) 60} [\href{https://arxiv.org/abs/cond-mat/0701379}{arXiv:cond-mat/0701379 [cond-mat.mes-hall]}].

\bibitem{hsieh2008topological}
D.~Hsieh, D.~Qian, L.~Wray, Y.~Xia, Y.\ S.~Hor, R.\ J.~Cava and M.\ Z.~Hasan, ``A topological Dirac insulator in a quantum spin Hall phase,''
\href{http://dx.doi.org/10.1038/nature06843}{Nature {\bf 452} (2008) 970} [\href{https://arxiv.org/abs/0910.2420}{arXiv:0910.2420 [cond-mat.mes-hall]}].

\bibitem{xia2009topological}
Y.~Xia, D.~Qian, D.~Hsieh, L.~Wray, A.~Pal, H.~Lin, A.~Bansil, D.~Grauer, Y.\ S.~Hor, R.\ J.~Cava and M.\ Z.~Hasan, 
``Observation of a large-gap topological-insulator class with a single Dirac cone on the surface,''
\href{http://dx.doi.org/10.1038/nphys1274}{Nature Physics {\bf 5} (2009) 398}.

\bibitem{RevModPhys.82.3045}
M.\ Z.~Hasan and C.\ L.~Kane, ``{\textit{Colloquium}: Topological insulators},''
\href{http://link.aps.org/doi/10.1103/RevModPhys.82.3045}{Rev.\ Mod.\ Phys.\ {\bf 82} (2010) 3045} [\href{https://arxiv.org/abs/1002.3895}{arXiv:1002.3895 [cond-mat.mes-hall]}].

\bibitem{polini2013artificial}
M.~Polini, F.~Guinea, M.~Lewenstein, H.\ C.~Manoharan and V.~Pellegrini, ``Artificial honeycomb lattices for electrons, atoms and photons,''
\href{http://dx.doi.org/10.1038/nnano.2013.161}{Nature Nanotechnology {\bf 8} (2013) 625}.

\bibitem{liu2014dirac3d}
Z.\ K.~Liu, B.~Zhou, Y.~Zhang, Z.\ J.~Wang, H.\ M.~Weng, D.~Prabhakaran, S.-K.~Mo, Z.\ X.~Shen, Z.~Fang, X.~Dai, Z.~Hussain, Y.\ L.~Chen, 
``Discovery of a Three-Dimensional Topological Dirac Semimetal, Na3Bi,''
\href{http://dx.doi.org/10.1126/science.1245085}{Science {\bf 343} (2014) 864-867} [\href{https://arxiv.org/abs/1310.0391}{arXiv:1310.0391 [cond-mat.mtrl-sci]}].

\bibitem{neupane2013dirac3d}
M.~Neupane, S.-Y.~Xu, R.~Sankar, N.~Alidoust, G.~Bian, C.~Liu, I.~Belopolski, T.-R.~Chang, H.-T.~Jeng, H.~Lin, A.~Bansil, F.~Chou and M.\ Z.~Hasan, 
``Observation of a three-dimensional topological Dirac semimetal phase in high-mobility Cd$_3$As$_2$,''
\href{http://dx.doi.org/10.1038/ncomms4786}{Nature Communications  {\bf 5} (2014) 3786} [\href{https://arxiv.org/abs/1309.7892}{arXiv:1309.7892 [cond-mat.mes-hall]}].

\bibitem{liu2013dirac3d}
Z.\ K.~Liu, J.~Jiang, B.~Zhou, Z.\ J.~Wang, Y.~Zhang, H.\ M.~Weng, D.~Prabhakaran, S.-K.~Mo, H.~Peng, P.~Dudin, T.~Kim, M.~Hoesch, Z.~Fang, X.~Dai, Z.\ X.~Shen, D.\ L.~Feng, Z.~Hussain and Y.\ L.~Chen,
``A stable three-dimensional topological Dirac semimetal Cd$_3$As$_2$,''
\href{http://dx.doi.org/10.1038/nmat3990}{Nature Materials {\bf 13} (2014) 677}.

\bibitem{PhysRevLett.113.027603}
S.~Borisenko, Q.~Gibson, D.~Evtushinsky, V.~Zabolotnyy, B.~B\"uchner, and R.\ J.~Cava, ``Experimental Realization of a Three-Dimensional Dirac Semimetal,''
\href{http://link.aps.org/doi/10.1103/PhysRevLett.113.027603}{Phys.\ Rev.\ Lett.\ {\bf 113} (2014) 027603} [\href{https://arxiv.org/abs/1309.7978}{arXiv:1309.7978 [cond-mat.mes-hall]}].

\bibitem{huang2014weyl}
S.-M.~Huang, S.-Y.~Xu, I.~Belopolski, C.-C.~Lee, G.~Chang, B.~Wang, N.~Alidoust, G.~Bian, M.~Neupane, C.~Zhang, S.~Jia, A.~Bansil, H.~Lin and M.\ Z.~Hasan, 
``A Weyl Fermion semimetal with surface Fermi arcs in the transition metal monopnictide TaAs class,''
\href{http://dx.doi.org/10.1038/ncomms8373}{Nature Communications {\bf 6} (2015) 7373}.

\bibitem{xu2015weyl}
S.-Y.~Xu et al., ``Discovery of a Weyl fermion state with Fermi arcs in niobium arsenide,''
\href{http://dx.doi.org/10.1038/nphys3437}{Nature Physics {\bf 11} (2015) 748}.

\bibitem{RevModPhys.88.021004}
A.~Bansil, H.~Lin and T.~Das, ``{\textit{Colloquium}  : Topological band theory},''
\href{http://link.aps.org/doi/10.1103/RevModPhys.88.021004}{Rev.\ Mod.\ Phys.\ {\bf 88} (2016) 021004} [\href{https://arxiv.org/abs/1603.03576}{arXiv:1603.03576 [cond-mat.mes-hall]}].

\bibitem{Wehling2014Dirac}
T.\ O.~Wehling, A.\ M.~Black-Schaffer and A.\ V.~Balatsky, ``Dirac materials,''
\href{http://dx.doi.org/10.1080/00018732.2014.927109}{Adv.\ Phys.\ {\bf 63} (2014) 1} [\href{http://arxiv.org/abs/1405.5774}{arXiv:1405.5774 [cond-mat.mtrl-sci]}].

\bibitem{vozmediano2010gauge}
M.\ A.\ H.~Vozmediano, M.\ I.~Katsnelson and F.~Guinea, ``Gauge fields in graphene,''
\href{http://dx.doi.org/10.1016/j.physrep.2010.07.003}{Physics Reports {\bf 496} (2010) 109} [\href{https://arxiv.org/abs/1003.5179}{arXiv:1003.5179 [cond-mat.mes-hall]}].

\bibitem{RevModPhys.84.1067}
V.\ N.~Kotov, B.~Uchoa, V.\ M.~Pereira, F.~Guinea and A.\ H.~Castro Neto, ``Electron-Electron Interactions in Graphene: Current Status and Perspectives,''
\href{http://link.aps.org/doi/10.1103/RevModPhys.84.1067}{Rev.\ Mod.\ Phys.\ {\bf 84} (2012) 1067} [\href{https://arxiv.org/abs/1012.3484}{arXiv:1012.3484 [cond-mat.str-el]}].
%\href{http://inspirehep.net/record/901457?ln=en}{Rev.\ Mod.\ Phys.\ {\bf 84} (2012) 1067}.

\bibitem{Coquand:2017wak}
  O.~Coquand, K.~Essafi, J.-P.~Kownacki and D.~Mouhanna,
 ``A glassy phase in quenched disordered graphene and crystalline membranes,''
\href{http://dx.doi.org/10.1103/PhysRevE.97.030102}{Phys.\ Rev.\ E {\bf 97} (2018) no.3,  030102}
[\href{https://arxiv.org/abs/1708.08364}{arXiv:1708.08364 [cond-mat.dis-nn]}].

\bibitem{Herbut:2006cs}
  I.~F.~Herbut, ``Interactions and phase transitions on graphene's honeycomb lattice,''
\href{http://dx.doi.org/10.1103/PhysRevLett.97.146401}{Phys.\ Rev.\ Lett.\  {\bf 97} (2006) 146401} [\href{https://arxiv.org/abs/cond-mat/0606195}{arXiv:cond-mat/0606195 [cond-mat.mes-hall]}].

\bibitem{PhysRevLett.114.126804}
C.~Faugeras, S.~Berciaud, P.~Leszczynski, Y.~Henni, K.~Nogajewski, M.~Orlita, T.~Taniguchi, K.~Watanabe, C.~ Forsythe, P.~Kim, R.~Jalil, A.\ K.~Geim, D.\ M.~Basko and M.~Potemski, 
``Landau Level Spectroscopy of Electron-Electron Interactions in Graphene,''
\href{http://link.aps.org/doi/10.1103/PhysRevLett.114.126804}{Phys.\ Rev.\ Lett.\ {\bf 114} 126804 (2015)} [\href{https://arxiv.org/abs/1412.0115}{arXiv:1412.0115 [cond-mat.mes-hall]}].

\bibitem{Hirata2017}
M.~Hirata, K.~Ishikawa, G.~Matsuno, A.~Kobayashi, K.~Miyagawa, M.~Tamura, C.~Berthier and K.~Kanoda, ``Extraordinary Coulomb correlations and incipient excitonic instability of Weyl fermions,''
\href{http://dx.doi.org/10.1126/science.aan5351}{Science {\bf 358} (2017) 1403}
[\href{https://arxiv.org/abs/1702.00097}{arXiv:1702.00097 [cond-mat.str-el]}].

\bibitem{Gonzalez:1993uz}
J.~Gonz\'alez, F.~Guinea and M.~A.~H.~Vozmediano, ``NonFermi liquid behavior of electrons in the half filled honeycomb lattice (A Renormalization group approach),''
\href{http://dx.doi.org/10.1016/0550-3213(94)90410-3}{Nucl.\ Phys.\ B {\bf 424} (1994) 595} [\href{https://arxiv.org/abs/hep-th/9311105}{arXiv:hep-th/9311105}].
  %%CITATION = doi:10.1016/0550-3213(94)90410-3;%%

\bibitem{bolotin2009fqhe}
K.\ I.~Bolotin, F.~Ghahari, M.\ D.~Shulman, H.\ L.~Stormer and P.~Kim, ``Observation of the fractional quantum Hall effect in graphene,''
\href{http://dx.doi.org/10.1038/nature08582}{Nature {\bf 462} (2009) 196} [\href{https://arxiv.org/abs/0910.2763}{arXiv:0910.2763 [cond-mat.mes-hall]}].
   
\bibitem{du2009fqhe}
X.~Du, I.~Skachko, F.~Duerr, A.~Luican, and E.\ Y.~Andrei, ``Fractional quantum Hall effect and insulating phase of Dirac electrons in graphene,''
\href{http://dx.doi.org/10.1038/nature08522}{Nature {\bf 462} (2209) 192} [\href{https://arxiv.org/abs/0910.2532}{arXiv:0910.2532 [cond-mat.mes-hall]}].

\bibitem{morpurgo2009fqhe}
A.\ F.~Morpurgo, ``Condensed-matter physics: Dirac electrons broken to pieces,''
\href{http://dx.doi.org/10.1038/462170a}{Nature {\bf 462} (2009) 170}.

\bibitem{elias2011cones}
D.\ C.~Elias, R.\ V.~Gorbachev, A.\ S.~Mayorov, S.\ V.~Morozov, A.\ A.~Zhukov, P.~Blake, L.\ A.~Ponomarenko, I.\ V.~Grigorieva, K.\ S.~Novoselov, F.~Guinea and A.\ K.~Geim,
``Dirac cones reshaped by interaction effects in suspended graphene,''
\href{http://dx.doi.org/10.1038/nphys2049}{Nature Physics {\bf 7} (2011) 701} [\href{https://arxiv.org/abs/1104.1396}{arXiv:1104.1396 [cond-mat.mes-hall]}].

\bibitem{siegel2011cones}
D.\ A.~Siegel, C.-H.~Parka, C.~Hwangb, J.~Deslippe, A.\ V.~Fedorovc, S.\ G.~Louie and A.~Lanzara, ``Many-body interactions in quasi-freestanding graphene,''
\href{http://dx.doi.org/10.1073/pnas.1100242108}{PNAS {\bf 108} (2011) 11365} [\href{https://arxiv.org/abs/1106.5822}{arXiv:1106.5822 [cond-mat.mtrl-sci]}].

\bibitem{bostwick2007}
A.~Bostwick, T.~Ohta, T.~Seyller, K.~Horn and E.~Rotenberg, ``Quasiparticle dynamics in graphene,''
\href{http://dx.doi.org/10.1038/nphys477}{Nature Physics {\bf 3} (2007) 36}.

\bibitem{PhysRevLett.102.176804}
G.~Li, A.~Luican and E.\ Y.~Andrei, ``Scanning Tunneling Spectroscopy of Graphene on Graphite,''
\href{http://dx.doi.org/10.1103/PhysRevLett.102.176804}{Phys.\ Rev.\ Lett.\ {\bf 102} (2009) 176804} [\href{https://arxiv.org/abs/0803.4016}{arXiv:0803.4016 [cond-mat.mes-hall]}].

\bibitem{PhysRevLett.77.3589}
J.~Gonz\'alez, F.~Guinea and M.~A.~H.~Vozmediano, ``Unconventional Quasiparticle Lifetime in Graphite,''
\href{http://link.aps.org/doi/10.1103/PhysRevLett.77.3589}{Phys.\ Rev.\ Lett.\ {\bf 77} (1996) 3589} [\href{https://arxiv.org/abs/cond-mat/9603113}{arXiv:cond-mat/9603113}].

\bibitem{PhysRevB.59.R2474}
J.~Gonz\'alez, F.~Guinea and M.~A.~H.~Vozmediano, ``Marginal-Fermi-liquid behavior from two-dimensional Coulomb interaction,''
\href{http://link.aps.org/doi/10.1103/PhysRevB.59.R2474}{Phys.\ Rev.\ B {\bf 59} (1999) R2474} [\href{https://arxiv.org/abs/cond-mat/9807130}{arXiv:cond-mat/9807130 [cond-mat.str-el]}].

\bibitem{PhysRevLett.113.105502}
J.~Hofmann, E.~Barnes and S.~Das~Sarma, ``Why Does Graphene Behave as a Weakly Interacting System?,''
\href{http://dx.doi.org/10.1103/PhysRevLett.113.105502}{Phys.\ Rev.\ Lett.\ {\bf 113} (2014) 105502} [\href{https://arxiv.org/abs/1405.7036}{arXiv:1405.7036 [cond-mat.mes-hall]}].

\bibitem{Kohn61PhysRev.123.1242}
W.~Kohn, ``Cyclotron Resonance and de Haas-van Alphen Oscillations of an Interacting Electron Gas,''
\href{http://dx.doi.org/10.1103/PhysRev.123.1242}{Phys.\ Rev.\ {\bf 123} (1961) 1242}.

\bibitem{PhysRevB.14.4439}
C.\ S.~Ting, S.\ C.~Ying and J.\ J.~Quinn, ``Theory of dynamical conductivity of interacting electrons,''
\href{http://link.aps.org/doi/10.1103/PhysRevB.14.4439}{Phys.\ Rev.\ B {\bf 14} (1976) 4439}.

\bibitem{Throckmorton:2018}
R.~E.\ Throckmorton and S.\ Das Sarma, ``Failure of Kohn's Theorem and f-sum-rule in intrinsic Dirac-Weyl materials in the presence of a filled Fermi sea,''
\href{http://dx.doi.org/10.1103/PhysRevB.98.155112}{Phys.\ Rev.\ B {\bf 98} (2018) 155112} [\href{https://arxiv.org/abs/1805.03650}{arXiv:1805.03650 [cond-mat.mes-hall]}].

\bibitem{Peres:2010mx}
N.~M.~R.~Peres, ``{\textit{Colloquium}  :  The Transport properties of graphene: An Introduction},''
\href{http://dx.doi.org/10.1103/RevModPhys.82.2673}{Rev.\ Mod.\ Phys.\  {\bf 82} (2010) 2673} [\href{https://arxiv.org/abs/1007.2849}{arXiv:1007.2849 [cond-mat.mes-hall]}].
%\href{http://inspirehep.net/record/868637}{Rev.\ Mod.\ Phys.\  {\bf 82} (2010) 2673}
  %%CITATION = doi:10.1103/RevModPhys.82.2673;%%

\bibitem{Gusynin:2006ym}
  V.~P.~Gusynin, S.~G.~Sharapov and J.~P.~Carbotte, ``Unusual microwave response of Dirac quasiparticles in graphene,''
\href{http://dx.doi.org/10.1103/PhysRevLett.96.256802}{Phys.\ Rev.\ Lett.\  {\bf 96} (2006) 256802} [\href{https://arxiv.org/abs/cond-mat/0603267}{arXiv:cond-mat/0603267 [cond-mat.str-el]}].

\bibitem{Gusynin:2009-ac}
V.~P.~Gusynin, S.\ G.~Sharapov and J.\ P.~Carbotte, ``On the universal ac optical background in graphene,''
\href{http://stacks.iop.org/1367-2630/11/i=9/a=095013}{New Journal of Physics {\bf 11} (2009) 095013} [\href{https://arxiv.org/abs/0908.2803}{arXiv:0908.2803 [cond-mat.str-el]}].

\bibitem{PhysRevB.83.195401}
A.~Giuliani, V.~Mastropietro and M.~Porta, ``Absence of interaction corrections in the optical conductivity of graphene,''
\href{http://link.aps.org/doi/10.1103/PhysRevB.83.195401}{Phys. Rev. B {\bf 83} (2011) 195401} [\href{https://arxiv.org/abs/1010.4461}{arXiv:1010.4461 [cond-mat.str-el]}].

\bibitem{Herbut08.PhysRevLett.100.046403}
I.\ F.~Herbut, V.~Juri\v{c}i\'c and O.~Vafek, ``Coulomb Interaction, Ripples, and the Minimal Conductivity of Graphene,''
\href{http://link.aps.org/doi/10.1103/PhysRevLett.100.046403}{Phys.\ Rev.\ Lett.\ {\bf 100} (2008) 046403} [\href{https://arxiv.org/abs/0707.4171}{arXiv:0707.4171 [cond-mat.mes-hall]}].

\bibitem{Mishchenko2008}
E.\ G.~Mishchenko, ``Minimal conductivity in graphene: Interaction corrections and ultraviolet anomaly,''
\href{http://stacks.iop.org/0295-5075/83/i=1/a=17005}{Europhys.\ Lett.\ {\bf 83} (2008) 17005} [\href{https://arxiv.org/abs/0709.4245}{arXiv:0709.4245 [cond-mat.mes-hall]}].

\bibitem{Juricic:2010dm}
V.~Juricic, O.~Vafek and I.~F.~Herbut, ``Conductivity of interacting massless Dirac particles in graphene: Collisionless regime,''
\href{http://dx.doi.org/10.1103/PhysRevB.82.235402}{Phys.\ Rev.\ B {\bf 82} (2010) 235402} [\href{https://arxiv.org/abs/1009.3269}{arXiv:1009.3269 [cond-mat.mes-hall]}].
  %%CITATION = doi:10.1103/PhysRevB.82.235402;%%

\bibitem{Sheehy09.PhysRevB.80.193411}
D.\ E.~Sheehy and J.~Schmalian, ``Optical transparency of graphene as determined by the fine-structure constant,''
\href{http://link.aps.org/doi/10.1103/PhysRevB.80.193411}{Phys.\ Rev.\ B {\bf 80} (2009) 193411} [\href{https://arxiv.org/abs/0906.5164}{arXiv:0906.5164 [cond-mat.mes-hall]}].

\bibitem{Abedinpour11.PhysRevB.84.045429}
S.\ H.~Abedinpour, G.~Vignale, A.~Principi, M.~Polini, W-K.~Tse and A.\ H.~MacDonald, ``Drude weight, plasmon dispersion, and ac conductivity in doped graphene sheets,''
\href{http://link.aps.org/doi/10.1103/PhysRevB.84.045429}{Phys. Rev. B {\bf 84} (2011) 045429} [\href{https://arxiv.org/abs/1101.4291}{arXiv:1101.4291 [cond-mat.mes-hall]}].

\bibitem{Sodemann12.PhysRevB.86.115408}
I.\ Sodemann and M.\ M.~Fogler, ``Interaction corrections to the polarization function of graphene,''
\href{http://link.aps.org/doi/10.1103/PhysRevB.86.115408}{Phys.\ Rev.\ B {\bf 86} (2012) 115408} [\href{https://arxiv.org/abs/1206.3519}{arXiv:1206.3519 [cond-mat.mes-hall]}].

\bibitem{Gazzola13.0295-5075-104-2-27002}
G.~Gazzola, A.\ L.~Cherchiglia, L.\ A.~Cabral, M.\ C.~Nemes and M.~Sampaio, ``Conductivity of Coulomb interacting massless Dirac particles in graphene: Regularization-dependent parameters and symmetry constraints,''
\href{http://stacks.iop.org/0295-5075/104/i=2/a=27002}{Europhys.\ Lett.\ {\bf 104} (2013) 27002} [\href{https://arxiv.org/abs/1305.6334}{arXiv:1305.6334 [cond-mat.mes-hall]}].

\bibitem{Rosenstein13.PhysRevLett.110.066602}
B.~Rosenstein, M.~Lewkowicz and T.~Maniv, ``Chiral Anomaly and Strength of the Electron-Electron Interaction in Graphene,''
\href{http://link.aps.org/doi/10.1103/PhysRevLett.110.066602}{Phys.\ Rev.\ Lett.\ {\bf 110} (2013) 066602} [\href{https://arxiv.org/abs/1210.3345}{arXiv:1210.3345 [cond-mat.str-el]}].

\bibitem{PhysRevB.87.205445}
I.\ F.~Herbut and V.~Mastropietro, ``Universal conductivity of graphene in the ultrarelativistic regime,''
\href{http://link.aps.org/doi/10.1103/PhysRevB.87.205445}{Phys.\ Rev.\ B {\bf 87} (2013) 205445} [\href{https://arxiv.org/abs/1304.1988}{arXiv:1304.1988 [cond-mat.str-el]}].

\bibitem{Link16.PhysRevB.93.235447}
J.~Link, P.\ P.~Orth, D.\ E.~Sheehy and J.~Schmalian, ``Universal collisionless transport of graphene,''
\href{http://link.aps.org/doi/10.1103/PhysRevB.93.235447}{Phys.\ Rev.\ B {\bf 93} (2016) 235447} [\href{https://arxiv.org/abs/1511.05984}{arXiv:1511.05984 [cond-mat.str-el]}].

\bibitem{Boyda:2016emg}
D.\ L.~Boyda, V.\ V.~Braguta, M.\ I.~Katsnelson and M.\ V.~Ulybyshev, ``Many-body effects on graphene conductivity: Quantum Monte Carlo calculations,''
\href{http://dx.doi.org/10.1103/PhysRevB.94.085421}{Phys.\ Rev.\ B {\bf 94} (2016) no.8,  085421} [\href{https://arxiv.org/abs/1601.05315}{arXiv:1601.05315 [cond-mat.str-el]}].

\bibitem{Stauber17.PhysRevLett.118.266801}
T.~Stauber, P.~Parida, M.~Trushin, M.\ V.~Ulybyshev, D.\ L.~Boyda and J.~Schliemann, ``Interacting Electrons in Graphene: Fermi Velocity Renormalization and Optical Response,''
\href{http://dx.doi.org/10.1103/PhysRevLett.118.266801}{Phys.\ Rev.\ Lett.\ {\bf 118} (2017) 266801} [\href{https://arxiv.org/abs/1704.03747}{arXiv:1704.03747 [cond-mat.mes-hall]}].

\bibitem{Fradkin86.PhysRevB.33.3263}
E.~Fradkin, ``Critical behavior of disordered degenerate semiconductors. II. Spectrum and transport properties in mean-field theory,''
\href{http://dx.doi.org/10.1103/PhysRevB.33.3263}{Phys.\ Rev.\ B {\bf 33} (1986) 3263}.

\bibitem{Lee93.PhysRevLett.71.1887}
P.\ A.~Lee, ``Localized states in a d-wave superconductor,''
\href{http://dx.doi.org/10.1103/PhysRevLett.71.1887}{Phys.\ Rev.\ Lett.\ {\bf 71} (1993) 1887}.

\bibitem{Ludwig94.PhysRevB.50.7526}
A.\ W.\ W.~Ludwig, M.\ P.\ A.~Fisher, R.~Shankar and G.~Grinstein, ``Integer quantum Hall transition: An alternative approach and exact results,''
\href{http://link.aps.org/doi/10.1103/PhysRevB.50.7526}{Phys.\ Rev.\ B {\bf 50} (1994) 7526}.

\bibitem{deJuan10.PhysRevB.82.125409}
F.\ de Juan, A.\ G.~Grushin and M.\ A.\ H.~Vozmediano, ``Renormalization of Coulomb interaction in graphene: Determining observable quantities,''
\href{http://dx.doi.org/10.1103/PhysRevB.82.125409}{Phys.\ Rev.\ B {\bf 82} (2010) 125409} [\href{https://arxiv.org/abs/1002.3111}{https://arxiv.org/abs/1002.3111}].

\bibitem{0268-1242-25-6-063001}
M.~Orlita and M.~Potemski, ``Dirac electronic states in graphene systems: optical spectroscopy studies,''
\href{http://stacks.iop.org/0268-1242/25/i=6/a=063001}{Semiconductor Science and Technology {\bf 25} (2010) 063001} [\href{https://arxiv.org/abs/1004.2949}{arXiv:1004.2949 [cond-mat.mes-hall]}].

\bibitem{Juricic:2009px}
  V.~Juricic, I.~F.~Herbut and G.~W.~Semenoff, ``Coulomb interaction at the metal-insulator critical point in graphene,''
\href{http://dx.doi.org/10.1103/PhysRevB.80.081405}{Phys.\ Rev.\ B {\bf 80} (2009) 081405} [\href{https://arxiv.org/abs/0906.3513}{arXiv:0906.3513 [cond-mat.str-el]}].

\bibitem{Semenoff:2011jf}
  G.~W.~Semenoff, ``Chiral Symmetry Breaking in Graphene,''
\href{http://dx.doi.org/10.1088/0031-8949/2012/T146/014016}{Phys.\ Scripta T {\bf 146} (2012) 014016} [\href{https://arxiv.org/abs/1108.2945}{arXiv:1108.2945 [hep-th]}].

\bibitem{Khveshchenko:2001zz}
D.~V.~Khveshchenko, ``Ghost Excitonic Insulator Transition in Layered Graphite,''
\href{http://dx.doi.org/10.1103/PhysRevLett.87.246802}{Phys.\ Rev.\ Lett.\  {\bf 87} (2001) 246802} [\href{https://arxiv.org/abs/cond-mat/0101306}{arXiv:cond-mat/0101306}].
  %%CITATION = doi:10.1103/PhysRevLett.87.246802;%%

\bibitem{Gorbar:2002iw}
E.~V.~Gorbar, V.~P.~Gusynin, V.~A.~Miransky and I.~A.~Shovkovy, ``Magnetic field driven metal insulator phase transition in planar systems,''
\href{http://dx.doi.org/10.1103/PhysRevB.66.045108}{Phys.\ Rev.\ B {\bf 66} (2002) 045108} [\href{https://arxiv.org/abs/cond-mat/0202422}{arXiv:cond-mat/0202422}].
  %%CITATION = doi:10.1103/PhysRevB.66.045108;%%

\bibitem{Leal:2003sg}
H.~Leal and D.~V.~Khveshchenko, ``Excitonic instability in two-dimensional degenerate semimetals,''
\href{http://dx.doi.org/10.1016/j.nuclphysb.2004.03.020}{Nucl.\ Phys.\ B {\bf 687} (2004) 323} [\href{https://arxiv.org/abs/cond-mat/0302164}{arXiv:cond-mat/0302164}].
  %%CITATION = doi:10.1016/j.nuclphysb.2004.03.020;%%

\bibitem{Son07.PhysRevB.75.235423}
D.\ T.~Son, ``Quantum critical point in graphene approached in the limit of infinitely strong Coulomb interaction,''
\href{http://link.aps.org/doi/10.1103/PhysRevB.75.235423}{Phys.\ Rev.\ B {\bf 75} (2007) 235423} [\href{https://arxiv.org/abs/cond-mat/0701501}{arXiv:cond-mat/0701501 [cond-mat.str-el]}].

\bibitem{VafekCase08}
O.~Vafek and M.~J.~Case, ``Renormalization group approach to two-dimensional Coulomb interacting Dirac fermions with random gauge potential,''
\href{http://dx.doi.org/10.1103/PhysRevB.77.033410}{Phys.\ Rev.\ B {\bf 77} (2008) 033410} [\href{https://arxiv.org/abs/0710.2907}{arXiv:0710.2907 [cond-mat.mes-hall]}].

\bibitem{Khveshchenko:2008ye}
D.~V.~Khveshchenko, ``Massive Dirac fermions in single-layer graphene,''
\href{http://dx.doi.org/10.1088/0953-8984/21/7/075303}{J.\ Phys.\ Condens.\ Matter {\bf 21} (2009) 075303} [\href{https://arxiv.org/abs/0807.0676}{arXiv:0807.0676 [cond-mat.str-el]}].
  %%CITATION = doi:10.1088/0953-8984/21/7/075303;%%

\bibitem{Liu09.PhysRevB.79.205429}
G.-Z.~Liu, W.~Li and G.~Cheng, ``Interaction and excitonic insulating transition in graphene,''
\href{http://link.aps.org/doi/10.1103/PhysRevB.79.205429}{Phys.\ Rev.\ B {\bf 79} (2009) 205429} [\href{https://arxiv.org/abs/0811.4471}{arXiv:0811.4471 [cond-mat.str-el]}].

\bibitem{Gamayun:2009em}
O.~V.~Gamayun, E.~V.~Gorbar and V.~P.~Gusynin, ``Gap generation and semimetal-insulator phase transition in graphene,''
\href{http://dx.doi.org/10.1103/PhysRevB.81.075429}{Phys.\ Rev.\ B {\bf 81} (2010) 075429} [\href{https://arxiv.org/abs/0911.4878}{arXiv:0911.4878 [cond-mat.str-el]}].
  %%CITATION = doi:10.1103/PhysRevB.81.075429;%%

\bibitem{Drut:2008rg}
J.~E.~Drut and T.~A.~Lahde, ``Is graphene in vacuum an insulator?,''
\href{http://dx.doi.org/10.1103/PhysRevLett.102.026802}{Phys.\ Rev.\ Lett.\  {\bf 102} (2009) 026802} [\href{https://arxiv.org/abs/0807.0834}{arXiv:0807.0834 [cond-mat.str-el]}].
  %%CITATION = doi:10.1103/PhysRevLett.102.026802;%%

\bibitem{Drut:2009aj}
J.~E.~Drut and T.~A.~Lahde, ``Lattice field theory simulations of graphene,''
\href{http://dx.doi.org/10.1103/PhysRevB.79.165425}{Phys.\ Rev.\ B {\bf 79} (2009) 165425} [\href{https://arxiv.org/abs/0901.0584}{arXiv:0901.0584 [cond-mat.str-el]}].
  %%CITATION = doi:10.1103/PhysRevB.79.165425;%%

\bibitem{Drut:2009zi}
J.~E.~Drut and T.~A.~Lahde, ``Critical exponents of the semimetal-insulator transition in graphene: A Monte Carlo study,''
\href{http://dx.doi.org/10.1103/PhysRevB.79.241405}{Phys.\ Rev.\ B {\bf 79} (2009) 241405} [\href{https://arxiv.org/abs/0905.1320}{arXiv:0905.1320 [cond-mat.str-el]}].
  %%CITATION = doi:10.1103/PhysRevB.79.241405;%%

\bibitem{WangLiu11a}
J.-R.~Wang and G.-Z.~Liu, ``Eliashberg theory of excitonic insulating transition in graphene,''
\href{http://stacks.iop.org/0953-8984/23/i=15/a=155602}{J. Phys.: Condens. Matter {\bf 23} (2011) 155602} [\href{https://arxiv.org/abs/1010.2880}{arXiv:1010.2880 [cond-mat.str-el]}].

\bibitem{WangLiu11b}
J.-R.~Wang and G.-Z.~Liu, ``Dynamic gap generation in graphene under the long-range Coulomb interaction,''
\href{http://stacks.iop.org/0953-8984/23/i=34/a=345601}{J. Phys.: Condens. Matter {\bf 23} (2011) 345601}.

\bibitem{Gonzalez12.PhysRevB.85.085420}
J.~Gonz\'alez, ``Electron self-energy effects on chiral symmetry breaking in graphene,''
\href{http://link.aps.org/doi/10.1103/PhysRevB.85.085420}{Phys.\ Rev.\ B {\bf 85} (2012) 085420} [\href{https://arxiv.org/abs/1202.0443}{arXiv:1202.0443 [cond-mat.mes-hall]}].

\bibitem{Wang2012}
J.-R.~Wang and G.-Z.~Liu, ``Absence of dynamical gap generation in suspended graphene,''
\href{http://dx.doi.org/10.1088/1367-2630/14/4/043036}{New J.\ Phys.\ {\bf 14} (2012) 043036} [\href{https://arxiv.org/abs/1202.1014}{arXiv:1202.1014 [cond-mat.str-el]}].

\bibitem{Buividovich12.PhysRevB.86.245117}
P.\ V.~Buividovich and M.\ I.~Polikarpov, ``Monte Carlo study of the electron transport properties of monolayer graphene within the tight-binding model,''
\href{http://link.aps.org/doi/10.1103/PhysRevB.86.245117}{Phys.\ Rev.\ B {\bf 86} (2012) 245117} [\href{https://arxiv.org/abs/1206.0619}{arXiv:1206.0619 [cond-mat.str-el]}].

\bibitem{Ulybyshev:2013swa}
M.~V.~Ulybyshev, P.~V.~Buividovich, M.~I.~Katsnelson and M.~I.~Polikarpov,
``Monte-Carlo study of the semimetal-insulator phase transition in monolayer graphene with realistic inter-electron interaction potential,''
\href{http://dx.doi.org/10.1103/PhysRevLett.111.056801}{Phys.\ Rev.\ Lett.\  {\bf 111} (2013) 056801} [\href{https://arxiv.org/abs/1304.3660}{arXiv:1304.3660 [cond-mat.str-el]}].
  %%CITATION = doi:10.1103/PhysRevLett.111.056801;%%

\bibitem{Popovic13.PhysRevB.88.205429}
C.~Popovici, C.\ S.~Fischer and L.\ von~Smekal, ``Fermi velocity renormalization and dynamical gap generation in graphene,''
\href{http://link.aps.org/doi/10.1103/PhysRevB.88.205429}{Phys.\ Rev.\ B {\bf 88} (2013) 205429} [\href{https://arxiv.org/abs/1308.6199}{arXiv:1308.6199 [hep-ph]}].

\bibitem{Gonzalez15.PhysRevB.92.125115}
J.~Gonz\'alez, ``Phase diagram of the quantum electrodynamics of two-dimensional and three-dimensional Dirac semimetals,''
\href{http://link.aps.org/doi/10.1103/PhysRevB.92.125115}{Phys.\ Rev.\ B {\bf 92} (2015) 125115} [\href{https://arxiv.org/abs/1502.07640}{arXiv:1502.07640 [cond-mat.mes-hall]}].

\bibitem{Katanin16.PhysRevB.93.035132}
A.\ A.~Katanin, ``Effect of vertex corrections on the possibility of chiral symmetry breaking induced by long-range Coulomb repulsion in graphene,''
\href{http://link.aps.org/doi/10.1103/PhysRevB.93.035132}{Phys.\ Rev.\ B {\bf 93} (2016) 035132} [\href{https://arxiv.org/abs/1508.07224}{arXiv:1508.07224 [cond-mat.str-el]}].

\bibitem{Carrington:2017hlc}
  M.~E.~Carrington, C.~S.~Fischer, L.~von Smekal and M.~H.~Thoma, ``The role of frequency dependence in dynamical gap generation in graphene,''
\href{http://doi.org/10.5506/APhysPolBSupp.10.519}{Acta Phys.\ Pol.\ B Proc.\ Suppl.\ {\bf 10} (2017) 519} [\href{https://arxiv.org/abs/1711.01962}{arXiv:1711.01962 [cond-mat.mes-hall]}].

\bibitem{nevius2015}
M.\ S.~Nevius, M.~Conrad, F.~Wang, A.~Celis, M.\ N.~Nair, A.~Taleb-Ibrahimi, A.~Tejeda and E.\ H.~Conrad, ``Semiconducting Graphene from Highly Ordered Substrate Interactions,''
\href{http://link.aps.org/doi/10.1103/PhysRevLett.115.136802}{Phys.\ Rev.\ Lett.\ {\bf 115} 136802 (2015)} [\href{https://arxiv.org/abs/1505.00435}{arXiv:1505.00435 [cond-mat.mtrl-sci]}].

\bibitem{neto2009}
A.\ H.~Castro Neto, ``Pauling's dreams for graphene,''
\href{http://link.aps.org/10.1103/Physics.2.30}{Physics Online Journal {\bf 2} (2009) 30}.

%\bibitem{Ulybyshev:2013swa}
%  M.~V.~Ulybyshev, P.~V.~Buividovich, M.~I.~Katsnelson and M.~I.~Polikarpov,
%``Monte-Carlo study of the semimetal-insulator phase transition in monolayer graphene with realistic inter-electron interaction potential,''
%\href{http://dx.doi.org/10.1103/PhysRevLett.111.056801}{Phys.\ Rev.\ Lett.\  {\bf 111} (2013) 056801} [\href{https://arxiv.org/abs/1304.3660}{arXiv:1304.3660 [cond-mat.str-el]}].

\bibitem{Miransky:2015ava}
V.~A.~Miransky and I.~A.~Shovkovy, ``Quantum field theory in a magnetic field: From quantum chromodynamics to graphene and Dirac semimetals,''
\href{http://dx.doi.org/10.1016/j.physrep.2015.02.003}{Phys.\ Rept.\  {\bf 576} (2015) 1} [\href{https://arxiv.org/abs/1503.00732}{arXiv:1503.00732 [hep-ph]}].
  %%CITATION = doi:10.1016/j.physrep.2015.02.003;%%

\bibitem{Gusynin:2013noa}
V.~P.~Gusynin, ``Graphene and quantum electrodynamics,''
\href{http://vant.kipt.kharkov.ua/ARTICLE/VANT_2013_3/article_2013_3_29.pdf}{Prob.\ Atomic Sci.\ Technol.\  {\bf 2013N3} (2013) 29}.
  %%CITATION = PASPF,2013N3,29;%%

\bibitem{DeTar:2016vhr}
  C.~DeTar, C.~Winterowd and S.~Zafeiropoulos,
``Magnetic Catalysis in Graphene Effective Field Theory,''
\href{http://dx.doi.org/10.1103/PhysRevLett.117.266802}{Phys.\ Rev.\ Lett.\  {\bf 117} (2016) no.26,  266802} [\href{https://arxiv.org/abs/1607.03137}{arXiv:1607.03137 [hep-lat]}].

\bibitem{DeTar:2016dmj}
  C.~DeTar, C.~Winterowd and S.~Zafeiropoulos,
``Lattice Field Theory Study of Magnetic Catalysis in Graphene,''
\href{http://dx.doi.org/10.1103/PhysRevB.95.165442}{Phys.\ Rev.\ B {\bf 95} (2017) no.16,  165442} [\href{https://arxiv.org/abs/1608.00666}{arXiv:1608.00666 [hep-lat]}].

\bibitem{PhysRevLett.116.116803}
H.~Isobe and N.~Nagaosa, ``Coulomb Interaction Effect in Weyl Fermions with Tilted Energy Dispersion in Two Dimensions,''
\href{http://link.aps.org/doi/10.1103/PhysRevLett.116.116803}{Phys.\ Rev.\ Lett.\ {\bf 116} (2016) 116803} [\href{https://arxiv.org/abs/1512.08704}{arXiv:1512.08704 [cond-mat.str-el]}].

\bibitem{PhysRevB.89.235431}
E.~Barnes, E.\ H.~Hwang, R.\ E.~Throckmorton and S.~Das Sarma, ``Effective field theory, three-loop perturbative expansion, and their experimental implications in graphene many-body effects,''
\href{http://link.aps.org/doi/10.1103/PhysRevB.89.235431}{Phys.\ Rev.\ B {\bf 89} (2014) 235431} [\href{https://arxiv.org/abs/1401.7011}{arXiv:1401.7011 [cond-mat.mes-hall]}].

\bibitem{Gorbar:2001qt}
E.~V.~Gorbar, V.~P.~Gusynin and V.~A.~Miransky, ``Dynamical chiral symmetry breaking on a brane in reduced QED,''
\href{http://dx.doi.org/10.1103/PhysRevD.64.105028}{Phys.\ Rev.\ D {\bf 64} (2001) 105028} [\href{https://arxiv.org/abs/hep-ph/0105059}{arXiv:hep-ph/0105059}].
  %%CITATION = doi:10.1103/PhysRevD.64.105028;%%

\bibitem{Marino:1992xi}
E.~C.~Marino, ``Quantum electrodynamics of particles on a plane and the Chern-Simons theory,''
\href{http://dx.doi.org/10.1016/0550-3213(93)90379-4}{Nucl.\ Phys.\ B {\bf 408} (1993) 551} [\href{https://arxiv.org/abs/hep-th/9301034}{arXiv:hep-th/9301034}].
  %%CITATION = doi:10.1016/0550-3213(93)90379-4;%%

\bibitem{Dorey:1991kp}
N.~Dorey and N.~E.~Mavromatos, ``QED in three-dimension and two-dimensional superconductivity without parity violation,''
\href{http://dx.doi.org/10.1016/0550-3213(92)90632-L}{Nucl.\ Phys.\ B {\bf 386} (1992) 614}.
  %%CITATION = doi:10.1016/0550-3213(92)90632-L;%%

\bibitem{Kovner:1990zz}
A.~Kovner and B.~Rosenstein, ``Kosterlitz-Thouless mechanism of two-dimensional superconductivity,''
\href{http://dx.doi.org/10.1103/PhysRevB.42.4748}{Phys.\ Rev.\ B {\bf 42} (1990) 4748}.
  %%CITATION = doi:10.1103/PhysRevB.42.4748;%%

\bibitem{Kaplan:2009kr}
  D.~B.~Kaplan, J.~W.~Lee, D.~T.~Son and M.~A.~Stephanov, ``Conformality Lost,''
\href{http://dx.doi.org/10.1103/PhysRevD.80.125005}{Phys.\ Rev.\ D {\bf 80} (2009) 125005} [\href{https://arxiv.org/abs/0905.4752}{arXiv:0905.4752 [hep-th]}].

\bibitem{Gracey:2006jc}
  J.~A.~Gracey, ``Practicalities of renormalizing quantum field theories,''
\href{https://arxiv.org/abs/hep-th/0605037}{hep-th/0605037}.

\bibitem{samko1993fractional}
S.~Samko, A.\ A.~Kilbas and O.~Marichev, ``Fractional Integrals and Derivatives,''
\newblock Taylor \& Francis (1993). 

\bibitem{Caffarelli07}
L.~Caffarelli and  L.~Silvestre, ``An Extension Problem Related to the Fractional Laplacian,''
\href{http://dx.doi.org/10.1080/03605300600987306}{Communications in Partial Differential Equations {\bf 32} (2007) 1245} [\href{https://arxiv.org/abs/math/0608640}{arXiv:math/0608640 [math.AP]}].

\bibitem{Rajabpour:2011qr}
  M.~A.~Rajabpour, ``Conformal symmetry in non-local field theories,''
\href{http://dx.doi.org/10.1007/JHEP06(2011)076}{JHEP {\bf 1106} (2011) 076} [\href{https://arxiv.org/abs/1103.3625}{arXiv:1103.3625 [hep-th]}].

\bibitem{Herzog:2017xha}
  C.~P.~Herzog and K.~W.~Huang, ``Boundary Conformal Field Theory and a Boundary Central Charge,''
\href{http://dx.doi.org/10.1007/JHEP10(2017)189}{JHEP {\bf 1710} (2017) 189} [\href{https://arxiv.org/abs/1707.06224}{arXiv:1707.06224 [hep-th]}].

\bibitem{Limtragool:2016gnl}
  K.~Limtragool and P.~W.~Phillips,
  ``Anomalous dimension of the electrical current in strange metals from the fractional Aharonov-Bohm effect,''
\href{http://dx.doi.org/10.1209/0295-5075/121/27003}{EPL {\bf 121} (2018) no.2,  27003} [\href{https://arxiv.org/abs/1601.02340}{arXiv:1601.02340 [cond-mat.str-el]}].


\bibitem{LaNave:2017lwf} 
  G.~La Nave and P.~Phillips, ``Exact Form of Boundary Operators Dual to Interacting Bulk Scalar Fields in the AdS/CFT Correspondence,''
\href{https://arxiv.org/abs/1702.00038}{arXiv:1702.00038 [hep-th]}.

\bibitem{LaNave:2017nex} 
  G.~La Nave and P.~Phillips,
 ``Anomalous Dimensions for Boundary Conserved Currents in Holography via the Caffarelli-Silvestri Mechanism for p-forms,''
\href{https://arxiv.org/abs/1708.00863}{arXiv:1708.00863 [hep-th]}.

%\bibitem{PhysRevB.95.075129}
%J.-R.~Wang, G.-Z.~Liu and C.-J.~Zhang, ``Excitonic pairing and insulating transition in two-dimensional semi-Dirac semimetals,''
%\href{http://dx.doi.org/10.1103/PhysRevB.95.075129}{Phys.\ Rev.\ B {\bf 95} (2017) 075129} [\href{https://arxiv.org/abs/1611.00145}{arXiv:1611.00145 [cond-mat.str-el]}].

\bibitem{Teber:2012de}
S.~Teber, ``Electromagnetic current correlations in reduced quantum electrodynamics,''
\href{http://dx.doi.org/10.1103/PhysRevD.86.025005}{Phys.\ Rev.\ D {\bf 86} (2012) 025005} [\href{https://arxiv.org/abs/1204.5664}{arXiv:1204.5664 [hep-ph]}].
  %%CITATION = doi:10.1103/PhysRevD.86.025005;%%

\bibitem{Kotikov:2013kcl}
A.~V.~Kotikov and S.~Teber, ``Note on an application of the method of uniqueness to reduced quantum electrodynamics,''
\href{http://dx.doi.org/10.1103/PhysRevD.87.087701}{Phys.\ Rev.\ D {\bf 87} (2013) 087701} [\href{https://arxiv.org/abs/1302.3939}{arXiv:1302.3939 [hep-ph]}].
  %%CITATION = doi:10.1103/PhysRevD.87.087701;%%

\bibitem{Kotikov:2013eha}
A.~V.~Kotikov and S.~Teber, ``Two-loop fermion self-energy in reduced quantum electrodynamics and application to the ultrarelativistic limit of graphene,''
\href{http://link.aps.org/10.1103/PhysRevD.89.065038}{Phys.\ Rev.\ D {\bf 89} (2014) 065038} [\href{https://arxiv.org/abs/1312.2430}{arXiv:1312.2430 [hep-ph]}].
  %%CITATION = doi:10.1103/PhysRevD.89.065038;%%

\bibitem{Teber:2014hna}
S.~Teber, ``Two-loop fermion self-energy and propagator in reduced QED$_{3,2}$,''
\href{http://link.aps.org/10.1103/PhysRevD.89.067702}{Phys.\ Rev.\ D {\bf 89} (2014)  067702} [\href{https://arxiv.org/abs/1402.5032}{arXiv:1402.5032 [hep-ph]}].
  %%CITATION = doi:10.1103/PhysRevD.89.067702;%%

\bibitem{Teber:2014ita}
S.~Teber and A.~V.~Kotikov, ``Interaction corrections to the minimal conductivity of graphene via dimensional regularization,''
\href{http://dx.doi.org/10.1209/0295-5075/107/57001}{Europhys.\ Lett.\  {\bf 107} (2014) 57001} [\href{https://arxiv.org/abs/1407.7501}{arXiv:1407.7501 [cond-mat.mes-hall]}].
  %%CITATION = doi:10.1209/0295-5075/107/57001;%%

\bibitem{Teber:2016unz}
S.~Teber and A.~V.~Kotikov, ``The method of uniqueness and the optical conductivity of graphene: new application of a powerful technique for multi-loop calculations,''
\href{http://dx.doi.org/10.1134/S004057791703014X}{Theoretical and Mathematical Physics {\bf 190} (2017) 446}
[\href{https://doi.org/10.4213/tmf9161}{Teor.\ Mat.\ Fiz.\  {\bf 190} (2017) no.3,  519}]
 %[\href{https://inspirehep.net/record/1419982}{arXiv:1602.01962 [hep-th]}]. 
 [\href{https://arxiv.org/abs/1602.01962}{arXiv:1602.01962 [hep-th]}].
  %%CITATION = ARXIV:1602.01962;%%

\bibitem{Kotikov:2016wrb}
A.~V.~Kotikov, V.~I.~Shilin and S.~Teber, ``Critical behaviour of ($2+1$)-dimensional QED: $1/N_f$-corrections in the Landau gauge,''
\href{http://dx.doi.org/10.1103/PhysRevD.94.056009}{Phys.\ Rev.\ D {\bf 94} (2016) 056009} [\href{https://arxiv.org/abs/1605.01911}{arXiv:1605.01911 [hep-th]}].
  %%CITATION = ARXIV:1605.01911;%%

\bibitem{Kotikov:2016prf}
  A.~V.~Kotikov and S.~Teber, ``Critical behavior of ($2+1$)-dimensional QED: $1/N_f$ corrections in an arbitrary nonlocal gauge,''
\href{http://dx.doi.org/10.1103/PhysRevD.94.114011}{Phys.\ Rev.\ D {\bf 94} (2016) no.11,  114011} [\href{https://arxiv.org/abs/1609.06912}{arXiv:1609.06912 [hep-th]}].

\bibitem{Kotikov:2016yrn}
  A.~V.~Kotikov and S.~Teber, ``Critical behaviour of reduced QED$_{4,3}$ and dynamical fermion gap generation in graphene,''
\href{http://dx.doi.org/10.1103/PhysRevD.94.114010}{Phys.\ Rev.\ D {\bf 94} (2016) no.11,  114010} [\href{https://arxiv.org/abs/1610.00934}{arXiv:1610.00934 [hep-th]}].

\bibitem{Kotikov:2016rgs}
  A.~V.~Kotikov and S.~Teber, ``New results for a two-loop massless propagator-type Feynman diagram,''
\href{http://dx.doi.org/10.1134/S0040577918020083}{Theor.\ Math.\ Phys.\  {\bf 194} (2018) no.2,  284}
[\href{http://dx.doi.org/10.4213/tmf9340}{Teor.\ Mat.\ Fiz.\  {\bf 194} (2018) no.2,  331}]
[\href{https://arxiv.org/abs/1611.07240}{arXiv:1611.07240 [hep-th]}].

\bibitem{KotikovT-unpublished}
A.~V.~Kotikov and S.~Teber (unpublished).

\bibitem{Teber:2018goo}
  S.~Teber and A.~V.~Kotikov,
``Field theoretic renormalization study of reduced quantum electrodynamics and applications to the ultrarelativistic limit of Dirac liquids,''
\href{http://dx.doi.org/10.1103/PhysRevD.97.074004}{Phys.\ Rev.\ D {\bf 97} (2018) no.7,  074004} [\href{https://arxiv.org/abs/1801.10385}{arXiv:1801.10385 [hep-th]}].

\bibitem{Teber:2018qcn}
  S.~Teber and A.~V.~Kotikov, ``Field theoretic renormalization study of interaction corrections to the universal ac conductivity of graphene,''
\href{http://dx.doi.org/10.1007/JHEP07(2018)082}{JHEP {\bf 1807} (2018) 082} [\href{https://arxiv.org/abs/1802.09898}{arXiv:1802.09898 [cond-mat.mes-hall]}].

\bibitem{Kotikov:2018wxe}
  A.~V.~Kotikov and S.~Teber, ``Multi-loop techniques for massless Feynman diagram calculations,''
\href{https://arxiv.org/abs/1805.05109}{arXiv:1805.05109 [hep-th]} (accepted for publication in Phys.\ Part.\ Nucl.).

\bibitem{Bogner:2017xhp}
  C.~Bogner, S.~Borowka, T.~Hahn, G.~Heinrich, S.\ P.~Jones, M.~Kerner, A.~von Manteuffel, M.~Michel, E.~Panzer and V.~Papara, 
 ``Loopedia, a Database for Loop Integrals,''
 \href{http://dx.doi.org/10.1016/j.cpc.2017.12.017}{Comput.\ Phys.\ Commun.\  {\bf 225} (2018) 1}
 [\href{https://arxiv.org/abs/1709.01266}{arXiv:1709.01266 [hep-ph]}].

\bibitem{Kazakov:1984bw}
  D.~I.~Kazakov, ``Analytical Methods For Multiloop Calculations: Two Lectures On The Method Of Uniqueness,''
\href{https://inspirehep.net/record/203305}{JINR-E2-84-410}.

\bibitem{Leibbrandt:1975dj}
  G.~Leibbrandt, ``Introduction to the Technique of Dimensional Regularization,''
\href{http://dx.doi.org/10.1103/RevModPhys.47.849}{Rev.\ Mod.\ Phys.\  {\bf 47} (1975) 849}.

\bibitem{Narison:1980ti}
  S.~Narison, ``Techniques of Dimensional Renormalization and Applications to the Two Point Functions of QCD and QED,''
\href{http://dx.doi.org/10.1016/0370-1573(82)90023-0}{Phys.\ Rept.\  {\bf 84} (1982) 263}.

\bibitem{Kilgore:2011ta}
  W.~B.~Kilgore, ``Regularization Schemes and Higher Order Corrections,''
\href{http://dx.doi.org/10.1103/PhysRevD.83.114005}{Phys.\ Rev.\ D {\bf 83} (2011) 114005} [\href{https://arxiv.org/abs/1102.5353}{arXiv:1102.5353 [hep-ph]}].

\bibitem{Chawdhry:2018awn}
  H.~A.~Chawdhry, M.~A.~Lim and A.~Mitov,
``Two-loop five-point massless QCD amplitudes within the IBP approach,''
\href{https://arxiv.org/abs/1805.09182}{arXiv:1805.09182 [hep-ph]}.

\bibitem{Passarino:1978jh}
  G.~Passarino and M.~J.~G.~Veltman, ``One Loop Corrections for e+ e- Annihilation Into mu+ mu- in the Weinberg Model,''
\href{http://dx.doi.org/10.1016/0550-3213(79)90234-7}{Nucl.\ Phys.\ B {\bf 160} (1979) 151.}

\bibitem{Denner:1991kt}
  A.~Denner, ``Techniques for calculation of electroweak radiative corrections at the one loop level and results for W physics at LEP-200,''
\href{http://dx.doi.org/10.1002/prop.2190410402}{Fortsch.\ Phys.\  {\bf 41} (1993) 307} [\href{https://arxiv.org/abs/0709.1075}{arXiv:0709.1075 [hep-ph]}].

\bibitem{Shtabovenko:2016sxi}
  V.~Shtabovenko, R.~Mertig and F.~Orellana, ``New Developments in FeynCalc 9.0,''
\href{http://dx.doi.org/10.1016/j.cpc.2016.06.008}{Comput.\ Phys.\ Commun.\  {\bf 207} (2016) 432} [\href{https://arxiv.org/abs/1601.01167}{arXiv:1601.01167 [hep-ph]}].

\bibitem{Mertig:1990an}
  R.~Mertig, M.~Bohm and A.~Denner, ``FEYN CALC: Computer algebraic calculation of Feynman amplitudes,''
\href{http://dx.doi.org/10.1016/10.1016/0010-4655(91)90130-D}{Comput.\ Phys.\ Commun.\  {\bf 64} (1991) 345}.

\bibitem{Gorishnii:1984te}
  S.~G.~Gorishnii and A.~P.~Isaev, ``On an Approach to the Calculation of Multiloop Massless Feynman Integrals,''
\href{http://dx.doi.org/10.1007/BF01018263}{Theor.\ Math.\ Phys.\  {\bf 62} (1985) 232 [Teor.\ Mat.\ Fiz.\  {\bf 62} (1985) 345]}.

\bibitem{Baikov:2010hf}
  P.~A.~Baikov and K.~G.~Chetyrkin, ``Four Loop Massless Propagators: An Algebraic Evaluation of All Master Integrals,''
\href{http://dx.doi.org/10.1016/j.nuclphysb.2010.05.004}{Nucl.\ Phys.\ B {\bf 837} (2010) 186} [\href{https://arxiv.org/abs/1004.1153}{arXiv:1004.1153 [hep-ph]}].

\bibitem{Grozin:2012xi}
  A.~G.~Grozin, ``Massless two-loop self-energy diagram: Historical review,''
\href{http://dx.doi.org/10.1142/S0217751X12300189}{Int.\ J.\ Mod.\ Phys.\ A {\bf 27} (2012) 1230018} [\href{https://arxiv.org/abs/1206.2572}{arXiv:1206.2572 [hep-ph]}].

\bibitem{Fleischer:1998nb}
  J.~Fleischer, A.~V.~Kotikov and O.~L.~Veretin, ``Analytic two loop results for selfenergy type and vertex type diagrams with one nonzero mass,''
\href{http://dx.doi.org/10.1016/S0550-3213(99)00078-4}{Nucl.\ Phys.\ B {\bf 547} (1999) 343} [\href{https://arxiv.org/abs/hep-ph/9808242}{hep-ph/9808242}].

\bibitem{Kotikov:2001sc}
  A.~V.~Kotikov and L.~N.~Lipatov, ``DGLAP and BFKL evolution equations in the N=4 supersymmetric gauge theory,'' [\href{https://arxiv.org/abs/hep-ph/0112346}{hep-ph/0112346}].

\bibitem{Kotikov:2002ab}
  A.~V.~Kotikov and L.~N.~Lipatov, ``DGLAP and BFKL equations in the $N=4$ supersymmetric gauge theory,''
\href{http://dx.doi.org/10.1016/S0550-3213(03)00264-5}{Nucl.\ Phys.\ B {\bf 661} (2003) 19} Erratum: \href{http://dx.doi.org/10.1016/j.nuclphysb.2004.02.032}{[Nucl.\ Phys.\ B {\bf 685} (2004) 405]} 
[\href{https://arxiv.org/abs/hep-ph/0208220}{hep-ph/0208220}].

\bibitem{Isaev:2003tk}
  A.~P.~Isaev, ``Multiloop Feynman integrals and conformal quantum mechanics,''
\href{http://dx.doi.org/10.1016/S0550-3213(03)00393-6}{Nucl.\ Phys.\ B {\bf 662} (2003) 461} [\href{https://arxiv.org/abs/hep-th/0303056}{hep-th/0303056}].

\bibitem{Weinberg:1959nj}
  S.~Weinberg, ``High-energy behavior in quantum field theory,''
\href{http://dx.doi.org/10.1103/PhysRev.118.838}{Phys.\ Rev.\  {\bf 118} (1960) 838}.
  %%CITATION = doi:10.1103/PhysRev.118.838;%%

\bibitem{Kissler:2016gxn}
H.~Kissler, ``Hopf-algebraic Renormalization of QED in the linear covariant Gauge,''
\href{http://dx.doi.org/10.1016/j.aop.2016.05.008}{Annals Phys.\  {\bf 372} (2016) 159} [\href{https://arxiv.org/abs/1602.07003}{arXiv:1602.07003 [hep-th]}.

\bibitem{PhysRevLett.80.5409}
J.~Ye and S.~Sachdev, ``Coulomb Interactions at Quantum Hall Critical Points of Systems in a Periodic Potential,''
\href{http://dx.doi.org/10.1103/PhysRevLett.80.5409}{Phys.\ Rev.\ Lett.\ {\bf 80} (1998) 5409} [\href{https://arxiv.org/abs/cond-mat/9712161}{arXiv:cond-mat/9712161 [cond-mat.mes-hall]}].

\bibitem{PhysRevLett.87.137004}
I.\ F.~Herbut, ``Quantum Critical Points with the Coulomb Interaction and the Dynamical Exponent: When and Why $\mathit{z}\phantom{\rule{0ex}{0ex}}=\phantom{\rule{0ex}{0ex}}1$,''
\href{http://dx.doi.org/10.1103/PhysRevLett.87.137004}{Phys.\ Rev.\ Lett.\ {\bf 87} (2001) 137004} [\href{https://arxiv.org/abs/cond-mat/0105544}{arXiv:cond-mat/0105544 [cond-mat.supr-con]}].

\bibitem{PhysRev.128.2425}
J.~Schwinger, ``Gauge Invariance and Mass. II,''
\href{http://dx.doi.org/10.1103/PhysRev.128.2425}{Phys.\ Rev.\ {\bf 128} (1962) 2425}.

\bibitem{Dimopoulos:2000iq}
  P.~Dimopoulos, K.~Farakos, C.~P.~Korthals-Altes, G.~Koutsoumbas and S.~Nicolis,
  ``Phase structure of the 5-D Abelian Higgs model with anisotropic couplings,''
\href{http://dx.doi.org/10.1088/1126-6708/2001/02/005}{JHEP {\bf 0102} (2001) 005} [\href{https://arxiv.org/abs/hep-lat/0012028}{arXiv:hep-lat/0012028}].

\bibitem{Shirkov:1989bp}
  D.~V.~Shirkov, ``Nonlocal Renormalization 'stopping' The Running Gauge,''
\href{http://dx.doi.org/10.1016/0550-3213(90)90103-K}{Nucl.\ Phys.\ B {\bf 332} (1990) 425}.

\bibitem{Jackiw:1980kv}
  R.~Jackiw and S.~Templeton, ``How Superrenormalizable Interactions Cure their Infrared Divergences,''
\href{http://dx.doi.org/10.1103/PhysRevD.23.2291}{Phys.\ Rev.\ D {\bf 23} (1981) 2291}.

\bibitem{Guendelman:1982fm}
  E.~I.~Guendelman and Z.~M.~Radulovic, ``LOOP EXPANSION IN MASSLESS QED in three-dimensions,''
\href{http://dx.doi.org/10.1103/PhysRevD.27.357}{Phys.\ Rev.\ D {\bf 27} (1983) 357}.

\bibitem{Guendelman:1983dt}
  E.~I.~Guendelman and Z.~M.~Radulovic, ``Infrared Divergences in Three-dimensional Gauge Theories,''
\href{http://dx.doi.org/10.1103/PhysRevD.30.1338}{Phys.\ Rev.\ D {\bf 30} (1984) 1338}.

\bibitem{King:1985hr}
  I.~D.~King and G.~Thompson,
``Nonperturbative Analysis Of Leading Logarithms In Three-dimensional Qed,''
\href{http://dx.doi.org/10.1103/PhysRevD.31.2148}{Phys.\ Rev.\ D {\bf 31} (1985) 2148}.

\bibitem{Gusynin:2000zb}
  V.~Gusynin, A.~Hams and M.~Reenders, ``Nonperturbative infrared dynamics of three-dimensional QED with four fermion interaction,''
\href{http://dx.doi.org/10.1103/PhysRevD.63.045025}{Phys.\ Rev.\ D {\bf 63} (2001) 045025} [\href{https://arxiv.org/abs/hep-ph/0005241}{hep-ph/0005241}].

\bibitem{Collins:1974bg}
  J.~C.~Collins, ``Structure of Counterterms in Dimensional Regularization,''
\href{http://dx.doi.org/10.1016/0550-3213(74)90521-5}{Nucl.\ Phys.\ B {\bf 80} (1974) 341}.

\bibitem{Grisaru:1986wj}
  M.~T.~Grisaru, D.~I.~Kazakov and D.~Zanon,
``Five Loop Divergences for the $N=2$ Supersymmetric Nonlinear $\sigma$ Model,''
\href{http://dx.doi.org/10.1016/0550-3213(87)90102-7}{Nucl.\ Phys.\ B {\bf 287} (1987) 189}.

\bibitem{Herbut:2013kca}
  I.~F.~Herbut and V.~Mastropietro, ``Universal conductivity of graphene in the ultrarelativistic regime,''
\href{http://dx.doi.org/10.1103/PhysRevB.87.205445}{Phys.\ Rev.\ B {\bf 87} (2013) no.20,  205445} [\href{https://arxiv.org/abs/1304.1988}{arXiv:1304.1988 [cond-mat.str-el]}].

\bibitem{Valenzuela:2014uia}
  D.~Valenzuela, S.~Hern\'andez-Ortiz, M.~Loewe and A.~Raya, ``Graphene transparency in weak magnetic fields,''
\href{http://dx.doi.org/10.1088/1751-8113/48/6/065402}{J.\ Phys.\ A {\bf 48} (2015) no.6,  065402} [\href{https://arxiv.org/abs/1410.5501}{arXiv:1410.5501 [cond-mat.mes-hall]}].

\bibitem{Hernandez-Ortiz:2015wua}
  S.~Hern\'andez-Ortiz, D.~Valenzuela, A.~Raya and S.~S\'anchez-Madrigal, ``Light absorption in distorted graphene,''
\href{http://dx.doi.org/10.1142/S0217979216500843}{Int.\ J.\ Mod.\ Phys.\  {\bf 30} (2016) no.14,  1650084} [\href{https://arxiv.org/abs/1509.06717}{arXiv:1509.06717 [cond-mat.mes-hall]}].

\bibitem{Marino:2015uda}
  E.~C.~Marino, L.~O.~Nascimento, V.~S.~Alves and C.~M.~Smith, ``Interaction Induced Quantum Valley Hall Effect in Graphene,''
\href{http://dx.doi.org/10.1103/PhysRevX.5.011040}{Phys.\ Rev.\ X {\bf 5} (2015) no.1,  011040} [\href{https://arxiv.org/abs/1309.5879}{arXiv:1309.5879 [cond-mat.str-el]}].

\bibitem{Kooi:2017ugi}
  S.~H.~Kooi, N.~Menezes, V.~S.~Alves and C.~M.~Smith, ``Quantum Valley Hall Effect in Massive Dirac Systems Coupled to a Scalar Field,''
\href{https://arxiv.org/abs/1702.02812}{arXiv:1702.02812 [cond-mat.str-el]}.

\bibitem{Alves:2017fij}
  V.~S.~Alves, R.~O.~C.~Junior, E.~C.~Marino and L.~O.~Nascimento, ``Dynamical Mass Generation in Pseudo Quantum Electrodynamics with Four-Fermion Interactions,''
\href{http://dx.doi.org/10.1103/PhysRevD.96.034005}{Phys.\ Rev.\ D {\bf 96} (2017) 034005} [\href{https://arxiv.org/abs/1704.00381}{arXiv:1704.00381 [cond-mat.str-el]}].

\bibitem{Barros:2017ygp}
  J.~C.~P.~Barros, M.~Dalmonte and A.~Trombettoni,
``Long-range interactions from $U\left(1\right)$ gauge fields via dimensional mismatch,''
\href{https://arxiv.org/abs/1708.06585}{arXiv:1708.06585 [cond-mat.stat-mech]}.

\bibitem{Marino:2014oba}
  E.~C.~Marino, L.~O.~Nascimento, V.~S.~Alves and C.~M.~Smith, ``Unitarity of theories containing fractional powers of the d?Alembertian operator,''
\href{http://dx.doi.org/10.1103/PhysRevD.90.105003}{Phys.\ Rev.\ D {\bf 90} (2014) no.10,  105003} [\href{https://arxiv.org/abs/1408.1637}{arXiv:1408.1637 [hep-th]}].

\bibitem{Ahmad:2016dsb}
  A.~Ahmad, J.~J.~Cobos-Mart\'inez, Y.~Concha-S\'anchez and A.~Raya, ``Landau-Khalatnikov-Fradkin transformations in Reduced Quantum Electrodynamics,''
\href{http://dx.doi.org/10.1103/PhysRevD.93.094035}{Phys.\ Rev.\ D {\bf 93} (2016) no.9,  094035} [\href{https://arxiv.org/abs/1604.03886}{arXiv:1604.03886 [hep-ph]}].

\bibitem{Karch:2018uft}
  A.~Karch and Y.~Sato,
``Conformal Manifolds with Boundaries or Defects,''
\href{http://dx.doi.org/10.1007/JHEP07(2018)156}{JHEP {\bf 1807} (2018) 156}  [\href{https://arxiv.org/abs/1805.10427}{arXiv:1805.10427 [hep-th]}].

\bibitem{Dudal:2018pta}
  D.~Dudal, A.~J.~Mizher and P.~Pais,
``On the exact quantum scale invariance of three-dimensional reduced QED theories,''
\href{https://arxiv.org/abs/1808.04709}{arXiv:1808.04709 [hep-th]}.

\bibitem{Herzog:2018lqz}
  C.~P.~Herzog, K.~W.~Huang, I.~Shamir and J.~Virrueta,
``Superconformal Models for Graphene and Boundary Central Charges,''
\href{http://dx.doi.org/10.1007/JHEP09(2018)161}{JHEP {\bf 1809} (2018) 161} [\href{https://arxiv.org/abs/1807.01700}{arXiv:1807.01700 [hep-th]}].

\bibitem{PhysRevLett.67.3852}
W.~Metzner and C.~Di Castro, ``Ward identities and the \ensuremath{\beta} function in the Luttinger liquid,''
\href{http://dx.doi.org/10.1103/PhysRevLett.67.3852}{Phys.\ Rev.\ Lett.\ {\bf 67} (1991) 3852}.

\bibitem{PhysRevB.47.16107}
W.~Metzner and C.~Di Castro, ``Conservation laws and correlation functions in the Luttinger liquid,''
\href{http://dx.doi.org/10.1103/PhysRevB.47.16107}{Phys.\ Rev.\ B {\bf 47} (1993) 16107}.

\bibitem{Johnson:1964da}
  K.~Johnson, M.~Baker and R.~Willey, ``Selfenergy of the electron,''
\href{http://dx.doi.org/10.1103/PhysRev.136.B1111}{Phys.\ Rev.\  {\bf 136} (1964) B1111}.

\bibitem{Roberts:1994dr}
  C.~D.~Roberts and A.~G.~Williams, ``Dyson-Schwinger equations and their application to hadronic physics,''
\href{http://dx.doi.org/10.1016/0146-6410(94)90049-3}{Prog.\ Part.\ Nucl.\ Phys.\  {\bf 33} (1994) 477} [\href{https://arxiv.org/abs/hep-ph/9403224}{hep-ph/9403224}].

\bibitem{Landau:1955zz}
  L.~D.~Landau and I.~M.~Khalatnikov, ``The gauge transformation of the Green function for charged particles,''
\href{http://www.jetp.ac.ru/cgi-bin/e/index/e/2/1/p69?a=list}{Sov.\ Phys.\ JETP {\bf 2} (1956) 69}  [Zh.\ Eksp.\ Teor.\ Fiz.\  {\bf 29} (1955) 89].
  %%CITATION = SPHJA,2,69;%%

\bibitem{Fradkin:1955jr}
  E.~S.~Fradkin, ``Concerning some general relations of quantum electrodynamics,''
\href{http://www.jetp.ac.ru/cgi-bin/e/index/e/2/2/p361?a=list}{Sov.\ Phys.\ JETP {\bf 2} (1956) 361}  [Zh.\ Eksp.\ Teor.\ Fiz.\  {\bf 29} (1955) 258].

\bibitem{Johnson:1959zz}
  K.~Johnson and B.~Zumino, ``Gauge Dependence of the Wave-Function Renormalization Constant in Quantum Electrodynamics,''
\href{http://dx.doi.org/10.1103/PhysRevLett.3.351}{Phys.\ Rev.\ Lett.\  {\bf 3} (1959) 351}.

\bibitem{Zumino:1959wt}
  B.~Zumino, ``Gauge properties of propagators in quantum electrodynamics,''
\href{http://dx.doi.org/10.1063/1.1703632}{J.\ Math.\ Phys.\  {\bf 1} (1960) 1}.

\bibitem{Sonoda:2000kn}
  H.~Sonoda, ``On the gauge parameter dependence of QED,''
\href{http://dx.doi.org/10.1016/S0370-2693(01)00030-2}{Phys.\ Lett.\ B {\bf 499} (2001) 253} [\href{https://arxiv.org/abs/hep-th/0008158}{hep-th/0008158}].

\bibitem{Ball:1980ay}
  J.~S.~Ball and T.~W.~Chiu, ``Analytic Properties of the Vertex Function in Gauge Theories. 1.,''
\href{http://dx.doi.org/10.1103/PhysRevD.22.2542}{Phys.\ Rev.\ D {\bf 22} (1980) 2542}.

\bibitem{Fischer:2004nq}
  C.~S.~Fischer, R.~Alkofer, T.~Dahm and P.~Maris, ``Dynamical chiral symmetry breaking in unquenched QED(3),''
\href{http://dx.doi.org/10.1103/PhysRevD.70.073007}{Phys.\ Rev.\ D {\bf 70} (2004) 073007} [\href{https://arxiv.org/abs/hep-ph/0407104}{hep-ph/0407104}].

\bibitem{Miransky:1996pd}
  V.~A.~Miransky and K.~Yamawaki, ``Conformal phase transition in gauge theories,''
\href{http://dx.doi.org/10.1103/PhysRevD.56.3768}{Phys.\ Rev.\ D {\bf 55} (1997) 5051};
\href{http://dx.doi.org/10.1103/PhysRevD.55.5051}{Erratum: [Phys.\ Rev.\ D {\bf 56} (1997) 3768]} [\href{https://arxiv.org/abs/hep-th/9611142}{hep-th/9611142}].

\bibitem{Fomin:1978rk}
  P.~I.~Fomin, V.~P.~Gusynin and V.~A.~Miransky, ``Vacuum Instability of Massless Electrodynamics and the Gell-mann-low Eigenvalue Condition for the Bare Coupling Constant,''
\href{http://dx.doi.org/10.1016/0370-2693(78)90366-0}{Phys.\ Lett.\  {\bf 78B} (1978) 136}.

\bibitem{Fomin:1984tv}
  P.~I.~Fomin, V.~P.~Gusynin, V.~A.~Miransky and Y.~A.~Sitenko, ``Dynamical Symmetry Breaking and Particle Mass Generation in Gauge Field Theories,''
\href{https://link.springer.com/content/pdf/10.1007/BF02740014.pdf}{Riv.\ Nuovo Cim.\  {\bf 6N5} (1983) 1}.

\bibitem{Fukuda:1976zb}
  R.~Fukuda and T.~Kugo, ``Schwinger-Dyson Equation for Massless Vector Theory and Absence of Fermion Pole,''
\href{http://dx.doi.org/10.1016/0550-3213(76)90572-1}{Nucl.\ Phys.\ B {\bf 117} (1976) 250}.

\bibitem{Bardeen:1985sm}
  W.~A.~Bardeen, C.~N.~Leung and S.~T.~Love, ``The Dilaton and Chiral Symmetry Breaking,''
\href{http://dx.doi.org/10.1103/PhysRevLett.56.1230}{Phys.\ Rev.\ Lett.\  {\bf 56} (1986) 1230}.

\bibitem{Atkinson:1993mz}
  D.~Atkinson, J.~C.~R.~Bloch, V.~P.~Gusynin, M.~R.~Pennington and M.~Reenders,
  ``Strong QED with weak gauge dependence: Critical coupling and anomalous dimension,''
\href{http://dx.doi.org/10.1016/0370-2693(94)90526-6}{Phys.\ Lett.\ B {\bf 329} (1994) 117}.

\bibitem{Gusynin:1998se}
  V.~P.~Gusynin, A.~W.~Schreiber, T.~Sizer and A.~G.~Williams, ``Chiral symmetry breaking in dimensionally regularized nonperturbative quenched QED,''
\href{http://dx.doi.org/10.1103/PhysRevD.60.065007}{Phys.\ Rev.\ D {\bf 60} (1999) 065007} [\href{https://arxiv.org/abs/hep-th/9811184}{hep-th/9811184}].
  %%CITATION = doi:10.1103/PhysRevD.60.065007

\bibitem{Jackiw:1999qq}
  R.~Jackiw, ``When radiative corrections are finite but undetermined,''
\href{http://dx.doi.org/10.1142/S021797920000114X}{Int.\ J.\ Mod.\ Phys.\ B {\bf 14} (2000) 2011} [\href{https://arxiv.org/abs/hep-th/9903044}{hep-th/9903044}].

\bibitem{Panzer:2014kia}
  E.~Panzer, ``Renormalization, Hopf algebras and Mellin transforms,''
\href{http://dx.doi.org/10.1090/conm/648/13003}{Contemp.\ Math.\  {\bf 648} (2015) 169} [\href{https://arxiv.org/abs/1407.4943}{arXiv:1407.4943 [hep-th]}].

\bibitem{szego1939orthogonal}
G.~Szeg{\"o}, ``Orthogonal Polynomials,''
\newblock American Mathematical Society colloquium publications (1939).

\bibitem{Rosner:1967zz}
  J.~L.~Rosner, ``Higher-order contributions to the divergent part of $Z(3)$ in a model quantum electrodynamics,''
\href{http://dx.doi.org/10.1016/0003-4916(67)90262-X}{Annals Phys.\  {\bf 44} (1967) 11}.

\bibitem{GuthLecturesEM}
A.~Guth, ``Electromagnetism II,''
\href{https://ocw.mit.edu/courses/physics/8-07-electromagnetism-ii-fall-2012/index.htm}{MIT Course (2012)}.

\bibitem{Carslon:1971}
B.\ C.~Carlson, ``New Proof of the Addition Theorem for Gegenbauer Polynomials,''
\href{http://dx.doi.org/10.1137/0502032}{SIAM J.\ Math.\ Anal.\ {\bf 2} (1971) 347}.

\bibitem{Celmaster:1980ji}
  W.~Celmaster and R.~J.~Gonsalves, ``Fourth Order QCD Contributions to the e+ e- Annihilation Cross-Section,''
\href{http://dx.doi.org/10.1103/PhysRevD.21.3112}{Phys.\ Rev.\ D {\bf 21} (1980) 3112}.

\end{thebibliography}\endgroup\newpage
